\documentclass{article}

\title
{
Definition and properties to assess multi-agent environments as social intelligence tests
}
\author
{
Javier Insa-Cabrera \ \ \ \ \ Jos\'e Hern\'andez-Orallo\\
DSIC, Universitat Polit\`ecnica de Val\`encia, Spain\\
{\tt jinsa@dsic.upv.es} \ \ \ \ \ {\tt jorallo@dsic.upv.es}
}
\date{\today}

\input{glyphtounicode} 
\usepackage{anysize} 
\usepackage{graphicx} 
\usepackage{amsthm} 
\usepackage{amsmath, array} 
\usepackage{amssymb} 
\usepackage{hyperref} 
\usepackage{colortbl} 
\usepackage{soul} 

\marginsize{2cm}{2cm}{2cm}{2cm}

\theoremstyle{definition}
\newtheorem{definition}{Definition}
\newtheorem{proposition}{Proposition}
\newtheorem{conjecture}{Conjecture}
\newtheorem{approximation}{Approximation}
\newtheorem{assumption}{Assumption}
\newtheorem{lemma}{Lemma}

\newcommand{\instantiation}[3]{\ensuremath{\dot{#1} \stackrel{#2}\leftarrow #3}}

\begin{document}
\maketitle

\begin{abstract}
Social intelligence in natural and artificial systems is usually measured by the evaluation of associated traits or tasks that are deemed to represent some facets of social behaviour. The amalgamation of these traits is then used to configure the intuitive notion of social intelligence. Instead, in this paper we start from a {\em parametrised} definition of social intelligence as the expected performance in a set of environments with several agents, and we assess and derive tests from it. This definition makes several dependencies explicit: (1) the definition depends on the choice (and weight) of environments and agents, (2) the definition may include both competitive and cooperative behaviours depending on how agents and rewards are arranged into teams, (3) the definition mostly depends on the abilities of other agents, and (4) the actual difference between social intelligence and general intelligence (or other abilities) depends on these choices.
As a result, we address the problem of converting this definition into a more precise one where some fundamental properties ensuring social behaviour (such as action and reward {\em dependency} and {\em anticipation} on competitive/cooperative behaviours) are met as well as some other more instrumental properties (such as {\em secernment}, {\em boundedness}, {\em symmetry}, {\em validity}, {\em reliability}, {\em efficiency}), which are convenient to convert the definition into a practical test.
From the definition and the formalised properties, we take a look at several representative multi-agent environments, tests and games to see whether they meet these properties.

\vspace{0.5cm}

\end{abstract}

\newpage

\setcounter{tocdepth}{2}
{ \small 
\renewcommand{\baselinestretch}{0.8}
\tableofcontents
\renewcommand{\baselinestretch}{1}
}

\section{Introduction}
\label{sec:introduction}
Evaluation tools are crucial in any discipline as a way to assess its progress and creations. Artificial intelligence, as a discipline, lacks general, well-grounded and universally accepted intelligence measurement tools. In fact, artificial intelligence is a paradigmatic case of how useful these tools would be and how impeding this lack is. There are, of course, some tools, benchmarks and contests, aimed at the measurement of humanoid intelligence or the performance in a particular set of tasks. However, the evolution and state of the art of artificial intelligence is now more focussed towards social abilities, and here the measuring tools are still rather incipient.

In the past two decades, the notion of agent and the area of multi-agent systems have shifted artificial intelligence to problems and solutions where `social' intelligence is more relevant. This shift towards a more social-oriented AI is related to the modern view of human intelligence as highly social, actually one of the most distinctive features of human intelligence over other kinds of animal intelligence. Some significant questions that appear here are then whether we are able to properly evaluate social intelligence in general (not only in AI, but universally) and whether we can develop measurement tools that distinguish between social intelligence and general intelligence.

In this paper, we address these questions by attempting a formalisation of social intelligence and some of its associated properties, such as social dependency and anticipation, and the properties a good test should have, such as discrimination, grading, reliability, boundedness, symmetry, efficiency, validity, etc.  
%
%
%
These properties help to: identify the components of social intelligence and its varieties; make clear that the mere appearance of other agents does not make a context social; pave the way for the analysis of whether many social environments, games and tests found in the literature are useful for measuring social intelligence. In particular, we will analyse a representative selection of environments, some usual in game theory, such as matching pennies and the prisoner's dilemma, and some more sophisticated (and realistic) social scenarios, such as the predator-prey (a pursuit game), Pac-Man or RoboCup Soccer. We will see whether they meet the properties a social intelligence testbed should have.


One side question that we will try to analyse here is whether social intelligence can be fully separated from general intelligence or, conversely, whether general intelligence can be seen as a special case of social intelligence where the presence and intelligence of other agents is not so relevant. 

The paper is organised as follows. 
Section \ref{sec:background} gives an overview of the approaches to social intelligence from several disciplines.  
Section \ref{sec:definition} presents a formal, {\em parametric} definition of social intelligence and specifies how a test can be derived from the definition. 
Section \ref{sec:social_properties} discusses several properties that are needed (or desirable) for a good test of social intelligence. 
Section \ref{sec:current_environments} discusses current social environments and games in AI, and their suitability to evaluate social intelligence following the definition and the properties. 
Finally, section \ref{sec:conclusions_and_future_work} closes the paper with some discussion and future work.

\section{Background}
\label{sec:background}
Social intelligence (and its true distinction from general intelligence) has been a matter of study for many years. Many definitions have been proposed such as the ``ability to understand and manage men and women, boys and girls -- to act wisely in human relations'' \cite{thorndike1920intelligence}, the ``ability to get along with others'' \cite{moss1927you}, the ``facility in dealing with human beings'' \cite{wechsler1958measurement}, or more specific definitions including ``[the] ability to get along with people in general, social technique or ease in society, knowledge of social matters, susceptibility to stimuli from other members of a group, as well as insight into the temporary moods or the underlying personality traits of friends and of strangers'' \cite{vernon1933some}. Nonetheless, none of these definitions is sufficiently formal and operational to provide a clear measurement procedure. In fact, all these definitions require the definition of many other new concepts that appear in the definition.

Despite the ambiguity of what social intelligence is, many tests have been proposed to measure social intelligence in humans (see \cite{weis2008theory} for a survey). Typically there are two kinds of tests: 1) some use paper and pencil tests to measure social {\em knowledge}, and 2) others use more real situation tests (such as viewing photographies) in order to find out how people react in social situations. Some examples of these tests propose storylines which must be ordered to make sense, find the best end to a given joke, or select a correct emotion to a given face. However, there is a third, but unusual way: to measure social intelligence in terms of the definitions above, by confronting a human against other humans and see whether the subject deals with them or get along well. Apart from practical reasons that make this approach more difficult for testing, there are important questions, such as the selection and role of the other humans in order to make the test objective and effective.

But social intelligence is not only present in humans. Other animal species have also demonstrated this kind of intelligence. The evaluation of animals is more difficult, as we cannot ask them to perform a test as we do with humans, so the third type of tests is more common in this setting. Also, tests include some food as rewards in order to motivate the animals to perform the tasks. This is the same configuration as in \emph{reinforcement learning} (RL) \cite{sutton1998reinforcement}, where rewards are provided in order to encourage agents to perform tasks. In social intelligence tests for animals, especially for those focussing on cooperation, animals must obtain some food or reward that cannot be obtained by one individual alone, but two or more animals interacting are needed to get their reward. Some of the capabilities evaluated with those tests are their predisposition to deal with others, and their selfishness or altruism.

Although these tests measure some aspects of social intelligence, many have been devised to evaluate social intelligence for a particular species and for very specific tasks. In these tests, it is highly debatable whether the tasks are representative of a broader view of social intelligence. Also, it is usually very difficult to compare the results with those of other species. Fortunately, there have been some exceptions to this (species) specialisation, and they are proliferating in the past decade. For instance, some recent work has shown that social abilities can be compared in a systematic way between human children and apes \cite{Herrmann-etal2007,Herrmann-etal2010}.

Even in the cases when the tests are generalised, they are still composed of a set of tasks that have been associated to social cognition indirectly, by observation or correlation, according to decades of experiments in the comparative cognition literature (see, e.g., \cite{wasserman2006comparative,shettleworth2013fundamentals}). As a result, they cannot directly relate to the common definitions of social intelligence. For instance, one typical task used in social intelligence is to establish eye contact or to recognise oneself in a mirror (see, e.g., \cite[pp. 452-453]{shettleworth2009cognition}). These tasks do not seem to be derived from any definition of social intelligence.

In order to elude this gap, many studies in ethology, comparative cognition and psychology just focus on specific issues, such as competition, cooperation, symbiosis, communication, group/swarm abilities, etc. 
However, in these scenarios, the emphasis is usually put on detecting and observing some of these phenomena, rather than properly evaluating abilities. For instance, prey-predator interaction and behaviour have been studied from many different points of view (including game theory \cite{osborne2004introduction}), but it is not clear how the ability of each subject can be objectively evaluated, especially because the interaction depends on the cognitive abilities of both prey and predator. For instance, lions are well prepared to predict zebra's movements and chase them, but they may be less able for other kinds of animals.

Despite these difficulties, comparative cognition \cite{wasserman2006comparative,shettleworth2013fundamentals} is more and more concerned about performing tests that compare the abilities of many different species and also the abilities of individuals of different species.
From this point of view, it should be more natural to provide a single test (with possibly many different customised interfaces and rewards) to assess every kind of species (or, in other words, a more general, or {\em universal} \cite{AIJ2010,upsychometrics2,hernandez2013howuniversal}, test). To achieve this, such a test should be able to evaluate any level and spectrum of social intelligence, instead of focussing on the specific range and particular abilities of a single species.

When thinking about social tests and making them more species-independent, we can take the most general perspective, which leads us to the consideration of machines as well. However, evaluating social intelligence in machines has been quite different to the assessment of human and animal social intelligence. Nonetheless, as occurs with animals, rewards or scores may be used as a measure of their performance and a way of giving feedback to make them show their abilities. Besides, environments must be presented in such a way that a machine can process the observations and perform a set of actions. This is done by providing them with sensors and actuators that interact with our physical world, or provide them with a logical or virtual environment.

The environments used in social tests for machines tend to represent tasks that the agents must perform by interacting with other agents, so the performance is calculated as their capability to successfully cooperate with and/or compete against them to achieve some goals (see \cite{benda1985optimal, Kitano:1997:RRW:267658.267738} for two testbeds in multi-agent environments). In this way, the evaluation is a simulation in a social context, which is more directly linked to the definition of social intelligence.

In the context of social sciences (stretching from economics to AI), game theory \cite{osborne2004introduction} has also studied the interaction of different agents in formalised structures (called games). For this purpose, game theory uses a formal approach to define a utility function, and the effort is made for finding the best strategy among all possible strategies, assuming that the rest of agents also try to obtain their best results following some kind of rational actions. Although game theory needs the interaction of several agents, the goal is not to evaluate social intelligence but rather to analyse how the agents (or just policies) behave in these games and whether they reach some kind of equilibrium. Several kinds of games try to represent or to analyse a variety of properties: cooperative or non-cooperative games, simultaneous or sequential games, normal-form or extensive-form games, zero-sum or general-sum games, and symmetric or asymmetric games \cite{osborne2004introduction,myerson2013game}. However, games that may have the most interesting properties or applications are not necessarily useful for testing. For instance, a game where equilibria are easy may be inappropriate if we want a discriminative test. Similarly, asymmetric games make it more difficult to assess agent performance, as they depend on the role each agent takes.

One important concept in game theory is the notion of zero-sum vs. non-zero-sum games.  
Zero-sum games are a particular set of games where a player's gains (or losses) are equally balanced by the other players' losses (or gains). These kinds of games are known as competitive games, since one's gains reduces the gains (or increases the losses) of the other player(s), making the players having opposed interests. When a zero-sum game only has two players it becomes a pure competitive game. But zero-sum games can also contain cooperation in games with three or more players. Two players can cooperate in order to compete against a third or more players. As the number of players increase, cooperation becomes more important. In contrast, general-sum games are those games where the payoffs sum more or less than zero, and games can be cooperative even for two players. Finally, another particular feature in game theory is that environments are generally simple (without objects or spaces) and it is just the continuous interaction between agents that matters.

Multi-agent systems (MAS), on the contrary, present richer and more diverse possibilities, both in competition \cite{rust1992behaviour, billings2000first, Kitano:1997:RRW:267658.267738, wellman2001designing, poundstone1993prisoner} and cooperation \cite{benda1985optimal, Tan93multi-agentreinforcement, Sen94learningto, robinson2002using, zhang199918, ostergaard2001emergent}. Agents are usually evaluated according to their performance in some tasks interacting with other agents. 
Environments are usually selected to represent some particular problems for which techniques are developed and evaluated. However, these evaluations lack some important features. They do not evaluate social intelligence in a general way, but they are typically designed to evaluate one kind of task. However, the most important problem is that they usually require very specific abilities, or when they require many, it is not clear how to disentangle them. For instance, if a MAS setting requires both competition and cooperation to solve a problem, it is not always easy to select or gauge the degree of relevance of each one in order to give more relevance to competition over cooperation, or vice versa. Nonetheless, the major issue is that many  capabilities other than social intelligence also contaminate the results, which makes many MAS scenarios unsuitable if we want to measure social abilities only.

As an alternative\footnote{Not only as an alternative to MAS scenarios, but also to the Turing Test, CAPTCHAs and IQ tests (see \cite{IQnotformachines} for a discussion).}, can we start from a formal definition of an ability and derive tests from it? This approach has been investigated for machine intelligence evaluation. Formal approaches started in the late 1990s using notions from Kolmogorov complexity, Solomonoff prediction and the MML principle \cite{DoweHajek98,HernandezOrallo-MinayaCollado98,HernandezOrallo00a,HernandezOrallo00b}. Dobrev \cite{dobrev2000} suggested that machine intelligence should be measured by evaluating agent performance in a range of worlds, an idea that was independently developed in \cite{LeggHutter07} under the name ``Universal Intelligence''. This definition treats intelligence as a general notion, calculating it as the performance of the agent in a wide range of environments. Following this definition, a framework to evaluate intelligence \cite{AIJ2010} and an environment following the framework \cite{hernandez2010hopefully} were proposed. In order to show their effectiveness, some experiments were performed \cite{CAEPIA2011Evaluating, AGI2011Comparing, AISB-AICAP2012a}, but their results suggested that the framework still has some limitations. One of the possible reasons is that these environments lacked the richness of interaction. From the formal definition, it is virtually impossible to randomly generate an environment that contains some kind of social behaviour. Therefore, in order to overcome some of the limitations of these tests, some other agents need to be included in the environment to generate social situations. 
This was the goal in \cite{Insa-Cabrera:2012:MSI:2437816.2437830}, where other agents were directly included in the environment. Some simple experiments were performed to evaluate machine social behaviour in environments where the agents were forced to compete and/or cooperate with other agents. The results of these experiments showed the impact on agents' performance when other agents are directly introduced in a test of general intelligence. These experiments were performed using the framework in \cite{AIJ2010}, which was originally designed to evaluate general intelligence, by simply including other agents in the environment. Nonetheless, a general environment such as this one does not seem enough to evaluate social intelligence, since some abilities other than social intelligence are also evaluated in these kinds of environments. In order to measure social intelligence in isolation, we need to provide an appropriate environment class where only social intelligence is needed (or at least, where the degree of social intelligence needed can be fine tuned).

In this context, the key issue is to determine what kind of agents we must include in the test to interact with and what their roles are. This boils down to choosing a distribution of agents. However, in order to provide an environment with some rich social situations, we need first to know the level of social intelligence of the agents provided by the distribution. This circular problem is turned into a recursive one in \cite{AGI2011DarwinWallace}, where different levels of distributions are recursively provided by constructing a new distribution of agents from a prior distribution by selecting (or increasing the probability of) those agents with higher performance. However, it is not easy to derive a definition of social intelligence from here or a procedure to create environments that would be the base for social intelligence tests.

Overall, there are many different approaches for the study and evaluation of social intelligence, but we lack a comprehensive theory, well-grounded tools and wide comparisons to better understand the problem and find better measurement devices.


\section{Defining Social Intelligence Universally}
\label{sec:definition}
One way of reaching a universal definition of social intelligence is to consider more specific definitions and generalising them for any kind of subject. Thorndike's definition of social intelligence refers to ``men, women, boys and girs'' \cite{thorndike1920intelligence}. So this approach would generalise this view with the variety of species in animal cognition, but also including machines, robots and other artificial systems. This is in the spirit of universal psychometrics \cite{upsychometrics2}, where we must consider any kind of agent (natural or artificial). Any of these systems can, in principle, be evaluated and can also be subject of interaction with the evaluee.

This can take us to definitions such as ``performance of an agent in a wide range of environments while interacting with other agents'' \cite{Insa-Cabrera:2012:MSI:2437816.2437830}. As a result, we see clearly that social intelligence is a \emph{relative} property, where we need to specify these other agents (and the range of environments). 

With this approach, we distinguish those traits that have positive consequences on the performance (rewards) from those that are associated to social intelligence but do not necessarily lead to better performance (such as being generous, open, extroverted, etc.). In other words, we understand that an agent is socially intelligent if it has the ability to perform better in a social environment, but not if it is very {\em sociable} but showing very poor performance. In the end, we want an operational definition such that its measurement can be directly linked to it, and not derived by some other traits that are usually associated to social intelligence in humans and animals.

So we must focus our attention on the specification of the set of environments used for measuring and, most especially, on the characteristics of other agents. Nonetheless, it is important to determine the {\em role} these agents take in the environment relative to the evaluated agent. 
%
%
%
%
%
For instance, the environment can be populated by very intelligent agents, but the possibilities of an evaluated agent to achieve its goals will depend on whether these agents are allies or enemies, or more generally if they are cooperative or competitive. The key issue is to establish whether the other agents goals and interests are compatible with one's goals.
 The concept is complex, as alliances can be created and broken even if no clear teams are established from the beginning (and this is an interesting property of social intelligence). Nonetheless we have to consider the notion of {\em role} from the beginning and make it visible at the top, jointly with the kind of environment and the kind of agents.

These roles or alliances determine  two major social behaviours: cooperation and competition. These are in fact linked to the issue that some agents share some goals while some other agents compete or are against other agents' goals. If we think of rewards (or any other kind of utility function) as a general way of expressing goals, interests and even resources they share or compete for, we can distinguish two major kinds of social intelligence:

\begin{definition}
\label{def:competitive}
Competitive social intelligence is the capability to obtain the best performance in an environment where other agents compete for the same rewards.
\end{definition}

\begin{definition}
\label{def:cooperative}
Cooperative social intelligence is the capability to obtain the best performance in an environment where other agents share the same rewards.
\end{definition}

Note that both definitions are not exclusive, as there are environments where both competitive and cooperative behaviours are possible. This is similar to the several degrees of general-sum games in game theory.
Nonetheless, it would be very useful to have some way to analyse competition and cooperation separately (as two main facets of social intelligence). How clear-cut this separation can be done is an open question, as both abilities are occasionally correlated. For instance, the creation of alliances in a purely competitive scenario leads to temporary or permanent cooperation, where the other agents are seen in an instrumental way.

In what follows, we will see how these informal definitions can be formalised and integrated.

\subsection{Multi-agent environments and team rewards}
\label{sec:multi-agent_environment_and_team}
\begin{figure}[!ht]
\centering

\includegraphics[width=0.45\textwidth]{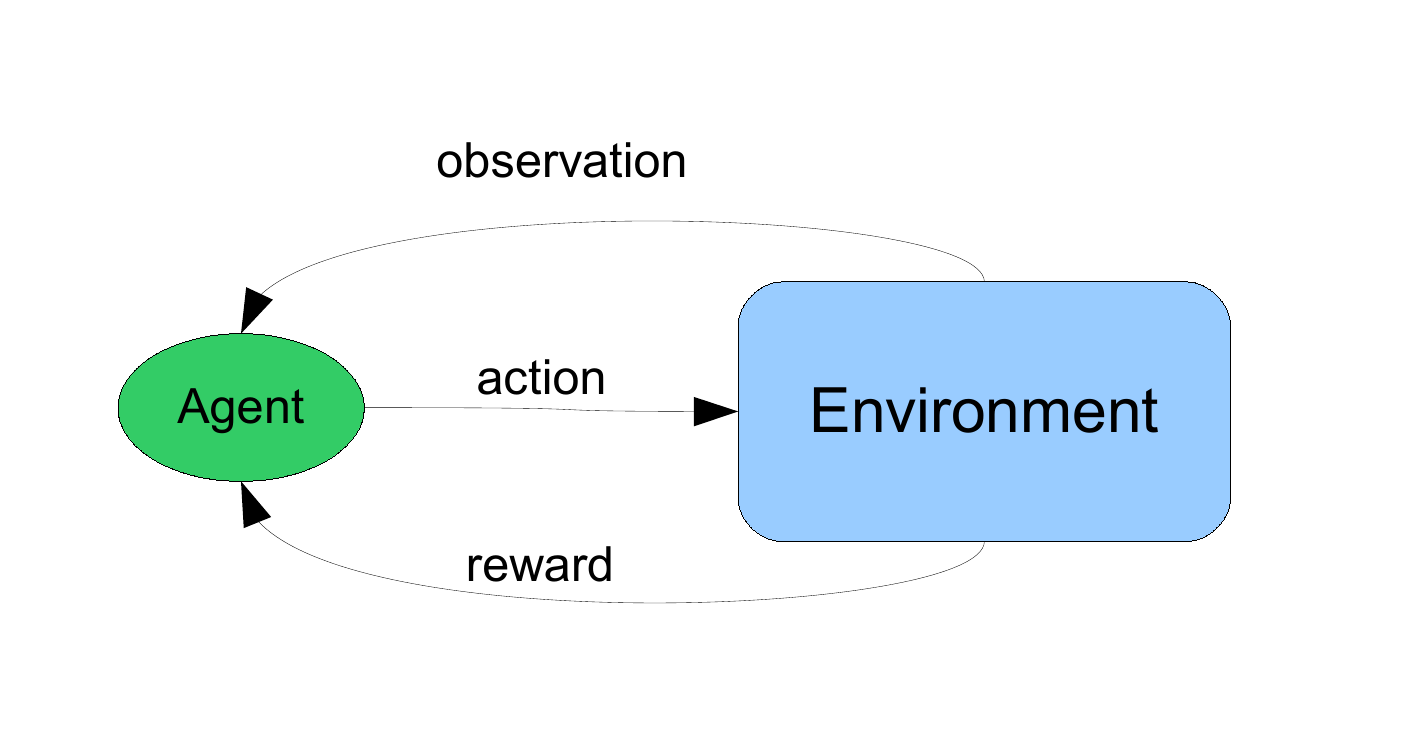}
\includegraphics[width=0.45\textwidth]{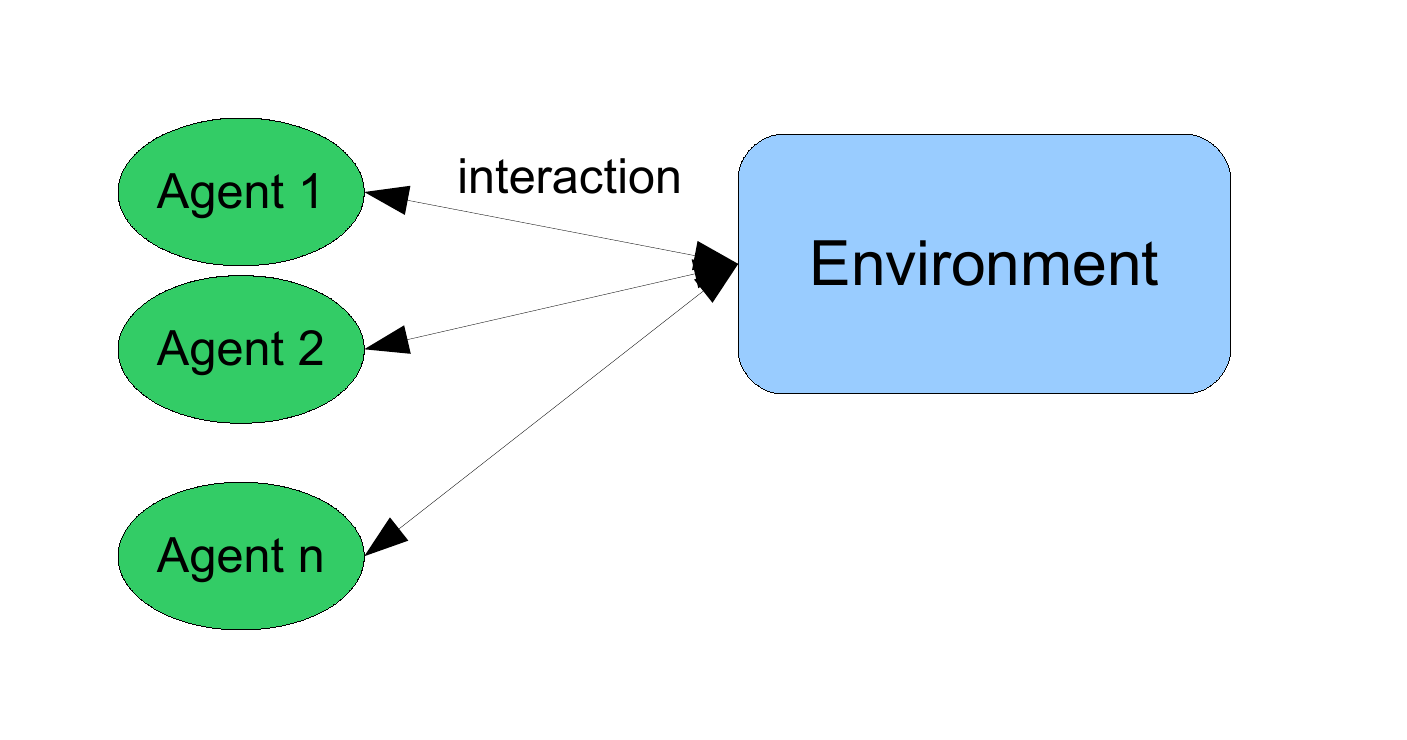}

\caption{Left: Interaction between a single agent and an environment. Right: Interaction of multiple agents in a multi-agent environment.}
\label{fig:multi-agent_environment}
\end{figure}

Before addressing a formal integration of definitions \ref{def:competitive} and \ref{def:cooperative}, we need to give a definition of (multi-agent) environment. An environment is a world where an agent can interact through actions, rewards and observations as seen in figure \ref{fig:multi-agent_environment} (left). This general view of the interaction between an agent and an environment can be extended to multi-agent systems by letting various agents interact simultaneously with the environment as seen in figure \ref{fig:multi-agent_environment} (right).

A multi-agent environment 
is an interactive scenario with several agents. An environment accepting $n$ agents defines $n$ parameters (one for each agent) denoted as {\em slots}. We use $i = 1, \dots, n$ to denote the agent slots. ${\cal A}_i$ is the action set for agent in slot $i$ and $a = (a_1, \dots, a_n) \in {\cal A}_1 \times \cdots \times {\cal A}_n$ is a joint action profile of the $n$ agents in the available discrete set of actions. ${\cal{O}}_i$ is the observation set that the agent in slot $i$ can perceive and ${\cal R}_i \subseteq \mathbb{Q}$ represents the possible rewards obtained by the agent in slot $i$. Both joint observation and reward profiles are denoted as $o = (o_1, \dots, o_n)$ and $r = (r_1, \dots, r_n)$ respectively, similarly as we did with actions. For each step, every agent must receive an observation $o_i \in {\cal O}_i$ and a reward $r_i \in {\cal R}_i$, and perform an action $a_i \in {\cal A}_i$. We will use $a_{i,k}$, $r_{i,k}$ and $o_{i,k}$ to respectively denote the action, reward and observation at step $k$ for the agent in slot $i$. We use $a_k$, $r_k$ and $o_k$ respectively to denote the joint actions, rewards and observations of all the agents at step $k$ (i.e., $a_1 = (a_{1,1}, \dots, a_{n,1})$ represents the joint actions at step $k = 1$). The order of events is always: observation, action and reward. For example, a sequence of two steps in a multi-agent environment is then a string such as $o_1a_1r_1o_2a_2r_2$ and the string $o_{1,1}a_{1,1}r_{1,1}o_{1,2}a_{1,2}r_{1,2}$ denotes the sequence of observations, actions and rewards for the agent in slot $1$.

Both the environment and the agents are defined as probabilistic measures. For the agent in slot $i$, the term $\pi(a_{i,k} | o_{i,1}a_{i,1}r_{i,1} \dots o_{i,k}) \rightarrow [0,1]$ denotes its probability to perform action $a_{i,k}$ after the sequence of events $o_{i,1}a_{i,1}r_{i,1} \dots o_{i,k}$. The observations provided by the environment to the agent in slot $i$ also have a probabilistic measure $\omega(o_{i,k} | o_1a_1r_1 \dots o_{k-1}a_{k-1}r_{k-1}) \rightarrow [0,1]$. As with observations, rewards are provided to the agent in slot $i$ depending on observations and actions on previous steps $\rho(r_{i,k} | o_1a_1r_1 \dots o_ka_k) \rightarrow [0,1]$. Note that the rewards obtained by each agent depend on the joint actions, observations and rewards of all the agents interacting in the environment, and not only on their own. A random agent (usually denoted by $\pi_r$) in slot $i$ is an agent that chooses its actions from ${\cal A}_i$ using a uniform distribution.

We use $\breve{A}^K_i(\pi, \mu)$ to denote the distribution (a probability measure) of strings representing the sequences of actions that $\pi$ performs in slot $i$ of $\mu$ during $K$ steps. If $K$ is omitted, we assume that $K$ is infinite, i.e., infinite sequences of actions for an endless episode. If agents and environment are deterministic (not stochastic) then this boils down to a probability measure giving probability 1 to one single string, the sequence of actions performed by $\pi$ on $\mu$.

Similarly, we use $\breve{R}^K_i(\pi, \mu)$ to denote the distribution of reward strings of $\pi$ in slot $i$ of $\mu$ during $K$ steps. If $K$ is omitted, we assume that $K$ is infinite. Again, if neither agents nor environment are stochastic, this is just a string. In the general case, we use $R_i^K(\pi, \mu)$ to denote the expected {\em average} reward (or value) with a discount factor $\gamma$. For instance, for $K=\infty$ this is 
$R_i(\pi, \mu) \triangleq 
\mathbb{E}(\lim\limits_{K \rightarrow \infty} \frac{\sum_{k=1}^K {\gamma^{k-1} r_{i,k}}}{\sum_{k=1}^K \gamma^{k-1}})$. 
Unless stated otherwise, we assume $\gamma = 1$.


\subsection{Teams}
We need to address a characterisation of slots, such that we can specify how agents participate in the environment. 
%
%
%
This actually means that we need to decide how the environment distributes rewards among the agents. An easy possibility will be to make each agent get its rewards without further constraints over other agents' rewards. With this configuration (e.g., general-sum games), both competition and cooperation may be completely useless for most environments, as the rewards are not limited or linked to the other agents. In contrast, if we set that the total set of rewards is limited in some way, we will foster competition, as happens in zero-sum games.
But in any of these two cases cooperation will hardly take place. Alliances could arise sporadically between at least two agents in order to bother (or defend against) a third agent. However, with low levels of social intelligence this seems unlikely to happen. For this reason we need to find a way to make agents cooperate, or at least to make it more likely before any (sophisticated) alliance can emerge on its own. One possible answer is the use of teams, defined as follows:

\begin{definition}
\label{def:team}
Agent slots $i$ and $j$ are in the same team iff $\forall k : r_{i,k} = r_{j,k}$
\end{definition}

\noindent which means that all agents in a team receive exactly the same rewards. This differs from alliances, where the agents could receive different rewards.
Note that teams are not alliances as usually understood in game theory. In fact, teams are fixed and cannot be changed by the agents. 
%
Also, we do not use the term alliance as we do not use any sophisticated mechanism to award rewards, related to the contribution of each agent in the team, as it is done with the Shapley Value \cite{roth1988shapley}. We just set rewards uniformly.

At this moment, we are ready to define an environment with parametrised agents by only specifying their slots and their team arrangement.

\begin{definition}
\label{def:environment}
A multi-agent environment $\mu$ accepting $N(\mu)$ agents (i.e., the number of slots in $\mu$) is a tuple $\left\langle {\cal A}, {\cal R}, {\cal O}, \omega, \rho, \tau \right\rangle$, where ${\cal A}$, ${\cal R}$, ${\cal O}$ represent the action sets, reward sets and observation sets respectively (i.e., ${\cal A} = {\cal A}_1 \times \cdots \times {\cal A}_{N(\mu)}$, ${\cal R} = {\cal R}_1 \times \cdots \times {\cal R}_{N(\mu)}$ and ${\cal O} = {\cal O}_1 \times \cdots \times {\cal O}_{N(\mu)}$) and $\omega$ and $\rho$ are the observation function and reward function respectively as defined in section \ref{sec:multi-agent_environment_and_team}. $\tau$ is a partition on the set of slots $\{1, \dots, N(\mu) \}$, where each set in $\tau$ represents a team.
\end{definition}

Note that with this definition the agents are not included in the environment. For instance, noughts and crosses could be defined as an environment $\mu_{nc}$ with two agents, where the partition set $\tau$ is defined as $\{ \{1\},  \{2\} \}$, which represents that this game allows two teams, and one agent in each. Another example is RoboCup Soccer \cite{Kitano:1997:RRW:267658.267738}, denoted by $\mu_{rc}$, whose $\tau$ would be $\{ \{1,2,3,4,5 \}, \{6,7,8,9,10 \} \}$, which represents that there are two teams, with slots  $\{1,2,3,4,5 \}$ in the first team and  slots $\{6,7,8,9,10 \}$ in the second team.

Once environments are defined, without including the agents, now we can define an {\em instantiation} for a particular agent setup. Formally, a {\em team line-up} $l$ is a list of agents. For instance, if we have a set of agents $\Pi = \{\pi_1,\pi_2,\pi_3,\pi_4\}$, a line-up from this set could be $l_1 = (\pi_2, \pi_3)$. The use of the same agent twice is allowed, so $l_2 = (\pi_1, \pi_1)$ is also a line-up. We denote by $\mu[l]$ the instantiation of an environment $\mu$ with a line-up $l$, provided that the length of $l$ is greater than or equal to the number of agents allowed by $\mu$ (if $l$ has more agents, the excess is ignored). The slots of the environment are then matched with the corresponding elements of $l$ following their order. For instance, for the noughts and crosses, an instantiation would be $\mu_{nc}[l_1]$. Note that different instantiations over the same environment would normally lead to different results.
A line-up pattern $\dot{l}$ is a list of agents where one or more elements are not instantiated. We can use instantiation to create more specific patterns or even to convert a pattern into a line-up. The instantiation of an agent $\pi$ at position $i$ on line-up pattern $\dot{l}$ of length $n$ is denoted by \instantiation{l}{i}{\pi}, which is exactly $\dot{l}_{1:(i-1)}\cdot\pi\cdot \dot{l}_{(i+1):n}$, where $l \cdot m$ denotes the concatenation of lists $l$ and $m$ and $l_{j:k}$ denotes the elements in $l$ from position $j$ to $k$. This notation can be extended to instantiate several agents simultaneously using \instantiation{l}{i,\dots,j}{\pi_1,\dots,\pi_n} to represent \instantiation{l}{i}{\pi_1} $\cdots \stackrel{j}\leftarrow \pi_n$. Once a line-up pattern $\dot{l}$ has all its elements instantiated becomes a line-up $l$. Note that environments can only be instantiated with line-ups, so first we need to instantiate all the elements from a line-up pattern to convert it to a line-up, and then use it to instantiate an environment. For instance, a line-up pattern for the set of agents $\Pi$ could be $\dot{l}_3 = (\pi_3, *)$, where $*$ represents an element that is not instantiated, and $\dot{l}_3\stackrel{2}\leftarrow \pi_4$ instantiates position 2 with agent $\pi_4$, converting the line-up pattern into the line-up $l_3 = (\pi_3, \pi_4)$.

We will use $L^n(\Pi)$ to specify the set of all the line-ups of length $n$ with agents of $\Pi$,
and $\dot{L}^n_{-i,\dots,j}(\Pi)$ to denote the set of all the line-up patterns of length $n$ with agents of $\Pi$ where positions $i, \dots, j$ are not instantiated. For instance, $\dot{L}^{n}_{-i}(\Pi)$ defines the set of all possible line-up patterns $\{\dot{l}_{1:i-1} \cdot \ast \cdot \dot{l}_{i+1:n}\}$ from the set of agents $\Pi$.

We will use line-up patterns with positions $i,\dots,j$ not being instantiated to evaluate agents in these positions, while the rest of the line-up pattern will be instantiated with the agents they will have to interact with.

$w_L$ denotes weights to line-ups formed with agents from a certain set $\Pi$, giving weights to the agents in the line-up and their positions. Similarly $w_{\dot{L}}$ denotes weights for line-up patterns formed with agents from $\Pi$. As line-ups and line-up patterns are clearly related, 
we will assume that $w_{\dot{L}}$ can be derived from $w_L$.

\begin{assumption}
\label{assum:line-up_weight}
The value of $w_{\dot{L}}$ for line-up patterns with only one element that is not instantiated is derived from $w_L$ as:

\begin{equation*}
\forall i,n,\Pi, \dot{l} \in \dot{L}^n_{-i}(\Pi) : w_{\dot{L}}(\dot{l}) = \sum_{\pi \in \Pi} w_L(\instantiation{l}{i}{\pi})  
\end{equation*}

\noindent and $w_{\dot{L}}$ for line-up patterns with 2 or more non-instantiated elements is recursively derived as:

\begin{equation*}
\forall i,\dots,j,n,\Pi, \dot{l} \in \dot{L}^n_{-i,\dots,j}(\Pi) : w_{\dot{L}}(\dot{l}) = \sum_{\pi \in \Pi} w_{\dot{L}}(\dot{l} \stackrel{i}\leftarrow \pi) = \cdots = \sum_{\pi \in \Pi} w_{\dot{L}}(\dot{l} \stackrel{j}\leftarrow \pi)  
\end{equation*}
\end{assumption}

Finally, note that we can calculate the expected average reward for any agent in line-up $l$ in an environment $\mu$.
We will use the notation $R_i(\mu[l])$, which gives us the expected average reward of the $i$th agent in line-up $l$ for environment $\mu$ (also in slot $i$). We can extend this notation to distributions as well (i.e. $\breve{A}_i(\mu[l])$ for distribution of action sequences and $\breve{R}_i(\mu[l])$ for distribution of reward sequences).

\subsection{A formal definition of social intelligence}
\label{sec:social_intelligence}
Having these ideas in mind we can now attempt a first definition of social intelligence. We first fix the line-up and vary on the possible environments.

\begin{definition}
We define the ability of an agent $\pi$ interacting in line-up $l$ which contains $\pi$ at position $i$, over a set of environments $M$ accepting at least $i$ agents and at most $|l|$ agents (being $|\cdot|$ the length of a list), weighting the environments by $w_M$ as:

\begin{equation}
\label{eq:social_intelligence_with_environments}
\Upsilon_i(l, M, w_M) \triangleq \sum_{\mu \in M} w_M(\mu) R_i(\mu[l_{1:N(\mu)}])
\end{equation}

\noindent where $|M| \geq 1$.
\end{definition}


Alternatively, we can think about a definition of the ability of an agent for a varying set of line-ups while fixing the environment.

\begin{definition}
We define the ability of an agent $\pi$ in slot $i$, interacting in an environment $\mu$ accepting at least $i$ agents, with a set of line-up patterns defined over agent set $\Pi$ and $w_{\dot{L}}$ as a weight for line-up patterns for the environment $\mu$:

\begin{equation}
\label{eq:social_intelligence_with_line-ups}
\Upsilon_i(\pi, \Pi, w_{\dot{L}}, \mu) \triangleq \sum_{\dot{l} \in \dot{L}^{N(\mu)}_{-i}(\Pi)} w_{\dot{L}}(\dot{l},\mu) R_i(\mu[\instantiation{l}{i}{\pi}])
\end{equation}

\noindent where $N(\mu) \geq 1$ and if $N(\mu) > 1$ then $|\Pi| \geq 1$ otherwise $|\Pi| \geq 0$.
\end{definition}



In the definition we have used a weight $w_{\dot{L}}$ that depends on $\mu$ (i.e., $w_{\dot{L}}(\dot{l},\mu)$). 
Note that $w_{\dot{L}}$ applies to $\dot{l}$, giving weights to the agents (and their positions) such that the evaluated agent $\pi$ interacts in the environment $\mu$ when it is located at position $i$. Note also that now $\dot{l}$ has always $N(\mu)$ elements when instantiated, so now no upper-restriction exists over the number of slots of $\mu$.

In fact, both $w_M$ and $w_{\dot{L}}$ could be integrated into a single weight for {\em instantiated environments}. However, we want to decouple agents and environments and use both of them as independent parameters. To make the agents and the environment independent we work with the next assumption.

\begin{assumption}
\label{assum:agent_environment_weight_independence}
If $w_L(l,\mu)$ and $w_{\dot{L}}(\dot{l},\mu)$, where $l \in L^{N(\mu)}(\Pi)$ and $\dot{l} \in \dot{L}^{N(\mu)}_{-i}(\Pi)$, are independent of $\mu$ then:

\begin{align*}
\forall l, \mu			& : w_L(l,\mu) = w_L(l)\\
\forall \dot{l}, \mu	& : w_{\dot{L}}(\dot{l},\mu) = w_{\dot{L}}(\dot{l})
\end{align*}
\end{assumption}

Assumption \ref{assum:agent_environment_weight_independence} gives the same weight to any line-up and line-up pattern whatever the environment. In what follows we will use $w_L(l)$ and $w_{\dot{L}}(\dot{l})$ as a weight for line-up $l$ and line-up pattern $\dot{l}$ respectively.

Under the assumption \ref{assum:agent_environment_weight_independence} we can integrate both equations \ref{eq:social_intelligence_with_environments} and \ref{eq:social_intelligence_with_line-ups} as follows:

\begin{definition}
\label{def:social_intelligence_for_a_slot}
The social intelligence of an agent $\pi$ in slot $i$, interacting with the class of agents $\Pi$ with a weight for line-up patterns $w_{\dot{L}}$, in a set of environments $M$ with a weight for environments $w_M$ is defined as:

\begin{equation}
\label{eq:social_intelligence_for_a_slot}
\Upsilon_i(\pi,\Pi,w_{\dot{L}},M,w_M) \triangleq \sum_{\mu \in M} w_M(\mu) \sum_{\dot{l} \in \dot{L}^{N(\mu)}_{-i}(\Pi)} w_{\dot{L}}(\dot{l}) R_i(\mu[\instantiation{l}{i}{\pi}])
\end{equation}

\noindent where $|M| \geq 1, \forall \mu \in M$ then $N(\mu) \geq 1$ and if $\exists \mu \in M | N(\mu) > 1$ then $|\Pi| \geq 1$ otherwise $|\Pi| \geq 0$.
\end{definition}

And summing the performance over all possible slots of the environments of $M$, weighting the slots of each environment by $w_S$, we have:

\begin{definition}
\label{def:social_intelligence}
The social intelligence of $\pi$ interacting with the class of agents $\Pi$ with a weight for line-up patterns $w_{\dot{L}}$, in a set of environments $M$ with a weight for environments $w_M$ and a weight for slots $w_S$ is defined as:

\begin{equation}
\label{eq:social_intelligence}
\Upsilon(\pi, \Pi, w_{\dot{L}}, M, w_M, w_S) \triangleq \sum_{\mu \in M} w_M(\mu) \sum_{i=1}^{N(\mu)} w_S(i,\mu) \sum_{\dot{l} \in \dot{L}^{N(\mu)}_{-i}(\Pi)} w_{\dot{L}}(\dot{l}) R_i(\mu[\instantiation{l}{i}{\pi}])
\end{equation}

\noindent where $|M| \geq 1, \forall \mu \in M$ then $N(\mu) \geq 1$ and if $\exists \mu \in M | N(\mu) > 1$ then $|\Pi| \geq 1$ otherwise $|\Pi| \geq 0$.
\end{definition}

This equation now removes the lower-restriction of the number of slots on the environments.

Finally the positions of the agents in the line-up patterns can be assumed independent:

\begin{assumption}
\label{assum:line-up_position_independence}
If $w_\Pi(\pi,i)$ defines the weight for the agent $\pi$ appearing at position $i$ in a line-up or line-up pattern, we assume $w_\Pi(\pi)$ to appear in all positions in a line-up or line-up pattern. Formally:

\begin{equation*}
\forall \pi,i : w_\Pi(\pi,i) = w_\Pi(\pi)
\end{equation*}
\end{assumption}

Under the assumption \ref{assum:line-up_position_independence}, $w_L$ and $w_{\dot{L}}$ can be derived as a function of terms from $w_\Pi$. Finally, we assume that the probability of an agent in a line-up and line-up pattern is independent of its position.

\begin{assumption}
\label{assum:line-up_weight_as_product_function}
We calculate $w_L$ as a product of agents weights $w_\Pi$ as:

\begin{equation*}
\forall{n,\Pi, l \in L^n(\Pi)} : w_{L}(l) = \prod_{1 \leq k \leq n} w_\Pi(l_{k:k})
\end{equation*}

\noindent and $w_{\dot{L}}$ is calculated as a product of agent weights $w_\Pi$ as:

\begin{equation}
\label{eq:line-up_as_product_function}
\forall{i,n,\Pi,\dot{l} \in \dot{L}^n_{-i}(\Pi)} : w_{\dot{L}}(\dot{l}) = \prod_{1 \leq k < i} w_\Pi(\dot{l}_{k:k}) \prod_{i < k \leq n} w_\Pi(\dot{l}_{k:k})
\end{equation}

This assumption is also extended to line-up patterns with 2 or more elements not instantiated as well.
\end{assumption}


\begin{proposition}
\label{prop:inside-out}
Under assumptions \ref{assum:line-up_position_independence} and \ref{assum:line-up_weight_as_product_function}, social intelligence as per equation \ref{eq:social_intelligence} is also defined as:

\begin{equation*}
\begin{aligned}
& \Upsilon(\pi,\Pi,w_{\dot{L}},M,w_M,w_S) = \Upsilon(\pi, \Pi, w_\Pi, M, w_M, w_S) =\\
& = \sum_{j=1}^\infty \sum_{i=1}^{j} \sum_{\dot{l} \in \dot{L}^j_{-i}(\Pi)} \left(\prod_{1 \leq k < i} w_\Pi(\dot{l}_{k:k}) \prod_{i < k \leq j} w_\Pi(\dot{l}_{k:k})\right) \sum_{\mu \in M^j} w_M(\mu) w_S(i,\mu) R_i(\mu[\instantiation{l}{i}{\pi}])
\end{aligned}
\end{equation*}

\noindent where $M^j$ denotes all the environments in $M$ with $j$ agent slots, $|M| \geq 1$, and if $\exists \mu \in M | N(\mu) > 1$ then $|\Pi| \geq 1$ otherwise $|\Pi| \geq 0$.

\begin{proof}
Definition \ref{def:social_intelligence} ranges over environments, their slots and then over line-up patterns, but we could express an equivalent equation by ranging over line-up patterns first and environments and their slots next:

\begin{equation*}
\begin{aligned}
\Upsilon(\pi,\Pi,w_{\dot{L}},M,w_M,w_S)	& = \sum_{\mu \in M} w_M(\mu) \sum_{i=1}^{N(\mu)} w_S(i,\mu) \sum_{\dot{l} \in \dot{L}^{N(\mu)}_{-i}(\Pi)} w_{\dot{L}}(\dot{l}) R_i(\mu[\instantiation{l}{i}{\pi}]) =\\
										& = \sum_{\mu \in M} w_M(\mu) \sum_{i=1}^{N(\mu)} w_S(i,\mu) \sum_{\dot{l} \in \dot{L}^{N(\mu)}_{-i}(\Pi)} \left(\prod_{1 \leq k < i} w_\Pi(\dot{l}_{k:k}) \prod_{i < k \leq N(\mu)} w_\Pi(\dot{l}_{k:k})\right) R_i(\mu[\instantiation{l}{i}{\pi}]) =\\
										& = \sum_{j=1}^\infty \sum_{\mu \in M^j} w_M(\mu) \sum_{i=1}^{j} w_S(i,\mu) \sum_{\dot{l} \in \dot{L}^j_{-i}(\Pi)} \left(\prod_{1 \leq k < i} w_\Pi(\dot{l}_{k:k}) \prod_{i < k \leq j} w_\Pi(\dot{l}_{k:k})\right) R_i(\mu[\instantiation{l}{i}{\pi}]) =\\
= \Upsilon(\pi,\Pi,w_\Pi,M,w_M,w_S)		& = \sum_{j=1}^\infty \sum_{i=1}^{j} \sum_{\dot{l} \in \dot{L}^j_{-i}(\Pi)} \left(\prod_{1 \leq k < i} w_\Pi(\dot{l}_{k:k}) \prod_{i < k \leq j} w_\Pi(\dot{l}_{k:k})\right) \sum_{\mu \in M^j} w_M(\mu) w_S(i,\mu) R_i(\mu[\instantiation{l}{i}{\pi}])
\end{aligned}
\end{equation*}
\end{proof}
\end{proposition}

This shows how we can parametrise the definition in terms of the weight of the other participants ($w_\Pi$) independently of their order in line-up patterns. For instance, the weight for each agent could depend on its (social) intelligence, provided we are able to estimate this value. The use of a product of weights makes sense if $w_\Pi$ is a unit measure.

{\bf The interpretation of the above definition is the expected performance of agent $\pi$ interacting with all possible line-up patterns generated using the set of agents $\Pi$, and in a set of environments $M$ with $\pi$ playing at all possible slots in each environment}\footnote{The above definitions could be simplified if for every environment $\mu_1$ and slot $i$ in it, there is always an environment $\mu_2$ with exactly the same behaviour where slot $i$ becomes $1$. That means that we could easily get rid of the summation over slots and work just with slot 1 for agent $\pi$. In other words, this would be like considering that the evaluated agent always plays with slot 1. As we will discuss later on, if the environments are symmetric, this problem also vanishes. Another option is to consider a uniform distribution for slots.}.

Proposition \ref{prop:inside-out} is not only useful for parametrising the definition in terms of the agents in isolation, but also because it decouples agents from environments. This makes sense in the context of (social) intelligence evaluation, as we want to consider other agents that are able to work in different environments, and not very specific agents that only work in one environment.

Definition \ref{def:social_intelligence} and its reformulation by proposition \ref{prop:inside-out} integrates all kinds of social behaviour, as it does not distinguish between agents appearing in the same team or opponent teams. For instance, if we consider a set $\Pi$ with very intelligent agents, some environments (and line-ups) will be easier if many of these agents appear in the same team, but will be harder if they appear in opponent teams. Also, the aggregation may consider many other environments where no social behaviour takes place. This means that the above equations are a skeleton for the definition, but we need to better analyse the pair ($\Pi$, $w_{\dot{L}}$) or ($\Pi$, $w_{\Pi}$) and the triplet ($M$, $w_M$, $w_S$).

\subsection{Tests}
\label{sec:tests}
A definition is not a test, most especially because many definitions range over infinite sets or an infinite number of steps. A test must be a finite procedure that can be feasibly applicable to an agent. For the moment, we will focus on non-adaptive tests, which are based on performing just a finite number of finite experiments or trials (episodes), which are independent of the previous ones.

Consequently, a test is defined using the definition of $\Upsilon(\pi,\Pi,w_{\dot{L}},M,w_M,w_S)$, where $\Pi$ is sampled with some distribution, $M$ is sampled with some distribution and the number of steps for each experiment is limited in some way. Sampling is understood to be without replacement when there is determinism (it does not make sense to repeat the same episode if the result is already known) but is understood to be with replacement for non-deterministic agents or environments.
We denote by $S \sim^n [A]_p$ a sample $S$ of $n$ elements from set $A$ using probability distribution $p$ for the powerset of $A$, i.e., for $2^A$.
The use of a distribution over samples instead of a distribution over exercises gives more flexibility about the conditions that we could establish on the sampling procedure. For instance, we could define a sample probability such that high diversity is ensured (apart from high accumulated relevance of the exercises that are chosen) or such that a range of difficulties is covered. Keep in mind that with this definition, the issues of with replacement or without replacement is re-understood as whether these samples allow repeated values or not.
With this notation, we can give the following definition of test:

\begin{definition}
\label{def:social_intelligence_test}
A {\em test} over $\Upsilon$ (definition \ref{def:social_intelligence} in the previous subsection), denoted by $\hat\Upsilon[p_\Pi, p_M, p_S, p_K, n_E]$, is a sample of $n_E$ episodes (or exercises) from all those summed in the definition, using agent distribution $p_\Pi$, an environment distribution $p_M$, a slot distribution $p_S$, and a distribution on the number of steps $p_K$.

\begin{equation*}
\hat{\Upsilon}[p_\Pi,p_M,p_S,p_K,n_E](\pi,\Pi,w_{\dot{L}},M,w_M,w_S) \triangleq \eta_{\mathcal{E}} \sum_{\left\langle\mu,i,\dot{l}\right\rangle \in \mathcal{E}} w_M(\mu) w_S(i,\mu) w_{\dot{L}}(\dot{l}) R_i^{K}(\mu[\instantiation{l}{i}{\pi}])
\end{equation*}

\noindent where $\eta_{\mathcal{E}}$ normalises the formula with $\eta_{\mathcal{E}} = \frac{1}{\sum_{\left\langle\mu,i,\dot{l}\right\rangle \in \mathcal{E}} w_M(\mu) w_S(i,\mu) w_{\dot{L}}(\dot{l})}$, $K$ is chosen using probability distribution $p_K$ and the episodes (or exercises) $\mathcal{E}$ are sampled as:


\begin{equation*}
\mathcal{E} \sim^{n_E} \left[\bigcup_{\mu \in M}\bigcup_{i = 1}^{N(\mu)}\left\{\left\langle\mu, i, \dot{l}\right\rangle : \dot{l} \in \dot{L}^{N(\mu)}_{-i}(\Pi)\right\}\right]_{p_{\mathcal{E}}}
\end{equation*}

with $p_{\mathcal{E}}$ being a distribution on the set of triplets $\left\langle\mu, i, \dot{l}\right\rangle$ based on $p_M$, $p_S$ and $p_\Pi$.

%


\end{definition}


Note that we use $p_\Pi$ for the line-up patterns, which could be the line-up pattern probability derived as the product of the probabilities of each agent in the line-up pattern following assumption \ref{assum:line-up_position_independence}, as in equation \ref{eq:line-up_as_product_function}. As we will see, if the environments are symmetric, we can get rid of $p_S$ and $w_S$ and just evaluate for slot 1.

It is important not to confuse the probabilities of sampling the line-up patterns, environments, slots and number of steps ($p_\Pi,p_M,p_S,p_K$) with any weight defined on them, in particular, the weights $w_{\dot{L}}$, $w_M$ and $w_S$ defined on line-up patterns, environments and slots respectively. Weights represent the relevance of an environment, their slots and line-up pattern for the definition (so it determines the abilities, roles and agents that have higher weight in the formula), while the distributions are just a way of sampling the usually large or infinite set of environments, slots and line-ups of agents. Weights and distributions might be related (or may be equal in order to ensure fast convergence to the actual value), but some other considerations may suggest that a less relevant case (low weight) can be sampled with high probability, as it may be highly representative or more robust, for instance. Actually, we want that a diversity of cases is sampled, rather than similar cases that will provide redundant information. This is why we use a distribution on $2^A$ and not on $A$ because otherwise we would not be able to measure how good (e.g., informative) a set of trials is. We will discuss this issue again in the following section in the context of reliability and efficiency.

\section{Properties about social intelligence testbeds}
\label{sec:social_properties}
In order to evaluate social intelligence and distinguish it from general intelligence, we need tests where social ability has to be used and, also, where we can perceive its consequences. This means that not every environment is useful for measuring social intelligence and not every subset of agents is also useful. We want tests such that the evaluated agent must use its social intelligence to understand and/or have influence over other agents' policies in such a way that this is useful to accomplish the evaluated agent's goals. We also need situations where common general intelligence is not enough. In a way, we want to {\em subtract} (from the summation of all environments and line-up patterns) those problems (as defined by classes of environments and agents) where general intelligence is enough (and social intelligence is useless) and those where intelligence (of any kind, social or non-social) is useless.

We will investigate some properties that are hence desirable (or necessary) for a testbed of environments\footnote{In what follows, we will explore some properties that can be applied to sets of environments rather than single environments. A definition or test can be composed of just one environment if $M$ has only one element (as for definition \ref{def:social_intelligence} having $|M| = 1$). Actually, we will give the definitions for one environment but they are easily extensible to a family or distribution of environments with a weight function used as a testbed.} and agents to actually measure social intelligence. Some other properties are more of a practical nature, such as the degree of discrimination and grading of the environment, the symmetry of slots, its reliability and efficiency.

Hereinafter, we will differentiate two kinds of set of agents: the set of agents we want to evaluate, denoted by $\Pi_e$, and the set of agents we want the environment to be populated with as opponents and team players, denoted by $\Pi_o$\footnote{Instead of using only one set for opponents and team players, it can also be extended by using two different sets; one for opponent players and another for team players.}. On many occasions both sets can be set equal, but it will be useful to keep them as separate sets.

\begin{figure}[!ht]
\centering

\includegraphics[width=0.9\textwidth]{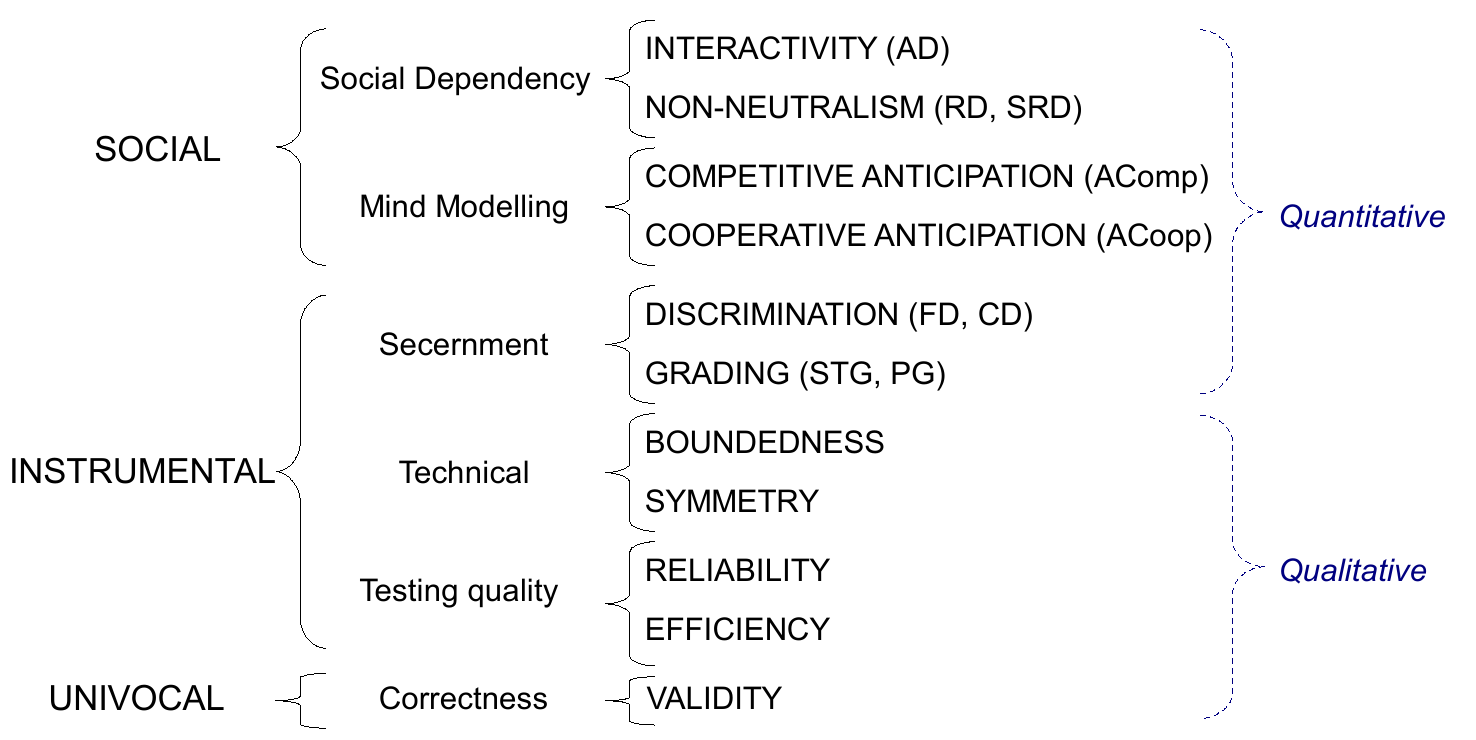}

\caption{Taxonomy of properties about social intelligence testbeds, grouped in three main categories (social, instrumental and univocal) and six subcategories (social dependency, mind modelling, secernment, technical, testing quality and correctness).}
\label{fig:properties_taxonomy}
\end{figure}

Figure \ref{fig:properties_taxonomy} gives a summary of the properties we will introduce in this section. They have different purposes and will reach different levels of formalisation. Actually, many of the properties (the quantitative ones) will be of the form $Prop(\Pi_e,w_{\Pi_e},\Pi_o,w_{\dot{L}},\mu,w_S)$, i.e., given these two sets $\Pi_e$ and $\Pi_o$, and the weights for them (in the form of agent weight $w_{\Pi_e}$ and line-up weight $w_{\dot{L}}$ respectively), they give a value for a given environment $\mu$ and slot weight $w_S$.

Finally, some of the properties below are presented for two slots but they could be extended to three or more slots as well. 
Similarly, all the following definitions are introduced for one environment, but they can be easily extended for a set of environments as well.

Next we will analyse and formalise these properties.

\subsection{Boundedness}
\label{sec:boundedness}
One property that we need to impose in order to make many of the previous definitions meaningful is that rewards must be bounded (otherwise, some summations will diverge). Any arbitrary choice of upper and lower bounds can be scaled to any other choice so, without loss of generality, we can assume that all of them are bounded between $-1$ and $1$. Formally:

\begin{equation*}
\forall i, k : -1 \leq r_{i,k} \leq 1
\end{equation*}

\noindent Note that they are bounded for every step. As $R_i(\cdot)$ is an average, then it is bounded as well if the rewards are bounded.

However, bounded rewards do not ensure that the measurement from definition \ref{def:social_intelligence} is bounded. In order to ensure a bounded result we also need to consider that weights are bounded, i.e., there are constants $c_M$, $c_S$, $c_{\dot{L}}$ and $c_\Pi$ such that:

\begin{align}
\forall M & : \sum_{\mu \in M} w_M(\mu) = c_M\\
\forall \mu & : \sum_{i=1}^{N(\mu)} w_S(i,\mu) = c_S\\
\forall i,n,\Pi_o & : \sum_{\dot{l} \in \dot{L}^n_{-i}(\Pi_o)} w_{\dot{L}}(\dot{l}) = c_{\dot{L}}\label{eq:pattern_weight_bounded}\\
\forall \Pi & : \sum_{\pi \in \Pi} w_\Pi(\pi) = c_\Pi\label{eq:agent_weight_bounded}
\end{align}

\noindent Equation \ref{eq:pattern_weight_bounded} can also be applied for two or more non-instantiated slots and equation \ref{eq:agent_weight_bounded} is used when assumption \ref{assum:line-up_weight_as_product_function} is made.

A convenient choice would be to have $c_M = c_S = c_{\dot{L}} = c_\Pi = 1$, and these weights would become unit measures (which should not be confused with the probabilities used in definition \ref{def:social_intelligence_test}). The expression of $w_{\dot{L}}$ in terms of a product of $w_\Pi$ make more sense if $w_\Pi$ is a unit measure.
With these conditions $\Upsilon$ and $\hat{\Upsilon}$ are bounded\footnote{Note that we are talking about the measure. For instance, $\Upsilon$ can be a measure that represents the, e.g., sigmoid function of an unbounded magnitude, easily recovered with a logit or probit function.}



An optional property that might be interesting occasionally is to consider environments whose reward sum is constant.
Without loss of generality, we can take this constant to be zero, which leads to the well-known notion of zero-sum games in game theory. 

\begin{definition}
An environment $\mu$ is zero-sum if and only if:

\begin{equation}
\label{eq:zero-sum}
\forall k : \sum^{N(\mu)}_{i=1} r_{i,k} = 0
\end{equation}
\end{definition}

The above definition may be too strict when we have environments with an episode goal at the end, but we want some positive or negative rewards to be given while agents approach the goal. A more convenient version follows:

\begin{definition}
An environment $\mu$ is zero-sum in the limit iff:

\begin{equation}
\label{eq:zero-sum_limit}
\lim\limits_{K \rightarrow \infty} \sum^K_{k=1} \sum^{N(\mu)}_{i=1} r_{i,k} = 0
\end{equation}
\end{definition}

\noindent Note that if we have teams, the previous definition could be changed in such a way that:

\begin{equation}
\label{eq:zero-sum_limit_teams}
\lim\limits_{K \rightarrow \infty} \sum^K_{k=1} \sum_{t \in \tau} \sum_{i \in t} r_{i,k} = 0
\end{equation}

So the sum of the agents' rewards in a team (or team's reward) does not need to be zero but the sum of all teams' rewards does.
For instance, if we have a team with agents $\{1,2\}$ and another team with agents $\{3,4,5\}$, then a reward (in the limit) of $1/4$ for agents $1$ and $2$ will imply $-1/6$ for each of the agents in the other team.

The zero-sum properties are appropriate for competition between teams. In fact, if teams have only one agent then we have {\em pure competitive} environments.
We can have both competition and cooperation by using teams in a zero-sum game, where agents in a team cooperate and agents in different teams compete.
If we want to evaluate {\em pure cooperation} (with one or more teams) then zero-sum games will not be appropriate.




\subsection{Interactivity}
\label{sec:interactivity}


By interactivity we mean the property that agents' actions have implications on the actions (and rewards) of the other agents. This is a key property as the existence of several agents in an environment does not ensure, per se, any social behaviour. In fact, it is important to realise that the use of several agents and their arrangement into teams does not ensure that some social behaviour can ever take place. Imagine a non-social environment, such as finding the way out of a maze with no other agents. Rewards depend on the agent finding the way out or not. While this is clearly non-social, we can use this environment as a building block and create an environment that takes two agents but makes them play separately on two equal mazes. We can generate rewards in at least four different ways: (1) we can give rewards separately without any modification on the outputs of the building blocks, (2) we can normalise them to a constant or a zero-sum (in the spirit of definition \ref{def:competitive}), (3) we can average both rewards and give them to both agents (in the spirit of definition \ref{def:cooperative}) or (4) any other combination of the rewards, including a stochastic (non-deterministic) combination. Note that none of these four options would contain or foster any kind of social behaviour. In fact, no agent is aware of the other (apart from the effect on rewards). However, the rewards can get highly correlated (as in ways 2 or 3 above) and do so in a non-additive or functional way.

The explanation of why there is no social behaviour in this case is that there is no influence between the actions of both agents. As a result, the big issue about choosing social contexts is how to determine that an agent has influence on other's actions. In fact, this is at the roots of definitions of interaction \cite{williams2011generalized,AGI2011Compression,ReteCogJavier2013,ReteCogDavid2013}, and the distinction between several kinds of interaction \cite{Keil-Goldin05}. Some other approaches have looked for some common information content between the peers. However, as pointed out by \cite{ReteCogDavid2013}, ``this may originate from a common source'', so common or mutual information is not sufficient for interaction to have taken place.


So we need a measure of interaction that is not based on correlation or common information content. However, the degree of influence that other agents may have on the actions of the agent we are evaluating is difficult to grasp; As environments and agents can be non-deterministic, changes can appear just randomly.

\subsubsection{Action Dependency}
\label{sec:AD}
We need to take a different approach. The key idea defines interaction in terms of sensitivity to other agents or, in other words, whether the inclusion of different agents in the environment has an effect on what the evaluated agent does. One formalisation of this idea goes as follows:

\begin{definition}
\label{def:AD_agent}
The action dependency degree for evaluated agent $\pi$ playing in slot $i$ in environment $\mu$ with a class of opponents and team players $\Pi_o$ with a weight of line-up patterns $w_{\dot{L}}$, is given by:

\begin{equation}
\label{eq:AD_agent}
AD_i(\pi,\Pi_o,w_{\dot{L}},\mu) \triangleq \eta_{\dot{L}^2} \sum_{\dot{u},\dot{v} \in \dot{L}^{N(\mu)}_{-i}(\Pi_o) | \dot{u} \neq \dot{v}} w_{\dot{L}}(\dot{u}) w_{\dot{L}}(\dot{v}) \Delta_S(\breve{A}_i(\mu[\instantiation{u}{i}{\pi}]), \breve{A}_i(\mu[\instantiation{v}{i}{\pi}]))
\end{equation}

\noindent where $\eta_{\dot{L}^2}$ normalises the formula with $\eta_{\dot{L}^2} = \frac{1}{\sum_{\dot{u},\dot{v} \in \dot{L}^{N(\mu)}_{-i}(\Pi_o) | \dot{u} \neq \dot{v}} w_{\dot{L}}(\dot{u}) w_{\dot{L}}(\dot{v})}$, $\Delta_S$ is a divergence function between string distributions, $|\Pi_o| \geq 2$, $N(\mu) \geq 2$ and $\exists \dot{u},\dot{v} \in \dot{L}^{N(\mu)}_{-i}(\Pi_o) | \dot{u} \neq \dot{v}, w_{\dot{L}}(\dot{u}) > 0$ and $w_{\dot{L}}(\dot{v}) > 0$.
\end{definition}

Note that $w_{\dot{L}}$ can be written in terms of $w_{\Pi_o}$ if we assume independence for the environment and agent positions (assumption \ref{assum:agent_environment_weight_independence} and \ref{assum:line-up_position_independence}), as we did in assumption \ref{assum:line-up_weight_as_product_function}.
Note also that $\breve{A}$ returns a distribution of action sequences if the environment or any of the agents is non-deterministic. If $AD_i$ is high, then the proportion of cases where two team line-up patterns lead to different sequences of actions for $\pi$ is high. This means that $\pi$ is highly sensitive in this environment about who else is in the environment. Conversely, if we have that for many pairs of line-up patterns the sequences of actions of $\pi$ are similar, this means that $\pi$'s actions are not affected by other agents.

The previous definition is relative to a distribution of line-up patterns on a population of agents, but it is given for a particular evaluated agent $\pi$. We may have that one evaluated agent can be very insensitive to line-up pattern changes, but other evaluated agents can be more sensitive in the same environment. If we want to generalise this for a class of evaluated agents $\Pi_e$ we have:

\begin{definition}
\label{def:AD_set}
The action dependency degree for a set of evaluated agents $\Pi_e$ with associated weight $w_{\Pi_e}$ playing in slot $i$ in environment $\mu$ with a class of opponents and team players $\Pi_o$ and a weight of line-up patterns $w_{\dot{L}}$ is given by:

\begin{equation}
\label{eq:AD_set}
AD_i(\Pi_e,w_{\Pi_e},\Pi_o,w_{\dot{L}},\mu) \triangleq \sum_{\pi \in \Pi_e} w_{\Pi_e}(\pi) AD_i(\pi,\Pi_o,w_{\dot{L}},\mu)
\end{equation}

\noindent where $|\Pi_e| \geq 1$.
\end{definition}


This definition is given only for slot $i$. Finally, we need to aggregate the action dependency degree for all slots:

\begin{definition}
\label{def:AD}
The action dependency degree for a set of evaluated agents $\Pi_e$ with associated weight $w_{\Pi_e}$ in environment $\mu$ with weight of slots $w_S$, a class of opponents and team players $\Pi_o$ and a weight of line-up patterns $w_{\dot{L}}$, is given by:

\begin{equation}
\label{eq:AD}
AD(\Pi_e,w_{\Pi_e},\Pi_o,w_{\dot{L}},\mu,w_S) \triangleq \sum_{i = 1}^{N(\mu)} w_S(i,\mu) AD_i(\Pi_e,w_{\Pi_e},\Pi_o,w_{\dot{L}},\mu)
\end{equation}

\noindent where $N(\mu) \geq 1$.
\end{definition}


It certainly remains to clarify what $\Delta_S$ can be. For deterministic cases an edit distance can be used. However, for non-deterministic cases we need to find alignments between the distributions or aggregate strings into some prototypes and compare them. One simple approach for both the deterministic and non-deterministic cases could be based on comparing action frequencies (independently of their order) or, alternatively, n-grams. We will see some specific examples in section \ref{sec:current_environments}. Note that different $\Delta_S$ functions may lead to different interpretations of action influence. For instance, there can be environments where a first few actions are interactive, but then no interaction takes places any more. In this case, the strings may be very different, but the degree, or more precisely, the timespan of interaction is small (like a butterfly effect).

\subsection{Non-neutralism}
\label{sec:non-neutralism}
The existence of interaction between agents does not ensure that these interactions are meaningful in terms of rewards. For instance, two agents can interact, but they may not affect each other's rewards. This, in ecological terms, is known as `neutralism'. In fact, in ecology, given two species, there are seven possible combinations of positive, negative or no effect between them, leading to six forms of symbiosis \cite{mac2003environmental}: neutralism (0,0), amensalism (0,-), commensalism (+,0), competition (-,-), mutualism (+,+), and predation/parasitism (+,-). In our case, as we want to characterise environments that may contain individuals (possibly more than two), we can simplify this to neutralism, cooperation (including commensalism and mutualism) and competition (including the rest). In other words, we want to analyse whether interaction has no effect on rewards, has a positive relation or a negative one.

\subsubsection{Reward Dependency}
\label{sec:RD}
So, the first thing that we need to determine is whether there is a dependency in rewards. This is very similar to the action dependency seen above:

\begin{definition}
\label{def:RD_agent}
The reward dependency degree for evaluated agent $\pi$ playing in slot $i$ in environment $\mu$ with a class of opponents and team players $\Pi_o$ and a weight of line-up patterns $w_{\dot{L}}$, is given by:

\begin{equation}
\label{eq:RD_agent}
RD_i(\pi,\Pi_o,w_{\dot{L}},\mu) \triangleq \eta_{\dot{L}^2} \sum_{\dot{u},\dot{v} \in \dot{L}^{N(\mu)}_{-i}(\Pi_o) | \dot{u} \neq \dot{v}} w_{\dot{L}}(\dot{u}) w_{\dot{L}}(\dot{v}) \Delta_{\mathbb{Q}}(R_i(\mu[\instantiation{u}{i}{\pi}]), R_i(\mu[\instantiation{v}{i}{\pi}]))
\end{equation}

\noindent where $\eta_{\dot{L}^2}$ normalises the formula with $\eta_{\dot{L}^2} = \frac{1}{\sum_{\dot{u},\dot{v} \in \dot{L}^{N(\mu)}_{-i}(\Pi_o) | \dot{u} \neq \dot{v}} w_{\dot{L}}(\dot{u}) w_{\dot{L}}(\dot{v})}$, $\Delta_{\mathbb{Q}}$ is a divergence function for rational numbers, $|\Pi_o| \geq 2$, $N(\mu) \geq 2$ and $\exists \dot{u},\dot{v} \in \dot{L}^{N(\mu)}_{-i}(\Pi_o) | \dot{u} \neq \dot{v}, w_{\dot{L}}(\dot{u}) > 0$ and $w_{\dot{L}}(\dot{v}) > 0$.
\end{definition}

Note that we use expected average rewards instead of a history of rewards. So, here the divergence compares numbers. For instance, we can use $\Delta_{\mathbb{Q}}(a,b) = 1 - \delta(a,b)$, where $\delta$ is the Kronecker delta function ($\delta(a,b) = 1$ if $a = b$ and 0 otherwise). With this choice, the previous function would boil down to the probability that by taking two team line-up patterns (using a weight or distribution $w_{\dot{L}}$), after instantiating both with $\pi$ in slot $i$, the expected average rewards of $\pi$ are different. Another option could be relative absolute difference, i.e., $\Delta_{\mathbb{Q}}(a,b) = \frac{|a-b|}{|a|+|b|}$.

Now, we can generalise this for a set of evaluated agents $\Pi_e$:

\begin{definition}
\label{def:RD_set}
The reward dependency degree for a set of evaluated agents $\Pi_e$ with associated weight $w_{\Pi_e}$ playing in slot $i$ in environment $\mu$ with a class of opponents and team players $\Pi_o$ and a weight of line-up patterns $w_{\dot{L}}$ is given by:

\begin{equation}
\label{eq:RD_set}
RD_i(\Pi_e,w_{\Pi_e},\Pi_o,w_{\dot{L}},\mu) \triangleq \sum_{\pi \in \Pi_e} w_{\Pi_e}(\pi) RD_i(\pi,\Pi_o,w_{\dot{L}},\mu)
\end{equation}

\noindent where $|\Pi_e| \geq 1$.
\end{definition}

So now we measure how dependent the rewards are in general (for any evaluated agent in $\Pi_e$).

The previous definition may slightly resemble the Shapley Value \cite{roth1988shapley} in cooperative game theory, but here we are not concerned about how relevant each agent is in a team (whether its contribution is higher than the contribution of its teammates), but to see whether there is effect on the rewards.

Finally, we aggregate for all slots:

\begin{definition}
\label{def:RD}
The reward dependency degree for a set of evaluated agents $\Pi_e$ with associated weight $w_{\Pi_e}$ in environment $\mu$ with weight of slots $w_S$, a class of opponents and team players $\Pi_o$ and a weight of line-up patterns $w_{\dot{L}}$ is given by:

\begin{equation}
\label{eq:RD}
RD(\Pi_e,w_{\Pi_e},\Pi_o,w_{\dot{L}},\mu,w_S) \triangleq \sum_{i = 1}^{N(\mu)} w_S(i,\mu) RD_i(\Pi_e,w_{\Pi_e},\Pi_o,w_{\dot{L}},\mu)
\end{equation}

\noindent where $N(\mu) \geq 1$.
\end{definition}

\subsubsection{Slot Reward Dependency}
\label{sec:SRD}
Both definitions \ref{def:AD} and \ref{def:RD} are necessary as we can have reward dependency without action dependency and
action dependency without reward dependency.

Now, we are interested in telling the {\em sign} of this dependency, i.e., how much cooperative or competitive this is.


\begin{definition}
\label{def:SRD_agent}
The slot reward dependency for evaluated agent $\pi$ playing in slot $i$ with slot $j$ (with i $\neq$ j) in environment $\mu$ with a class of opponents and team players $\Pi_o$ and a weight of line-up patterns $w_{\dot{L}}$ is given by:

\begin{equation}
\label{eq:SRD_agent}
SRD_{i,j}(\pi,\Pi_o,w_{\dot{L}},\mu) \triangleq corr_{\dot{l} \in \dot{L}^{N(\mu)}_{-i}(\Pi_o)}[w_{\dot{L}}(\dot{l})](R_i(\mu[\instantiation{l}{i}{\pi}]), R_j(\mu[\instantiation{l}{i}{\pi}]))
\end{equation}

\noindent where $corr_{x \in X}[w](a,b)$ is a weighted ($w$) correlation function between $a$ and $b$ for the elements generated by $X$, $|\Pi_o| \geq 1$ and $N(\mu) \geq 2$.
\end{definition}

Any correlation function can be used here (Pearson, Spearman, etc.). Clearly, if two slots are in the same team, from definition \ref{def:team}, we have that $SRD$ is $1$. In the case of a zero-sum game with only two teams, any two slots of different teams have a $SRD$ of $-1$ (provided there is at least one `match' which is not a tie). Note that as usual with correlation measures, if we have that two slots are reward independent, then $SRD$ is $0$. However, having $SRD=0$ does not necessarily imply independency. This means that we need to calculate the reward dependency degree and then ask whether this dependency is positive or negative for pairs of slots.

Now, we can generalise this for a set of evaluated agents $\Pi_e$:

\begin{definition}
\label{def:SRD_set}
The slot reward dependency for a set of evaluated agents $\Pi_e$ with associated weight $w_{\Pi_e}$ playing in slot $i$ with slot $j$ (with i $\neq$ j) in environment $\mu$ with a class of opponents and team players $\Pi_o$ and a weight of line-up patterns $w_{\dot{L}}$ is given by:

\begin{equation}
\label{eq:SRD_set}
SRD_{i,j}(\Pi_e,w_{\Pi_e},\Pi_o,w_{\dot{L}},\mu) \triangleq \sum_{\pi \in \Pi_e} w_{\Pi_e}(\pi) SRD_{i,j}(\pi,\Pi_o,w_{\dot{L}},\mu)
\end{equation}

\noindent where $|\Pi_e| \geq 1$.
\end{definition}

Finally we aggregate all combinations of pairs of slots:

%

\begin{definition}
\label{def:SRD}
The slot reward dependency for a set of evaluated agents $\Pi_e$ with associated weight $w_{\Pi_e}$ in environment $\mu$ with weight of slots $w_S$, with a class of opponents and team players $\Pi_o$ and a weight of line-up patterns $w_{\dot{L}}$ is given by:

\begin{equation}
\label{eq:SRD}
\begin{aligned}
& SRD(\Pi_e,w_{\Pi_e},\Pi_o,w_{\dot{L}},\mu,w_S) \triangleq \eta_{S_1^2} \sum_{i=1}^{N(\mu)} w_S(i,\mu) \times\\
& \times \left(\sum_{j=1}^{i-1} w_S(j,\mu) SRD_{i,j}(\Pi_e,w_{\Pi_e},\Pi_o,w_{\dot{L}},\mu) + \sum_{j=i+1}^{N(\mu)} w_S(j,\mu) SRD_{i,j}(\Pi_e,w_{\Pi_e},\Pi_o,w_{\dot{L}},\mu)\right)
\end{aligned}
\end{equation}

\noindent where $\eta_{S_1^2}$ normalises the formula with $\eta_{S_1^2} = \frac{1}{\sum_{i=1}^{N(\mu)} w_S(i,\mu) \left(\sum_{j=1}^{i-1} w_S(j,\mu) + \sum_{j=i+1}^{N(\mu)} w_S(j,\mu)\right)}$, $N(\mu) \geq 2$ and $\exists i,j | 1 \leq i \leq N(\mu), 1 \leq j \leq N(\mu), i \neq j, w_S(i,\mu) > 0$ and $w_S(j,\mu) > 0$.
\end{definition}

In practice, in order to evaluate social abilities, we require environments with high $RD$. Then, depending on the use of teams and normalisations, we can gauge whether we want to evaluate competition or cooperation, and have some positive $SRD$ with some slots and some negative $SRD$ with some other slots. This is easily obtained by using teams.


\subsection{Secernment}
\label{sec:secernment}
It is an important characteristic for a test to be able to give different values for different evaluated agents. Otherwise, if the results are the same (or very similar) for most evaluated agents, we get little information. In other words, we want tests (i.e., environment and set of agents populating it) to secern, to be discriminative. Although there are many approaches to the idea of discriminating power (see e.g., \cite{orallo2014JAAMAS}), one simple notion that accounts for this concept quite well is the variance of results.

\subsubsection{Fine and Coarse Discrimination}
\label{sec:FD}
\label{sec:CD}
In order to formalise this notion of variance (or number of different values) of the expected average rewards of the set of evaluated agents, we can just compare pairs of values as follows:

\begin{definition}
\label{def:FD_agents}
The fine discriminating power for evaluated agents $\pi_1$ and $\pi_2$ playing in slot $i$ in environment $\mu$ with a class of opponents and team players $\Pi_o$ and a weight of line-up patterns $w_{\dot{L}}$ is given by:

\begin{equation}
\label{eq:FD_agents}
FD_i(\pi_1,\pi_2,\Pi_o,w_{\dot{L}},\mu) \triangleq \sum_{\dot{l} \in \dot{L}^{N(\mu)}_{-i}(\Pi_o)} w_{\dot{L}}(\dot{l}) \Delta_{\mathbb{Q}}(R_i(\mu[\instantiation{l}{i}{\pi_1}]), R_i(\mu[\instantiation{l}{i}{\pi_2}]))
\end{equation}

\noindent where $\Delta_{\mathbb{Q}}$ is a divergence function for rational numbers, $N(\mu) \geq 1$ and if $N(\mu) > 1$ then $|\Pi_o| \geq 1$ otherwise $|\Pi_o| \geq 0$.
\end{definition}

This measures the expected average reward divergence of two evaluated agents placed both in slot $i$ in the same line-up patterns. If $\Delta_{\mathbb{Q}}$ is some kind of numeric difference (e.g., the absolute difference or the squared difference), then this measure would be similar to some kind of dispersion of expected average rewards (like a variance). If $\Delta_{\mathbb{Q}}$ equals $1-\delta$ (with $\delta$ being the Kronecker delta function) we have that this measures the number of times two different evaluated agents score differently. 

We can generalise this for a set of evaluated agents $\Pi_e$:

\begin{definition}
\label{def:FD_set}
The fine discriminating power for a set of evaluated agents $\Pi_e$ with associated weight $w_{\Pi_e}$ playing in slot $i$ in environment $\mu$ with a class of opponents and team players $\Pi_o$ and a weight of line-up patterns $w_{\dot{L}}$ is given by:

\begin{equation}
\label{eq:FD_set}
FD_i(\Pi_e,w_{\Pi_e},\Pi_o,w_{\dot{L}},\mu) \triangleq \eta_{\Pi^2} \sum_{\pi_1,\pi_2 \in \Pi_e | \pi_1 \neq \pi_2} w_{\Pi_e}(\pi_1) w_{\Pi_e}(\pi_2) FD_i(\pi_1,\pi_2,\Pi_o,w_{\dot{L}},\mu)
\end{equation}
\end{definition}

\noindent where $\eta_{\Pi^2}$ normalises the formula with $\eta_{\Pi^2} = \frac{1}{\sum_{\pi_1,\pi_2 \in \Pi_e | \pi_1 \neq \pi_2} w_{\Pi_e}(\pi_1) w_{\Pi_e}(\pi_2)}$, $|\Pi_e| \geq 2$ and $\exists \pi_1,\pi_2 \in \Pi_e | \pi_1 \neq \pi_2$, $w_{\Pi_e}(\pi_1) > 0$ and $w_{\Pi_e}(\pi_2) > 0$.

Again, we can sum over all slots:

\begin{definition}
\label{def:FD}
The fine discriminating power for a set of evaluated agents $\Pi_e$ with associated weight $w_{\Pi_e}$ in environment $\mu$ with weight of slots $w_S$, with a class of opponents and team players $\Pi_o$ and a weight of line-up patterns $w_{\dot{L}}$ is given by:

\begin{equation}
\label{eq:FD}
FD(\Pi_e,w_{\Pi_e},\Pi_o,w_{\dot{L}},\mu,w_S) \triangleq \sum_{i=1}^{N(\mu)} w_S(i,\mu) FD_i(\Pi_e,w_{\Pi_e},\Pi_o,w_{\dot{L}},\mu)
\end{equation}

\noindent where $N(\mu) \geq 1$.
\end{definition}

Being able to discriminate in terms of pair of agents for each line-up pattern in an environment can be generalised with the overall result of a measure $\Upsilon$ by considering the aggregated values on this measure, namely:

\begin{definition}
\label{def:CD_agents}
The coarse discriminating power for evaluated agents $\pi_1$ and $\pi_2$ playing in slot $i$ in environment $\mu$ with a class of opponents and team players $\Pi_o$ and a weight of line-up patterns $w_{\dot{L}}$ is given by:

\begin{equation}
\label{eq:CD_agents}
CD_i(\pi_1,\pi_2,\Pi_o,w_{\dot{L}},\mu) \triangleq \Delta_{\mathbb{Q}}(\Upsilon_i(\pi_1,\Pi_o,w_{\dot{L}},\mu), \Upsilon_i(\pi_2,\Pi_o,w_{\dot{L}},\mu))
\end{equation}

\noindent where $\Upsilon_i(\pi,\Pi,w_{\dot{L}},\mu)$ is defined in equation \ref{eq:social_intelligence_with_line-ups} and $\Delta_{\mathbb{Q}}$ is a divergence function for rational numbers.
\end{definition}

We generalise this for a set of evaluated agents $\Pi_e$:

\begin{definition}
\label{def:CD_set}
The coarse discriminating power for a set of evaluated agents $\Pi_e$ with associated weight $w_{\Pi_e}$ playing in slot $i$ in environment $\mu$ with a class of opponents and team players $\Pi_o$ and a weight of line-up patterns $w_{\dot{L}}$ is given by:

\begin{equation}
\label{eq:CD_set}
CD_i(\Pi_e,w_{\Pi_e},\Pi_o,w_{\dot{L}},\mu) \triangleq \eta_{\Pi^2} \sum_{\pi_1,\pi_2 \in \Pi_e | \pi_1 \neq \pi_2} w_{\Pi_e}(\pi_1) w_{\Pi_e}(\pi_2) CD_i(\pi_1,\pi_2,\Pi_o,w_{\dot{L}},\mu)
\end{equation}
\end{definition}

\noindent where $\eta_{\Pi^2}$ normalises the formula with $\eta_{\Pi^2} = \frac{1}{\sum_{\pi_1,\pi_2 \in \Pi_e | \pi_1 \neq \pi_2} w_{\Pi_e}(\pi_1) w_{\Pi_e}(\pi_2)}$, $|\Pi_e| \geq 2$ and $\exists \pi_1,\pi_2 \in \Pi_e | \pi_1 \neq \pi_2$, $w_{\Pi_e}(\pi_1) > 0$ and $w_{\Pi_e}(\pi_2) > 0$.

And summing over all slots:

\begin{definition}
\label{def:CD}
The coarse discriminating power for a set of evaluated agents $\Pi_e$ with associated weight $w_{\Pi_e}$ in environment $\mu$ with weight of slots $w_S$, with a class of opponents and team players $\Pi_o$ and a weight of line-ups pattern $w_{\dot{L}}$ is given by:

\begin{equation}
\label{eq:CD}
CD(\Pi_e,w_{\Pi_e},\Pi_o,w_{\dot{L}},\mu,w_S) \triangleq \sum_{i=1}^{N(\mu)} w_S(i,\mu) CD_i(\Pi_e,w_{\Pi_e},\Pi_o,w_{\dot{L}},\mu)
\end{equation}

\noindent where $N(\mu) \geq 1$.
\end{definition}

In both {\em fine} and {\em coarse} discrimination the goal is to check if two evaluated agents obtain similar results. 
The difference resides at the level we analyse their results. While the {\em fine} checks the similarities for each line-up pattern, the {\em coarse} is more oriented to seeing the overall similarity.

\subsubsection{Strict Total and Partial Grading}
\label{sec:STG}
\label{sec:PG}
An environment and set of agents populating it being discriminative when comparing evaluated agents does not mean that there is a gradation or order between the results for a set of evaluated agents. For instance, if we have three agents $\pi_1$, $\pi_2$ and $\pi_3$ that we want to evaluate in a competitive environment with two agent slots and two teams, and we get that $\pi_1$ scores better when interacts with $\pi_2$, $\pi_2$ scores better when interacts with $\pi_3$ and $\pi_3$ scores better when interacts with $\pi_1$, then there is no way to establish a gradation for these three agents.
Idealistically, we would like to have a strict total order, but this is unrealistic for many environments.

So the idea we will pursue is to evaluate how close an environment and set of agents populating it are to this ideal situation from the expected average rewards of the evaluated agents (without an aggregated rating system\footnote{A common approach is to create a rating when we have many experiments, as done with sport ratings, such as the ELO rating \cite{elo1978rating} in chess.}):

\begin{definition}
\label{def:STG_agents}
The strict total grading quality for evaluated agents $\pi_1, \pi_2$ and $\pi_3$ playing in slots $i$ and $j$ (with i $\neq$ j) in environment $\mu$ with a class of opponents and team players $\Pi_o$ and a weight of line-up patterns $w_{\dot{L}}$ is given by:

\begin{equation}
\label{eq:STG_agents}
STG_{i,j}(\pi_1,\pi_2,\pi_3,\Pi_o,w_{\dot{L}},\mu) \triangleq \sum_{\dot{l} \in \dot{L}^{N(\mu)}_{-i,j}(\Pi_o)} w_{\dot{L}}(\dot{l}) STO_{i,j}(\pi_1,\pi_2,\pi_3,\dot{l},\mu)
\end{equation}

\noindent where $N(\mu) \geq 2$, if $N(\mu) > 2$ then $|\Pi_o| \geq 1$ otherwise $|\Pi_o| \geq 0$ and $STO_{i,j}(\pi_1,\pi_2,\pi_3,\dot{l},\mu)$ (where $\dot{l}$ has all its elements instantiated except positions $i$ and $j$) is 1 if there is a permutation of the three evaluated agents such that there is a strict total order in their expected average rewards when placed by pairs in $\dot{l}$ interacting in environment $\mu$ in slots $i$ and $j$, and 0 otherwise. Formally, it is 1 iff there is a permutation $(\pi_1',\pi_2',\pi_3')$ such that:
$R_i(\mu[\instantiation{l}{i,j}{\pi_1',\pi_2'}]) < R_j(\mu[\instantiation{l}{i,j}{\pi_1',\pi_2'}])$,
$R_i(\mu[\instantiation{l}{i,j}{\pi_2',\pi_3'}]) < R_j(\mu[\instantiation{l}{i,j}{\pi_2',\pi_3'}])$ and 
$R_i(\mu[\instantiation{l}{i,j}{\pi_1',\pi_3'}]) < R_j(\mu[\instantiation{l}{i,j}{\pi_1',\pi_3'}])$. 
For instance, if we have three agents $a$, $b$ and $c$ in an environment $\mu$ and line-up pattern $\dot{l}$ for slots $i$ and $j$, and their expected average rewards when placed by pairs in $\dot{l}$ shows us that $b < a$, $a < c$ and $b < c$, then we have $STO_{i,j}(a,b,c,\dot{l},\mu) = 1$ with the permutation $(b,a,c)$.
This property is related to reliability, which we will see later on.
\end{definition}

We generalise this for a set of evaluated agents $\Pi_e$:

\begin{definition}
\label{def:STG_set}
The strict total grading quality for a set of evaluated agents $\Pi_e$ with associated weight $w_{\Pi_e}$ playing in slots $i$ and $j$ (with i $\neq$ j) in environment $\mu$ with a class of opponents and team players $\Pi_o$ and a weight of line-up patterns $w_{\dot{L}}$ is given by:

\begin{equation}
\label{eq:STG_set}
STG_{i,j}(\Pi_e,w_{\Pi_e},\Pi_o,w_{\dot{L}},\mu) \triangleq \eta_{\Pi^3} \sum_{\pi_1,\pi_2,\pi_3 \in \Pi_e | \pi_1 \neq \pi_2 \neq \pi_3} w_{\Pi_e}(\pi_1) w_{\Pi_e}(\pi_2) w_{\Pi_e}(\pi_3) STG_{i,j}(\pi_1,\pi_2,\pi_3,\Pi_o,w_{\dot{L}},\mu)
\end{equation}

\noindent where $\eta_{\Pi^3}$ normalises the formula with $\eta_{\Pi^3} = \frac{1}{\sum_{\pi_1,\pi_2,\pi_3 \in \Pi_e | \pi_1 \neq \pi_2 \neq \pi_3} w_{\Pi_e}(\pi_1) w_{\Pi_e}(\pi_2) w_{\Pi_e}(\pi_3)}$, $|\Pi_e| \geq 3$ and $\exists \pi_1,\pi_2,\pi_3 \in \Pi_e | \pi_1 \neq \pi_2 \neq \pi_3, w_{\Pi_e}(\pi_1) > 0, w_{\Pi_e}(\pi_2) > 0$ and $w_{\Pi_e}(\pi_3) > 0$.
\end{definition}

Now, if we aggregate all combinations of pairs of slots, we have:

\begin{definition}
\label{def:STG}
The strict total grading quality for a set of evaluated agents $\Pi_e$ with associated weight $w_{\Pi_e}$ in environment $\mu$ with weight of slots $w_S$, with a class of opponents and team players $\Pi_o$ and a weight of line-up patterns $w_{\dot{L}}$ is given by:

\begin{equation}
\begin{aligned}
& STG(\Pi_e,w_{\Pi_e},\Pi_o,w_{\dot{L}},\mu,w_S) \triangleq \eta_{S_1^2} \sum_{i=1}^{N(\mu)} w_S(i,\mu) \times\\
& \times \left(\sum_{j=1}^{i-1} w_S(j,\mu) STG_{i,j}(\Pi_e,w_{\Pi_e},\Pi_o,w_{\dot{L}},\mu) + \sum_{j=i+1}^{N(\mu)} w_S(j,\mu) STG_{i,j}(\Pi_e,w_{\Pi_e},\Pi_o,w_{\dot{L}},\mu)\right)
\end{aligned}
\end{equation}

\noindent where $\eta_{S_1^2}$ normalises the formula with $\eta_{S_1^2} = \frac{1}{\sum_{i=1}^{N(\mu)} w_S(i,\mu) \left(\sum_{j=1}^{i-1} w_S(j,\mu) + \sum_{j=i+1}^{N(\mu)} w_S(j,\mu)\right)}$, $N(\mu) \geq 2$ and $\exists i,j | 1 \leq i \leq N(\mu), 1 \leq j \leq N(\mu), i \neq j, w_S(i,\mu) > 0$ and $w_S(j,\mu) > 0$.
\end{definition}

The previous definition only considers strict total orders, and is useful to determine whether we can find a strict total order for the evaluated agents. However, this does not say much about the existence of grading `inconsistencies', such as non-discriminative cases such as $\pi_1 = \pi_2$, $\pi_2 = \pi_3$ and $\pi_1 = \pi_3$ which, for the above definition, are considered in the same way as not ordering cases such as $\pi_1 > \pi_2$, $\pi_2 > \pi_3$ and $\pi_1 < \pi_3$. In order to distinguish these cases, we can give a new definition as follows:

\begin{definition}
\label{def:PG}
The partial grading quality for a set of evaluated agents $\Pi_e$ with associated weight $w_{\Pi_e}$ in environment $\mu$ with weight of slots $w_S$, with a class of opponents and team players $\Pi_o$ and a weight of line-up patterns $w_{\dot{L}}$ is defined as in definition \ref{def:STG} with the use of a partial order with $\leq$ instead of a strict total order with $<$. It is denoted by $PG$.
\end{definition}

If $STG$ and $PG$ are high, this means that a derivation of a rating is highly consistent to what we see when using evaluated agents from $\Pi_e$ on slots $i$ and $j$. A very similar property is known as monotonicity in \cite[sec. 5]{Turing100}\cite{AISB-AICAP2012b}, showing an agent set for the matching pennies game that is non-monotonic. Nonetheless, a partial order can still be constructed for the agent set of all finite state machines for this game \cite{hibbard2008adversarial}. 

The existence of a meaningful rating allows for subselections of $\Pi_o$ according to this rating, which can be used to furbish the tests with high-rank agents that can lead to more sophisticated social environments (which can be detrimental or beneficial, so making it more or less difficult, respectively, depending on whether it is used for the same team or for opponents).

\subsection{Anticipation}
\label{sec:anticipation}
One crucial property that is related to social intelligence is anticipation, which means that in both competition and cooperation, evaluated agents can benefit from anticipating other agents' moves or, in more general terms, by having a theory of other's minds. While a formalisation of this concept is very elusive, we can at least introduce an approximation.

\subsubsection{Competitive Anticipation}
\label{sec:AComp}
The first thing we need to do is to distinguish between competition and cooperation. 
In competitive anticipation  
we usually expect that evaluated agents can perform better if their opponents can be well anticipated. An example of this is a predator-prey situation. 
This phenomenon is difficult to define in general, but we can introduce a simplified approach based on the idea that one evaluated agent anticipates {\em competitively} if its expected average reward competing against a (generally) non-random agent is higher than its expected average reward competing against a random agent. This can be generalised as follows:


\begin{definition}
\label{def:AComp_agents}
The anticipation benefit for evaluated agent $\pi_1$ against agent $\pi_2$, playing in slots $i$ and $j$ respectively (with $i$ and $j$ in different teams) when competing in environment $\mu$ with a class of opponents and team players $\Pi_o$ and a weight of line-up patterns $w_{\dot{L}}$ is given by:

\begin{equation}
\label{eq:AComp_agents}
AComp_{i,j}(\pi_1,\pi_2,\Pi_o,w_{\dot{L}},\mu) \triangleq \sum_{\dot{l} \in \dot{L}^{N(\mu)}_{-i,j}(\Pi_o)} w_{\dot{L}}(\dot{l}) \frac{1}{2} \left(R_i(\mu[\instantiation{l}{i,j}{\pi_1,\pi_2}]) - R_i(\mu[\instantiation{l}{i,j}{\pi_1,\pi_r}])\right)
\end{equation}

\noindent where $\pi_r$ is a random agent, $\frac{1}{2}$ normalises the result of $AComp_{i,j}$ to be between $-1$ and $1$, $N(\mu) \geq 2$ and if $N(\mu) > 2$ then $|\Pi_o| \geq 1$ otherwise $|\Pi_o| \geq 0$.
\end{definition}

We generalise this for a set of evaluated agents $\Pi_e$:

\begin{definition}
\label{def:AComp_set}
The anticipation benefit for a set of evaluated agents $\Pi_e$ with associated weight $w_{\Pi_e}$ playing in slot $i$ when competing against slot $j$ (with $i$ and $j$ in different teams) in environment $\mu$ with a class of opponents and team players $\Pi_o$ and a weight of line-up patterns $w_{\dot{L}}$ is given by:

\begin{equation}
\label{eq:AComp_set}
AComp_{i,j}(\Pi_e, w_{\Pi_e},\Pi_o,w_{\dot{L}},\mu) \triangleq \sum_{\pi_1 \in \Pi_e} w_{\Pi_e}(\pi_1) \max_{\pi_2 \in \Pi_o} AComp_{i,j}(\pi_1,\pi_2,\Pi_o,w_{\dot{L}},\mu)
\end{equation}

\noindent where $|\Pi_e| \geq 1$ and $|\Pi_o| \geq 1$.
\end{definition}

By aggregating all combinations of pairs of slots, we have:

\begin{definition}
\label{def:AComp}
The anticipation benefit for a set of evaluated agents $\Pi_e$ with associated weight $w_{\Pi_e}$ when competing in environment $\mu$ with weight of slots $w_S$, with a class of opponents and team players $\Pi_o$ and a weight of line-up patterns $w_{\dot{L}}$ is given by:
%
%
%

\begin{equation}
\label{eq:AComp}
AComp(\Pi_e,w_{\Pi_e},\Pi_o,w_{\dot{L}},\mu,w_S) \triangleq \eta_{S_2^2} \sum_{t_1,t_2 \in \tau | t_1 \neq t_2} \sum_{i \in t_1} w_S(i,\mu) \sum_{j \in t_2} w_S(j,\mu) AComp_{i,j}(\Pi_e,w_{\Pi_e},\Pi_o,w_{\dot{L}},\mu)
\end{equation}

\noindent where $\eta_{S_2^2}$ normalises the formula with $\eta_{S_2^2} = \frac{1}{\sum_{t_1,t_2 \in \tau | t_1 \neq t_2} \sum_{i \in t_1} w_S(i,\mu) \sum_{j \in t_2} w_S(j,\mu)}$, $\tau$ is the partition of slots on teams for environment $\mu$, $N(\mu) \geq 2$ and $\exists t_1,t_2 \in \tau | t_1 \neq t_2, \exists i \in t_1, j \in t_2 | w_S(i,\mu) > 0$ and $w_S(j,\mu) > 0$.
\end{definition}

If $AComp$ is positive this means that evaluated agents behave better against non-random agents (in general) than against random agents.
One good example of the above definition is when slot $i$ is a predator and $j$ is a prey. If, in general, evaluated agents in a class perform better with non-random preys than with random preys then the environment shows a benefit for this evaluated agent class (for these slots). Of course, this depends on $\Pi_o$, but if we include non-random opponents with some movement patterns and/or some degree of intelligence, the definition becomes more meaningful.

\subsubsection{Cooperative Anticipation}
\label{sec:ACoop}
On the other hand, in cooperative anticipation (or coordination) good rewards can only be obtained with both agents performing some actions together. Again a general definition accounting for all possible coordination situations is difficult, but we can introduce a simplified approach based on the following intuitive definition: two agents $\pi_1$ and $\pi_2$ coordinate cooperatively if the sum of the expected average rewards of $\pi_1$ and $\pi_2$ are higher when they interact together than the sum of each one interacting with a random agent.

\begin{definition}
\label{def:ACoop_agents}
The anticipation benefit for evaluated agent $\pi_1$ and agent $\pi_2$ playing in slots $i$ and $j$ respectively (with $i \neq j$ but in the same team) when cooperating in environment $\mu$ with a class of opponents and team players $\Pi_o$ and a weight of line-up patterns $w_{\dot{L}}$ is given by:

\begin{equation}
\label{eq:ACoop_agents}
\begin{aligned}
& ACoop_{i,j}(\pi_1,\pi_2,\Pi_o,w_{\dot{L}},\mu) \triangleq \sum_{\dot{l} \in \dot{L}^{N(\mu)}_{-i,j}(\Pi_o)} w_{\dot{L}}(\dot{l}) \frac{1}{4} \times\\
& \times \left(R_i(\mu[\instantiation{l}{i,j}{\pi_1,\pi_2}]) + R_j(\mu[\instantiation{l}{i,j}{\pi_1,\pi_2}]) - R_i(\mu[\instantiation{l}{i,j}{\pi_1,\pi_r}]) - R_j(\mu[\instantiation{l}{i,j}{\pi_r,\pi_2}])\right)
\end{aligned}
\end{equation}

\noindent where $\pi_r$ is a random agent, $\frac{1}{4}$ normalises the result of $ACoop_{i,j}$ to be between $-1$ and $1$, $N(\mu) \geq 2$ and if $N(\mu) > 2$ then $|\Pi_o| \geq 1$ otherwise $|\Pi_o| \geq 0$.
\end{definition}

The use of random agents for the above definition may not work in some cases if there are more than two elements in a team, as coordination may only take place when all of them coordinate and not only a pair (if a random agent is included, this can be very disruptive). In this case, the above definition could be extended to reach the number of elements in the team instead.

We can generalise this for a set of evaluated agents $\Pi_e$:



\begin{definition}
\label{def:ACoop_set}
The anticipation benefit for a set of evaluated agents $\Pi_e$ with associated weight $w_{\Pi_e}$ playing in slot $i$ when cooperating with slot $j$ (with $i \neq j$ but in the same team) in environment $\mu$ with a class of opponents and team players $\Pi_o$ and a weight of line-up patterns $w_{\dot{L}}$ is given by:

\begin{equation}
\label{eq:ACoop_set}
ACoop_{i,j}(\Pi_e,w_{\Pi_e},\Pi_o,w_{\dot{L}},\mu) \triangleq \sum_{\pi_1 \in \Pi_e} w_{\Pi_e}(\pi_1) \max_{\pi_2 \in \Pi_o} ACoop_{i,j}(\pi_1,\pi_2,\Pi_o,w_{\dot{L}},\mu)
\end{equation}

\noindent where $|\Pi_e| \geq 1$ and $|\Pi_o| \geq 1$.
\end{definition}

Finally, if we aggregate all combinations of pairs of slots, we have:


\begin{definition}
\label{def:ACoop}
The anticipation benefit for a set of evaluated agents $\Pi_e$ with associated weight $w_{\Pi_e}$ when cooperating in environment $\mu$ with weight of slots $w_S$, with a class of opponents and team players $\Pi_o$ and a weight of line-up patterns $w_{\dot{L}}$ is given by:

%
%

\begin{equation}
\label{eq:ACoop}
\begin{aligned}
ACoop(\Pi_e,w_{\Pi_e},\Pi_o,w_{\dot{L}},\mu,w_S)	& \triangleq \eta_{S_3^2} \sum_{t \in \tau} \sum_{i,j \in t | i \neq j} w_S(i,\mu) w_S(j,\mu) ACoop_{i,j}(\Pi_e,w_{\Pi_e},\Pi_o,w_{\dot{L}},\mu) +\\
													& + \sum_{t_1,t_2,t_3 \in \tau | t_1 \neq t_2 \neq t_3} \sum_{i \in t_1} w_S(i,\mu) \sum_{j \in t_2} w_S(j,\mu) ACoop_{i,j}(\Pi_e,w_{\Pi_e},\Pi_o,w_{\dot{L}},\mu)
\end{aligned}
\end{equation}

\noindent where $\eta_{S_3^2}$ normalises the formula with $\eta_{S_3^2} = \frac{1}{\sum_{t \in \tau} \sum_{i,j \in t | i \neq j} w_S(i,\mu) w_S(j,\mu) + \sum_{t_1,t_2,t_3 \in \tau | t_1 \neq t_2 \neq t_3} \sum_{i \in t_1} w_S(i,\mu) \sum_{j \in t_2} w_S(j,\mu)}$, $\tau$ is the partition of slots on teams for environment $\mu$, $N(\mu) \geq 2$ and $\exists i,j \in t \in \tau | i \neq j, w_S(i,\mu) > 0$ and $w_S(j,\mu) > 0$ or $\exists t_1,t_2,t_3 \in \tau | t_1 \neq t_2 \neq t_3, \exists i \in t_1, j \in t_2 | w_S(i,\mu) > 0$ and $w_S(j,\mu) > 0$.
\end{definition}

The previous definition includes slots in two different teams when the environment contains three or more teams. This is because two agents from different teams can cooperate against another agent in a third team.

We have defined the anticipation with only two slots, but competition and cooperation can also appear with three or more agents.



\subsection{Symmetry}
\label{sec:symmetry}
In game theory, a symmetric game is a game where the payoffs for playing a particular strategy depend only on the other strategies employed by the rest of agents, not on who is playing them. This property is very useful for evaluating purposes. Since we can change the positions of the agents and they will maintain their results, we only need to evaluate the agent playing in one position of the environment.

For our definition of multi-agent environments this definition of symmetry must be reconsidered. The previous definition means that for each pair of line-ups with the same agents but in different order, the agents obtain exactly the same results. But with the inclusion of teams this definition is not appropriate. For example, using an environment with the partition of slots on teams $\tau = \{\{1,2\},\{3,4\}\}$ and line-up $l = (\pi_1, \pi_2, \pi_3, \pi_4$), we have that agents $\pi_1$ and $\pi_2$ must both obtain the same result, as $\pi_3$ and $\pi_4$ as well. Following the definition and switching the positions of $\pi_2$ and $\pi_3$ we obtain line-up $l' = (\pi_1, \pi_3, \pi_2, \pi_4)$, which now means that agents $\pi_1$ and $\pi_3$ must have the same rewards (since they are now in the same team) while maintaining their previous results, as $\pi_2$ and $\pi_4$ as well. This situation can only occur when all slots (and therefore teams) obtain equal results.

Instead, we extend this definition of symmetry to include teams. First, we denote by $\sigma(l)$ the set of line-ups permuting the agent positions in line-up $l$. This set will correspond with the one used in game theory to define symmetry. To adapt this set to our definition, we must select a subset of line-ups from $\sigma(l)$ respecting the teams defined in $\tau$. We denote this subset with $\sigma(l,\tau)$, where we only select line-ups from $\sigma(l)$ if original teams are maintained. Following the example, line-up $l'$ is not included in $\sigma(l,\tau)$ since $\pi_1$ and $\pi_3$ from $l'$ were not in the same team in $l$ (as $\pi_2$ and $\pi_4$ as well). However, $l'' = (\pi_3, \pi_4, \pi_2, \pi_1)$ is included in $\sigma(l,\tau)$, since both pair of agents $(\pi_1, \pi_2)$ and $(\pi_3, \pi_4)$ are still in the same team. From here, we define symmetry for a multi-agent environment as follows:

\begin{definition}
\label{def:symmetry}
We say a multi-agent environment $\mu$ is symmetric if and only if every team in $\tau$ has the same number of elements and:

\begin{equation}
\forall i,K,\Pi,l\in L^{N(\mu)}(\Pi),l'\in\sigma(l,\tau) : \breve{R}^K_i(\mu[l]) = \breve{R}^K_{i'}(\mu[l'])
\end{equation}

\noindent where $i'$ represents the slot of agent $l_{i:i}$ in $l'$.
\end{definition}

Note that we impose that each set in $\tau$ must have the same number of elements. This is because we only consider environments to be symmetric if we can evaluate an agent in every slot and obtain the same result. Having teams with different number of elements will not allow us to do this.

This definition now fits our goal of symmetry. But too few environments will fit this definition of symmetry because it is too restrictive. However, we could divide this definition of symmetry of teams into two parts depending on the relation between the slots:

For the first part we look at the relation between the slots within each team:

\begin{definition}
\label{def:intra_team_symmetry}
We say an environment is \emph{Intra-Team Symmetric} when the agents within every team can be swapped without affecting their results.
\end{definition}

This kind of symmetry will allow us to evaluate a team in an environment without taking into account their positions within every set of $\tau$.

For the second part we look at the relation between the slots in different teams. This can be also divided into two parts:

\begin{definition}
\label{def:partial_inter_team_symmetry}
We say an environment is \emph{Partial Inter-Team Symmetric} if every pair of teams having the same number of elements can be swapped without affecting their results.
\end{definition}

\begin{definition}
\label{def:total_inter_team_symmetry}
We say an environment is \emph{Total Inter-Team Symmetric} if every pair of teams can be swapped without affecting their results.
\end{definition}

This kind of symmetry will allow us to evaluate a team in an environment without taking into account in which set of $\tau$ they are situated.

Definition \ref{def:symmetry} will correspond with an Intra-Team and Total Inter-Team Symmetry, where every team of agents can be located in every set of $\tau$ and in different order, maintaining their performance expectation.

Symmetry is not a necessary condition for social behaviours, but it is a very practical one for measurement as the result does not depend on the slot we use to evaluate and all slots are useful for evaluating the same ability. This simplifies all previous calculations (for social intelligence and properties) as we can obtain the same results by calculating them for only one slot or pair of slots.

\subsection{Validity}
\label{sec:validity}
Validity is the most important property of a cognitive test in psychometrics. In our context, the validity of a definition is that it accounts for the notion we expect it to grasp. For instance, if we say that a given definition of $\Upsilon$ measures social intelligence but it actually measures arithmetic abilities then the definition is not valid. In our case, this depends on the choice of $\Pi_o$ ($\Pi$) and $M$ as the core of definition \ref{def:social_intelligence}.

Poor validity may have two sources (or may appear in two different variants): a definition may be too specific (it does not account for all the abilities the notion is thought to consider) or it is too general (it includes some abilities that are not part of the notion to be measured). In other words, the measure should account for {\em all, but not more,} of the concept it tries to represent. We refer to these two issues of validity as the generality of the measure and the specificity of the measure.

Regarding generality, we should be careful about the use of very restrictive choices for $\Pi_o$ and $M$. It could be possible to find a single environment that meets all the properties seen in the previous subsections. However, using just one environment is prone to specialisation, as usual in many AI benchmarks. For instance, if we use a particular maze as an environment with a set of particular agents, then we can have good scores by using a very specialised agent for this situation, which may be unable to succeed in other social contexts. For instance, chess with current chess players is an example where a specialised system (e.g., Deep Blue) is able to score well, while it is clearly useless for other problems. A similar over-specialisation may happen if the agent class is too small. This is usual in biology, where some species specialise for predating (or establishing a symbiosis) with other species.

Consequently, the environment class and the agent class must be general enough to avoid that some predefined or hardwired policies could be optimal for these classes. This is the key issue of a (social) {\em intelligence} test; it must be as general as possible. We need to choose a diverse environment class. One possibility is to consider all environments (as done by \cite{LeggHutter07,AIJ2010}), and another is to find an environment class that is sufficiently representative (as attempted in \cite{hernandez2010hopefully}).



Similarly, as happens with the environment diversity and most especially while evaluating social intelligence, we need to consider a class of agents that leads to a diversity in line-up patterns. This class of agents should incorporate many different types of agents: random agents, agents with some predetermined policies, agents that are able to learn, human agents, agents with low social intelligence, agents with high social intelligence, etc. The set of all possible agents (either artificial or biological) is known as {\em machine kingdom} in \cite{upsychometrics2,hernandez2013potential} and raises many questions about the feasibility of any test considering this astronomically large set. Also, there are doubts about what the weight for this universal set should be. Instead, some representative kinds of agents could be chosen. In this way, we could aim at social intelligence relative to a smaller (and well-defined) set of agents, possibly specialising the definition by limiting the resources, the program size \cite{agentpolicies2013} or the intelligence of the agents \cite{AGI2011DarwinWallace}.
 


Regarding specificity, it is equally important for a measurement to only include those environments and agents that really reflect what we want to measure. For instance, it is desirable that the evaluation of an ability is done in an environment where no other abilities are required, or in other words, we want that the environment evaluates the ability in isolation. Otherwise, it will not be clear which part of the result comes from the ability to be evaluated, and which part comes from other abilities. Although it is very difficult to avoid any contamination, the idea is to ensure that the role of these other abilities are minor, or are taken for granted for all agents.

The properties of dependency (interactivity, non-neutralism) and anticipation (both competitive and cooperative) seen in previous subsections have been included for the sake of specificity. We are certainly not interested in non-social environments as this would contaminate the measure with other abilities. In fact, one of the recurrent issues in defining and measuring social intelligence is to be specific enough to distinguish it from general intelligence.

Unlike all other properties in this section, validity is not a property for which a formal account can be given, as it precisely accounts for how well the definition reflects the natural or intuitive notion that is to be measured. 

The assurement about validity must come then from the use of a formal definition (as for definition \ref{def:social_intelligence}) with a meaningful instantiation for the agent and environment classes, and also from the experimental results that can be obtained through the tests derived from the definition. Note that in psychometrics there is usually a lack of a proper definition of the cognitive abilities of interest (e.g., psychometrics has not presented an unambiguous definition of intelligence), so validity is applied to a test and not to the definition of a cognitive ability. In fact, the concept is frequently derived from the test, as has happened with the modern view of intelligence, as ``the ability measured by IQ tests''. 

As discussed in the introduction, we are interested in a formal definition whose validity we can discuss and appraise, and from which tests are derived, and not the other way round.

\subsection{Reliability}
\label{sec:reliability}
Another key issue in psychometric tests is the notion of reliability, which means that the measurement is close to the actual value. Note that this is different to validity, which refers about the true identification or definition of the actual value. In other words, if we assume validity, i.e., that our definition is correct\footnote{The lack of a proper definition for many abilities makes reliability refer to the quality of the result of a single application of the test in comparison to the idealised average result if the test could be repeated indefinitely.}, reliability refers to the quality of the measurement with respect to the actual value. 
More technically, if the actual value of $\pi$ for an ability $\phi$ is $v$ then we want a test to give a value which is close to $v$. The cause of the divergence may be systematic (bias), non-systematic (variance) or both.

First, we need to consider that reliability applies to tests, as introduced in section \ref{sec:tests}. Reliability is then defined by considering that a test can be repeated many times, so becoming a random variable that we can compare to the true value. Formally:

\begin{definition}
Given a definition of a cognitive ability $\Upsilon$ and a test $\hat{\Upsilon}$, the test error is given by:

\begin{equation}
TE(\hat{\Upsilon}) \triangleq  Mean((\hat{\Upsilon} - \Upsilon)^2)
\end{equation}

\noindent where the mean is calculated over the repeated application of the test (to one subject or more subjects).
\end{definition}

The reliability $Rel(\hat{\Upsilon})$ can be defined as a decreasing function over $TE(\hat{\Upsilon})$, such as $Rel(\hat{\Upsilon}) = e^{-TE(\hat{\Upsilon})}$.

The reason for defining test error as the mean {\em squared} error (and not an absolute error) is a customary choice in many measures of error, as we can decompose it into the squared bias: $(Mean(\hat{\Upsilon}) - \Upsilon)^2$ and the variance of the error $Var(\hat{\Upsilon} - \Upsilon)$. These values can be calculated for just one evaluated agent or for all evaluated agents.


If the bias is not zero this means that the mechanism to sample the exercises and/or the number of iterations is inappropriate, and the choices for $p_{\Pi_o}$ ($p_\Pi$), $p_M$, $p_S$ and $p_K$ in definition \ref{def:social_intelligence_test} must be revised. If there is a high variance, this suggests that the number of episodes $n_E$ is too small, and we need more (i.e., exercises in a tests) to get a less volatile result.

In this sense, note that some of the properties studied in previous subsections can hold for $\Upsilon$ but may be significantly different for unreliable tests (i.e., approximations) $\hat{\Upsilon}$.

The estimation of $TE(\hat{\Upsilon})$ or $Rel(\hat{\Upsilon})$ depends on knowing the true value of $\Upsilon$. This is not possible in practice for most environments, so $\Upsilon$ will need to be estimated for large samples and compared with an actual test (working with a small sample). Because of the difficulties of estimating this, in what follows we will just give a qualitative assessment.

\subsection{Efficiency}
\label{sec:efficiency}
This property refers to how efficient a test is in terms of the (computational) time required to get a reliable score. 
It is easy to see that efficiency and reliability are opposed. If we were able to perform an infinitely number of infinite episodes, then we would have $\hat{\Upsilon} = \Upsilon$, with perfect reliability, as we would exhaust $\Pi_o$ and $M$. However, as we try to make tests not only finite but more efficient, we lose reliability because of the sampling procedure. If done properly, it is usually the variance component of the reliability decomposition that is affected if it is possible to keep the bias close to 0 even with very low values of the number of episodes $n_E$ in definition \ref{def:social_intelligence_test}.

The definition of efficiency then needs to be defined as a ratio between the reliability and the time taken by the test (depending mostly on $n_E$ and $p_K$, but also on $p_{\Pi_o}$ ($p_\Pi$), $p_M$ and $p_S$).

\begin{definition}
Given a definition of a cognitive ability $\Upsilon$ and a test $\hat{\Upsilon}$, the efficiency is given by:

\begin{equation}
Eff(\hat{\Upsilon}) \triangleq  Rel(\hat{\Upsilon}) / Time(\hat{\Upsilon}) 
\end{equation}

\noindent where $Time$ is the average time taken by test $\hat{\Upsilon}$. Time can be measured as physical (real) time or as computational time (steps).
\end{definition}

While this is the way it should be measured, the big issue is how to choose environments and agents such that a high efficiency is attained. Clearly, if the selected environments are insensitive to agents actions or require too many actions to affect rewards, then this will negatively affect efficiency. As we are interested in social abilities, interactivity and non-neutralism must be high, as otherwise most steps will be useless to get information about the evaluated agent. This of course includes cases where the evaluated agent is stuck or bored because their opponents (or teammates) are too good or too bad, or the environment leads the evaluated agent to heaven or hell situations where actions are almost irrelevant, 

Naturally, a way of making tests more efficient is by the use of adaptive tests, as in computerised adaptive testing. We will not explore this possibility in this paper as our definition of test in section \ref{sec:tests} is not adaptive (for adaptive versions of universal tests, the reader is referred to \cite{AIJ2010,upsychometrics2}).

Similarly to reliability, we will just give a qualitative assessment of efficiency.

\subsection{Summary of properties}
\label{sec:summary_properties}
In tables \ref{table:summary_quantitative_properties} and \ref{table:summary_qualitative_properties} we can see a summary of all previous properties. Table \ref{table:summary_quantitative_properties} shows the {\em quantitative} properties, while table \ref{table:summary_qualitative_properties} shows the {\em qualitative}\footnote{Some of them can in principle be quantified, but we only give a qualitative assessment in this paper.} properties. This completes our picture jointly with figure \ref{fig:properties_taxonomy}.

\begin{table}
\centering

\begin{tabular}{|p{5.8cm}|p{3cm}|c|p{3cm}|p{3cm}|}
\hline
Property & Meaning & Range & Lowest value & Highest value\\
\hline
$AD(\Pi_e,w_{\Pi_e},\Pi_o,w_{\dot{L}},\mu,w_S)$ \mbox{\small Social:Social dependency:Interactivity} & The {\bf action dependency} of the evaluated agents on the line-up pattern they encounter. & $[0,1]$ & The evaluated agents do not take into account other agents' actions in their behaviour. & The evaluated agents behave completely depending on the other agents' actions.\\
\hline
$RD(\Pi_e,w_{\Pi_e},\Pi_o,w_{\dot{L}},\mu,w_S)$ \mbox{\small Social:Social dependency:Non-neutralism} & The {\bf reward dependency} of the evaluated agents on the line-up pattern they encounter. & $[0,1]$ & Each evaluated agent obtains the same expected average reward independently of the line-up pattern. & The agents in the line-up pattern directly exercise influence on the expected average rewards of each evaluated agent.\\
\hline
$SRD(\Pi_e,w_{\Pi_e},\Pi_o,w_{\dot{L}},\mu,w_S)$ \mbox{\small Social:Social dependency:Non-neutralism} & The {\bf slot reward dependency} measures how competitive or cooperative the environment is. & $[-1,1]$ & The environment is completely competitive. & The environment is completely cooperative.\\
\hline
$FD(\Pi_e,w_{\Pi_e},\Pi_o,w_{\dot{L}},\mu,w_S)$ \mbox{\small Instrumental:Secernment:Discrimination} & The {\bf fine discrimination} between pairs of evaluated agents when interacting with the same line-up patterns. & $[0,1]$ & Every evaluated agent obtains the same expected average reward for each line-up pattern. & Every evaluated agent obtains different expected average rewards for each line-up pattern.\\
\hline
$CD(\Pi_e,w_{\Pi_e},\Pi_o,w_{\dot{L}},\mu,w_S)$ \mbox{\small Instrumental:Secernment:Discrimination} & The {\bf coarse discrimination} between pairs of evaluated agents. & $[0,1]$ & Every evaluated agent obtains the same (social) intelligence value. & Every evaluated agent obtains different (social) intelligence values.\\
\hline
$STG(\Pi_e,w_{\Pi_e},\Pi_o,w_{\dot{L}},\mu,w_S)$ \mbox{\small Instrumental:Secernment:Grading} & The {\bf strict total grading} measures the level of strict grading ($<$) between the evaluated agents. & $[0,1]$ & There is no strict total order between the evaluated agents. & There is a strict total order between all the evaluated agents.\\
\hline
$PG(\Pi_e,w_{\Pi_e},\Pi_o,w_{\dot{L}},\mu,w_S)$ \mbox{\small Instrumental:Secernment:Grading} & The {\bf partial grading} measures the level of partial grading ($\leq$) between the evaluated agents. & $[0,1]$ & There is no partial order between the evaluated agents. & There is a partial order between all the evaluated agents.\\
\hline
$AComp(\Pi_e,w_{\Pi_e},\Pi_o,w_{\dot{L}},\mu,w_S)$ \mbox{\small Social:Mind modelling:Competitive} & The benefit of {\bf anticipating competitive} agents. & $[-1,1]$ & Every evaluated agent completely fails at anticipating competitive agents' behaviour. & Every evaluated agent perfectly anticipates competitive agents' behaviour.\\
\hline
$ACoop(\Pi_e,w_{\Pi_e},\Pi_o,w_{\dot{L}},\mu,w_S)$ \mbox{\small Social:Mind modelling:Cooperative} & The benefit of {\bf anticipating cooperative} agents. & $[-1,1]$ & Every evaluated agent completely fails at anticipating cooperative agents' behaviour. & Every evaluated agent perfectly anticipates cooperative agents' behaviour.\\
\hline
\end{tabular}

\caption{Summary of the quantitative properties about a social intelligence testbed $\mu$ with slot probability $w_S$, evaluated agent set $\Pi_e$ with weights $w_{\Pi_e}$, class of opponents and team players $\Pi_o$ and line-up probability $w_{\dot{L}}$. For each property the table shows its arguments, a brief description, the range of values and a description of the situations when their lowest and highest values occur.}
\label{table:summary_quantitative_properties}
\end{table}

\begin{table}
\centering

\begin{tabular}{|p{4cm}|p{10cm}|}
\hline
Property & Meaning\\
\hline
{\bf Boundedness} \mbox{\small Instrumental:Technical} & Rewards are bounded, so all quantitative properties are also bounded.\\
\hline
{\bf Symmetry} \mbox{\small Instrumental:Technical} & Desirable condition to simplify the measurement in such a way that only one slot has to be used to evaluate an agent.\\
\hline
{\bf Validity} \mbox{\small Univocal:Correctness} & The test evaluates what it is supposed to evaluate and nothing more.\\
\hline
{\bf Reliability} \mbox{\small Instrumental:Testing quality} & The result of an agent's ability in the test is close to its real value.\\
\hline
{\bf Efficiency} \mbox{\small Instrumental:Testing quality} & Decreases with the amount of (computational) time used in the test to obtain a reliable score.\\
\hline
\end{tabular}

\caption{Summary of the qualitative properties (or for which we will give a qualitative assessment) about a social intelligence testbed, providing a brief description of what each property represents.}
\label{table:summary_qualitative_properties}
\end{table}

With these properties we managed to represent how appropriate an environment $\mu$ and the set of agents $\Pi_o$ we use are in order to evaluate the social intelligence of a given set of evaluated agents $\Pi_e$. The set of properties we propose provides key information about the testbed we are analysing. First, we can measure the influence that a set of agents $\Pi_o$ produces in a set of evaluated agents $\Pi_e$. Second, we can analyse to what extent the anticipation abilities are useful for a set of evaluated agents $\Pi_e$ interacting with a set of agents $\Pi_o$. Third, we can determine whether cooperation or competition is given more importance in the testbed. Fourth, we estimate the discriminative power that the testbed has for the evaluation of different agents. Fifth, the grading power of the testbed indicates how effective it is to rank agents. And sixth, we have some instrumental properties that are convenient to convert the definition into a practical test.

We think that these properties can be useful to characterise a testbed.

For the quantitative properties we find two kinds of properties. The properties whose values range from $0$ to $1$ determine the percentage of fulfilment that the environment $\mu$ and the set of agents $\Pi_o$ have about this property when evaluating a set of agents $\Pi_e$. Therefore, the lower the value the worse the system is in regard to this property, and the higher the value the better. On the contrary, the kind of properties whose values range from $-1$ to $1$ must not be interpreted in the same way. Instead, these properties measure to which kind of type the system is more focussed on, and not a level of accomplishment or quality.

\section{Degree of compliance of several multi-agent and social scenarios}
\label{sec:current_environments}
Many games and environments have been proposed as testbeds to evaluate performance in a multi-agent environment \cite{Pell96,Weyns-etal05,shoham2008multiagent,Zatuchna-Bagnall09}. Typically these games and environments are created or selected to represent a specific problem or family to analyse or solve. 
Since we are interested in developing social intelligence tests, it is first mandatory to evaluate whether these other previous testbeds could be valid as they are (or with minor modifications). If not, they can still be a good source of inspiration to figure out new environment classes by reusing some of their ideas or hybridising some of their features.

We would have liked to explore many games and environments, but we can just practically do a selection of some of the most common and representative in the area of multi-agent systems, game theory and (social) computer games. We will focus on some testbeds whose specification is complete, so we can analyse the level of compliance of these testbeds for the properties seen in the previous section. In particular, the testbeds that we are going to analyse in this section are: matching pennies, prisoner's dilemma, predator-prey (a pursuit game), Pac-Man and RoboCup Soccer.



\subsection{Graphical analysis for the properties}
Before starting with the games and environments testbeds, we are going to introduce some indicators and a graphical representation that will be illustrated on a figurative environment. In table \ref{table:indicators_summary} we show a summary with the most important elements we are using in this section.

\begin{table}[!ht]
\centering

\begin{tabular}{|p{3cm}|p{2.5cm}|p{5cm}|p{4cm}|}
\hline
Terminology & Notation & Explanation & Example\\
\hline
Social environment (or game) & $\mu$ & The rules of the game. & Chess.\\
\hline
Social setup & $\left\langle \Pi_o,w_{\dot{L}},\mu,w_S \right\rangle$ &  This is more specific than just a game or environment and determines all the players and positions and their alignment probabilities $w_{\dot{L}}$ and slots. & Playing chess against the ten 
best players in Spain with uniform probability, starting with white pieces.\\
\hline
Evaluated agents and distribution &  $\left\langle \Pi_e, w_{\Pi_e} \right\rangle$ & The set of agents to be evaluated and an associated distribution. & Evaluate the students in a chess club with uniform distribution.\\
\hline
\end{tabular}

\caption{Summary of the elements used for the evaluation of agents in a multi-agent environment.}
\label{table:indicators_summary}
\end{table}

In order to assess compliance with interactivity, non-neutralism, anticipation and other properties for an environment $\mu$ we need to specify the evaluated agent class $\Pi_e$ with associated weight $w_{\Pi_e}$, the agent class $\Pi_o$ which populates the environment, line-up pattern weights $w_{\dot{L}}$ and slot weights $w_S$. One choice for $\Pi_e$, $w_{\Pi_e}$ and $\Pi_o$ would be to consider any possible agent that is expressible using a given policy language. This, however, would make the calculation of most properties difficult (if not impossible). A better approach would be to use a (representative) sample of all agents or a sample of a meaningful class. Instead of that, and in order to give a more general picture of the environment itself, we will show the range of values that each property can have (independently of the agents), and how much this range can be restricted (for better or worse) depending on which $\Pi_e, w_{\Pi_e}$ and $\Pi_o$ we select. In fact, when evaluating a set of agents $\Pi_e$ in a certain setup, we should provide which set of agents $\Pi_o$ will populate the environment, but we could also let this set unfixed in order to measure the environment itself. Finally, the use of different weights can lead to different ranges for the properties, but in what follows we will assume uniform weights for line-up patterns $w_{\dot{L}}$ and slots $w_S$.

We divided the properties into three types. In the first type we have the properties which have a quantitative value that can range between $0$ and $1$. In the second type we have the properties which have a quantitative value that can range between $-1$ and $1$. And in the third type are the properties for which we provide a qualitative value.

For the first two types of properties we calculate the range that each property can have in an environment. For this, we need to calculate the lowest and highest values that this range can have for each quantitative property $Prop$. To achieve this, we will select $\Pi_e,w_{\Pi_e}$ and $\Pi_o$ (from the set of all possible $\Pi_e,w_{\Pi_e}$ and $\Pi_o$ such that $Prop$ is defined) that obtain the lowest and highest values respectively. We define $General$ as follows:

\begin{definition}
\label{def:general_range}
We denote $General$ to be the range of values from $General_{\min}(Prop,\mu)$ to $General_{\max}(Prop,\mu)$, where:

\begin{align}
General_{\min}(Prop,\mu) & \triangleq \min_{\Pi_e,w_{\Pi_e},\Pi_o} Prop(\Pi_e,w_{\Pi_e},\Pi_o,w_{\dot{L}},\mu,w_S)\label{eq:general_min}\\
General_{\max}(Prop,\mu) & \triangleq \max_{\Pi_e,w_{\Pi_e},\Pi_o} Prop(\Pi_e,w_{\Pi_e},\Pi_o,w_{\dot{L}},\mu,w_S)\label{eq:general_max}
\end{align}

\noindent where the weight for slots $w_S$ and weight for line-up patterns $w_{\dot{L}}$ are uniform weights and the triplet $\left\langle\Pi_e,w_{\Pi_e},\Pi_o\right\rangle$ is selected (from the set of all possible $\left\langle\Pi_e,w_{\Pi_e},\Pi_o\right\rangle$ such that $Prop$ is defined) to minimise/maximise the values of a quantitative property $Prop$ for environment $\mu$.
\end{definition}

For the first type of properties $Prop$ we can select some set of agents $\Pi_o$ to obtain a situation where $General$ is restricted in such a way that $General_{\max}(Prop,\mu)$ decreases. In particular, we are interested in the setup with the ``lowest maximum'', i.e., we consider those $\Pi_o$ that minimise this maximum. We define $Left$ as follows:

\begin{definition}
\label{def:left_range}
We denote $Left$ to be the most restricted range of values from $Left_{\min}(Prop,\mu)$ to $Left_{\max}(Prop,\mu)$ that we can obtain when $\Pi_o$ is selected (from the set of all possible $\Pi_o$ such that $Prop$ is defined) to decrease the values of a quantitative property $Prop$ which range is between $0$ and $1$ for environment $\mu$, where:

\begin{align}
Left_{\min}(Prop,\mu) & \triangleq General_{\min}(Prop,\mu)\label{eq:left_min}\\
Left_{\max}(Prop,\mu) & \triangleq \min_{\Pi_o} \max_{\Pi_e,w_{\Pi_e}} Prop(\Pi_e,w_{\Pi_e},\Pi_o,w_{\dot{L}},\mu,w_S)\label{eq:left_max}
\end{align}
\end{definition}

In the same way, for the first type of properties $Prop$ we can select some set of agents $\Pi_o$ to obtain a situation where $General$ is restricted in such a way that $General_{\min}(Prop,\mu)$ increases. In particular, we are interested in the setup with the ``highest minimum'', i.e., we consider those $\Pi_o$ that maximise this minimum. We define $Right$ as follows:

\begin{definition}
\label{def:right_range}
We denote $Right$ to be the most restricted range of values from $Right_{\min}(Prop,\mu)$ to $Right_{\max}(Prop,\mu)$ that we can obtain when $\Pi_o$ is selected (from the set of all possible $\Pi_o$ such that $Prop$ is defined) to increase the values of a quantitative property $Prop$ which range is between $0$ and $1$ for environment $\mu$, where:

\begin{align}
Right_{\min}(Prop,\mu) & \triangleq \max_{\Pi_o} \min_{\Pi_e,w_{\Pi_e}} Prop(\Pi_e,w_{\Pi_e},\Pi_o,w_{\dot{L}},\mu,w_S)\label{eq:right_min}\\
Right_{\max}(Prop,\mu) & \triangleq General_{\max}(Prop,\mu)\label{eq:right_max}
\end{align}
\end{definition}

For the first type of properties, the $General$, $Left$ and $Right$ ranges become better as long as their minimum and maximum values become higher. If the $Left$ range values are lower, this would mean that a bad selection of $\Pi_o$ is disastrous for the quality of the testbed. If $Right$ range values are higher would mean that there is a good selection of $\Pi_o$ which improves the quality of the testbed. The comparison between $Left$ and $Right$ with $General$ shows us the importance that a good selection for the set of opponents and team players $\Pi_o$ has for a property $Prop$ in an environment $\mu$. As these three ranges are more different, the selection of agents $\Pi_o$ becomes more important in order to provide a better quality for the testbed.

\begin{figure}[!ht]
\centering

\includegraphics[width=0.85\textwidth]{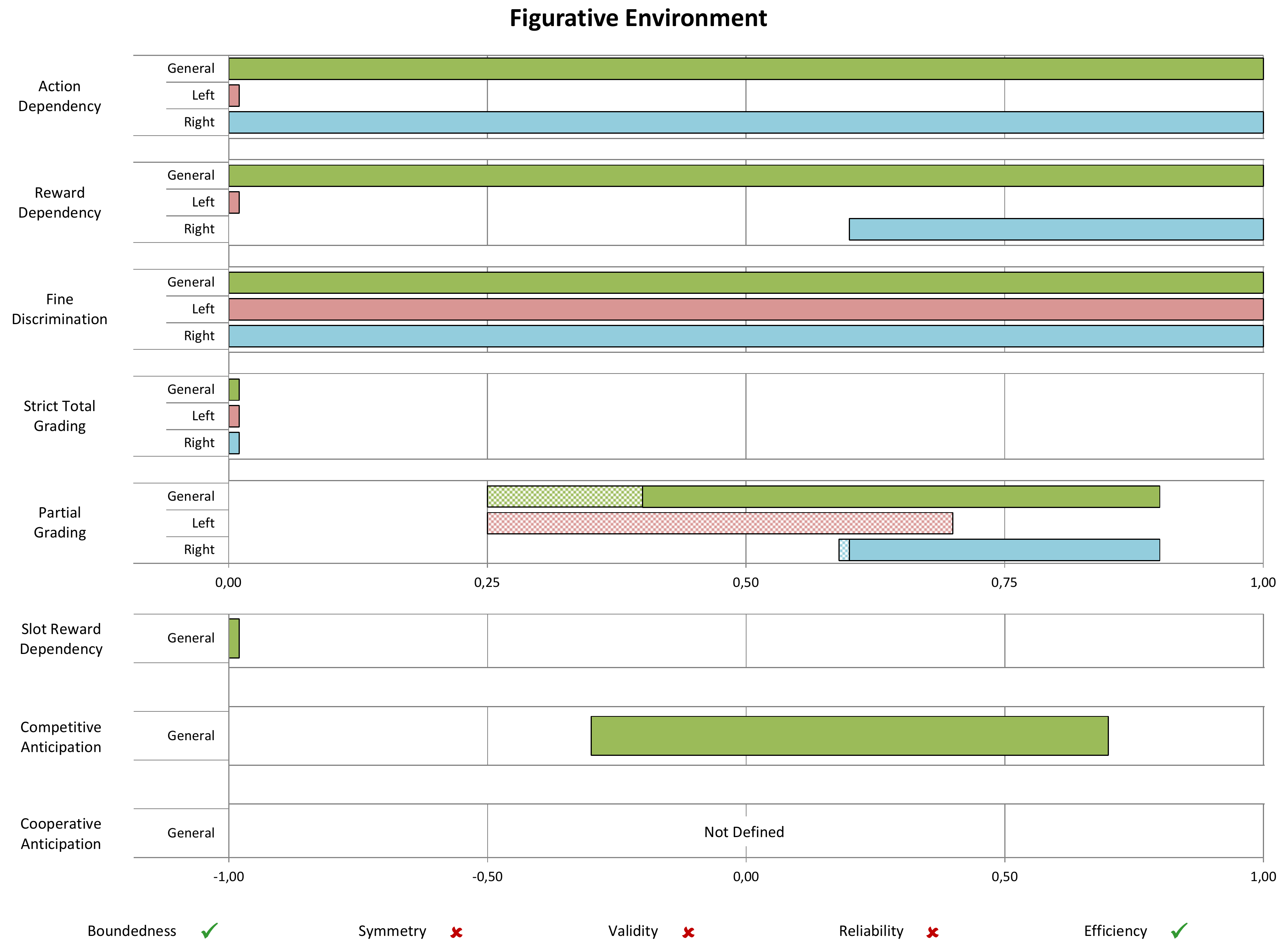}

\caption{Social properties for a figurative environment. Lighter bands mean the values are not formally calculated, but an estimation is given.}
\label{fig:figurative_environment}
\end{figure}

In figure \ref{fig:figurative_environment} we present the properties of a figurative environment divided in three sections. The top section represents five quantitative properties whose range can be between $0$ and $1$\footnote{Since FD and CD are similar properties, we decided to just calculate the FD property in order to simplify our analysis.}. The middle section represents the three quantitative properties whose range can be between $-1$ and $1$. Finally, the bottom section represents the five qualitative properties.

We can see that each property of the first type has the early mentioned $General$, $Left$ and $Right$ ranges represented with three bands. The first property (Action Dependency) has a $General$ range from $0$ to $1$, represented with the first band. This is the broadest range that this kind of property can have. This means that this environment can have any value for this property depending on the set of agents $\Pi_o$, the set of evaluated agents $\Pi_e$ and its weight distribution $w_{\Pi_e}$. The second band represents its $Left$ range, which is equal to $[0,0]$. In this case, there exists a set of agents $\Pi_o$ that restricts this range to the minimum possible range. The third band represents its $Right$ range, which remains from $0$ to $1$. Now, no set of agents $\Pi_o$ can be selected to restrict this range. In the next four properties we see some other examples for the three ranges that this type of property can have. As we can see in the last property of this type (PG), we use a lighter color to represent that (part of) a range is not formally calculated, but instead we provide an estimation.

Next we arrive at the second type of properties. Here the values for the $General$ range can be between $-1$ and $1$. Unlike the previous case, we do not provide $Left$ and $Right$ ranges for this type of properties. This is because these properties represent for which kind of social intelligence the environment is more oriented, so there are not really good or bad ranges for this property.

In the first property of this second type (Slot Reward Dependency), we see that the $General$ range for this property is equal to $[-1,-1]$, indicating that this environment is purely competitive. The next property shows another example for this type of properties.
Meanwhile, the last property has a label with the text ``Not Defined''. This is because for this environment the property is not defined, so we cannot represent it.

Finally, we arrived at the last type of properties. Here, we denote whether the environment meets the properties by using a tick ($\checkmark$) or cross ($\times$) mark respectively.

Now that we have explained how we will represent the properties, let us start analysing some true environments.

\subsection{Matching pennies}
\label{sec:matching_pennies}
Matching pennies \cite{weibull1995evolutionary} can be considered the simplest game in game theory featuring competition. This game consists of two players (or agents) each flipping a coin. If both coins match player 1 wins, otherwise player 2 wins.

This game is played as a repeated game, which means that the game is played on a single iteration and the game is repeated for several iterations. Each player can see the actions performed by the other player. The game is usually repeated during $K$ steps (i.e., it is the iterated matching pennies), so players can use past steps in order to predict the other player's strategy. Following definition \ref{def:environment}, for agent slot $i$ this environment only allows two actions ${\cal A}_i = \{$Head, Tail$\}$ and only provides two rewards ${\cal R}_i = \{-1, 1\}$, which correspond to lose and win respectively. Clearly, in this game, $\tau = \{\{1\},\{2\}\}$ represents the partition of slots in teams, i.e., has two teams and only one slot in each. For agent in slot $i$ the environment provides an observation set ${\cal O}_i = {\cal A}_j \cup \{null\}$ (where agent slot $j$ represents the slot of the other agent) and the observation function $\omega$ returns to each agent the action performed by the other agent in the previous iteration or $null$ if it is the first iteration. Figure \ref{fig:matching_pennies_payoff} shows the reward function $\rho$ as a reward matrix, which has the actions of both agents as input and their rewards as outputs.

\begin{figure}[!ht]
\centering

\begin{tabular}[c]{|c|c|c|}
\hline
		& Head		& Tail\\
\hline
Head	& ($1, -1$)	& ($-1, 1$)\\
\hline
Tail	& ($-1, 1$)	& ($1, -1$)\\
\hline
\end{tabular}

\caption{Matching pennies' payoff matrix. Rows and columns represent the actions of agent 1 and 2 respectively. Cells content $(X, Y)$ corresponds to the obtained rewards for agent 1 and agent 2 respectively when the actions leading to that cell are performed.}
\label{fig:matching_pennies_payoff}
\end{figure}

Next we discuss the level of compliance of the matching pennies environment with respect to the properties seen in section \ref{sec:social_properties}. We can see a summary of the social properties for the matching pennies in figure \ref{fig:matching_pennies_properties}. In appendix \ref{appen:matching_pennies_properties} we prove how we obtained these values.

\begin{figure}[!ht]
\centering

\includegraphics[width=0.85\textwidth]{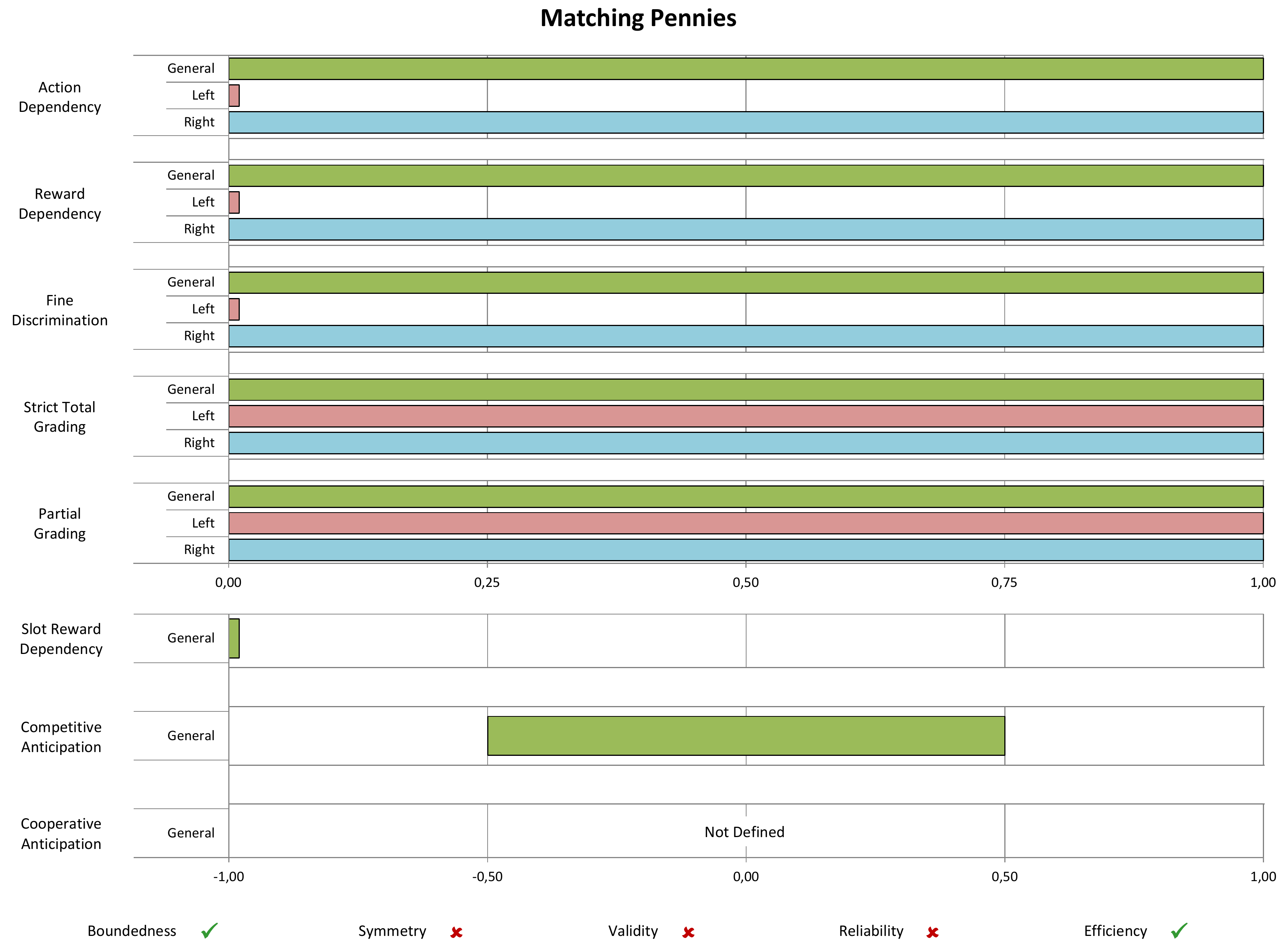}

\caption{Social properties for the matching pennies using uniform weights for $w_S$ and $w_{\dot{L}}$, and $\Delta_S(a,b)$ and $\Delta_{\mathbb{Q}}(a,b)$ return $1$ if $a$ and $b$ are equal and $0$ otherwise.}
\label{fig:matching_pennies_properties}
\end{figure}

If we start with the properties, we see that this game is {\em bounded}, as rewards are always between $-1$ and $1$. This means that if the weight functions $w_S$ and $w_{\dot{L}}$ are bounded, the value of $\Upsilon$ (and many other properties) will be bounded. Also, matching pennies is a well known zero-sum game, which means that the payoffs of the agents always sum zero (as we can see in figure \ref{fig:matching_pennies_payoff}) and, therefore, agents have totally opposed interests.

We next move to the symmetry property. This game has only two teams with one agent on each team. So, in order to prove that this environment is not symmetric, we only need to find a pair of line-ups $l_1 = (\pi_1, \pi_2)$ and $l_2 = (\pi_2, \pi_1)$ where the sequence of rewards for $\pi_1$ and/or $\pi_2$ differs in both line-ups. This becomes trivial by using the same agent $\pi_t$ (which always performs Tail) as both $\pi_1$ and $\pi_2$ (i.e., $l_1 = l_2 = (\pi_t,\pi_t)$) and check whether the agent obtains the same result in both slots. Since $\pi_t$ gets a result of $1$ in slot $1$ and a result of $-1$ in slot $2$, we can conclude that this environment is {\em not symmetric}. This forces us to calculate some other properties for all the slots.

If we look to the $General$ range for the action dependency (AD) property, we see that it goes from $0$ to $1$ (propositions \ref{prop:matching_pennies_AD_general_min} and \ref{prop:matching_pennies_AD_general_max}). That means that the evaluated agents can either interact without noticing the other agent or can perform actions depending on which agent they encounter. But some particular selection of $\Pi_o$ could make this environment to have a too restrictive $Left$ range with respect to this property, making it equal to $[0,0]$ (proposition \ref{prop:matching_pennies_AD_left_max}), so no evaluated agent could perform different actions depending on which agent it interacts with. In addition, we see that no particular selection of $\Pi_o$ can restrict the $Right$ range, which remains from 0 to 1 (proposition \ref{prop:matching_pennies_AD_right_min}).

When we look at the $General$ range for the reward dependency (RD) property, we see that it goes from $0$ to $1$ (propositions \ref{prop:matching_pennies_RD_general_min} and \ref{prop:matching_pennies_RD_general_max}). This means that the evaluated agents can either obtain the same expected average reward or can obtain different expected average rewards depending on which agent they encounter. But some particular selection of $\Pi_o$ could make this environment have a too restrictive $Left$ range with respect to this property, making it equal to $[0,0]$ (proposition \ref{prop:matching_pennies_RD_left_max}). In addition, we see that no particular selection of $\Pi_o$ can restrict the $Right$ range, which remains from $0$ to $1$ (proposition \ref{prop:matching_pennies_RD_right_min}).

The $General$ range for the fine discrimination (FD) property goes from $0$ to $1$ (propositions \ref{prop:matching_pennies_FD_general_min} and \ref{prop:matching_pennies_FD_general_max}). This means that two different evaluated agents can either obtain the same expected average reward or can obtain different expected average rewards. The $Left$ range can be restricted to be equal to $[0,0]$ (proposition \ref{prop:matching_pennies_FD_left_max}), and the $Right$ range goes from $0$ to $1$ (proposition \ref{prop:matching_pennies_FD_right_min}). This means that, with a bad selection of $\Pi_o$, no pair of evaluated agents can be differentiated in terms of performance, and it does not exist a $\Pi_o$ to always differentiate any pair of evaluated agents.

The $General$ range for the strict total grading (STG) and partial grading (PG) properties has, as in all previous properties, a minimum value of $0$ (propositions \ref{prop:matching_pennies_STG_general_min} and \ref{prop:matching_pennies_PG_general_min}) and a maximum value of $1$ (propositions \ref{prop:matching_pennies_STG_general_max} and \ref{prop:matching_pennies_PG_general_max}). This means that we cannot provide an ordering for some sets of evaluated agents, but we can provide it for some other sets of evaluated agents. In both STG and PG, $Left$ cannot be restricted by any $\Pi_o$, remaining from $0$ to $1$ (propositions \ref{prop:matching_pennies_STG_left_max} and \ref{prop:matching_pennies_PG_left_max}), and the same occurs with $Right$, which cannot be restricted by any $\Pi_o$, remaining from $0$ to $1$ (propositions \ref{prop:matching_pennies_STG_right_min} and \ref{prop:matching_pennies_PG_right_min}).

This environment always has a $General$ range for the slot reward dependency (SRD) property equal to $[-1,-1]$ (proposition \ref{prop:matching_pennies_SRD_general_range}). This means that both agent slots have opposed interests, i.e., it is entirely competitive.

The $General$ range for the competitive anticipation (AComp) property goes from $-\frac{1}{2}$ to $\frac{1}{2}$ (propositions \ref{prop:matching_pennies_AComp_general_min} and \ref{prop:matching_pennies_AComp_general_max}). These values tell us that an evaluated agent can improve its results by correctly anticipating the actions of the other agent, but an incorrect anticipation will worsen its results.

On the contrary, we cannot evaluate the cooperative anticipation (ACoop) property in this environment. This is because each of the two teams has only one slot (in addition, this is also a zero-sum game), so the formula cannot be applied.

We now discuss the validity property. Matching pennies is a very simple game. As a result, it seems clear that it is not general enough to be used (alone) as the basis of a social intelligent test. Nonetheless, there are different opinions about this, as it has been suggested that matching pennies {\em could} be an intelligence test on its own, under the name `Adversarial Sequence Prediction' \cite{hibbard2008adversarial,hibbard2011measuring}. In fact, a tournament was organised in 2011 where computer algorithms competed\footnote{See \url{http://matchingpennies.com/tournament/}.} and very interesting emergence phenomena were observed. 
Only strategies that were able to see patterns in the other players scored well (better than random). There are of course some counter-intuitive things about this game. Actually, random agents score exactly the same with any opponent, even with very intelligent ones. This raises concerns about the validity and reliability of this environment for a test since, in an intelligence test, the average score of a random policy should not obtain a good result, since random agents are clearly not intelligent. Also, matching pennies results are non-monotonic for a set of agents. In \cite{Turing100} there is an example of an agent set for matching pennies that is non-monotonic (so $PG < 1$). Nonetheless, partial orders can still be constructed for the agent set of all finite state machines \cite{hibbard2011measuring}. 
Another important problem about matching pennies is that it only evaluates pure competition (it is a zero-sum game), and no form of cooperation can be found (although some versions, or the ternary extension, rock-paper-scissors, could allow for cooperation).
Finally, another strong argument against the validity of this game as a good environment (alone) for a social intelligence test is that some current systems may score better than humans, even though these systems are not (socially) intelligent at all and they are designed to play matching pennies only.

Finally, efficiency is a property where matching pennies excels, since every action has immediate consequences. This means that in most cases, it can be enough to design a test where only a few dozen steps are performed in order to have a good approximation. Of course, there might be agents that may change their behaviour in step 10,000 so their assessment up to this point will be completely different to their assessment in the limit (if no discounting factor is used).
For some agents, with our definition of reliability, we get that some pairs of agents are stable from step 2, so high degrees of reliability (low test error) can be achieved very soon. When a random agent is involved, the approximation is slower, but not much slower, depending on a binomial distribution.

Overall, matching pennies is an interesting game, but it lacks the generality that a social intelligence test should have. Nonetheless, it is a very simple game that illustrates how the range values of several properties can be calculated and provide very useful information about how an environment or game behaves. In the end, it has been a useful exercise before analysing more sophisticated scenarios below.

\subsection{Prisoner's dilemma}
\label{sec:prisoner_dilemma}
In the prisoner's dilemma \cite{poundstone1993prisoner} two prisoners (or agents) are suspects of a crime, and are asked if the other prisoner is guilty of that crime. If both cooperate and do not blame the other, both spend a short time in prison. If one cooperates but not the other, the one who blamed reduces its time in prison to the minimum sentence, but the other prisoner obtains the maximum sentence. Finally, if both prisoners blame the other, both spend a long time in prison.

As happens with the matching pennies, this game is played as a repeated game, which means that the game is played on a single iteration and the game is repeated for several iterations. Each player can see the actions performed by the other player. The game is usually repeated during $K$ steps (i.e., it is the iterated prisoner's dilemma), so players can use past steps in order to predict the other player's strategy. Following definition \ref{def:environment}, for agent slot $i$ this environment only allows two actions ${\cal A}_i = \{$Cooperate, Blame$\}$ and provides four rewards ${\cal R}_i = \{1, 2, 3, 4\}$, which correspond to the time spent in prison. In this game, $\tau = \{\{1\},\{2\}\}$ represents the partition of slots in teams, which has two teams and only one slot in each. For agent in slot $i$ the environment provides an observation set ${\cal O}_i = {\cal A}_j \cup \{null\}$ (where agent slot $j$ represents the slot of the other agent) and the observation function $\omega$ returns to each agent the action performed by the other agent in the previous iteration or $null$ if it is the first iteration. Figure \ref{fig:prisoner_dilemma_payoff} shows the reward function $\rho$ as a reward matrix, which has the actions of both agents as input and their rewards as outputs.

\begin{figure}[!ht]
\centering

\begin{tabular}[c]{|c|c|c|}
\hline
			& Cooperate	& Blame\\
\hline
Cooperate	& $(2, 2)$	& $(4, 1)$\\
\hline
Blame		& $(1, 4)$	& $(3, 3)$\\
\hline
\end{tabular}

\caption{Prisoner's dilemma's payoff matrix. Rows and columns represent the actions of agent 1 and 2 respectively.}
\label{fig:prisoner_dilemma_payoff}
\end{figure}

The payoff matrix in figure \ref{fig:prisoner_dilemma_payoff} is not normalised. We can normalise this matrix to be between $-1$ and $1$, giving the lowest reward to the highest period in prison and vice versa. Once rewards are normalised, for agent slot $i$ they are set to ${\cal R}_i = \{-1, -0.33, 0.33, 1\}$. In figure \ref{fig:prisoner_dilemma_payoff_normalised} we can see this payoff matrix normalised.

\begin{figure}[!ht]
\centering

\begin{tabular}[c]{|c|c|c|}
\hline
			& Cooperate			& Blame\\
\hline
Cooperate	& $(0.33, 0.33)$	& $(-1, 1)$\\
\hline
Blame		& $(1, -1)$			& $(-0.33, -0.33)$\\
\hline
\end{tabular}

\caption{Prisoner's dilemma's payoff matrix normalised. Maximum sentence in prison is represented by the lowest reward, while minimum sentence is represented with the highest reward.}
\label{fig:prisoner_dilemma_payoff_normalised}
\end{figure}

Prisoner's dilemma differs from matching pennies by including some cooperation, since both agents can choose to cooperate to spend a relatively little time in prison.

We can see a summary of the social properties for the prisoner's dilemma in figure \ref{fig:prisoner_dilemma_properties}. In appendix \ref{appen:prisoner_dilemma_properties} we prove how we obtained these values.

\begin{figure}[!ht]
\centering

\includegraphics[width=0.85\textwidth]{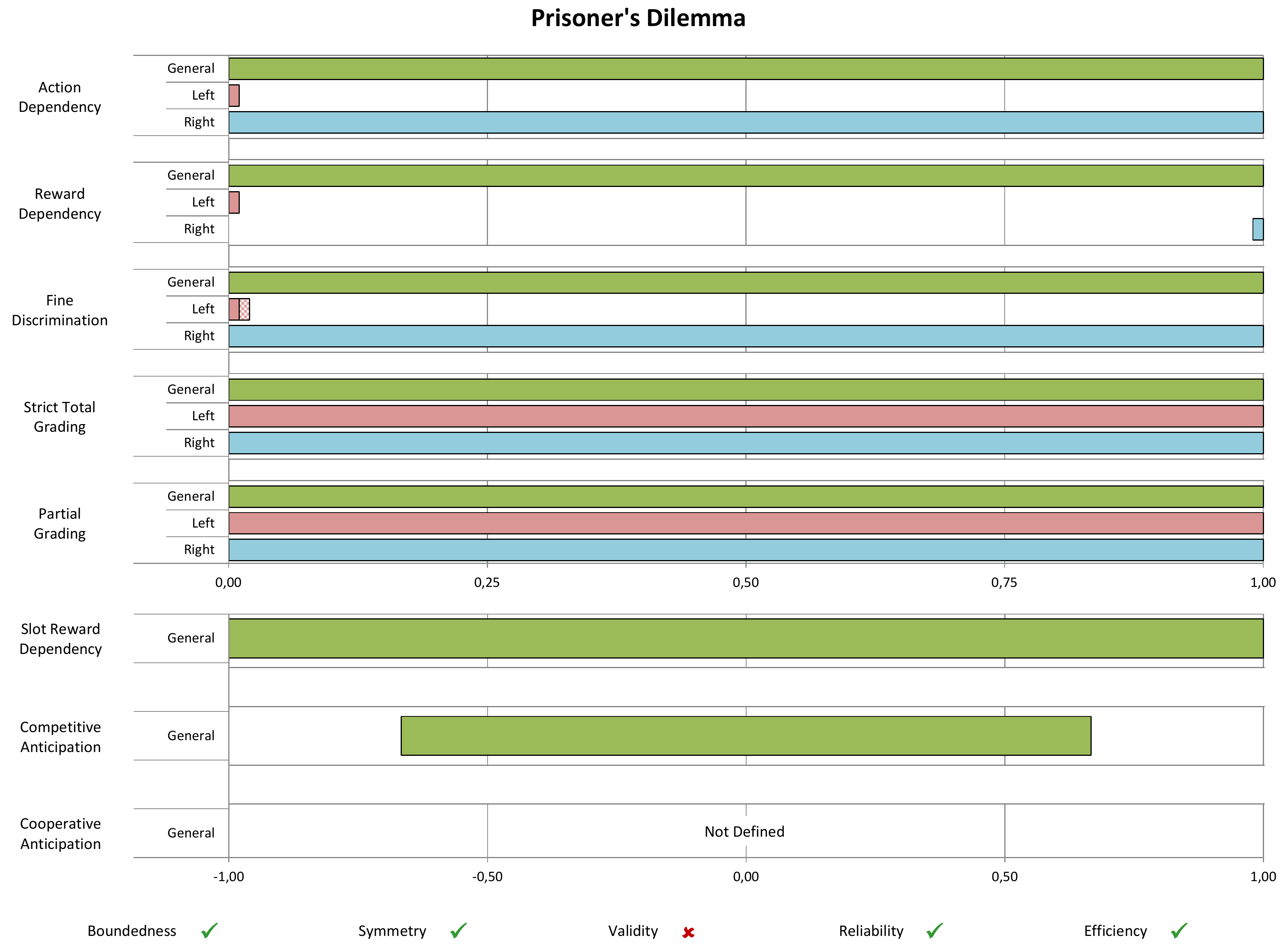}

\caption{Social properties for the prisoner's dilemma using uniform weights for $w_S$ and $w_{\dot{L}}$, and $\Delta_S(a,b)$ and $\Delta_{\mathbb{Q}}(a,b)$ return $1$ if $a$ and $b$ are equal and $0$ otherwise. Lighter bands mean the values are not formally calculated, but an estimation is given.}
\label{fig:prisoner_dilemma_properties}
\end{figure}

With the normalised payoff matrix we can see that this game is {\em bounded}, since $\frac{1}{K \cdot n}\sum^K_{k=1} \sum^n_{i=1} r_{i,k} = c$ where $-0.33 \leq c \leq 0.33$. In this game some cooperation can appear, as for example, when both agents always decide to cooperate, so they obtain the maximum joint reward ($0.33$). But blaming can provide the best reward to one player if the other player still cooperates, so cooperating now provides the worst reward. Finally, if both decide to blame, both obtain the worst joint reward ($-0.33$). As in matching pennies, if the weight functions $w_S$ and $w_{\dot{L}}$ are bounded, the value of $\Upsilon$ (and many properties) will be bounded.

Analysing the symmetry property, we can see that the payoff matrix is clearly symmetric for both players. This makes that the payoffs of any strategies made by the agents do not depend on which slots they are, since they will obtain the same rewards. From this observation we can conclude that this environment is {\em symmetric}, which allows us to calculate some other properties only for one slot and assume that it is maintained for the other.

If we look at the ranges for the action dependency (AD) property, we encounter exactly the same scenario than in the matching pennies. That means that evaluated agents can either interact without noticing the other agent, obtaining a value of $0$ (proposition \ref{prop:prisoner_dilemma_AD_general_min}), or can always perform actions depending on the agent they encounter, obtaining a value of $1$ (proposition \ref{prop:prisoner_dilemma_AD_general_max}). But some particular selections of $\Pi_o$ can provide a too restricted $Left$ range, forcing this value to be equal to $[0,0]$ (proposition \ref{prop:prisoner_dilemma_AD_left_max}), so no evaluated agent from $\Pi_e$ can behave differently depending on the agents of $\Pi_o$. In addition, no particular selection of $\Pi_o$ can restrict the $Right$ range, remaining from $0$ to $1$ (proposition \ref{prop:prisoner_dilemma_AD_right_min}).

We start to find some differences with the matching pennies when we analyse the reward dependency (RD) property. As in matching pennies, the $General$ range for this property goes from 0 to 1 (propositions \ref{prop:prisoner_dilemma_RD_general_min} and \ref{prop:prisoner_dilemma_RD_general_max}), so the expected average rewards of the evaluated agents can either depend or not on which agent they encounter. Also, some particular selection of $\Pi_o$ could make this environment to have a too restrictive $Left$ range with respect to this property, making it equal to $[0,0]$ (proposition \ref{prop:prisoner_dilemma_RD_left_max}), so no evaluated agent obtains different expected average rewards depending on which agent it interacts with. But, a good selection of $\Pi_o$ can restrict the $Right$ range making it equal to $[1,1]$ (proposition \ref{prop:prisoner_dilemma_RD_right_min}). This means that the evaluated agents will obtain different expected average rewards depending on which agent they interact with.

We now move to the fine discrimination (FD) property. The $General$ range for this property goes from $0$ to $1$ (propositions \ref{prop:prisoner_dilemma_FD_general_min} and \ref{prop:prisoner_dilemma_FD_general_max}). This means that two evaluated agents can either obtain the same expected average reward or different expected average rewards depending on the set of agents $\Pi_o$ we select. It exists a particular $\Pi_o$ which restricts the $Left$ range to be equal to $[0,0]$ (conjecture \ref{conj:prisoner_dilemma_FD_left_max}), meaning that the environment will not be discriminating the evaluated agents, since every evaluated agent will obtain the same expected average reward. It is not possible to restrict the $Right$ range in such a way that we can always discriminate every pair of evaluated agents, so it remains from $0$ to $1$ (proposition \ref{prop:prisoner_dilemma_FD_right_min}), so it is possible to find two different evaluated agents from some particular $\Pi_e$ obtaining the same result.

The $General$ range for the strict total grading (STG) and partial grading (PG) properties can, as in all previous properties, reach a minimum value of 0 (propositions \ref{prop:prisoner_dilemma_STG_general_min} and \ref{prop:prisoner_dilemma_PG_general_min}) and a maximum value of 1 (propositions \ref{prop:prisoner_dilemma_STG_general_max} and \ref{prop:prisoner_dilemma_PG_general_max}) for this environment. This means that we cannot provide an ordering for some sets of evaluated agents, but we can provide it for some other sets of evaluated agents. In both STG and PG, the $Left$ range cannot be restricted by any $\Pi_o$, remaining from $0$ to $1$ (propositions \ref{prop:prisoner_dilemma_STG_left_max} and \ref{prop:prisoner_dilemma_PG_left_max}), and the same occurs with the $Right$ range, which cannot be restricted by any $\Pi_o$, remaining from $0$ to $1$ (propositions \ref{prop:prisoner_dilemma_STG_right_min} and \ref{prop:prisoner_dilemma_PG_right_min}).

The $General$ range for the slot reward dependency (SRD) property goes from $-1$ to $1$ (propositions \ref{prop:prisoner_dilemma_SRD_general_min} and \ref{prop:prisoner_dilemma_SRD_general_max}). This provides very different distributions of expected average rewards depending on the strategies used by the agents.

The $General$ range for the competitive anticipation (AComp) property goes from $-\frac{2}{3}$ to $\frac{2}{3}$ (propositions \ref{prop:prisoner_dilemma_AComp_general_min} and \ref{prop:prisoner_dilemma_AComp_general_max}). Anticipating the strategy of the other agent can be really useful to obtain a good expected average reward, but it can also provide a really bad expected average reward if the strategy is not correctly anticipated.

We cannot evaluate the cooperative anticipation (ACoop) property in this environment. This is because, as in the matching pennies environment, each of the two teams has only one slot, so the formula cannot be applied. This seems counter-intuitive since one of the actions is named to cooperate, but this cooperation is not meant to improve the agent's own rewards, but to improve the other agent's rewards. Indeed, if we reframe the game by using only one team and calculating the team reward as the mean of the agents' rewards, then both agents can cooperate to obtain the best joint reward. This will lead us to a situation where a bad cooperative anticipation between the agents' actions can negatively affect the team reward, but also, a good cooperative anticipation will have good benefits for the team.

We now discuss the validity property. The prisoner's dilemma is similar in simplicity to the matching pennies game. But in this game, competition is not so strong, providing some cooperation between the two teams and making this game more general than the matching pennies. But in this game we cannot evaluate cooperation within a team, so it is not general enough to evaluate social intelligence. Actually, some simple strategies can clearly make the adversary's results get stuck, forcing it to obtain bad rewards independently of its strategy. This raises concerns about the validity of this environment for a test, since an intelligence test should not give bad results to intelligent agents.

Finally, the efficiency property is almost as good as in the matching pennies environment. Every action has immediate consequences on the agents' results, having good approximations for their results in few steps. Therefore, reliable values for the ability measured will be reached in short time.

The prisoner's dilemma resembles the matching pennies in many aspects, but it is slightly a more complex environment. As many similarities exists between both environments many properties remain equal.

Let us see a more complex environment.

\subsection{Predator-prey (Pursuit game)}
\label{sec:predator-prey}
\begin{figure}[!ht]
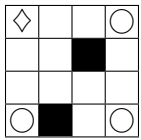

\newcolumntype{C}{>{\centering\arraybackslash}p{12px}}
\def \block {\cellcolor{black}}
\def \hunter {$\bigcirc$}
\def \prey {$\diamondsuit$}

\centering
\setlength{\tabcolsep}{0px}
\begin{tabular}[c]{|C|C|C|C|}
\hline
\prey   &         &         & \hunter \\
\hline
        &         & \block  &         \\
\hline
        &         &         &         \\
\hline
\hunter & \block  &         & \hunter \\
\hline
\end{tabular}

\caption{A predator-prey scenario with a 4x4 grid space. \hunter{} denotes a predator, \prey{} denotes the prey and a black cell denotes a block. At each step, agents can stay or move one cell horizontally or vertically, but blocks cannot be crossed. The prey is chased once it shares a cell with a predator.}
\label{fig:predator-prey}
\end{figure}

One typical environment for cooperation that uses a 2D discrete space is a pursuit game called Predator-prey \cite{benda1985optimal}, where the evaluee acts as a predator and has to cooperate/coordinate with other two predators in order to chase a prey. If they succeed chasing the prey, the goal is achieved. Figure \ref{fig:predator-prey} shows an example of a predator-prey environment.

%
%

Many variants have been proposed about this scenario, which provides a high diversity of environments. Some examples include spaces with and without obstacles or boundaries, and many variants about the parameters have been considered: the distance of the scenario that the agents can perceive, the number of predators or preys, the speed of the agents, etc. Even the definition of how the prey is chased has been modified, e.g., the prey is surrounded by the predators, or one predator chases the prey by occupying the same position. Some of these variants add a variety of social complexity to the game, such as different levels of cooperation/competition by having to interact with different numbers of predators or preys, or having faster preys.

These and other pursuit games have been widely studied and used in multi-agent systems (i.e., \cite{Tan93multi-agentreinforcement, denzinger1996experiments, denzinger2000evolutionary, frankl1987pursuit, ahmadabadi2002expertness}), but, in our opinion, no thorough study about their properties has been developed so far.

Since we cannot analyse all the variants we just select one of them to analyse it, but different variants could have different properties' values. For this analysis we will use the environment shown in figure \ref{fig:predator-prey}, which also shows its initial observation. The game is typically performed in episodes. We will make an episode to end after six iterations are performed.

Following definition \ref{def:environment}, for agent slot $i$ this environment allows four actions ${\cal A}_i = \{$Up, Right, Down, Left$\}$, which leads the agent to the cell facing this direction (when an agent performs an action leading to a block or boundary, the agent does not move). For agent slot $i$ the environment provides three rewards ${\cal R}_i = \{0,-6,6\}$, which correspond to `the episode is not finished', `lose the chase' and `win the chase' respectively. $\tau = \{\{1\},\{2,3,4\}\}$ represents the partition of slots in teams. The first team $\{1\}$ contains the prey (which is located in the upper left corner) and the second team $\{2,3,4\}$ contains three predators (which are located in the upper right, bottom left and bottom right corners respectively). For agent in slot $i$ the environment provides an observation set ${\cal O}_i$ which corresponds to the set of spaces with any possible location of the agents, and the observation function $\omega$ returns for every agent a description of the space as, for example, figure \ref{fig:predator-prey}. For the analysis of this environment we will follow the same procedure as for previous environments. This is the reason why we allow the evaluated agent to play in every slot, even as the prey. But, as mentioned above, if we only let the agent play as a predator, the values of the properties will be different. Figure \ref{fig:predator-prey_payoff} shows the reward function $\rho$ as a payoff matrix which has the current iteration and the chasing situation as input and the agents' rewards as output. Note that after the six iterations, the average reward will be $-1$ or $1$ depending on whether the agent wins the chase or not.

%

\begin{figure}[!ht]
\centering

\begin{tabular}[c]{|c|c|c|}
\hline
				& Prey has been chased	& Prey has not been chased\\
\hline
Iteration $1$	& $(0, 0)$				& $(0, 0)$\\
\hline
Iteration $2$	& $(0, 0)$				& $(0, 0)$\\
\hline
Iteration $3$	& $(0, 0)$				& $(0, 0)$\\
\hline
Iteration $4$	& $(0, 0)$				& $(0, 0)$\\
\hline
Iteration $5$	& $(0, 0)$				& $(0, 0)$\\
\hline
Iteration $6$	& $(-6, 6)$				& $(6, -6)$\\
\hline
\end{tabular}

\caption{Predator-prey's payoff matrix. Rows represent the iteration, while columns represents whether the prey has been chased or not. Cells content $(X, Y)$ corresponds to the obtained rewards for the prey and each predator respectively.}
\label{fig:predator-prey_payoff}
\end{figure}

We can see a summary of the social properties for the predator-prey in figure \ref{fig:predator-prey_properties}. In appendix \ref{appen:predator-prey_properties} we prove how we obtained these values.

\begin{figure}[!ht]
\centering

\includegraphics[width=0.85\textwidth]{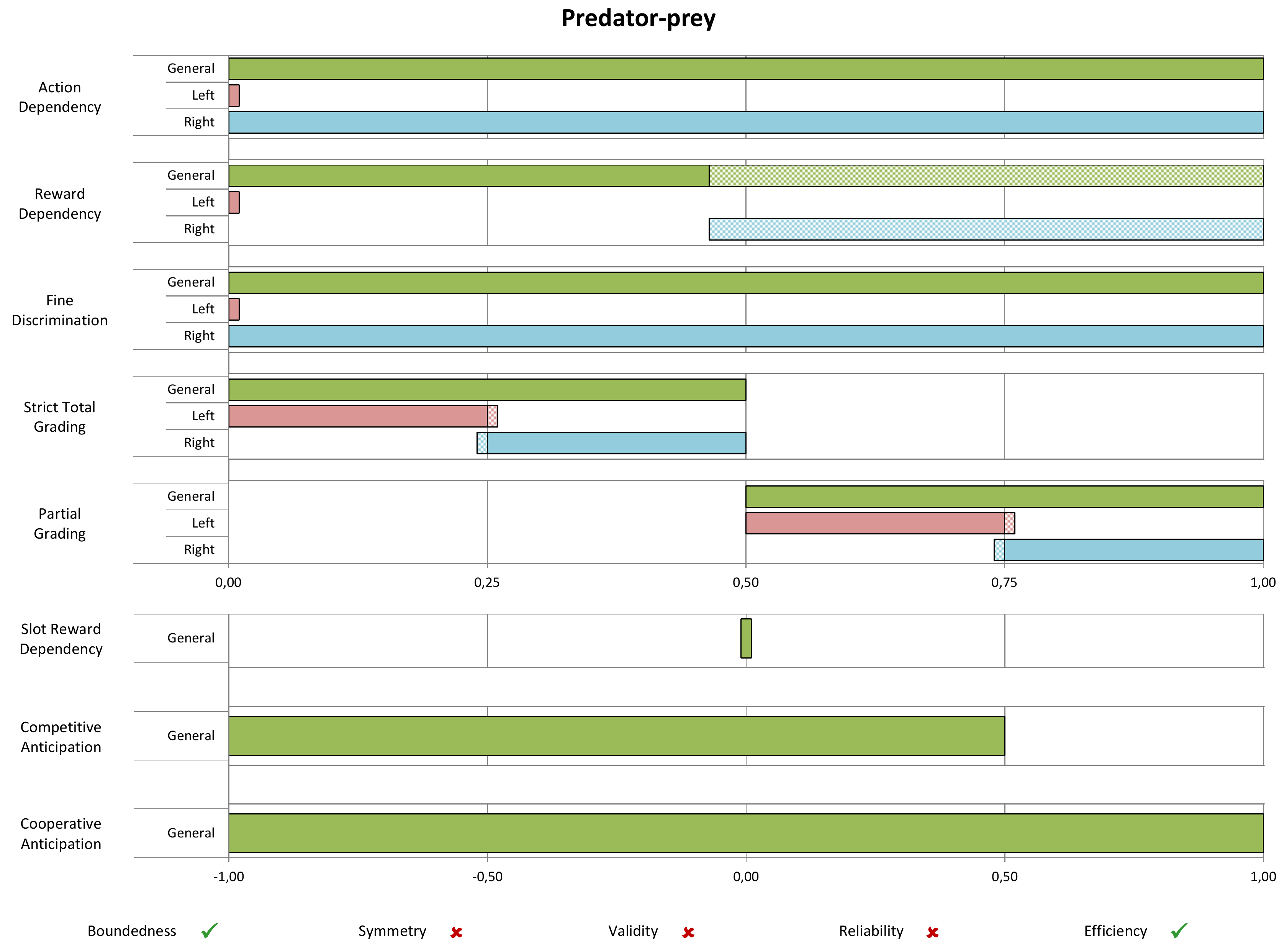}

\caption{Social properties for the predator-prey using uniform weights for $w_S$ and $w_{\dot{L}}$, and $\Delta_S(a,b)$ and $\Delta_{\mathbb{Q}}(a,b)$ return $1$ if $a$ and $b$ are equal and $0$ otherwise. Lighter bands mean the values are not formally calculated, but an estimation is given.}
\label{fig:predator-prey_properties}
\end{figure}

When we start with the properties we see that this game is {\em bounded} since its rewards are between $-6$ and $6$, so average rewards are between $-1$ and $1$. As with previous environments, if weight functions $w_S$ and $w_{\dot{L}}$ are bounded, the value of $\Upsilon$ and many properties will be bounded.

Analysing the symmetry property, we clearly see that this game is {\em not symmetric}. First, prey and predator teams do not have the same number of agents. And second, changing the slots of the agents within the predator team does not provide the same sequences of rewards for the agents, owing to they start in different positions. This forces us to calculate all the other properties for all the slots.

If we look at the action dependency (AD) property, we cannot see any difference with previous environments, even having a so different environment. For the $General$ range, evaluated agents can either interact without noticing the other agents, providing a minimum value of $0$ (proposition \ref{prop:predator-prey_AD_general_min}), or can perform different actions depending on the agents they encounter, providing a maximum value of $1$ (proposition \ref{prop:predator-prey_AD_general_max}). But some particular bad selections of $\Pi_o$ could restrict $Left$ range to be equal to $[0,0]$ (proposition \ref{prop:predator-prey_AD_left_max}). On the contrary, no particular selection of $\Pi_o$ can restrict the $Right$ range for this property, remaining from $0$ to $1$ (proposition \ref{prop:predator-prey_AD_right_min}).

When we look at the ranges for the reward dependency (RD) property, we can see that the $General$ range goes from $0$ to $1$ (proposition \ref{prop:predator-prey_RD_general_min} and conjecture \ref{conj:predator-prey_RD_general_max}) so the rewards of the evaluated agents can either differ or not depending on the agents they encounter. A really bad selection of $\Pi_o$ can make the $Left$ range too restrictive, staying on $[0,0]$ (proposition \ref{prop:predator-prey_RD_left_max}) no matter which $\Pi_e$ we are evaluating. But we can restrict the $Right$ range by selecting a properly $\Pi_o$, obtaining a range from $\frac{13}{28}$ to $1$ (approximation \ref{approx:predator-prey_RD_right_min}). This would force that almost half of the results of the evaluated agents will be different depending on the line-up pattern they interact with.

We now move to the ranges for the fine discrimination (FD) property. The $General$ range goes from $0$ to $1$ (propositions \ref{prop:predator-prey_FD_general_min} and \ref{prop:predator-prey_FD_general_max}), i.e., the result of one evaluated agent can be different or not from the result of another evaluated agent depending on which set of agents $\Pi_o$ is selected. But a bad selection of $\Pi_o$ can restrict too much the $Left$ range, making this value equal to $[0,0]$ (proposition \ref{prop:predator-prey_FD_left_max}), so every evaluated agent will obtain the same result. However, the $Right$ range cannot be restricted, remaining from $0$ to $1$ (proposition \ref{prop:predator-prey_FD_right_min}), since there are always two different evaluated agents that can obtain the same expected average reward, independently of the set of agents $\Pi_o$.

Strict total grading (STG) and partial grading (PG) properties are clearly different from previous environments. The $General$ range for the strict total grading property can only go from $0$ to $\frac{1}{2}$ (propositions \ref{prop:predator-prey_STG_general_min} and \ref{prop:predator-prey_STG_general_max}), so we cannot even have a strict total ordering for all the evaluated agents. In addition, its $Left$ range can be restricted to be from $0$ to $\frac{1}{4}$ (approximation \ref{approx:predator-prey_STG_left_max}), so it would be possible to obtain some strict total ordering. At least, its $Right$ range can be restricted to be from $\frac{1}{4}$ to $\frac{1}{2}$ (approximation \ref{approx:predator-prey_STG_right_min}), which somehow alleviates this situation. But instead, its partial grading has really good ranges. Its $General$ range goes from $\frac{1}{2}$ to $1$ (propositions \ref{prop:predator-prey_PG_general_min} and \ref{prop:predator-prey_PG_general_max}), its $Left$ range can be restricted to be from $\frac{1}{2}$ to $\frac{3}{4}$ (approximation \ref{approx:predator-prey_PG_left_max}) and its $Right$ range can be restricted to be from $\frac{3}{4}$ to $1$ (approximation \ref{approx:predator-prey_PG_right_min}). This makes this environment good for grading the evaluated agents, so even with a bad selection of $\Pi_o$ we will still be able to obtain some partial orders between some of the evaluated agents, and provides a promising partial grading for the evaluated agents if $\Pi_o$ is well selected.

The slot reward dependency (SRD) has a $General$ range equal to $[0,0]$ (proposition \ref{prop:predator-prey_SRD_general_range}). This particular value comes from the opposite results of preys and predators. As early mentioned, if we had only used the slots of the predators for evaluating the agents, we would have had another range.

The $General$ range for the competitive anticipation (AComp) property goes from $-1$ to $\frac{1}{2}$ (propositions \ref{prop:predator-prey_AComp_general_min} and \ref{prop:predator-prey_AComp_general_max}). Anticipating the strategy of the other team can be useful to improve the expected average reward, but only when playing in the predator team, since playing as the prey does not have this benefit. But conversely, a bad anticipation does really penalise the expected average reward of an evaluated agent either playing in the prey or predator team.

In this environment we can find cooperation between the agents within the predator team. The $General$ range for the cooperative anticipation (ACoop) goes from $-1$ to $1$ (propositions \ref{prop:predator-prey_ACoop_general_min} and \ref{prop:predator-prey_ACoop_general_max}). This makes a good coordination within the predator team to always chase the prey, but a bad coordination can let the prey to escape.

We now discuss the validity property. In the predator-prey environment we can encounter both competition between the prey and predators, and cooperation among the predators, which makes it a complex game to evaluate social intelligence. Both competition and cooperation seem important in this game, giving more importance to competition rather than cooperation as we can see from AComp and ACoop, where a bad result can come from incorrectly anticipating while an incorrect anticipation in cooperation does not necessarily provide bad rewards. But this game gives us a general situation to evaluate social intelligence in a broad way. Also, agents that are evaluated in this environment can have good orderings as properties STG and PG reflect, making it a good environment to classify the agents. However, the abilities that the agents need to accomplish their goals are not balanced. It is easier for the predator team to win the chase if they cooperate adequately, where the prey will not have a chance to survive. Also, the $Left$ ranges of the properties are usually very restrictive when $\Pi_o$ is not selected carefully. So it is really necessary to select a correct set of agents, but still, once a certain level of intelligence is reached we cannot evaluate higher levels of intelligence, since their result will remain equal. From here, we can say that the social intelligence that this environment is evaluating is clearly limited to a certain level, and once this level is reached the results will not vary as happens in the $Left$ range for the RD property.
Summarizing, this environment allows us to evaluate both competition and cooperation which makes it a good environment to evaluate social intelligence, but this can only be evaluated until a certain level of intelligence. Over this level, the results will not reflect the true intelligence, so the environment will not be evaluating its actual value.
As a result, the game is {\em not valid} for interesting levels of social intelligence.

Finally, this environment is {\em not reliable} for just one exercise. It has a high variability as only 6 iteration are allowed and the space is small, so small changes in the first movements may have very different consequences. With respect to its efficiency, we just need a few interactions to provide us a fast result for the evaluated agent, which makes this environment {\em efficient}. This means that reliability can be obtained in reasonable time by the repetition of many episodes.

The predator-prey gives us a more complex environment than previous environments. However, it can still provide us a good environment to evaluate social intelligence if $\Pi_o$ is wisely chosen. But we must be careful, since we could obtain a really poor environment if they are not properly selected. The results we have obtained came from our choice to include the possibility to evaluate the agent also playing as the prey. A more classical approach could have given us a different picture for this environment.

\subsection{Pac-Man}
\label{sec:pac-man}
\begin{figure}[!ht]
\centering

\includegraphics[width=0.30\textwidth]{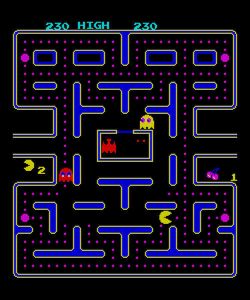}

\caption{A Pac-Man game screenshot. Pac-Man is represented by a yellow circle with a mouth, which must avoid enemies represented by ghosts. Small pills represent the food that Pac-Man must collect, and big pills change chasing agents' roles for a limited period of time. \textit{Taken from \url{http://www.freepik.com/} with permission.}}
\label{fig:pac-man}
\end{figure}

Computer games are also used as mainstream environments to evaluate AI systems. One example of the use of games for evaluating AI is the ALE (Arcade Learning Environment) \cite{bellemare2012arcade}, a framework where a set of arcade computer games are used to evaluate the performance of current AI algorithms. Here, we will analyse Pac-Man, a simple and well known game, but still complex enough to the state of the art in AI, which uses a 2D maze. The AI community has used this environment as a testbed in order to evaluate their algorithms (e.g., \cite{veness2011monte, gallagher2003learning}). This game resembles a pursuit game, but this time the player represents the prey role (most of the time), so it must avoid being caught by the enemies (represented by ghosts). In order to win, Pac-Man must also collect all the pills that are present in the environment, which also provide some points. On the other hand, ghosts are appearing in the environment one by one over time, and they win if at least one of them is able to chase Pac-Man. If Pac-Man is able to reach certain locations in the environment and eat specific pills, it becomes invulnerable for a short period of time, and receives additional points by chasing the ghosts. Figure \ref{fig:pac-man} shows a Pac-Man game screenshot.

From the huge diversity of possible situations that can occur in this environment, it is difficult to formally analyse some of the properties as we did with previous environments. As long as the systems are more complex, it becomes more difficult to determine their actual levels of cooperation and competition, and more effort is needed to formalise them and find some $\Pi_e$, $w_{\Pi_e}$ and $\Pi_o$ to find the environment ranges. Instead, we will analyse this environment in an informal way. As in previous environments we will assume uniform weights for $w_{\dot{L}}$ and $w_S$.

\begin{figure}[!ht]
\centering

\includegraphics[width=0.85\textwidth]{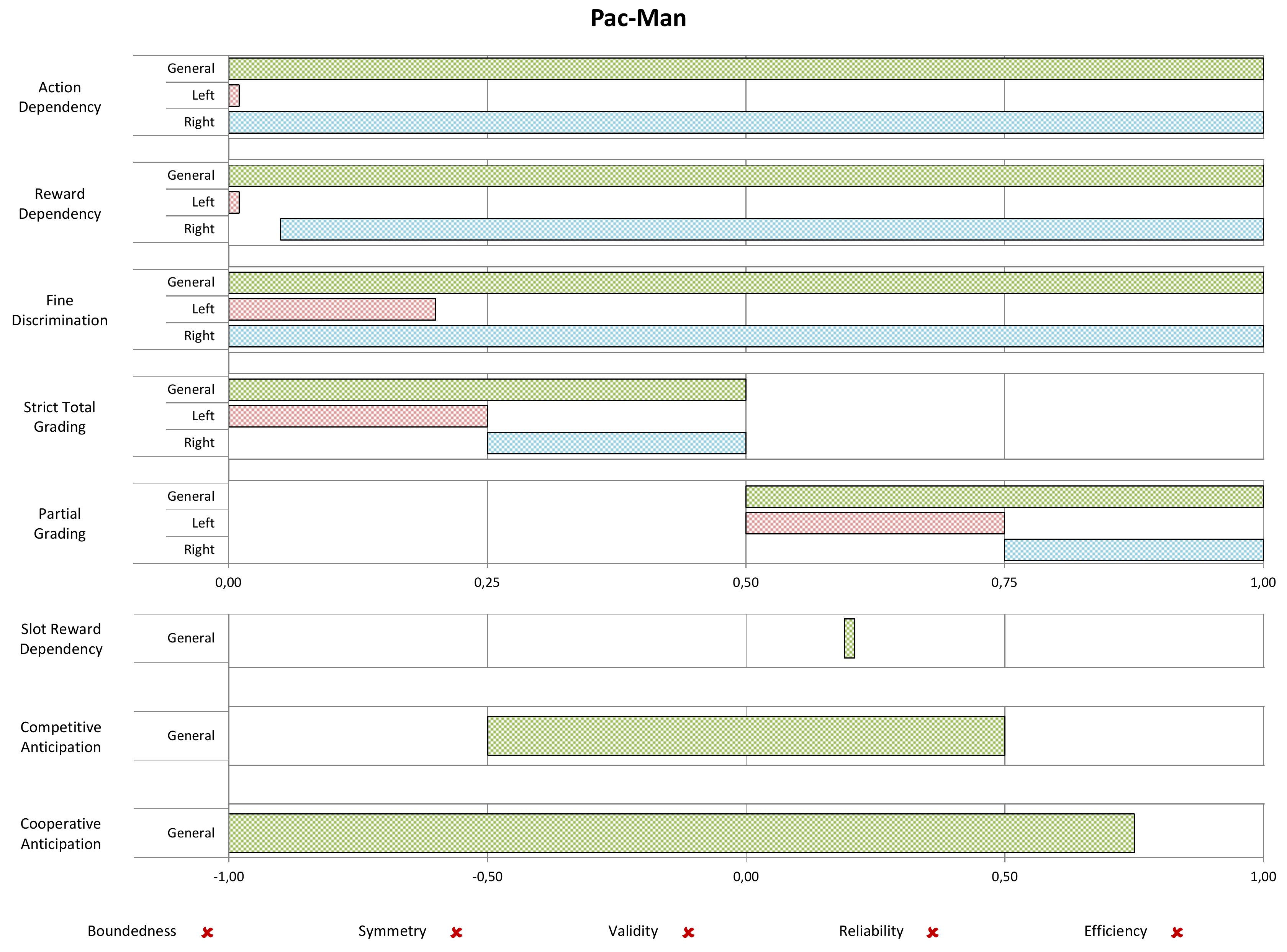}

\caption{Estimation of the social properties for the Pac-Man environment using uniform weights for $w_S$ and $w_{\dot{L}}$, and $\Delta_S(a,b)$ and $\Delta_{\mathbb{Q}}(a,b)$ return $1$ if $a$ and $b$ are equal and $0$ otherwise.}
\label{fig:pac-man_properties}
\end{figure}

In figure \ref{fig:pac-man_properties} we show a summary with an estimation of the properties for Pac-Man.

When we start with the properties we see that this game is {\em not bounded}, since Pac-Man can obtain more and more points (or rewards) as long as it continues surpassing levels. This makes that $\Upsilon$ will not be bounded for this environment. Reframing the game by calculating an average of points by time as rewards in order to make it bounded will change the goal of the game significantly.

Also, the game is {\em not symmetric}. On one hand, both teams do not have the same number of slots, which makes the game {\em not Total Inter-Team Symmetric}. On the other hand, we could say that the environment is Intra-Team Symmetric, since every ghost has the same probability to chase Pac-Man, but this is not exact, since each ghost appears in different moments of the game, so swapping their behaviour could not provide exactly the same results, making the environment {\em not Intra-Team Symmetric}.

The action dependency (AD) property seems to be as in previous environments. All agents have the possibility to ignore the actions of the other agents or act according to what they did in previous interactions.

When we look at the reward dependency (RD) property, it could obtain some different values. Indeed, it is too easy to chase Pac-Man if the four ghosts cooperate coherently, but a bad behaviour for the ghosts can facilitate the game for Pac-Man. In addition, small differences in the behaviour of the agents can provide very different results as, for example, a ghost passes near Pac-Man and decides to chase or to avoid it. This small difference in behaviour will provide high differences in their results. Also, when a ghost is far from Pac-Man, small differences in its behaviour will probably lead to similar results.

The fine discrimination (FD) property also has a huge range of values. As mentioned above, the behaviour of two different evaluated agents can both obtain the same or very different results, highly depending on the behaviour of the other agents.

It seems difficult to know whether we can establish a grading between the evaluated agents in this environment. But we venture that the grading properties could be similar to the ones provided in the predator-prey environment, since both environments have many similarities.

It is also difficult to provide a slot reward dependency, since rewards obtained by one team typically do not reflect on the other team. For example, every point obtained by Pac-Man does not directly have influence on the ghost team's rewards, and chasing Pac-Man only prevents it from obtaining more points. But just assuming that the rewards of each team are always different, we can obtain a value as we did in the SRD for the predator-prey environment (which has a similar configuration) to obtain an approximated value, meaning that the slot reward dependency is more focused on cooperation than competition.

If we look at the anticipation properties, it is possible that competitive anticipation does not have a huge reflect on rewards, but still anticipating competitors will provide some good rewards. In cooperative anticipation, it is possible that one ghost can do worse than a random agent, leading to really bad values in this case, but a good anticipation can make chasing Pac-Man easier.

This game is {\em not very reliable} as it depends on many small details. Also, the game is {\em not efficient}. We will need to run the game at least for dozens of iterations to get some stability in the agents' expected average rewards, since many of the first actions obtain the same rewards but the crucial part of the game comes when the pills become scattered.

Finally, with this game we can both evaluate competition and cooperation. We can find competition, since each team can only gain rewards by making the other team lose rewards. Additionally, in the ghost team cooperation is also needed to properly chase Pac-Man. For this game, the selection of agents is crucial, a set of predators with high level of intelligence can make Pac-Man efforts useless, which will always obtain bad results. This makes the game somewhat valid, but only for lower levels of (social) intelligence if the selection of agents is wisely chosen, but not for agents with high levels of (social) intelligence where the game becomes {\em not valid}.

\subsection{RoboCup Soccer}
\label{sec:robocup_soccer}
\begin{figure}[!ht]
\centering

\includegraphics[width=0.45\textwidth]{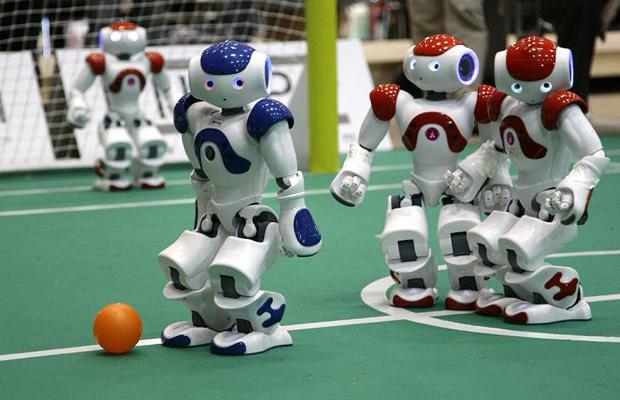}

\caption{A standard RoboCup Soccer platform with robots playing a match. \textit{Used with permission from \url{http://www.robocup2013.org/}, photograph by: Bart van Overbeeke.}}
\label{fig:robocup_soccer}
\end{figure}

As an example of a 3D space game we find the RoboCup Soccer competition \cite{Kitano:1997:RRW:267658.267738}. Here, two artificial multi-agent systems (or teams) have to compete against each other in order to win a soccer match. The agents in each team must cooperate to make the ball reach the adversary's goal, while cooperate to avoid the adversary to score a goal. The game follows the rules of a typical soccer match. Figure \ref{fig:robocup_soccer} shows a RoboCup Soccer match.

As happens with the previous environment, this game has a huge diversity of possible situations that can occur (including physical and virtual versions), which makes difficult to formally analyse it. Again, in such complex scenario, it becomes more difficult to determine the levels of cooperation and competition. But also, due to the high level of complexity of this game, teams tend to some specialisation, with each player focussing on some specific aspects of the game instead of focusing on the problem in a general way. Again, we will analyse this environment in an informal way. As in previous environments we will assume uniform weights for $w_{\dot{L}}$ and $w_S$.

\begin{figure}[!ht]
\centering

\includegraphics[width=0.85\textwidth]{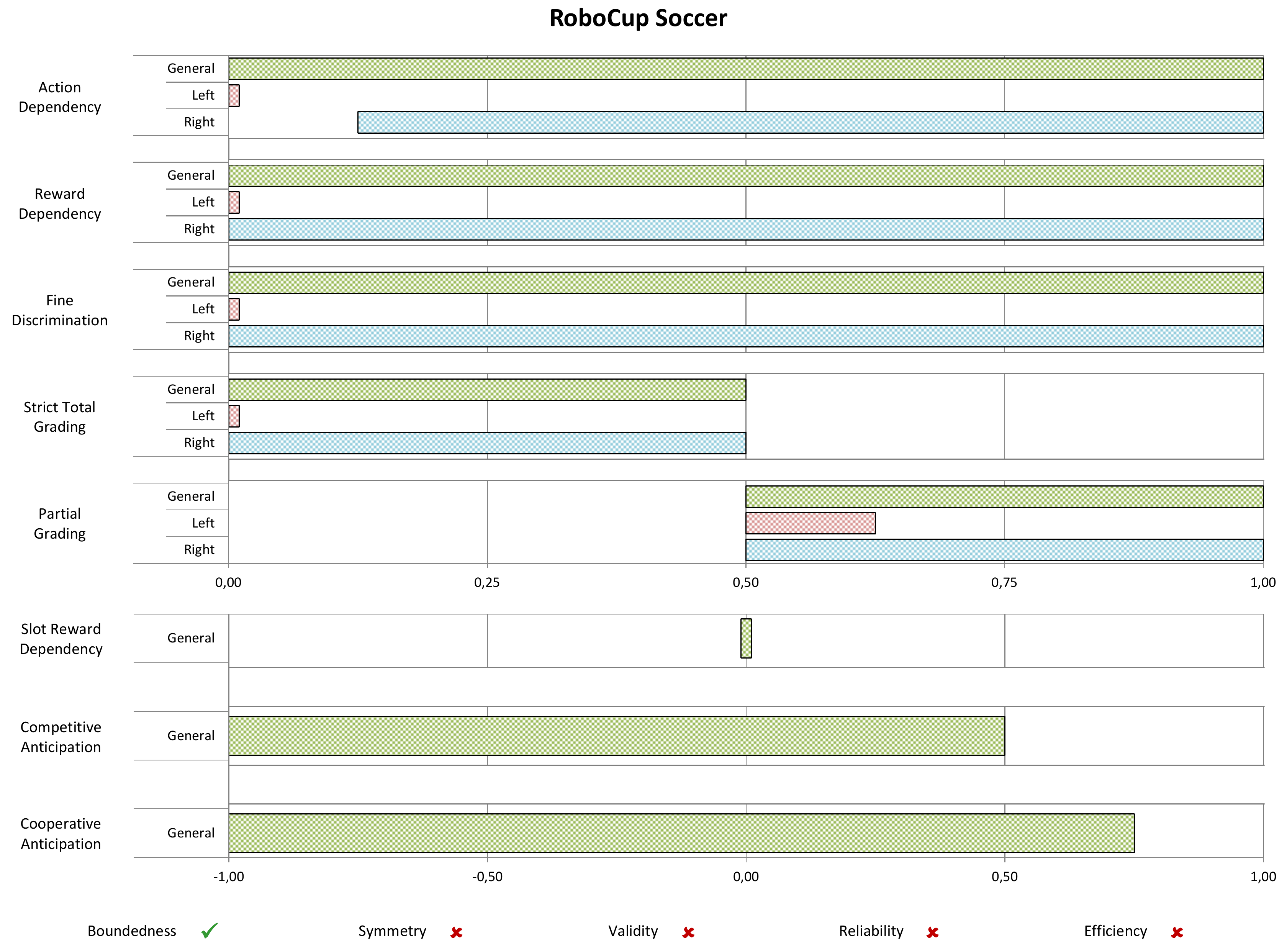}

\caption{Estimation of the social properties for the RoboCup Soccer using uniform weights for $w_S$ and $w_{\dot{L}}$, and $\Delta_S(a,b)$ and $\Delta_{\mathbb{Q}}(a,b)$ return $1$ if $a$ and $b$ are equal and $0$ otherwise.}
\label{fig:robocup_soccer_properties}
\end{figure}

As in the previous environment, in figure \ref{fig:robocup_soccer_properties} we show a summary with an estimation of the social properties for the RoboCup Soccer.

When we start with the properties we see that this game is {\em bounded}, since rewards ($1$ for win, $0$ for a tie and $-1$ for lose) are bounded. This makes that $\Upsilon$ will be bounded if weights are also bounded.

When we look at the symmetry property, both teams have the same number of slots and, if we ignore which team starts with the possession of the ball on each half, it makes the environment {\em Total Inter-Team Symmetric}, so both teams can swap their slots and the result will remain the same. But, if we want to swap the slots of two players within the same team, they will not obtain the same results (as for example the goalkeeper has different rules), so the environment is {\em not Intra-Team Symmetric}. Since for the symmetry condition we need the environment to be both {\em Total Inter-Team Symmetric} and {\em Intra-Team Symmetric}, we can conclude that the RoboCup Soccer is {\em not symmetric}.

The action dependency (AD) property is similar to previous environments. All the evaluated agents have the possibility to act differently depending on the agents they encounter, but, at least, in this game an agent can affect the actions of the evaluated agents. For example, one agent can knock the evaluated agent down to the ground, so now it will only be able to stand up.

We now see the reward dependency (RD) property, which can have a huge range. A change in the line-up of course can change the result of the match.  
Conversely, only changing one agent in the line-up can completely change the match result to make a team lose. changing the agents' rewards. And we can reason in the same way for the fine discrimination (FD) property, since the behaviour of only one agent (the agent to be evaluated) can also make its team to either win/lose the game, or obtain the same result.

It is not easy to determine if there exists some grading between the evaluated agents in the RoboCup Soccer. Instead, we can take a look at some professional (human) Soccer leagues. It is not unusual to see situations where two teams repeatedly tie, and a third team beats one of them while loses against the other one repeatedly. This situation shows us that there is no strict grading between teams (and neither is  between their players) for this game.

The slot reward dependency is straightforward for this environment. We have two teams with five players on each team. Every agent within a team will obtain the same reward, while the other agents within the other team will obtain the opposite reward. Using a correlation function over these rewards and over all slots, we obtain a slot reward dependency equal to 0.

If we look at the anticipation properties, correctly anticipating both competitive and cooperative can provide a high advantage for each of the teams, so the team of the evaluated agent can score more goals and win the game more easily. A bad competitive anticipation can lead the opponent to win the game, while a good one can provide good results. In cooperative anticipation, the evaluated agent could play worse if it is not correctly anticipating its teammate, but a good anticipation can provide them really good results.

Let us now consider the validity of this game. First, with this game we can both evaluate competition and cooperation. We can find competition, since both teams must compete to win the game. Additionally, the agents within each team can cooperate to mislead the other team and score more goals. Second, increasing the social intelligence of the agents will typically increase the difficulty of the match, since more skilled agents will score goals more easily, and also defend better, preventing the other team to score. This makes the game useful to match a high variety of skill levels. But conversely, this game also evaluates some other skills than social intelligence, such as for example, their ability to predict the movement that the ball will do when it is kicked. In principle, there are reasons to consider this game a valid scenario to evaluate social intelligence. However, since the agents will need more than their social intelligence in order to play the game, we think that an evaluation using this kind of environment will not only evaluate social intelligence, but also other abilities such as motion understanding. For this reason we consider this environment {\em not valid} to evaluate social intelligence.

This game is {\em not reliable}. In this game, it is widely known that two teams that are facing each other do not necessarily obtain the same result in every match. This depends highly on the decisions made by the agents, since a small decision change can lead to a very different result on the match. Also, the result of the game depends on luck (at least for the physical version), since the players cannot always predict the correct movement of the ball.

\subsection{Summary}
\label{sec:environments_summary}
So far, we have analysed some ranges of values that the properties presented in this paper can have for the five environments we have selected. Next, let us group the environment properties by the different ranges in order to make a comparison between the environments through their properties.

\begin{figure}[!ht]
\centering

\includegraphics[width=0.85\textwidth]{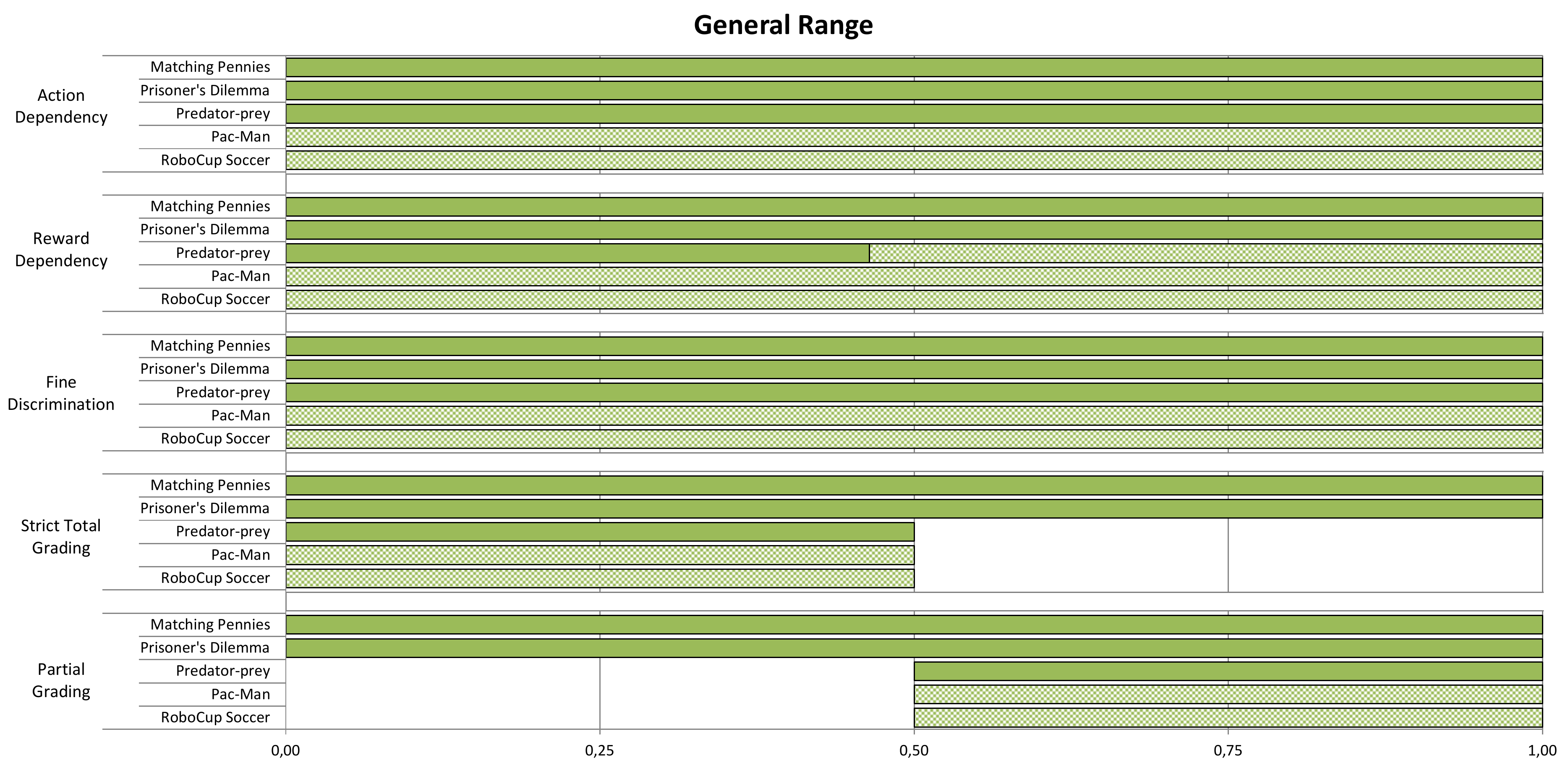}
\includegraphics[width=0.85\textwidth]{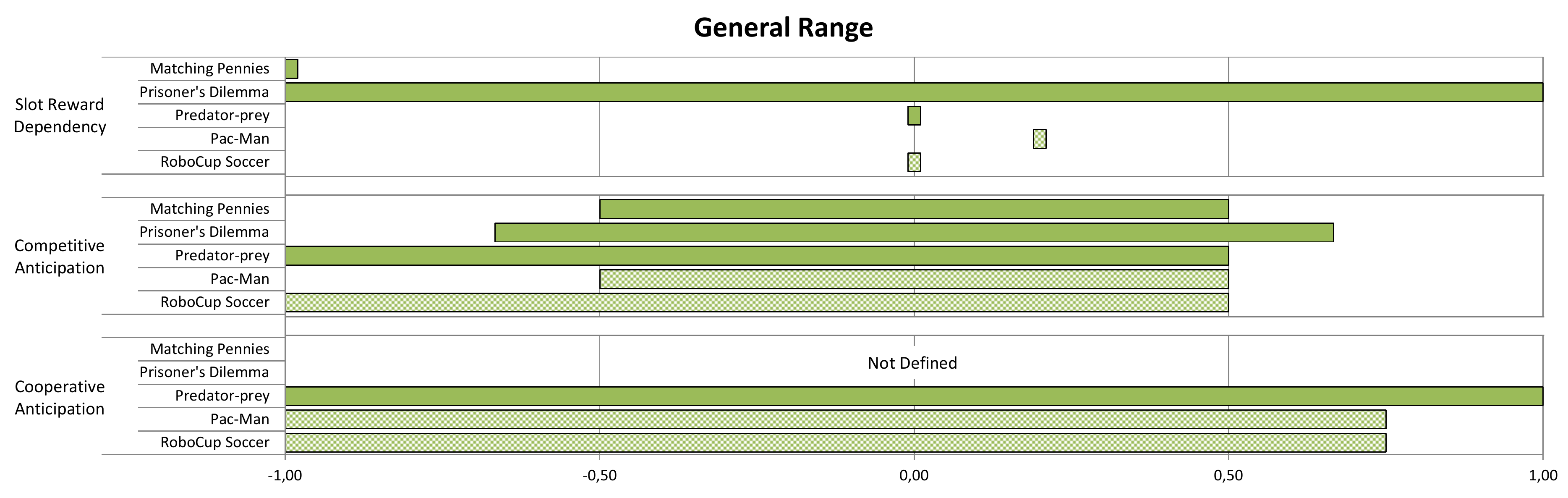}

\caption{General range for the quantitative properties of the five environments analysed in this paper using uniform weights for $w_S$ and $w_{\dot{L}}$, and $\Delta_S(a,b)$ and $\Delta_{\mathbb{Q}}(a,b)$ return $1$ if $a$ and $b$ are equal and $0$ otherwise. Lighter bands mean that an estimation is given.}
\label{fig:environments_general_range}
\end{figure}

In figure \ref{fig:environments_general_range} we see the $General$ range of the quantitative properties for the five environments.

As we can see in the top part of figure \ref{fig:environments_general_range}, there are few differences with respect to this range for the environments. In the first three properties all the environments have the broadest possible range. We only see some difference for the grading properties in the last three environments. While the first two environments have the broadest possible range, the last three environments seems that as long as the strict total grading gets worse, the partial grading gets a better range. This is simply explained because agents in the same team obtain the same rewards, so it is not possible to obtain a strict total ordering for them while it is still possible for agents in different teams. Obviously, the agents in the same team always have a partial ordering, making this range higher.

In the bottom part of figure \ref{fig:environments_general_range} we can see more differences. In the slot reward dependency property we can see the first big difference between the environments. The majority of the environments have a unique (and usually different) value for this property. While the matching pennies is completely competitive, Pac-Man is slightly more oriented to cooperation, and predator-prey and RoboCup Soccer are neutral (i.e., provides both competition and cooperation to the same extent). But, prisoner's dilemma is the only one that has the broadest possible range instead of having a predetermined configuration, so this environment allows the agents to dynamically cooperate and compete with agents in the other team depending on which actions they perform.

Finally we have the two anticipation properties. In both of them, the agents obtain better rewards when they correctly anticipate, but their rewards get worse when they incorrectly anticipate.

In the competitive anticipation we see some small differences between the environments. While a correct competitive anticipation provides good rewards, some environments usually provide much worse rewards when incorrectly anticipating competitive agents, as we can see for the predator-prey and RoboCup Soccer.
For the cooperative anticipation we see another different picture. In this case two of the environments do not have this property, since they do not provide teams where the agents will be able to cooperate. However, the last three environments provide teams where cooperative anticipation can be useful between cooperative agents, and also they (almost) provide the broadest range for this property.

\begin{figure}[!ht]
\centering

\includegraphics[width=0.85\textwidth]{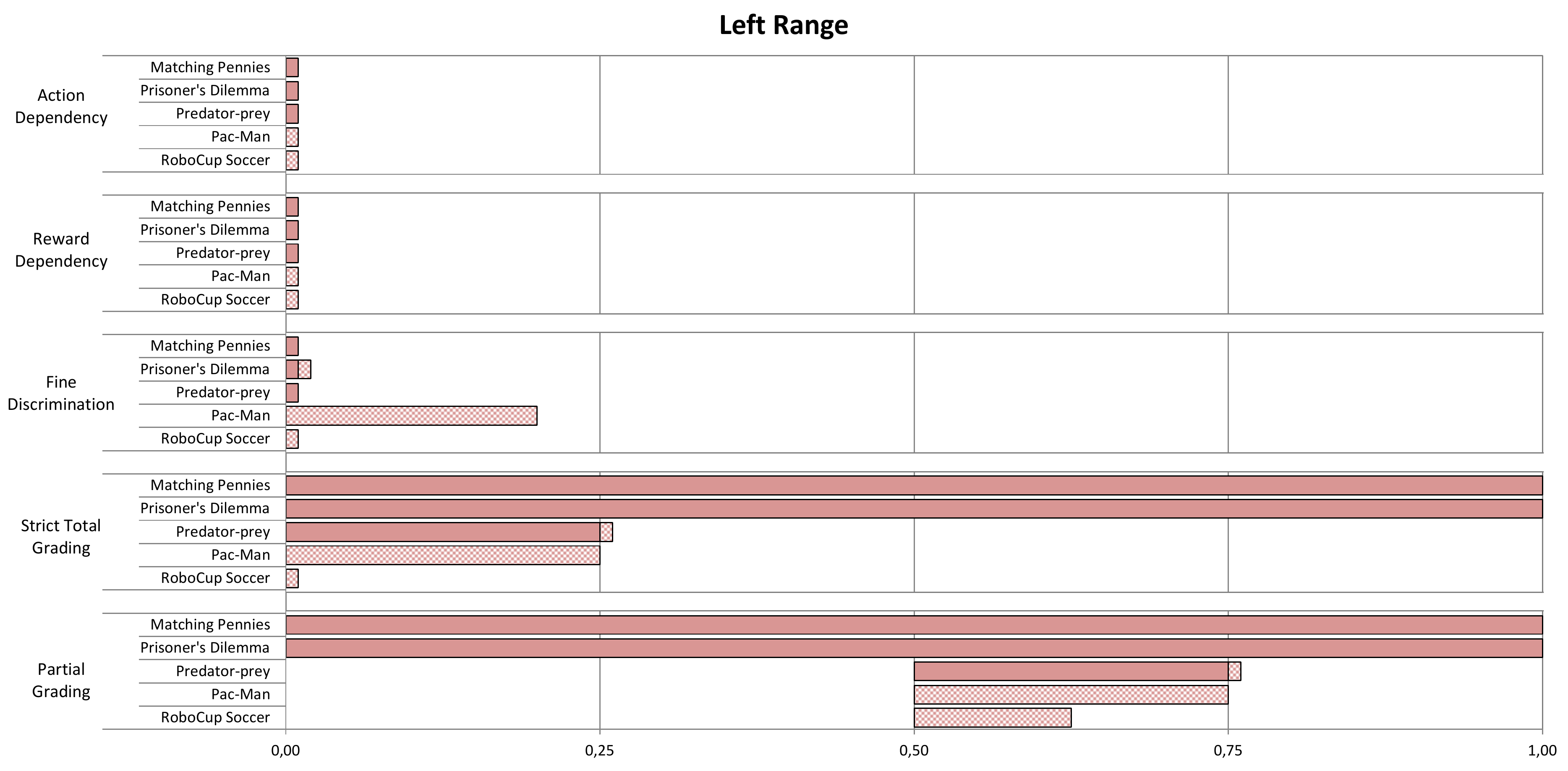}

\caption{Left range for the quantitative properties which range is between $0$ and $1$ of the five environments analysed in this paper using uniform weights for $w_S$ and $w_{\dot{L}}$, and $\Delta_S(a,b)$ and $\Delta_{\mathbb{Q}}(a,b)$ return $1$ if $a$ and $b$ are equal and $0$ otherwise. Lighter bands mean that an estimation is given.}
\label{fig:environments_left_range}
\end{figure}

It is much more interesting to see what happens with these environments if we make a bad selection of $\Pi_o$. In figure \ref{fig:environments_left_range} we can see the $Left$ ranges of the quantitative properties which range is between $0$ and $1$ for the five environments.

We can see that both the action and reward dependency properties have the worst possible range for all the environments. This means that a bad selection of $\Pi_o$ will be disastrous with respect to these properties. In fact, this is not surprising, since $\Pi_o$ could be populated only with agents having the same exact behaviour, so the evaluated agents will not be able to behave or obtain different rewards depending on which agents from $\Pi_o$ they are interacting with.

For the fine discrimination property, we can see that (almost) all the environments have a very low range, so the evaluated agents can hardly be discriminated.

The strict total grading property clearly gives us an order of their compliance for the environments. RoboCup Soccer is clearly the worst environment, while predator-prey and Pac-Man follow it. Meanwhile, matching pennies and prisoner's dilemma cannot be restricted with any particular $\Pi_o$, obtaining the best range for this property.

Finally, the partial grading has different ranges for the environments. Predator-prey and Pac-Man have the same range, and they also have a clearly better range than the RoboCup Soccer. Matching pennies and prisoner's dilemma have the broadest possible range. At this point, it is not clear which pair matching pennies and prisoner's dilemma, or predator-prey and Pac-Man has a better range. From one side, the range for matching pennies and prisoner's dilemma goes from $0$ to $1$, so the selection of $\Pi_o$ does not necessarily worsens the property, but still it is possible to have a bad value. From the other side, the worst value for the predator-prey and Pac-Man cannot be as bad as in matching pennies and prisoner's dilemma. However, the selection of $\Pi_o$ does worsen the values for this property.

\begin{figure}[!ht]
\centering

\includegraphics[width=0.85\textwidth]{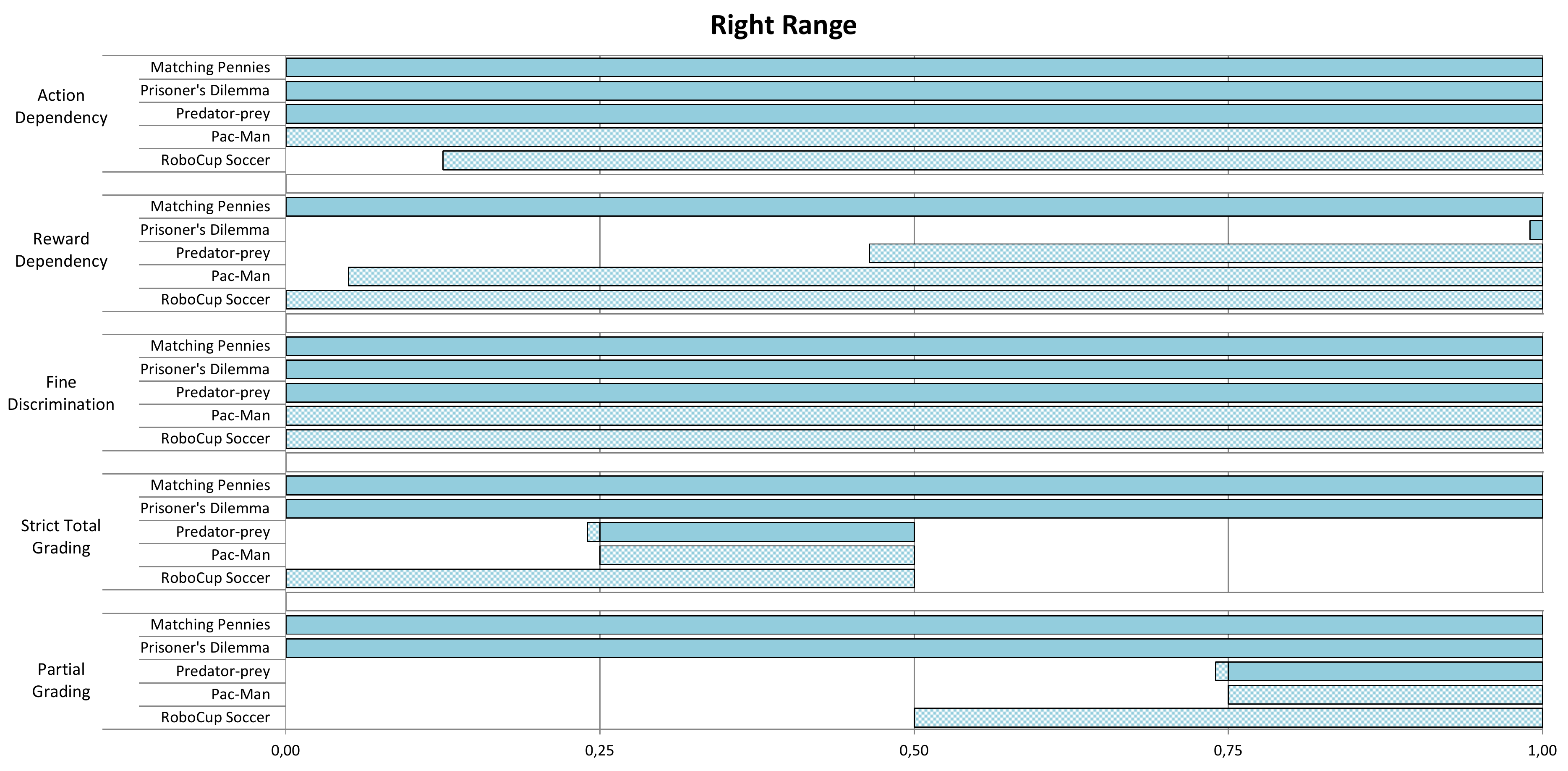}

\caption{Right range for the quantitative properties which range is between $0$ and $1$ of the five environments analysed in this paper using uniform weights for $w_S$ and $w_{\dot{L}}$, and $\Delta_S(a,b)$ and $\Delta_{\mathbb{Q}}(a,b)$ return $1$ if $a$ and $b$ are equal and $0$ otherwise. Lighter bands mean that an estimation is given.}
\label{fig:environments_right_range}
\end{figure}

Let us next see the effect that a good selection of $\Pi_o$ can have on some quantitative properties for the environments. This is possibly the most interesting picture, because it gives us the best we can do with a right choice of $\Pi_o$. In figure \ref{fig:environments_right_range} we can see the $Right$ ranges of the quantitative properties which range is between $0$ and $1$ for the five environments.

We can see that the action dependency and fine discrimination properties cannot improve much by selecting an appropriate $\Pi_o$. At least RoboCup Soccer can slightly restrict its action dependency range, which means that even with the best possible choice of $\Pi_o$ we cannot ensure that there will be a high action dependency.

For the reward dependency property the ranges vary. Matching pennies and RoboCup Soccer cannot restrict their respective ranges. Pac-Man can slightly restrict this range and predator-prey does not have a bad range, so depending on which line-up pattern the evaluated agents encounter, they will certainly obtain some differences in their expected average rewards. But prisoner's dilemma can ace this property, making the expected average rewards different for every evaluated agent depending on the line-up pattern they interact with.

For the strict total grading we find more differences. It is not clear which environment has a better range, but at least we can say that RoboCup Soccer has the worst range among the five environments. However, it is not as clear which of the other four environment has a better range. From one side, the range of matching pennies and prisoners' dilemma goes from $0$ to $1$, so the selection of $\Pi_o$ does not necessarily improve their property, but still it is possible to have a good value. On the other hand, the best value for the predator-prey and Pac-Man cannot be as good as in matching pennies and prisoner's dilemma. However, the selection of $\Pi_o$ does improve the values for this property.

Finally, the partial grading gives us more information about the orderings. RoboCup Soccer has improved its range, but it is still worse than the predator-prey and Pac-Man, which now are clearly the best to obtain an ordering. However matching pennies and prisoner's dilemma cannot restrict this range, making them the worst of the five environments.

Overall, the {\em Right} range is more informative. Note that the $\Pi_o$ that we use to calculate the values for each environment's property is not necessarily the same. We just obtain the values locally for each property. That means that some points could not be achievable at the same time.

\begin{figure}[!ht]
\centering

\includegraphics[width=0.85\textwidth]{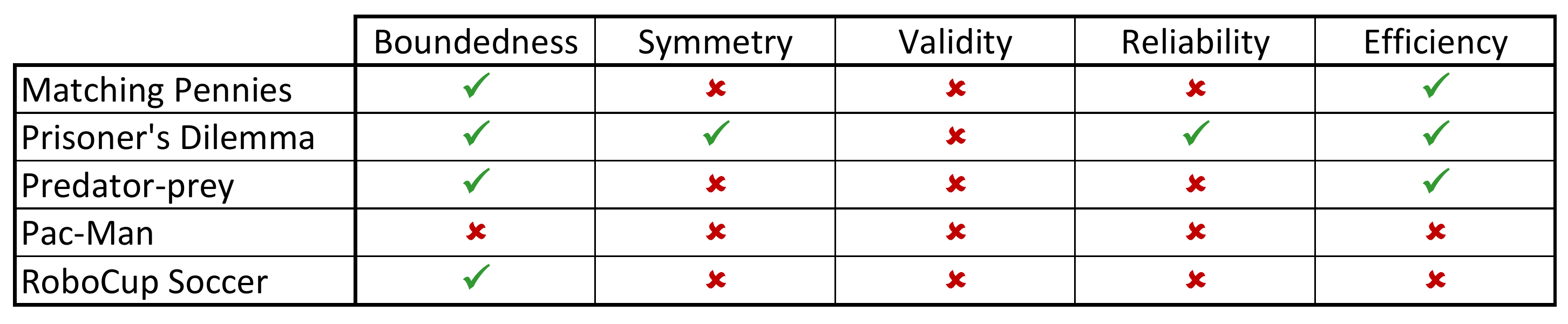}

\caption{Qualitative properties of the five environments analysed in this paper.}
\label{fig:environments_qualitative}
\end{figure}

Lastly, in figure \ref{fig:environments_qualitative} we see a summary of the qualitative properties to obtain a practical test for the environments.

As we can see, almost all the environments have bounded rewards. This provides a bounded value for $\Upsilon$. But only the prisoner's dilemma is symmetric, so in order to evaluate an agent in the other environments, we will need to evaluate them in all the slots. With respect to the validity property, no environment is correctly evaluating social intelligence. Some of them are not sufficiently general, as happens with the matching pennies or the prisoners' dilemma. We have the opposite situation with RoboCup Soccer, where more abilities are evaluated  and it seems difficult to isolate social intelligence from these other abilities. Finally, other environments can only evaluate the social intelligence to a certain degree, as happens in the predator-prey or Pac-Man. With respect to the reliable property, the prisoners' dilemma can evaluate every level of performance with a good result, but we cannot say the same for the rest of the environments. In matching pennies every agent interacting against a random agent will obtain the same result independently of their intelligence, while the rest of environments will not be able to provide a correct value for this property over certain level of intelligence. Lastly, due to its simplicity, matching pennies, prisoners' dilemma and predator-prey are really efficient to provide a result for the agents, but the rest of environments will need to run the game during several iterations and episodes to converge to their results.

From this analysis and comparison between the properties of the environments we made in this section, we can provide some insights. First, we give some findings about the five environments.

\begin{itemize}
\item As we have seen in our analysis, these environments are typically covering anticipation well. Competitive anticipation is well covered in all the environments, while cooperative anticipation is not defined for the first two environments, but the last three are covering it very well. It also seems that the partial grading is generally well covered, so we can find some partial orderings between the evaluated agents.
\item We can find some other properties that the environments are not covering well. One example is the action dependency, where (almost) every environment analysed in this paper obtained the same poor ranges. This property is something which is not usually thought about when designing an environment, but the possibility of having influence on the actions that other players can do is an interesting thing to consider when designing a multi-agent environment and, in particular, if we want to evaluate social intelligence. Another property which is not usually well covered is the slot reward dependency, where these environments are typically only giving one value. We do not mean that the environments do not have good values, but instead, an environment having a broader range of values will provide us a more interesting scenario, where the relations of competition and cooperation between agents can change dynamically.
\item As we can see from their qualitative properties, four of the five environments we selected have some difficulties to be used as a practical test. None of them provides symmetry to simplify the evaluation. The range of abilities required to succeed in these environments are not appropriate to be a valid test to evaluate social intelligence accurately, as well as their reliability is compromised depending on the agents we use to populate the environment. At least some of them are efficient enough to obtain results in a short period of time, and they usually provide bounded rewards, so we can calculate a bounded value of the (social) intelligence of the agents.
\item With these properties, we obtained different ranges of $General$, $Left$ and $Right$ for each of the five environments. In addition, we could see that some little changes over the definition of an environment (as occurs with the matching pennies and prisoner's dilemma) are clearly reflected with these properties. In fact, every kind of environment will have particular ranges of values for these properties, with which we will be able to select the (social) environment(s) that best fits our goals (e.g., select an environment focused on anticipating other agents).
\item As we have seen, a good selection of $\Pi_o$ is crucial in order to obtain appropriate social environments. But, how could we provide a $\Pi_o$ which is appropriate to evaluate the social ability of an agent in a large number of environments? This is a difficult task. When starting the evaluation, since the intelligence of the evaluated agent is not known, it would be appropriate to use one $\Pi_o$ which agents are not too smart during the first environments. As long as the evaluation goes ahead and the intelligence of the evaluee is better known, it would be better to use another $\Pi_o$ whose agents are conforming to this level of intelligence. In order to solve this problem, we could provide a unique $\Pi_o$ and use some kind of distribution which is continuously evolving, giving more probability to the agents which are obtaining better results on these environments (e.g., as in the spirit of the Darwin-Wallace distribution \cite{AGI2011DarwinWallace}).
\end{itemize}

\noindent From the previous analysis we can now distinguish the features of the environments that could be reused for the design of better environments to measure social intelligence more effectively. The first environment we saw is matching pennies, but it does not seem to have any particular useful feature from the properties we analysed. Next we saw the prisoner's dilemma environment, which is similar to the matching pennies with some little modifications. This prisoner's dilemma offers some nice features to include in a social intelligence test. First, we notice its capability to dynamically change the relation between the slots, providing a competitive and cooperative environment at the same time depending on the agents' actions. Second, the evaluated agent can obtain drastically different results when it interacts with very different agents from $\Pi_o$. And third, the symmetry of the payoff matrix, its reliability and efficiency makes this environment a good candidate to provide a simple test. The third environment was the predator-prey. This is the first environment that we analysed providing several agents in (at least) one team. From this team of agents, it is possible to anticipate cooperative agents in such a way that really good performance can be achieved when it is done correctly, and an incorrect anticipation can provide really bad rewards. The same occurs while anticipating competitive agents, but in this case, both teams can anticipate the agents in the other team. Also, we can obtain good partial gradings for the evaluated agents and it is a really efficient game, providing a result in a short period of time. However, Pac-Man and RoboCup Soccer do not provide significant features beyond those provided by predator-prey. At least, in RoboCup Soccer it is possible to exert a slight influence on other agents' actions, but only to some extent.

Conversely, we also distinguish those features that we do not want to appear in environments for social intelligent tests. The first feature we distinguish is that none of these environments are valid to evaluate social intelligence since they are evaluating: 1) more abilities than necessary, as in RoboCup Soccer where the agents need their motion understanding to play the game, 2) not enough abilities, as in matching pennies and prisoner's dilemma where the agents cannot cooperate with agents in the same team, or 3) is only valid for lower levels of intelligence, as in predator-prey and Pac-Man where the predators and ghost can easily chase the prey and Pac-Man respectively once they reach a certain level of (social) intelligence. Also, the majority of the environments do not provide a reliable result due to many reasons. For example, the results of the agents can be easily restricted to always obtain the same results (as occurs in matching pennies by using a random agent). Also, little changes in the behaviour of the agents can create a butterfly effect, making the agents to obtain very different results (as occurs in the last three environments). Also, the environments are typically not symmetric, which will force to evaluate the agents in all slots, and complex environments usually need a lot of iterations to provide a ``reliable'' result. In these five environments, it is weird (if not impossible) to find a situation where an agent can directly influence on which actions are available for one (or more) of the other agents. The capability to directly influence on the available actions of the rest of agents could provide us a richer social environment. Also, some environments (matching pennies and prisoner's dilemma) are not suitable to let the agents anticipate cooperation within a team, since they do not provide the agents a team of agents to cooperate with them. Finally, when we see in more detail some environments, we notice that predator-prey and Pac-Man provide a really difficult/hostile environment, where the predators and ghosts respectively have an enormous advantage to win the game.

Even if it is not the goal of this paper (but a future work), we consider that a good environment measuring social intelligence would have (at least) these characteristics: 1) It provides two or more teams to interact with, and two or more agents on each team. By having this, the agents will be able to compete against the other team(s), cooperate with the agents within the same team and, in the case where more than two teams are presented in the environment, cooperate with other teams. This will provide anticipation to the environment, so the agents will be able to competitive and cooperative anticipate other agents. 2) The agents should influence in some way the rewards obtained by the other agents, providing reward dependency to the environment. 3) There should be limited rewards that the agents can obtain and the payoff of the agents will only depend on the actions they perform. This will provide us a bounded and symmetric environment, which will be ideal to create a practical test. 4) There should not be easy equilibria in the environment. If such circumstance occurs, most of the agents (the intelligent ones) will always perform the same actions, which will limit the results obtained by the agents. Avoiding easy equilibria will provide to the environment more discriminative power, reward dependency and grading for the agents. 5) The environment should provide different kind of spaces where the agents can move. This will avoid the agents to specialise to a particular space. This will make the environment more valid to evaluate (social) intelligence in a more broader way. 6) The agents must be able to influence in some extent the actions that the other agents can perform, creating richer social situations and providing some action dependency to the environment.

Finally, what can we say about the properties? Are they sufficient to characterise any environment? How should they be used? Are the plots useful? Is the $Left$ and $Right$ ranges more meaningful than the $General$ ranges? Some insights below.

\begin{itemize}
\item With these properties we are able to obtain different values for each environment. This gives us some idea about the strengths and weakness of each environment.
\item Here, we only evaluated one agent from a set of evaluated agents $\Pi_e$ interacting with a set of opponents and team players $\Pi_o$. But more specialised properties (and even a definition of social collective intelligence) can be easily extended by, for example, dividing the set of agents $\Pi_o$ into two sets (i.e., one for opponent players and one for team players) or, instead of evaluating the agents in isolation, evaluating together a group (or collective) of agents.
\item There are also other issues which may not be covered on these properties, as for example communication among teammates. They do not provide information about misleading opponents, or the possibility of the agents to influence the actions of other agents on its benefit. Also, the properties do not show us the contribution of the agents to their teams' rewards, as well as the impact that their inclusion in the line-up has in the results of the other teams.
\item We proposed some useful properties to measure the appropriateness of the environments to evaluate social intelligence, which are also useful to characterise these environments. These properties provide us some interesting information about the environments such as the fine and coarse discrimination, which give us a measure of their discriminative power. Other interesting properties are the action, reward and slot reward dependency, giving us an idea about the existing dependency between the actions/rewards of the agents, and which relation between the slots is given more importance in the environment (i.e., it is a more competitive- or cooperative-oriented environment), and if this relation is static or can change during the evaluation.
\item We used the $Left$ and $Right$ ranges in order to compare for each property how a particular good or bad selection of $\Pi_o$ can affect that property in an environment. Conversely, a real test shall provide a unique $\Pi_o$ to evaluate the agents, obtaining a unique range for each property. This selection of $\Pi_o$ will (most probably) make the properties to barely look like the $Left$ or $Right$ ranges we calculated for this five environments, providing instead more varied ranges. Therefore, a comparison between some testbeds with their $\Pi_o$ fixed will give us a more clear idea about their differences.
\end{itemize}

\section{Conclusions and Future work}
\label{sec:conclusions_and_future_work}
Social intelligence has been an important area of study in psychology, comparative cognition and economics for more than a century, and more recently, in artificial intelligence. However, despite the fact that other tests have been created to evaluate other cognitive abilities, nowadays it is still difficult to find a proper test to evaluate social intelligence. Also, current tests tend to be focussed on evaluating the ability of a single species and it is even more complicated to find a test to evaluate social intelligence that is applicable to machines. In fact, tests designed to succeed on a task that requires social intelligence usually also require other abilities to succeed in the task, making them not appropriate for this purpose. This lack of general socially-oriented tests may be due to the absence of a precise (and formal) definition of social intelligence.

In this paper we formalise a definition of social intelligence and some useful social properties for multi-agent environments. In particular, the contributions are:

\begin{itemize}
\item We analysed what social behaviour implies, reviewing what it means from several disciplines.
\item We have considered various options for a definition, and we finally proposed a parametrised definition that formalises the notion of performing well in an environment with other (social) agents. We also indicate how a test can be constructed using this definition.
\item We proposed some properties along with their formal definitions in order to better analyse the appropriateness of an environment to evaluate social intelligence.
\item With this definition of social intelligence and the properties proposed, we analysed and compared several tests and games from artificial intelligence and game theory where social intelligence has an important role to see which properties they meet and which can be improved in order to evaluate social intelligence.
\end{itemize}

\noindent This definition of social intelligence along with the properties proposed here are a first attempt in order to determine whether a testbed is useful for the assessment of social intelligence. As far as we know, this is the first approximation to provide a formal definition of social intelligence along with some useful properties to judge a certain testbed. More research will provide us more information about which properties can be improved and information about other properties to complement the ones presented here.


The characterisation of social intelligence proposed here along with the properties we propose to determine whether an environment is useful to evaluate social intelligence has some open features to be solved.

\begin{itemize}
\item First, it is not clear which utility function the agents must have. Should they be regulated using a discount factor as usual in reinforcement learning? Should we give more importance to later rewards, when the agents are supposed to understand how to behave? Or is it better to use an average (as we did in this paper), giving the same importance to all the rewards?
\item Second, every test should provide which level of complexity it is evaluating. We postulated that the level of complexity should be determined by the agents included in the environment (and their intelligence), the agent setting that determines how teams are formed and the environment where the agent is evaluated. We determined how the first two parameters should influence the formula, but without indicating the formula itself. Should the level of intelligence of the agents weight more than the agent setting, or should it be otherwise? How can we consider the environment in this formula?
\item Third, we noticed that every space used to define an environment necessarily evaluates some spatial intelligence. In theory, we should calculate which part of the result comes from spatial intelligence and subtract it, but this seems very difficult. Alternatively, we could figure out an environment class where no other abilities needed to interact will be really useful. Or at least with a wide variety such that, on average, these other abilities do not bias the result.
\item Fourth, we could also relax the definition of social intelligence by letting the agents cooperate without placing them in a team. Life has taught us that alliances can arise from several agents, even when they do not share the same objectives, to improve the chances of success. It would be interesting to analyse whether a test evaluating only competition can indirectly evaluate this spontaneous cooperation.
\item Fifth, social intelligence is linked to communication and language. We have not included any property or feature in the definitions to account for the presence of communication and language, or to facilitate that. Clearly, communication is possible through actions. Even language can be transmitted by the agent actions with a proper coding. However, this could be rendered more easily to agents. Nonetheless, any particular  communication protocol can make the test non-universal. Instead we think that some extra actions that could be observed immediately by other agents (or by a subset of them) could be basic enough as a signal.
\end{itemize}

\noindent This article goes beyond the simple properties of game theory in many ways, opening a number of possibilities for the evaluation of multi-agent environments. Of course, further research is needed to clarify all these questions.
We have set some formal principles and made the difficulties arise. In fact, the evaluation of non-social intelligence itself has not been fully achieved and it is still being investigated. The evaluation of social intelligence is still more convoluted.
This article provides the basis of how we can evaluate social intelligence in a formal way following these principles.


\section*{Acknowledgements}
This work was supported by the MEC projects EXPLORA-INGENIO TIN 2009-06078-E, CONSOLIDER-INGENIO 26706 and TINs 2010-21062-C02-02, 2013-45732-C4-1-P and GVA projects PROMETEO/2008/051 and PROMETEO/2011/052. Javier Insa-Cabrera was sponsored by Spanish MEC-FPU grant AP2010-4389.


\bibliography{biblio}

\section*{Appendix}
\appendix

Before starting with each of the environments, we will prove a lemma that will be helpful for the $Left$ and $Right$ ranges.

We could calculate $Left$ and $Right$ using $\Pi_e$ and $\Pi_o$ with a high number of agents. However, the more agents we include the more difficult the calculation becomes. Instead of this, and in order to simplify calculations, we can just use the minimum necessary number of agents in $\Pi_e$ and $\Pi_o$ for that property to obtain the maximum/minimum value following the idea on lemma \ref{lemma:minimum_necessary_agent_for_properties}:

\begin{lemma}
\label{lemma:minimum_necessary_agent_for_properties}
In order to calculate $Left$/$Right$ maximum/minimum value for a property $Prop$, the length of the set of evaluated agents $|\Pi_e|$ and the length of the set of opponents and team players $|\Pi_o|$ can be respectively equal to the minimum number of evaluated agents $n$ and agents to populate the environment $m$ needed to calculate $Prop$.

\begin{proof}
Let $\Pi_e = \{\pi_1,\dots,\pi_n,\dots,\pi_p\}$ be the set of agents evaluated with weight $w_{\Pi_e}$ in an environment $\mu$ with weight of slots $w_S$ using a set of opponents and team players $\Pi_o$ and $w_{\dot{L}}$ as a weight for line-up patterns.

Let us suppose that we want to calculate the value for a property $Prop$ which needs $n$ evaluated agents to be defined, its definition calculates first the value for each evaluated agent $\pi$ and then these values are weighted using $w_{\Pi_e}(\pi)$ to provide the property value. Following this definition we obtain a list of values $(v_1,\dots,v_n,\dots,v_p)$ (one for each evaluated agent) that will be weighted with $w_{\Pi_e}$ to obtain the property value $v$. If we get rid of the evaluated agent which obtains the maximum value for $Prop$ and we rearrange $w_{\Pi_e}$ to sum $1$, then the new property value $v'$ will always be lower than $v$. We can repeat this process until $n$ agents remain in $\Pi_e$ (i.e., $|\Pi_e| = n$) to obtain $v_{min}$ for this set of evaluated agents. An analogous process applies to obtaining $v_{max}$ by getting rid of the evaluated agent which obtains the minimum value for $Prop$.

The same reasoning applies to the properties that calculate the values for pairs of evaluated agents, but in this case we get rid of the agent whose sum of values (the values of the pairs where this agent is used) is highest/lowest. Also, the same reasoning applies for $\Pi_o$.
\end{proof}
\end{lemma}

\section{Matching Pennies properties}
\label{appen:matching_pennies_properties}
In this section we prove how we obtained the values for the properties for the matching pennies environment (section \ref{sec:matching_pennies}). To calculate some of the values for the properties, we make use of lemma \ref{lemma:matching_pennies_random_agent}.

\begin{lemma}
\label{lemma:matching_pennies_random_agent}
In the matching pennies environment and for any slot, introducing a random agent $\pi_r$ in a line-up will always provide an expected average reward equal to $0$ for both agents.

\begin{proof}
A random agent $\pi_r$ has a probability of $p^j_{r,h} = p^j_{r,t} = \frac{1}{2}$ to perform both Head and Tail at iteration $j$. Let us denote with $\pi_s$ the agent that $\pi_r$ is interacting with, and denote with $p^1_{s,h}$ the probability of performing Head and $p^1_{s,t} = 1 - p^1_{s,h}$ the probability of performing Tail at the first iteration for $\pi_s$.

To calculate the expected reward of an agent, we sum the possible rewards that this agent can obtain multiplied by the probability that these rewards occurs. When we calculate the expected reward for $\pi_r$ for the first iteration in the matching pennies environment $\mu$ in any slot $i$, we obtain:

\begin{equation*}
\forall i : R^1_i(\pi_r,\mu) = p^1_{r,h} \left(p^1_{s,h} \times r^1_{h,h,i} + p^1_{s,t} \times r^1_{h,t,i}\right) + p^1_{r,t} \left(p^1_{s,h} \times r^1_{t,h,i} + p^1_{s,t} \times r^1_{t,t,i}\right)
\end{equation*}

\noindent where $r^j_{a_1,a_2,i}$ is the reward that the agent in slot $i$ obtains at iteration $j$ when one agent performs $a_1$ and the other agent performs $a_2$.

From the matching pennies' payoff matrix (figure \ref{fig:matching_pennies_payoff}), we can see that for every slot $i$, $r_{h,h,i} = r_{t,t,i}$ and $r_{h,t,i} = r_{t,h,i}$, so we name them $r_{e,i}$ and $r_{d,i}$ respectively. We can also see that the reward values are the inverse of each other, having $r_{d,i} = -r_{e,i}$. Renaming the rewards in the formula and rearranging it we obtain:

\begin{equation*}
\begin{aligned}
\forall i : R^1_i(\pi_r,\mu)	& = p^1_{r,h} \left(p^1_{s,h} \times r_{e,i} + p^1_{s,t} \times r_{d,i}\right) + p^1_{r,t} \left(p^1_{s,h} \times r_{d,i} + p^1_{s,t} \times r_{e,i}\right) =\\
								& = p^1_{r,h} \left(p^1_{s,h} \times r_{e,i} + p^1_{s,t} \times (-r_{e,i})\right) + p^1_{r,t} \left(p^1_{s,h} \times (-r_{e,i}) + p^1_{s,t} \times r_{e,i}\right) =\\
								& = p^1_{r,h} \left(r_{e,i} \times \left(p^1_{s,h} - p^1_{s,t}\right)\right) + p^1_{r,t} \left((-r_{e,i}) \times \left(p^1_{s,h} - p^1_{s,t}\right)\right) =\\
								& = p^1_{r,h} \left(r_{e,i} \times \left(p^1_{s,h} - p^1_{s,t}\right)\right) - p^1_{r,t} \left(r_{e,i} \times \left(p^1_{s,h} - p^1_{s,t}\right)\right) =\\
								& = \left(p^1_{r,h} - p^1_{r,t}\right) \times \left(r_{e,i} \times \left(p^1_{s,h} - p^1_{s,t}\right)\right)
\end{aligned}
\end{equation*}

And since $\pi_r$ gives the same probability to both Head and Tail ($p^1_{r,h} = p^1_{r,t} = \frac{1}{2}$) we obtain the following expected reward:

\begin{equation*}
\forall i : R^1_i(\pi_r,\mu) = \left(\frac{1}{2} - \frac{1}{2}\right) \times \left(r_{e,i} \times \left(p^1_{s,h} - p^1_{s,t}\right)\right) = 0
\end{equation*}

We calculated the expected reward for the first iteration. At this point $\pi_s$ could change its behaviour depending on what happened on the previous iteration, using different probabilities $p^2_{s,h}$ and $p^2_{s,t}$ for iteration 2. But note that it does not matter which probabilities $p^n_{s,h}$ and $p^n_{s,t}$ we use, the result will still be $0$, and averaging over the iterations we obtain an expected average reward equal to $0$. Obviously, since this is a zero-sum game, when $\pi_r$ obtains an expected average reward of $0$, $\pi_s$ obtains the same expected average reward of $0$.
\end{proof}
\end{lemma}

\subsection{Action Dependency}
We start with the action dependency (AD) property. As given in section \ref{sec:AD}, we want to know if the evaluated agents behave differently depending on which line-up they interact with. We use $\Delta_S(a,b) = 1$ if distributions $a$ and $b$ are equal and $0$ otherwise.

\begin{proposition}
\label{prop:matching_pennies_AD_general_min}
$General_{min}$ for the action dependency (AD) property is equal to $0$ for the matching pennies environment.

\begin{proof}
To find $General_{min}$ (equation \ref{eq:general_min}), we need to find a trio $\left\langle\Pi_e,w_{\Pi_e},\Pi_o\right\rangle$ which minimises the property as much as possible. We can have this situation by selecting $\Pi_e = \{\pi_t\}$ with $w_{\Pi_e}(\pi_t) = 1$ and $\Pi_o = \{{\pi_h}_1, {\pi_h}_2\}$ (a $\pi_h$ agent always performs Head and a $\pi_t$ agent always performs Tail).

Following definition \ref{def:AD}, we obtain the AD value for this $\left\langle\Pi_e,w_{\Pi_e},\Pi_o\right\rangle$. Since the environment is not symmetric, we need to calculate this property for every slot. Following definition \ref{def:AD_set}, we could calculate its AD value for each slot but, since $\Pi_e$ has only one agent and its weight is equal to $1$, it is equivalent to use directly definition \ref{def:AD_agent}. We start with slot 1:

\begin{equation*}
\begin{aligned}
AD_1(\pi_t,\Pi_o,w_{\dot{L}},\mu)	& = \eta_{\dot{L}^2} \sum_{\dot{u},\dot{v} \in \dot{L}^{N(\mu)}_{-1}(\Pi_o) | \dot{u} \neq \dot{v}} w_{\dot{L}}(\dot{u}) w_{\dot{L}}(\dot{v}) \Delta_S(\breve{A}_1(\mu[\instantiation{u}{1}{\pi_t}]), \breve{A}_1(\mu[\instantiation{v}{1}{\pi_t}])) =\\
									& = 2 \frac{2}{1} \frac{1}{2} \frac{1}{2} \Delta_S(\breve{A}_1(\mu[\pi_t,{\pi_h}_1]), \breve{A}_1(\mu[\pi_t,{\pi_h}_2]))
\end{aligned}
\end{equation*}

\noindent Note that we avoided to calculate both $\Delta_S(a,b)$ and $\Delta_S(b,a)$ since they provide the same result, by calculating only $\Delta_S(a,b)$ and multiplying the result by $2$.

In this case, we only need to calculate $\Delta_S(\breve{A}_1(\mu[\pi_t,\pi_{h_1}]), \breve{A}_1(\mu[\pi_t,\pi_{h_2}]))$, where the agent in both slots 1 ($\pi_t$) will perform the same sequence of actions (always Tail) independently of the line-up. So:

\begin{equation*}
AD_1(\pi_t,\Pi_o,w_{\dot{L}},\mu) = 2 \frac{2}{1} \frac{1}{2} \frac{1}{2} 0 = 0
\end{equation*}

And for slot 2, the agent in both slots 2 ($\pi_t$) will also perform the same sequence of actions (always Tail) independently of the line-up. So:

\begin{equation*}
\begin{aligned}
AD_2(\pi_t,\Pi_o,w_{\dot{L}},\mu)	& = \eta_{\dot{L}^2} \sum_{\dot{u},\dot{v} \in \dot{L}^{N(\mu)}_{-2}(\Pi_o) | \dot{u} \neq \dot{v}} w_{\dot{L}}(\dot{u}) w_{\dot{L}}(\dot{v}) \Delta_S(\breve{A}_2(\mu[\instantiation{u}{2}{\pi_t}]), \breve{A}_2(\mu[\instantiation{v}{2}{\pi_t}])) =\\
									& = 2 \frac{2}{1} \frac{1}{2} \frac{1}{2} \Delta_S(\breve{A}_2(\mu[{\pi_h}_1,\pi_t]), \breve{A}_2(\mu[{\pi_h}_2,\pi_t])) =\\
									& = 2 \frac{2}{1} \frac{1}{2} \frac{1}{2} 0 = 0
\end{aligned}
\end{equation*}

And finally, we weight over the slots:

\begin{equation*}
\begin{aligned}
AD(\Pi_e,w_{\Pi_e},\Pi_o,w_{\dot{L}},\mu,w_S)	& =	\sum_{i = 1}^{N(\mu)} w_S(i,\mu) AD_i(\Pi_e,w_{\Pi_e},\Pi_o,w_{\dot{L}},\mu) =\\
												& =	\sum_{i = 1}^{N(\mu)} w_S(i,\mu) AD_i(\pi_t,\Pi_o,w_{\dot{L}},\mu) =\\
												& = \frac{1}{2} \{AD_1(\pi_t,\Pi_o,w_{\dot{L}},\mu) + AD_2(\pi_t,\Pi_o,w_{\dot{L}},\mu)\} =\\
												& = \frac{1}{2} \left\{0 + 0\right\} = 0
\end{aligned}
\end{equation*}

Since $0$ is the lowest possible value for the action dependency property, therefore matching pennies has $General_{min} = 0$ for this property.
\end{proof}
\end{proposition}

\begin{proposition}
\label{prop:matching_pennies_AD_general_max}
$General_{max}$ for the action dependency (AD) property is equal to $1$ for the matching pennies environment.

\begin{proof}
To find $General_{max}$ (equation \ref{eq:general_max}), we need to find a trio $\left\langle\Pi_e,w_{\Pi_e},\Pi_o\right\rangle$ which maximises the property as much as possible. We can have this situation by selecting $\Pi_e = \{\pi_m\}$ with $w_{\Pi_e}(\pi_m) = 1$ and $\Pi_o = \{\pi_h, \pi_t\}$ (a $\pi_m$ agent first acts randomly and then always mimics the other agent's last action, a $\pi_h$ agent always performs Head and a $\pi_t$ agent always performs Tail).

Following definition \ref{def:AD}, we obtain the AD value for this $\left\langle\Pi_e,w_{\Pi_e},\Pi_o\right\rangle$. Since the environment is not symmetric, we need to calculate this property for every slot. Following definition \ref{def:AD_set}, we could calculate its AD value for each slot but, since $\Pi_e$ has only one agent and its weight is equal to $1$, it is equivalent to use directly definition \ref{def:AD_agent}. We start with slot 1:

\begin{equation*}
\begin{aligned}
AD_1(\pi_m,\Pi_o,w_{\dot{L}},\mu)	& = \eta_{\dot{L}^2} \sum_{\dot{u},\dot{v} \in \dot{L}^{N(\mu)}_{-1}(\Pi_o) | \dot{u} \neq \dot{v}} w_{\dot{L}}(\dot{u}) w_{\dot{L}}(\dot{v}) \Delta_S(\breve{A}_1(\mu[\instantiation{u}{1}{\pi_m}]), \breve{A}_1(\mu[\instantiation{v}{1}{\pi_m}])) =\\
									& = 2 \frac{2}{1} \frac{1}{2} \frac{1}{2} \Delta_S(\breve{A}_1(\mu[\pi_m,\pi_h]), \breve{A}_1(\mu[\pi_m,\pi_t]))
\end{aligned}
\end{equation*}

\noindent Note that we avoided to calculate both $\Delta_S(a,b)$ and $\Delta_S(b,a)$ since they provide the same result, by calculating only $\Delta_S(a,b)$ and multiplying the result by $2$.

From iteration 2, $\pi_m$ will mimic the last action of the agent in slot $2$, and since $\pi_h$ will always perform Head and $\pi_t$ will always perform Tail, the agent in both slots 1 ($\pi_m$) will perform different sequences of actions on each line-up. So:

\begin{equation*}
AD_1(\pi_m,\Pi_o,w_{\dot{L}},\mu) = 2 \frac{2}{1} \frac{1}{2} \frac{1}{2} 1 = 1
\end{equation*}

And for slot 2, the agent in both slots 2 ($\pi_m$) will also perform different sequences of actions on each line-up. So:

\begin{equation*}
\begin{aligned}
AD_2(\pi_m,\Pi_o,w_{\dot{L}},\mu)	& = \eta_{\dot{L}^2} \sum_{\dot{u},\dot{v} \in \dot{L}^{N(\mu)}_{-2}(\Pi_o) | \dot{u} \neq \dot{v}} w_{\dot{L}}(\dot{u}) w_{\dot{L}}(\dot{v}) \Delta_S(\breve{A}_2(\mu[\instantiation{u}{2}{\pi_m}]), \breve{A}_2(\mu[\instantiation{v}{2}{\pi_m}])) =\\
									& = 2 \frac{2}{1} \frac{1}{2} \frac{1}{2} \Delta_S(\breve{A}_2(\mu[\pi_h,\pi_m]), \breve{A}_2(\mu[\pi_t,\pi_m])) =\\
									& = 2 \frac{2}{1} \frac{1}{2} \frac{1}{2} 1 = 1
\end{aligned}
\end{equation*}

And finally, we weight over the slots:

\begin{equation*}
\begin{aligned}
AD(\Pi_e,w_{\Pi_e},\Pi_o,w_{\dot{L}},\mu,w_S)	& =	\sum_{i = 1}^{N(\mu)} w_S(i,\mu) AD_i(\Pi_e,w_{\Pi_e},\Pi_o,w_{\dot{L}},\mu) =\\
												& =	\sum_{i = 1}^{N(\mu)} w_S(i,\mu) AD_i(\pi_m,\Pi_o,w_{\dot{L}},\mu) =\\
												& = \frac{1}{2} \{AD_1(\pi_m,\Pi_o,w_{\dot{L}},\mu) + AD_2(\pi_m,\Pi_o,w_{\dot{L}},\mu)\} =\\
												& = \frac{1}{2} \left\{1 + 1\right\} = 1
\end{aligned}
\end{equation*}

Since $1$ is the highest possible value for the action dependency property, therefore matching pennies has $General_{max} = 1$ for this property.
\end{proof}
\end{proposition}

\begin{proposition}
\label{prop:matching_pennies_AD_left_max}
$Left_{max}$ for the action dependency (AD) property is equal to $0$ for the matching pennies environment.

\begin{proof}
To find $Left_{max}$ (equation \ref{eq:left_max}), we need to find a pair $\left\langle\Pi_e,w_{\Pi_e}\right\rangle$ which maximises the property as much as possible while $\Pi_o$ minimises it. Using $\Pi_o = \{{\pi_h}_1,{\pi_h}_2\}$ (a $\pi_h$ agent always performs Head) we find this situation no matter which pair $\left\langle\Pi_e,w_{\Pi_e}\right\rangle$ we use.

Following definition \ref{def:AD}, we obtain the AD value for this $\left\langle\Pi_e,w_{\Pi_e},\Pi_o\right\rangle$ (where $\Pi_e$ and $w_{\Pi_e}$ are instantiated with any permitted values). Since the environment is not symmetric, we need to calculate this property for every slot. Following definition \ref{def:AD_set}, we can calculate its AD value for each slot. We start with slot 1:

\begin{equation*}
AD_1(\Pi_e,w_{\Pi_e},\Pi_o,w_{\dot{L}},\mu) = \sum_{\pi \in \Pi_e} w_{\Pi_e}(\pi) AD_1(\pi,\Pi_o,w_{\dot{L}},\mu)
\end{equation*}

We do not know which $\Pi_e$ we have, but we know that we will need to evaluate $AD_1(\pi,\Pi_o,w_{\dot{L}},\mu)$ for all evaluated agent $\pi \in \Pi_e$. We follow definition \ref{def:AD_agent} to calculate this value for a figurative evaluated agent $\pi$ from $\Pi_e$:

\begin{equation*}
\begin{aligned}
AD_1(\pi,\Pi_o,w_{\dot{L}},\mu)	& = \eta_{\dot{L}^2} \sum_{\dot{u},\dot{v} \in \dot{L}^{N(\mu)}_{-1}(\Pi_o) | \dot{u} \neq \dot{v}} w_{\dot{L}}(\dot{u}) w_{\dot{L}}(\dot{v}) \Delta_S(\breve{A}_1(\mu[\instantiation{u}{1}{\pi}]), \breve{A}_1(\mu[\instantiation{v}{1}{\pi}])) =\\
								& = 2 \frac{2}{1} \frac{1}{2} \frac{1}{2} \Delta_S(\breve{A}_1(\mu[\pi,{\pi_h}_1]), \breve{A}_1(\mu[\pi,{\pi_h}_2]))
\end{aligned}
\end{equation*}

\noindent Note that we avoided to calculate both $\Delta_S(a,b)$ and $\Delta_S(b,a)$ since they provide the same result, by calculating only $\Delta_S(a,b)$ and multiplying the result by $2$.

A $\pi_h$ agent will always perform Head, so we obtain a situation where the agent in both slots 1 (any $\pi$) will be able to differentiate with which agent is interacting, so it will not be able to change its distribution of action sequences depending on the opponent's behaviour. So:

\begin{equation*}
AD_1(\pi,\Pi_o,w_{\dot{L}},\mu) = 2 \frac{2}{1} \frac{1}{2} \frac{1}{2} 0 = 0
\end{equation*}

Therefore, no matter which agents are in $\Pi_e$ and their weights $w_{\Pi_e}$ we obtain:

\begin{equation*}
AD_1(\Pi_e,w_{\Pi_e},\Pi_o,w_{\dot{L}},\mu) = 0
\end{equation*}

And for slot 2:

\begin{equation*}
AD_2(\Pi_e,w_{\Pi_e},\Pi_o,w_{\dot{L}},\mu) = \sum_{\pi \in \Pi_e} w_{\Pi_e}(\pi) AD_2(\pi,\Pi_o,w_{\dot{L}},\mu)
\end{equation*}

Again, we do not know which $\Pi_e$ we have, but we know that we will need to evaluate $AD_2(\pi,\Pi_o,w_{\dot{L}},\mu)$ for all evaluated agent $\pi \in \Pi_e$. We follow definition \ref{def:AD_agent} to calculate this value for a figurative evaluated agent $\pi$ from $\Pi_e$:

\begin{equation*}
\begin{aligned}
AD_2(\pi,\Pi_o,w_{\dot{L}},\mu)	& = \eta_{\dot{L}^2} \sum_{\dot{u},\dot{v} \in \dot{L}^{N(\mu)}_{-2}(\Pi_o) | \dot{u} \neq \dot{v}} w_{\dot{L}}(\dot{u}) w_{\dot{L}}(\dot{v}) \Delta_S(\breve{A}_2(\mu[\instantiation{u}{2}{\pi}]), \breve{A}_2(\mu[\instantiation{v}{2}{\pi}])) =\\
								& = 2 \frac{2}{1} \frac{1}{2} \frac{1}{2} \Delta_S(\breve{A}_2(\mu[{\pi_h}_1,\pi]), \breve{A}_2(\mu[{\pi_h}_2,\pi]))
\end{aligned}
\end{equation*}

Again, a $\pi_h$ agent will always perform Head, so we obtain a situation where the agent in both slots 2 (any $\pi$) will be able to differentiate with which agent is interacting, so it will not be able to change its distribution of action sequences depending on the opponent's behaviour. So:

\begin{equation*}
AD_2(\pi,\Pi_o,w_{\dot{L}},\mu) = 2 \frac{2}{1} \frac{1}{2} \frac{1}{2} 0 = 0
\end{equation*}

Therefore, no matter which agents are in $\Pi_e$ and their weights $w_{\Pi_e}$ we obtain:

\begin{equation*}
AD_2(\Pi_e,w_{\Pi_e},\Pi_o,w_{\dot{L}},\mu) = 0
\end{equation*}

And finally, we weight over the slots:

\begin{equation*}
\begin{aligned}
AD(\Pi_e,w_{\Pi_e},\Pi_o,w_{\dot{L}},\mu,w_S)	& =	\sum_{i = 1}^{N(\mu)} w_S(i,\mu) AD_i(\Pi_e,w_{\Pi_e},\Pi_o,w_{\dot{L}},\mu) =\\
												& = \frac{1}{2} \{AD_1(\Pi_e,w_{\Pi_e},\Pi_o,w_{\dot{L}},\mu) + AD_2(\Pi_e,w_{\Pi_e},\Pi_o,w_{\dot{L}},\mu)\} =\\
												& = \frac{1}{2} \left\{0 + 0\right\} = 0
\end{aligned}
\end{equation*}

So, for every pair $\left\langle\Pi_e,w_{\Pi_e}\right\rangle$ we obtain the same result:

\begin{equation*}
\forall \Pi_e,w_{\Pi_e} : AD(\Pi_e,w_{\Pi_e},\Pi_o,w_{\dot{L}},\mu,w_S) = 0
\end{equation*}

Therefore, matching pennies has $Left_{max} = 0$ for this property.
\end{proof}
\end{proposition}

\begin{proposition}
\label{prop:matching_pennies_AD_right_min}
$Right_{min}$ for the action dependency (AD) property is equal to $0$ for the matching pennies environment.

\begin{proof}
To find $Right_{min}$ (equation \ref{eq:right_min}), we need to find a pair $\left\langle\Pi_e,w_{\Pi_e}\right\rangle$ which minimises the property as much as possible while $\Pi_o$ maximises it. Using $\Pi_e = \{\pi_h\}$ with $w_{\Pi_e}(\pi_h) = 1$ (a $\pi_h$ agent always performs Head) we find this situation no matter which $\Pi_o$ we use.

Following definition \ref{def:AD}, we obtain the AD value for this $\left\langle\Pi_e,w_{\Pi_e},\Pi_o\right\rangle$ (where $\Pi_o$ is instantiated with any permitted value). Since the environment is not symmetric, we need to calculate this property for every slot. Following definition \ref{def:AD_set}, we could calculate its AD value for each slot but, since $\Pi_e$ has only one agent and its weight is equal to $1$, it is equivalent to use directly definition \ref{def:AD_agent}. We start with slot 1:

\begin{equation*}
AD_1(\pi_h,\Pi_o,w_{\dot{L}},\mu) = \eta_{\dot{L}^2} \sum_{\dot{u},\dot{v} \in \dot{L}^{N(\mu)}_{-1}(\Pi_o) | \dot{u} \neq \dot{v}} w_{\dot{L}}(\dot{u}) w_{\dot{L}}(\dot{v}) \Delta_S(\breve{A}_1(\mu[\instantiation{u}{1}{\pi_h}]), \breve{A}_1(\mu[\instantiation{v}{1}{\pi_h}]))
\end{equation*}

We do not know which $\Pi_o$ we have, but we know that we will need to obtain two different line-up patterns $\dot{u}$ and $\dot{v}$ from $\dot{L}^{N(\mu)}_{-1}(\Pi_o)$ to calculate $\Delta_S(\breve{A}_1(\mu[\instantiation{u}{1}{\pi_h}]), \breve{A}_1(\mu[\instantiation{v}{1}{\pi_h}]))$. We calculate this value for two figurative line-up patterns $\dot{u} = (*,\pi_1)$ and $\dot{v} = (*,\pi_2)$ from $\dot{L}^{N(\mu)}_{-1}(\Pi_o)$:

\begin{equation*}
\Delta_S(\breve{A}_1(\mu[\instantiation{u}{1}{\pi_h}]), \breve{A}_1(\mu[\instantiation{v}{1}{\pi_h}])) = \Delta_S(\breve{A}_1(\mu[\pi_h,\pi_1]), \breve{A}_1(\mu[\pi_h,\pi_2]))
\end{equation*}

Here, the agent in both slots 1 ($\pi_h$) will perform the same sequence of actions (always Head) independently of the line-up. So no matter which agents are in $\Pi_o$ we obtain:

\begin{equation*}
AD_1(\pi_h,\Pi_o,w_{\dot{L}},\mu) = 0
\end{equation*}

And for slot 2:

\begin{equation*}
AD_2(\pi_h,\Pi_o,w_{\dot{L}},\mu) = \eta_{\dot{L}^2} \sum_{\dot{u},\dot{v} \in \dot{L}^{N(\mu)}_{-2}(\Pi_o) | \dot{u} \neq \dot{v}} w_{\dot{L}}(\dot{u}) w_{\dot{L}}(\dot{v}) \Delta_S(\breve{A}_2(\mu[\instantiation{u}{2}{\pi_h}]), \breve{A}_2(\mu[\instantiation{v}{2}{\pi_h}]))
\end{equation*}

Again, we do not know which $\Pi_o$ we have, but we know that we will need to obtain two different line-up patterns $\dot{u}$ and $\dot{v}$ from $\dot{L}^{N(\mu)}_{-2}(\Pi_o)$ to calculate $\Delta_S(\breve{A}_2(\mu[\instantiation{u}{2}{\pi_h}]), \breve{A}_2(\mu[\instantiation{v}{2}{\pi_h}]))$. We calculate this value for two figurative line-up patterns $\dot{u} = (*,\pi_1)$ and $\dot{v} = (*,\pi_2)$ from $\dot{L}^{N(\mu)}_{-2}(\Pi_o)$:

\begin{equation*}
\Delta_S(\breve{A}_2(\mu[\instantiation{u}{2}{\pi_h}]), \breve{A}_2(\mu[\instantiation{v}{2}{\pi_h}])) = \Delta_S(\breve{A}_2(\mu[\pi_1,\pi_h]), \breve{A}_2(\mu[\pi_2,\pi_h]))
\end{equation*}

Again, the agent in both slots 2 ($\pi_h$) will perform the same sequence of actions (always Head) independently of the line-up. So no matter which agents are in $\Pi_o$ we obtain:

\begin{equation*}
AD_2(\pi_h,\Pi_o,w_{\dot{L}},\mu) = 0
\end{equation*}

And finally, we weight over the slots:

\begin{equation*}
\begin{aligned}
AD(\Pi_e,w_{\Pi_e},\Pi_o,w_{\dot{L}},\mu,w_S)	& =	\sum_{i = 1}^{N(\mu)} w_S(i,\mu) AD_i(\Pi_e,w_{\Pi_e},\Pi_o,w_{\dot{L}},\mu) =\\
												& =	\sum_{i = 1}^{N(\mu)} w_S(i,\mu) AD_i(\pi_h,\Pi_o,w_{\dot{L}},\mu) =\\
												& = \frac{1}{2} \{AD_1(\pi_h,\Pi_o,w_{\dot{L}},\mu) + AD_2(\pi_h,\Pi_o,w_{\dot{L}},\mu)\} =\\
												& = \frac{1}{2} \left\{0 + 0\right\} = 0
\end{aligned}
\end{equation*}

So, for every $\Pi_o$ we obtain the same result:

\begin{equation*}
\forall \Pi_o : AD(\Pi_e,w_{\Pi_e},\Pi_o,w_{\dot{L}},\mu,w_S) = 0
\end{equation*}

Therefore, matching pennies has $Right_{min} = 0$ for this property.
\end{proof}
\end{proposition}

\subsection{Reward Dependency}
We continue with the reward dependency (RD) property. As given in section \ref{sec:RD}, we want to know if the evaluated agents obtain different expected average rewards depending on which line-up they interact with. We use $\Delta_{\mathbb{Q}}(a,b) = 1$ if numbers $a$ and $b$ are equal and $0$ otherwise.

\begin{proposition}
\label{prop:matching_pennies_RD_general_min}
$General_{min}$ for the reward dependency (RD) property is equal to $0$ for the matching pennies environment.

\begin{proof}
To find $General_{min}$ (equation \ref{eq:general_min}), we need to find a trio $\left\langle\Pi_e,w_{\Pi_e},\Pi_o\right\rangle$ which minimises the property as much as possible. We can have this situation by selecting $\Pi_e = \{\pi_t\}$ with $w_{\Pi_e}(\pi_t) = 1$ and $\Pi_o = \{{\pi_h}_1, {\pi_h}_2\}$ (a $\pi_h$ agent always performs Head and a $\pi_t$ agent always performs Tail).

Following definition \ref{def:RD}, we obtain the RD value for this $\left\langle\Pi_e,w_{\Pi_e},\Pi_o\right\rangle$. Since the environment is not symmetric, we need to calculate this property for every slot. Following definition \ref{def:RD_set}, we could calculate its RD value for each slot but, since $\Pi_e$ has only one agent and its weight is equal to $1$, it is equivalent to use directly definition \ref{def:RD_agent}. We start with slot 1:

\begin{equation*}
\begin{aligned}
RD_1(\pi_t,\Pi_o,w_{\dot{L}},\mu)	& = \eta_{\dot{L}^2} \sum_{\dot{u},\dot{v} \in \dot{L}^{N(\mu)}_{-1}(\Pi_o) | \dot{u} \neq \dot{v}} w_{\dot{L}}(\dot{u}) w_{\dot{L}}(\dot{v}) \Delta_{\mathbb{Q}}(R_1(\mu[\instantiation{u}{1}{\pi_t}]), R_1(\mu[\instantiation{v}{1}{\pi_t}])) =\\
									& = 2 \frac{2}{1} \frac{1}{2} \frac{1}{2} \Delta_{\mathbb{Q}}(R_1(\mu[\pi_t,{\pi_h}_1]), R_1(\mu[\pi_t,{\pi_h}_2]))
\end{aligned}
\end{equation*}

\noindent Note that we avoided to calculate both $\Delta_{\mathbb{Q}}(a,b)$ and $\Delta_{\mathbb{Q}}(b,a)$ since they provide the same result, by calculating only $\Delta_{\mathbb{Q}}(a,b)$ and multiplying the result by $2$.

In this case, we only need to calculate $\Delta_{\mathbb{Q}}(R_1(\mu[\pi_t,{\pi_h}_1]), R_1(\mu[\pi_t,{\pi_h}_2]))$, where $\pi_t$ will always perform Tail and a $\pi_h$ agent will always perform Head, so the agent in both slots 1 ($\pi_t$) will obtain the same expected average reward ($-1$) independently of the line-up. So:

\begin{equation*}
RD_1(\pi_t,\Pi_o,w_{\dot{L}},\mu) = 2 \frac{2}{1} \frac{1}{2} \frac{1}{2} 0 = 0
\end{equation*}

And for slot 2, the agent in both slots 2 ($\pi_t$) will also obtain the same expected average reward ($1$) independently of the line-up. So:

\begin{equation*}
\begin{aligned}
RD_2(\pi_t,\Pi_o,w_{\dot{L}},\mu)	& = \eta_{\dot{L}^2} \sum_{\dot{u},\dot{v} \in \dot{L}^{N(\mu)}_{-2}(\Pi_o) | \dot{u} \neq \dot{v}} w_{\dot{L}}(\dot{u}) w_{\dot{L}}(\dot{v}) \Delta_{\mathbb{Q}}(R_2(\mu[\instantiation{u}{2}{\pi_t}]), R_2(\mu[\instantiation{v}{2}{\pi_t}])) =\\
									& = 2 \frac{2}{1} \frac{1}{2} \frac{1}{2} \Delta_{\mathbb{Q}}(R_2(\mu[{\pi_h}_1,\pi_t]), R_2(\mu[{\pi_h}_2,\pi_t])) =\\
									& = 2 \frac{2}{1} \frac{1}{2} \frac{1}{2} 0 = 0
\end{aligned}
\end{equation*}

And finally, we weight over the slots:

\begin{equation*}
\begin{aligned}
RD(\Pi_e,w_{\Pi_e},\Pi_o,w_{\dot{L}},\mu,w_S)	& =	\sum_{i = 1}^{N(\mu)} w_S(i,\mu) RD_i(\Pi_e,w_{\Pi_e},\Pi_o,w_{\dot{L}},\mu) =\\
												& =	\sum_{i = 1}^{N(\mu)} w_S(i,\mu) RD_i(\pi_t,\Pi_o,w_{\dot{L}},\mu) =\\
												& = \frac{1}{2} \{RD_1(\pi_t,\Pi_o,w_{\dot{L}},\mu) + RD_2(\pi_t,\Pi_o,w_{\dot{L}},\mu)\} =\\
												& = \frac{1}{2} \left\{0 + 0\right\} = 0
\end{aligned}
\end{equation*}

Since $0$ is the lowest possible value for the reward dependency property, therefore matching pennies has $General_{min} = 0$ for this property.
\end{proof}
\end{proposition}

\begin{proposition}
\label{prop:matching_pennies_RD_general_max}
$General_{max}$ for the reward dependency (RD) property is equal to $1$ for the matching pennies environment.

\begin{proof}
To find $General_{max}$ (equation \ref{eq:general_max}), we need to find a trio $\left\langle\Pi_e,w_{\Pi_e},\Pi_o\right\rangle$ which maximises the property as much as possible. We can have this situation by selecting $\Pi_e = \{\pi_t\}$ with $w_{\Pi_e}(\pi_t) = 1$ and $\Pi_o = \{\pi_h, \pi_t\}$ (a $\pi_h$ agent always performs Head and a $\pi_t$ agent always performs Tail).

Following definition \ref{def:RD}, we obtain the RD value for this $\left\langle\Pi_e,w_{\Pi_e},\Pi_o\right\rangle$. Since the environment is not symmetric, we need to calculate this property for every slot. Following definition \ref{def:RD_set}, we could calculate its RD value for each slot but, since $\Pi_e$ has only one agent and its weight is equal to $1$, it is equivalent to use directly definition \ref{def:RD_agent}. We start with slot 1:

\begin{equation*}
\begin{aligned}
RD_1(\pi_t,\Pi_o,w_{\dot{L}},\mu)	& = \eta_{\dot{L}^2} \sum_{\dot{u},\dot{v} \in \dot{L}^{N(\mu)}_{-1}(\Pi_o) | \dot{u} \neq \dot{v}} w_{\dot{L}}(\dot{u}) w_{\dot{L}}(\dot{v}) \Delta_{\mathbb{Q}}(R_1(\mu[\instantiation{u}{1}{\pi_t}]), R_1(\mu[\instantiation{v}{1}{\pi_t}])) =\\
									& = 2 \frac{2}{1} \frac{1}{2} \frac{1}{2} \Delta_{\mathbb{Q}}(R_1(\mu[\pi_t,\pi_h]), R_1(\mu[\pi_t,\pi_t]))
\end{aligned}
\end{equation*}

\noindent Note that we avoided to calculate both $\Delta_{\mathbb{Q}}(a,b)$ and $\Delta_{\mathbb{Q}}(b,a)$ since they provide the same result, by calculating only $\Delta_{\mathbb{Q}}(a,b)$ and multiplying the result by $2$.

In line-up $(\pi_t,\pi_h)$, where $\pi_t$ will always perform Tail and $\pi_h$ will always perform Head, the agent in slot 1 ($\pi_t$) will obtain one expected average reward ($-1$), while in line-up $(\pi_t,\pi_t)$, where both $\pi_t$ will always perform Tail, the agent in slot 1 ($\pi_t$) will obtain a different expected average reward ($1$). So:

\begin{equation*}
RD_1(\pi_t,\Pi_o,w_{\dot{L}},\mu) = 2 \frac{2}{1} \frac{1}{2} \frac{1}{2} 1 = 1
\end{equation*}

And for slot 2, the agent in slot 2 ($\pi_t$) will also obtain different expected average rewards depending on the line-up. So:

\begin{equation*}
\begin{aligned}
RD_2(\pi_t,\Pi_o,w_{\dot{L}},\mu)	& = \eta_{\dot{L}^2} \sum_{\dot{u},\dot{v} \in \dot{L}^{N(\mu)}_{-2}(\Pi_o) | \dot{u} \neq \dot{v}} w_{\dot{L}}(\dot{u}) w_{\dot{L}}(\dot{v}) \Delta_{\mathbb{Q}}(R_2(\mu[\instantiation{u}{2}{\pi_t}]), R_2(\mu[\instantiation{v}{2}{\pi_t}])) =\\
									& = 2 \frac{2}{1} \frac{1}{2} \frac{1}{2} \Delta_{\mathbb{Q}}(R_2(\mu[\pi_h,\pi_t]), R_2(\mu[\pi_t,\pi_t]))\\
									& = 2 \frac{2}{1} \frac{1}{2} \frac{1}{2} 1 = 1
\end{aligned}
\end{equation*}

And finally, we weight over the slots:

\begin{equation*}
\begin{aligned}
RD(\Pi_e,w_{\Pi_e},\Pi_o,w_{\dot{L}},\mu,w_S)	& =	\sum_{i = 1}^{N(\mu)} w_S(i,\mu) RD_i(\Pi_e,w_{\Pi_e},\Pi_o,w_{\dot{L}},\mu) =\\
												& =	\sum_{i = 1}^{N(\mu)} w_S(i,\mu) RD_i(\pi_t,\Pi_o,w_{\dot{L}},\mu) =\\
												& = \frac{1}{2} \{RD_1(\pi_t,\Pi_o,w_{\dot{L}},\mu) + RD_2(\pi_t,\Pi_o,w_{\dot{L}},\mu)\} =\\
												& = \frac{1}{2} \left\{1 + 1\right\} = 1
\end{aligned}
\end{equation*}

Since $1$ is the highest possible value for the reward dependency property, therefore matching pennies has $General_{max} = 1$ for this property.
\end{proof}
\end{proposition}

\begin{proposition}
\label{prop:matching_pennies_RD_left_max}
$Left_{max}$ for the reward dependency (RD) property is equal to $0$ for the matching pennies environment.

\begin{proof}
To find $Left_{max}$ (equation \ref{eq:left_max}), we need to find a pair $\left\langle\Pi_e,w_{\Pi_e}\right\rangle$ which maximises the property as much as possible while $\Pi_o$ minimises it. Using $\Pi_o = \{{\pi_h}_1,{\pi_h}_2\}$ (a $\pi_h$ agent always performs Head) we find this situation no matter which pair $\left\langle\Pi_e,w_{\Pi_e}\right\rangle$ we use.

Following definition \ref{def:RD}, we obtain the RD value for this $\left\langle\Pi_e,w_{\Pi_e},\Pi_o\right\rangle$ (where $\Pi_e$ and $w_{\Pi_e}$ are instantiated with any permitted values). Since the environment is not symmetric, we need to calculate this property for every slot. Following definition \ref{def:RD_set}, we can calculate its RD value for each slot. We start with slot 1:

\begin{equation*}
RD_1(\Pi_e,w_{\Pi_e},\Pi_o,w_{\dot{L}},\mu) = \sum_{\pi \in \Pi_e} w_{\Pi_e}(\pi) RD_1(\pi,\Pi_o,w_{\dot{L}},\mu)
\end{equation*}

We do not know which $\Pi_e$ we have, but we know that we will need to evaluate $RD_1(\pi,\Pi_o,w_{\dot{L}},\mu)$ for all evaluated agent $\pi \in \Pi_e$. We follow definition \ref{def:RD_agent} to calculate this value for a figurative evaluated agent $\pi$ from $\Pi_e$:

\begin{equation*}
\begin{aligned}
RD_1(\pi,\Pi_o,w_{\dot{L}},\mu)	& = \eta_{\dot{L}^2} \sum_{\dot{u},\dot{v} \in \dot{L}^{N(\mu)}_{-1}(\Pi_o) | \dot{u} \neq \dot{v}} w_{\dot{L}}(\dot{u}) w_{\dot{L}}(\dot{v}) \Delta_{\mathbb{Q}}(R_1(\mu[\instantiation{u}{1}{\pi}]), R_1(\mu[\instantiation{v}{1}{\pi}])) =\\
								& = 2 \frac{2}{1} \frac{1}{2} \frac{1}{2} \Delta_{\mathbb{Q}}(R_1(\mu[\pi,{\pi_h}_1]), R_1(\mu[\pi,{\pi_h}_2]))
\end{aligned}
\end{equation*}

\noindent Note that we avoided to calculate both $\Delta_{\mathbb{Q}}(a,b)$ and $\Delta_{\mathbb{Q}}(b,a)$ since they provide the same result, by calculating only $\Delta_{\mathbb{Q}}(a,b)$ and multiplying the result by $2$.

A $\pi_h$ agent will always perform Head, so we obtain a situation where the agent in both slots 1 (any $\pi$) will be able to differentiate with which agent is interacting, so it will not be able to change its distribution of action sequences depending on the opponent's behaviour, obtaining agent in both slots 1 (any $\pi$) the same expected average reward. So:

\begin{equation*}
RD_1(\pi,\Pi_o,w_{\dot{L}},\mu) = 2 \frac{2}{1} \frac{1}{2} \frac{1}{2} 0 = 0
\end{equation*}

Therefore, no matter which agents are in $\Pi_e$ and their weights $w_{\Pi_e}$ we obtain:

\begin{equation*}
RD_1(\Pi_e,w_{\Pi_e},\Pi_o,w_{\dot{L}},\mu) = 0
\end{equation*}

And for slot 2:

\begin{equation*}
RD_2(\Pi_e,w_{\Pi_e},\Pi_o,w_{\dot{L}},\mu) = \sum_{\pi \in \Pi_e} w_{\Pi_e}(\pi) RD_2(\pi,\Pi_o,w_{\dot{L}},\mu)
\end{equation*}

Again, we do not know which $\Pi_e$ we have, but we know that we will need to evaluate $RD_2(\pi,\Pi_o,w_{\dot{L}},\mu)$ for all evaluated agent $\pi \in \Pi_e$. We follow definition \ref{def:RD_agent} to calculate this value for a figurative evaluated agent $\pi$ from $\Pi_e$:

\begin{equation*}
\begin{aligned}
RD_2(\pi,\Pi_o,w_{\dot{L}},\mu)	& = \eta_{\dot{L}^2} \sum_{\dot{u},\dot{v} \in \dot{L}^{N(\mu)}_{-2}(\Pi_o) | \dot{u} \neq \dot{v}} w_{\dot{L}}(\dot{u}) w_{\dot{L}}(\dot{v}) \Delta_{\mathbb{Q}}(R_2(\mu[\instantiation{u}{2}{\pi}]), R_2(\mu[\instantiation{v}{2}{\pi}])) =\\
								& = 2 \frac{2}{1} \frac{1}{2} \frac{1}{2} \Delta_{\mathbb{Q}}(R_2(\mu[{\pi_h}_1,\pi]), R_2(\mu[{\pi_h}_2,\pi]))
\end{aligned}
\end{equation*}

Again, a $\pi_h$ agent will always perform Head, so we obtain a situation where the agent in both slots 2 (any $\pi$) will be able to differentiate with which agent is interacting, so it will not be able to change its distribution of action sequences depending on the opponent's behaviour, obtaining agent in both slots 2 (any $\pi$) the same expected average reward. So:

\begin{equation*}
RD_2(\pi,\Pi_o,w_{\dot{L}},\mu) = 2 \frac{2}{1} \frac{1}{2} \frac{1}{2} 0 = 0
\end{equation*}

Therefore, no matter which agents are in $\Pi_e$ and their weights $w_{\Pi_e}$ we obtain:

\begin{equation*}
RD_2(\Pi_e,w_{\Pi_e},\Pi_o,w_{\dot{L}},\mu) = 0
\end{equation*}

And finally, we weight over the slots:

\begin{equation*}
\begin{aligned}
RD(\Pi_e,w_{\Pi_e},\Pi_o,w_{\dot{L}},\mu,w_S)	& =	\sum_{i = 1}^{N(\mu)} w_S(i,\mu) RD_i(\Pi_e,w_{\Pi_e},\Pi_o,w_{\dot{L}},\mu) =\\
												& = \frac{1}{2} \{RD_1(\Pi_e,w_{\Pi_e},\Pi_o,w_{\dot{L}},\mu) + RD_2(\Pi_e,w_{\Pi_e},\Pi_o,w_{\dot{L}},\mu)\} =\\
												& = \frac{1}{2} \left\{0 + 0\right\} = 0
\end{aligned}
\end{equation*}

So, for every pair $\left\langle\Pi_e,w_{\Pi_e}\right\rangle$ we obtain the same result:

\begin{equation*}
\forall \Pi_e,w_{\Pi_e} : RD(\Pi_e,w_{\Pi_e},\Pi_o,w_{\dot{L}},\mu,w_S) = 0
\end{equation*}

Therefore, matching pennies has $Left_{max} = 0$ for this property.
\end{proof}
\end{proposition}

\begin{proposition}
\label{prop:matching_pennies_RD_right_min}
$Right_{min}$ for the reward dependency (RD) property is equal to $0$ for the matching pennies environment.

\begin{proof}
To find $Right_{min}$ (equation \ref{eq:right_min}), we need to find a pair $\left\langle\Pi_e,w_{\Pi_e}\right\rangle$ which minimises the property as much as possible while $\Pi_o$ maximises it. Using $\Pi_e = \{\pi_r\}$ with $w_{\Pi_e}(\pi_r) = 1$ (a $\pi_r$ agent always acts randomly) we find this situation no matter which $\Pi_o$ we use.

Following definition \ref{def:RD}, we obtain the RD value for this $\left\langle\Pi_e,w_{\Pi_e},\Pi_o\right\rangle$ (where $\Pi_o$ is instantiated with any permitted value). Since the environment is not symmetric, we need to calculate this property for every slot. Following definition \ref{def:RD_set}, we could calculate its RD value for each slot but, since $\Pi_e$ has only one agent and its weight is equal to $1$, it is equivalent to use directly definition \ref{def:RD_agent}. We start with slot 1:

\begin{equation*}
RD_1(\pi_r,\Pi_o,w_{\dot{L}},\mu) = \eta_{\dot{L}^2} \sum_{\dot{u},\dot{v} \in \dot{L}^{N(\mu)}_{-1}(\Pi_o) | \dot{u} \neq \dot{v}} w_{\dot{L}}(\dot{u}) w_{\dot{L}}(\dot{v}) \Delta_{\mathbb{Q}}(R_1(\mu[\instantiation{u}{1}{\pi_r}]), R_1(\mu[\instantiation{v}{1}{\pi_r}]))
\end{equation*}

We do not know which $\Pi_o$ we have, but we know that we will need to obtain two different line-up patterns $\dot{u}$ and $\dot{v}$ from $\dot{L}^{N(\mu)}_{-1}(\Pi_o)$ to calculate $\Delta_{\mathbb{Q}}(R_1(\mu[\instantiation{u}{1}{\pi_r}]), R_1(\mu[\instantiation{v}{1}{\pi_r}]))$. We calculate this value for two figurative line-up patterns $\dot{u} = (*,\pi_1)$ and $\dot{v} = (*,\pi_2)$ from $\dot{L}^{N(\mu)}_{-1}(\Pi_o)$:

\begin{equation*}
\Delta_{\mathbb{Q}}(R_1(\mu[\instantiation{u}{1}{\pi_r}]), R_1(\mu[\instantiation{v}{1}{\pi_r}])) = \Delta_{\mathbb{Q}}(R_1(\mu[\pi_r,\pi_1]), R_1(\mu[\pi_r,\pi_2]))
\end{equation*}

Here, the agent in both slots 1 ($\pi_r$) makes its expected average reward equal to its opponent expected average reward (both have an expected average reward of $0$ as proved in lemma \ref{lemma:matching_pennies_random_agent}). So no matter which agents are in $\Pi_o$ we obtain:

\begin{equation*}
RD_1(\pi_r,\Pi_o,w_{\dot{L}},\mu) = 0
\end{equation*}

And for slot 2:

\begin{equation*}
RD_2(\pi_r,\Pi_o,w_{\dot{L}},\mu) = \eta_{\dot{L}^2} \sum_{\dot{u},\dot{v} \in \dot{L}^{N(\mu)}_{-2}(\Pi_o) | \dot{u} \neq \dot{v}} w_{\dot{L}}(\dot{u}) w_{\dot{L}}(\dot{v}) \Delta_{\mathbb{Q}}(R_2(\mu[\instantiation{u}{2}{\pi_r}]), R_2(\mu[\instantiation{v}{2}{\pi_r}]))
\end{equation*}

Again, we do not know which $\Pi_o$ we have, but we know that we will need to obtain two different line-up patterns $\dot{u}$ and $\dot{v}$ from $\dot{L}^{N(\mu)}_{-2}(\Pi_o)$ to calculate $\Delta_{\mathbb{Q}}(R_2(\mu[\instantiation{u}{2}{\pi_r}]), R_2(\mu[\instantiation{v}{2}{\pi_r}]))$. We calculate this value for two figurative line-up patterns $\dot{u} = (*,\pi_1)$ and $\dot{v} = (*,\pi_2)$ from $\dot{L}^{N(\mu)}_{-2}(\Pi_o)$:

\begin{equation*}
\Delta_{\mathbb{Q}}(R_2(\mu[\instantiation{u}{2}{\pi_r}]), R_2(\mu[\instantiation{v}{2}{\pi_r}])) = \Delta_{\mathbb{Q}}(R_2(\mu[\pi_1,\pi_r]), R_2(\mu[\pi_2,\pi_r]))
\end{equation*}

Again, the agent in both slots 2 ($\pi_r$) makes its expected average reward equal to its opponent expected average reward (both have an expected average reward of $0$ as proved in lemma \ref{lemma:matching_pennies_random_agent}). So no matter which agents are in $\Pi_o$ we obtain:

\begin{equation*}
RD_2(\pi_r,\Pi_o,w_{\dot{L}},\mu) = 0
\end{equation*}

And finally, we weight over the slots:

\begin{equation*}
\begin{aligned}
RD(\Pi_e,w_{\Pi_e},\Pi_o,w_{\dot{L}},\mu,w_S)	& =	\sum_{i = 1}^{N(\mu)} w_S(i,\mu) RD_i(\Pi_e,w_{\Pi_e},\Pi_o,w_{\dot{L}},\mu) =\\
												& =	\sum_{i = 1}^{N(\mu)} w_S(i,\mu) RD_i(\pi_r,\Pi_o,w_{\dot{L}},\mu) =\\
												& = \frac{1}{2} \{RD_1(\pi_r,\Pi_o,w_{\dot{L}},\mu) + RD_2(\pi_r,\Pi_o,w_{\dot{L}},\mu)\} =\\
												& = \frac{1}{2} \left\{0 + 0\right\} = 0
\end{aligned}
\end{equation*}

So, for every $\Pi_o$ we obtain the same result:

\begin{equation*}
\forall \Pi_o : RD(\Pi_e,w_{\Pi_e},\Pi_o,w_{\dot{L}},\mu,w_S) = 0
\end{equation*}

Therefore, matching pennies has $Right_{min} = 0$ for this property.
\end{proof}
\end{proposition}

\subsection{Fine Discrimination}
Now we move to the fine discrimination (FD) property. As given in section \ref{sec:FD}, we want to know if different evaluated agents obtain different expected average rewards when interacting in the environment. We use $\Delta_{\mathbb{Q}}(a,b) = 1$ if numbers $a$ and $b$ are equal and $0$ otherwise.

\begin{proposition}
\label{prop:matching_pennies_FD_general_min}
$General_{min}$ for the fine discrimination (FD) property is equal to $0$ for the matching pennies environment.

\begin{proof}
To find $General_{min}$ (equation \ref{eq:general_min}), we need to find a trio $\left\langle\Pi_e,w_{\Pi_e},\Pi_o\right\rangle$ which minimises the property as much as possible. We can have this situation by selecting $\Pi_e = \{{\pi_t}_1,{\pi_t}_2\}$ with uniform weight for $w_{\Pi_e}$ and $\Pi_o = \{\pi_h\}$ (a $\pi_h$ agent always performs Head and a $\pi_t$ agent always performs Tail).

Following definition \ref{def:FD}, we obtain the FD value for this $\left\langle\Pi_e,w_{\Pi_e},\Pi_o\right\rangle$. Since the environment is not symmetric, we need to calculate this property for every slot. Following definition \ref{def:FD_set}, we can calculate its FD value for each slot. We start with slot 1:

\begin{equation*}
\begin{aligned}
FD_1(\Pi_e,w_{\Pi_e},\Pi_o,w_{\dot{L}},\mu)	& = \eta_{\Pi^2} \sum_{\pi_1,\pi_2 \in \Pi_e | \pi_1 \neq \pi_2} w_{\Pi_e}(\pi_1) w_{\Pi_e}(\pi_2) FD_1(\pi_1,\pi_2,\Pi_o,w_{\dot{L}},\mu) =\\
											& = 2 \frac{2}{1} \frac{1}{2} \frac{1}{2} FD_1({\pi_t}_1,{\pi_t}_2,\Pi_o,w_{\dot{L}},\mu)
\end{aligned}
\end{equation*}

\noindent Note that we avoided to calculate both $FD_i(\pi_1,\pi_2,\Pi_o,w_{\dot{L}},\mu)\}$ and $FD_i(\pi_2,\pi_1,\Pi_o,w_{\dot{L}},\mu)\}$ since they provide the same result, by calculating only $FD_i(\pi_1,\pi_2,\Pi_o,w_{\dot{L}},\mu)\}$ and multiplying the result by $2$.

In this case, we only need to calculate $FD_1({\pi_t}_1,{\pi_t}_2,\Pi_o,w_{\dot{L}},\mu)$. We follow definition \ref{def:FD_agents} to calculate this value:

\begin{equation*}
\begin{aligned}
FD_1({\pi_t}_1,{\pi_t}_2,\Pi_o,w_{\dot{L}},\mu)	& = \sum_{\dot{l} \in \dot{L}^{N(\mu)}_{-1}(\Pi_o)} w_{\dot{L}}(\dot{l}) \Delta_{\mathbb{Q}}(R_1(\mu[\instantiation{l}{1}{{\pi_t}_1}]), R_1(\mu[\instantiation{l}{1}{{\pi_t}_2}])) =\\
												& = \Delta_{\mathbb{Q}}(R_1(\mu[{\pi_t}_1,\pi_h]), R_1(\mu[{\pi_t}_2,\pi_h]))
\end{aligned}
\end{equation*}

Here, a $\pi_t$ agent will always perform Tail and $\pi_h$ will always perform Head, so both agents in slot 1 (${\pi_t}_1$ and ${\pi_t}_2$) will obtain the same expected average reward ($-1$). So:

\begin{equation*}
FD_1({\pi_t}_1,{\pi_t}_2,\Pi_o,w_{\dot{L}},\mu) = 0
\end{equation*}

Therefore:

\begin{equation*}
FD_1(\Pi_e,w_{\Pi_e},\Pi_o,w_{\dot{L}},\mu) = 2 \frac{2}{1} \frac{1}{2} \frac{1}{2} 0 = 0
\end{equation*}

And for slot 2:

\begin{equation*}
\begin{aligned}
FD_2(\Pi_e,w_{\Pi_e},\Pi_o,w_{\dot{L}},\mu)	& = \eta_{\Pi^2} \sum_{\pi_1,\pi_2 \in \Pi_e | \pi_1 \neq \pi_2} w_{\Pi_e}(\pi_1) w_{\Pi_e}(\pi_2) FD_2(\pi_1,\pi_2,\Pi_o,w_{\dot{L}},\mu) =\\
											& = 2 \frac{2}{1} \frac{1}{2} \frac{1}{2} FD_2({\pi_t}_1,{\pi_t}_2,\Pi_o,w_{\dot{L}},\mu)
\end{aligned}
\end{equation*}

In this case, we only need to calculate $FD_2({\pi_t}_1,{\pi_t}_2,\Pi_o,w_{\dot{L}},\mu)$. We follow definition \ref{def:FD_agents} to calculate this value:

\begin{equation*}
\begin{aligned}
FD_2({\pi_t}_1,{\pi_t}_2,\Pi_o,w_{\dot{L}},\mu)	& = \sum_{\dot{l} \in \dot{L}^{N(\mu)}_{-2}(\Pi_o)} w_{\dot{L}}(\dot{l}) \Delta_{\mathbb{Q}}(R_2(\mu[\instantiation{l}{2}{{\pi_t}_1}]), R_2(\mu[\instantiation{l}{2}{{\pi_t}_2}])) =\\
												& = \Delta_{\mathbb{Q}}(R_2(\mu[\pi_h,{\pi_t}_1]), R_2(\mu[\pi_h,{\pi_t}_2]))
\end{aligned}
\end{equation*}

Here, $\pi_h$ will always perform Head and a $\pi_t$ agent will always perform Tail, so both agents in slot 2 (${\pi_t}_1$ and ${\pi_t}_2$) will obtain the same expected average reward ($1$). So:

\begin{equation*}
FD_2({\pi_t}_1,{\pi_t}_2,\Pi_o,w_{\dot{L}},\mu) = 0
\end{equation*}

Therefore:

\begin{equation*}
FD_2(\Pi_e,w_{\Pi_e},\Pi_o,w_{\dot{L}},\mu) = 2 \frac{2}{1} \frac{1}{2} \frac{1}{2} 0 = 0
\end{equation*}

And finally, we weight over the slots:

\begin{equation*}
\begin{aligned}
FD(\Pi_e,w_{\Pi_e},\Pi_o,w_{\dot{L}},\mu,w_S)	& =	\sum_{i = 1}^{N(\mu)} w_S(i,\mu) FD_i(\Pi_e,w_{\Pi_e},\Pi_o,w_{\dot{L}},\mu) =\\
												& = \frac{1}{2} \{FD_1(\Pi_e,w_{\Pi_e},\Pi_o,w_{\dot{L}},\mu) + FD_2(\Pi_e,w_{\Pi_e},\Pi_o,w_{\dot{L}},\mu)\} =\\
												& = \frac{1}{2} \left\{0 + 0\right\} = 0
\end{aligned}
\end{equation*}

Since $0$ is the lowest possible value for the fine discriminative property, therefore matching pennies has $General_{min} = 0$ for this property.
\end{proof}
\end{proposition}

\begin{proposition}
\label{prop:matching_pennies_FD_general_max}
$General_{max}$ for the fine discrimination (FD) property is equal to $1$ for the matching pennies environment.

\begin{proof}
To find $General_{max}$ (equation \ref{eq:general_max}), we need to find a trio $\left\langle\Pi_e,w_{\Pi_e},\Pi_o\right\rangle$ which maximises the property as much as possible. We can have this situation by selecting $\Pi_e = \{\pi_h,\pi_t\}$ with uniform weight for $w_{\Pi_e}$ and $\Pi_o = \{\pi_h\}$ (a $\pi_h$ agent always performs Head and a $\pi_t$ agent always performs Tail).

Following definition \ref{def:FD}, we obtain the FD value for this $\left\langle\Pi_e,w_{\Pi_e},\Pi_o\right\rangle$. Since the environment is not symmetric, we need to calculate this property for every slot. Following definition \ref{def:FD_set}, we can calculate its FD value for each slot. We start with slot 1:

\begin{equation*}
\begin{aligned}
FD_1(\Pi_e,w_{\Pi_e},\Pi_o,w_{\dot{L}},\mu)	& = \eta_{\Pi^2} \sum_{\pi_1,\pi_2 \in \Pi_e | \pi_1 \neq \pi_2} w_{\Pi_e}(\pi_1) w_{\Pi_e}(\pi_2) FD_1(\pi_1,\pi_2,\Pi_o,w_{\dot{L}},\mu) =\\
											& = 2 \frac{2}{1} \frac{1}{2} \frac{1}{2} FD_1(\pi_h,\pi_t,\Pi_o,w_{\dot{L}},\mu)
\end{aligned}
\end{equation*}

\noindent Note that we avoided to calculate both $FD_i(\pi_1,\pi_2,\Pi_o,w_{\dot{L}},\mu)\}$ and $FD_i(\pi_2,\pi_1,\Pi_o,w_{\dot{L}},\mu)\}$ since they provide the same result, by calculating only $FD_i(\pi_1,\pi_2,\Pi_o,w_{\dot{L}},\mu)\}$ and multiplying the result by $2$.

In this case, we only need to calculate $FD_1(\pi_h,\pi_t,\Pi_o,w_{\dot{L}},\mu)$. We follow definition \ref{def:FD_agents} to calculate this value:

\begin{equation*}
\begin{aligned}
FD_1(\pi_h,\pi_t,\Pi_o,w_{\dot{L}},\mu)	& = \sum_{\dot{l} \in \dot{L}^{N(\mu)}_{-1}(\Pi_o)} w_{\dot{L}}(\dot{l}) \Delta_{\mathbb{Q}}(R_1(\mu[\instantiation{l}{1}{\pi_h}]), R_1(\mu[\instantiation{l}{1}{\pi_t}])) =\\
										& = \Delta_{\mathbb{Q}}(R_1(\mu[\pi_h,\pi_h]), R_1(\mu[\pi_t,\pi_h]))
\end{aligned}
\end{equation*}

In line-up $(\pi_h,\pi_h)$, where both $\pi_h$ will always perform Head, the agent in slot 1 ($\pi_h$) will obtain one expected average reward ($1$), while in line-up $(\pi_t,\pi_h)$, where $\pi_t$ will always perform Tail and $\pi_h$ will always perform Head, the agent in slot 1 ($\pi_t$) will obtain a different expected average reward ($-1$). So:

\begin{equation*}
FD_1(\pi_h,\pi_t,\Pi_o,w_{\dot{L}},\mu) = 1
\end{equation*}

Therefore:

\begin{equation*}
FD_1(\Pi_e,w_{\Pi_e},\Pi_o,w_{\dot{L}},\mu) = 2 \frac{2}{1} \frac{1}{2} \frac{1}{2} 1 = 1
\end{equation*}

And for slot 2:

\begin{equation*}
\begin{aligned}
FD_2(\Pi_e,w_{\Pi_e},\Pi_o,w_{\dot{L}},\mu)	& = \eta_{\Pi^2} \sum_{\pi_1,\pi_2 \in \Pi_e | \pi_1 \neq \pi_2} w_{\Pi_e}(\pi_1) w_{\Pi_e}(\pi_2) FD_2(\pi_1,\pi_2,\Pi_o,w_{\dot{L}},\mu) =\\
											& = 2 \frac{2}{1} \frac{1}{2} \frac{1}{2} FD_2(\pi_h,\pi_t,\Pi_o,w_{\dot{L}},\mu)
\end{aligned}
\end{equation*}

In this case, we only need to calculate $FD_2(\pi_h,\pi_t,\Pi_o,w_{\dot{L}},\mu)$. We follow definition \ref{def:FD_agents} to calculate this value:

\begin{equation*}
\begin{aligned}
FD_2(\pi_h,\pi_t,\Pi_o,w_{\dot{L}},\mu)	& = \sum_{\dot{l} \in \dot{L}^{N(\mu)}_{-2}(\Pi_o)} w_{\dot{L}}(\dot{l}) \Delta_{\mathbb{Q}}(R_2(\mu[\instantiation{l}{2}{\pi_h}]), R_2(\mu[\instantiation{l}{2}{\pi_t}])) =\\
										& = \Delta_{\mathbb{Q}}(R_2(\mu[\pi_h,\pi_h]), R_2(\mu[\pi_h,\pi_t]))
\end{aligned}
\end{equation*}

In line-up $(\pi_h,\pi_h)$, where both $\pi_h$ will always perform Head, the agent in slot 2 ($\pi_h$) will obtain one expected average reward ($-1$), while in line-up $(\pi_h,\pi_t)$, where $\pi_h$ will always perform Head and $\pi_t$ will always perform Head, the agent in slot 2 ($\pi_t$) will obtain a different expected average reward ($1$). So:

\begin{equation*}
FD_2(\pi_h,\pi_t,\Pi_o,w_{\dot{L}},\mu) = 1
\end{equation*}

Therefore:

\begin{equation*}
FD_2(\Pi_e,w_{\Pi_e},\Pi_o,w_{\dot{L}},\mu) = 2 \frac{2}{1} \frac{1}{2} \frac{1}{2} 1 = 1
\end{equation*}

And finally, we weight over the slots:

\begin{equation*}
\begin{aligned}
FD(\Pi_e,w_{\Pi_e},\Pi_o,w_{\dot{L}},\mu,w_S)	& =	\sum_{i = 1}^{N(\mu)} w_S(i,\mu) FD_i(\Pi_e,w_{\Pi_e},\Pi_o,w_{\dot{L}},\mu) =\\
												& = \frac{1}{2} \{FD_1(\Pi_e,w_{\Pi_e},\Pi_o,w_{\dot{L}},\mu) + FD_2(\Pi_e,w_{\Pi_e},\Pi_o,w_{\dot{L}},\mu)\} =\\
												& = \frac{1}{2} \left\{1 + 1\right\} = 1
\end{aligned}
\end{equation*}

Since $1$ is the highest possible value for the fine discriminative property, therefore matching pennies has $General_{max} = 1$ for this property.
\end{proof}
\end{proposition}

\begin{proposition}
\label{prop:matching_pennies_FD_left_max}
$Left_{max}$ for the fine discrimination (FD) property is equal to $0$ for the matching pennies environment.

\begin{proof}
To find $Left_{max}$ (equation \ref{eq:left_max}), we need to find a pair $\left\langle\Pi_e,w_{\Pi_e}\right\rangle$ which maximises the property as much as possible while $\Pi_o$ minimises it. Using $\Pi_o = \{\pi_r\}$ (a $\pi_r$ agent always acts randomly) we find this situation no matter which pair $\left\langle\Pi_e,w_{\Pi_e}\right\rangle$ we use.

Following definition \ref{def:FD}, we obtain the FD value for this $\left\langle\Pi_e,w_{\Pi_e},\Pi_o\right\rangle$ (where $\Pi_e$ and $w_{\Pi_e}$ are instantiated with any permitted values). Since the environment is not symmetric, we need to calculate this property for every slot. Following definition \ref{def:FD_set}, we can calculate its FD value for each slot. We start with slot 1:

\begin{equation*}
FD_1(\Pi_e,w_{\Pi_e},\Pi_o,w_{\dot{L}},\mu) = \eta_{\Pi^2} \sum_{\pi_1,\pi_2 \in \Pi_e | \pi_1 \neq \pi_2} w_{\Pi_e}(\pi_1) w_{\Pi_e}(\pi_2) FD_1(\pi_1,\pi_2,\Pi_o,w_{\dot{L}},\mu)
\end{equation*}

We do not know which $\Pi_e$ we have, but we know that we will need to evaluate $FD_1(\pi_1,\pi_2,\Pi_o,w_{\dot{L}},\mu)$ for all pair of evaluated agents $\pi_1,\pi_2 \in \Pi_e | \pi_1 \neq \pi_2$. We follow definition \ref{def:FD_agents} to calculate this value for two figurative evaluated agents $\pi_1$ and $\pi_2$ from $\Pi_e$ such that $\pi_1 \neq \pi_2$:

\begin{equation*}
\begin{aligned}
FD_1(\pi_1,\pi_2,\Pi_o,w_{\dot{L}},\mu)	& = \sum_{\dot{l} \in \dot{L}^{N(\mu)}_{-1}(\Pi_o)} w_{\dot{L}}(\dot{l}) \Delta_{\mathbb{Q}}(R_1(\mu[\instantiation{l}{1}{\pi_1}]), R_1(\mu[\instantiation{l}{1}{\pi_2}])) =\\
										& = \Delta_{\mathbb{Q}}(R_1(\mu[\pi_1,\pi_r]), R_1(\mu[\pi_2,\pi_r]))
\end{aligned}
\end{equation*}

Here, the agent in both slots 2 ($\pi_r$) makes its expected average reward equal to its opponent expected average reward (both have an expected average reward of $0$ as proved in lemma \ref{lemma:matching_pennies_random_agent}). So no matter which agents $\pi_1$ and $\pi_2$ are we obtain:

\begin{equation*}
FD_1(\pi_1,\pi_2,\Pi_o,w_{\dot{L}},\mu) = 0
\end{equation*}

Therefore:

\begin{equation*}
FD_1(\Pi_e,w_{\Pi_e},\Pi_o,w_{\dot{L}},\mu) = 0
\end{equation*}

And for slot 2:

\begin{equation*}
FD_2(\Pi_e,w_{\Pi_e},\Pi_o,w_{\dot{L}},\mu) = \eta_{\Pi^2} \sum_{\pi_1,\pi_2 \in \Pi_e | \pi_1 \neq \pi_2} w_{\Pi_e}(\pi_1) w_{\Pi_e}(\pi_2) FD_2(\pi_1,\pi_2,\Pi_o,w_{\dot{L}},\mu)
\end{equation*}

Again, we do not know which $\Pi_e$ we have, but we know that we will need to evaluate $FD_2(\pi_1,\pi_2,\Pi_o,w_{\dot{L}},\mu)$ for all pair of evaluated agents $\pi_1,\pi_2 \in \Pi_e | \pi_1 \neq \pi_2$. We follow definition \ref{def:FD_agents} to calculate this value for two figurative evaluated agents $\pi_1$ and $\pi_2$ from $\Pi_e$ such that $\pi_1 \neq \pi_2$:

\begin{equation*}
\begin{aligned}
FD_2(\pi_1,\pi_2,\Pi_o,w_{\dot{L}},\mu)	& = \sum_{\dot{l} \in \dot{L}^{N(\mu)}_{-2}(\Pi_o)} w_{\dot{L}}(\dot{l}) \Delta_{\mathbb{Q}}(R_2(\mu[\instantiation{l}{2}{\pi_1}]), R_1(\mu[\instantiation{l}{2}{\pi_2}])) =\\
										& = \Delta_{\mathbb{Q}}(R_2(\mu[\pi_r,\pi_1]), R_2(\mu[\pi_r,\pi_2]))
\end{aligned}
\end{equation*}

Again, the agent in both slots 1 ($\pi_r$) makes its expected average reward equal to its opponent expected average reward (both have an expected average reward of $0$ as proved in lemma \ref{lemma:matching_pennies_random_agent}). So no matter which agents $\pi_1$ and $\pi_2$ are we obtain:

\begin{equation*}
FD_2(\pi_1,\pi_2,\Pi_o,w_{\dot{L}},\mu) = 0
\end{equation*}

Therefore:

\begin{equation*}
FD_2(\Pi_e,w_{\Pi_e},\Pi_o,w_{\dot{L}},\mu) = 0
\end{equation*}

And finally, we weight over the slots:

\begin{equation*}
\begin{aligned}
FD(\Pi_e,w_{\Pi_e},\Pi_o,w_{\dot{L}},\mu,w_S)	& = \sum_{i=1}^{N(\mu)} w_S(i,\mu) FD_i(\Pi_e,w_{\Pi_e},\Pi_o,w_{\dot{L}},\mu)\\
												& = \frac{1}{2} \{FD_1(\Pi_e,w_{\Pi_e},\Pi_o,w_{\dot{L}},\mu) + FD_2(\Pi_e,w_{\Pi_e},\Pi_o,w_{\dot{L}},\mu)\} =\\
												& = \frac{1}{2} \left\{0 + 0\right\} = 0
\end{aligned}
\end{equation*}

So, for every pair $\left\langle\Pi_e,w_{\Pi_e}\right\rangle$ we obtain the same result:

\begin{equation*}
\forall \Pi_e,w_{\Pi_e} : FD(\Pi_e,w_{\Pi_e},\Pi_o,w_{\dot{L}},\mu,w_S) = 0
\end{equation*}

Therefore, matching pennies has $Left_{max} = 0$ for this property.
\end{proof}
\end{proposition}

\begin{proposition}
\label{prop:matching_pennies_FD_right_min}
$Right_{min}$ for the fine discrimination (FD) property is equal to $0$ for the matching pennies environment.

\begin{proof}
To find $Right_{min}$ (equation \ref{eq:right_min}), we need to find a pair $\left\langle\Pi_e,w_{\Pi_e}\right\rangle$ which minimises the property as much as possible while $\Pi_o$ maximises it. Using $\Pi_e = \{{\pi_t}_1,{\pi_t}_2\}$ with uniform weight for $w_{\Pi_e}$ (a $\pi_t$ agent always performs Tail) we find this situation no matter which $\Pi_o$ we use.

Following definition \ref{def:FD}, we obtain the FD value for this $\left\langle\Pi_e,w_{\Pi_e},\Pi_o\right\rangle$ (where $\Pi_o$ is instantiated with any permitted values). Since the environment is not symmetric, we need to calculate this property for every slot. Following definition \ref{def:FD_set}, we can calculate its FD value for each slot. We start with slot 1:

\begin{equation*}
\begin{aligned}
FD_1(\Pi_e,w_{\Pi_e},\Pi_o,w_{\dot{L}},\mu)	& = \eta_{\Pi^2} \sum_{\pi_1,\pi_2 \in \Pi_e | \pi_1 \neq \pi_2} w_{\Pi_e}(\pi_1) w_{\Pi_e}(\pi_2) FD_1(\pi_1,\pi_2,\Pi_o,w_{\dot{L}},\mu) =\\
											& = 2 \frac{2}{1} \frac{1}{2} \frac{1}{2} FD_1({\pi_t}_1,{\pi_t}_2,\Pi_o,w_{\dot{L}},\mu)
\end{aligned}
\end{equation*}

\noindent Note that we avoided to calculate both $FD_i(\pi_1,\pi_2,\Pi_o,w_{\dot{L}},\mu)\}$ and $FD_i(\pi_2,\pi_1,\Pi_o,w_{\dot{L}},\mu)\}$ since they provide the same result, by calculating only $FD_i(\pi_1,\pi_2,\Pi_o,w_{\dot{L}},\mu)\}$ and multiplying the result by $2$.

In this case, we only need to calculate $FD_1({\pi_t}_1,{\pi_t}_2,\Pi_o,w_{\dot{L}},\mu)$. We follow definition \ref{def:FD_agents} to calculate this value:

\begin{equation*}
FD_1({\pi_t}_1,{\pi_t}_2,\Pi_o,w_{\dot{L}},\mu) = \sum_{\dot{l} \in \dot{L}^{N(\mu)}_{-1}(\Pi_o)} w_{\dot{L}}(\dot{l}) \Delta_{\mathbb{Q}}(R_1(\mu[\instantiation{l}{1}{{\pi_t}_1}]), R_1(\mu[\instantiation{l}{1}{{\pi_t}_2}]))
\end{equation*}

We do not know which $\Pi_o$ we have, but we know that we will need to obtain a line-up pattern $\dot{l}$ from $\dot{L}^{N(\mu)}_{-1}(\Pi_o)$ to calculate $\Delta_{\mathbb{Q}}(R_1(\mu[\instantiation{l}{1}{{\pi_t}_1}]), R_1(\mu[\instantiation{l}{1}{{\pi_t}_2}]))$. We calculate this value for a figurative line-up pattern $\dot{l} = (*,\pi)$ from $\dot{L}^{N(\mu)}_{-1}(\Pi_o)$:

\begin{equation*}
\Delta_{\mathbb{Q}}(R_1(\mu[\instantiation{l}{1}{{\pi_t}_1}]), R_1(\mu[\instantiation{l}{1}{{\pi_t}_2}])) = \Delta_{\mathbb{Q}}(R_1(\mu[{\pi_t}_1,\pi]), R_1(\mu[{\pi_t}_2,\pi]))
\end{equation*}

A $\pi_t$ agent will always perform Tail, so we obtain a situation where the agent in both slots 2 (any $\pi$) will be able to differentiate with which agent is interacting, so it will not be able to change its distribution of action sequences depending on the opponent's behaviour, obtaining both agents in slot 1 (${\pi_t}_1$ and ${\pi_t}_2$) the same expected average reward. So:

\begin{equation*}
FD_1({\pi_t}_1,{\pi_t}_2,\Pi_o,w_{\dot{L}},\mu) = 0
\end{equation*}

Therefore:

\begin{equation*}
FD_1(\Pi_e,w_{\Pi_e},\Pi_o,w_{\dot{L}},\mu) = 2 \frac{2}{1} \frac{1}{2} \frac{1}{2} 0 = 0
\end{equation*}

And for slot 2:

\begin{equation*}
\begin{aligned}
FD_2(\Pi_e,w_{\Pi_e},\Pi_o,w_{\dot{L}},\mu)	& = \eta_{\Pi^2} \sum_{\pi_1,\pi_2 \in \Pi_e | \pi_1 \neq \pi_2} w_{\Pi_e}(\pi_1) w_{\Pi_e}(\pi_2) FD_2(\pi_1,\pi_2,\Pi_o,w_{\dot{L}},\mu) =\\
											& = 2 \frac{2}{1} \frac{1}{2} \frac{1}{2} FD_2({\pi_t}_1,{\pi_t}_2,\Pi_o,w_{\dot{L}},\mu)
\end{aligned}
\end{equation*}

In this case, we only need to calculate $FD_2({\pi_t}_1,{\pi_t}_2,\Pi_o,w_{\dot{L}},\mu)$. We follow definition \ref{def:FD_agents} to calculate this value:

\begin{equation*}
FD_2({\pi_t}_1,{\pi_t}_2,\Pi_o,w_{\dot{L}},\mu) = \sum_{\dot{l} \in \dot{L}^{N(\mu)}_{-2}(\Pi_o)} w_{\dot{L}}(\dot{l}) \Delta_{\mathbb{Q}}(R_2(\mu[\instantiation{l}{2}{{\pi_t}_1}]), R_2(\mu[\instantiation{l}{2}{{\pi_t}_2}]))
\end{equation*}

Again, we do not know which $\Pi_o$ we have, but we know that we will need to obtain a line-up pattern $\dot{l}$ from $\dot{L}^{N(\mu)}_{-2}(\Pi_o)$ to calculate $\Delta_{\mathbb{Q}}(R_2(\mu[\instantiation{l}{2}{{\pi_t}_1}]), R_2(\mu[\instantiation{l}{2}{{\pi_t}_2}]))$. We calculate this value for a figurative line-up pattern $\dot{l} = (\pi,*)$ from $\dot{L}^{N(\mu)}_{-2}(\Pi_o)$:

\begin{equation*}
\Delta_{\mathbb{Q}}(R_2(\mu[\instantiation{l}{2}{{\pi_t}_1}]), R_2(\mu[\instantiation{l}{2}{{\pi_t}_2}])) = \Delta_{\mathbb{Q}}(R_2(\mu[\pi,{\pi_t}_1]), R_2(\mu[\pi,{\pi_t}_2]))
\end{equation*}

Again, a $\pi_t$ agent will always perform Tail, so we obtain a situation where the agent in both slots 1 (any $\pi$) will be able to differentiate with which agent is interacting, so it will not be able to change its distribution of action sequences depending on the opponent's behaviour, obtaining both agents in slot 2 (${\pi_t}_1$ and ${\pi_t}_2$) the same expected average reward. So:

\begin{equation*}
FD_2({\pi_t}_1,{\pi_t}_2,\Pi_o,w_{\dot{L}},\mu) = 0
\end{equation*}

Therefore:

\begin{equation*}
FD_2(\Pi_e,w_{\Pi_e},\Pi_o,w_{\dot{L}},\mu) = 2 \frac{2}{1} \frac{1}{2} \frac{1}{2} 0 = 0
\end{equation*}

And finally, we weight over the slots:

\begin{equation*}
\begin{aligned}
FD(\Pi_e,w_{\Pi_e},\Pi_o,w_{\dot{L}},\mu,w_S)	& = \sum_{i=1}^{N(\mu)} w_S(i,\mu) FD_i(\Pi_e,w_{\Pi_e},\Pi_o,w_{\dot{L}},\mu)\\
												& = \frac{1}{2} \{FD_1(\Pi_e,w_{\Pi_e},\Pi_o,w_{\dot{L}},\mu) + FD_2(\Pi_e,w_{\Pi_e},\Pi_o,w_{\dot{L}},\mu)\} =\\
												& = \frac{1}{2} \{0 + 0\} = 0
\end{aligned}
\end{equation*}

So, for every $\Pi_o$ we obtain the same result:

\begin{equation*}
\forall \Pi_o : FD(\Pi_e,w_{\Pi_e},\Pi_o,w_{\dot{L}},\mu,w_S) = 0
\end{equation*}

Therefore, matching pennies has $Right_{min} = 0$ for this property.
\end{proof}
\end{proposition}

\subsection{Strict Total Grading}
We arrive to the strict total grading (STG) property. As given in section \ref{sec:STG}, we want to know if there exists a strict ordering between the evaluated agents when interacting in the environment.

To simplify the notation, we use the next table to represent the STO:
$R_i(\mu[\instantiation{l}{i,j}{\pi_1,\pi_2}]) < R_j(\mu[\instantiation{l}{i,j}{\pi_1,\pi_2}])$,
$R_i(\mu[\instantiation{l}{i,j}{\pi_2,\pi_3}]) < R_j(\mu[\instantiation{l}{i,j}{\pi_2,\pi_3}])$ and
$R_i(\mu[\instantiation{l}{i,j}{\pi_1,\pi_3}]) < R_j(\mu[\instantiation{l}{i,j}{\pi_1,\pi_3}])$.

\begin{center}
\begin{tabular}{c c c}
Slot i & & Slot j\\
\hline
$\pi_1$ & $<$ & $\pi_2$\\
$\pi_2$ & $<$ & $\pi_3$\\
$\pi_1$ & $<$ & $\pi_3$
\end{tabular}
\end{center}

\begin{proposition}
\label{prop:matching_pennies_STG_general_min}
$General_{min}$ for the strict total grading (STG) property is equal to $0$ for the matching pennies environment.

\begin{proof}
To find $General_{min}$ (equation \ref{eq:general_min}), we need to find a trio $\left\langle\Pi_e,w_{\Pi_e},\Pi_o\right\rangle$ which minimises the property as much as possible. We can have this situation by selecting $\Pi_e = \{{\pi_r}_1,{\pi_r}_2,{\pi_r}_3\}$ with uniform weight for $w_{\Pi_e}$ and $\Pi_o = \emptyset$ (a $\pi_r$ agent always acts randomly).

Following definition \ref{def:STG}, we obtain the STG value for this $\left\langle\Pi_e,w_{\Pi_e},\Pi_o\right\rangle$. Since the environment is not symmetric, we need to calculate this property for every pair of slots. Following definition \ref{def:STG_set}, we can calculate its STG value for each pair of slots. We start with slots 1 and 2:

\begin{equation*}
\begin{aligned}
STG_{1,2}(\Pi_e,w_{\Pi_e},\Pi_o,w_{\dot{L}},\mu)	& = \eta_{\Pi^3} \sum_{\pi_1,\pi_2,\pi_3 \in \Pi_e | \pi_1 \neq \pi_2 \neq \pi_3} w_{\Pi_e}(\pi_1) w_{\Pi_e}(\pi_2) w_{\Pi_e}(\pi_3) STG_{1,2}(\pi_1,\pi_2,\pi_3,\Pi_o,w_{\dot{L}},\mu) =\\
													& = 6 \frac{9}{2} \frac{1}{3} \frac{1}{3} \frac{1}{3} STG_{1,2}({\pi_r}_1,{\pi_r}_2,{\pi_r}_3,\Pi_o,w_{\dot{L}},\mu)
\end{aligned}
\end{equation*}

\noindent Note that we avoided to calculate all the permutations of $\pi_1,\pi_2,\pi_3$ for $STG_{i,j}(\pi_1,\pi_2,\pi_3,\Pi_o,w_{\dot{L}},\mu)$ since they provide the same result, by calculating only one permutation and multiplying the result by the number of permutations $6$.

In this case, we only need to calculate $STG_{1,2}({\pi_r}_1,{\pi_r}_2,{\pi_r}_3,\Pi_o,w_{\dot{L}},\mu)$. We follow definition \ref{def:STG_agents} to calculate this value:

\begin{equation*}
\begin{aligned}
STG_{1,2}({\pi_r}_1,{\pi_r}_2,{\pi_r}_3,\Pi_o,w_{\dot{L}},\mu)	& = \sum_{\dot{l} \in \dot{L}^{N(\mu)}_{-1,2}(\Pi_o)} w_{\dot{L}}(\dot{l}) STO_{1,2}({\pi_r}_1,{\pi_r}_2,{\pi_r}_3,\dot{l},\mu) =\\
																& = STO_{1,2}({\pi_r}_1,{\pi_r}_2,{\pi_r}_3,(*,*),\mu)
\end{aligned}
\end{equation*}

The following table shows us $STO_{1,2}$ for all the permutations of ${\pi_r}_1,{\pi_r}_2,{\pi_r}_3$.

\begin{center}
\begin{tabular}{c c c | c c c | c c c}
Slot 1 & & Slot 2				& Slot 1 & & Slot 2					& Slot 1 & & Slot 2\\
\hline
${\pi_r}_1$ & $<$ & ${\pi_r}_2$	& ${\pi_r}_1$ & $<$ & ${\pi_r}_3$	& ${\pi_r}_2$ & $<$ & ${\pi_r}_1$\\
${\pi_r}_2$ & $<$ & ${\pi_r}_3$	& ${\pi_r}_3$ & $<$ & ${\pi_r}_2$	& ${\pi_r}_1$ & $<$ & ${\pi_r}_3$\\
${\pi_r}_1$ & $<$ & ${\pi_r}_3$	& ${\pi_r}_1$ & $<$ & ${\pi_r}_2$	& ${\pi_r}_2$ & $<$ & ${\pi_r}_3$
\end{tabular}
\begin{tabular}{c c c | c c c | c c c}
Slot 1 & & Slot 2				& Slot 1 & & Slot 2					& Slot 1 & & Slot 2\\
\hline
${\pi_r}_2$ & $<$ & ${\pi_r}_3$	& ${\pi_r}_3$ & $<$ & ${\pi_r}_1$	& ${\pi_r}_3$ & $<$ & ${\pi_r}_2$\\
${\pi_r}_3$ & $<$ & ${\pi_r}_1$	& ${\pi_r}_1$ & $<$ & ${\pi_r}_2$	& ${\pi_r}_2$ & $<$ & ${\pi_r}_1$\\
${\pi_r}_2$ & $<$ & ${\pi_r}_1$	& ${\pi_r}_3$ & $<$ & ${\pi_r}_2$	& ${\pi_r}_3$ & $<$ & ${\pi_r}_1$
\end{tabular}
\end{center}

But, it is not possible to find a STO, since for every permutation we have at least one random agent making its expected average reward equal to its opponent expected average reward (both have an expected average reward of $0$ as proved in lemma \ref{lemma:matching_pennies_random_agent}). So:

\begin{equation*}
STG_{1,2}({\pi_r}_1,{\pi_r}_2,{\pi_r}_3,\Pi_o,w_{\dot{L}},\mu) = 0
\end{equation*}

Therefore:

\begin{equation*}
STG_{1,2}(\Pi_e,w_{\Pi_e},\Pi_o,w_{\dot{L}},\mu) = 6 \frac{9}{2} \frac{1}{3} \frac{1}{3} \frac{1}{3} 0 = 0
\end{equation*}

And for slots 2 and 1:

\begin{equation*}
\begin{aligned}
STG_{2,1}(\Pi_e,w_{\Pi_e},\Pi_o,w_{\dot{L}},\mu)	& = \eta_{\Pi^3} \sum_{\pi_1,\pi_2,\pi_3 \in \Pi_e | \pi_1 \neq \pi_2 \neq \pi_3} w_{\Pi_e}(\pi_1) w_{\Pi_e}(\pi_2) w_{\Pi_e}(\pi_3) STG_{2,1}(\pi_1,\pi_2,\pi_3,\Pi_o,w_{\dot{L}},\mu) =\\
													& = 6 \frac{9}{2} \frac{1}{3} \frac{1}{3} \frac{1}{3} STG_{2,1}({\pi_r}_1,{\pi_r}_2,{\pi_r}_3,\Pi_o,w_{\dot{L}},\mu)
\end{aligned}
\end{equation*}

Again, we only need to calculate $STG_{2,1}({\pi_r}_1,{\pi_r}_2,{\pi_r}_3,\Pi_o,w_{\dot{L}},\mu)$. We follow definition \ref{def:STG_agents} to calculate this value:

\begin{equation*}
\begin{aligned}
STG_{2,1}({\pi_r}_1,{\pi_r}_2,{\pi_r}_3,\Pi_o,w_{\dot{L}},\mu)	& = \sum_{\dot{l} \in \dot{L}^{N(\mu)}_{-2,1}(\Pi_o)} w_{\dot{L}}(\dot{l}) STO_{2,1}({\pi_r}_1,{\pi_r}_2,{\pi_r}_3,\dot{l},\mu) =\\
																& = STO_{2,1}({\pi_r}_1,{\pi_r}_2,{\pi_r}_3,(*,*),\mu)
\end{aligned}
\end{equation*}

The following table shows us $STO_{2,1}$ for all the permutations of ${\pi_r}_1,{\pi_r}_2,{\pi_r}_3$.

\begin{center}
\begin{tabular}{c c c | c c c | c c c}
Slot 2 & & Slot 1				& Slot 2 & & Slot 1					& Slot 2 & & Slot 1\\
\hline
${\pi_r}_1$ & $<$ & ${\pi_r}_2$	& ${\pi_r}_1$ & $<$ & ${\pi_r}_3$	& ${\pi_r}_2$ & $<$ & ${\pi_r}_1$\\
${\pi_r}_2$ & $<$ & ${\pi_r}_3$	& ${\pi_r}_3$ & $<$ & ${\pi_r}_2$	& ${\pi_r}_1$ & $<$ & ${\pi_r}_3$\\
${\pi_r}_1$ & $<$ & ${\pi_r}_3$	& ${\pi_r}_1$ & $<$ & ${\pi_r}_2$	& ${\pi_r}_2$ & $<$ & ${\pi_r}_3$
\end{tabular}
\begin{tabular}{c c c | c c c | c c c}
Slot 2 & & Slot 1				& Slot 2 & & Slot 1					& Slot 2 & & Slot 1\\
\hline
${\pi_r}_2$ & $<$ & ${\pi_r}_3$	& ${\pi_r}_3$ & $<$ & ${\pi_r}_1$	& ${\pi_r}_3$ & $<$ & ${\pi_r}_2$\\
${\pi_r}_3$ & $<$ & ${\pi_r}_1$	& ${\pi_r}_1$ & $<$ & ${\pi_r}_2$	& ${\pi_r}_2$ & $<$ & ${\pi_r}_1$\\
${\pi_r}_2$ & $<$ & ${\pi_r}_1$	& ${\pi_r}_3$ & $<$ & ${\pi_r}_2$	& ${\pi_r}_3$ & $<$ & ${\pi_r}_1$
\end{tabular}
\end{center}

Again, it is not possible to find a STO, since for every permutation we have at least one random agent making its expected average reward equal to its opponent expected average reward (both have an expected average reward of $0$ as proved in lemma \ref{lemma:matching_pennies_random_agent}). So:

\begin{equation*}
STG_{2,1}({\pi_r}_1,{\pi_r}_2,{\pi_r}_3,\Pi_o,w_{\dot{L}},\mu) = 0
\end{equation*}

Therefore:

\begin{equation*}
STG_{2,1}(\Pi_e,w_{\Pi_e},\Pi_o,w_{\dot{L}},\mu) = 6 \frac{9}{2} \frac{1}{3} \frac{1}{3} \frac{1}{3} 0 = 0
\end{equation*}

And finally, we weight over the slots:

\begin{equation*}
\begin{aligned}
& STG(\Pi_e,w_{\Pi_e},\Pi_o,w_{\dot{L}},\mu,w_S) = \eta_{S_1^2} \sum_{i=1}^{N(\mu)} w_S(i,\mu) \times\\
& \times \left(\sum_{j=1}^{i-1} w_S(j,\mu) STG_{i,j}(\Pi_e,w_{\Pi_e},\Pi_o,w_{\dot{L}},\mu) + \sum_{j=i+1}^{N(\mu)} w_S(j,\mu) STG_{i,j}(\Pi_e,w_{\Pi_e},\Pi_o,w_{\dot{L}},\mu)\right) =\\
& \ \ \ \ \ \ \ \ \ \ \ \ \ \ \ \ \ \ \ \ \ \ \ \ \ \ \ \ \ \ \ \ \ \ \ \ \ \ = \frac{2}{1} \frac{1}{2} \frac{1}{2} \{STG_{1,2}(\Pi_e,w_{\Pi_e},\Pi_o,w_{\dot{L}},\mu) + STG_{2,1}(\Pi_e,w_{\Pi_e},\Pi_o,w_{\dot{L}},\mu)\} =\\
& \ \ \ \ \ \ \ \ \ \ \ \ \ \ \ \ \ \ \ \ \ \ \ \ \ \ \ \ \ \ \ \ \ \ \ \ \ \ = \frac{2}{1} \frac{1}{2} \frac{1}{2} \left\{0 + 0\right\} = 0
\end{aligned}
\end{equation*}

Since $0$ is the lowest possible value for the strict total grading property, therefore matching pennies has $General_{min} = 0$ for this property.
\end{proof}
\end{proposition}

\begin{proposition}
\label{prop:matching_pennies_STG_general_max}
$General_{max}$ for the strict total grading (STG) property is equal to $1$ for the matching pennies environment.

\begin{proof}
To find $General_{max}$ (equation \ref{eq:general_max}), we need to find a trio $\left\langle\Pi_e,w_{\Pi_e},\Pi_o\right\rangle$ which maximises the property as much as possible. We can have this situation by selecting $\Pi_e = \{\pi_h,\pi_{h/t},\pi_{m/o}\}$ with uniform weight for $w_{\Pi_e}$ and $\Pi_o = \emptyset$ (a $\pi_h$ agent always performs Head, a $\pi_{h/t}$ agent always performs Head when playing in slot 1 and always performs Tail when playing in slot 2, and a $\pi_{m/o}$ agent always mimics the last action of its opponent when playing in slot 1 and always performs the opposite of this action when playing in slot 2)\footnote{$\pi_{h/t}$ and $\pi_{m/o}$ have to know in which slot they are playing. To infer this, they start with a random action at the first iteration and then look at the action of their opponent and the reward they obtain.}.

Following definition \ref{def:STG}, we obtain the STG value for this $\left\langle\Pi_e,w_{\Pi_e},\Pi_o\right\rangle$. Since the environment is not symmetric, we need to calculate this property for every pair of slots. Following definition \ref{def:STG_set}, we can calculate its STG value for each pair of slots. We start with slots 1 and 2:

\begin{equation*}
\begin{aligned}
STG_{1,2}(\Pi_e,w_{\Pi_e},\Pi_o,w_{\dot{L}},\mu)	& = \eta_{\Pi^3} \sum_{\pi_1,\pi_2,\pi_3 \in \Pi_e | \pi_1 \neq \pi_2 \neq \pi_3} w_{\Pi_e}(\pi_1) w_{\Pi_e}(\pi_2) w_{\Pi_e}(\pi_3) STG_{1,2}(\pi_1,\pi_2,\pi_3,\Pi_o,w_{\dot{L}},\mu) =\\
													& = 6 \frac{9}{2} \frac{1}{3} \frac{1}{3} \frac{1}{3} STG_{1,2}(\pi_h,\pi_{h/t},\pi_{m/o},\Pi_o,w_{\dot{L}},\mu)
\end{aligned}
\end{equation*}

\noindent Note that we avoided to calculate all the permutations of $\pi_1,\pi_2,\pi_3$ for $STG_{i,j}(\pi_1,\pi_2,\pi_3,\Pi_o,w_{\dot{L}},\mu)$ since they provide the same result, by calculating only one permutation and multiplying the result by the number of permutations $6$.

In this case, we only need to calculate $STG_{1,2}(\pi_h,\pi_{h/t},\pi_{m/o},\Pi_o,w_{\dot{L}},\mu)$. We follow definition \ref{def:STG_agents} to calculate this value:

\begin{equation*}
\begin{aligned}
STG_{1,2}(\pi_h,\pi_{h/t},\pi_{m/o},\Pi_o,w_{\dot{L}},\mu)	& = \sum_{\dot{l} \in \dot{L}^{N(\mu)}_{-1,2}(\Pi_o)} w_{\dot{L}}(\dot{l}) STO_{1,2}(\pi_h,\pi_{h/t},\pi_{m/o},\dot{l},\mu) =\\
															& = STO_{1,2}(\pi_h,\pi_{h/t},\pi_{m/o},(*,*),\mu)
\end{aligned}
\end{equation*}

The following table shows us $STO_{1,2}$ for all the permutations of $\pi_h,\pi_{h/t},\pi_{m/o}$.

\begin{center}
\begin{tabular}{c c c | c c c | c c c}
Slot 1 & & Slot 2				& Slot 1 & & Slot 2					& Slot 1 & & Slot 2\\
\hline
$\pi_h$ & $<$ & $\pi_{h/t}$		& $\pi_h$ & $<$ & $\pi_{m/o}$		& $\pi_{h/t}$ & $<$ & $\pi_h$\\
$\pi_{h/t}$ & $<$ & $\pi_{m/o}$	& $\pi_{m/o}$ & $<$ & $\pi_{h/t}$	& $\pi_h$ & $<$ & $\pi_{m/o}$\\
$\pi_h$ & $<$ & $\pi_{m/o}$		& $\pi_h$ & $<$ & $\pi_{h/t}$		& $\pi_{h/t}$ & $<$ & $\pi_{m/o}$
\end{tabular}
\begin{tabular}{c c c | c c c | c c c}
Slot 1 & & Slot 2				& Slot 1 & & Slot 2					& Slot 1 & & Slot 2\\
\hline
$\pi_{h/t}$ & $<$ & $\pi_{m/o}$	& $\pi_{m/o}$ & $<$ & $\pi_h$		& $\pi_{m/o}$ & $<$ & $\pi_{h/t}$\\
$\pi_{m/o}$ & $<$ & $\pi_h$		& $\pi_h$ & $<$ & $\pi_{h/t}$		& $\pi_{h/t}$ & $<$ & $\pi_h$\\
$\pi_{h/t}$ & $<$ & $\pi_h$		& $\pi_{m/o}$ & $<$ & $\pi_{h/t}$	& $\pi_{m/o}$ & $<$ & $\pi_h$
\end{tabular}
\end{center}

It is possible to find a STO for the first permutation. In $\pi_h < \pi_{h/t}$, $\pi_h$ will always perform Head and $\pi_{h/t}$ will always perform Tail, so they will obtain an expected average reward of $-1$ and $1$ respectively. In $\pi_{h/t} < \pi_{m/o}$, $\pi_{h/t}$ will always perform Head and $\pi_{m/o}$ will always perform Tail, so they will obtain an expected average reward of $-1$ and $1$ respectively. In $\pi_h < \pi_{m/o}$, $\pi_h$ will always perform Head and $\pi_{m/o}$ will always perform Tail, so they will obtain an expected average reward of $-1$ and $1$ respectively. So:

\begin{equation*}
STG_{1,2}(\pi_h,\pi_{h/t},\pi_{m/o},\Pi_o,w_{\dot{L}},\mu) = 1
\end{equation*}

Therefore:

\begin{equation*}
STG_{1,2}(\Pi_e,w_{\Pi_e},\Pi_o,w_{\dot{L}},\mu) = 6 \frac{9}{2} \frac{1}{3} \frac{1}{3} \frac{1}{3} 1 = 1
\end{equation*}

And for slots 2 and 1:

\begin{equation*}
\begin{aligned}
STG_{2,1}(\Pi_e,w_{\Pi_e},\Pi_o,w_{\dot{L}},\mu)	& = \eta_{\Pi^3} \sum_{\pi_1,\pi_2,\pi_3 \in \Pi_e | \pi_1 \neq \pi_2 \neq \pi_3} w_{\Pi_e}(\pi_1) w_{\Pi_e}(\pi_2) w_{\Pi_e}(\pi_3) STG_{2,1}(\pi_1,\pi_2,\pi_3,\Pi_o,w_{\dot{L}},\mu) =\\
													& = 6 \frac{9}{2} \frac{1}{3} \frac{1}{3} \frac{1}{3} STG_{2,1}(\pi_h,\pi_{h/t},\pi_{m/o},\Pi_o,w_{\dot{L}},\mu)
\end{aligned}
\end{equation*}

Again, we only need to calculate $STG_{2,1}(\pi_h,\pi_{h/t},\pi_{m/o},\Pi_o,w_{\dot{L}},\mu)$. We follow definition \ref{def:STG_agents} to calculate this value:

\begin{equation*}
\begin{aligned}
STG_{2,1}(\pi_h,\pi_{h/t},\pi_{m/o},\Pi_o,w_{\dot{L}},\mu)	& = \sum_{\dot{l} \in \dot{L}^{N(\mu)}_{-2,1}(\Pi_o)} w_{\dot{L}}(\dot{l}) STO_{2,1}(\pi_h,\pi_{h/t},\pi_{m/o},\dot{l},\mu) =\\
															& = STO_{2,1}(\pi_h,\pi_{h/t},\pi_{m/o},(*,*),\mu)
\end{aligned}
\end{equation*}

The following table shows us $STO_{2,1}$ for all the permutations of $\pi_h,\pi_{h/t},\pi_{m/o}$.

\begin{center}
\begin{tabular}{c c c | c c c | c c c}
Slot 2 & & Slot 1				& Slot 2 & & Slot 1					& Slot 2 & & Slot 1\\
\hline
$\pi_h$ & $<$ & $\pi_{h/t}$		& $\pi_h$ & $<$ & $\pi_{m/o}$		& $\pi_{h/t}$ & $<$ & $\pi_h$\\
$\pi_{h/t}$ & $<$ & $\pi_{m/o}$	& $\pi_{m/o}$ & $<$ & $\pi_{h/t}$	& $\pi_h$ & $<$ & $\pi_{m/o}$\\
$\pi_h$ & $<$ & $\pi_{m/o}$		& $\pi_h$ & $<$ & $\pi_{h/t}$		& $\pi_{h/t}$ & $<$ & $\pi_{m/o}$
\end{tabular}
\begin{tabular}{c c c | c c c | c c c}
Slot 2 & & Slot 1				& Slot 2 & & Slot 1					& Slot 2 & & Slot 1\\
\hline
$\pi_{h/t}$ & $<$ & $\pi_{m/o}$	& $\pi_{m/o}$ & $<$ & $\pi_h$		& $\pi_{m/o}$ & $<$ & $\pi_{h/t}$\\
$\pi_{m/o}$ & $<$ & $\pi_h$		& $\pi_h$ & $<$ & $\pi_{h/t}$		& $\pi_{h/t}$ & $<$ & $\pi_h$\\
$\pi_{h/t}$ & $<$ & $\pi_h$		& $\pi_{m/o}$ & $<$ & $\pi_{h/t}$	& $\pi_{m/o}$ & $<$ & $\pi_h$
\end{tabular}
\end{center}

Again, it is possible to find a STO for the first permutation. In $\pi_h < \pi_{h/t}$, $\pi_h$ will always perform Head and $\pi_{h/t}$ will always perform Head, so they will obtain an expected average reward of $-1$ and $1$ respectively. In $\pi_{h/t} < \pi_{m/o}$, $\pi_{h/t}$ will always perform Tail and $\pi_{m/o}$ will always perform Tail, so they will obtain an expected average reward of $-1$ and $1$ respectively. In $\pi_h < \pi_{m/o}$, $\pi_h$ will always perform Head and $\pi_{m/o}$ will always perform Head, so they will obtain an expected average reward of $-1$ and $1$ respectively. So:

\begin{equation*}
STG_{2,1}(\pi_h,\pi_{h/t},\pi_{m/o},\Pi_o,w_{\dot{L}},\mu) = 1
\end{equation*}

Therefore:

\begin{equation*}
STG_{2,1}(\Pi_e,w_{\Pi_e},\Pi_o,w_{\dot{L}},\mu) = 6 \frac{9}{2} \frac{1}{3} \frac{1}{3} \frac{1}{3} 1 = 1
\end{equation*}

And finally, we weight over the slots:

\begin{equation*}
\begin{aligned}
& STG(\Pi_e,w_{\Pi_e},\Pi_o,w_{\dot{L}},\mu,w_S) = \eta_{S_1^2} \sum_{i=1}^{N(\mu)} w_S(i,\mu) \times\\
& \times \left(\sum_{j=1}^{i-1} w_S(j,\mu) STG_{i,j}(\Pi_e,w_{\Pi_e},\Pi_o,w_{\dot{L}},\mu) + \sum_{j=i+1}^{N(\mu)} w_S(j,\mu) STG_{i,j}(\Pi_e,w_{\Pi_e},\Pi_o,w_{\dot{L}},\mu)\right) =\\
& \ \ \ \ \ \ \ \ \ \ \ \ \ \ \ \ \ \ \ \ \ \ \ \ \ \ \ \ \ \ \ \ \ \ \ \ \ \ = \frac{2}{1} \frac{1}{2} \frac{1}{2} \{STG_{1,2}(\Pi_e,w_{\Pi_e},\Pi_o,w_{\dot{L}},\mu) + STG_{2,1}(\Pi_e,w_{\Pi_e},\Pi_o,w_{\dot{L}},\mu)\} =\\
& \ \ \ \ \ \ \ \ \ \ \ \ \ \ \ \ \ \ \ \ \ \ \ \ \ \ \ \ \ \ \ \ \ \ \ \ \ \ = \frac{2}{1} \frac{1}{2} \frac{1}{2} \left\{1 + 1\right\} = 1
\end{aligned}
\end{equation*}

Since $1$ is the highest possible value for the strict total grading property, therefore matching pennies has $General_{max} = 1$ for this property.
\end{proof}
\end{proposition}

\begin{proposition}
\label{prop:matching_pennies_STG_left_max}
$Left_{max}$ for the strict total grading (STG) property is equal to $1$ for the matching pennies environment.

\begin{proof}
To find $Left_{max}$ (equation \ref{eq:left_max}), we need to find a pair $\left\langle\Pi_e,w_{\Pi_e}\right\rangle$ which maximises the property as much as possible while $\Pi_o$ minimises it. Using $\Pi_e = \{\pi_h,\pi_{h/t},\pi_{m/o}\}$ with uniform weight for $w_{\Pi_e}$ (a $\pi_h$ agent always performs Head, a $\pi_{h/t}$ agent always performs Head when playing in slot 1 and always performs Tail when playing in slot 2, and a $\pi_{m/o}$ agent always mimics the last action of its opponent when playing in slot 1 and always performs the opposite of this action when playing in slot 2)\footnote{$\pi_{h/t}$ and $\pi_{m/o}$ have to know in which slot they are playing. To infer this, they start with a random action at the first iteration and then look at the action of their opponent and the reward they obtain.} we find this situation no matter which $\Pi_o$ we use.

Following definition \ref{def:STG}, we obtain the STG value for this $\left\langle\Pi_e,w_{\Pi_e},\Pi_o\right\rangle$ (where $\Pi_o$ is instantiated with any permitted value). Since the environment is not symmetric, we need to calculate this property for every pair of slots. Following definition \ref{def:STG_set}, we can calculate its STG value for each pair of slots. We start with slots 1 and 2:

\begin{equation*}
\begin{aligned}
STG_{1,2}(\Pi_e,w_{\Pi_e},\Pi_o,w_{\dot{L}},\mu)	& = \eta_{\Pi^3} \sum_{\pi_1,\pi_2,\pi_3 \in \Pi_e | \pi_1 \neq \pi_2 \neq \pi_3} w_{\Pi_e}(\pi_1) w_{\Pi_e}(\pi_2) w_{\Pi_e}(\pi_3) STG_{1,2}(\pi_1,\pi_2,\pi_3,\Pi_o,w_{\dot{L}},\mu) =\\
													& = 6 \frac{9}{2} \frac{1}{3} \frac{1}{3} \frac{1}{3} STG_{1,2}(\pi_h,\pi_{h/t},\pi_{m/o},\Pi_o,w_{\dot{L}},\mu)
\end{aligned}
\end{equation*}

\noindent Note that we avoided to calculate all the permutations of $\pi_1,\pi_2,\pi_3$ for $STG_{i,j}(\pi_1,\pi_2,\pi_3,\Pi_o,w_{\dot{L}},\mu)$ since they provide the same result, by calculating only one permutation and multiplying the result by the number of permutations $6$.

In this case, we only need to calculate $STG_{1,2}(\pi_h,\pi_{h/t},\pi_{m/o},\Pi_o,w_{\dot{L}},\mu)$. We follow definition \ref{def:STG_agents} to calculate this value:

\begin{equation*}
\begin{aligned}
STG_{1,2}(\pi_h,\pi_{h/t},\pi_{m/o},\Pi_o,w_{\dot{L}},\mu)	& = \sum_{\dot{l} \in \dot{L}^{N(\mu)}_{-1,2}(\Pi_o)} w_{\dot{L}}(\dot{l}) STO_{1,2}(\pi_h,\pi_{h/t},\pi_{m/o},\dot{l},\mu) =\\
															& = STO_{1,2}(\pi_h,\pi_{h/t},\pi_{m/o},(*,*),\mu)
\end{aligned}
\end{equation*}

\noindent Note that the choice of $\Pi_o$ does not affect the result of $STG_{1,2}$.

The following table shows us $STO_{1,2}$ for all the permutations of $\pi_h,\pi_{h/t},\pi_{m/o}$.

\begin{center}
\begin{tabular}{c c c | c c c | c c c}
Slot 1 & & Slot 2				& Slot 1 & & Slot 2					& Slot 1 & & Slot 2\\
\hline
$\pi_h$ & $<$ & $\pi_{h/t}$		& $\pi_h$ & $<$ & $\pi_{m/o}$		& $\pi_{h/t}$ & $<$ & $\pi_h$\\
$\pi_{h/t}$ & $<$ & $\pi_{m/o}$	& $\pi_{m/o}$ & $<$ & $\pi_{h/t}$	& $\pi_h$ & $<$ & $\pi_{m/o}$\\
$\pi_h$ & $<$ & $\pi_{m/o}$		& $\pi_h$ & $<$ & $\pi_{h/t}$		& $\pi_{h/t}$ & $<$ & $\pi_{m/o}$
\end{tabular}
\begin{tabular}{c c c | c c c | c c c}
Slot 1 & & Slot 2				& Slot 1 & & Slot 2					& Slot 1 & & Slot 2\\
\hline
$\pi_{h/t}$ & $<$ & $\pi_{m/o}$	& $\pi_{m/o}$ & $<$ & $\pi_h$		& $\pi_{m/o}$ & $<$ & $\pi_{h/t}$\\
$\pi_{m/o}$ & $<$ & $\pi_h$		& $\pi_h$ & $<$ & $\pi_{h/t}$		& $\pi_{h/t}$ & $<$ & $\pi_h$\\
$\pi_{h/t}$ & $<$ & $\pi_h$		& $\pi_{m/o}$ & $<$ & $\pi_{h/t}$	& $\pi_{m/o}$ & $<$ & $\pi_h$
\end{tabular}
\end{center}

It is possible to find a STO for the first permutation. In $\pi_h < \pi_{h/t}$, $\pi_h$ will always perform Head and $\pi_{h/t}$ will always perform Tail, so they will obtain an expected average reward of $-1$ and $1$ respectively. In $\pi_{h/t} < \pi_{m/o}$, $\pi_{h/t}$ will always perform Head and $\pi_{m/o}$ will always perform Tail, so they will obtain an expected average reward of $-1$ and $1$ respectively. In $\pi_h < \pi_{m/o}$, $\pi_h$ will always perform Head and $\pi_{m/o}$ will always perform Tail, so they will obtain an expected average reward of $-1$ and $1$ respectively. So:

\begin{equation*}
STG_{1,2}(\pi_h,\pi_{h/t},\pi_{m/o},\Pi_o,w_{\dot{L}},\mu) = 1
\end{equation*}

Therefore:

\begin{equation*}
STG_{1,2}(\Pi_e,w_{\Pi_e},\Pi_o,w_{\dot{L}},\mu) = 6 \frac{9}{2} \frac{1}{3} \frac{1}{3} \frac{1}{3} 1 = 1
\end{equation*}

And for slots 2 and 1:

\begin{equation*}
\begin{aligned}
STG_{2,1}(\Pi_e,w_{\Pi_e},\Pi_o,w_{\dot{L}},\mu)	& = \eta_{\Pi^3} \sum_{\pi_1,\pi_2,\pi_3 \in \Pi_e | \pi_1 \neq \pi_2 \neq \pi_3} w_{\Pi_e}(\pi_1) w_{\Pi_e}(\pi_2) w_{\Pi_e}(\pi_3) STG_{2,1}(\pi_1,\pi_2,\pi_3,\Pi_o,w_{\dot{L}},\mu) =\\
													& = 6 \frac{9}{2} \frac{1}{3} \frac{1}{3} \frac{1}{3} STG_{2,1}(\pi_h,\pi_{h/t},\pi_{m/o},\Pi_o,w_{\dot{L}},\mu)
\end{aligned}
\end{equation*}

Again, we only need to calculate $STG_{2,1}(\pi_h,\pi_{h/t},\pi_{m/o},\Pi_o,w_{\dot{L}},\mu)$. We follow definition \ref{def:STG_agents} to calculate this value:

\begin{equation*}
\begin{aligned}
STG_{2,1}(\pi_h,\pi_{h/t},\pi_{m/o},\Pi_o,w_{\dot{L}},\mu)	& = \sum_{\dot{l} \in \dot{L}^{N(\mu)}_{-2,1}(\Pi_o)} w_{\dot{L}}(\dot{l}) STO_{2,1}(\pi_h,\pi_{h/t},\pi_{m/o},\dot{l},\mu) =\\
															& = STO_{2,1}(\pi_h,\pi_{h/t},\pi_{m/o},(*,*),\mu)
\end{aligned}
\end{equation*}

\noindent Note again that the choice of $\Pi_o$ does not affect the result of $STG_{2,1}$.

The following table shows us $STO_{2,1}$ for all the permutations of $\pi_h,\pi_{h/t},\pi_{m/o}$.

\begin{center}
\begin{tabular}{c c c | c c c | c c c}
Slot 2 & & Slot 1				& Slot 2 & & Slot 1					& Slot 2 & & Slot 1\\
\hline
$\pi_h$ & $<$ & $\pi_{h/t}$		& $\pi_h$ & $<$ & $\pi_{m/o}$		& $\pi_{h/t}$ & $<$ & $\pi_h$\\
$\pi_{h/t}$ & $<$ & $\pi_{m/o}$	& $\pi_{m/o}$ & $<$ & $\pi_{h/t}$	& $\pi_h$ & $<$ & $\pi_{m/o}$\\
$\pi_h$ & $<$ & $\pi_{m/o}$		& $\pi_h$ & $<$ & $\pi_{h/t}$		& $\pi_{h/t}$ & $<$ & $\pi_{m/o}$
\end{tabular}
\begin{tabular}{c c c | c c c | c c c}
Slot 2 & & Slot 1				& Slot 2 & & Slot 1					& Slot 2 & & Slot 1\\
\hline
$\pi_{h/t}$ & $<$ & $\pi_{m/o}$	& $\pi_{m/o}$ & $<$ & $\pi_h$		& $\pi_{m/o}$ & $<$ & $\pi_{h/t}$\\
$\pi_{m/o}$ & $<$ & $\pi_h$		& $\pi_h$ & $<$ & $\pi_{h/t}$		& $\pi_{h/t}$ & $<$ & $\pi_h$\\
$\pi_{h/t}$ & $<$ & $\pi_h$		& $\pi_{m/o}$ & $<$ & $\pi_{h/t}$	& $\pi_{m/o}$ & $<$ & $\pi_h$
\end{tabular}
\end{center}

Again, it is possible to find a STO for the first permutation. In $\pi_h < \pi_{h/t}$, $\pi_h$ will always perform Head and $\pi_{h/t}$ will always perform Head, so they will obtain an expected average reward of $-1$ and $1$ respectively. In $\pi_{h/t} < \pi_{m/o}$, $\pi_{h/t}$ will always perform Tail and $\pi_{m/o}$ will always perform Tail, so they will obtain an expected average reward of $-1$ and $1$ respectively. In $\pi_h < \pi_{m/o}$, $\pi_h$ will always perform Head and $\pi_{m/o}$ will always perform Head, so they will obtain an expected average reward of $-1$ and $1$ respectively. So:

\begin{equation*}
STG_{2,1}(\pi_h,\pi_{h/t},\pi_{m/o},\Pi_o,w_{\dot{L}},\mu) = 1
\end{equation*}

Therefore:

\begin{equation*}
STG_{2,1}(\Pi_e,w_{\Pi_e},\Pi_o,w_{\dot{L}},\mu) = 6 \frac{9}{2} \frac{1}{3} \frac{1}{3} \frac{1}{3} 1 = 1
\end{equation*}

And finally, we weight over the slots:

\begin{equation*}
\begin{aligned}
& STG(\Pi_e,w_{\Pi_e},\Pi_o,w_{\dot{L}},\mu,w_S) = \eta_{S_1^2} \sum_{i=1}^{N(\mu)} w_S(i,\mu) \times\\
& \times \left(\sum_{j=1}^{i-1} w_S(j,\mu) STG_{i,j}(\Pi_e,w_{\Pi_e},\Pi_o,w_{\dot{L}},\mu) + \sum_{j=i+1}^{N(\mu)} w_S(j,\mu) STG_{i,j}(\Pi_e,w_{\Pi_e},\Pi_o,w_{\dot{L}},\mu)\right) =\\
& \ \ \ \ \ \ \ \ \ \ \ \ \ \ \ \ \ \ \ \ \ \ \ \ \ \ \ \ \ \ \ \ \ \ \ \ \ \ = \frac{2}{1} \frac{1}{2} \frac{1}{2} \{STG_{1,2}(\Pi_e,w_{\Pi_e},\Pi_o,w_{\dot{L}},\mu) + STG_{2,1}(\Pi_e,w_{\Pi_e},\Pi_o,w_{\dot{L}},\mu)\} =\\
& \ \ \ \ \ \ \ \ \ \ \ \ \ \ \ \ \ \ \ \ \ \ \ \ \ \ \ \ \ \ \ \ \ \ \ \ \ \ = \frac{2}{1} \frac{1}{2} \frac{1}{2} \left\{1 + 1\right\} = 1
\end{aligned}
\end{equation*}

Since $1$ is the highest possible value for the strict total grading property and no $\Pi_o$ is able to influence this result, therefore matching pennies has $Left_{max} = 1$ for this property.
\end{proof}
\end{proposition}

\begin{proposition}
\label{prop:matching_pennies_STG_right_min}
$Right_{min}$ for the strict total grading (STG) property is equal to $0$ for the matching pennies environment.

\begin{proof}
To find $Right_{min}$ (equation \ref{eq:right_min}), we need to find a pair $\left\langle\Pi_e,w_{\Pi_e}\right\rangle$ which minimises the property as much as possible while $\Pi_o$ maximises it. Using $\Pi_e = \{{\pi_r}_1,{\pi_r}_2,{\pi_r}_3\}$ with uniform weight for $w_{\Pi_e}$ (a $\pi_r$ agent always acts randomly) we find this situation no matter which $\Pi_o$ we use.

Following definition \ref{def:STG}, we obtain the STG value for this $\left\langle\Pi_e,w_{\Pi_e},\Pi_o\right\rangle$ (where $\Pi_o$ is instantiated with any permitted value). Since the environment is not symmetric, we need to calculate this property for every pair of slots. Following definition \ref{def:STG_set}, we can calculate its STG value for each pair of slots. We start with slots 1 and 2:

\begin{equation*}
\begin{aligned}
STG_{1,2}(\Pi_e,w_{\Pi_e},\Pi_o,w_{\dot{L}},\mu)	& = \eta_{\Pi^3} \sum_{\pi_1,\pi_2,\pi_3 \in \Pi_e | \pi_1 \neq \pi_2 \neq \pi_3} w_{\Pi_e}(\pi_1) w_{\Pi_e}(\pi_2) w_{\Pi_e}(\pi_3) STG_{1,2}(\pi_1,\pi_2,\pi_3,\Pi_o,w_{\dot{L}},\mu) =\\
													& = 6 \frac{9}{2} \frac{1}{3} \frac{1}{3} \frac{1}{3} STG_{1,2}({\pi_r}_1,{\pi_r}_2,{\pi_r}_3,\Pi_o,w_{\dot{L}},\mu)
\end{aligned}
\end{equation*}

\noindent Note that we avoided to calculate all the permutations of $\pi_1,\pi_2,\pi_3$ for $STG_{i,j}(\pi_1,\pi_2,\pi_3,\Pi_o,w_{\dot{L}},\mu)$ since they provide the same result, by calculating only one permutation and multiplying the result by the number of permutations $6$.

In this case, we only need to calculate $STG_{1,2}({\pi_r}_1,{\pi_r}_2,{\pi_r}_3,\Pi_o,w_{\dot{L}},\mu)$. We follow definition \ref{def:STG_agents} to calculate this value:

\begin{equation*}
\begin{aligned}
STG_{1,2}({\pi_r}_1,{\pi_r}_2,{\pi_r}_3,\Pi_o,w_{\dot{L}},\mu)	& = \sum_{\dot{l} \in \dot{L}^{N(\mu)}_{-1,2}(\Pi_o)} w_{\dot{L}}(\dot{l}) STO_{1,2}({\pi_r}_1,{\pi_r}_2,{\pi_r}_3,\dot{l},\mu) =\\
																& = STO_{1,2}({\pi_r}_1,{\pi_r}_2,{\pi_r}_3,(*,*),\mu)
\end{aligned}
\end{equation*}

\noindent Note that the choice of $\Pi_o$ does not affect the result of $STG_{1,2}$.

The following table shows us $STO_{1,2}$ for all the permutations of ${\pi_r}_1,{\pi_r}_2,{\pi_r}_3$.

\begin{center}
\begin{tabular}{c c c | c c c | c c c}
Slot 1 & & Slot 2				& Slot 1 & & Slot 2					& Slot 1 & & Slot 2\\
\hline
${\pi_r}_1$ & $<$ & ${\pi_r}_2$	& ${\pi_r}_1$ & $<$ & ${\pi_r}_3$	& ${\pi_r}_2$ & $<$ & ${\pi_r}_1$\\
${\pi_r}_2$ & $<$ & ${\pi_r}_3$	& ${\pi_r}_3$ & $<$ & ${\pi_r}_2$	& ${\pi_r}_1$ & $<$ & ${\pi_r}_3$\\
${\pi_r}_1$ & $<$ & ${\pi_r}_3$	& ${\pi_r}_1$ & $<$ & ${\pi_r}_2$	& ${\pi_r}_2$ & $<$ & ${\pi_r}_3$
\end{tabular}
\begin{tabular}{c c c | c c c | c c c}
Slot 1 & & Slot 2				& Slot 1 & & Slot 2					& Slot 1 & & Slot 2\\
\hline
${\pi_r}_2$ & $<$ & ${\pi_r}_3$	& ${\pi_r}_3$ & $<$ & ${\pi_r}_1$	& ${\pi_r}_3$ & $<$ & ${\pi_r}_2$\\
${\pi_r}_3$ & $<$ & ${\pi_r}_1$	& ${\pi_r}_1$ & $<$ & ${\pi_r}_2$	& ${\pi_r}_2$ & $<$ & ${\pi_r}_1$\\
${\pi_r}_2$ & $<$ & ${\pi_r}_1$	& ${\pi_r}_3$ & $<$ & ${\pi_r}_2$	& ${\pi_r}_3$ & $<$ & ${\pi_r}_1$
\end{tabular}
\end{center}

But, it is not possible to find a STO, since for every permutation we have at least one random agent making its expected average reward equal to its opponent expected average reward (both have an expected average reward of $0$ as proved in lemma \ref{lemma:matching_pennies_random_agent}). So:

\begin{equation*}
STG_{1,2}({\pi_r}_1,{\pi_r}_2,{\pi_r}_3,\Pi_o,w_{\dot{L}},\mu) = 0
\end{equation*}

Therefore:

\begin{equation*}
STG_{1,2}(\Pi_e,w_{\Pi_e},\Pi_o,w_{\dot{L}},\mu) = 6 \frac{9}{2} \frac{1}{3} \frac{1}{3} \frac{1}{3} 0 = 0
\end{equation*}

And for slots 2 and 1:

\begin{equation*}
\begin{aligned}
STG_{2,1}(\Pi_e,w_{\Pi_e},\Pi_o,w_{\dot{L}},\mu)	& = \eta_{\Pi^3} \sum_{\pi_1,\pi_2,\pi_3 \in \Pi_e | \pi_1 \neq \pi_2 \neq \pi_3} w_{\Pi_e}(\pi_1) w_{\Pi_e}(\pi_2) w_{\Pi_e}(\pi_3) STG_{2,1}(\pi_1,\pi_2,\pi_3,\Pi_o,w_{\dot{L}},\mu) =\\
													& = 6 \frac{9}{2} \frac{1}{3} \frac{1}{3} \frac{1}{3} STG_{2,1}({\pi_r}_1,{\pi_r}_2,{\pi_r}_3,\Pi_o,w_{\dot{L}},\mu)
\end{aligned}
\end{equation*}

Again, we only need to calculate $STG_{2,1}({\pi_r}_1,{\pi_r}_2,{\pi_r}_3,\Pi_o,w_{\dot{L}},\mu)$. We follow definition \ref{def:STG_agents} to calculate this value:

\begin{equation*}
\begin{aligned}
STG_{2,1}({\pi_r}_1,{\pi_r}_2,{\pi_r}_3,\Pi_o,w_{\dot{L}},\mu)	& = \sum_{\dot{l} \in \dot{L}^{N(\mu)}_{-2,1}(\Pi_o)} w_{\dot{L}}(\dot{l}) STO_{2,1}({\pi_r}_1,{\pi_r}_2,{\pi_r}_3,\dot{l},\mu) =\\
																& = STO_{2,1}({\pi_r}_1,{\pi_r}_2,{\pi_r}_3,(*,*),\mu)
\end{aligned}
\end{equation*}

\noindent Note again that the choice of $\Pi_o$ does not affect the result of $STG_{2,1}$.

The following table shows us $STO_{2,1}$ for all the permutations of ${\pi_r}_1,{\pi_r}_2,{\pi_r}_3$.

\begin{center}
\begin{tabular}{c c c | c c c | c c c}
Slot 2 & & Slot 1				& Slot 2 & & Slot 1					& Slot 2 & & Slot 1\\
\hline
${\pi_r}_1$ & $<$ & ${\pi_r}_2$	& ${\pi_r}_1$ & $<$ & ${\pi_r}_3$	& ${\pi_r}_2$ & $<$ & ${\pi_r}_1$\\
${\pi_r}_2$ & $<$ & ${\pi_r}_3$	& ${\pi_r}_3$ & $<$ & ${\pi_r}_2$	& ${\pi_r}_1$ & $<$ & ${\pi_r}_3$\\
${\pi_r}_1$ & $<$ & ${\pi_r}_3$	& ${\pi_r}_1$ & $<$ & ${\pi_r}_2$	& ${\pi_r}_2$ & $<$ & ${\pi_r}_3$
\end{tabular}
\begin{tabular}{c c c | c c c | c c c}
Slot 2 & & Slot 1				& Slot 2 & & Slot 1					& Slot 2 & & Slot 1\\
\hline
${\pi_r}_2$ & $<$ & ${\pi_r}_3$	& ${\pi_r}_3$ & $<$ & ${\pi_r}_1$	& ${\pi_r}_3$ & $<$ & ${\pi_r}_2$\\
${\pi_r}_3$ & $<$ & ${\pi_r}_1$	& ${\pi_r}_1$ & $<$ & ${\pi_r}_2$	& ${\pi_r}_2$ & $<$ & ${\pi_r}_1$\\
${\pi_r}_2$ & $<$ & ${\pi_r}_1$	& ${\pi_r}_3$ & $<$ & ${\pi_r}_2$	& ${\pi_r}_3$ & $<$ & ${\pi_r}_1$
\end{tabular}
\end{center}

Again, it is not possible to find a STO, since for every permutation we have at least one random agent making its expected average reward equal to its opponent expected average reward (both have an expected average reward of $0$ as proved in lemma \ref{lemma:matching_pennies_random_agent}). So:

\begin{equation*}
STG_{2,1}({\pi_r}_1,{\pi_r}_2,{\pi_r}_3,\Pi_o,w_{\dot{L}},\mu) = 0
\end{equation*}

Therefore:

\begin{equation*}
STG_{2,1}(\Pi_e,w_{\Pi_e},\Pi_o,w_{\dot{L}},\mu) = 6 \frac{9}{2} \frac{1}{3} \frac{1}{3} \frac{1}{3} 0 = 0
\end{equation*}

And finally, we weight over the slots:

\begin{equation*}
\begin{aligned}
& STG(\Pi_e,w_{\Pi_e},\Pi_o,w_{\dot{L}},\mu,w_S) = \eta_{S_1^2} \sum_{i=1}^{N(\mu)} w_S(i,\mu) \times\\
& \times \left(\sum_{j=1}^{i-1} w_S(j,\mu) STG_{i,j}(\Pi_e,w_{\Pi_e},\Pi_o,w_{\dot{L}},\mu) + \sum_{j=i+1}^{N(\mu)} w_S(j,\mu) STG_{i,j}(\Pi_e,w_{\Pi_e},\Pi_o,w_{\dot{L}},\mu)\right) =\\
& \ \ \ \ \ \ \ \ \ \ \ \ \ \ \ \ \ \ \ \ \ \ \ \ \ \ \ \ \ \ \ \ \ \ \ \ \ \ = \frac{2}{1} \frac{1}{2} \frac{1}{2} \{STG_{1,2}(\Pi_e,w_{\Pi_e},\Pi_o,w_{\dot{L}},\mu) + STG_{2,1}(\Pi_e,w_{\Pi_e},\Pi_o,w_{\dot{L}},\mu)\} =\\
& \ \ \ \ \ \ \ \ \ \ \ \ \ \ \ \ \ \ \ \ \ \ \ \ \ \ \ \ \ \ \ \ \ \ \ \ \ \ = \frac{2}{1} \frac{1}{2} \frac{1}{2} \left\{0 + 0\right\} = 0
\end{aligned}
\end{equation*}

Since $0$ is the lowest possible value for the strict total grading property and no $\Pi_o$ is able to influence this result, therefore matching pennies has $Right_{min} = 0$ for this property.
\end{proof}
\end{proposition}

\subsection{Partial Grading}
Now we arrive to the partial grading (PG) property. As given in section \ref{sec:PG}, we want to know if there exists a partial ordering between the evaluated agents when interacting in the environment.

To simplify the notation, we use the next table to represent the PO:
$R_i(\mu[\instantiation{l}{i,j}{\pi_1,\pi_2}]) \leq R_j(\mu[\instantiation{l}{i,j}{\pi_1,\pi_2}])$,
$R_i(\mu[\instantiation{l}{i,j}{\pi_2,\pi_3}]) \leq R_j(\mu[\instantiation{l}{i,j}{\pi_2,\pi_3}])$ and
$R_i(\mu[\instantiation{l}{i,j}{\pi_1,\pi_3}]) \leq R_j(\mu[\instantiation{l}{i,j}{\pi_1,\pi_3}])$.

\begin{center}
\begin{tabular}{c c c}
Slot i & & Slot j\\
\hline
$\pi_1$ & $\leq$ & $\pi_2$\\
$\pi_2$ & $\leq$ & $\pi_3$\\
$\pi_1$ & $\leq$ & $\pi_3$
\end{tabular}
\end{center}

\begin{proposition}
\label{prop:matching_pennies_PG_general_min}
$General_{min}$ for the partial grading (PG) property is equal to $0$ for the matching pennies environment.

\begin{proof}
To find $General_{min}$ (equation \ref{eq:general_min}), we need to find a trio $\left\langle\Pi_e,w_{\Pi_e},\Pi_o\right\rangle$ which minimises the property as much as possible. We can have this situation by selecting $\Pi_e = \{{\pi_h}_1,{\pi_h}_2,\pi_t\}$ with uniform weight for $w_{\Pi_e}$ and $\Pi_o = \emptyset$ (a $\pi_h$ agent always performs Head and a $\pi_t$ agent always performs Tail).

Following definition \ref{def:PG}, we obtain the PG value for this $\left\langle\Pi_e,w_{\Pi_e},\Pi_o\right\rangle$. Since the environment is not symmetric, we need to calculate this property for every pair of slots. Following definition \ref{def:STG_set} (for PG), we can calculate its PG value for each pair of slots. We start with slots 1 and 2:

\begin{equation*}
\begin{aligned}
PG_{1,2}(\Pi_e,w_{\Pi_e},\Pi_o,w_{\dot{L}},\mu)	& = \eta_{\Pi^3} \sum_{\pi_1,\pi_2,\pi_3 \in \Pi_e | \pi_1 \neq \pi_2 \neq \pi_3} w_{\Pi_e}(\pi_1) w_{\Pi_e}(\pi_2) w_{\Pi_e}(\pi_3) PG_{1,2}(\pi_1,\pi_2,\pi_3,\Pi_o,w_{\dot{L}},\mu) =\\
												& = 6 \frac{9}{2} \frac{1}{3} \frac{1}{3} \frac{1}{3} PG_{1,2}({\pi_h}_1,{\pi_h}_2,\pi_t,\Pi_o,w_{\dot{L}},\mu)
\end{aligned}
\end{equation*}

\noindent Note that we avoided to calculate all the permutations of $\pi_1,\pi_2,\pi_3$ for $PG_{i,j}(\pi_1,\pi_2,\pi_3,\Pi_o,w_{\dot{L}},\mu)$ since they provide the same result, by calculating only one permutation and multiplying the result by the number of permutations $6$.

In this case, we only need to calculate $PG_{1,2}({\pi_h}_1,{\pi_h}_2,\pi_t,\Pi_o,w_{\dot{L}},\mu)$. We follow definition \ref{def:STG_agents} (for PG) to calculate this value:

\begin{equation*}
\begin{aligned}
PG_{1,2}({\pi_h}_1,{\pi_h}_2,\pi_t,\Pi_o,w_{\dot{L}},\mu)	& = \sum_{\dot{l} \in \dot{L}^{N(\mu)}_{-1,2}(\Pi_o)} w_{\dot{L}}(\dot{l}) PO_{1,2}({\pi_h}_1,{\pi_h}_2,\pi_t,\dot{l},\mu) =\\
															& = PO_{1,2}({\pi_h}_1,{\pi_h}_2,\pi_t,(*,*),\mu)
\end{aligned}
\end{equation*}

The following table shows us $PO_{1,2}$ for all the permutations of ${\pi_h}_1,{\pi_h}_2,\pi_t$.

\begin{center}
\begin{tabular}{c c c | c c c | c c c}
Slot 1 & & Slot 2					& Slot 1 & & Slot 2						& Slot 1 & & Slot 2\\
\hline
${\pi_h}_1$ & $\leq$ & ${\pi_h}_2$	& ${\pi_h}_1$ & $\leq$ & $\pi_t$		& ${\pi_h}_2$ & $\leq$ & ${\pi_h}_1$\\
${\pi_h}_2$ & $\leq$ & $\pi_t$		& $\pi_t$ & $\leq$ & ${\pi_h}_2$		& ${\pi_h}_1$ & $\leq$ & $\pi_t$\\
${\pi_h}_1$ & $\leq$ & $\pi_t$		& ${\pi_h}_1$ & $\leq$ & ${\pi_h}_2$	& ${\pi_h}_2$ & $\leq$ & $\pi_t$
\end{tabular}
\begin{tabular}{c c c | c c c | c c c}
Slot 1 & & Slot 2					& Slot 1 & & Slot 2						& Slot 1 & & Slot 2\\
\hline
${\pi_h}_2$ & $\leq$ & $\pi_t$		& $\pi_t$ & $\leq$ & ${\pi_h}_1$		& $\pi_t$ & $\leq$ & ${\pi_h}_2$\\
$\pi_t$ & $\leq$ & ${\pi_h}_1$		& ${\pi_h}_1$ & $\leq$ & ${\pi_h}_2$	& ${\pi_h}_2$ & $\leq$ & ${\pi_h}_1$\\
${\pi_h}_2$ & $\leq$ & ${\pi_h}_1$	& $\pi_t$ & $\leq$ & ${\pi_h}_2$		& $\pi_t$ & $\leq$ & ${\pi_h}_1$
\end{tabular}
\end{center}

But, it is not possible to find a PO, since for every permutation we have either ${\pi_h}_1 \leq {\pi_h}_2$ or ${\pi_h}_2 \leq {\pi_h}_1$. In both cases, a $\pi_h$ agent will always perform Head, so they will obtain an expected average reward of $1$ and $-1$ respectively. So:

\begin{equation*}
PG_{1,2}({\pi_h}_1,{\pi_h}_2,\pi_t,\Pi_o,w_{\dot{L}},\mu) = 0
\end{equation*}

Therefore:

\begin{equation*}
PG_{1,2}(\Pi_e,w_{\Pi_e},\Pi_o,w_{\dot{L}},\mu) = 6 \frac{9}{2} \frac{1}{3} \frac{1}{3} \frac{1}{3} 0 = 0
\end{equation*}

And for slots 2 and 1:

\begin{equation*}
\begin{aligned}
PG_{2,1}(\Pi_e,w_{\Pi_e},\Pi_o,w_{\dot{L}},\mu)	& = \eta_{\Pi^3} \sum_{\pi_1,\pi_2,\pi_3 \in \Pi_e | \pi_1 \neq \pi_2 \neq \pi_3} w_{\Pi_e}(\pi_1) w_{\Pi_e}(\pi_2) w_{\Pi_e}(\pi_3) PG_{2,1}(\pi_1,\pi_2,\pi_3,\Pi_o,w_{\dot{L}},\mu) =\\
												& = 6 \frac{9}{2} \frac{1}{3} \frac{1}{3} \frac{1}{3} PG_{2,1}({\pi_h}_1,{\pi_h}_2,\pi_t,\Pi_o,w_{\dot{L}},\mu)
\end{aligned}
\end{equation*}

Again, we only need to calculate $PG_{2,1}({\pi_h}_1,{\pi_h}_2,\pi_t,\Pi_o,w_{\dot{L}},\mu)$. We follow definition \ref{def:STG_agents} (for PG) to calculate this value:

\begin{equation*}
\begin{aligned}
PG_{2,1}({\pi_h}_1,{\pi_h}_2,\pi_t,\Pi_o,w_{\dot{L}},\mu)	& = \sum_{\dot{l} \in \dot{L}^{N(\mu)}_{-2,1}(\Pi_o)} w_{\dot{L}}(\dot{l}) PO_{2,1}({\pi_h}_1,{\pi_h}_2,\pi_t,\dot{l},\mu) =\\
															& = PO_{2,1}({\pi_h}_1,{\pi_h}_2,\pi_t,(*,*),\mu)
\end{aligned}
\end{equation*}

The following table shows us $PO_{2,1}$ for all the permutations of ${\pi_h}_1,{\pi_h}_2,\pi_t$.

\begin{center}
\begin{tabular}{c c c | c c c | c c c}
Slot 2 & & Slot 1					& Slot 2 & & Slot 1						& Slot 2 & & Slot 1\\
\hline
${\pi_h}_1$ & $\leq$ & ${\pi_h}_2$	& ${\pi_h}_1$ & $\leq$ & $\pi_t$		& ${\pi_h}_2$ & $\leq$ & ${\pi_h}_1$\\
${\pi_h}_2$ & $\leq$ & $\pi_t$		& $\pi_t$ & $\leq$ & ${\pi_h}_2$		& ${\pi_h}_1$ & $\leq$ & $\pi_t$\\
${\pi_h}_1$ & $\leq$ & $\pi_t$		& ${\pi_h}_1$ & $\leq$ & ${\pi_h}_2$	& ${\pi_h}_2$ & $\leq$ & $\pi_t$
\end{tabular}
\begin{tabular}{c c c | c c c | c c c}
Slot 2 & & Slot 1					& Slot 2 & & Slot 1						& Slot 2 & & Slot 1\\
\hline
${\pi_h}_2$ & $\leq$ & $\pi_t$		& $\pi_t$ & $\leq$ & ${\pi_h}_1$		& $\pi_t$ & $\leq$ & ${\pi_h}_2$\\
$\pi_t$ & $\leq$ & ${\pi_h}_1$		& ${\pi_h}_1$ & $\leq$ & ${\pi_h}_2$	& ${\pi_h}_2$ & $\leq$ & ${\pi_h}_1$\\
${\pi_h}_2$ & $\leq$ & ${\pi_h}_1$	& $\pi_t$ & $\leq$ & ${\pi_h}_2$		& $\pi_t$ & $\leq$ & ${\pi_h}_1$
\end{tabular}
\end{center}

Again, it is not possible to find a PO, since for every permutation we have either ${\pi_h}_1 \leq \pi_t$ or $\pi_t \leq {\pi_h}_1$. In ${\pi_h}_1 \leq \pi_t$, ${\pi_h}_1$ will always perform Head and $\pi_t$ will always perform Tail, so they will obtain an expected average reward of $1$ and $-1$ respectively. In $\pi_t \leq {\pi_h}_1$, $\pi_t$ will always perform Tail and ${\pi_h}_1$ will always perform Head, so they will obtain an expected average reward of $1$ and $-1$ respectively. So:

\begin{equation*}
PG_{2,1}({\pi_h}_1,{\pi_h}_2,\pi_t,\Pi_o,w_{\dot{L}},\mu) = 0
\end{equation*}

Therefore:

\begin{equation*}
PG_{2,1}(\Pi_e,w_{\Pi_e},\Pi_o,w_{\dot{L}},\mu) = 6 \frac{9}{2} \frac{1}{3} \frac{1}{3} \frac{1}{3} 0 = 0
\end{equation*}

And finally, we weight over the slots:

\begin{equation*}
\begin{aligned}
& PG(\Pi_e,w_{\Pi_e},\Pi_o,w_{\dot{L}},\mu,w_S) = \eta_{S_1^2} \sum_{i=1}^{N(\mu)} w_S(i,\mu) \times\\
& \times \left(\sum_{j=1}^{i-1} w_S(j,\mu) PG_{i,j}(\Pi_e,w_{\Pi_e},\Pi_o,w_{\dot{L}},\mu) + \sum_{j=i+1}^{N(\mu)} w_S(j,\mu) PG_{i,j}(\Pi_e,w_{\Pi_e},\Pi_o,w_{\dot{L}},\mu)\right) =\\
& \ \ \ \ \ \ \ \ \ \ \ \ \ \ \ \ \ \ \ \ \ \ \ \ \ \ \ \ \ \ \ \ \ \ \ \ = \frac{2}{1} \frac{1}{2} \frac{1}{2} \{PG_{1,2}(\Pi_e,w_{\Pi_e},\Pi_o,w_{\dot{L}},\mu) + PG_{2,1}(\Pi_e,w_{\Pi_e},\Pi_o,w_{\dot{L}},\mu)\} =\\
& \ \ \ \ \ \ \ \ \ \ \ \ \ \ \ \ \ \ \ \ \ \ \ \ \ \ \ \ \ \ \ \ \ \ \ \ = \frac{2}{1} \frac{1}{2} \frac{1}{2} \left\{0 + 0\right\} = 0
\end{aligned}
\end{equation*}

Since $0$ is the lowest possible value for the partial grading property, therefore matching pennies has $General_{min} = 0$ for this property.
\end{proof}
\end{proposition}

\begin{proposition}
\label{prop:matching_pennies_PG_general_max}
$General_{max}$ for the partial grading (PG) property is equal to $1$ for the matching pennies environment.

\begin{proof}
To find $General_{max}$ (equation \ref{eq:general_max}), we need to find a trio $\left\langle\Pi_e,w_{\Pi_e},\Pi_o\right\rangle$ which maximises the property as much as possible. We can have this situation by selecting $\Pi_e = \{{\pi_r}_1,{\pi_r}_2,{\pi_r}_3\}$ with uniform weight for $w_{\Pi_e}$ and $\Pi_o = \emptyset$ (a $\pi_r$ agent always acts randomly).

Following definition \ref{def:PG}, we obtain the PG value for this $\left\langle\Pi_e,w_{\Pi_e},\Pi_o\right\rangle$. Since the environment is not symmetric, we need to calculate this property for every pair of slots. Following definition \ref{def:STG_set} (for PG), we can calculate its PG value for each pair of slots. We start with slots 1 and 2:

\begin{equation*}
\begin{aligned}
PG_{1,2}(\Pi_e,w_{\Pi_e},\Pi_o,w_{\dot{L}},\mu)	& = \eta_{\Pi^3} \sum_{\pi_1,\pi_2,\pi_3 \in \Pi_e | \pi_1 \neq \pi_2 \neq \pi_3} w_{\Pi_e}(\pi_1) w_{\Pi_e}(\pi_2) w_{\Pi_e}(\pi_3) PG_{1,2}(\pi_1,\pi_2,\pi_3,\Pi_o,w_{\dot{L}},\mu) =\\
												& = 6 \frac{9}{2} \frac{1}{3} \frac{1}{3} \frac{1}{3} PG_{1,2}({\pi_r}_1,{\pi_r}_2,{\pi_r}_3,\Pi_o,w_{\dot{L}},\mu)
\end{aligned}
\end{equation*}

\noindent Note that we avoided to calculate all the permutations of $\pi_1,\pi_2,\pi_3$ for $PG_{i,j}(\pi_1,\pi_2,\pi_3,\Pi_o,w_{\dot{L}},\mu)$ since they provide the same result, by calculating only one permutation and multiplying the result by the number of permutations $6$.

In this case, we only need to calculate $PG_{1,2}({\pi_r}_1,{\pi_r}_2,{\pi_r}_3,\Pi_o,w_{\dot{L}},\mu)$. We follow definition \ref{def:STG_agents} (for PG) to calculate this value:

\begin{equation*}
\begin{aligned}
PG_{1,2}({\pi_r}_1,{\pi_r}_2,{\pi_r}_3,\Pi_o,w_{\dot{L}},\mu)	& = \sum_{\dot{l} \in \dot{L}^{N(\mu)}_{-1,2}(\Pi_o)} w_{\dot{L}}(\dot{l}) PO_{1,2}(\pi_h,\pi_{h/t},\pi_{m/o},\dot{l},\mu) =\\
																& = PO_{1,2}({\pi_r}_1,{\pi_r}_2,{\pi_r}_3,(*,*),\mu)
\end{aligned}
\end{equation*}

The following table shows us $PO_{1,2}$ for all the permutations of ${\pi_r}_1,{\pi_r}_2,{\pi_r}_3$.

\begin{center}
\begin{tabular}{c c c | c c c | c c c}
Slot 1 & & Slot 2					& Slot 1 & & Slot 2						& Slot 1 & & Slot 2\\
\hline
${\pi_r}_1$ & $\leq$ & ${\pi_r}_2$	& ${\pi_r}_1$ & $\leq$ & ${\pi_r}_3$	& ${\pi_r}_2$ & $\leq$ & ${\pi_r}_1$\\
${\pi_r}_2$ & $\leq$ & ${\pi_r}_3$	& ${\pi_r}_3$ & $\leq$ & ${\pi_r}_2$	& ${\pi_r}_1$ & $\leq$ & ${\pi_r}_3$\\
${\pi_r}_1$ & $\leq$ & ${\pi_r}_3$	& ${\pi_r}_1$ & $\leq$ & ${\pi_r}_2$	& ${\pi_r}_2$ & $\leq$ & ${\pi_r}_3$
\end{tabular}
\begin{tabular}{c c c | c c c | c c c}
Slot 1 & & Slot 2					& Slot 1 & & Slot 2						& Slot 1 & & Slot 2\\
\hline
${\pi_r}_2$ & $\leq$ & ${\pi_r}_3$	& ${\pi_r}_3$ & $\leq$ & ${\pi_r}_1$	& ${\pi_r}_3$ & $\leq$ & ${\pi_r}_2$\\
${\pi_r}_3$ & $\leq$ & ${\pi_r}_1$	& ${\pi_r}_1$ & $\leq$ & ${\pi_r}_2$	& ${\pi_r}_2$ & $\leq$ & ${\pi_r}_1$\\
${\pi_r}_2$ & $\leq$ & ${\pi_r}_1$	& ${\pi_r}_3$ & $\leq$ & ${\pi_r}_2$	& ${\pi_r}_3$ & $\leq$ & ${\pi_r}_1$
\end{tabular}
\end{center}

It is possible to find a PO for every permutation, since we have at least one random agent making its expected average reward equal to its opponent expected average reward (both have an expected average reward of $0$ as proved in lemma \ref{lemma:matching_pennies_random_agent}). So:

\begin{equation*}
PG_{1,2}({\pi_r}_1,{\pi_r}_2,{\pi_r}_3,\Pi_o,w_{\dot{L}},\mu) = 1
\end{equation*}

Therefore:

\begin{equation*}
PG_{1,2}(\Pi_e,w_{\Pi_e},\Pi_o,w_{\dot{L}},\mu) = 6 \frac{9}{2} \frac{1}{3} \frac{1}{3} \frac{1}{3} 1 = 1
\end{equation*}

And for slots 2 and 1:

\begin{equation*}
\begin{aligned}
PG_{2,1}(\Pi_e,w_{\Pi_e},\Pi_o,w_{\dot{L}},\mu)	& = \eta_{\Pi^3} \sum_{\pi_1,\pi_2,\pi_3 \in \Pi_e | \pi_1 \neq \pi_2 \neq \pi_3} w_{\Pi_e}(\pi_1) w_{\Pi_e}(\pi_2) w_{\Pi_e}(\pi_3) PG_{2,1}(\pi_1,\pi_2,\pi_3,\Pi_o,w_{\dot{L}},\mu) =\\
												& = 6 \frac{9}{2} \frac{1}{3} \frac{1}{3} \frac{1}{3} PG_{2,1}({\pi_r}_1,{\pi_r}_2,{\pi_r}_3,\Pi_o,w_{\dot{L}},\mu)
\end{aligned}
\end{equation*}

Again, we only need to calculate $PG_{2,1}({\pi_r}_1,{\pi_r}_2,{\pi_r}_3,\Pi_o,w_{\dot{L}},\mu)$. We follow definition \ref{def:STG_agents} (for PG) to calculate this value:

\begin{equation*}
\begin{aligned}
PG_{2,1}({\pi_r}_1,{\pi_r}_2,{\pi_r}_3,\Pi_o,w_{\dot{L}},\mu)	& = \sum_{\dot{l} \in \dot{L}^{N(\mu)}_{-2,1}(\Pi_o)} w_{\dot{L}}(\dot{l}) PO_{2,1}({\pi_r}_1,{\pi_r}_2,{\pi_r}_3,\dot{l},\mu) =\\
																& = PO_{2,1}({\pi_r}_1,{\pi_r}_2,{\pi_r}_3,(*,*),\mu)
\end{aligned}
\end{equation*}

The following table shows us $PO_{2,1}$ for all the permutations of ${\pi_r}_1,{\pi_r}_2,{\pi_r}_3$.

\begin{center}
\begin{tabular}{c c c | c c c | c c c}
Slot 2 & & Slot 1					& Slot 2 & & Slot 1						& Slot 2 & & Slot 1\\
\hline
${\pi_r}_1$ & $\leq$ & ${\pi_r}_2$	& ${\pi_r}_1$ & $\leq$ & ${\pi_r}_3$	& ${\pi_r}_2$ & $\leq$ & ${\pi_r}_1$\\
${\pi_r}_2$ & $\leq$ & ${\pi_r}_3$	& ${\pi_r}_3$ & $\leq$ & ${\pi_r}_2$	& ${\pi_r}_1$ & $\leq$ & ${\pi_r}_3$\\
${\pi_r}_1$ & $\leq$ & ${\pi_r}_3$	& ${\pi_r}_1$ & $\leq$ & ${\pi_r}_2$	& ${\pi_r}_2$ & $\leq$ & ${\pi_r}_3$
\end{tabular}
\begin{tabular}{c c c | c c c | c c c}
Slot 2 & & Slot 1					& Slot 2 & & Slot 1						& Slot 2 & & Slot 1\\
\hline
${\pi_r}_2$ & $\leq$ & ${\pi_r}_3$	& ${\pi_r}_3$ & $\leq$ & ${\pi_r}_1$	& ${\pi_r}_3$ & $\leq$ & ${\pi_r}_2$\\
${\pi_r}_3$ & $\leq$ & ${\pi_r}_1$	& ${\pi_r}_1$ & $\leq$ & ${\pi_r}_2$	& ${\pi_r}_2$ & $\leq$ & ${\pi_r}_1$\\
${\pi_r}_2$ & $\leq$ & ${\pi_r}_1$	& ${\pi_r}_3$ & $\leq$ & ${\pi_r}_2$	& ${\pi_r}_3$ & $\leq$ & ${\pi_r}_1$
\end{tabular}
\end{center}

Again, it is possible to find a PO for every permutation, since we have at least one random agent making its expected average reward equal to its opponent expected average reward (both have an expected average reward of $0$ as proved in lemma \ref{lemma:matching_pennies_random_agent}). So:

\begin{equation*}
PG_{2,1}({\pi_r}_1,{\pi_r}_2,{\pi_r}_3,\Pi_o,w_{\dot{L}},\mu) = 1
\end{equation*}

Therefore:

\begin{equation*}
PG_{2,1}(\Pi_e,w_{\Pi_e},\Pi_o,w_{\dot{L}},\mu) = 6 \frac{9}{2} \frac{1}{3} \frac{1}{3} \frac{1}{3} 1 = 1
\end{equation*}

And finally, we weight over the slots:

\begin{equation*}
\begin{aligned}
& PG(\Pi_e,w_{\Pi_e},\Pi_o,w_{\dot{L}},\mu,w_S) = \eta_{S_1^2} \sum_{i=1}^{N(\mu)} w_S(i,\mu) \times\\
& \times \left(\sum_{j=1}^{i-1} w_S(j,\mu) PG_{i,j}(\Pi_e,w_{\Pi_e},\Pi_o,w_{\dot{L}},\mu) + \sum_{j=i+1}^{N(\mu)} w_S(j,\mu) PG_{i,j}(\Pi_e,w_{\Pi_e},\Pi_o,w_{\dot{L}},\mu)\right) =\\
& \ \ \ \ \ \ \ \ \ \ \ \ \ \ \ \ \ \ \ \ \ \ \ \ \ \ \ \ \ \ \ \ \ \ \ \ = \frac{2}{1} \frac{1}{2} \frac{1}{2} \{PG_{1,2}(\Pi_e,w_{\Pi_e},\Pi_o,w_{\dot{L}},\mu) + PG_{2,1}(\Pi_e,w_{\Pi_e},\Pi_o,w_{\dot{L}},\mu)\} =\\
& \ \ \ \ \ \ \ \ \ \ \ \ \ \ \ \ \ \ \ \ \ \ \ \ \ \ \ \ \ \ \ \ \ \ \ \ = \frac{2}{1} \frac{1}{2} \frac{1}{2} \left\{1 + 1\right\} = 1
\end{aligned}
\end{equation*}

Since $1$ is the highest possible value for the partial grading property, therefore matching pennies has $General_{max} = 1$ for this property.
\end{proof}
\end{proposition}

\begin{proposition}
\label{prop:matching_pennies_PG_left_max}
$Left_{max}$ for the partial grading (PG) property is equal to $1$ for the matching pennies environment.

\begin{proof}
To find $Left_{max}$ (equation \ref{eq:left_max}), we need to find a pair $\left\langle\Pi_e,w_{\Pi_e}\right\rangle$ which maximises the property as much as possible while $\Pi_o$ minimises it. Using $\Pi_e = \{{\pi_r}_1,{\pi_r}_2,{\pi_r}_3\}$ with uniform weight for $w_{\Pi_e}$ (a $\pi_r$ agent always acts randomly) we find this situation no matter which $\Pi_o$ we use.

Following definition \ref{def:PG}, we obtain the PG value for this $\left\langle\Pi_e,w_{\Pi_e},\Pi_o\right\rangle$ (where $\Pi_o$ is instantiated with any permitted value). Since the environment is not symmetric, we need to calculate this property for every pair of slots. Following definition \ref{def:STG_set} (for PG), we can calculate its PG value for each pair of slots. We start with slots 1 and 2:

\begin{equation*}
\begin{aligned}
PG_{1,2}(\Pi_e,w_{\Pi_e},\Pi_o,w_{\dot{L}},\mu)	& = \eta_{\Pi^3} \sum_{\pi_1,\pi_2,\pi_3 \in \Pi_e | \pi_1 \neq \pi_2 \neq \pi_3} w_{\Pi_e}(\pi_1) w_{\Pi_e}(\pi_2) w_{\Pi_e}(\pi_3) PG_{1,2}(\pi_1,\pi_2,\pi_3,\Pi_o,w_{\dot{L}},\mu) =\\
												& = 6 \frac{9}{2} \frac{1}{3} \frac{1}{3} \frac{1}{3} PG_{1,2}({\pi_r}_1,{\pi_r}_2,{\pi_r}_3,\Pi_o,w_{\dot{L}},\mu)
\end{aligned}
\end{equation*}

\noindent Note that we avoided to calculate all the permutations of $\pi_1,\pi_2,\pi_3$ for $PG_{i,j}(\pi_1,\pi_2,\pi_3,\Pi_o,w_{\dot{L}},\mu)$ since they provide the same result, by calculating only one permutation and multiplying the result by the number of permutations $6$.

In this case, we only need to calculate $PG_{1,2}({\pi_r}_1,{\pi_r}_2,{\pi_r}_3,\Pi_o,w_{\dot{L}},\mu)$. We follow definition \ref{def:STG_agents} (for PG) to calculate this value:

\begin{equation*}
\begin{aligned}
PG_{1,2}({\pi_r}_1,{\pi_r}_2,{\pi_r}_3,\Pi_o,w_{\dot{L}},\mu)	& = \sum_{\dot{l} \in \dot{L}^{N(\mu)}_{-1,2}(\Pi_o)} w_{\dot{L}}(\dot{l}) PO_{1,2}(\pi_h,\pi_{h/t},\pi_{m/o},\dot{l},\mu) =\\
																& = PO_{1,2}({\pi_r}_1,{\pi_r}_2,{\pi_r}_3,(*,*),\mu)
\end{aligned}
\end{equation*}

\noindent Note that the choice of $\Pi_o$ does not affect the result of $PG_{1,2}$.

The following table shows us $PO_{1,2}$ for all the permutations of ${\pi_r}_1,{\pi_r}_2,{\pi_r}_3$.

\begin{center}
\begin{tabular}{c c c | c c c | c c c}
Slot 1 & & Slot 2					& Slot 1 & & Slot 2						& Slot 1 & & Slot 2\\
\hline
${\pi_r}_1$ & $\leq$ & ${\pi_r}_2$	& ${\pi_r}_1$ & $\leq$ & ${\pi_r}_3$	& ${\pi_r}_2$ & $\leq$ & ${\pi_r}_1$\\
${\pi_r}_2$ & $\leq$ & ${\pi_r}_3$	& ${\pi_r}_3$ & $\leq$ & ${\pi_r}_2$	& ${\pi_r}_1$ & $\leq$ & ${\pi_r}_3$\\
${\pi_r}_1$ & $\leq$ & ${\pi_r}_3$	& ${\pi_r}_1$ & $\leq$ & ${\pi_r}_2$	& ${\pi_r}_2$ & $\leq$ & ${\pi_r}_3$
\end{tabular}
\begin{tabular}{c c c | c c c | c c c}
Slot 1 & & Slot 2					& Slot 1 & & Slot 2						& Slot 1 & & Slot 2\\
\hline
${\pi_r}_2$ & $\leq$ & ${\pi_r}_3$	& ${\pi_r}_3$ & $\leq$ & ${\pi_r}_1$	& ${\pi_r}_3$ & $\leq$ & ${\pi_r}_2$\\
${\pi_r}_3$ & $\leq$ & ${\pi_r}_1$	& ${\pi_r}_1$ & $\leq$ & ${\pi_r}_2$	& ${\pi_r}_2$ & $\leq$ & ${\pi_r}_1$\\
${\pi_r}_2$ & $\leq$ & ${\pi_r}_1$	& ${\pi_r}_3$ & $\leq$ & ${\pi_r}_2$	& ${\pi_r}_3$ & $\leq$ & ${\pi_r}_1$
\end{tabular}
\end{center}

It is possible to find a PO for every permutation, since we have at least one random agent making its expected average reward equal to its opponent expected average reward (both have an expected average reward of $0$ as proved in lemma \ref{lemma:matching_pennies_random_agent}). So:

\begin{equation*}
PG_{1,2}({\pi_r}_1,{\pi_r}_2,{\pi_r}_3,\Pi_o,w_{\dot{L}},\mu) = 1
\end{equation*}

Therefore:

\begin{equation*}
PG_{1,2}(\Pi_e,w_{\Pi_e},\Pi_o,w_{\dot{L}},\mu) = 6 \frac{9}{2} \frac{1}{3} \frac{1}{3} \frac{1}{3} 1 = 1
\end{equation*}

And for slots 2 and 1:

\begin{equation*}
\begin{aligned}
PG_{2,1}(\Pi_e,w_{\Pi_e},\Pi_o,w_{\dot{L}},\mu)	& = \eta_{\Pi^3} \sum_{\pi_1,\pi_2,\pi_3 \in \Pi_e | \pi_1 \neq \pi_2 \neq \pi_3} w_{\Pi_e}(\pi_1) w_{\Pi_e}(\pi_2) w_{\Pi_e}(\pi_3) PG_{2,1}(\pi_1,\pi_2,\pi_3,\Pi_o,w_{\dot{L}},\mu) =\\
												& = 6 \frac{9}{2} \frac{1}{3} \frac{1}{3} \frac{1}{3} PG_{2,1}({\pi_r}_1,{\pi_r}_2,{\pi_r}_3,\Pi_o,w_{\dot{L}},\mu)
\end{aligned}
\end{equation*}

Again, we only need to calculate $PG_{2,1}({\pi_r}_1,{\pi_r}_2,{\pi_r}_3,\Pi_o,w_{\dot{L}},\mu)$. We follow definition \ref{def:STG_agents} (for PG) to calculate this value:

\begin{equation*}
\begin{aligned}
PG_{2,1}({\pi_r}_1,{\pi_r}_2,{\pi_r}_3,\Pi_o,w_{\dot{L}},\mu)	& = \sum_{\dot{l} \in \dot{L}^{N(\mu)}_{-2,1}(\Pi_o)} w_{\dot{L}}(\dot{l}) PO_{2,1}({\pi_r}_1,{\pi_r}_2,{\pi_r}_3,\dot{l},\mu) =\\
																& = PO_{2,1}({\pi_r}_1,{\pi_r}_2,{\pi_r}_3,(*,*),\mu)
\end{aligned}
\end{equation*}

\noindent Note again that the choice of $\Pi_o$ does not affect the result of $PG_{2,1}$.

The following table shows us $PO_{2,1}$ for all the permutations of ${\pi_r}_1,{\pi_r}_2,{\pi_r}_3$.

\begin{center}
\begin{tabular}{c c c | c c c | c c c}
Slot 2 & & Slot 1					& Slot 2 & & Slot 1						& Slot 2 & & Slot 1\\
\hline
${\pi_r}_1$ & $\leq$ & ${\pi_r}_2$	& ${\pi_r}_1$ & $\leq$ & ${\pi_r}_3$	& ${\pi_r}_2$ & $\leq$ & ${\pi_r}_1$\\
${\pi_r}_2$ & $\leq$ & ${\pi_r}_3$	& ${\pi_r}_3$ & $\leq$ & ${\pi_r}_2$	& ${\pi_r}_1$ & $\leq$ & ${\pi_r}_3$\\
${\pi_r}_1$ & $\leq$ & ${\pi_r}_3$	& ${\pi_r}_1$ & $\leq$ & ${\pi_r}_2$	& ${\pi_r}_2$ & $\leq$ & ${\pi_r}_3$
\end{tabular}
\begin{tabular}{c c c | c c c | c c c}
Slot 2 & & Slot 1					& Slot 2 & & Slot 1						& Slot 2 & & Slot 1\\
\hline
${\pi_r}_2$ & $\leq$ & ${\pi_r}_3$	& ${\pi_r}_3$ & $\leq$ & ${\pi_r}_1$	& ${\pi_r}_3$ & $\leq$ & ${\pi_r}_2$\\
${\pi_r}_3$ & $\leq$ & ${\pi_r}_1$	& ${\pi_r}_1$ & $\leq$ & ${\pi_r}_2$	& ${\pi_r}_2$ & $\leq$ & ${\pi_r}_1$\\
${\pi_r}_2$ & $\leq$ & ${\pi_r}_1$	& ${\pi_r}_3$ & $\leq$ & ${\pi_r}_2$	& ${\pi_r}_3$ & $\leq$ & ${\pi_r}_1$
\end{tabular}
\end{center}

Again, it is possible to find a PO for every permutation, since we have at least one random agent making its expected average reward equal to its opponent expected average reward (both have an expected average reward of $0$ as proved in lemma \ref{lemma:matching_pennies_random_agent}). So:

\begin{equation*}
PG_{2,1}({\pi_r}_1,{\pi_r}_2,{\pi_r}_3,\Pi_o,w_{\dot{L}},\mu) = 1
\end{equation*}

Therefore:

\begin{equation*}
PG_{2,1}(\Pi_e,w_{\Pi_e},\Pi_o,w_{\dot{L}},\mu) = 6 \frac{9}{2} \frac{1}{3} \frac{1}{3} \frac{1}{3} 1 = 1
\end{equation*}

And finally, we weight over the slots:

\begin{equation*}
\begin{aligned}
& PG(\Pi_e,w_{\Pi_e},\Pi_o,w_{\dot{L}},\mu,w_S) = \eta_{S_1^2} \sum_{i=1}^{N(\mu)} w_S(i,\mu) \times\\
& \times \left(\sum_{j=1}^{i-1} w_S(j,\mu) PG_{i,j}(\Pi_e,w_{\Pi_e},\Pi_o,w_{\dot{L}},\mu) + \sum_{j=i+1}^{N(\mu)} w_S(j,\mu) PG_{i,j}(\Pi_e,w_{\Pi_e},\Pi_o,w_{\dot{L}},\mu)\right) =\\
& \ \ \ \ \ \ \ \ \ \ \ \ \ \ \ \ \ \ \ \ \ \ \ \ \ \ \ \ \ \ \ \ \ \ \ \ = \frac{2}{1} \frac{1}{2} \frac{1}{2} \{PG_{1,2}(\Pi_e,w_{\Pi_e},\Pi_o,w_{\dot{L}},\mu) + PG_{2,1}(\Pi_e,w_{\Pi_e},\Pi_o,w_{\dot{L}},\mu)\} =\\
& \ \ \ \ \ \ \ \ \ \ \ \ \ \ \ \ \ \ \ \ \ \ \ \ \ \ \ \ \ \ \ \ \ \ \ \ = \frac{2}{1} \frac{1}{2} \frac{1}{2} \left\{1 + 1\right\} = 1
\end{aligned}
\end{equation*}

Since $1$ is the highest possible value for the partial grading property and no $\Pi_o$ is able to influence this result, therefore matching pennies has $Left_{max} = 1$ for this property.
\end{proof}
\end{proposition}

\begin{proposition}
\label{prop:matching_pennies_PG_right_min}
$Right_{min}$ for the partial grading (PG) property is equal to $0$ for the matching pennies environment.

\begin{proof}
To find $Right_{min}$ (equation \ref{eq:right_min}), we need to find a pair $\left\langle\Pi_e,w_{\Pi_e}\right\rangle$ which minimises the property as much as possible while $\Pi_o$ maximises it. Using $\Pi_e = \{{\pi_h}_1,{\pi_h}_2,\pi_t\}$ with uniform weight for $w_{\Pi_e}$ (a $\pi_h$ agent always perform Head and a $\pi_t$ agent always perform Tail) we find this situation no matter which $\Pi_o$ we use.

Following definition \ref{def:PG}, we obtain the PG value for this $\left\langle\Pi_e,w_{\Pi_e},\Pi_o\right\rangle$ (where $\Pi_o$ is instantiated with any permitted value). Since the environment is not symmetric, we need to calculate this property for every pair of slots. Following definition \ref{def:STG_set} (for PG), we can calculate its PG value for each pair of slots. We start with slots 1 and 2:

\begin{equation*}
\begin{aligned}
PG_{1,2}(\Pi_e,w_{\Pi_e},\Pi_o,w_{\dot{L}},\mu)	& = \eta_{\Pi^3} \sum_{\pi_1,\pi_2,\pi_3 \in \Pi_e | \pi_1 \neq \pi_2 \neq \pi_3} w_{\Pi_e}(\pi_1) w_{\Pi_e}(\pi_2) w_{\Pi_e}(\pi_3) PG_{1,2}(\pi_1,\pi_2,\pi_3,\Pi_o,w_{\dot{L}},\mu) =\\
												& = 6 \frac{9}{2} \frac{1}{3} \frac{1}{3} \frac{1}{3} PG_{1,2}({\pi_h}_1,{\pi_h}_2,\pi_t,\Pi_o,w_{\dot{L}},\mu)
\end{aligned}
\end{equation*}

\noindent Note that we avoided to calculate all the permutations of $\pi_1,\pi_2,\pi_3$ for $PG_{i,j}(\pi_1,\pi_2,\pi_3,\Pi_o,w_{\dot{L}},\mu)$ since they provide the same result, by calculating only one permutation and multiplying the result by the number of permutations $6$.

In this case, we only need to calculate $PG_{1,2}({\pi_h}_1,{\pi_h}_2,\pi_t,\Pi_o,w_{\dot{L}},\mu)$. We follow definition \ref{def:STG_agents} (for PG) to calculate this value:

\begin{equation*}
\begin{aligned}
PG_{1,2}({\pi_h}_1,{\pi_h}_2,\pi_t,\Pi_o,w_{\dot{L}},\mu)	& = \sum_{\dot{l} \in \dot{L}^{N(\mu)}_{-1,2}(\Pi_o)} w_{\dot{L}}(\dot{l}) PO_{1,2}({\pi_h}_1,{\pi_h}_2,\pi_t,\dot{l},\mu) =\\
															& = PO_{1,2}({\pi_h}_1,{\pi_h}_2,\pi_t,(*,*),\mu)
\end{aligned}
\end{equation*}

\noindent Note that the choice of $\Pi_o$ does not affect the result of $PG_{1,2}$.

The following table shows us $PO_{1,2}$ for all the permutations of ${\pi_h}_1,{\pi_h}_2,\pi_t$.

\begin{center}
\begin{tabular}{c c c | c c c | c c c}
Slot 1 & & Slot 2					& Slot 1 & & Slot 2						& Slot 1 & & Slot 2\\
\hline
${\pi_h}_1$ & $\leq$ & ${\pi_h}_2$	& ${\pi_h}_1$ & $\leq$ & $\pi_t$		& ${\pi_h}_2$ & $\leq$ & ${\pi_h}_1$\\
${\pi_h}_2$ & $\leq$ & $\pi_t$		& $\pi_t$ & $\leq$ & ${\pi_h}_2$		& ${\pi_h}_1$ & $\leq$ & $\pi_t$\\
${\pi_h}_1$ & $\leq$ & $\pi_t$		& ${\pi_h}_1$ & $\leq$ & ${\pi_h}_2$	& ${\pi_h}_2$ & $\leq$ & $\pi_t$
\end{tabular}
\begin{tabular}{c c c | c c c | c c c}
Slot 1 & & Slot 2					& Slot 1 & & Slot 2						& Slot 1 & & Slot 2\\
\hline
${\pi_h}_2$ & $\leq$ & $\pi_t$		& $\pi_t$ & $\leq$ & ${\pi_h}_1$		& $\pi_t$ & $\leq$ & ${\pi_h}_2$\\
$\pi_t$ & $\leq$ & ${\pi_h}_1$		& ${\pi_h}_1$ & $\leq$ & ${\pi_h}_2$	& ${\pi_h}_2$ & $\leq$ & ${\pi_h}_1$\\
${\pi_h}_2$ & $\leq$ & ${\pi_h}_1$	& $\pi_t$ & $\leq$ & ${\pi_h}_2$		& $\pi_t$ & $\leq$ & ${\pi_h}_1$
\end{tabular}
\end{center}

But, it is not possible to find a PO, since for every permutation we have either ${\pi_h}_1 \leq {\pi_h}_2$ or ${\pi_h}_2 \leq {\pi_h}_1$. In both cases, $\pi_h$ will always perform Head, so they will obtain an expected average reward of $1$ and $-1$ respectively. So:

\begin{equation*}
PG_{1,2}({\pi_h}_1,{\pi_h}_2,\pi_t,\Pi_o,w_{\dot{L}},\mu) = 0
\end{equation*}

Therefore:

\begin{equation*}
PG_{1,2}(\Pi_e,w_{\Pi_e},\Pi_o,w_{\dot{L}},\mu) = 6 \frac{9}{2} \frac{1}{3} \frac{1}{3} \frac{1}{3} 0 = 0
\end{equation*}

And for slots 2 and 1:

\begin{equation*}
\begin{aligned}
PG_{2,1}(\Pi_e,w_{\Pi_e},\Pi_o,w_{\dot{L}},\mu)	& = \eta_{\Pi^3} \sum_{\pi_1,\pi_2,\pi_3 \in \Pi_e | \pi_1 \neq \pi_2 \neq \pi_3} w_{\Pi_e}(\pi_1) w_{\Pi_e}(\pi_2) w_{\Pi_e}(\pi_3) PG_{2,1}(\pi_1,\pi_2,\pi_3,\Pi_o,w_{\dot{L}},\mu) =\\
												& = 6 \frac{9}{2} \frac{1}{3} \frac{1}{3} \frac{1}{3} PG_{2,1}({\pi_h}_1,{\pi_h}_2,\pi_t,\Pi_o,w_{\dot{L}},\mu)
\end{aligned}
\end{equation*}

Again, we only need to calculate $PG_{2,1}({\pi_h}_1,{\pi_h}_2,\pi_t,\Pi_o,w_{\dot{L}},\mu)$. We follow definition \ref{def:STG_agents} (for PG) to calculate this value:

\begin{equation*}
\begin{aligned}
PG_{2,1}({\pi_h}_1,{\pi_h}_2,\pi_t,\Pi_o,w_{\dot{L}},\mu)	& = \sum_{\dot{l} \in \dot{L}^{N(\mu)}_{-2,1}(\Pi_o)} w_{\dot{L}}(\dot{l}) PO_{2,1}({\pi_h}_1,{\pi_h}_2,\pi_t,\dot{l},\mu) =\\
															& = PO_{2,1}({\pi_h}_1,{\pi_h}_2,\pi_t,(*,*),\mu)
\end{aligned}
\end{equation*}

\noindent Note again that the choice of $\Pi_o$ does not affect the result of $PG_{2,1}$.

The following table shows us $PO_{2,1}$ for all the permutations of ${\pi_h}_1,{\pi_h}_2,\pi_t$.

\begin{center}
\begin{tabular}{c c c | c c c | c c c}
Slot 2 & & Slot 1					& Slot 2 & & Slot 1						& Slot 2 & & Slot 1\\
\hline
${\pi_h}_1$ & $\leq$ & ${\pi_h}_2$	& ${\pi_h}_1$ & $\leq$ & $\pi_t$		& ${\pi_h}_2$ & $\leq$ & ${\pi_h}_1$\\
${\pi_h}_2$ & $\leq$ & $\pi_t$		& $\pi_t$ & $\leq$ & ${\pi_h}_2$		& ${\pi_h}_1$ & $\leq$ & $\pi_t$\\
${\pi_h}_1$ & $\leq$ & $\pi_t$		& ${\pi_h}_1$ & $\leq$ & ${\pi_h}_2$	& ${\pi_h}_2$ & $\leq$ & $\pi_t$
\end{tabular}
\begin{tabular}{c c c | c c c | c c c}
Slot 2 & & Slot 1					& Slot 2 & & Slot 1						& Slot 2 & & Slot 1\\
\hline
${\pi_h}_2$ & $\leq$ & $\pi_t$		& $\pi_t$ & $\leq$ & ${\pi_h}_1$		& $\pi_t$ & $\leq$ & ${\pi_h}_2$\\
$\pi_t$ & $\leq$ & ${\pi_h}_1$		& ${\pi_h}_1$ & $\leq$ & ${\pi_h}_2$	& ${\pi_h}_2$ & $\leq$ & ${\pi_h}_1$\\
${\pi_h}_2$ & $\leq$ & ${\pi_h}_1$	& $\pi_t$ & $\leq$ & ${\pi_h}_2$		& $\pi_t$ & $\leq$ & ${\pi_h}_1$
\end{tabular}
\end{center}

Again, it is not possible to find a PO, since for every permutation we have either ${\pi_h}_1 \leq \pi_t$ or $\pi_t \leq {\pi_h}_1$. In ${\pi_h}_1 \leq \pi_t$, ${\pi_h}_1$ will always perform Head and $\pi_t$ will always perform Tail, so they will obtain an expected average reward of $1$ and $-1$ respectively. In $\pi_t \leq {\pi_h}_1$, $\pi_t$ will always perform Tail and ${\pi_h}_1$ will always perform Head, so they will obtain an expected average reward of $1$ and $-1$ respectively. So:

\begin{equation*}
PG_{2,1}({\pi_h}_1,{\pi_h}_2,\pi_t,\Pi_o,w_{\dot{L}},\mu) = 0
\end{equation*}

Therefore:

\begin{equation*}
PG_{2,1}(\Pi_e,w_{\Pi_e},\Pi_o,w_{\dot{L}},\mu) = 6 \frac{9}{2} \frac{1}{3} \frac{1}{3} \frac{1}{3} 0 = 0
\end{equation*}

And finally, we weight over the slots:

\begin{equation*}
\begin{aligned}
& PG(\Pi_e,w_{\Pi_e},\Pi_o,w_{\dot{L}},\mu,w_S) = \eta_{S_1^2} \sum_{i=1}^{N(\mu)} w_S(i,\mu) \times\\
& \times \left(\sum_{j=1}^{i-1} w_S(j,\mu) PG_{i,j}(\Pi_e,w_{\Pi_e},\Pi_o,w_{\dot{L}},\mu) + \sum_{j=i+1}^{N(\mu)} w_S(j,\mu) PG_{i,j}(\Pi_e,w_{\Pi_e},\Pi_o,w_{\dot{L}},\mu)\right) =\\
& \ \ \ \ \ \ \ \ \ \ \ \ \ \ \ \ \ \ \ \ \ \ \ \ \ \ \ \ \ \ \ \ \ \ \ \ = \frac{2}{1} \frac{1}{2} \frac{1}{2} \{PG_{1,2}(\Pi_e,w_{\Pi_e},\Pi_o,w_{\dot{L}},\mu) + PG_{2,1}(\Pi_e,w_{\Pi_e},\Pi_o,w_{\dot{L}},\mu)\} =\\
& \ \ \ \ \ \ \ \ \ \ \ \ \ \ \ \ \ \ \ \ \ \ \ \ \ \ \ \ \ \ \ \ \ \ \ \ = \frac{2}{1} \frac{1}{2} \frac{1}{2} \left\{0 + 0\right\} = 0
\end{aligned}
\end{equation*}

Since $0$ is the lowest possible value for the partial grading property and no $\Pi_o$ is able to influence this result, therefore matching pennies has $Right_{min} = 0$ for this property.
\end{proof}
\end{proposition}

\subsection{Slot Reward Dependency}
Next we see the slot reward dependency (SRD) property. As given in section \ref{sec:SRD}, we want to know how much competitive or cooperative the environment is.

\begin{proposition}
\label{prop:matching_pennies_SRD_general_range}
$General$ range for the slot reward dependency (SRD) property is equal to $[-1,-1]$ for the matching pennies environment.

\begin{proof}
Following definition \ref{def:SRD}, we obtain the SRD value for any $\left\langle\Pi_e,w_{\Pi_e},\Pi_o\right\rangle$ (where $\Pi_e, w_{\Pi_e}$ and $\Pi_o$ are instantiated with any permitted values). Since the environment is not symmetric, we need to calculate this property for every pair of slots. Following definition \ref{def:SRD_set}, we can calculate its SRD value for each pair of slots. We start with slots 1 and 2:

\begin{equation*}
SRD_{1,2}(\Pi_e,w_{\Pi_e},\Pi_o,w_{\dot{L}},\mu) = \sum_{\pi \in \Pi_e} w_{\Pi_e}(\pi) SRD_{1,2}(\pi,\Pi_o,w_{\dot{L}},\mu)
\end{equation*}

We do not know which $\Pi_e$ we have, but we know that we will need to evaluate $SRD_{1,2}(\pi,\Pi_o,w_{\dot{L}},\mu)$ for all evaluated agent $\pi \in \Pi_e$. We follow definition \ref{def:SRD_agent} to calculate this value for a figurative evaluated agent $\pi_1$ from $\Pi_e$:

\begin{equation*}
SRD_{1,2}(\pi_1,\Pi_o,w_{\dot{L}},\mu) = corr_{\dot{l} \in \dot{L}^{N(\mu)}_{-1}(\Pi_o)}[w_{\dot{L}}(\dot{l})](R_1(\mu[\instantiation{l}{1}{\pi_1}]), R_2(\mu[\instantiation{l}{1}{\pi_1}]))
\end{equation*}

We do not know which $\Pi_o$ we have, but we know that we will need to obtain a line-up pattern $\dot{l}$ from $\dot{L}^{N(\mu)}_{-1}(\Pi_o)$ to calculate $corr(R_1(\mu[\instantiation{l}{1}{\pi_1}]), R_2(\mu[\instantiation{l}{1}{\pi_1}]))$. We calculate this value for a figurative line-up pattern $\dot{l} = (*,\pi_2)$ from $\dot{L}^{N(\mu)}_{-1}(\Pi_o)$:

\begin{equation*}
corr(R_1(\mu[\instantiation{l}{1}{\pi_1}]), R_2(\mu[\instantiation{l}{1}{\pi_1}])) = corr(R_1(\mu[\pi_1,\pi_2]), R_2(\mu[\pi_1,\pi_2]))
\end{equation*}

This game is a zero-sum game with two agents. That means that, in every game, the sum of both agents' rewards will always be zero, or in other words, when the agent in slot 1 (any $\pi_1$) obtains a reward $r$ the agent in slot 2 (any $\pi_2$) obtains $-r$ as reward, and this relation is propagated to expected average rewards as well. Since we use a correlation function between the expected average rewards, and the agents in slots 1 and 2 always obtain opposite expected average reward, then the correlation function will always obtain the same value\footnote{Provided there is at least one game which is not a tie.} of $-1$. So:

\begin{equation*}
SRD_{1,2}(\pi_1,\Pi_o,w_{\dot{L}},\mu) = -1
\end{equation*}

Therefore:

\begin{equation*}
SRD_{1,2}(\Pi_e,w_{\Pi_e},\Pi_o,w_{\dot{L}},\mu) = -1
\end{equation*}

And for slots 2 and 1:

\begin{equation*}
SRD_{2,1}(\Pi_e,w_{\Pi_e},\Pi_o,w_{\dot{L}},\mu) = \sum_{\pi \in \Pi_e} w_{\Pi_e}(\pi) SRD_{2,1}(\pi,\Pi_o,w_{\dot{L}},\mu)
\end{equation*}

Again, we do not know which $\Pi_e$ we have, but we know that we will need to evaluate $SRD_{2,1}(\pi,\Pi_o,w_{\dot{L}},\mu)$ for all evaluated agent $\pi \in \Pi_e$. We follow definition \ref{def:SRD_agent} to calculate this value for a figurative evaluated agent $\pi_1$ from $\Pi_e$:

\begin{equation*}
SRD_{2,1}(\pi_1,\Pi_o,w_{\dot{L}},\mu) = corr_{\dot{l} \in \dot{L}^{N(\mu)}_{-2}(\Pi_o)}[w_{\dot{L}}(\dot{l})](R_2(\mu[\instantiation{l}{2}{\pi_1}]), R_1(\mu[\instantiation{l}{2}{\pi_1}]))
\end{equation*}

Again, we do not know which $\Pi_o$ we have, but we know that we will need to obtain a line-up pattern $\dot{l}$ from $\dot{L}^{N(\mu)}_{-2}(\Pi_o)$ to calculate $corr(R_2(\mu[\instantiation{l}{2}{\pi_1}]), R_1(\mu[\instantiation{l}{2}{\pi_1}]))$. We calculate this value for a figurative line-up pattern $\dot{l} = (\pi_2,*)$ from $\dot{L}^{N(\mu)}_{-2}(\Pi_o)$:

\begin{equation*}
corr(R_2(\mu[\instantiation{l}{2}{\pi_1}]), R_1(\mu[\instantiation{l}{2}{\pi_1}])) = corr(R_2(\mu[\pi_2,\pi_1]), R_1(\mu[\pi_2,\pi_1]))
\end{equation*}

Again, this game is a zero-sum game with two agents. That means that, in every game, the sum of both agents' rewards will always be zero, or in other words, when the agent in slot 2 (any $\pi_1$) obtains a reward $r$ the agent in slot 1 (any $\pi_2$) obtains $-r$ as reward, and this relation is propagated to expected average rewards as well. Since we use a correlation function between the expected average rewards, and the agents in slots 2 and 1 always obtain opposite expected average reward, then the correlation function will always obtain the same value of $-1$. So:

\begin{equation*}
SRD_{2,1}(\pi_1,\Pi_o,w_{\dot{L}},\mu) = -1
\end{equation*}

Therefore:

\begin{equation*}
SRD_{2,1}(\Pi_e,w_{\Pi_e},\Pi_o,w_{\dot{L}},\mu) = -1
\end{equation*}

And finally, we weight over the slots:

\begin{equation*}
\begin{aligned}
& SRD(\Pi_e,w_{\Pi_e},\Pi_o,w_{\dot{L}},\mu,w_S) = \eta_{S_1^2} \sum_{i=1}^{N(\mu)} w_S(i,\mu) \times\\
& \times \left(\sum_{j=1}^{i-1} w_S(j,\mu) SRD_{i,j}(\Pi_e,w_{\Pi_e},\Pi_o,w_{\dot{L}},\mu) + \sum_{j=i+1}^{N(\mu)} w_S(j,\mu) SRD_{i,j}(\Pi_e,w_{\Pi_e},\Pi_o,w_{\dot{L}},\mu)\right) =\\
& \ \ \ \ \ \ \ \ \ \ \ \ \ \ \ \ \ \ \ \ \ \ \ \ \ \ \ \ \ \ \ \ \ \ \ \ \ \ = \frac{2}{1} \frac{1}{2} \frac{1}{2} \{SRD_{1,2}(\Pi_e,w_{\Pi_e},\Pi_o,w_{\dot{L}},\mu) + SRD_{2,1}(\Pi_e,w_{\Pi_e},\Pi_o,w_{\dot{L}},\mu)\} =\\
& \ \ \ \ \ \ \ \ \ \ \ \ \ \ \ \ \ \ \ \ \ \ \ \ \ \ \ \ \ \ \ \ \ \ \ \ \ \ = \frac{2}{1} \frac{1}{2} \frac{1}{2} \left\{-1 + (-1)\right\} = -1
\end{aligned}
\end{equation*}

So, for every trio $\left\langle\Pi_e,w_{\Pi_e},\Pi_o\right\rangle$ we obtain the same result:

\begin{equation*}
\forall \Pi_e,w_{\Pi_e},\Pi_o : SRD(\Pi_e,w_{\Pi_e},\Pi_o,w_{\dot{L}},\mu,w_S) = -1
\end{equation*}

Therefore, matching pennies has $General = [-1,-1]$ for this property.
\end{proof}
\end{proposition}

\subsection{Competitive Anticipation}
Finally, we follow with the competitive anticipation (AComp) property. As given in section \ref{sec:AComp}, we want to know how much benefit the evaluated agents obtain when they anticipate competing agents.

\begin{proposition}
\label{prop:matching_pennies_AComp_general_min}
$General_{min}$ for the competitive anticipation (AComp) property is equal to $-\frac{1}{2}$ for the matching pennies environment.

\begin{proof}
To find $General_{min}$ (equation \ref{eq:general_min}), we need to find a trio $\left\langle\Pi_e,w_{\Pi_e},\Pi_o\right\rangle$ which minimises the property as much as possible. We can have this situation by selecting $\Pi_e = \{\pi_{t/h}\}$ with $w_{\Pi_e}(\pi_{t/h}) = 1$ and $\Pi_o = \{\pi_h\}$ (a $\pi_h$ agent always performs Head and a $\pi_{t/h}$ agent always performs Tail when playing in slot 1 and always performs Head when playing in slot 2)\footnote{$\pi_{t/h}$ has to know in which slot it is playing. To infer this, it starts with a random action at the first iteration and then look at the action of its opponent and the reward it obtains.}.

Following definition \ref{def:AComp}, we obtain the AComp value for this $\left\langle\Pi_e,w_{\Pi_e},\Pi_o\right\rangle$. Since the environment is not symmetric, we need to calculate this property for every pair of slots in different teams. Following definition \ref{def:AComp_set}, we could calculate its AComp value for each pair of slots but, since $\Pi_e$ has only one agent, its weight is equal to $1$ and $\Pi_o$ also has only one agent, it is equivalent to use directly definition \ref{def:AComp_agents}. We start with slots 1 and 2:

\begin{equation*}
\begin{aligned}
AComp_{1,2}(\pi_{t/h},\pi_h,\Pi_o,w_{\dot{L}},\mu)	& = \sum_{\dot{l} \in \dot{L}^{N(\mu)}_{-1,2}(\Pi_o)} w_{\dot{L}}(\dot{l}) \frac{1}{2} \left(R_1(\mu[\instantiation{l}{1,2}{\pi_{t/h},\pi_h}]) - R_1(\mu[\instantiation{l}{1,2}{\pi_{t/h},\pi_r}])\right) =\\
													& = \frac{1}{2} \left(R_1(\mu[\pi_{t/h},\pi_h]) - R_1(\mu[\pi_{t/h},\pi_r])\right)
\end{aligned}
\end{equation*}

We know from lemma \ref{lemma:matching_pennies_random_agent} that $R_1(\mu[\pi_{t/h},\pi_r]) = 0$, so we only need to calculate $R_1(\mu[\pi_{t/h},\pi_h])$, where $\pi_{t/h}$ will always perform Tail and $\pi_h$ will always perform Head, so the agent in slot 1 ($\pi_{t/h}$) will obtain an expected average reward of $-1$. So:

\begin{equation*}
AComp_{1,2}(\pi_{t/h},\pi_h,\Pi_o,w_{\dot{L}},\mu) = \frac{1}{2} \left((-1) - 0\right) = -\frac{1}{2}
\end{equation*}

\noindent Note that this is the minimum possible value for $AComp_{1,2}(\pi_1,\pi_2,\Pi_o,w_{\dot{L}},\mu)$ since $R_1(\mu[\pi_1,\pi_r])$ will always be equal to $0$ for this environment.

And for slots 2 and 1:

\begin{equation*}
\begin{aligned}
AComp_{2,1}(\pi_{t/h},\pi_h,\Pi_o,w_{\dot{L}},\mu)	& = \sum_{\dot{l} \in \dot{L}^{N(\mu)}_{-2,1}(\Pi_o)} w_{\dot{L}}(\dot{l}) \frac{1}{2} \left(R_2(\mu[\instantiation{l}{2,1}{\pi_{t/h},\pi_h}]) - R_2(\mu[\instantiation{l}{2,1}{\pi_{t/h},\pi_r}])\right) =\\
													& = \frac{1}{2} \left(R_2(\mu[\pi_h,\pi_{t/h}]) - R_2(\mu[\pi_r,\pi_{t/h}])\right)
\end{aligned}
\end{equation*}

Again, we know from lemma \ref{lemma:matching_pennies_random_agent} that $R_2(\mu[\pi_r,\pi_{t/h}]) = 0$, so we only need to calculate $R_2(\mu[\pi_h,\pi_{t/h}])$, where $\pi_h$ and $\pi_{t/h}$ will always perform Head, so the agent in slot 2 ($\pi_{t/h}$) will obtain an expected average reward of $-1$. So:

\begin{equation*}
AComp_{2,1}(\pi_{t/h},\pi_h,\Pi_o,w_{\dot{L}},\mu) = \frac{1}{2} \left((-1) - 0\right) = -\frac{1}{2}
\end{equation*}

\noindent Note again that this is the minimum possible value for $AComp_{2,1}(\pi_1,\pi_2,\Pi_o,w_{\dot{L}},\mu)$ since $R_2(\mu[\pi_r,\pi_1])$ will always be equal to $0$ for this environment.

And finally, we weight over the slots:


\begin{equation*}
\begin{aligned}
AComp(\Pi_e,w_{\Pi_e},\Pi_o,w_{\dot{L}},\mu,w_S)	& = \eta_{S_2^2} \sum_{t_1,t_2 \in \tau | t_1 \neq t_2} \sum_{i \in t_1} w_S(i,\mu) \sum_{j \in t_2} w_S(j,\mu) AComp_{i,j}(\Pi_e,w_{\Pi_e},\Pi_o,w_{\dot{L}},\mu) =\\
													& = \frac{2}{1} \frac{1}{2} \frac{1}{2} \{AComp_{1,2}(\Pi_e,w_{\Pi_e},\Pi_o,w_{\dot{L}},\mu) + AComp_{2,1}(\Pi_e,w_{\Pi_e},\Pi_o,w_{\dot{L}},\mu)\} =\\
													& = \frac{2}{1} \frac{1}{2} \frac{1}{2} \{AComp_{1,2}(\pi_{t/h},\pi_h,\Pi_o,w_{\dot{L}},\mu) + AComp_{2,1}(\pi_{t/h},\pi_h,\Pi_o,w_{\dot{L}},\mu)\} =\\
													& = \frac{2}{1} \frac{1}{2} \frac{1}{2} \left\{-\frac{1}{2} + \left(-\frac{1}{2}\right)\right\} = -\frac{1}{2}
\end{aligned}
\end{equation*}

Since $-\frac{1}{2}$ is the lowest possible value that we can obtain for the competitive anticipation property, therefore matching pennies has $General_{min} = -\frac{1}{2}$ for this property.
\end{proof}
\end{proposition}

\begin{proposition}
\label{prop:matching_pennies_AComp_general_max}
$General_{max}$ for the competitive anticipation (AComp) property is equal to $\frac{1}{2}$ for the matching pennies environment.

\begin{proof}
To find $General_{max}$ (equation \ref{eq:general_max}), we need to find a trio $\left\langle\Pi_e,w_{\Pi_e},\Pi_o\right\rangle$ which maximises the property as much as possible. We can have this situation by selecting $\Pi_e = \{\pi_{h/t}\}$ with $w_{\Pi_e}(\pi_{h/t}) = 1$ and $\Pi_o = \{\pi_h\}$ (a $\pi_h$ agent always performs Head and a $\pi_{h/t}$ agent always performs Head when playing in slot 1 and always performs Tail when playing in slot 2)\footnote{$\pi_{h/t}$ has to know in which slot it is playing. To infer this, it starts with a random action at the first iteration and then look at the action of its opponent and the reward it obtains.}.

Following definition \ref{def:AComp}, we obtain the AComp value for this $\left\langle\Pi_e,w_{\Pi_e},\Pi_o\right\rangle$. Since the environment is not symmetric, we need to calculate this property for every pair of slots in different teams. Following definition \ref{def:AComp_set}, we could calculate its AComp value for each pair of slots but, since $\Pi_e$ has only one agent, its weight is equal to $1$ and $\Pi_o$ also has only one agent, it is equivalent to use directly definition \ref{def:AComp_agents}. We start with slots 1 and 2:

\begin{equation*}
\begin{aligned}
AComp_{1,2}(\pi_{h/t},\pi_h,\Pi_o,w_{\dot{L}},\mu)	& = \sum_{\dot{l} \in \dot{L}^{N(\mu)}_{-1,2}(\Pi_o)} w_{\dot{L}}(\dot{l}) \frac{1}{2} \left(R_1(\mu[\instantiation{l}{1,2}{\pi_{h/t},\pi_h}]) - R_1(\mu[\instantiation{l}{1,2}{\pi_{h/t},\pi_r}])\right) =\\
													& = \frac{1}{2} \left(R_1(\mu[\pi_{h/t},\pi_h]) - R_1(\mu[\pi_{h/t},\pi_r])\right)
\end{aligned}
\end{equation*}

We know from lemma \ref{lemma:matching_pennies_random_agent} that $R_1(\mu[\pi_{h/t},\pi_r]) = 0$, so we only need to calculate $R_1(\mu[\pi_{h/t},\pi_h])$, where $\pi_{h/t}$ and $\pi_h$ will always perform Head, so the agent in slot 1 ($\pi_{h/t}$) will obtain an expected average reward of $1$. So:

\begin{equation*}
AComp_{1,2}(\pi_{h/t},\pi_h,\Pi_o,w_{\dot{L}},\mu) = \frac{1}{2} \left(1 - 0\right) = \frac{1}{2}
\end{equation*}

\noindent Note that this is the maximum possible value for $AComp_{1,2}(\pi_1,\pi_2,\Pi_o,w_{\dot{L}},\mu)$ since $R_1(\mu[\pi_1,\pi_r])$ will always be equal to $0$ for this environment.

And for slots 2 and 1:

\begin{equation*}
\begin{aligned}
AComp_{2,1}(\pi_{h/t},\pi_h,\Pi_o,w_{\dot{L}},\mu)	& = \sum_{\dot{l} \in \dot{L}^{N(\mu)}_{-2,1}(\Pi_o)} w_{\dot{L}}(\dot{l}) \frac{1}{2} \left(R_2(\mu[\instantiation{l}{2,1}{\pi_{h/t},\pi_h}]) - R_2(\mu[\instantiation{l}{2,1}{\pi_{h/t},\pi_r}])\right) =\\
													& = \frac{1}{2} \left(R_2(\mu[\pi_h,\pi_{h/t}]) - R_2(\mu[\pi_r,\pi_{h/t}])\right)
\end{aligned}
\end{equation*}

Again, we know from lemma \ref{lemma:matching_pennies_random_agent} that $R_2(\mu[\pi_r,\pi_{h/t}]) = 0$, so we only need to calculate $R_2(\mu[\pi_h,\pi_{h/t}])$, where $\pi_h$ will always perform Head and $\pi_{h/t}$ will always perform Tail, so the agent in slot 2 ($\pi_{h/t}$) will obtain an expected average reward of $1$. So:

\begin{equation*}
AComp_{2,1}(\pi_{h/t},\pi_h,\Pi_o,w_{\dot{L}},\mu) = \frac{1}{2} \left(1 - 0\right) = \frac{1}{2}
\end{equation*}

\noindent Note again that this is the maximum possible value for $AComp_{2,1}(\pi_1,\pi_2,\Pi_o,w_{\dot{L}},\mu)$ since $R_2(\mu[\pi_r,\pi_1])$ will always be equal to $0$ for this environment.

And finally, we weight over the slots:


\begin{equation*}
\begin{aligned}
AComp(\Pi_e,w_{\Pi_e},\Pi_o,w_{\dot{L}},\mu,w_S)	& = \eta_{S_2^2} \sum_{t_1,t_2 \in \tau | t_1 \neq t_2} \sum_{i \in t_1} w_S(i,\mu) \sum_{j \in t_2} w_S(j,\mu) AComp_{i,j}(\Pi_e,w_{\Pi_e},\Pi_o,w_{\dot{L}},\mu) =\\
													& = \frac{2}{1} \frac{1}{2} \frac{1}{2} \{AComp_{1,2}(\Pi_e,w_{\Pi_e},\Pi_o,w_{\dot{L}},\mu) + AComp_{2,1}(\Pi_e,w_{\Pi_e},\Pi_o,w_{\dot{L}},\mu)\} =\\
													& = \frac{2}{1} \frac{1}{2} \frac{1}{2} \{AComp_{1,2}(\pi_{h/t},\pi_h,\Pi_o,w_{\dot{L}},\mu) + AComp_{2,1}(\pi_{h/t},\pi_h,\Pi_o,w_{\dot{L}},\mu)\} =\\
													& = \frac{2}{1} \frac{1}{2} \frac{1}{2} \left\{\frac{1}{2} + \frac{1}{2}\right\} = \frac{1}{2}
\end{aligned}
\end{equation*}

Since $\frac{1}{2}$ is the highest possible value that we can obtain for the competitive anticipation property, therefore matching pennies has $General_{max} = \frac{1}{2}$ for this property.
\end{proof}
\end{proposition}

\section{Prisoner's Dilemma properties}
\label{appen:prisoner_dilemma_properties}
In this section we prove how we obtained the values for the properties for the prisoner's dilemma environment (section \ref{sec:prisoner_dilemma}).

\subsection{Action Dependency}
We start with the action dependency (AD) property. As given in section \ref{sec:AD}, we want to know if the evaluated agents behave differently depending on which line-up they interact with. We use $\Delta_S(a,b) = 1$ if distributions $a$ and $b$ are equal and $0$ otherwise.

\begin{proposition}
\label{prop:prisoner_dilemma_AD_general_min}
$General_{min}$ for the action dependency (AD) property is equal to $0$ for the prisoner's dilemma environment.

\begin{proof}
To find $General_{min}$ (equation \ref{eq:general_min}), we need to find a trio $\left\langle\Pi_e,w_{\Pi_e},\Pi_o\right\rangle$ which minimises the property as much as possible. We can have this situation by selecting $\Pi_e = \{\pi_b\}$ with $w_{\Pi_e}(\pi_b) = 1$ and $\Pi_o = \{{\pi_c}_1, {\pi_c}_2\}$ (a $\pi_c$ agent always performs Cooperate and a $\pi_b$ agent always performs Blame).

Following definition \ref{def:AD}, we obtain the AD value for this $\left\langle\Pi_e,w_{\Pi_e},\Pi_o\right\rangle$. Since the environment is symmetric, we just need to calculate this property for one slot and generalise its result to all slots. Following definition \ref{def:AD_set}, we could calculate its AD value for slot 1 but, since $\Pi_e$ has only one agent and its weight is equal to $1$, it is equivalent to use directly definition \ref{def:AD_agent} for slot 1:

\begin{equation*}
\begin{aligned}
AD_1(\pi_b,\Pi_o,w_{\dot{L}},\mu)	& = \eta_{\dot{L}^2} \sum_{\dot{u},\dot{v} \in \dot{L}^{N(\mu)}_{-1}(\Pi_o) | \dot{u} \neq \dot{v}} w_{\dot{L}}(\dot{u}) w_{\dot{L}}(\dot{v}) \Delta_S(\breve{A}_1(\mu[\instantiation{u}{1}{\pi_b}]), \breve{A}_1(\mu[\instantiation{v}{1}{\pi_b}])) =\\
									& = 2 \frac{2}{1} \frac{1}{2} \frac{1}{2} \Delta_S(\breve{A}_1(\mu[\pi_b,{\pi_c}_1]), \breve{A}_1(\mu[\pi_b,{\pi_c}_2]))
\end{aligned}
\end{equation*}

\noindent Note that we avoided to calculate both $\Delta_S(a,b)$ and $\Delta_S(b,a)$ since they provide the same result, by calculating only $\Delta_S(a,b)$ and multiplying the result by $2$.

In this case, we only need to calculate $\Delta_S(\breve{A}_1(\mu[\pi_b,\pi_{c_1}]), \breve{A}_1(\mu[\pi_b,\pi_{c_2}]))$, where the agent in both slots 1 ($\pi_b$) will perform the same sequence of actions (always Blame) independently of the line-up. So:

\begin{equation*}
AD_1(\pi_b,\Pi_o,w_{\dot{L}},\mu) = 2 \frac{2}{1} \frac{1}{2} \frac{1}{2} 0 = 0
\end{equation*}

And finally, generalising for all slots:

\begin{equation*}
\begin{aligned}
AD(\Pi_e,w_{\Pi_e},\Pi_o,w_{\dot{L}},\mu,w_S)	& =	\sum_{i = 1}^{N(\mu)} w_S(i,\mu) AD_i(\Pi_e,w_{\Pi_e},\Pi_o,w_{\dot{L}},\mu) =\\
												& =	AD_1(\Pi_e,w_{\Pi_e},\Pi_o,w_{\dot{L}},\mu) =\\
												& =	AD_1(\pi_b,\Pi_o,w_{\dot{L}},\mu) =\\
												& = 0
\end{aligned}
\end{equation*}

Since $0$ is the lowest possible value for the action dependency property, therefore prisoner's dilemma has $General_{min} = 0$ for this property.
\end{proof}
\end{proposition}

\begin{proposition}
\label{prop:prisoner_dilemma_AD_general_max}
$General_{max}$ for the action dependency (AD) property is equal to $1$ for the prisoner's dilemma environment.

\begin{proof}
To find $General_{max}$ (equation \ref{eq:general_max}), we need to find a trio $\left\langle\Pi_e,w_{\Pi_e},\Pi_o\right\rangle$ which maximises the property as much as possible. We can have this situation by selecting $\Pi_e = \{\pi_m\}$ with $w_{\Pi_e}(\pi_m) = 1$ and $\Pi_o = \{\pi_c, \pi_b\}$ (a $\pi_m$ agent first acts randomly and then always mimics the other agent's last action, a $\pi_c$ agent always performs Cooperate and a $\pi_b$ agent always performs Blame).

Following definition \ref{def:AD}, we obtain the AD value for this $\left\langle\Pi_e,w_{\Pi_e},\Pi_o\right\rangle$. Since the environment is symmetric, we just need to calculate this property for one slot and generalise its result to all slots. Following definition \ref{def:AD_set}, we could calculate its AD value for slot 1 but, since $\Pi_e$ has only one agent and its weight is equal to $1$, it is equivalent to use directly definition \ref{def:AD_agent} for slot 1:

\begin{equation*}
\begin{aligned}
AD_1(\pi_m,\Pi_o,w_{\dot{L}},\mu)	& = \eta_{\dot{L}^2} \sum_{\dot{u},\dot{v} \in \dot{L}^{N(\mu)}_{-1}(\Pi_o) | \dot{u} \neq \dot{v}} w_{\dot{L}}(\dot{u}) w_{\dot{L}}(\dot{v}) \Delta_S(\breve{A}_1(\mu[\instantiation{u}{1}{\pi_m}]), \breve{A}_1(\mu[\instantiation{v}{1}{\pi_m}])) =\\
									& = 2 \frac{2}{1} \frac{1}{2} \frac{1}{2} \Delta_S(\breve{A}_1(\mu[\pi_m,\pi_c]), \breve{A}_1(\mu[\pi_m,\pi_b]))
\end{aligned}
\end{equation*}

\noindent Note that we avoided to calculate both $\Delta_S(a,b)$ and $\Delta_S(b,a)$ since they provide the same result, by calculating only $\Delta_S(a,b)$ and multiplying the result by $2$.

From iteration 2, $\pi_m$ will mimic the last action of the agent in slot $2$, and since $\pi_c$ will always perform Cooperate and $\pi_b$ will always perform Blame, the agent in both slots 1 ($\pi_m$) will perform different sequences of actions on each line-up. So:

\begin{equation*}
AD_1(\pi_m,\Pi_o,w_{\dot{L}},\mu) = 2 \frac{2}{1} \frac{1}{2} \frac{1}{2} 1 = 1
\end{equation*}

And finally, generalising for all slots:

\begin{equation*}
\begin{aligned}
AD(\Pi_e,w_{\Pi_e},\Pi_o,w_{\dot{L}},\mu,w_S)	& =	\sum_{i = 1}^{N(\mu)} w_S(i,\mu) AD_i(\Pi_e,w_{\Pi_e},\Pi_o,w_{\dot{L}},\mu) =\\
												& =	AD_1(\Pi_e,w_{\Pi_e},\Pi_o,w_{\dot{L}},\mu) =\\
												& =	AD_1(\pi_m,\Pi_o,w_{\dot{L}},\mu) =\\
												& = 1
\end{aligned}
\end{equation*}

Since $1$ is the highest possible value for the action dependency property, therefore prisoner's dilemma has $General_{max} = 1$ for this property.
\end{proof}
\end{proposition}

\begin{proposition}
\label{prop:prisoner_dilemma_AD_left_max}
$Left_{max}$ for the action dependency (AD) property is equal to $0$ for the prisoner's dilemma environment.

\begin{proof}
To find $Left_{max}$ (equation \ref{eq:left_max}), we need to find a pair $\left\langle\Pi_e,w_{\Pi_e}\right\rangle$ which maximises the property as much as possible while $\Pi_o$ minimises it. Using $\Pi_o = \{{\pi_c}_1,{\pi_c}_2\}$ (a $\pi_c$ agent always performs Cooperate) we find this situation no matter which pair $\left\langle\Pi_e,w_{\Pi_e}\right\rangle$ we use.

Following definition \ref{def:AD}, we obtain the AD value for this $\left\langle\Pi_e,w_{\Pi_e},\Pi_o\right\rangle$ (where $\Pi_e$ and $w_{\Pi_e}$ are instantiated with any permitted values). Since the environment is symmetric, we just need to calculate this property for one slot and generalise its result to all slots. Following definition \ref{def:AD_set}, we can calculate its AD value for slot 1:

\begin{equation*}
AD_1(\Pi_e,w_{\Pi_e},\Pi_o,w_{\dot{L}},\mu) = \sum_{\pi \in \Pi_e} w_{\Pi_e}(\pi) AD_1(\pi,\Pi_o,w_{\dot{L}},\mu)
\end{equation*}

We do not know which $\Pi_e$ we have, but we know that we will need to evaluate $AD_1(\pi,\Pi_o,w_{\dot{L}},\mu)$ for all evaluated agent $\pi \in \Pi_e$. We follow definition \ref{def:AD_agent} to calculate this value for a figurative evaluated agent $\pi$ from $\Pi_e$:

\begin{equation*}
\begin{aligned}
AD_1(\pi,\Pi_o,w_{\dot{L}},\mu)	& = \eta_{\dot{L}^2} \sum_{\dot{u},\dot{v} \in \dot{L}^{N(\mu)}_{-1}(\Pi_o) | \dot{u} \neq \dot{v}} w_{\dot{L}}(\dot{u}) w_{\dot{L}}(\dot{v}) \Delta_S(\breve{A}_1(\mu[\instantiation{u}{1}{\pi}]), \breve{A}_1(\mu[\instantiation{v}{1}{\pi}])) =\\
								& = 2 \frac{2}{1} \frac{1}{2} \frac{1}{2} \Delta_S(\breve{A}_1(\mu[\pi,{\pi_c}_1]), \breve{A}_1(\mu[\pi,{\pi_c}_2]))
\end{aligned}
\end{equation*}

\noindent Note that we avoided to calculate both $\Delta_S(a,b)$ and $\Delta_S(b,a)$ since they provide the same result, by calculating only $\Delta_S(a,b)$ and multiplying the result by $2$.

A $\pi_c$ agent will always perform Cooperate, so we obtain a situation where the agent in both slots 1 (any $\pi$) will be able to differentiate with which agent is interacting, so it will not be able to change its distribution of action sequences depending on the opponent's behaviour. So:

\begin{equation*}
AD_1(\pi,\Pi_o,w_{\dot{L}},\mu) = 2 \frac{2}{1} \frac{1}{2} \frac{1}{2} 0 = 0
\end{equation*}

Therefore, no matter which agents are in $\Pi_e$ and their weights $w_{\Pi_e}$ we obtain:

\begin{equation*}
AD_1(\Pi_e,w_{\Pi_e},\Pi_o,w_{\dot{L}},\mu) = 0
\end{equation*}

And finally, generalising for all slots:

\begin{equation*}
\begin{aligned}
AD(\Pi_e,w_{\Pi_e},\Pi_o,w_{\dot{L}},\mu,w_S)	& =	\sum_{i = 1}^{N(\mu)} w_S(i,\mu) AD_i(\Pi_e,w_{\Pi_e},\Pi_o,w_{\dot{L}},\mu) =\\
												& = AD_1(\Pi_e,w_{\Pi_e},\Pi_o,w_{\dot{L}},\mu) =\\
												& = 0
\end{aligned}
\end{equation*}

So, for every pair $\left\langle\Pi_e,w_{\Pi_e}\right\rangle$ we obtain the same result:

\begin{equation*}
\forall \Pi_e,w_{\Pi_e} : AD(\Pi_e,w_{\Pi_e},\Pi_o,w_{\dot{L}},\mu,w_S) = 0
\end{equation*}

Therefore, prisoner's dilemma has $Left_{max} = 0$ for this property.
\end{proof}
\end{proposition}

\begin{proposition}
\label{prop:prisoner_dilemma_AD_right_min}
$Right_{min}$ for the action dependency (AD) property is equal to $0$ for the prisoner's dilemma environment.

\begin{proof}
To find $Right_{min}$ (equation \ref{eq:right_min}), we need to find a pair $\left\langle\Pi_e,w_{\Pi_e}\right\rangle$ which minimises the property as much as possible while $\Pi_o$ maximises it. Using $\Pi_e = \{\pi_c\}$ with $w_{\Pi_e}(\pi_c) = 1$ (a $\pi_c$ agent always performs Cooperate) we find this situation no matter which $\Pi_o$ we use.

Following definition \ref{def:AD}, we obtain the AD value for this $\left\langle\Pi_e,w_{\Pi_e},\Pi_o\right\rangle$ (where $\Pi_o$ is instantiated with any permitted value). Since the environment is symmetric, we just need to calculate this property for one slot and generalise its result to all slots. Following definition \ref{def:AD_set}, we could calculate its AD value for slot 1 but, since $\Pi_e$ has only one agent and its weight is equal to $1$, it is equivalent to use directly definition \ref{def:AD_agent} for slot 1:

\begin{equation*}
AD_1(\pi_c,\Pi_o,w_{\dot{L}},\mu) = \eta_{\dot{L}^2} \sum_{\dot{u},\dot{v} \in \dot{L}^{N(\mu)}_{-1}(\Pi_o) | \dot{u} \neq \dot{v}} w_{\dot{L}}(\dot{u}) w_{\dot{L}}(\dot{v}) \Delta_S(\breve{A}_1(\mu[\instantiation{u}{1}{\pi_c}]), \breve{A}_1(\mu[\instantiation{v}{1}{\pi_c}]))
\end{equation*}

We do not know which $\Pi_o$ we have, but we know that we will need to obtain two different line-up patterns $\dot{u}$ and $\dot{v}$ from $\dot{L}^{N(\mu)}_{-1}(\Pi_o)$ to calculate $\Delta_S(\breve{A}_1(\mu[\instantiation{u}{1}{\pi_c}]), \breve{A}_1(\mu[\instantiation{v}{1}{\pi_c}]))$. We calculate this value for two figurative line-up patterns $\dot{u} = (*,\pi_1)$ and $\dot{v} = (*,\pi_2)$ from $\dot{L}^{N(\mu)}_{-1}(\Pi_o)$:

\begin{equation*}
\Delta_S(\breve{A}_1(\mu[\instantiation{u}{1}{\pi_c}]), \breve{A}_1(\mu[\instantiation{v}{1}{\pi_c}])) = \Delta_S(\breve{A}_1(\mu[\pi_c,\pi_1]), \breve{A}_1(\mu[\pi_c,\pi_2]))
\end{equation*}

Here, the agent in both slots 1 ($\pi_c$) will perform the same sequence of actions (always Cooperate) independently of the line-up. So no matter which agents are in $\Pi_o$ we obtain:

\begin{equation*}
AD_1(\pi_c,\Pi_o,w_{\dot{L}},\mu) = 0
\end{equation*}

And finally, generalising for all slots:

\begin{equation*}
\begin{aligned}
AD(\Pi_e,w_{\Pi_e},\Pi_o,w_{\dot{L}},\mu,w_S)	& =	\sum_{i = 1}^{N(\mu)} w_S(i,\mu) AD_i(\Pi_e,w_{\Pi_e},\Pi_o,w_{\dot{L}},\mu) =\\
												& =	AD_1(\Pi_e,w_{\Pi_e},\Pi_o,w_{\dot{L}},\mu) =\\
												& = AD_1(\pi_c,\Pi_o,w_{\dot{L}},\mu) =\\
												& = 0
\end{aligned}
\end{equation*}

So, for every $\Pi_o$ we obtain the same result:

\begin{equation*}
\forall \Pi_o : AD(\Pi_e,w_{\Pi_e},\Pi_o,w_{\dot{L}},\mu,w_S) = 0
\end{equation*}

Therefore, prisoner's dilemma has $Right_{min} = 0$ for this property.
\end{proof}
\end{proposition}

\subsection{Reward Dependency}
We continue with the reward dependency (RD) property. As given in section \ref{sec:RD}, we want to know if the evaluated agents obtain different expected average rewards depending on which line-up they interact with. We use $\Delta_{\mathbb{Q}}(a,b) = 1$ if numbers $a$ and $b$ are equal and $0$ otherwise.

\begin{proposition}
\label{prop:prisoner_dilemma_RD_general_min}
$General_{min}$ for the reward dependency (RD) property is equal to $0$ for the prisoner's dilemma environment.

\begin{proof}
To find $General_{min}$ (equation \ref{eq:general_min}), we need to find a trio $\left\langle\Pi_e,w_{\Pi_e},\Pi_o\right\rangle$ which minimises the property as much as possible. We can have this situation by selecting $\Pi_e = \{\pi_b\}$ with $w_{\Pi_e}(\pi_b) = 1$ and $\Pi_o = \{{\pi_c}_1, {\pi_c}_2\}$ (a $\pi_c$ agent always performs Cooperate and a $\pi_b$ agent always performs Blame).

Following definition \ref{def:RD}, we obtain the RD value for this $\left\langle\Pi_e,w_{\Pi_e},\Pi_o\right\rangle$. Since the environment is symmetric, we just need to calculate this property for one slot and generalise its result to all slots. Following definition \ref{def:RD_set}, we could calculate its RD value for slot 1 but, since $\Pi_e$ has only one agent and its weight is equal to $1$, it is equivalent to use directly definition \ref{def:RD_agent} for slot 1:

\begin{equation*}
\begin{aligned}
RD_1(\pi_b,\Pi_o,w_{\dot{L}},\mu)	& = \eta_{\dot{L}^2} \sum_{\dot{u},\dot{v} \in \dot{L}^{N(\mu)}_{-1}(\Pi_o) | \dot{u} \neq \dot{v}} w_{\dot{L}}(\dot{u}) w_{\dot{L}}(\dot{v}) \Delta_{\mathbb{Q}}(R_1(\mu[\instantiation{u}{1}{\pi_b}]), R_1(\mu[\instantiation{v}{1}{\pi_b}])) =\\
									& = 2 \frac{2}{1} \frac{1}{2} \frac{1}{2} \Delta_{\mathbb{Q}}(R_1(\mu[\pi_b,{\pi_c}_1]), R_1(\mu[\pi_b,{\pi_c}_2]))
\end{aligned}
\end{equation*}

\noindent Note that we avoided to calculate both $\Delta_{\mathbb{Q}}(a,b)$ and $\Delta_{\mathbb{Q}}(b,a)$ since they provide the same result, by calculating only $\Delta_{\mathbb{Q}}(a,b)$ and multiplying the result by $2$.

In this case, we only need to calculate $\Delta_{\mathbb{Q}}(R_1(\mu[\pi_b,{\pi_c}_1]), R_1(\mu[\pi_b,{\pi_c}_2]))$, where $\pi_b$ will always perform Blame and $\pi_c$ will always perform Cooperate, so the agent in both slots 1 ($\pi_b$) will obtain the same expected average reward ($1$) independently of the line-up. So:

\begin{equation*}
RD_1(\pi_b,\Pi_o,w_{\dot{L}},\mu) = 2 \frac{2}{1} \frac{1}{2} \frac{1}{2} 0 = 0
\end{equation*}

And finally, generalising for all slots:

\begin{equation*}
\begin{aligned}
RD(\Pi_e,w_{\Pi_e},\Pi_o,w_{\dot{L}},\mu,w_S)	& =	\sum_{i = 1}^{N(\mu)} w_S(i,\mu) RD_i(\Pi_e,w_{\Pi_e},\Pi_o,w_{\dot{L}},\mu) =\\
												& =	RD_1(\Pi_e,w_{\Pi_e},\Pi_o,w_{\dot{L}},\mu) =\\
												& =	RD_1(\pi_b,\Pi_o,w_{\dot{L}},\mu) =\\
												& = 0
\end{aligned}
\end{equation*}

Since $0$ is the lowest possible value for the reward dependency property, therefore prisoner's dilemma has $General_{min} = 0$ for this property.
\end{proof}
\end{proposition}

\begin{proposition}
\label{prop:prisoner_dilemma_RD_general_max}
$General_{max}$ for the reward dependency (RD) property is equal to $1$ for the prisoner's dilemma environment.

\begin{proof}
To find $General_{max}$ (equation \ref{eq:general_max}), we need to find a trio $\left\langle\Pi_e,w_{\Pi_e},\Pi_o\right\rangle$ which maximises the property as much as possible. We can have this situation by selecting $\Pi_e = \{\pi_b\}$ with $w_{\Pi_e}(\pi_b) = 1$ and $\Pi_o = \{\pi_c, \pi_b\}$ (a $\pi_c$ agent always performs Cooperate and a $\pi_b$ agent always performs Blame).

Following definition \ref{def:RD}, we obtain the RD value for this $\left\langle\Pi_e,w_{\Pi_e},\Pi_o\right\rangle$. Since the environment is symmetric, we just need to calculate this property for one slot and generalise its result to all slots. Following definition \ref{def:RD_set}, we could calculate its RD value for slot 1 but, since $\Pi_e$ has only one agent and its weight is equal to $1$, it is equivalent to use directly definition \ref{def:RD_agent} for slot 1:

\begin{equation*}
\begin{aligned}
RD_1(\pi_b,\Pi_o,w_{\dot{L}},\mu)	& = \eta_{\dot{L}^2} \sum_{\dot{u},\dot{v} \in \dot{L}^{N(\mu)}_{-1}(\Pi_o) | \dot{u} \neq \dot{v}} w_{\dot{L}}(\dot{u}) w_{\dot{L}}(\dot{v}) \Delta_{\mathbb{Q}}(R_1(\mu[\instantiation{u}{1}{\pi_b}]), R_1(\mu[\instantiation{v}{1}{\pi_b}])) =\\
									& = 2 \frac{2}{1} \frac{1}{2} \frac{1}{2} \Delta_{\mathbb{Q}}(R_1(\mu[\pi_b,\pi_c]), R_1(\mu[\pi_b,\pi_b]))
\end{aligned}
\end{equation*}

\noindent Note that we avoided to calculate both $\Delta_{\mathbb{Q}}(a,b)$ and $\Delta_{\mathbb{Q}}(b,a)$ since they provide the same result, by calculating only $\Delta_{\mathbb{Q}}(a,b)$ and multiplying the result by $2$.

In line-up $(\pi_b,\pi_c)$, where $\pi_b$ will always perform Blame and $\pi_c$ will always perform Cooperate, the agent in slot 1 ($\pi_b$) will obtain one expected average reward ($1$), while in line-up $(\pi_b,\pi_b)$, where both $\pi_b$ will always perform Blame, the agent in slot 1 ($\pi_b$) will obtain a different expected average reward ($-\frac{1}{3}$). So:

\begin{equation*}
RD_1(\pi_b,\Pi_o,w_{\dot{L}},\mu) = 2 \frac{2}{1} \frac{1}{2} \frac{1}{2} 1 = 1
\end{equation*}

And finally, generalising for all slots:

\begin{equation*}
\begin{aligned}
RD(\Pi_e,w_{\Pi_e},\Pi_o,w_{\dot{L}},\mu,w_S)	& =	\sum_{i = 1}^{N(\mu)} w_S(i,\mu) RD_i(\Pi_e,w_{\Pi_e},\Pi_o,w_{\dot{L}},\mu) =\\
												& =	RD_1(\Pi_e,w_{\Pi_e},\Pi_o,w_{\dot{L}},\mu) =\\
												& =	RD_1(\pi_b,\Pi_o,w_{\dot{L}},\mu) =\\
												& = 1
\end{aligned}
\end{equation*}

Since $1$ is the highest possible value for the reward dependency property, therefore prisoner's dilemma has $General_{max} = 1$ for this property.
\end{proof}
\end{proposition}

\begin{proposition}
\label{prop:prisoner_dilemma_RD_left_max}
$Left_{max}$ for the reward dependency (RD) property is equal to $0$ for the prisoner's dilemma environment.

\begin{proof}
To find $Left_{max}$ (equation \ref{eq:left_max}), we need to find a pair $\left\langle\Pi_e,w_{\Pi_e}\right\rangle$ which maximises the property as much as possible while $\Pi_o$ minimises it. Using $\Pi_o = \{{\pi_c}_1,{\pi_c}_2\}$ (a $\pi_c$ agent always performs Cooperate) we find this situation no matter which pair $\left\langle\Pi_e,w_{\Pi_e}\right\rangle$ we use.

Following definition \ref{def:RD}, we obtain the RD value for this $\left\langle\Pi_e,w_{\Pi_e},\Pi_o\right\rangle$ (where $\Pi_e$ and $w_{\Pi_e}$ are instantiated with any permitted values). Since the environment is symmetric, we just need to calculate this property for one slot and generalise its result to all slots. Following definition \ref{def:RD_set}, we can calculate its RD value for slot 1:

\begin{equation*}
RD_1(\Pi_e,w_{\Pi_e},\Pi_o,w_{\dot{L}},\mu) = \sum_{\pi \in \Pi_e} w_{\Pi_e}(\pi) RD_1(\pi,\Pi_o,w_{\dot{L}},\mu)
\end{equation*}

We do not know which $\Pi_e$ we have, but we know that we will need to evaluate $RD_1(\pi,\Pi_o,w_{\dot{L}},\mu)$ for all evaluated agent $\pi \in \Pi_e$. We follow definition \ref{def:RD_agent} to calculate this value for a figurative evaluated agent $\pi$ from $\Pi_e$:

\begin{equation*}
\begin{aligned}
RD_1(\pi,\Pi_o,w_{\dot{L}},\mu)	& = \eta_{\dot{L}^2} \sum_{\dot{u},\dot{v} \in \dot{L}^{N(\mu)}_{-1}(\Pi_o) | \dot{u} \neq \dot{v}} w_{\dot{L}}(\dot{u}) w_{\dot{L}}(\dot{v}) \Delta_{\mathbb{Q}}(R_1(\mu[\instantiation{u}{1}{\pi}]), R_1(\mu[\instantiation{v}{1}{\pi}])) =\\
								& = 2 \frac{2}{1} \frac{1}{2} \frac{1}{2} \Delta_{\mathbb{Q}}(R_1(\mu[\pi,{\pi_c}_1]), R_1(\mu[\pi,{\pi_c}_2]))
\end{aligned}
\end{equation*}

\noindent Note that we avoided to calculate both $\Delta_{\mathbb{Q}}(a,b)$ and $\Delta_{\mathbb{Q}}(b,a)$ since they provide the same result, by calculating only $\Delta_{\mathbb{Q}}(a,b)$ and multiplying the result by $2$.

A $\pi_c$ agent will always perform Cooperate, so we obtain a situation where the agent in both slots 1 (any $\pi$) will be able to differentiate with which agent is interacting, so it will not be able to change its distribution of action sequences depending on the opponent's behaviour, obtaining agent in both slots 1 (any $\pi$) the same expected average reward. So:

\begin{equation*}
RD_1(\pi,\Pi_o,w_{\dot{L}},\mu) = 2 \frac{2}{1} \frac{1}{2} \frac{1}{2} 0 = 0
\end{equation*}

Therefore, no matter which agents are in $\Pi_e$ and their weights $w_{\Pi_e}$ we obtain:

\begin{equation*}
RD_1(\Pi_e,w_{\Pi_e},\Pi_o,w_{\dot{L}},\mu) = 0
\end{equation*}

And finally, generalising for all slots:

\begin{equation*}
\begin{aligned}
RD(\Pi_e,w_{\Pi_e},\Pi_o,w_{\dot{L}},\mu,w_S)	& =	\sum_{i = 1}^{N(\mu)} w_S(i,\mu) RD_i(\Pi_e,w_{\Pi_e},\Pi_o,w_{\dot{L}},\mu) =\\
												& =	RD_1(\Pi_e,w_{\Pi_e},\Pi_o,w_{\dot{L}},\mu) =\\
												& = 0
\end{aligned}
\end{equation*}

So, for every pair $\left\langle\Pi_e,w_{\Pi_e}\right\rangle$ we obtain the same result:

\begin{equation*}
\forall \Pi_e,w_{\Pi_e} : RD(\Pi_e,w_{\Pi_e},\Pi_o,w_{\dot{L}},\mu,w_S) = 0
\end{equation*}

Therefore, prisoner's dilemma has $Left_{max} = 0$ for this property.
\end{proof}
\end{proposition}

\begin{proposition}
\label{prop:prisoner_dilemma_RD_right_min}
$Right_{min}$ for the reward dependency (RD) property is equal to $1$ for the prisoner's dilemma environment.

\begin{proof}
To find $Right_{min}$ (equation \ref{eq:right_min}), we need to find a pair $\left\langle\Pi_e,w_{\Pi_e}\right\rangle$ which minimises the property as much as possible while $\Pi_o$ maximises it. Using $\Pi_o = \{\pi_c,\pi_b\}$ (a $\pi_c$ agent always performs Cooperate and a $\pi_b$ agent always performs Blame) we find this situation no matter which pair $\left\langle\Pi_e,w_{\Pi_e}\right\rangle$ we use.

Following definition \ref{def:RD}, we obtain the RD value for this $\left\langle\Pi_e,w_{\Pi_e},\Pi_o\right\rangle$ (where $\Pi_e$ and $w_{\Pi_e}$ are instantiated with any permitted values). Since the environment is symmetric, we just need to calculate this property for one slot and generalise its result to all slots. Following definition \ref{def:RD_set}, we can calculate its RD value for slot 1:

\begin{equation*}
RD_1(\Pi_e,w_{\Pi_e},\Pi_o,w_{\dot{L}},\mu) = \sum_{\pi \in \Pi_e} w_{\Pi_e}(\pi) RD_1(\pi,\Pi_o,w_{\dot{L}},\mu)
\end{equation*}

We do not know which $\Pi_e$ we have, but we know that we will need to evaluate $RD_1(\pi,\Pi_o,w_{\dot{L}},\mu)$ for all evaluated agent $\pi \in \Pi_e$. We follow definition \ref{def:RD_agent} to calculate this value for a figurative evaluated agent $\pi$ from $\Pi_e$:

\begin{equation*}
\begin{aligned}
RD_1(\pi,\Pi_o,w_{\dot{L}},\mu)	& = \eta_{\dot{L}^2} \sum_{\dot{u},\dot{v} \in \dot{L}^{N(\mu)}_{-1}(\Pi_o) | \dot{u} \neq \dot{v}} w_{\dot{L}}(\dot{u}) w_{\dot{L}}(\dot{v}) \Delta_{\mathbb{Q}}(R_1(\mu[\instantiation{u}{1}{\pi}]), R_1(\mu[\instantiation{v}{1}{\pi}])) =\\
								& = 2 \frac{2}{1} \frac{1}{2} \frac{1}{2} \Delta_{\mathbb{Q}}(R_1(\mu[\pi,\pi_c]), R_1(\mu[\pi,\pi_b]))
\end{aligned}
\end{equation*}

\noindent Note that we avoided to calculate both $\Delta_{\mathbb{Q}}(a,b)$ and $\Delta_{\mathbb{Q}}(b,a)$ since they provide the same result, by calculating only $\Delta_{\mathbb{Q}}(a,b)$ and multiplying the result by $2$.
From the matching pennies' payoff matrix (figure \ref{fig:matching_pennies_payoff})

In line-up $(\pi,\pi_c)$, where $\pi_c$ will always perform Cooperate, the agent in slot 1 ($\pi$) will obtain an expected average reward between $\frac{1}{3}$ and $1$, while in line-up $(\pi,\pi_b)$, where $\pi_b$ will always perform Blame, the agent in slot 1 ($\pi$) will obtain another expected average reward between $-1$ and $-\frac{1}{3}$. So:

\begin{equation*}
RD_1(\pi,\Pi_o,w_{\dot{L}},\mu) = 2 \frac{2}{1} \frac{1}{2} \frac{1}{2} 1 = 1
\end{equation*}

Therefore, no matter which agents are in $\Pi_e$ and their weights $w_{\Pi_e}$ we obtain:

\begin{equation*}
RD_1(\Pi_e,w_{\Pi_e},\Pi_o,w_{\dot{L}},\mu) = 1
\end{equation*}

And finally, generalising for all slots:

\begin{equation*}
\begin{aligned}
RD(\Pi_e,w_{\Pi_e},\Pi_o,w_{\dot{L}},\mu,w_S)	& =	\sum_{i = 1}^{N(\mu)} w_S(i,\mu) RD_i(\Pi_e,w_{\Pi_e},\Pi_o,w_{\dot{L}},\mu) =\\
												& =	RD_1(\Pi_e,w_{\Pi_e},\Pi_o,w_{\dot{L}},\mu) =\\
												& = 1
\end{aligned}
\end{equation*}

So, for every pair $\left\langle\Pi_e,w_{\Pi_e}\right\rangle$ we obtain the same result:

\begin{equation*}
\forall \Pi_e,w_{\Pi_e} : RD(\Pi_e,w_{\Pi_e},\Pi_o,w_{\dot{L}},\mu,w_S) = 1
\end{equation*}

Therefore, prisoner's dilemma has $Right_{min} = 1$ for this property.
\end{proof}
\end{proposition}

\subsection{Fine Discrimination}
Now we move to the fine discrimination (FD) property. As given in section \ref{sec:FD}, we want to know if different evaluated agents obtain different expected average rewards when interacting in the environment. We use $\Delta_{\mathbb{Q}}(a,b) = 1$ if numbers $a$ and $b$ are equal and $0$ otherwise.

\begin{proposition}
\label{prop:prisoner_dilemma_FD_general_min}
$General_{min}$ for the fine discrimination (FD) property is equal to $0$ for the prisoner's dilemma environment.

\begin{proof}
To find $General_{min}$ (equation \ref{eq:general_min}), we need to find a trio $\left\langle\Pi_e,w_{\Pi_e},\Pi_o\right\rangle$ which minimises the property as much as possible. We can have this situation by selecting $\Pi_e = \{{\pi_b}_1,{\pi_b}_2\}$ with uniform weight for $w_{\Pi_e}$ and $\Pi_o = \{\pi_c\}$ (a $\pi_c$ agent always performs Cooperate and a $\pi_b$ agent always performs Blame).

Following definition \ref{def:FD}, we obtain the FD value for this $\left\langle\Pi_e,w_{\Pi_e},\Pi_o\right\rangle$. Since the environment is symmetric, we just need to calculate this property for one slot and generalise its result to all slots. Following definition \ref{def:FD_set}, we can calculate its FD value for slot 1:

\begin{equation*}
\begin{aligned}
FD_1(\Pi_e,w_{\Pi_e},\Pi_o,w_{\dot{L}},\mu)	& = \eta_{\Pi^2} \sum_{\pi_1,\pi_2 \in \Pi_e | \pi_1 \neq \pi_2} w_{\Pi_e}(\pi_1) w_{\Pi_e}(\pi_2) FD_1(\pi_1,\pi_2,\Pi_o,w_{\dot{L}},\mu) =\\
											& = 2 \frac{2}{1} \frac{1}{2} \frac{1}{2} FD_1({\pi_b}_1,{\pi_b}_2,\Pi_o,w_{\dot{L}},\mu)
\end{aligned}
\end{equation*}

\noindent Note that we avoided to calculate both $FD_i(\pi_1,\pi_2,\Pi_o,w_{\dot{L}},\mu)\}$ and $FD_i(\pi_2,\pi_1,\Pi_o,w_{\dot{L}},\mu)\}$ since they provide the same result, by calculating only $FD_i(\pi_1,\pi_2,\Pi_o,w_{\dot{L}},\mu)\}$ and multiplying the result by $2$.

In this case, we only need to calculate $FD_1({\pi_b}_1,{\pi_b}_2,\Pi_o,w_{\dot{L}},\mu)$. We follow definition \ref{def:FD_agents} to calculate this value:

\begin{equation*}
\begin{aligned}
FD_1({\pi_b}_1,{\pi_b}_2,\Pi_o,w_{\dot{L}},\mu)	& = \sum_{\dot{l} \in \dot{L}^{N(\mu)}_{-1}(\Pi_o)} w_{\dot{L}}(\dot{l}) \Delta_{\mathbb{Q}}(R_1(\mu[\instantiation{l}{1}{{\pi_b}_1}]), R_1(\mu[\instantiation{l}{1}{{\pi_b}_2}])) =\\
												& = \Delta_{\mathbb{Q}}(R_1(\mu[{\pi_b}_1,\pi_c]), R_1(\mu[{\pi_b}_2,\pi_c]))
\end{aligned}
\end{equation*}

Here, a $\pi_b$ agent will always perform Blame and $\pi_c$ will always perform Cooperate, so both agents in slot 1 (${\pi_b}_1$ and ${\pi_b}_2$) will obtain the same expected average reward ($1$). So:

\begin{equation*}
FD_1({\pi_b}_1,{\pi_b}_2,\Pi_o,w_{\dot{L}},\mu) = 0
\end{equation*}

Therefore:

\begin{equation*}
FD_1(\Pi_e,w_{\Pi_e},\Pi_o,w_{\dot{L}},\mu) = 2 \frac{2}{1} \frac{1}{2} \frac{1}{2} 0 = 0
\end{equation*}

And finally, generalising for all slots:

\begin{equation*}
\begin{aligned}
FD(\Pi_e,w_{\Pi_e},\Pi_o,w_{\dot{L}},\mu,w_S)	& =	\sum_{i = 1}^{N(\mu)} w_S(i,\mu) FD_i(\Pi_e,w_{\Pi_e},\Pi_o,w_{\dot{L}},\mu) =\\
												& =	FD_1(\Pi_e,w_{\Pi_e},\Pi_o,w_{\dot{L}},\mu) =\\
												& = 0
\end{aligned}
\end{equation*}

Since $0$ is the lowest possible value for the fine discriminative property, therefore prisoner's dilemma has $General_{min} = 0$ for this property.
\end{proof}
\end{proposition}

\begin{proposition}
\label{prop:prisoner_dilemma_FD_general_max}
$General_{max}$ for the fine discrimination (FD) property is equal to $1$ for the prisoner's dilemma environment.

\begin{proof}
To find $General_{max}$ (equation \ref{eq:general_max}), we need to find a trio $\left\langle\Pi_e,w_{\Pi_e},\Pi_o\right\rangle$ which maximises the property as much as possible. We can have this situation by selecting $\Pi_e = \{\pi_c,\pi_b\}$ with uniform weight for $w_{\Pi_e}$ and $\Pi_o = \{\pi_c\}$ (a $\pi_c$ agent always performs Cooperate and a $\pi_b$ agent always performs Blame).

Following definition \ref{def:FD}, we obtain the FD value for this $\left\langle\Pi_e,w_{\Pi_e},\Pi_o\right\rangle$. Since the environment is symmetric, we just need to calculate this property for one slot and generalise its result to all slots. Following definition \ref{def:FD_set}, we can calculate its FD value for slot 1:

\begin{equation*}
\begin{aligned}
FD_1(\Pi_e,w_{\Pi_e},\Pi_o,w_{\dot{L}},\mu)	& = \eta_{\Pi^2} \sum_{\pi_1,\pi_2 \in \Pi_e | \pi_1 \neq \pi_2} w_{\Pi_e}(\pi_1) w_{\Pi_e}(\pi_2) FD_1(\pi_1,\pi_2,\Pi_o,w_{\dot{L}},\mu) =\\
											& = 2 \frac{2}{1} \frac{1}{2} \frac{1}{2} FD_1(\pi_c,\pi_b,\Pi_o,w_{\dot{L}},\mu)
\end{aligned}
\end{equation*}

\noindent Note that we avoided to calculate both $FD_i(\pi_1,\pi_2,\Pi_o,w_{\dot{L}},\mu)\}$ and $FD_i(\pi_2,\pi_1,\Pi_o,w_{\dot{L}},\mu)\}$ since they provide the same result, by calculating only $FD_i(\pi_1,\pi_2,\Pi_o,w_{\dot{L}},\mu)\}$ and multiplying the result by $2$.

In this case, we only need to calculate $FD_1(\pi_c,\pi_b,\Pi_o,w_{\dot{L}},\mu)$. We follow definition \ref{def:FD_agents} to calculate this value:

\begin{equation*}
\begin{aligned}
FD_1(\pi_c,\pi_b,\Pi_o,w_{\dot{L}},\mu)	& = \sum_{\dot{l} \in \dot{L}^{N(\mu)}_{-1}(\Pi_o)} w_{\dot{L}}(\dot{l}) \Delta_{\mathbb{Q}}(R_1(\mu[\instantiation{l}{1}{\pi_c}]), R_1(\mu[\instantiation{l}{1}{\pi_b}])) =\\
										& = \Delta_{\mathbb{Q}}(R_1(\mu[\pi_c,\pi_c]), R_1(\mu[\pi_b,\pi_c]))
\end{aligned}
\end{equation*}

In line-up $(\pi_c,\pi_c)$, where both $\pi_c$ will always perform Cooperate, the agent in slot 1 ($\pi_c$) will obtain one expected average reward ($\frac{1}{3}$), while in line-up $(\pi_b,\pi_c)$, where $\pi_b$ will always perform Blame and $\pi_c$ will always perform Cooperate, the agent in slot 1 ($\pi_b$) will obtain a different expected average reward ($1$). So:

\begin{equation*}
FD_1(\pi_c,\pi_b,\Pi_o,w_{\dot{L}},\mu) = 1
\end{equation*}

Therefore:

\begin{equation*}
FD_1(\Pi_e,w_{\Pi_e},\Pi_o,w_{\dot{L}},\mu) = 2 \frac{2}{1} \frac{1}{2} \frac{1}{2} 1 = 1
\end{equation*}

And finally, generalising for all slots:

\begin{equation*}
\begin{aligned}
FD(\Pi_e,w_{\Pi_e},\Pi_o,w_{\dot{L}},\mu,w_S)	& =	\sum_{i = 1}^{N(\mu)} w_S(i,\mu) FD_i(\Pi_e,w_{\Pi_e},\Pi_o,w_{\dot{L}},\mu) =\\
												& =	FD_1(\Pi_e,w_{\Pi_e},\Pi_o,w_{\dot{L}},\mu) =\\
												& = 1
\end{aligned}
\end{equation*}

Since $1$ is the highest possible value for the fine discriminative property, therefore prisoner's dilemma has $General_{max} = 1$ for this property.
\end{proof}
\end{proposition}

\begin{conjecture}
\label{conj:prisoner_dilemma_FD_left_max}
$Left_{max}$ for the fine discrimination (FD) property is equal to $0$ for the prisoner's dilemma environment.

An agent $\pi \in \Pi_o$ can force every evaluated agent to obtain an expected average reward equal to $0$ (in the limit). The procedure is simple. While the evaluated agent has an expected average reward lower/greater than $0$, $\pi$ performs Cooperate/Blame forcing the evaluated agent to increase/decrease its expected average reward. If this procedure is repeated indefinitely, the expected average reward of any evaluated agent will tend to $0$. So:

\begin{equation*}
\forall \pi_1,\pi_2 \exists \Pi_o : FD(\pi_1,\pi_2,\Pi_o,w_{\dot{L}},\mu,w_S) = 0
\end{equation*}

Therefore, prisoner's dilemma has $Left_{max} = 0$ for this property.
\end{conjecture}

\begin{proposition}
\label{prop:prisoner_dilemma_FD_right_min}
$Right_{min}$ for the fine discrimination (FD) property is equal to $0$ for the prisoner's dilemma environment.

\begin{proof}
To find $Right_{min}$ (equation \ref{eq:right_min}), we need to find a pair $\left\langle\Pi_e,w_{\Pi_e}\right\rangle$ which minimises the property as much as possible while $\Pi_o$ maximises it. Using $\Pi_e = \{{\pi_b}_1,{\pi_b}_2\}$ with uniform weight for $w_{\Pi_e}$ (a $\pi_b$ agent always performs Blame) we find this situation no matter which $\Pi_o$ we use.

Following definition \ref{def:FD}, we obtain the FD value for this $\left\langle\Pi_e,w_{\Pi_e},\Pi_o\right\rangle$ (where $\Pi_o$ is instantiated with any permitted values). Since the environment is symmetric, we just need to calculate this property for one slot and generalise its result to all slots. Following definition \ref{def:FD_set}, we can calculate its FD value for slot 1:

\begin{equation*}
\begin{aligned}
FD_1(\Pi_e,w_{\Pi_e},\Pi_o,w_{\dot{L}},\mu)	& = \eta_{\Pi^2} \sum_{\pi_1,\pi_2 \in \Pi_e | \pi_1 \neq \pi_2} w_{\Pi_e}(\pi_1) w_{\Pi_e}(\pi_2) FD_1(\pi_1,\pi_2,\Pi_o,w_{\dot{L}},\mu) =\\
											& = 2 \frac{2}{1} \frac{1}{2} \frac{1}{2} FD_1({\pi_b}_1,{\pi_b}_2,\Pi_o,w_{\dot{L}},\mu)
\end{aligned}
\end{equation*}

\noindent Note that we avoided to calculate both $FD_i(\pi_1,\pi_2,\Pi_o,w_{\dot{L}},\mu)\}$ and $FD_i(\pi_2,\pi_1,\Pi_o,w_{\dot{L}},\mu)\}$ since they provide the same result, by calculating only $FD_i(\pi_1,\pi_2,\Pi_o,w_{\dot{L}},\mu)\}$ and multiplying the result by $2$.

In this case, we only need to calculate $FD_1({\pi_b}_1,{\pi_b}_2,\Pi_o,w_{\dot{L}},\mu)$. We follow definition \ref{def:FD_agents} to calculate this value:

\begin{equation*}
FD_1({\pi_b}_1,{\pi_b}_2,\Pi_o,w_{\dot{L}},\mu) = \sum_{\dot{l} \in \dot{L}^{N(\mu)}_{-1}(\Pi_o)} w_{\dot{L}}(\dot{l}) \Delta_{\mathbb{Q}}(R_1(\mu[\instantiation{l}{1}{{\pi_b}_1}]), R_1(\mu[\instantiation{l}{1}{{\pi_b}_2}]))
\end{equation*}

We do not know which $\Pi_o$ we have, but we know that we will need to obtain a line-up pattern $\dot{l}$ from $\dot{L}^{N(\mu)}_{-1}(\Pi_o)$ to calculate $\Delta_{\mathbb{Q}}(R_1(\mu[\instantiation{l}{1}{{\pi_b}_1}]), R_1(\mu[\instantiation{l}{1}{{\pi_b}_2}]))$. We calculate this value for a figurative line-up pattern $\dot{l} = (*,\pi)$ from $\dot{L}^{N(\mu)}_{-1}(\Pi_o)$:

\begin{equation*}
\Delta_{\mathbb{Q}}(R_1(\mu[\instantiation{l}{1}{{\pi_b}_1}]), R_1(\mu[\instantiation{l}{1}{{\pi_b}_2}])) = \Delta_{\mathbb{Q}}(R_1(\mu[{\pi_b}_1,\pi]), R_1(\mu[{\pi_b}_2,\pi]))
\end{equation*}

A $\pi_b$ agent will always perform Blame, so we obtain a situation where the agent in both slots 2 (any $\pi$) will be able to differentiate with which agent is interacting, so it will not be able to change its distribution of action sequences depending on the opponent's behaviour, obtaining both agents in slot 1 (${\pi_b}_1$ and ${\pi_b}_2$) the same expected average reward. So:

\begin{equation*}
FD_1({\pi_b}_1,{\pi_b}_2,\Pi_o,w_{\dot{L}},\mu) = 0
\end{equation*}

Therefore:

\begin{equation*}
FD_1(\Pi_e,w_{\Pi_e},\Pi_o,w_{\dot{L}},\mu) = 2 \frac{2}{1} \frac{1}{2} \frac{1}{2} 0 = 0
\end{equation*}

And finally, generalising for all slots:

\begin{equation*}
\begin{aligned}
FD(\Pi_e,w_{\Pi_e},\Pi_o,w_{\dot{L}},\mu,w_S)	& = \sum_{i=1}^{N(\mu)} w_S(i,\mu) FD_i(\Pi_e,w_{\Pi_e},\Pi_o,w_{\dot{L}},\mu)\\
												& = FD_1(\Pi_e,w_{\Pi_e},\Pi_o,w_{\dot{L}},\mu)\\
												& = 0
\end{aligned}
\end{equation*}

So, for every $\Pi_o$ we obtain the same result:

\begin{equation*}
\forall \Pi_o : FD(\Pi_e,w_{\Pi_e},\Pi_o,w_{\dot{L}},\mu,w_S) = 0
\end{equation*}

Therefore, prisoner's dilemma has $Right_{min} = 0$ for this property.
\end{proof}
\end{proposition}

\subsection{Strict Total Grading}
We arrive to the strict total grading (STG) property. As given in section \ref{sec:STG}, we want to know if there exists a strict ordering between the evaluated agents when interacting in the environment.

To simplify the notation, we use the next table to represent the STO:
$R_i(\mu[\instantiation{l}{i,j}{\pi_1,\pi_2}]) < R_j(\mu[\instantiation{l}{i,j}{\pi_1,\pi_2}])$,
$R_i(\mu[\instantiation{l}{i,j}{\pi_2,\pi_3}]) < R_j(\mu[\instantiation{l}{i,j}{\pi_2,\pi_3}])$ and
$R_i(\mu[\instantiation{l}{i,j}{\pi_1,\pi_3}]) < R_j(\mu[\instantiation{l}{i,j}{\pi_1,\pi_3}])$.

\begin{center}
\begin{tabular}{c c c}
Slot i & & Slot j\\
\hline
$\pi_1$ & $<$ & $\pi_2$\\
$\pi_2$ & $<$ & $\pi_3$\\
$\pi_1$ & $<$ & $\pi_3$
\end{tabular}
\end{center}

\begin{proposition}
\label{prop:prisoner_dilemma_STG_general_min}
$General_{min}$ for the strict total grading (STG) property is equal to $0$ for the prisoner's dilemma environment.

\begin{proof}
To find $General_{min}$ (equation \ref{eq:general_min}), we need to find a trio $\left\langle\Pi_e,w_{\Pi_e},\Pi_o\right\rangle$ which minimises the property as much as possible. We can have this situation by selecting $\Pi_e = \{{\pi_b}_1,{\pi_b}_2,{\pi_b}_3\}$ with uniform weight for $w_{\Pi_e}$ and $\Pi_o = \emptyset$ (a $\pi_b$ agent always performs Blame).

Following definition \ref{def:STG}, we obtain the STG value for this $\left\langle\Pi_e,w_{\Pi_e},\Pi_o\right\rangle$. Since the environment is symmetric, we just need to calculate this property for one pair of slots and generalise its result to all pair of slots. Following definition \ref{def:STG_set}, we can calculate its STG value for slots 1 and 2:

\begin{equation*}
\begin{aligned}
STG_{1,2}(\Pi_e,w_{\Pi_e},\Pi_o,w_{\dot{L}},\mu)	& = \eta_{\Pi^3} \sum_{\pi_1,\pi_2,\pi_3 \in \Pi_e | \pi_1 \neq \pi_2 \neq \pi_3} w_{\Pi_e}(\pi_1) w_{\Pi_e}(\pi_2) w_{\Pi_e}(\pi_3) STG_{1,2}(\pi_1,\pi_2,\pi_3,\Pi_o,w_{\dot{L}},\mu) =\\
													& = 6 \frac{9}{2} \frac{1}{3} \frac{1}{3} \frac{1}{3} STG_{1,2}({\pi_b}_1,{\pi_b}_2,{\pi_b}_3,\Pi_o,w_{\dot{L}},\mu)
\end{aligned}
\end{equation*}

\noindent Note that we avoided to calculate all the permutations of $\pi_1,\pi_2,\pi_3$ for $STG_{i,j}(\pi_1,\pi_2,\pi_3,\Pi_o,w_{\dot{L}},\mu)$ since they provide the same result, by calculating only one permutation and multiplying the result by the number of permutations $6$.

In this case, we only need to calculate $STG_{1,2}({\pi_b}_1,{\pi_b}_2,{\pi_b}_3,\Pi_o,w_{\dot{L}},\mu)$. We follow definition \ref{def:STG_agents} to calculate this value:

\begin{equation*}
\begin{aligned}
STG_{1,2}({\pi_b}_1,{\pi_b}_2,{\pi_b}_3,\Pi_o,w_{\dot{L}},\mu)	& = \sum_{\dot{l} \in \dot{L}^{N(\mu)}_{-1,2}(\Pi_o)} w_{\dot{L}}(\dot{l}) STO_{1,2}({\pi_b}_1,{\pi_b}_2,{\pi_b}_3,\dot{l},\mu) =\\
																& = STO_{1,2}({\pi_b}_1,{\pi_b}_2,{\pi_b}_3,(*,*),\mu)
\end{aligned}
\end{equation*}

The following table shows us $STO_{1,2}$ for all the permutations of ${\pi_b}_1,{\pi_b}_2,{\pi_b}_3$.

\begin{center}
\begin{tabular}{c c c | c c c | c c c}
Slot 1 & & Slot 2				& Slot 1 & & Slot 2					& Slot 1 & & Slot 2\\
\hline
${\pi_b}_1$ & $<$ & ${\pi_b}_2$	& ${\pi_b}_1$ & $<$ & ${\pi_b}_3$	& ${\pi_b}_2$ & $<$ & ${\pi_b}_1$\\
${\pi_b}_2$ & $<$ & ${\pi_b}_3$	& ${\pi_b}_3$ & $<$ & ${\pi_b}_2$	& ${\pi_b}_1$ & $<$ & ${\pi_b}_3$\\
${\pi_b}_1$ & $<$ & ${\pi_b}_3$	& ${\pi_b}_1$ & $<$ & ${\pi_b}_2$	& ${\pi_b}_2$ & $<$ & ${\pi_b}_3$
\end{tabular}
\begin{tabular}{c c c | c c c | c c c}
Slot 1 & & Slot 2				& Slot 1 & & Slot 2					& Slot 1 & & Slot 2\\
\hline
${\pi_b}_2$ & $<$ & ${\pi_b}_3$	& ${\pi_b}_3$ & $<$ & ${\pi_b}_1$	& ${\pi_b}_3$ & $<$ & ${\pi_b}_2$\\
${\pi_b}_3$ & $<$ & ${\pi_b}_1$	& ${\pi_b}_1$ & $<$ & ${\pi_b}_2$	& ${\pi_b}_2$ & $<$ & ${\pi_b}_1$\\
${\pi_b}_2$ & $<$ & ${\pi_b}_1$	& ${\pi_b}_3$ & $<$ & ${\pi_b}_2$	& ${\pi_b}_3$ & $<$ & ${\pi_b}_1$
\end{tabular}
\end{center}

But, it is not possible to find a STO, since for every permutation we always have ${\pi_b}_i < {\pi_b}_j$, where a $\pi_b$ agent will always perform Blame, so they will both obtain an expected average reward of $-\frac{1}{3}$. So:

\begin{equation*}
STG_{1,2}({\pi_b}_1,{\pi_b}_2,{\pi_b}_3,\Pi_o,w_{\dot{L}},\mu) = 0
\end{equation*}

Therefore:

\begin{equation*}
STG_{1,2}(\Pi_e,w_{\Pi_e},\Pi_o,w_{\dot{L}},\mu) = 6 \frac{9}{2} \frac{1}{3} \frac{1}{3} \frac{1}{3} 0 = 0
\end{equation*}

And finally, generalising for all slots:

\begin{equation*}
\begin{aligned}
& STG(\Pi_e,w_{\Pi_e},\Pi_o,w_{\dot{L}},\mu,w_S) = \eta_{S_1^2} \sum_{i=1}^{N(\mu)} w_S(i,\mu) \times\\
& \times \left(\sum_{j=1}^{i-1} w_S(j,\mu) STG_{i,j}(\Pi_e,w_{\Pi_e},\Pi_o,w_{\dot{L}},\mu) + \sum_{j=i+1}^{N(\mu)} w_S(j,\mu) STG_{i,j}(\Pi_e,w_{\Pi_e},\Pi_o,w_{\dot{L}},\mu)\right) =\\
& \ \ \ \ \ \ \ \ \ \ \ \ \ \ \ \ \ \ \ \ \ \ \ \ \ \ \ \ \ \ \ \ \ \ \ \ \ \ = STG_{1,2}(\Pi_e,w_{\Pi_e},\Pi_o,w_{\dot{L}},\mu) =\\
& \ \ \ \ \ \ \ \ \ \ \ \ \ \ \ \ \ \ \ \ \ \ \ \ \ \ \ \ \ \ \ \ \ \ \ \ \ \ = 0
\end{aligned}
\end{equation*}

Since $0$ is the lowest possible value for the strict total grading property, therefore prisoner's dilemma has $General_{min} = 0$ for this property.
\end{proof}
\end{proposition}

\begin{proposition}
\label{prop:prisoner_dilemma_STG_general_max}
$General_{max}$ for the strict total grading (STG) property is equal to $1$ for the prisoner's dilemma environment.

\begin{proof}
To find $General_{max}$ (equation \ref{eq:general_max}), we need to find a trio $\left\langle\Pi_e,w_{\Pi_e},\Pi_o\right\rangle$ which maximises the property as much as possible. We can have this situation by selecting $\Pi_e = \{\pi_c,\pi_b,\pi_r\}$ with uniform weight for $w_{\Pi_e}$ and $\Pi_o = \emptyset$ (a $\pi_c$ agent always performs Cooperate, a $\pi_b$ agent always performs Blame and a $\pi_r$ agent always acts randomly).

Following definition \ref{def:STG}, we obtain the STG value for this $\left\langle\Pi_e,w_{\Pi_e},\Pi_o\right\rangle$. Since the environment is symmetric, we just need to calculate this property for one pair of slots and generalise its result to all pair of slots. Following definition \ref{def:STG_set}, we can calculate its STG value for slots 1 and 2:

\begin{equation*}
\begin{aligned}
STG_{1,2}(\Pi_e,w_{\Pi_e},\Pi_o,w_{\dot{L}},\mu)	& = \eta_{\Pi^3} \sum_{\pi_1,\pi_2,\pi_3 \in \Pi_e | \pi_1 \neq \pi_2 \neq \pi_3} w_{\Pi_e}(\pi_1) w_{\Pi_e}(\pi_2) w_{\Pi_e}(\pi_3) STG_{1,2}(\pi_1,\pi_2,\pi_3,\Pi_o,w_{\dot{L}},\mu) =\\
													& = 6 \frac{9}{2} \frac{1}{3} \frac{1}{3} \frac{1}{3} STG_{1,2}(\pi_c,\pi_b,\pi_r,\Pi_o,w_{\dot{L}},\mu)
\end{aligned}
\end{equation*}

\noindent Note that we avoided to calculate all the permutations of $\pi_1,\pi_2,\pi_3$ for $STG_{i,j}(\pi_1,\pi_2,\pi_3,\Pi_o,w_{\dot{L}},\mu)$ since they provide the same result, by calculating only one permutation and multiplying the result by the number of permutations $6$.

In this case, we only need to calculate $STG_{1,2}(\pi_c,\pi_b,\pi_r,\Pi_o,w_{\dot{L}},\mu)$. We follow definition \ref{def:STG_agents} to calculate this value:

\begin{equation*}
\begin{aligned}
STG_{1,2}(\pi_c,\pi_b,\pi_r,\Pi_o,w_{\dot{L}},\mu)	& = \sum_{\dot{l} \in \dot{L}^{N(\mu)}_{-1,2}(\Pi_o)} w_{\dot{L}}(\dot{l}) STO_{1,2}(\pi_c,\pi_b,\pi_r,\dot{l},\mu) =\\
													& = STO_{1,2}(\pi_c,\pi_b,\pi_r,(*,*),\mu)
\end{aligned}
\end{equation*}

The following table shows us $STO_{1,2}$ for all the permutations of $\pi_c,\pi_b,\pi_r$.

\begin{center}
\begin{tabular}{c c c | c c c | c c c}
Slot 1 & & Slot 2		& Slot 1 & & Slot 2			& Slot 1 & & Slot 2\\
\hline
$\pi_c$ & $<$ & $\pi_b$	& $\pi_c$ & $<$ & $\pi_r$	& $\pi_b$ & $<$ & $\pi_c$\\
$\pi_b$ & $<$ & $\pi_r$	& $\pi_r$ & $<$ & $\pi_b$	& $\pi_c$ & $<$ & $\pi_r$\\
$\pi_c$ & $<$ & $\pi_r$	& $\pi_c$ & $<$ & $\pi_b$	& $\pi_b$ & $<$ & $\pi_r$
\end{tabular}
\begin{tabular}{c c c | c c c | c c c}
Slot 1 & & Slot 2		& Slot 1 & & Slot 2			& Slot 1 & & Slot 2\\
\hline
$\pi_b$ & $<$ & $\pi_r$	& $\pi_r$ & $<$ & $\pi_c$	& $\pi_r$ & $<$ & $\pi_b$\\
$\pi_r$ & $<$ & $\pi_c$	& $\pi_c$ & $<$ & $\pi_b$	& $\pi_b$ & $<$ & $\pi_c$\\
$\pi_b$ & $<$ & $\pi_c$	& $\pi_r$ & $<$ & $\pi_b$	& $\pi_r$ & $<$ & $\pi_c$
\end{tabular}
\end{center}

It is possible to find a STO for the second permutation. In $\pi_c < \pi_r$, $\pi_c$ will always perform Cooperate and $\pi_r$ will always act randomly, so they will obtain an expected average reward of $-\frac{1}{3}$ and $\frac{2}{3}$ respectively. In $\pi_r < \pi_b$, $\pi_r$ will always act randomly and $\pi_b$ will always perform Blame, so they will obtain an expected average reward of $-\frac{2}{3}$ and $\frac{1}{3}$ respectively. In $\pi_c < \pi_b$, $\pi_c$ will always perform Cooperate and $\pi_b$ will always perform Blame, so they will obtain an expected average reward of $-1$ and $1$ respectively. So:

\begin{equation*}
STG_{1,2}(\pi_c,\pi_b,\pi_r,\Pi_o,w_{\dot{L}},\mu) = 1
\end{equation*}

Therefore:

\begin{equation*}
STG_{1,2}(\Pi_e,w_{\Pi_e},\Pi_o,w_{\dot{L}},\mu) = 6 \frac{9}{2} \frac{1}{3} \frac{1}{3} \frac{1}{3} 1 = 1
\end{equation*}

And finally, generalising for all slots:

\begin{equation*}
\begin{aligned}
& STG(\Pi_e,w_{\Pi_e},\Pi_o,w_{\dot{L}},\mu,w_S) = \eta_{S_1^2} \sum_{i=1}^{N(\mu)} w_S(i,\mu) \times\\
& \times \left(\sum_{j=1}^{i-1} w_S(j,\mu) STG_{i,j}(\Pi_e,w_{\Pi_e},\Pi_o,w_{\dot{L}},\mu) + \sum_{j=i+1}^{N(\mu)} w_S(j,\mu) STG_{i,j}(\Pi_e,w_{\Pi_e},\Pi_o,w_{\dot{L}},\mu)\right) =\\
& \ \ \ \ \ \ \ \ \ \ \ \ \ \ \ \ \ \ \ \ \ \ \ \ \ \ \ \ \ \ \ \ \ \ \ \ \ \ = STG_{1,2}(\Pi_e,w_{\Pi_e},\Pi_o,w_{\dot{L}},\mu) =\\
& \ \ \ \ \ \ \ \ \ \ \ \ \ \ \ \ \ \ \ \ \ \ \ \ \ \ \ \ \ \ \ \ \ \ \ \ \ \ = 1
\end{aligned}
\end{equation*}

Since $1$ is the highest possible value for the strict total grading property, therefore prisoner's dilemma has $General_{max} = 1$ for this property.
\end{proof}
\end{proposition}

\begin{proposition}
\label{prop:prisoner_dilemma_STG_left_max}
$Left_{max}$ for the strict total grading (STG) property is equal to $1$ for the prisoner's dilemma environment.

\begin{proof}
To find $Left_{max}$ (equation \ref{eq:left_max}), we need to find a pair $\left\langle\Pi_e,w_{\Pi_e}\right\rangle$ which maximises the property as much as possible while $\Pi_o$ minimises it. Using $\Pi_e = \{\pi_c,\pi_b,\pi_r\}$ with uniform weight for $w_{\Pi_e}$ (a $\pi_c$ agent always performs Cooperate, a $\pi_b$ agent always performs Blame and a $\pi_r$ agent always acts randomly) we find this situation no matter which $\Pi_o$ we use.

Following definition \ref{def:STG}, we obtain the STG value for this $\left\langle\Pi_e,w_{\Pi_e},\Pi_o\right\rangle$ (where $\Pi_o$ is instantiated with any permitted values). Since the environment is symmetric, we just need to calculate this property for one pair of slots and generalise its result to all pair of slots. Following definition \ref{def:STG_set}, we can calculate its STG value for slots 1 and 2:

\begin{equation*}
\begin{aligned}
STG_{1,2}(\Pi_e,w_{\Pi_e},\Pi_o,w_{\dot{L}},\mu)	& = \eta_{\Pi^3} \sum_{\pi_1,\pi_2,\pi_3 \in \Pi_e | \pi_1 \neq \pi_2 \neq \pi_3} w_{\Pi_e}(\pi_1) w_{\Pi_e}(\pi_2) w_{\Pi_e}(\pi_3) STG_{1,2}(\pi_1,\pi_2,\pi_3,\Pi_o,w_{\dot{L}},\mu) =\\
													& = 6 \frac{9}{2} \frac{1}{3} \frac{1}{3} \frac{1}{3} STG_{1,2}(\pi_c,\pi_b,\pi_r,\Pi_o,w_{\dot{L}},\mu)
\end{aligned}
\end{equation*}

\noindent Note that we avoided to calculate all the permutations of $\pi_1,\pi_2,\pi_3$ for $STG_{i,j}(\pi_1,\pi_2,\pi_3,\Pi_o,w_{\dot{L}},\mu)$ since they provide the same result, by calculating only one permutation and multiplying the result by the number of permutations $6$.

In this case, we only need to calculate $STG_{1,2}(\pi_c,\pi_b,\pi_r,\Pi_o,w_{\dot{L}},\mu)$. We follow definition \ref{def:STG_agents} to calculate this value:

\begin{equation*}
\begin{aligned}
STG_{1,2}(\pi_c,\pi_b,\pi_r,\Pi_o,w_{\dot{L}},\mu)	& = \sum_{\dot{l} \in \dot{L}^{N(\mu)}_{-1,2}(\Pi_o)} w_{\dot{L}}(\dot{l}) STO_{1,2}(\pi_c,\pi_b,\pi_r,\dot{l},\mu) =\\
													& = STO_{1,2}(\pi_c,\pi_b,\pi_r,(*,*),\mu)
\end{aligned}
\end{equation*}

\noindent Note that the choice of $\Pi_o$ does not affect the result of $STG_{1,2}$.

The following table shows us $STO_{1,2}$ for all the permutations of $\pi_c,\pi_b,\pi_r$.

\begin{center}
\begin{tabular}{c c c | c c c | c c c}
Slot 1 & & Slot 2		& Slot 1 & & Slot 2			& Slot 1 & & Slot 2\\
\hline
$\pi_c$ & $<$ & $\pi_b$	& $\pi_c$ & $<$ & $\pi_r$	& $\pi_b$ & $<$ & $\pi_c$\\
$\pi_b$ & $<$ & $\pi_r$	& $\pi_r$ & $<$ & $\pi_b$	& $\pi_c$ & $<$ & $\pi_r$\\
$\pi_c$ & $<$ & $\pi_r$	& $\pi_c$ & $<$ & $\pi_b$	& $\pi_b$ & $<$ & $\pi_r$
\end{tabular}
\begin{tabular}{c c c | c c c | c c c}
Slot 1 & & Slot 2		& Slot 1 & & Slot 2			& Slot 1 & & Slot 2\\
\hline
$\pi_b$ & $<$ & $\pi_r$	& $\pi_r$ & $<$ & $\pi_c$	& $\pi_r$ & $<$ & $\pi_b$\\
$\pi_r$ & $<$ & $\pi_c$	& $\pi_c$ & $<$ & $\pi_b$	& $\pi_b$ & $<$ & $\pi_c$\\
$\pi_b$ & $<$ & $\pi_c$	& $\pi_r$ & $<$ & $\pi_b$	& $\pi_r$ & $<$ & $\pi_c$
\end{tabular}
\end{center}

It is possible to find a STO for the second permutation. In $\pi_c < \pi_r$, $\pi_c$ will always perform Cooperate and $\pi_r$ will always act randomly, so they will obtain an expected average reward of $-\frac{1}{3}$ and $\frac{2}{3}$ respectively. In $\pi_r < \pi_b$, $\pi_r$ will always act randomly and $\pi_b$ will always perform Blame, so they will obtain an expected average reward of $-\frac{2}{3}$ and $\frac{1}{3}$ respectively. In $\pi_c < \pi_b$, $\pi_c$ will always perform Cooperate and $\pi_b$ will always perform Blame, so they will obtain an expected average reward of $-1$ and $1$ respectively. So:

\begin{equation*}
STG_{1,2}(\pi_c,\pi_b,\pi_r,\Pi_o,w_{\dot{L}},\mu) = 1
\end{equation*}

Therefore:

\begin{equation*}
STG_{1,2}(\Pi_e,w_{\Pi_e},\Pi_o,w_{\dot{L}},\mu) = 6 \frac{9}{2} \frac{1}{3} \frac{1}{3} \frac{1}{3} 1 = 1
\end{equation*}

And finally, generalising for all slots:

\begin{equation*}
\begin{aligned}
& STG(\Pi_e,w_{\Pi_e},\Pi_o,w_{\dot{L}},\mu,w_S) = \eta_{S_1^2} \sum_{i=1}^{N(\mu)} w_S(i,\mu) \times\\
& \times \left(\sum_{j=1}^{i-1} w_S(j,\mu) STG_{i,j}(\Pi_e,w_{\Pi_e},\Pi_o,w_{\dot{L}},\mu) + \sum_{j=i+1}^{N(\mu)} w_S(j,\mu) STG_{i,j}(\Pi_e,w_{\Pi_e},\Pi_o,w_{\dot{L}},\mu)\right) =\\
& \ \ \ \ \ \ \ \ \ \ \ \ \ \ \ \ \ \ \ \ \ \ \ \ \ \ \ \ \ \ \ \ \ \ \ \ \ \ = STG_{1,2}(\Pi_e,w_{\Pi_e},\Pi_o,w_{\dot{L}},\mu) =\\
& \ \ \ \ \ \ \ \ \ \ \ \ \ \ \ \ \ \ \ \ \ \ \ \ \ \ \ \ \ \ \ \ \ \ \ \ \ \ = 1
\end{aligned}
\end{equation*}

So, for every $\Pi_o$ we obtain the same result:

\begin{equation*}
\forall \Pi_o : STG(\Pi_e,w_{\Pi_e},\Pi_o,w_{\dot{L}},\mu,w_S) = 1
\end{equation*}

Therefore, prisoner's dilemma has $Left_{max} = 1$ for this property.
\end{proof}
\end{proposition}

\begin{proposition}
\label{prop:prisoner_dilemma_STG_right_min}
$Right_{min}$ for the strict total grading (STG) property is equal to $0$ for the prisoner's dilemma environment.

\begin{proof}
To find $Right_{min}$ (equation \ref{eq:right_min}), we need to find a pair $\left\langle\Pi_e,w_{\Pi_e}\right\rangle$ which minimises the property as much as possible while $\Pi_o$ maximises it. Using $\Pi_e = \{{\pi_b}_1,{\pi_b}_2,{\pi_b}_3\}$ with uniform weight for $w_{\Pi_e}$ (a $\pi_b$ agent always performs Blame) we find this situation no matter which $\Pi_o$ we use.

Following definition \ref{def:STG}, we obtain the STG value for this $\left\langle\Pi_e,w_{\Pi_e},\Pi_o\right\rangle$ (where $\Pi_o$ is instantiated with any permitted values). Since the environment is symmetric, we just need to calculate this property for one pair of slots and generalise its result to all pair of slots. Following definition \ref{def:STG_set}, we can calculate its STG value for slots 1 and 2:

\begin{equation*}
\begin{aligned}
STG_{1,2}(\Pi_e,w_{\Pi_e},\Pi_o,w_{\dot{L}},\mu)	& = \eta_{\Pi^3} \sum_{\pi_1,\pi_2,\pi_3 \in \Pi_e | \pi_1 \neq \pi_2 \neq \pi_3} w_{\Pi_e}(\pi_1) w_{\Pi_e}(\pi_2) w_{\Pi_e}(\pi_3) STG_{1,2}(\pi_1,\pi_2,\pi_3,\Pi_o,w_{\dot{L}},\mu) =\\
													& = 6 \frac{9}{2} \frac{1}{3} \frac{1}{3} \frac{1}{3} STG_{1,2}({\pi_b}_1,{\pi_b}_2,{\pi_b}_3,\Pi_o,w_{\dot{L}},\mu)
\end{aligned}
\end{equation*}

\noindent Note that we avoided to calculate all the permutations of $\pi_1,\pi_2,\pi_3$ for $STG_{i,j}(\pi_1,\pi_2,\pi_3,\Pi_o,w_{\dot{L}},\mu)$ since they provide the same result, by calculating only one permutation and multiplying the result by the number of permutations $6$.

In this case, we only need to calculate $STG_{1,2}({\pi_b}_1,{\pi_b}_2,{\pi_b}_3,\Pi_o,w_{\dot{L}},\mu)$. We follow definition \ref{def:STG_agents} to calculate this value:

\begin{equation*}
\begin{aligned}
STG_{1,2}({\pi_b}_1,{\pi_b}_2,{\pi_b}_3,\Pi_o,w_{\dot{L}},\mu)	& = \sum_{\dot{l} \in \dot{L}^{N(\mu)}_{-1,2}(\Pi_o)} w_{\dot{L}}(\dot{l}) STO_{1,2}({\pi_b}_1,{\pi_b}_2,{\pi_b}_3,\dot{l},\mu) =\\
																& = STO_{1,2}({\pi_b}_1,{\pi_b}_2,{\pi_b}_3,(*,*),\mu)
\end{aligned}
\end{equation*}

\noindent Note that the choice of $\Pi_o$ does not affect the result of $STG_{1,2}$.

The following table shows us $STO_{1,2}$ for all the permutations of ${\pi_b}_1,{\pi_b}_2,{\pi_b}_3$.

\begin{center}
\begin{tabular}{c c c | c c c | c c c}
Slot 1 & & Slot 2				& Slot 1 & & Slot 2					& Slot 1 & & Slot 2\\
\hline
${\pi_b}_1$ & $<$ & ${\pi_b}_2$	& ${\pi_b}_1$ & $<$ & ${\pi_b}_3$	& ${\pi_b}_2$ & $<$ & ${\pi_b}_1$\\
${\pi_b}_2$ & $<$ & ${\pi_b}_3$	& ${\pi_b}_3$ & $<$ & ${\pi_b}_2$	& ${\pi_b}_1$ & $<$ & ${\pi_b}_3$\\
${\pi_b}_1$ & $<$ & ${\pi_b}_3$	& ${\pi_b}_1$ & $<$ & ${\pi_b}_2$	& ${\pi_b}_2$ & $<$ & ${\pi_b}_3$
\end{tabular}
\begin{tabular}{c c c | c c c | c c c}
Slot 1 & & Slot 2				& Slot 1 & & Slot 2					& Slot 1 & & Slot 2\\
\hline
${\pi_b}_2$ & $<$ & ${\pi_b}_3$	& ${\pi_b}_3$ & $<$ & ${\pi_b}_1$	& ${\pi_b}_3$ & $<$ & ${\pi_b}_2$\\
${\pi_b}_3$ & $<$ & ${\pi_b}_1$	& ${\pi_b}_1$ & $<$ & ${\pi_b}_2$	& ${\pi_b}_2$ & $<$ & ${\pi_b}_1$\\
${\pi_b}_2$ & $<$ & ${\pi_b}_1$	& ${\pi_b}_3$ & $<$ & ${\pi_b}_2$	& ${\pi_b}_3$ & $<$ & ${\pi_b}_1$
\end{tabular}
\end{center}

But, it is not possible to find a STO, since for every permutation we always have ${\pi_b}_i < {\pi_b}_j$, where a $\pi_b$ agent will always perform Blame, so they will both obtain an expected average reward of $-\frac{1}{3}$. So:

\begin{equation*}
STG_{1,2}({\pi_b}_1,{\pi_b}_2,{\pi_b}_3,\Pi_o,w_{\dot{L}},\mu) = 0
\end{equation*}

Therefore:

\begin{equation*}
STG_{1,2}(\Pi_e,w_{\Pi_e},\Pi_o,w_{\dot{L}},\mu) = 6 \frac{9}{2} \frac{1}{3} \frac{1}{3} \frac{1}{3} 0 = 0
\end{equation*}

And finally, generalising for all slots:

\begin{equation*}
\begin{aligned}
& STG(\Pi_e,w_{\Pi_e},\Pi_o,w_{\dot{L}},\mu,w_S) = \eta_{S_1^2} \sum_{i=1}^{N(\mu)} w_S(i,\mu) \times\\
& \times \left(\sum_{j=1}^{i-1} w_S(j,\mu) STG_{i,j}(\Pi_e,w_{\Pi_e},\Pi_o,w_{\dot{L}},\mu) + \sum_{j=i+1}^{N(\mu)} w_S(j,\mu) STG_{i,j}(\Pi_e,w_{\Pi_e},\Pi_o,w_{\dot{L}},\mu)\right) =\\
& \ \ \ \ \ \ \ \ \ \ \ \ \ \ \ \ \ \ \ \ \ \ \ \ \ \ \ \ \ \ \ \ \ \ \ \ \ \ = STG_{1,2}(\Pi_e,w_{\Pi_e},\Pi_o,w_{\dot{L}},\mu) =\\
& \ \ \ \ \ \ \ \ \ \ \ \ \ \ \ \ \ \ \ \ \ \ \ \ \ \ \ \ \ \ \ \ \ \ \ \ \ \ = 0
\end{aligned}
\end{equation*}

So, for every $\Pi_o$ we obtain the same result:

\begin{equation*}
\forall \Pi_o : STG(\Pi_e,w_{\Pi_e},\Pi_o,w_{\dot{L}},\mu,w_S) = 0
\end{equation*}

Therefore, prisoner's dilemma has $Right_{min} = 0$ for this property.
\end{proof}
\end{proposition}

\subsection{Partial Grading}
Now we arrive to the partial grading (PG) property. As given in section \ref{sec:PG}, we want to know if there exists a partial ordering between the evaluated agents when interacting in the environment.

To simplify the notation, we use the next table to represent the PO:
$R_i(\mu[\instantiation{l}{i,j}{\pi_1,\pi_2}]) \leq R_j(\mu[\instantiation{l}{i,j}{\pi_1,\pi_2}])$,
$R_i(\mu[\instantiation{l}{i,j}{\pi_2,\pi_3}]) \leq R_j(\mu[\instantiation{l}{i,j}{\pi_2,\pi_3}])$ and
$R_i(\mu[\instantiation{l}{i,j}{\pi_1,\pi_3}]) \leq R_j(\mu[\instantiation{l}{i,j}{\pi_1,\pi_3}])$.

\begin{center}
\begin{tabular}{c c c}
Slot i & & Slot j\\
\hline
$\pi_1$ & $\leq$ & $\pi_2$\\
$\pi_2$ & $\leq$ & $\pi_3$\\
$\pi_1$ & $\leq$ & $\pi_3$
\end{tabular}
\end{center}

\begin{proposition}
\label{prop:prisoner_dilemma_PG_general_min}
$General_{min}$ for the partial grading (PG) property is equal to $0$ for the prisoner's dilemma environment.

\begin{proof}
To find $General_{min}$ (equation \ref{eq:general_min}), we need to find a trio $\left\langle\Pi_e,w_{\Pi_e},\Pi_o\right\rangle$ which minimises the property as much as possible. We can have this situation by selecting $\Pi_e = \{\pi_b^{-ccb},\pi_b^{-bbc},\pi_m\}$ with uniform weight for $w_{\Pi_e}$ and $\Pi_o = \emptyset$ (a $\pi_m$ agent first acts randomly and then always mimics the other agent's last action, a $\pi_b^{-ccb}$ agent always performs Blame except for the last three actions where it performs Cooperate twice and finalises performing Blame, and a $\pi_b^{-bbc}$ agent always performs Blame except for the last three actions where it performs Blame twice and finalises performing Cooperate).

Following definition \ref{def:PG}, we obtain the PG value for this $\left\langle\Pi_e,w_{\Pi_e},\Pi_o\right\rangle$. Since the environment is symmetric, we just need to calculate this property for one pair of slots and generalise its result to all pair of slots. Following definition \ref{def:STG_set} (for PG), we can calculate its PG value for slots 1 and 2:

\begin{equation*}
\begin{aligned}
PG_{1,2}(\Pi_e,w_{\Pi_e},\Pi_o,w_{\dot{L}},\mu)	& = \eta_{\Pi^3} \sum_{\pi_1,\pi_2,\pi_3 \in \Pi_e | \pi_1 \neq \pi_2 \neq \pi_3} w_{\Pi_e}(\pi_1) w_{\Pi_e}(\pi_2) w_{\Pi_e}(\pi_3) PG_{1,2}(\pi_1,\pi_2,\pi_3,\Pi_o,w_{\dot{L}},\mu) =\\
												& = 6 \frac{9}{2} \frac{1}{3} \frac{1}{3} \frac{1}{3} PG_{1,2}(\pi_b^{-ccb},\pi_b^{-bbc},\pi_m,\Pi_o,w_{\dot{L}},\mu)
\end{aligned}
\end{equation*}

\noindent Note that we avoided to calculate all the permutations of $\pi_1,\pi_2,\pi_3$ for $PG_{i,j}(\pi_1,\pi_2,\pi_3,\Pi_o,w_{\dot{L}},\mu)$ since they provide the same result, by calculating only one permutation and multiplying the result by the number of permutations $6$.

In this case, we only need to calculate $PG_{1,2}(\pi_b^{-ccb},\pi_b^{-bbc},\pi_m,\Pi_o,w_{\dot{L}},\mu)$. We follow definition \ref{def:STG_agents} (for PG) to calculate this value:

\begin{equation*}
\begin{aligned}
PG_{1,2}(\pi_b^{-ccb},\pi_b^{-bbc},\pi_m,\Pi_o,w_{\dot{L}},\mu)	& = \sum_{\dot{l} \in \dot{L}^{N(\mu)}_{-1,2}(\Pi_o)} w_{\dot{L}}(\dot{l}) PO_{1,2}(\pi_b^{-ccb},\pi_b^{-bbc},\pi_m,\dot{l},\mu) =\\
																& = PO_{1,2}(\pi_b^{-ccb},\pi_b^{-bbc},\pi_m,(*,*),\mu)
\end{aligned}
\end{equation*}

The following table shows us $PO_{1,2}$ for all the permutations of $\pi_b^{-ccb},\pi_b^{-bbc},\pi_m$.

\begin{center}
\begin{tabular}{c c c | c c c | c c c}
Slot 1 & & Slot 2							& Slot 1 & & Slot 2							& Slot 1 & & Slot 2\\
\hline
$\pi_b^{-ccb}$ & $\leq$ & $\pi_b^{-bbc}$	& $\pi_b^{-ccb}$ & $\leq$ & $\pi_m$			& $\pi_b^{-bbc}$ & $\leq$ & $\pi_b^{-ccb}$\\
$\pi_b^{-bbc}$ & $\leq$ & $\pi_m$			& $\pi_m$ & $\leq$ & $\pi_b^{-bbc}$			& $\pi_b^{-ccb}$ & $\leq$ & $\pi_m$\\
$\pi_b^{-ccb}$ & $\leq$ & $\pi_m$			& $\pi_b^{-ccb}$ & $\leq$ & $\pi_b^{-bbc}$	& $\pi_b^{-bbc}$ & $\leq$ & $\pi_m$
\end{tabular}
\begin{tabular}{c c c | c c c | c c c}
Slot 1 & & Slot 2							& Slot 1 & & Slot 2							& Slot 1 & & Slot 2\\
\hline
$\pi_b^{-bbc}$ & $\leq$ & $\pi_m$			& $\pi_m$ & $\leq$ & $\pi_b^{-ccb}$			& $\pi_m$ & $\leq$ & $\pi_b^{-bbc}$\\
$\pi_m$ & $\leq$ & $\pi_b^{-ccb}$			& $\pi_b^{-ccb}$ & $\leq$ & $\pi_b^{-bbc}$	& $\pi_b^{-bbc}$ & $\leq$ & $\pi_b^{-ccb}$\\
$\pi_b^{-bbc}$ & $\leq$ & $\pi_b^{-ccb}$	& $\pi_m$ & $\leq$ & $\pi_b^{-bbc}$			& $\pi_m$ & $\leq$ & $\pi_b^{-ccb}$
\end{tabular}
\end{center}

But, it is not possible to find a PO for any permutation, since $\pi_b^{-ccb} > \pi_m$, $\pi_b^{-bbc} > \pi_b^{-ccb}$ and $\pi_m > \pi_b^{-bbc}$. So:

\begin{equation*}
PG_{1,2}(\pi_b^{-ccb},\pi_b^{-bbc},\pi_m,\Pi_o,w_{\dot{L}},\mu) = 0
\end{equation*}

Therefore:

\begin{equation*}
PG_{1,2}(\Pi_e,w_{\Pi_e},\Pi_o,w_{\dot{L}},\mu) = 6 \frac{9}{2} \frac{1}{3} \frac{1}{3} \frac{1}{3} 0 = 0
\end{equation*}

And finally, generalising for all slots:

\begin{equation*}
\begin{aligned}
& PG(\Pi_e,w_{\Pi_e},\Pi_o,w_{\dot{L}},\mu,w_S) = \eta_{S_1^2} \sum_{i=1}^{N(\mu)} w_S(i,\mu) \times\\
& \times \left(\sum_{j=1}^{i-1} w_S(j,\mu) PG_{i,j}(\Pi_e,w_{\Pi_e},\Pi_o,w_{\dot{L}},\mu) + \sum_{j=i+1}^{N(\mu)} w_S(j,\mu) PG_{i,j}(\Pi_e,w_{\Pi_e},\Pi_o,w_{\dot{L}},\mu)\right) =\\
& \ \ \ \ \ \ \ \ \ \ \ \ \ \ \ \ \ \ \ \ \ \ \ \ \ \ \ \ \ \ \ \ \ \ \ \ = PG_{1,2}(\Pi_e,w_{\Pi_e},\Pi_o,w_{\dot{L}},\mu) =\\
& \ \ \ \ \ \ \ \ \ \ \ \ \ \ \ \ \ \ \ \ \ \ \ \ \ \ \ \ \ \ \ \ \ \ \ \ = 0
\end{aligned}
\end{equation*}

Since $0$ is the lowest possible value for the partial grading property, therefore prisoner's dilemma has $General_{min} = 0$ for this property.
\end{proof}
\end{proposition}

\begin{proposition}
\label{prop:prisoner_dilemma_PG_general_max}
$General_{max}$ for the partial grading (PG) property is equal to $1$ for the prisoner's dilemma environment.

\begin{proof}
To find $General_{max}$ (equation \ref{eq:general_max}), we need to find a trio $\left\langle\Pi_e,w_{\Pi_e},\Pi_o\right\rangle$ which maximises the property as much as possible. We can have this situation by selecting $\Pi_e = \{{\pi_b}_1,{\pi_b}_2,{\pi_b}_3\}$ with uniform weight for $w_{\Pi_e}$ and $\Pi_o = \emptyset$ (a $\pi_b$ agent always performs Blame).

Following definition \ref{def:PG}, we obtain the PG value for this $\left\langle\Pi_e,w_{\Pi_e},\Pi_o\right\rangle$. Since the environment is symmetric, we just need to calculate this property for one pair of slots and generalise its result to all pair of slots. Following definition \ref{def:STG_set} (for PG), we can calculate its PG value for slots 1 and 2:

\begin{equation*}
\begin{aligned}
PG_{1,2}(\Pi_e,w_{\Pi_e},\Pi_o,w_{\dot{L}},\mu)	& = \eta_{\Pi^3} \sum_{\pi_1,\pi_2,\pi_3 \in \Pi_e | \pi_1 \neq \pi_2 \neq \pi_3} w_{\Pi_e}(\pi_1) w_{\Pi_e}(\pi_2) w_{\Pi_e}(\pi_3) PG_{1,2}(\pi_1,\pi_2,\pi_3,\Pi_o,w_{\dot{L}},\mu) =\\
												& = 6 \frac{9}{2} \frac{1}{3} \frac{1}{3} \frac{1}{3} PG_{1,2}({\pi_b}_1,{\pi_b}_2,{\pi_b}_3,\Pi_o,w_{\dot{L}},\mu)
\end{aligned}
\end{equation*}

\noindent Note that we avoided to calculate all the permutations of $\pi_1,\pi_2,\pi_3$ for $PG_{i,j}(\pi_1,\pi_2,\pi_3,\Pi_o,w_{\dot{L}},\mu)$ since they provide the same result, by calculating only one permutation and multiplying the result by the number of permutations $6$.

In this case, we only need to calculate $PG_{1,2}({\pi_b}_1,{\pi_b}_2,{\pi_b}_3,\Pi_o,w_{\dot{L}},\mu)$. We follow definition \ref{def:STG_agents} (for PG) to calculate this value:

\begin{equation*}
\begin{aligned}
PG_{1,2}({\pi_b}_1,{\pi_b}_2,{\pi_b}_3,\Pi_o,w_{\dot{L}},\mu)	& = \sum_{\dot{l} \in \dot{L}^{N(\mu)}_{-1,2}(\Pi_o)} w_{\dot{L}}(\dot{l}) PO_{1,2}({\pi_b}_1,{\pi_b}_2,{\pi_b}_3,\dot{l},\mu) =\\
																& = PO_{1,2}({\pi_b}_1,{\pi_b}_2,{\pi_b}_3,(*,*),\mu)
\end{aligned}
\end{equation*}

The following table shows us $PO_{1,2}$ for all the permutations of ${\pi_b}_1,{\pi_b}_2,{\pi_b}_3$.

\begin{center}
\begin{tabular}{c c c | c c c | c c c}
Slot 1 & & Slot 2					& Slot 1 & & Slot 2						& Slot 1 & & Slot 2\\
\hline
${\pi_b}_1$ & $\leq$ & ${\pi_b}_2$	& ${\pi_b}_1$ & $\leq$ & ${\pi_b}_3$	& ${\pi_b}_2$ & $\leq$ & ${\pi_b}_1$\\
${\pi_b}_2$ & $\leq$ & ${\pi_b}_3$	& ${\pi_b}_3$ & $\leq$ & ${\pi_b}_2$	& ${\pi_b}_1$ & $\leq$ & ${\pi_b}_3$\\
${\pi_b}_1$ & $\leq$ & ${\pi_b}_3$	& ${\pi_b}_1$ & $\leq$ & ${\pi_b}_2$	& ${\pi_b}_2$ & $\leq$ & ${\pi_b}_3$
\end{tabular}
\begin{tabular}{c c c | c c c | c c c}
Slot 1 & & Slot 2					& Slot 1 & & Slot 2						& Slot 1 & & Slot 2\\
\hline
${\pi_b}_2$ & $\leq$ & ${\pi_b}_3$	& ${\pi_b}_3$ & $\leq$ & ${\pi_b}_1$	& ${\pi_b}_3$ & $\leq$ & ${\pi_b}_2$\\
${\pi_b}_3$ & $\leq$ & ${\pi_b}_1$	& ${\pi_b}_1$ & $\leq$ & ${\pi_b}_2$	& ${\pi_b}_2$ & $\leq$ & ${\pi_b}_1$\\
${\pi_b}_2$ & $\leq$ & ${\pi_b}_1$	& ${\pi_b}_3$ & $\leq$ & ${\pi_b}_2$	& ${\pi_b}_3$ & $\leq$ & ${\pi_b}_1$
\end{tabular}
\end{center}

It is possible to find a PO for every permutation, since we always have ${\pi_b}_i \leq {\pi_b}_j$, where a $\pi_b$ agent will always perform Blame, so they will both obtain an expected average reward of $-\frac{1}{3}$. So:

\begin{equation*}
PG_{1,2}({\pi_b}_1,{\pi_b}_2,{\pi_b}_3,\Pi_o,w_{\dot{L}},\mu) = 1
\end{equation*}

Therefore:

\begin{equation*}
PG_{1,2}(\Pi_e,w_{\Pi_e},\Pi_o,w_{\dot{L}},\mu) = 6 \frac{9}{2} \frac{1}{3} \frac{1}{3} \frac{1}{3} 1 = 1
\end{equation*}

And finally, generalising for all slots:

\begin{equation*}
\begin{aligned}
& PG(\Pi_e,w_{\Pi_e},\Pi_o,w_{\dot{L}},\mu,w_S) = \eta_{S_1^2} \sum_{i=1}^{N(\mu)} w_S(i,\mu) \times\\
& \times \left(\sum_{j=1}^{i-1} w_S(j,\mu) PG_{i,j}(\Pi_e,w_{\Pi_e},\Pi_o,w_{\dot{L}},\mu) + \sum_{j=i+1}^{N(\mu)} w_S(j,\mu) PG_{i,j}(\Pi_e,w_{\Pi_e},\Pi_o,w_{\dot{L}},\mu)\right) =\\
& \ \ \ \ \ \ \ \ \ \ \ \ \ \ \ \ \ \ \ \ \ \ \ \ \ \ \ \ \ \ \ \ \ \ \ \ = PG_{1,2}(\Pi_e,w_{\Pi_e},\Pi_o,w_{\dot{L}},\mu) =\\
& \ \ \ \ \ \ \ \ \ \ \ \ \ \ \ \ \ \ \ \ \ \ \ \ \ \ \ \ \ \ \ \ \ \ \ \ = 1
\end{aligned}
\end{equation*}

Since $1$ is the highest possible value for the partial grading property, therefore prisoner's dilemma has $General_{max} = 1$ for this property.
\end{proof}
\end{proposition}

\begin{proposition}
\label{prop:prisoner_dilemma_PG_left_max}
$Left_{max}$ for the partial grading (PG) property is equal to $1$ for the prisoner's dilemma environment.

\begin{proof}
To find $Left_{max}$ (equation \ref{eq:left_max}), we need to find a pair $\left\langle\Pi_e,w_{\Pi_e}\right\rangle$ which maximises the property as much as possible while $\Pi_o$ minimises it. Using $\Pi_e = \{{\pi_b}_1,{\pi_b}_2,{\pi_b}_3\}$ with uniform weight for $w_{\Pi_e}$ (a $\pi_b$ agent always performs Blame) we find this situation no matter which $\Pi_o$ we use.

Following definition \ref{def:PG}, we obtain the PG value for this $\left\langle\Pi_e,w_{\Pi_e},\Pi_o\right\rangle$ (where $\Pi_o$ is instantiated with any permitted values). Since the environment is symmetric, we just need to calculate this property for one pair of slots and generalise its result to all pair of slots. Following definition \ref{def:STG_set} (for PG), we can calculate its PG value for slots 1 and 2:

\begin{equation*}
\begin{aligned}
PG_{1,2}(\Pi_e,w_{\Pi_e},\Pi_o,w_{\dot{L}},\mu)	& = \eta_{\Pi^3} \sum_{\pi_1,\pi_2,\pi_3 \in \Pi_e | \pi_1 \neq \pi_2 \neq \pi_3} w_{\Pi_e}(\pi_1) w_{\Pi_e}(\pi_2) w_{\Pi_e}(\pi_3) PG_{1,2}(\pi_1,\pi_2,\pi_3,\Pi_o,w_{\dot{L}},\mu) =\\
												& = 6 \frac{9}{2} \frac{1}{3} \frac{1}{3} \frac{1}{3} PG_{1,2}({\pi_b}_1,{\pi_b}_2,{\pi_b}_3,\Pi_o,w_{\dot{L}},\mu)
\end{aligned}
\end{equation*}

\noindent Note that we avoided to calculate all the permutations of $\pi_1,\pi_2,\pi_3$ for $PG_{i,j}(\pi_1,\pi_2,\pi_3,\Pi_o,w_{\dot{L}},\mu)$ since they provide the same result, by calculating only one permutation and multiplying the result by the number of permutations $6$.

In this case, we only need to calculate $PG_{1,2}({\pi_b}_1,{\pi_b}_2,{\pi_b}_3,\Pi_o,w_{\dot{L}},\mu)$. We follow definition \ref{def:STG_agents} (for PG) to calculate this value:

\begin{equation*}
\begin{aligned}
PG_{1,2}({\pi_b}_1,{\pi_b}_2,{\pi_b}_3,\Pi_o,w_{\dot{L}},\mu)	& = \sum_{\dot{l} \in \dot{L}^{N(\mu)}_{-1,2}(\Pi_o)} w_{\dot{L}}(\dot{l}) PO_{1,2}({\pi_b}_1,{\pi_b}_2,{\pi_b}_3,\dot{l},\mu) =\\
																& = PO_{1,2}({\pi_b}_1,{\pi_b}_2,{\pi_b}_3,(*,*),\mu)
\end{aligned}
\end{equation*}

\noindent Note that the choice of $\Pi_o$ does not affect the result of $PG_{1,2}$.

The following table shows us $PO_{1,2}$ for all the permutations of ${\pi_b}_1,{\pi_b}_2,{\pi_b}_3$.

\begin{center}
\begin{tabular}{c c c | c c c | c c c}
Slot 1 & & Slot 2					& Slot 1 & & Slot 2						& Slot 1 & & Slot 2\\
\hline
${\pi_b}_1$ & $\leq$ & ${\pi_b}_2$	& ${\pi_b}_1$ & $\leq$ & ${\pi_b}_3$	& ${\pi_b}_2$ & $\leq$ & ${\pi_b}_1$\\
${\pi_b}_2$ & $\leq$ & ${\pi_b}_3$	& ${\pi_b}_3$ & $\leq$ & ${\pi_b}_2$	& ${\pi_b}_1$ & $\leq$ & ${\pi_b}_3$\\
${\pi_b}_1$ & $\leq$ & ${\pi_b}_3$	& ${\pi_b}_1$ & $\leq$ & ${\pi_b}_2$	& ${\pi_b}_2$ & $\leq$ & ${\pi_b}_3$
\end{tabular}
\begin{tabular}{c c c | c c c | c c c}
Slot 1 & & Slot 2					& Slot 1 & & Slot 2						& Slot 1 & & Slot 2\\
\hline
${\pi_b}_2$ & $\leq$ & ${\pi_b}_3$	& ${\pi_b}_3$ & $\leq$ & ${\pi_b}_1$	& ${\pi_b}_3$ & $\leq$ & ${\pi_b}_2$\\
${\pi_b}_3$ & $\leq$ & ${\pi_b}_1$	& ${\pi_b}_1$ & $\leq$ & ${\pi_b}_2$	& ${\pi_b}_2$ & $\leq$ & ${\pi_b}_1$\\
${\pi_b}_2$ & $\leq$ & ${\pi_b}_1$	& ${\pi_b}_3$ & $\leq$ & ${\pi_b}_2$	& ${\pi_b}_3$ & $\leq$ & ${\pi_b}_1$
\end{tabular}
\end{center}

It is possible to find a PO for every permutation, since we always have ${\pi_b}_i \leq {\pi_b}_j$, where a $\pi_b$ agent will always perform Blame, so they will both obtain an expected average reward of $-\frac{1}{3}$. So:

\begin{equation*}
PG_{1,2}({\pi_b}_1,{\pi_b}_2,{\pi_b}_3,\Pi_o,w_{\dot{L}},\mu) = 1
\end{equation*}

Therefore:

\begin{equation*}
PG_{1,2}(\Pi_e,w_{\Pi_e},\Pi_o,w_{\dot{L}},\mu) = 6 \frac{9}{2} \frac{1}{3} \frac{1}{3} \frac{1}{3} 1 = 1
\end{equation*}

And finally, generalising for all slots:

\begin{equation*}
\begin{aligned}
& PG(\Pi_e,w_{\Pi_e},\Pi_o,w_{\dot{L}},\mu,w_S) = \eta_{S_1^2} \sum_{i=1}^{N(\mu)} w_S(i,\mu) \times\\
& \times \left(\sum_{j=1}^{i-1} w_S(j,\mu) PG_{i,j}(\Pi_e,w_{\Pi_e},\Pi_o,w_{\dot{L}},\mu) + \sum_{j=i+1}^{N(\mu)} w_S(j,\mu) PG_{i,j}(\Pi_e,w_{\Pi_e},\Pi_o,w_{\dot{L}},\mu)\right) =\\
& \ \ \ \ \ \ \ \ \ \ \ \ \ \ \ \ \ \ \ \ \ \ \ \ \ \ \ \ \ \ \ \ \ \ \ \ = PG_{1,2}(\Pi_e,w_{\Pi_e},\Pi_o,w_{\dot{L}},\mu) =\\
& \ \ \ \ \ \ \ \ \ \ \ \ \ \ \ \ \ \ \ \ \ \ \ \ \ \ \ \ \ \ \ \ \ \ \ \ = 1
\end{aligned}
\end{equation*}

So, for every $\Pi_o$ we obtain the same result:

\begin{equation*}
\forall \Pi_o : PG(\Pi_e,w_{\Pi_e},\Pi_o,w_{\dot{L}},\mu,w_S) = 1
\end{equation*}

Therefore, prisoner's dilemma has $Left_{max} = 1$ for this property.
\end{proof}
\end{proposition}

\begin{proposition}
\label{prop:prisoner_dilemma_PG_right_min}
$Right_{min}$ for the partial grading (PG) property is equal to $0$ for the prisoner's dilemma environment.

\begin{proof}
To find $Right_{min}$ (equation \ref{eq:right_min}), we need to find a pair $\left\langle\Pi_e,w_{\Pi_e}\right\rangle$ which minimises the property as much as possible while $\Pi_o$ maximises it. Using $\Pi_e = \{\pi_b^{-ccb},\pi_b^{-bbc},\pi_m\}$ with uniform weight for $w_{\Pi_e}$ (a $\pi_m$ agent first acts randomly and then always mimics the other agent's last action, a $\pi_b^{-ccb}$ agent always performs Blame except for the last three actions where it performs Cooperate twice and finalises performing Blame, and a $\pi_b^{-bbc}$ agent always performs Blame except for the last three actions where it performs Blame twice and finalises performing Cooperate) we find this situation no matter which $\Pi_o$ we use.

Following definition \ref{def:PG}, we obtain the PG value for this $\left\langle\Pi_e,w_{\Pi_e},\Pi_o\right\rangle$ (where $\Pi_o$ is instantiated with any permitted values). Since the environment is symmetric, we just need to calculate this property for one pair of slots and generalise its result to all pair of slots. Following definition \ref{def:STG_set} (for PG), we can calculate its PG value for slots 1 and 2:

\begin{equation*}
\begin{aligned}
PG_{1,2}(\Pi_e,w_{\Pi_e},\Pi_o,w_{\dot{L}},\mu)	& = \eta_{\Pi^3} \sum_{\pi_1,\pi_2,\pi_3 \in \Pi_e | \pi_1 \neq \pi_2 \neq \pi_3} w_{\Pi_e}(\pi_1) w_{\Pi_e}(\pi_2) w_{\Pi_e}(\pi_3) PG_{1,2}(\pi_1,\pi_2,\pi_3,\Pi_o,w_{\dot{L}},\mu) =\\
												& = 6 \frac{9}{2} \frac{1}{3} \frac{1}{3} \frac{1}{3} PG_{1,2}(\pi_b^{-ccb},\pi_b^{-bbc},\pi_m,\Pi_o,w_{\dot{L}},\mu)
\end{aligned}
\end{equation*}

\noindent Note that we avoided to calculate all the permutations of $\pi_1,\pi_2,\pi_3$ for $PG_{i,j}(\pi_1,\pi_2,\pi_3,\Pi_o,w_{\dot{L}},\mu)$ since they provide the same result, by calculating only one permutation and multiplying the result by the number of permutations $6$.

In this case, we only need to calculate $PG_{1,2}(\pi_b^{-ccb},\pi_b^{-bbc},\pi_m,\Pi_o,w_{\dot{L}},\mu)$. We follow definition \ref{def:STG_agents} (for PG) to calculate this value:

\begin{equation*}
\begin{aligned}
PG_{1,2}(\pi_b^{-ccb},\pi_b^{-bbc},\pi_m,\Pi_o,w_{\dot{L}},\mu)	& = \sum_{\dot{l} \in \dot{L}^{N(\mu)}_{-1,2}(\Pi_o)} w_{\dot{L}}(\dot{l}) PO_{1,2}(\pi_b^{-ccb},\pi_b^{-bbc},\pi_m,\dot{l},\mu) =\\
																& = PO_{1,2}(\pi_b^{-ccb},\pi_b^{-bbc},\pi_m,(*,*),\mu)
\end{aligned}
\end{equation*}

\noindent Note that the choice of $\Pi_o$ does not affect the result of $PG_{1,2}$.

The following table shows us $PO_{1,2}$ for all the permutations of $\pi_b^{-ccb},\pi_b^{-bbc},\pi_m$.

\begin{center}
\begin{tabular}{c c c | c c c | c c c}
Slot 1 & & Slot 2							& Slot 1 & & Slot 2							& Slot 1 & & Slot 2\\
\hline
$\pi_b^{-ccb}$ & $\leq$ & $\pi_b^{-bbc}$	& $\pi_b^{-ccb}$ & $\leq$ & $\pi_m$			& $\pi_b^{-bbc}$ & $\leq$ & $\pi_b^{-ccb}$\\
$\pi_b^{-bbc}$ & $\leq$ & $\pi_m$			& $\pi_m$ & $\leq$ & $\pi_b^{-bbc}$			& $\pi_b^{-ccb}$ & $\leq$ & $\pi_m$\\
$\pi_b^{-ccb}$ & $\leq$ & $\pi_m$			& $\pi_b^{-ccb}$ & $\leq$ & $\pi_b^{-bbc}$	& $\pi_b^{-bbc}$ & $\leq$ & $\pi_m$
\end{tabular}
\begin{tabular}{c c c | c c c | c c c}
Slot 1 & & Slot 2							& Slot 1 & & Slot 2							& Slot 1 & & Slot 2\\
\hline
$\pi_b^{-bbc}$ & $\leq$ & $\pi_m$			& $\pi_m$ & $\leq$ & $\pi_b^{-ccb}$			& $\pi_m$ & $\leq$ & $\pi_b^{-bbc}$\\
$\pi_m$ & $\leq$ & $\pi_b^{-ccb}$			& $\pi_b^{-ccb}$ & $\leq$ & $\pi_b^{-bbc}$	& $\pi_b^{-bbc}$ & $\leq$ & $\pi_b^{-ccb}$\\
$\pi_b^{-bbc}$ & $\leq$ & $\pi_b^{-ccb}$	& $\pi_m$ & $\leq$ & $\pi_b^{-bbc}$			& $\pi_m$ & $\leq$ & $\pi_b^{-ccb}$
\end{tabular}
\end{center}

But, it is not possible to find a PO for any permutation, since $\pi_b^{-ccb} < \pi_b^{-bbc}$, $\pi_b^{-bbc} < \pi_m$ but $\pi_b^{-ccb} > \pi_m$. So:

\begin{equation*}
PG_{1,2}(\pi_b^{-ccb},\pi_b^{-bbc},\pi_m,\Pi_o,w_{\dot{L}},\mu) = 0
\end{equation*}

Therefore:

\begin{equation*}
PG_{1,2}(\Pi_e,w_{\Pi_e},\Pi_o,w_{\dot{L}},\mu) = 6 \frac{9}{2} \frac{1}{3} \frac{1}{3} \frac{1}{3} 0 = 0
\end{equation*}

And finally, generalising for all slots:

\begin{equation*}
\begin{aligned}
& PG(\Pi_e,w_{\Pi_e},\Pi_o,w_{\dot{L}},\mu,w_S) = \eta_{S_1^2} \sum_{i=1}^{N(\mu)} w_S(i,\mu) \times\\
& \times \left(\sum_{j=1}^{i-1} w_S(j,\mu) PG_{i,j}(\Pi_e,w_{\Pi_e},\Pi_o,w_{\dot{L}},\mu) + \sum_{j=i+1}^{N(\mu)} w_S(j,\mu) PG_{i,j}(\Pi_e,w_{\Pi_e},\Pi_o,w_{\dot{L}},\mu)\right) =\\
& \ \ \ \ \ \ \ \ \ \ \ \ \ \ \ \ \ \ \ \ \ \ \ \ \ \ \ \ \ \ \ \ \ \ \ \ = PG_{1,2}(\Pi_e,w_{\Pi_e},\Pi_o,w_{\dot{L}},\mu) =\\
& \ \ \ \ \ \ \ \ \ \ \ \ \ \ \ \ \ \ \ \ \ \ \ \ \ \ \ \ \ \ \ \ \ \ \ \ = 0
\end{aligned}
\end{equation*}

So, for every $\Pi_o$ we obtain the same result:

\begin{equation*}
\forall \Pi_o : PG(\Pi_e,w_{\Pi_e},\Pi_o,w_{\dot{L}},\mu,w_S) = 0
\end{equation*}

Therefore, prisoner's dilemma has $Right_{min} = 0$ for this property.
\end{proof}
\end{proposition}

\subsection{Slot Reward Dependency}
Next we see the slot reward dependency (SRD) property. As given in section \ref{sec:SRD}, we want to know how much competitive or cooperative the environment is.

\begin{proposition}
\label{prop:prisoner_dilemma_SRD_general_min}
$General_{min}$ for the slot reward dependency (SRD) property is equal to $-1$ for the prisoner's dilemma environment.

\begin{proof}
To find $General_{min}$ (equation \ref{eq:general_min}), we need to find a trio $\left\langle\Pi_e,w_{\Pi_e},\Pi_o\right\rangle$ which minimises the property as much as possible. We can have this situation by selecting $\Pi_e = \{\pi_b\}$ with $w_{\Pi_e}(\pi_b) = 1$ and $\Pi_o = \{\pi_c\}$ (a $\pi_c$ agent always performs Cooperate and a $\pi_b$ agent always performs Blame).

Following definition \ref{def:SRD}, we obtain the SRD value for this $\left\langle\Pi_e,w_{\Pi_e},\Pi_o\right\rangle$. Since the environment is symmetric, we just need to calculate this property for one pair of slots and generalise its result to all pair of slots. Following definition \ref{def:SRD_set}, we could calculate its SRD value for slots 1 and 2 but, since $\Pi_e$ has only one agent and its weight is equal to $1$, it is equivalent to use directly definition \ref{def:SRD_agent} for slots 1 and 2:

\begin{equation*}
\begin{aligned}
SRD_{1,2}(\pi_b,\Pi_o,w_{\dot{L}},\mu)	& = corr_{\dot{l} \in \dot{L}^{N(\mu)}_{-1}(\Pi_o)}[w_{\dot{L}}(\dot{l})](R_1(\mu[\instantiation{l}{1}{\pi_b}]), R_2(\mu[\instantiation{l}{1}{\pi_b}])) =\\
										& = corr(R_1(\mu[\pi_b,\pi_c]), R_2(\mu[\pi_b,\pi_c]))
\end{aligned}
\end{equation*}

In line-up $(\pi_b,\pi_c)$, where $\pi_b$ will always perform Blame and $\pi_c$ will always perform Cooperate, they will obtain an expected average reward of $1$ and $-1$ respectively. Since we use a correlation function between the expected average rewards, and the agents in slots 1 and 2 obtain an expected average reward of $-1$ and $1$ respectively, then the correlation function will obtain the value of $-1$. So:

\begin{equation*}
SRD_{1,2}(\pi_b,\Pi_o,w_{\dot{L}},\mu) = -1
\end{equation*}

And finally:

\begin{equation*}
\begin{aligned}
& SRD(\Pi_e,w_{\Pi_e},\Pi_o,w_{\dot{L}},\mu,w_S) = \eta_{S_1^2} \sum_{i=1}^{N(\mu)} w_S(i,\mu) \times\\
& \times \left(\sum_{j=1}^{i-1} w_S(j,\mu) SRD_{i,j}(\Pi_e,w_{\Pi_e},\Pi_o,w_{\dot{L}},\mu) + \sum_{j=i+1}^{N(\mu)} w_S(j,\mu) SRD_{i,j}(\Pi_e,w_{\Pi_e},\Pi_o,w_{\dot{L}},\mu)\right) =\\
& \ \ \ \ \ \ \ \ \ \ \ \ \ \ \ \ \ \ \ \ \ \ \ \ \ \ \ \ \ \ \ \ \ \ \ \ \ \ = SRD_{1,2}(\Pi_e,w_{\Pi_e},\Pi_o,w_{\dot{L}},\mu) =\\
& \ \ \ \ \ \ \ \ \ \ \ \ \ \ \ \ \ \ \ \ \ \ \ \ \ \ \ \ \ \ \ \ \ \ \ \ \ \ = SRD_{1,2}(\pi_b,\Pi_o,w_{\dot{L}},\mu) =\\
& \ \ \ \ \ \ \ \ \ \ \ \ \ \ \ \ \ \ \ \ \ \ \ \ \ \ \ \ \ \ \ \ \ \ \ \ \ \ = -1
\end{aligned}
\end{equation*}

Since $-1$ is the lowest possible value for the slot reward dependency property, therefore prisoner's dilemma has $General_{min} = -1$ for this property.
\end{proof}
\end{proposition}

\begin{proposition}
\label{prop:prisoner_dilemma_SRD_general_max}
$General_{max}$ for the slot reward dependency (SRD) property is equal to $1$ for the prisoner's dilemma environment.

\begin{proof}
To find $General_{max}$ (equation \ref{eq:general_max}), we need to find a trio $\left\langle\Pi_e,w_{\Pi_e},\Pi_o\right\rangle$ which minimises the property as much as possible. We can have this situation by selecting $\Pi_e = \{\pi_b\}$ with $w_{\Pi_e}(\pi_b) = 1$ and $\Pi_o = \{\pi_b\}$ (a $\pi_b$ agent always performs Blame).

Following definition \ref{def:SRD}, we obtain the SRD value for this $\left\langle\Pi_e,w_{\Pi_e},\Pi_o\right\rangle$. Since the environment is symmetric, we just need to calculate this property for one pair of slots and generalise its result to all pair of slots. Following definition \ref{def:SRD_set}, we could calculate its SRD value for slots 1 and 2 but, since $\Pi_e$ has only one agent and its weight is equal to $1$, it is equivalent to use directly definition \ref{def:SRD_agent} for slots 1 and 2:

\begin{equation*}
\begin{aligned}
SRD_{1,2}(\pi_b,\Pi_o,w_{\dot{L}},\mu)	& = corr_{\dot{l} \in \dot{L}^{N(\mu)}_{-1}(\Pi_o)}[w_{\dot{L}}(\dot{l})](R_1(\mu[\instantiation{l}{1}{\pi_b}]), R_2(\mu[\instantiation{l}{1}{\pi_b}])) =\\
										& = corr(R_1(\mu[\pi_b,\pi_b]), R_2(\mu[\pi_b,\pi_b]))
\end{aligned}
\end{equation*}

In line-up $(\pi_b,\pi_b)$, where both $\pi_b$ will always perform Blame, they will both obtain an expected average reward of $-\frac{1}{3}$. Since we use a correlation function between the expected average rewards, and the agents in slots 1 and 2 obtain the same expected average reward of $-\frac{1}{3}$, then the correlation function will obtain the value of $1$. So:

\begin{equation*}
SRD_{1,2}(\pi_b,\Pi_o,w_{\dot{L}},\mu) = 1
\end{equation*}

And finally:

\begin{equation*}
\begin{aligned}
& SRD(\Pi_e,w_{\Pi_e},\Pi_o,w_{\dot{L}},\mu,w_S) = \eta_{S_1^2} \sum_{i=1}^{N(\mu)} w_S(i,\mu) \times\\
& \times \left(\sum_{j=1}^{i-1} w_S(j,\mu) SRD_{i,j}(\Pi_e,w_{\Pi_e},\Pi_o,w_{\dot{L}},\mu) + \sum_{j=i+1}^{N(\mu)} w_S(j,\mu) SRD_{i,j}(\Pi_e,w_{\Pi_e},\Pi_o,w_{\dot{L}},\mu)\right) =\\
& \ \ \ \ \ \ \ \ \ \ \ \ \ \ \ \ \ \ \ \ \ \ \ \ \ \ \ \ \ \ \ \ \ \ \ \ \ \ = SRD_{1,2}(\Pi_e,w_{\Pi_e},\Pi_o,w_{\dot{L}},\mu) =\\
& \ \ \ \ \ \ \ \ \ \ \ \ \ \ \ \ \ \ \ \ \ \ \ \ \ \ \ \ \ \ \ \ \ \ \ \ \ \ = SRD_{1,2}(\pi_b,\Pi_o,w_{\dot{L}},\mu) =\\
& \ \ \ \ \ \ \ \ \ \ \ \ \ \ \ \ \ \ \ \ \ \ \ \ \ \ \ \ \ \ \ \ \ \ \ \ \ \ = 1
\end{aligned}
\end{equation*}

Since $1$ is the highest possible value for the slot reward dependency property, therefore prisoner's dilemma has $General_{max} = 1$ for this property.
\end{proof}
\end{proposition}

\subsection{Competitive Anticipation}
Finally, we follow with the competitive anticipation (AComp) property. As given in section \ref{sec:AComp}, we want to know how much benefit the evaluated agents obtain when they anticipate competing agents.

\begin{proposition}
\label{prop:prisoner_dilemma_AComp_general_min}
$General_{min}$ for the competitive anticipation (AComp) property is equal to $-\frac{2}{3}$ for the prisoner's dilemma environment.

\begin{proof}
To find $General_{min}$ (equation \ref{eq:general_min}), we need to find a trio $\left\langle\Pi_e,w_{\Pi_e},\Pi_o\right\rangle$ which minimises the property as much as possible. We can have this situation by selecting $\Pi_e = \{\pi_{c/b}\}$ with $w_{\Pi_e}(\pi_{c/b}) = 1$ and $\Pi_o = \{\pi_b\}$ (a $\pi_{c/b}$ agent performs Cooperate until the other agent also performs Cooperate, then it starts to perform Blame, and a $\pi_b$ agent always performs Blame).

Following definition \ref{def:AComp}, we obtain the AComp value for this $\left\langle\Pi_e,w_{\Pi_e},\Pi_o\right\rangle$. Since the environment is symmetric, we just need to calculate this property for one pair of slots in different teams, and generalise its result to all pair of slots. Following definition \ref{def:AComp_set}, we could calculate its AComp value for slots 1 and 2 but, since $\Pi_e$ has only one agent, its weight is equal to $1$ and $\Pi_o$ also has only one agent, it is equivalent to use directly definition \ref{def:AComp_agents} for slots 1 and 2:

\begin{equation*}
\begin{aligned}
AComp_{1,2}(\pi_{c/b},\pi_b,\Pi_o,w_{\dot{L}},\mu)	& = \sum_{\dot{l} \in \dot{L}^{N(\mu)}_{-1,2}(\Pi_o)} w_{\dot{L}}(\dot{l}) \frac{1}{2} \left(R_1(\mu[\instantiation{l}{1,2}{\pi_{c/b},\pi_b}]) - R_1(\mu[\instantiation{l}{1,2}{\pi_{c/b},\pi_r}])\right) =\\
													& = \frac{1}{2} \left(R_1(\mu[\pi_{c/b},\pi_b]) - R_1(\mu[\pi_{c/b},\pi_r])\right)
\end{aligned}
\end{equation*}

In line-up $(\pi_{c/b},\pi_b)$, where $\pi_{c/b}$ will always perform Cooperate and $\pi_b$ will always perform Blame, the agent in slot 1 ($\pi_{c/b}$) will obtain an expected average reward of $-1$. In line-up $(\pi_{c/b},\pi_r)$, where $\pi_{c/b}$ will start with Cooperate and then will continue with Blame once $\pi_r$ performs Cooperate, and $\pi_r$ will always act randomly, the agent in slot 1 ($\pi_{c/b}$) will obtain an expected average reward of $\frac{1}{3}$ (in the limit). So:

\begin{equation*}
AComp_{1,2}(\pi_{c/b},\pi_b,\Pi_o,w_{\dot{L}},\mu) = \frac{1}{2} \left((-1) - \frac{1}{3}\right) = -\frac{2}{3}
\end{equation*}

\noindent Note that this is the minimum possible value for $AComp_{1,2}(\pi_1,\pi_2,\Pi_o,w_{\dot{L}},\mu)$ since interacting with a random agent cannot provide a greater result than $\frac{1}{3}$ for this environment.

And finally, generalising for all slots:


\begin{equation*}
\begin{aligned}
AComp(\Pi_e,w_{\Pi_e},\Pi_o,w_{\dot{L}},\mu,w_S)	& = \eta_{S_2^2} \sum_{t_1,t_2 \in \tau | t_1 \neq t_2} \sum_{i \in t_1} w_S(i,\mu) \sum_{j \in t_2} w_S(j,\mu) AComp_{i,j}(\Pi_e,w_{\Pi_e},\Pi_o,w_{\dot{L}},\mu) =\\
													& = AComp_{1,2}(\Pi_e,w_{\Pi_e},\Pi_o,w_{\dot{L}},\mu) =\\
													& = AComp_{1,2}(\pi_{c/b},\pi_b,\Pi_o,w_{\dot{L}},\mu) =\\
													& = -\frac{2}{3}
\end{aligned}
\end{equation*}

Since $-\frac{2}{3}$ is the lowest possible value that we can obtain for the competitive anticipation property, therefore prisoner's dilemma has $General_{min} = -\frac{2}{3}$ for this property.
\end{proof}
\end{proposition}

\begin{proposition}
\label{prop:prisoner_dilemma_AComp_general_max}
$General_{max}$ for the competitive anticipation (AComp) property is equal to $\frac{2}{3}$ for the prisoner's dilemma environment.

\begin{proof}
To find $General_{max}$ (equation \ref{eq:general_max}), we need to find a trio $\left\langle\Pi_e,w_{\Pi_e},\Pi_o\right\rangle$ which maximises the property as much as possible. We can have this situation by selecting $\Pi_e = \{\pi_{b/c}\}$ with $w_{\Pi_e}(\pi_{b/c}) = 1$ and $\Pi_o = \{\pi_c\}$ (a $\pi_{b/c}$ agent performs Blame until the other agent also performs Blame, then it starts to perform Cooperate, and a $\pi_c$ agent always performs Cooperate).

Following definition \ref{def:AComp}, we obtain the AComp value for this $\left\langle\Pi_e,w_{\Pi_e},\Pi_o\right\rangle$. Since the environment is symmetric, we just need to calculate this property for one pair of slots in different teams, and generalise its result to all pair of slots. Following definition \ref{def:AComp_set}, we could calculate its AComp value for slots 1 and 2 but, since $\Pi_e$ has only one agent, its weight is equal to $1$ and $\Pi_o$ also has only one agent, it is equivalent to use directly definition \ref{def:AComp_agents} for slots 1 and 2:

\begin{equation*}
\begin{aligned}
AComp_{1,2}(\pi_{b/c},\pi_c,\Pi_o,w_{\dot{L}},\mu)	& = \sum_{\dot{l} \in \dot{L}^{N(\mu)}_{-1,2}(\Pi_o)} w_{\dot{L}}(\dot{l}) \frac{1}{2} \left(R_1(\mu[\instantiation{l}{1,2}{\pi_{b/c},\pi_c}]) - R_1(\mu[\instantiation{l}{1,2}{\pi_{b/c},\pi_r}])\right) =\\
													& = \frac{1}{2} \left(R_1(\mu[\pi_{b/c},\pi_c]) - R_1(\mu[\pi_{b/c},\pi_r])\right)
\end{aligned}
\end{equation*}

In line-up $(\pi_{b/c},\pi_c)$, where $\pi_{b/c}$ will always perform Blame and $\pi_c$ will always perform Cooperate, the agent in slot 1 ($\pi_{b/c}$) will obtain an expected average reward of $1$. In line-up $(\pi_{b/c},\pi_r)$, where $\pi_{b/c}$ will start with Blame and then will continue with Cooperate once $\pi_r$ performs Blame, and $\pi_r$ will always act randomly, the agent in slot 1 ($\pi_{b/c}$) will obtain an expected average reward of $-\frac{1}{3}$ (in the limit). So:

\begin{equation*}
AComp_{1,2}(\pi_{b/c},\pi_c,\Pi_o,w_{\dot{L}},\mu) = \frac{1}{2} \left(1 - \left(-\frac{1}{3}\right)\right) = \frac{2}{3}
\end{equation*}

\noindent Note that this is the maximum possible value for $AComp_{1,2}(\pi_1,\pi_2,\Pi_o,w_{\dot{L}},\mu)$ since interacting with a random agent cannot provide a lower result than $-\frac{1}{3}$ for this environment.

And finally, generalising for all slots:


\begin{equation*}
\begin{aligned}
AComp(\Pi_e,w_{\Pi_e},\Pi_o,w_{\dot{L}},\mu,w_S)	& = \eta_{S_2^2} \sum_{t_1,t_2 \in \tau | t_1 \neq t_2} \sum_{i \in t_1} w_S(i,\mu) \sum_{j \in t_2} w_S(j,\mu) AComp_{i,j}(\Pi_e,w_{\Pi_e},\Pi_o,w_{\dot{L}},\mu) =\\
													& = AComp_{1,2}(\Pi_e,w_{\Pi_e},\Pi_o,w_{\dot{L}},\mu) =\\
													& = AComp_{1,2}(\pi_{b/c},\pi_c,\Pi_o,w_{\dot{L}},\mu) =\\
													& = \frac{2}{3}
\end{aligned}
\end{equation*}

Since $\frac{2}{3}$ is the highest possible value that we can obtain for the competitive anticipation property, therefore prisoner's dilemma has $General_{max} = \frac{2}{3}$ for this property.
\end{proof}
\end{proposition}

\section{Predator-prey properties}
\label{appen:predator-prey_properties}
In this section we prove how we obtained the values for the properties for the predator-prey environment (section \ref{sec:predator-prey}). To calculate some of the values for the properties, we make use of lemma \ref{lemma:predator-prey_always_chase}.

\begin{lemma}
\label{lemma:predator-prey_always_chase}
When three well coordinated predators are trying to chase the prey, it will always be chased in 5 iterations or less no matter the behaviour of the prey.

Since there exists a lot of variants to chase the prey, we cannot show them all. Instead, here we show one of the largest sequences of actions to chase the prey in 5 iterations when the prey is trying to escape and the predators are well coordinated.

\begin{figure}[!ht]
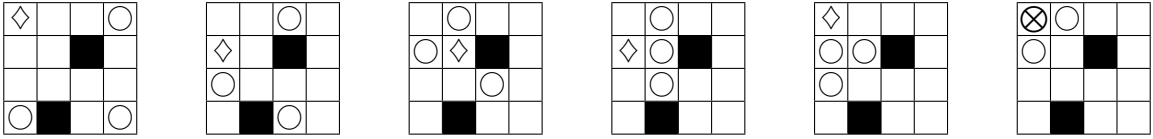

\newcolumntype{C}{>{\centering\arraybackslash}p{12px}}
\def \block {\cellcolor{black}}
\def \hunter {$\bigcirc$}
\def \prey {$\diamondsuit$}
\def \chased {$\bigotimes$}

\centering

\begin{minipage}{0.15\textwidth}
\centering
\setlength{\tabcolsep}{0px}

\begin{tabular}[c]{|C|C|C|C|}
\hline
\prey   &         &         & \hunter \\
\hline
        &         & \block  &         \\
\hline
        &         &         &         \\
\hline
\hunter & \block  &         & \hunter \\
\hline
\end{tabular}
\end{minipage}
\begin{minipage}{0.15\textwidth}
\centering
\setlength{\tabcolsep}{0px}

\begin{tabular}[c]{|C|C|C|C|}
\hline
        &         & \hunter &         \\
\hline
\prey   &         & \block  &         \\
\hline
\hunter &         &         &         \\
\hline
        & \block  & \hunter &         \\
\hline
\end{tabular}
\end{minipage}
\begin{minipage}{0.15\textwidth}
\centering
\setlength{\tabcolsep}{0px}

\begin{tabular}[c]{|C|C|C|C|}
\hline
        & \hunter &         &         \\
\hline
\hunter & \prey   & \block  &         \\
\hline
        &         & \hunter &         \\
\hline
        & \block  &         &         \\
\hline
\end{tabular}
\end{minipage}
\begin{minipage}{0.15\textwidth}
\centering
\setlength{\tabcolsep}{0px}

\begin{tabular}[c]{|C|C|C|C|}
\hline
        & \hunter &         &         \\
\hline
\prey   & \hunter & \block  &         \\
\hline
        & \hunter &         &         \\
\hline
        & \block  &         &         \\
\hline
\end{tabular}
\end{minipage}
\begin{minipage}{0.15\textwidth}
\centering
\setlength{\tabcolsep}{0px}

\begin{tabular}[c]{|C|C|C|C|}
\hline
\prey   &         &         &         \\
\hline
\hunter & \hunter & \block  &         \\
\hline
\hunter &         &         &         \\
\hline
        & \block  &         &         \\
\hline
\end{tabular}
\end{minipage}
\begin{minipage}{0.15\textwidth}
\centering
\setlength{\tabcolsep}{0px}

\begin{tabular}[c]{|C|C|C|C|}
\hline
\chased & \hunter &         &         \\
\hline
\hunter &         & \block  &         \\
\hline
        &         &         &         \\
\hline
        & \block  &         &         \\
\hline
\end{tabular}
\end{minipage}

\caption{One of the largest sequences of actions to chase the prey in 5 iterations with three well coordinated predators. $\bigotimes$ represents a chased prey.}
\end{figure}

Other behaviours of the prey will lead it closer to the boundaries, which would be easier for the predators to chase it.
\end{lemma}

\subsection{Action Dependency}
We start with the action dependency (AD) property. As given in section \ref{sec:AD}, we want to know if the evaluated agents behave differently depending on which line-up they interact with. We use $\Delta_S(a,b) = 1$ if distributions $a$ and $b$ are equal and $0$ otherwise.

\begin{proposition}
\label{prop:predator-prey_AD_general_min}
$General_{min}$ for the action dependency (AD) property is equal to $0$ for the predator-prey environment.

\begin{proof}
To find $General_{min}$ (equation \ref{eq:general_min}), we need to find a trio $\left\langle\Pi_e,w_{\Pi_e},\Pi_o\right\rangle$ which minimises the property as much as possible. We can have this situation by selecting $\Pi_e = \{\pi_u\}$ with $w_{\Pi_e}(\pi_u) = 1$ and $\Pi_o = \{{\pi_d}_1, {\pi_d}_2\}$ (a $\pi_d$ agent always performs Down and a $\pi_u$ agent always performs Up).

Following definition \ref{def:AD}, we obtain the AD value for this $\left\langle\Pi_e,w_{\Pi_e},\Pi_o\right\rangle$. Since the environment is not symmetric, we need to calculate this property for every slot. Following definition \ref{def:AD_set}, we could calculate its AD value for each slot but, since $\Pi_e$ has only one agent and its weight is equal to $1$, it is equivalent to use directly definition \ref{def:AD_agent}. We start with slot 1:

\begin{equation*}
\begin{aligned}
AD_1(\pi_u,\Pi_o,w_{\dot{L}},\mu)	& = \eta_{\dot{L}^2} \sum_{\dot{u},\dot{v} \in \dot{L}^{N(\mu)}_{-1}(\Pi_o) | \dot{u} \neq \dot{v}} w_{\dot{L}}(\dot{u}) w_{\dot{L}}(\dot{v}) \Delta_S(\breve{A}_1(\mu[\instantiation{u}{1}{\pi_u}]), \breve{A}_1(\mu[\instantiation{v}{1}{\pi_u}])) =\\
									& = 2 \frac{8}{7} \frac{1}{8} \frac{1}{8} \{\Delta_S(\breve{A}_1(\mu[\pi_u,{\pi_d}_1,{\pi_d}_1,{\pi_d}_1]), \breve{A}_1(\mu[\pi_u,{\pi_d}_1,{\pi_d}_1,{\pi_d}_2])) +\\
									& + \Delta_S(\breve{A}_1(\mu[\pi_u,{\pi_d}_1,{\pi_d}_1,{\pi_d}_1]), \breve{A}_1(\mu[\pi_u,{\pi_d}_1,{\pi_d}_2,{\pi_d}_1])) +\\
									& \ \ \ \ \ \ \ \ \ \ \ \ \ \ \ \ \ \ \ \ \ \ \ \ \ \ \ \ \ \ \ \ \ \ \ \ \ \ \ \ \vdots\\
									& + \Delta_S(\breve{A}_1(\mu[\pi_u,{\pi_d}_2,{\pi_d}_2,{\pi_d}_1]), \breve{A}_1(\mu[\pi_u,{\pi_d}_2,{\pi_d}_2,{\pi_d}_2]))\}
\end{aligned}
\end{equation*}

\noindent Note that we avoided to calculate both $\Delta_S(a,b)$ and $\Delta_S(b,a)$ since they provide the same result, by calculating only $\Delta_S(a,b)$ and multiplying the result by $2$.

In this case, we have 28 possible pairs of line-ups where the agent in both slots 1 ($\pi_u$) will perform the same sequence of actions (always Up) independently of the line-up. So:

\begin{equation*}
AD_1(\pi_u,\Pi_o,w_{\dot{L}},\mu) = 2 \frac{8}{7} \frac{1}{8} \frac{1}{8} \left\{28 \times 0\right\} = 0
\end{equation*}

For slot 2, the agent in both slots 2 ($\pi_u$) will also perform the same sequence of actions (always Up) independently of the line-up. So:

\begin{equation*}
\begin{aligned}
AD_2(\pi_u,\Pi_o,w_{\dot{L}},\mu)	& = \eta_{\dot{L}^2} \sum_{\dot{u},\dot{v} \in \dot{L}^{N(\mu)}_{-2}(\Pi_o) | \dot{u} \neq \dot{v}} w_{\dot{L}}(\dot{u}) w_{\dot{L}}(\dot{v}) \Delta_S(\breve{A}_2(\mu[\instantiation{u}{2}{\pi_u}]), \breve{A}_2(\mu[\instantiation{v}{2}{\pi_u}])) =\\
									& = 2 \frac{8}{7} \frac{1}{8} \frac{1}{8} \{\Delta_S(\breve{A}_2(\mu[{\pi_d}_1,\pi_u,{\pi_d}_1,{\pi_d}_1]), \breve{A}_2(\mu[{\pi_d}_1,\pi_u,{\pi_d}_1,{\pi_d}_2])) +\\
									& + \Delta_S(\breve{A}_2(\mu[{\pi_d}_1,\pi_u,{\pi_d}_1,{\pi_d}_1]), \breve{A}_2(\mu[{\pi_d}_1,\pi_u,{\pi_d}_2,{\pi_d}_1])) +\\
									& \ \ \ \ \ \ \ \ \ \ \ \ \ \ \ \ \ \ \ \ \ \ \ \ \ \ \ \ \ \ \ \ \ \ \ \ \ \ \ \ \vdots\\
									& + \Delta_S(\breve{A}_2(\mu[{\pi_d}_2,\pi_u,{\pi_d}_2,{\pi_d}_1]), \breve{A}_2(\mu[{\pi_d}_2,\pi_u,{\pi_d}_2,{\pi_d}_2]))\} =\\
									& = 2 \frac{8}{7} \frac{1}{8} \frac{1}{8} \left\{28 \times 0\right\} = 0
\end{aligned}
\end{equation*}

For slot 3, the agent in both slots 3 ($\pi_u$) will also perform the same sequence of actions (always Up) independently of the line-up. So:

\begin{equation*}
\begin{aligned}
AD_3(\pi_u,\Pi_o,w_{\dot{L}},\mu)	& = \eta_{\dot{L}^2} \sum_{\dot{u},\dot{v} \in \dot{L}^{N(\mu)}_{-3}(\Pi_o) | \dot{u} \neq \dot{v}} w_{\dot{L}}(\dot{u}) w_{\dot{L}}(\dot{v}) \Delta_S(\breve{A}_3(\mu[\instantiation{u}{3}{\pi_u}]), \breve{A}_3(\mu[\instantiation{v}{3}{\pi_u}])) =\\
									& = 2 \frac{8}{7} \frac{1}{8} \frac{1}{8} \{\Delta_S(\breve{A}_3(\mu[{\pi_d}_1,{\pi_d}_1,\pi_u,{\pi_d}_1]), \breve{A}_3(\mu[{\pi_d}_1,{\pi_d}_1,\pi_u,{\pi_d}_2])) +\\
									& + \Delta_S(\breve{A}_3(\mu[{\pi_d}_1,{\pi_d}_1,\pi_u,{\pi_d}_1]), \breve{A}_3(\mu[{\pi_d}_1,{\pi_d}_2,\pi_u,{\pi_d}_1])) +\\
									& \ \ \ \ \ \ \ \ \ \ \ \ \ \ \ \ \ \ \ \ \ \ \ \ \ \ \ \ \ \ \ \ \ \ \ \ \ \ \ \ \vdots\\
									& + \Delta_S(\breve{A}_3(\mu[{\pi_d}_2,{\pi_d}_2,\pi_u,{\pi_d}_1]), \breve{A}_3(\mu[{\pi_d}_2,{\pi_d}_2,\pi_u,{\pi_d}_2]))\} =\\
									& = 2 \frac{8}{7} \frac{1}{8} \frac{1}{8} \left\{28 \times 0\right\} = 0
\end{aligned}
\end{equation*}

And for slot 4, the agent in both slots 4 ($\pi_u$) will also perform the same sequence of actions (always Up) independently of the line-up. So:

\begin{equation*}
\begin{aligned}
AD_4(\pi_u,\Pi_o,w_{\dot{L}},\mu)	& = \eta_{\dot{L}^2} \sum_{\dot{u},\dot{v} \in \dot{L}^{N(\mu)}_{-4}(\Pi_o) | \dot{u} \neq \dot{v}} w_{\dot{L}}(\dot{u}) w_{\dot{L}}(\dot{v}) \Delta_S(\breve{A}_4(\mu[\instantiation{u}{4}{\pi_u}]), \breve{A}_4(\mu[\instantiation{v}{4}{\pi_u}])) =\\
									& = 2 \frac{8}{7} \frac{1}{8} \frac{1}{8} \{\Delta_S(\breve{A}_4(\mu[{\pi_d}_1,{\pi_d}_1,{\pi_d}_1,\pi_u]), \breve{A}_4(\mu[{\pi_d}_1,{\pi_d}_1,{\pi_d}_2,\pi_u])) +\\
									& + \Delta_S(\breve{A}_4(\mu[{\pi_d}_1,{\pi_d}_1,{\pi_d}_1,\pi_u]), \breve{A}_4(\mu[{\pi_d}_1,{\pi_d}_2,{\pi_d}_1,\pi_u])) +\\
									& \ \ \ \ \ \ \ \ \ \ \ \ \ \ \ \ \ \ \ \ \ \ \ \ \ \ \ \ \ \ \ \ \ \ \ \ \ \ \ \ \vdots\\
									& + \Delta_S(\breve{A}_4(\mu[{\pi_d}_2,{\pi_d}_2,{\pi_d}_1,\pi_u]), \breve{A}_4(\mu[{\pi_d}_2,{\pi_d}_2,{\pi_d}_2,\pi_u]))\} =\\
									& = 2 \frac{8}{7} \frac{1}{8} \frac{1}{8} \left\{28 \times 0\right\} = 0
\end{aligned}
\end{equation*}

And finally, we weight over the slots:

\begin{equation*}
\begin{aligned}
AD(\Pi_e,w_{\Pi_e},\Pi_o,w_{\dot{L}},\mu,w_S)	& =	\sum_{i = 1}^{N(\mu)} w_S(i,\mu) AD_i(\Pi_e,w_{\Pi_e},\Pi_o,w_{\dot{L}},\mu) =\\
												& =	\sum_{i = 1}^{N(\mu)} w_S(i,\mu) AD_i(\pi_u,\Pi_o,w_{\dot{L}},\mu) =\\
												& = \frac{1}{4} \{AD_1(\pi_u,\Pi_o,w_{\dot{L}},\mu) + AD_2(\pi_u,\Pi_o,w_{\dot{L}},\mu) +\\
												& + AD_3(\pi_u,\Pi_o,w_{\dot{L}},\mu) + AD_4(\pi_u,\Pi_o,w_{\dot{L}},\mu)\} =\\
												& = \frac{1}{4} \left\{0 + 0 + 0 + 0\right\} = 0
\end{aligned}
\end{equation*}

Since $0$ is the lowest possible value for the action dependency property, therefore predator-prey has $General_{min} = 0$ for this property.
\end{proof}
\end{proposition}

\begin{proposition}
\label{prop:predator-prey_AD_general_max}
$General_{max}$ for the action dependency (AD) property is equal to $1$ for the predator-prey environment.

\begin{proof}
To find $General_{max}$ (equation \ref{eq:general_max}), we need to find a trio $\left\langle\Pi_e,w_{\Pi_e},\Pi_o\right\rangle$ which maximises the property as much as possible. We can have this situation by selecting $\Pi_e = \{\pi_m\}$ with $w_{\Pi_e}(\pi_m) = 1$ and $\Pi_o = \{\pi_u, \pi_d\}$ (a $\pi_m$ agent first acts randomly and then always mimics sequentially the other agents' last action, a $\pi_u$ agent always performs Up and a $\pi_d$ agent always performs Down).

Following definition \ref{def:AD}, we obtain the AD value for this $\left\langle\Pi_e,w_{\Pi_e},\Pi_o\right\rangle$. Since the environment is not symmetric, we need to calculate this property for every slot. Following definition \ref{def:AD_set}, we could calculate its AD value for each slot but, since $\Pi_e$ has only one agent and its weight is equal to $1$, it is equivalent to use directly definition \ref{def:AD_agent}. We start with slot 1:

\begin{equation*}
\begin{aligned}
AD_1(\pi_m,\Pi_o,w_{\dot{L}},\mu)	& = \eta_{\dot{L}^2} \sum_{\dot{u},\dot{v} \in \dot{L}^{N(\mu)}_{-1}(\Pi_o) | \dot{u} \neq \dot{v}} w_{\dot{L}}(\dot{u}) w_{\dot{L}}(\dot{v}) \Delta_S(\breve{A}_1(\mu[\instantiation{u}{1}{\pi_m}]), \breve{A}_1(\mu[\instantiation{v}{1}{\pi_m}])) =\\
									& = 2 \frac{8}{7} \frac{1}{8} \frac{1}{8} \{\Delta_S(\breve{A}_1(\mu[\pi_m,\pi_u,\pi_u,\pi_u]), \breve{A}_1(\mu[\pi_m,\pi_u,\pi_u,\pi_d])) +\\
									& + \Delta_S(\breve{A}_1(\mu[\pi_m,\pi_u,\pi_u,\pi_u]), \breve{A}_1(\mu[\pi_m,\pi_u,\pi_d,\pi_u])) +\\
									& \ \ \ \ \ \ \ \ \ \ \ \ \ \ \ \ \ \ \ \ \ \ \ \ \ \ \ \ \ \ \ \ \ \ \ \ \ \vdots\\
									& + \Delta_S(\breve{A}_1(\mu[\pi_m,\pi_d,\pi_d,\pi_u]), \breve{A}_1(\mu[\pi_m,\pi_d,\pi_d,\pi_d]))\}
\end{aligned}
\end{equation*}

\noindent Note that we avoided to calculate both $\Delta_S(a,b)$ and $\Delta_S(b,a)$ since they provide the same result, by calculating only $\Delta_S(a,b)$ and multiplying the result by $2$.

In this case, we have 28 possible pairs of line-ups where the agent in both slots 1 ($\pi_m$) will perform different sequences of actions depending on the line-up. So:

\begin{equation*}
AD_1(\pi_m,\Pi_o,w_{\dot{L}},\mu) = 2 \frac{8}{7} \frac{1}{8} \frac{1}{8} \left\{28 \times 1\right\} = 1
\end{equation*}

For slot 2, the agent in both slots 2 ($\pi_m$) will also perform different sequences of actions depending on the line-up. So:

\begin{equation*}
\begin{aligned}
AD_2(\pi_m,\Pi_o,w_{\dot{L}},\mu)	& = \eta_{\dot{L}^2} \sum_{\dot{u},\dot{v} \in \dot{L}^{N(\mu)}_{-2}(\Pi_o) | \dot{u} \neq \dot{v}} w_{\dot{L}}(\dot{u}) w_{\dot{L}}(\dot{v}) \Delta_S(\breve{A}_2(\mu[\instantiation{u}{2}{\pi_m}]), \breve{A}_2(\mu[\instantiation{v}{2}{\pi_m}])) =\\
									& = 2 \frac{8}{7} \frac{1}{8} \frac{1}{8} \{\Delta_S(\breve{A}_2(\mu[\pi_u,\pi_m,\pi_u,\pi_u]), \breve{A}_2(\mu[\pi_u,\pi_m,\pi_u,\pi_d])) +\\
									& + \Delta_S(\breve{A}_2(\mu[\pi_u,\pi_m,\pi_u,\pi_u]), \breve{A}_2(\mu[\pi_u,\pi_m,\pi_d,\pi_u])) +\\
									& \ \ \ \ \ \ \ \ \ \ \ \ \ \ \ \ \ \ \ \ \ \ \ \ \ \ \ \ \ \ \ \ \ \ \ \ \ \vdots\\
									& + \Delta_S(\breve{A}_2(\mu[\pi_d,\pi_m,\pi_d,\pi_u]), \breve{A}_2(\mu[\pi_d,\pi_m,\pi_d,\pi_d]))\} =\\
									& = 2 \frac{8}{7} \frac{1}{8} \frac{1}{8} \left\{28 \times 1\right\} = 1
\end{aligned}
\end{equation*}

For slot 3, the agent in both slots 3 ($\pi_m$) will also perform different sequences of actions depending on the line-up. So:

\begin{equation*}
\begin{aligned}
AD_3(\pi_m,\Pi_o,w_{\dot{L}},\mu)	& = \eta_{\dot{L}^2} \sum_{\dot{u},\dot{v} \in \dot{L}^{N(\mu)}_{-3}(\Pi_o) | \dot{u} \neq \dot{v}} w_{\dot{L}}(\dot{u}) w_{\dot{L}}(\dot{v}) \Delta_S(\breve{A}_3(\mu[\instantiation{u}{3}{\pi_m}]), \breve{A}_3(\mu[\instantiation{v}{3}{\pi_m}])) =\\
									& = 2 \frac{8}{7} \frac{1}{8} \frac{1}{8} \{\Delta_S(\breve{A}_3(\mu[\pi_u,\pi_u,\pi_m,\pi_u]), \breve{A}_3(\mu[\pi_u,\pi_u,\pi_m,\pi_d])) +\\
									& + \Delta_S(\breve{A}_3(\mu[\pi_u,\pi_u,\pi_m,\pi_u]), \breve{A}_3(\mu[\pi_u,\pi_d,\pi_m,\pi_u])) +\\
									& \ \ \ \ \ \ \ \ \ \ \ \ \ \ \ \ \ \ \ \ \ \ \ \ \ \ \ \ \ \ \ \ \ \ \ \ \ \vdots\\
									& + \Delta_S(\breve{A}_3(\mu[\pi_d,\pi_d,\pi_m,\pi_u]), \breve{A}_3(\mu[\pi_d,\pi_d,\pi_m,\pi_d]))\} =\\
									& = 2 \frac{8}{7} \frac{1}{8} \frac{1}{8} \left\{28 \times 1\right\} = 1
\end{aligned}
\end{equation*}

And for slot 4, the agent in both slots 4 ($\pi_m$) will also perform different sequences of actions depending on the line-up. So:

\begin{equation*}
\begin{aligned}
AD_4(\pi_m,\Pi_o,w_{\dot{L}},\mu)	& = \eta_{\dot{L}^2} \sum_{\dot{u},\dot{v} \in \dot{L}^{N(\mu)}_{-4}(\Pi_o) | \dot{u} \neq \dot{v}} w_{\dot{L}}(\dot{u}) w_{\dot{L}}(\dot{v}) \Delta_S(\breve{A}_4(\mu[\instantiation{u}{4}{\pi_m}]), \breve{A}_4(\mu[\instantiation{v}{4}{\pi_m}])) =\\
									& = 2 \frac{8}{7} \frac{1}{8} \frac{1}{8} \{\Delta_S(\breve{A}_4(\mu[\pi_u,\pi_u,\pi_u,\pi_m]), \breve{A}_4(\mu[\pi_u,\pi_u,\pi_d,\pi_m])) +\\
									& + \Delta_S(\breve{A}_4(\mu[\pi_u,\pi_u,\pi_u,\pi_m]), \breve{A}_4(\mu[\pi_u,\pi_d,\pi_u,\pi_m])) +\\
									& \ \ \ \ \ \ \ \ \ \ \ \ \ \ \ \ \ \ \ \ \ \ \ \ \ \ \ \ \ \ \ \ \ \ \ \ \ \vdots\\
									& + \Delta_S(\breve{A}_4(\mu[\pi_d,\pi_d,\pi_u,\pi_m]), \breve{A}_4(\mu[\pi_d,\pi_d,\pi_d,\pi_m]))\} =\\
									& = 2 \frac{8}{7} \frac{1}{8} \frac{1}{8} \left\{28 \times 1\right\} = 1
\end{aligned}
\end{equation*}

And finally, we weight over the slots:

\begin{equation*}
\begin{aligned}
AD(\Pi_e,w_{\Pi_e},\Pi_o,w_{\dot{L}},\mu,w_S)	& =	\sum_{i = 1}^{N(\mu)} w_S(i,\mu) AD_i(\Pi_e,w_{\Pi_e},\Pi_o,w_{\dot{L}},\mu) =\\
												& =	\sum_{i = 1}^{N(\mu)} w_S(i,\mu) AD_i(\pi_m,\Pi_o,w_{\dot{L}},\mu) =\\
												& = \frac{1}{4} \{AD_1(\pi_m,\Pi_o,w_{\dot{L}},\mu) + AD_2(\pi_m,\Pi_o,w_{\dot{L}},\mu) +\\
												& + AD_3(\pi_m,\Pi_o,w_{\dot{L}},\mu) + AD_4(\pi_m,\Pi_o,w_{\dot{L}},\mu)\} =\\
												& = \frac{1}{4} \left\{1 + 1 + 1 + 1\right\} = 1
\end{aligned}
\end{equation*}

Since $1$ is the highest possible value for the action dependency property, therefore predator-prey has $General_{max} = 1$ for this property.
\end{proof}
\end{proposition}

\begin{proposition}
\label{prop:predator-prey_AD_left_max}
$Left_{max}$ for the action dependency (AD) property is equal to $0$ for the predator-prey environment.

\begin{proof}
To find $Left_{max}$ (equation \ref{eq:left_max}), we need to find a pair $\left\langle\Pi_e,w_{\Pi_e}\right\rangle$ which maximises the property as much as possible while $\Pi_o$ minimises it. Using $\Pi_o = \{{\pi_d}_1,{\pi_d}_2\}$ (a $\pi_d$ agent always performs Down) we find this situation no matter which pair $\left\langle\Pi_e,w_{\Pi_e}\right\rangle$ we use.

Following definition \ref{def:AD}, we obtain the AD value for this $\left\langle\Pi_e,w_{\Pi_e},\Pi_o\right\rangle$ (where $\Pi_e$ and $w_{\Pi_e}$ are instantiated with any permitted values). Since the environment is not symmetric, we need to calculate this property for every slot. Following definition \ref{def:AD_set}, we can calculate its AD value for each slot. We start with slot 1:

\begin{equation*}
AD_1(\Pi_e,w_{\Pi_e},\Pi_o,w_{\dot{L}},\mu) = \sum_{\pi \in \Pi_e} w_{\Pi_e}(\pi) AD_1(\pi,\Pi_o,w_{\dot{L}},\mu)
\end{equation*}

We do not know which $\Pi_e$ we have, but we know that we will need to evaluate $AD_1(\pi,\Pi_o,w_{\dot{L}},\mu)$ for all evaluated agent $\pi \in \Pi_e$. We follow definition \ref{def:AD_agent} to calculate this value for a figurative evaluated agent $\pi$ from $\Pi_e$:

\begin{equation*}
\begin{aligned}
AD_1(\pi,\Pi_o,w_{\dot{L}},\mu)	& = \eta_{\dot{L}^2} \sum_{\dot{u},\dot{v} \in \dot{L}^{N(\mu)}_{-1}(\Pi_o) | \dot{u} \neq \dot{v}} w_{\dot{L}}(\dot{u}) w_{\dot{L}}(\dot{v}) \Delta_S(\breve{A}_1(\mu[\instantiation{u}{1}{\pi}]), \breve{A}_1(\mu[\instantiation{v}{1}{\pi}])) =\\
								& = 2 \frac{8}{7} \frac{1}{8} \frac{1}{8} \{\Delta_S(\breve{A}_1(\mu[\pi,{\pi_d}_1,{\pi_d}_1,{\pi_d}_1]), \breve{A}_1(\mu[\pi,{\pi_d}_1,{\pi_d}_1,{\pi_d}_2])) +\\
								& + \Delta_S(\breve{A}_1(\mu[\pi,{\pi_d}_1,{\pi_d}_1,{\pi_d}_1]), \breve{A}_1(\mu[\pi,{\pi_d}_1,{\pi_d}_2,{\pi_d}_1])) +\\
								& \ \ \ \ \ \ \ \ \ \ \ \ \ \ \ \ \ \ \ \ \ \ \ \ \ \ \ \ \ \ \ \ \ \ \ \ \ \ \ \vdots\\
								& + \Delta_S(\breve{A}_1(\mu[\pi,{\pi_d}_2,{\pi_d}_2,{\pi_d}_1]), \breve{A}_1(\mu[\pi,{\pi_d}_2,{\pi_d}_2,{\pi_d}_2]))\}
\end{aligned}
\end{equation*}

\noindent Note that we avoided to calculate both $\Delta_S(a,b)$ and $\Delta_S(b,a)$ since they provide the same result, by calculating only $\Delta_S(a,b)$ and multiplying the result by $2$.

A $\pi_d$ agent will always perform Down, so for the 28 possible pairs of line-ups we obtain a situation where the agent in both slots 1 (any $\pi$) will be able to differentiate with which agents is interacting, so it will not be able to change its distribution of action sequences depending on the line-up. So:

\begin{equation*}
AD_1(\pi,\Pi_o,w_{\dot{L}},\mu) = 2 \frac{8}{7} \frac{1}{8} \frac{1}{8} \left\{28 \times 0\right\} = 0
\end{equation*}

Therefore, no matter which agents are in $\Pi_e$ and their weights $w_{\Pi_e}$ we obtain:

\begin{equation*}
AD_1(\Pi_e,w_{\Pi_e},\Pi_o,w_{\dot{L}},\mu) = 0
\end{equation*}

For slot 2:

\begin{equation*}
AD_2(\Pi_e,w_{\Pi_e},\Pi_o,w_{\dot{L}},\mu) = \sum_{\pi \in \Pi_e} w_{\Pi_e}(\pi) AD_2(\pi,\Pi_o,w_{\dot{L}},\mu)
\end{equation*}

We do not know which $\Pi_e$ we have, but we know that we will need to evaluate $AD_2(\pi,\Pi_o,w_{\dot{L}},\mu)$ for all evaluated agent $\pi \in \Pi_e$. We follow definition \ref{def:AD_agent} to calculate this value for a figurative evaluated agent $\pi$ from $\Pi_e$:

\begin{equation*}
\begin{aligned}
AD_2(\pi,\Pi_o,w_{\dot{L}},\mu)	& = \eta_{\dot{L}^2} \sum_{\dot{u},\dot{v} \in \dot{L}^{N(\mu)}_{-2}(\Pi_o) | \dot{u} \neq \dot{v}} w_{\dot{L}}(\dot{u}) w_{\dot{L}}(\dot{v}) \Delta_S(\breve{A}_2(\mu[\instantiation{u}{2}{\pi}]), \breve{A}_2(\mu[\instantiation{v}{2}{\pi}])) =\\
								& = 2 \frac{8}{7} \frac{1}{8} \frac{1}{8} \{\Delta_S(\breve{A}_2(\mu[{\pi_d}_1,\pi,{\pi_d}_1,{\pi_d}_1]), \breve{A}_2(\mu[{\pi_d}_1,\pi,{\pi_d}_1,{\pi_d}_2])) +\\
								& + \Delta_S(\breve{A}_2(\mu[{\pi_d}_1,\pi,{\pi_d}_1,{\pi_d}_1]), \breve{A}_2(\mu[{\pi_d}_1,\pi,{\pi_d}_2,{\pi_d}_1])) +\\
								& \ \ \ \ \ \ \ \ \ \ \ \ \ \ \ \ \ \ \ \ \ \ \ \ \ \ \ \ \ \ \ \ \ \ \ \ \ \ \ \vdots\\
								& + \Delta_S(\breve{A}_2(\mu[{\pi_d}_2,\pi,{\pi_d}_2,{\pi_d}_1]), \breve{A}_2(\mu[{\pi_d}_2,\pi,{\pi_d}_2,{\pi_d}_2]))\}
\end{aligned}
\end{equation*}

A $\pi_d$ agent will always perform Down, so for the 28 possible pairs of line-ups we obtain a situation where the agent in both slots 2 (any $\pi$) will be able to differentiate with which agents is interacting, so it will not be able to change its distribution of action sequences depending on the line-up. So:

\begin{equation*}
AD_2(\pi,\Pi_o,w_{\dot{L}},\mu) = 2 \frac{8}{7} \frac{1}{8} \frac{1}{8} \left\{28 \times 0\right\} = 0
\end{equation*}

Therefore, no matter which agents are in $\Pi_e$ and their weights $w_{\Pi_e}$ we obtain:

\begin{equation*}
AD_2(\Pi_e,w_{\Pi_e},\Pi_o,w_{\dot{L}},\mu) = 0
\end{equation*}

For slot 3:

\begin{equation*}
AD_3(\Pi_e,w_{\Pi_e},\Pi_o,w_{\dot{L}},\mu) = \sum_{\pi \in \Pi_e} w_{\Pi_e}(\pi) AD_3(\pi,\Pi_o,w_{\dot{L}},\mu)
\end{equation*}

We do not know which $\Pi_e$ we have, but we know that we will need to evaluate $AD_3(\pi,\Pi_o,w_{\dot{L}},\mu)$ for all evaluated agent $\pi \in \Pi_e$. We follow definition \ref{def:AD_agent} to calculate this value for a figurative evaluated agent $\pi$ from $\Pi_e$:

\begin{equation*}
\begin{aligned}
AD_3(\pi,\Pi_o,w_{\dot{L}},\mu)	& = \eta_{\dot{L}^2} \sum_{\dot{u},\dot{v} \in \dot{L}^{N(\mu)}_{-3}(\Pi_o) | \dot{u} \neq \dot{v}} w_{\dot{L}}(\dot{u}) w_{\dot{L}}(\dot{v}) \Delta_S(\breve{A}_3(\mu[\instantiation{u}{3}{\pi}]), \breve{A}_3(\mu[\instantiation{v}{3}{\pi}])) =\\
								& = 2 \frac{8}{7} \frac{1}{8} \frac{1}{8} \{\Delta_S(\breve{A}_3(\mu[{\pi_d}_1,{\pi_d}_1,\pi,{\pi_d}_1]), \breve{A}_3(\mu[{\pi_d}_1,{\pi_d}_1,\pi,{\pi_d}_2])) +\\
								& + \Delta_S(\breve{A}_3(\mu[{\pi_d}_1,{\pi_d}_1,\pi,{\pi_d}_1]), \breve{A}_3(\mu[{\pi_d}_1,{\pi_d}_2,\pi,{\pi_d}_1])) +\\
								& \ \ \ \ \ \ \ \ \ \ \ \ \ \ \ \ \ \ \ \ \ \ \ \ \ \ \ \ \ \ \ \ \ \ \ \ \ \ \ \vdots\\
								& + \Delta_S(\breve{A}_3(\mu[{\pi_d}_2,{\pi_d}_2,\pi,{\pi_d}_1]), \breve{A}_3(\mu[{\pi_d}_2,{\pi_d}_2,\pi,{\pi_d}_2]))\}
\end{aligned}
\end{equation*}

A $\pi_d$ agent will always perform Down, so for the 28 possible pairs of line-ups we obtain a situation where the agent in both slots 3 (any $\pi$) will be able to differentiate with which agents is interacting, so it will not be able to change its distribution of action sequences depending on the line-up. So:

\begin{equation*}
AD_3(\pi,\Pi_o,w_{\dot{L}},\mu) = 2 \frac{8}{7} \frac{1}{8} \frac{1}{8} \left\{28 \times 0\right\} = 0
\end{equation*}

Therefore, no matter which agents are in $\Pi_e$ and their weights $w_{\Pi_e}$ we obtain:

\begin{equation*}
AD_3(\Pi_e,w_{\Pi_e},\Pi_o,w_{\dot{L}},\mu) = 0
\end{equation*}

And for slot 4:

\begin{equation*}
AD_4(\Pi_e,w_{\Pi_e},\Pi_o,w_{\dot{L}},\mu) = \sum_{\pi \in \Pi_e} w_{\Pi_e}(\pi) AD_4(\pi,\Pi_o,w_{\dot{L}},\mu)
\end{equation*}

We do not know which $\Pi_e$ we have, but we know that we will need to evaluate $AD_4(\pi,\Pi_o,w_{\dot{L}},\mu)$ for all evaluated agent $\pi \in \Pi_e$. We follow definition \ref{def:AD_agent} to calculate this value for a figurative evaluated agent $\pi$ from $\Pi_e$:

\begin{equation*}
\begin{aligned}
AD_4(\pi,\Pi_o,w_{\dot{L}},\mu)	& = \eta_{\dot{L}^2} \sum_{\dot{u},\dot{v} \in \dot{L}^{N(\mu)}_{-4}(\Pi_o) | \dot{u} \neq \dot{v}} w_{\dot{L}}(\dot{u}) w_{\dot{L}}(\dot{v}) \Delta_S(\breve{A}_4(\mu[\instantiation{u}{4}{\pi}]), \breve{A}_4(\mu[\instantiation{v}{4}{\pi}])) =\\
								& = 2 \frac{8}{7} \frac{1}{8} \frac{1}{8} \{\Delta_S(\breve{A}_4(\mu[{\pi_d}_1,{\pi_d}_1,{\pi_d}_1,\pi]), \breve{A}_4(\mu[{\pi_d}_1,{\pi_d}_1,{\pi_d}_2,\pi])) +\\
								& + \Delta_S(\breve{A}_4(\mu[{\pi_d}_1,{\pi_d}_1,{\pi_d}_1,\pi]), \breve{A}_4(\mu[{\pi_d}_1,{\pi_d}_2,{\pi_d}_1,\pi])) +\\
								& \ \ \ \ \ \ \ \ \ \ \ \ \ \ \ \ \ \ \ \ \ \ \ \ \ \ \ \ \ \ \ \ \ \ \ \ \ \ \ \vdots\\
								& + \Delta_S(\breve{A}_4(\mu[{\pi_d}_2,{\pi_d}_2,{\pi_d}_1,\pi]), \breve{A}_4(\mu[{\pi_d}_2,{\pi_d}_2,{\pi_d}_2,\pi]))\}
\end{aligned}
\end{equation*}

A $\pi_d$ agent will always perform Down, so for the 28 possible pairs of line-ups we obtain a situation where the agent in both slots 4 (any $\pi$) will be able to differentiate with which agents is interacting, so it will not be able to change its distribution of action sequences depending on the line-up. So:

\begin{equation*}
AD_4(\pi,\Pi_o,w_{\dot{L}},\mu) = 2 \frac{8}{7} \frac{1}{8} \frac{1}{8} \left\{28 \times 0\right\} = 0
\end{equation*}

Therefore, no matter which agents are in $\Pi_e$ and their weights $w_{\Pi_e}$ we obtain:

\begin{equation*}
AD_4(\Pi_e,w_{\Pi_e},\Pi_o,w_{\dot{L}},\mu) = 0
\end{equation*}

And finally, we weight over the slots:

\begin{equation*}
\begin{aligned}
AD(\Pi_e,w_{\Pi_e},\Pi_o,w_{\dot{L}},\mu,w_S)	& =	\sum_{i = 1}^{N(\mu)} w_S(i,\mu) AD_i(\Pi_e,w_{\Pi_e},\Pi_o,w_{\dot{L}},\mu) =\\
												& = \frac{1}{4} \{AD_1(\Pi_e,w_{\Pi_e},\Pi_o,w_{\dot{L}},\mu) + AD_2(\Pi_e,w_{\Pi_e},\Pi_o,w_{\dot{L}},\mu) +\\
												& + AD_3(\Pi_e,w_{\Pi_e},\Pi_o,w_{\dot{L}},\mu) + AD_4(\Pi_e,w_{\Pi_e},\Pi_o,w_{\dot{L}},\mu)\} =\\
												& = \frac{1}{4} \left\{0 + 0 + 0 + 0\right\} = 0
\end{aligned}
\end{equation*}

So, for every pair $\left\langle\Pi_e,w_{\Pi_e}\right\rangle$ we obtain the same result:

\begin{equation*}
\forall \Pi_e,w_{\Pi_e} : AD(\Pi_e,w_{\Pi_e},\Pi_o,w_{\dot{L}},\mu,w_S) = 0
\end{equation*}

Therefore, predator-prey has $Left_{max} = 0$ for this property.
\end{proof}
\end{proposition}

\begin{proposition}
\label{prop:predator-prey_AD_right_min}
$Right_{min}$ for the action dependency (AD) property is equal to $0$ for the predator-prey environment.

\begin{proof}
To find $Right_{min}$ (equation \ref{eq:right_min}), we need to find a pair $\left\langle\Pi_e,w_{\Pi_e}\right\rangle$ which minimises the property as much as possible while $\Pi_o$ maximises it. Using $\Pi_e = \{\pi_u\}$ with $w_{\Pi_e}(\pi_u) = 1$ (a $\pi_u$ agent always performs Up) we find this situation no matter which $\Pi_o$ we use.

Following definition \ref{def:AD}, we obtain the AD value for this $\left\langle\Pi_e,w_{\Pi_e},\Pi_o\right\rangle$ (where $\Pi_o$ is instantiated with any permitted value). Since the environment is not symmetric, we need to calculate this property for every slot. Following definition \ref{def:AD_set}, we could calculate its AD value for each slot but, since $\Pi_e$ has only one agent and its weight is equal to $1$, it is equivalent to use directly definition \ref{def:AD_agent}. We start with slot 1:

\begin{equation*}
AD_1(\pi_u,\Pi_o,w_{\dot{L}},\mu) = \eta_{\dot{L}^2} \sum_{\dot{u},\dot{v} \in \dot{L}^{N(\mu)}_{-1}(\Pi_o) | \dot{u} \neq \dot{v}} w_{\dot{L}}(\dot{u}) w_{\dot{L}}(\dot{v}) \Delta_S(\breve{A}_1(\mu[\instantiation{u}{1}{\pi_u}]), \breve{A}_1(\mu[\instantiation{v}{1}{\pi_u}]))
\end{equation*}

We do not know which $\Pi_o$ we have, but we know that we will need to obtain two different line-up patterns $\dot{u}$ and $\dot{v}$ from $\dot{L}^{N(\mu)}_{-1}(\Pi_o)$ to calculate $\Delta_S(\breve{A}_1(\mu[\instantiation{u}{1}{\pi_u}]), \breve{A}_1(\mu[\instantiation{v}{1}{\pi_u}]))$. We calculate this value for two figurative line-up patterns $\dot{u} = (*,\pi_1,\pi_2,\pi_3)$ and $\dot{v} = (*,\pi_4,\pi_5,\pi_6)$ from $\dot{L}^{N(\mu)}_{-1}(\Pi_o)$:

\begin{equation*}
\Delta_S(\breve{A}_1(\mu[\instantiation{u}{1}{\pi_u}]), \breve{A}_1(\mu[\instantiation{v}{1}{\pi_u}])) = \Delta_S(\breve{A}_1(\mu[\pi_u,\pi_1,\pi_2,\pi_3]), \breve{A}_1(\mu[\pi_u,\pi_4,\pi_5,\pi_6]))
\end{equation*}

Here, the agent in both slots 1 ($\pi_u$) will perform the same sequence of actions (always Up) independently of the line-up. So no matter which agents are in $\Pi_o$ we obtain:

\begin{equation*}
AD_1(\pi_u,\Pi_o,w_{\dot{L}},\mu) = 0
\end{equation*}

For slot 2:

\begin{equation*}
AD_2(\pi_u,\Pi_o,w_{\dot{L}},\mu) = \eta_{\dot{L}^2} \sum_{\dot{u},\dot{v} \in \dot{L}^{N(\mu)}_{-2}(\Pi_o) | \dot{u} \neq \dot{v}} w_{\dot{L}}(\dot{u}) w_{\dot{L}}(\dot{v}) \Delta_S(\breve{A}_2(\mu[\instantiation{u}{2}{\pi_u}]), \breve{A}_2(\mu[\instantiation{v}{2}{\pi_u}]))
\end{equation*}

We do not know which $\Pi_o$ we have, but we know that we will need to obtain two different line-up patterns $\dot{u}$ and $\dot{v}$ from $\dot{L}^{N(\mu)}_{-2}(\Pi_o)$ to calculate $\Delta_S(\breve{A}_2(\mu[\instantiation{u}{2}{\pi_u}]), \breve{A}_2(\mu[\instantiation{v}{2}{\pi_u}]))$. We calculate this value for two figurative line-up patterns $\dot{u} = (\pi_1,*,\pi_2,\pi_3)$ and $\dot{v} = (\pi_4,*,\pi_5,\pi_6)$ from $\dot{L}^{N(\mu)}_{-2}(\Pi_o)$:

\begin{equation*}
\Delta_S(\breve{A}_2(\mu[\instantiation{u}{2}{\pi_u}]), \breve{A}_2(\mu[\instantiation{v}{2}{\pi_u}])) = \Delta_S(\breve{A}_2(\mu[\pi_1,\pi_u,\pi_2,\pi_3]), \breve{A}_2(\mu[\pi_4,\pi_u,\pi_5,\pi_6]))
\end{equation*}

Here, the agent in both slots 2 ($\pi_u$) will perform the same sequence of actions (always Up) independently of the line-up. So no matter which agents are in $\Pi_o$ we obtain:

\begin{equation*}
AD_2(\pi_u,\Pi_o,w_{\dot{L}},\mu) = 0
\end{equation*}

For slot 3:

\begin{equation*}
AD_3(\pi_u,\Pi_o,w_{\dot{L}},\mu) = \eta_{\dot{L}^2} \sum_{\dot{u},\dot{v} \in \dot{L}^{N(\mu)}_{-3}(\Pi_o) | \dot{u} \neq \dot{v}} w_{\dot{L}}(\dot{u}) w_{\dot{L}}(\dot{v}) \Delta_S(\breve{A}_3(\mu[\instantiation{u}{3}{\pi_u}]), \breve{A}_3(\mu[\instantiation{v}{3}{\pi_u}]))
\end{equation*}

We do not know which $\Pi_o$ we have, but we know that we will need to obtain two different line-up patterns $\dot{u}$ and $\dot{v}$ from $\dot{L}^{N(\mu)}_{-3}(\Pi_o)$ to calculate $\Delta_S(\breve{A}_3(\mu[\instantiation{u}{3}{\pi_u}]), \breve{A}_3(\mu[\instantiation{v}{3}{\pi_u}]))$. We calculate this value for two figurative line-up patterns $\dot{u} = (\pi_1,\pi_2,*,\pi_3)$ and $\dot{v} = (\pi_4,\pi_5,*,\pi_6)$ from $\dot{L}^{N(\mu)}_{-3}(\Pi_o)$:

\begin{equation*}
\Delta_S(\breve{A}_3(\mu[\instantiation{u}{3}{\pi_u}]), \breve{A}_3(\mu[\instantiation{v}{3}{\pi_u}])) = \Delta_S(\breve{A}_3(\mu[\pi_1,\pi_2,\pi_u,\pi_3]), \breve{A}_3(\mu[\pi_4,\pi_5,\pi_u,\pi_6]))
\end{equation*}

Here, the agent in both slots 3 ($\pi_u$) will perform the same sequence of actions (always Up) independently of the line-up. So no matter which agents are in $\Pi_o$ we obtain:

\begin{equation*}
AD_3(\pi_u,\Pi_o,w_{\dot{L}},\mu) = 0
\end{equation*}

And for slot 4:

\begin{equation*}
AD_4(\pi_u,\Pi_o,w_{\dot{L}},\mu) = \eta_{\dot{L}^2} \sum_{\dot{u},\dot{v} \in \dot{L}^{N(\mu)}_{-4}(\Pi_o) | \dot{u} \neq \dot{v}} w_{\dot{L}}(\dot{u}) w_{\dot{L}}(\dot{v}) \Delta_S(\breve{A}_4(\mu[\instantiation{u}{4}{\pi_u}]), \breve{A}_4(\mu[\instantiation{v}{4}{\pi_u}]))
\end{equation*}

We do not know which $\Pi_o$ we have, but we know that we will need to obtain two different line-up patterns $\dot{u}$ and $\dot{v}$ from $\dot{L}^{N(\mu)}_{-4}(\Pi_o)$ to calculate $\Delta_S(\breve{A}_4(\mu[\instantiation{u}{4}{\pi_u}]), \breve{A}_4(\mu[\instantiation{v}{4}{\pi_u}]))$. We calculate this value for two figurative line-up patterns $\dot{u} = (\pi_1,\pi_2,\pi_3,*)$ and $\dot{v} = (\pi_4,\pi_5,\pi_6,*)$ from $\dot{L}^{N(\mu)}_{-4}(\Pi_o)$:

\begin{equation*}
\Delta_S(\breve{A}_4(\mu[\instantiation{u}{4}{\pi_u}]), \breve{A}_4(\mu[\instantiation{v}{4}{\pi_u}])) = \Delta_S(\breve{A}_4(\mu[\pi_1,\pi_2,\pi_3,\pi_u]), \breve{A}_4(\mu[\pi_4,\pi_5,\pi_6,\pi_u]))
\end{equation*}

Here, the agent in both slots 4 ($\pi_u$) will perform the same sequence of actions (always Up) independently of the line-up. So no matter which agents are in $\Pi_o$ we obtain:

\begin{equation*}
AD_4(\pi_u,\Pi_o,w_{\dot{L}},\mu) = 0
\end{equation*}

And finally, we weight over the slots:

\begin{equation*}
\begin{aligned}
AD(\Pi_e,w_{\Pi_e},\Pi_o,w_{\dot{L}},\mu,w_S)	& =	\sum_{i = 1}^{N(\mu)} w_S(i,\mu) AD_i(\Pi_e,w_{\Pi_e},\Pi_o,w_{\dot{L}},\mu) =\\
												& =	\sum_{i = 1}^{N(\mu)} w_S(i,\mu) AD_i(\pi_u,\Pi_o,w_{\dot{L}},\mu) =\\
												& = \frac{1}{4} \{ AD_1(\pi_u,\Pi_o,w_{\dot{L}},\mu) + AD_2(\pi_u,\Pi_o,w_{\dot{L}},\mu) +\\
												& + AD_3(\pi_u,\Pi_o,w_{\dot{L}},\mu) + AD_4(\pi_u,\Pi_o,w_{\dot{L}},\mu)\} =\\
												& = \frac{1}{4} \left\{0 + 0 + 0 + 0\right\} = 0
\end{aligned}
\end{equation*}

So, for every $\Pi_o$ we obtain the same result:

\begin{equation*}
\forall \Pi_o : AD(\Pi_e,w_{\Pi_e},\Pi_o,w_{\dot{L}},\mu,w_S) = 0
\end{equation*}

Therefore, predator-prey has $Right_{min} = 0$ for this property.
\end{proof}
\end{proposition}

\subsection{Reward Dependency}
We continue with the reward dependency (RD) property. As given in section \ref{sec:RD}, we want to know if the evaluated agents obtain different expected average rewards depending on which line-up they interact with. We use $\Delta_{\mathbb{Q}}(a,b) = 1$ if numbers $a$ and $b$ are equal and $0$ otherwise.

\begin{proposition}
\label{prop:predator-prey_RD_general_min}
$General_{min}$ for the reward dependency (RD) property is equal to $0$ for the predator-prey environment.

\begin{proof}
To find $General_{min}$ (equation \ref{eq:general_min}), we need to find a trio $\left\langle\Pi_e,w_{\Pi_e},\Pi_o\right\rangle$ which minimises the property as much as possible. We can have this situation by selecting $\Pi_e = \{\pi_u\}$ with $w_{\Pi_e}(\pi_u) = 1$ and $\Pi_o = \{{\pi_d}_1, {\pi_d}_2\}$ (a $\pi_d$ agent always performs Down and a $\pi_u$ agent always performs Up).

Following definition \ref{def:RD}, we obtain the RD value for this $\left\langle\Pi_e,w_{\Pi_e},\Pi_o\right\rangle$. Since the environment is not symmetric, we need to calculate this property for every slot. Following definition \ref{def:RD_set}, we could calculate its RD value for each slot but, since $\Pi_e$ has only one agent and its weight is equal to $1$, it is equivalent to use directly definition \ref{def:RD_agent}. We start with slot 1:

\begin{equation*}
\begin{aligned}
RD_1(\pi_u,\Pi_o,w_{\dot{L}},\mu)	& = \eta_{\dot{L}^2} \sum_{\dot{u},\dot{v} \in \dot{L}^{N(\mu)}_{-1}(\Pi_o) | \dot{u} \neq \dot{v}} w_{\dot{L}}(\dot{u}) w_{\dot{L}}(\dot{v}) \Delta_{\mathbb{Q}}(R_1(\mu[\instantiation{u}{1}{\pi_u}]), R_1(\mu[\instantiation{v}{1}{\pi_u}])) =\\
									& = 2 \frac{8}{7} \frac{1}{8} \frac{1}{8} \{\Delta_{\mathbb{Q}}(R_1(\mu[\pi_u,{\pi_d}_1,{\pi_d}_1,{\pi_d}_1]), R_1(\mu[\pi_u,{\pi_d}_1,{\pi_d}_1,{\pi_d}_2])) +\\
									& + \Delta_{\mathbb{Q}}(R_1(\mu[\pi_u,{\pi_d}_1,{\pi_d}_1,{\pi_d}_1]), R_1(\mu[\pi_u,{\pi_d}_1,{\pi_d}_2,{\pi_d}_1])) +\\
									& \ \ \ \ \ \ \ \ \ \ \ \ \ \ \ \ \ \ \ \ \ \ \ \ \ \ \ \ \ \ \ \ \ \ \ \ \ \ \ \ \vdots\\
									& + \Delta_{\mathbb{Q}}(R_1(\mu[\pi_u,{\pi_d}_2,{\pi_d}_2,{\pi_d}_1]), R_1(\mu[\pi_u,{\pi_d}_2,{\pi_d}_2,{\pi_d}_2]))\}
\end{aligned}
\end{equation*}

\noindent Note that we avoided to calculate both $\Delta_{\mathbb{Q}}(a,b)$ and $\Delta_{\mathbb{Q}}(b,a)$ since they provide the same result, by calculating only $\Delta_{\mathbb{Q}}(a,b)$ and multiplying the result by $2$.

In this case, we have 28 possible pairs of line-ups, where $\pi_u$ will always perform Up and a $\pi_d$ agent will always perform Down, so the agent in both slots 1 ($\pi_u$) will obtain the same expected average reward ($1$) independently of the line-up. So:

\begin{equation*}
RD_1(\pi_u,\Pi_o,w_{\dot{L}},\mu) = 2 \frac{8}{7} \frac{1}{8} \frac{1}{8} \left\{28 \times 0\right\} = 0
\end{equation*}

For slot 2, the agent in both slots 2 ($\pi_u$) will also obtain the same expected average reward ($1$) independently of the line-up. So:

\begin{equation*}
\begin{aligned}
RD_2(\pi_u,\Pi_o,w_{\dot{L}},\mu)	& = \eta_{\dot{L}^2} \sum_{\dot{u},\dot{v} \in \dot{L}^{N(\mu)}_{-2}(\Pi_o) | \dot{u} \neq \dot{v}} w_{\dot{L}}(\dot{u}) w_{\dot{L}}(\dot{v}) \Delta_{\mathbb{Q}}(R_2(\mu[\instantiation{u}{2}{\pi_u}]), R_2(\mu[\instantiation{v}{2}{\pi_u}])) =\\
									& = 2 \frac{8}{7} \frac{1}{8} \frac{1}{8} \{\Delta_{\mathbb{Q}}(R_2(\mu[{\pi_d}_1,\pi_u,{\pi_d}_1,{\pi_d}_1]), R_2(\mu[{\pi_d}_1,\pi_u,{\pi_d}_1,{\pi_d}_2])) +\\
									& + \Delta_{\mathbb{Q}}(R_2(\mu[{\pi_d}_1,\pi_u,{\pi_d}_1,{\pi_d}_1]), R_2(\mu[{\pi_d}_1,\pi_u,{\pi_d}_2,{\pi_d}_1])) +\\
									& \ \ \ \ \ \ \ \ \ \ \ \ \ \ \ \ \ \ \ \ \ \ \ \ \ \ \ \ \ \ \ \ \ \ \ \ \ \ \ \ \vdots\\
									& + \Delta_{\mathbb{Q}}(R_2(\mu[{\pi_d}_2,\pi_u,{\pi_d}_2,{\pi_d}_1]), R_2(\mu[{\pi_d}_2,\pi_u,{\pi_d}_2,{\pi_d}_2]))\} =\\
									& = 2 \frac{8}{7} \frac{1}{8} \frac{1}{8} \left\{28 \times 0\right\} = 0
\end{aligned}
\end{equation*}

For slot 3, the agent in both slots 3 ($\pi_u$) will also obtain the same expected average reward ($-1$) independently of the line-up. So:

\begin{equation*}
\begin{aligned}
RD_3(\pi_u,\Pi_o,w_{\dot{L}},\mu)	& = \eta_{\dot{L}^2} \sum_{\dot{u},\dot{v} \in \dot{L}^{N(\mu)}_{-3}(\Pi_o) | \dot{u} \neq \dot{v}} w_{\dot{L}}(\dot{u}) w_{\dot{L}}(\dot{v}) \Delta_{\mathbb{Q}}(R_3(\mu[\instantiation{u}{3}{\pi_u}]), R_3(\mu[\instantiation{v}{3}{\pi_u}])) =\\
									& = 2 \frac{8}{7} \frac{1}{8} \frac{1}{8} \{\Delta_{\mathbb{Q}}(R_3(\mu[{\pi_d}_1,{\pi_d}_1,\pi_u,{\pi_d}_1]), R_3(\mu[{\pi_d}_1,{\pi_d}_1,\pi_u,{\pi_d}_2])) +\\
									& + \Delta_{\mathbb{Q}}(R_3(\mu[{\pi_d}_1,{\pi_d}_1,\pi_u,{\pi_d}_1]), R_3(\mu[{\pi_d}_1,{\pi_d}_2,\pi_u,{\pi_d}_1])) +\\
									& \ \ \ \ \ \ \ \ \ \ \ \ \ \ \ \ \ \ \ \ \ \ \ \ \ \ \ \ \ \ \ \ \ \ \ \ \ \ \ \ \vdots\\
									& + \Delta_{\mathbb{Q}}(R_3(\mu[{\pi_d}_2,{\pi_d}_2,\pi_u,{\pi_d}_1]), R_3(\mu[{\pi_d}_2,{\pi_d}_2,\pi_u,{\pi_d}_2]))\} =\\
									& = 2 \frac{8}{7} \frac{1}{8} \frac{1}{8} \left\{28 \times 0\right\} = 0
\end{aligned}
\end{equation*}

And for slot 4, the agent in both slots 4 ($\pi_u$) will also obtain the same expected average reward ($1$) independently of the line-up. So:

\begin{equation*}
\begin{aligned}
RD_4(\pi_u,\Pi_o,w_{\dot{L}},\mu)	& = \eta_{\dot{L}^2} \sum_{\dot{u},\dot{v} \in \dot{L}^{N(\mu)}_{-4}(\Pi_o) | \dot{u} \neq \dot{v}} w_{\dot{L}}(\dot{u}) w_{\dot{L}}(\dot{v}) \Delta_{\mathbb{Q}}(R_4(\mu[\instantiation{u}{4}{\pi_u}]), R_4(\mu[\instantiation{v}{4}{\pi_u}])) =\\
									& = 2 \frac{8}{7} \frac{1}{8} \frac{1}{8} \{\Delta_{\mathbb{Q}}(R_4(\mu[{\pi_d}_1,{\pi_d}_1,{\pi_d}_1,\pi_u]), R_4(\mu[{\pi_d}_1,{\pi_d}_1,{\pi_d}_2,\pi_u])) +\\
									& + \Delta_{\mathbb{Q}}(R_4(\mu[{\pi_d}_1,{\pi_d}_1,{\pi_d}_1,\pi_u]), R_4(\mu[{\pi_d}_1,{\pi_d}_2,{\pi_d}_1,\pi_u])) +\\
									& \ \ \ \ \ \ \ \ \ \ \ \ \ \ \ \ \ \ \ \ \ \ \ \ \ \ \ \ \ \ \ \ \ \ \ \ \ \ \ \ \vdots\\
									& + \Delta_{\mathbb{Q}}(R_4(\mu[{\pi_d}_2,{\pi_d}_2,{\pi_d}_1,\pi_u]), R_4(\mu[{\pi_d}_2,{\pi_d}_2,{\pi_d}_2,\pi_u]))\} =\\
									& = 2 \frac{8}{7} \frac{1}{8} \frac{1}{8} \left\{28 \times 0\right\} = 0
\end{aligned}
\end{equation*}

And finally, we weight over the slots:

\begin{equation*}
\begin{aligned}
RD(\Pi_e,w_{\Pi_e},\Pi_o,w_{\dot{L}},\mu,w_S)	& =	\sum_{i = 1}^{N(\mu)} w_S(i,\mu) RD_i(\Pi_e,w_{\Pi_e},\Pi_o,w_{\dot{L}},\mu) =\\
												& =	\sum_{i = 1}^{N(\mu)} w_S(i,\mu) RD_i(\pi_u,\Pi_o,w_{\dot{L}},\mu) =\\
												& = \frac{1}{4} \{RD_1(\pi_u,\Pi_o,w_{\dot{L}},\mu) + RD_2(\pi_u,\Pi_o,w_{\dot{L}},\mu) +\\
												& + RD_3(\pi_u,\Pi_o,w_{\dot{L}},\mu) + RD_4(\pi_u,\Pi_o,w_{\dot{L}},\mu)\} =\\
												& = \frac{1}{4} \left\{0 + 0 + 0 + 0\right\} = 0
\end{aligned}
\end{equation*}

Since $0$ is the lowest possible value for the reward dependency property, therefore predator-prey has $General_{min} = 0$ for this property.
\end{proof}
\end{proposition}

\begin{conjecture}
\label{conj:predator-prey_RD_general_max}
$General_{max}$ for the reward dependency (RD) property is equal to $1$ for the predator-prey environment.

To find $General_{max}$ (equation \ref{eq:general_max}), we need to find a trio $\left\langle\Pi_e,w_{\Pi_e},\Pi_o\right\rangle$ which maximises the property as much as possible. We can have this situation by selecting $\Pi_e = \{\pi_r\}$ with $w_{\Pi_e}(\pi_r) = 1$ and $\Pi_o = \{\pi_s, \pi_r\}$ (a $\pi_r$ agent always acts randomly and a $\pi_s$ agent always stays in the same cell\footnote{Note that every cell has an action which is blocked by a block or a boundary, therefore an agent performing this action will stay at its current cell.}).

Following definition \ref{def:RD}, we obtain the RD value for this $\left\langle\Pi_e,w_{\Pi_e},\Pi_o\right\rangle$. Since the environment is not symmetric, we need to calculate this property for every slot. Following definition \ref{def:RD_set}, we could calculate its RD value for each slot but, since $\Pi_e$ has only one agent and its weight is equal to $1$, it is equivalent to use directly definition \ref{def:RD_agent}. We start with slot 1:

\begin{equation*}
\begin{aligned}
RD_1(\pi_r,\Pi_o,w_{\dot{L}},\mu)	& = \eta_{\dot{L}^2} \sum_{\dot{u},\dot{v} \in \dot{L}^{N(\mu)}_{-1}(\Pi_o) | \dot{u} \neq \dot{v}} w_{\dot{L}}(\dot{u}) w_{\dot{L}}(\dot{v}) \Delta_{\mathbb{Q}}(R_1(\mu[\instantiation{u}{1}{\pi_r}]), R_1(\mu[\instantiation{v}{1}{\pi_r}])) =\\
									& = 2 \frac{8}{7} \frac{1}{8} \frac{1}{8} \{\Delta_{\mathbb{Q}}(R_1(\mu[\pi_r,\pi_s,\pi_s,\pi_s]), R_1(\mu[\pi_r,\pi_s,\pi_s,\pi_r])) +\\
									& + \Delta_{\mathbb{Q}}(R_1(\mu[\pi_r,\pi_s,\pi_s,\pi_s]), R_1(\mu[\pi_r,\pi_s,\pi_r,\pi_s])) +\\
									& \ \ \ \ \ \ \ \ \ \ \ \ \ \ \ \ \ \ \ \ \ \ \ \ \ \ \ \ \ \ \ \ \ \ \ \ \vdots\\
									& + \Delta_{\mathbb{Q}}(R_1(\mu[\pi_r,\pi_r,\pi_r,\pi_s]), R_1(\mu[\pi_r,\pi_r,\pi_r,\pi_r]))\}
\end{aligned}
\end{equation*}

\noindent Note that we avoided to calculate both $\Delta_{\mathbb{Q}}(a,b)$ and $\Delta_{\mathbb{Q}}(b,a)$ since they provide the same result, by calculating only $\Delta_{\mathbb{Q}}(a,b)$ and multiplying the result by $2$.

The expected average reward of these line-ups highly depends on the agents' positions, obtaining an expected average reward from $-1$ to $1$ (exclusive, since there always exists some probability that the prey will either be chased or not) to the agent in slot 1 ($\pi_r$). One reason is the stochastic behaviour of the $\pi_r$ agents, which makes that no pair of line-ups can obtain exactly the same result. Another reason is that the positions of the slots where the random agents play as a predator do not have a symmetric place in the space (blocks are not symmetrically located in the space) which, for each $\pi_r$ in a different slot, provides (most likely) different probabilities to chase the prey\footnote{It is more likely that the prey will be chased by the lower left predator than the upper right predator, and the lower right predator will have the lowest chance to chase the prey.}. This makes every pair of line-up to have different expected average rewards, making its reward dependency equal to 1.

\begin{equation*}
RD_1(\pi_r,\Pi_o,w_{\dot{L}},\mu) = 2 \frac{8}{7} \frac{1}{8} \frac{1}{8} \{28 \times 1\} = 1
\end{equation*}

For slot 2, also the result of these line-ups highly depends on the agents' positions. So:

\begin{equation*}
\begin{aligned}
RD_2(\pi_r,\Pi_o,w_{\dot{L}},\mu)	& = \eta_{\dot{L}^2} \sum_{\dot{u},\dot{v} \in \dot{L}^{N(\mu)}_{-2}(\Pi_o) | \dot{u} \neq \dot{v}} w_{\dot{L}}(\dot{u}) w_{\dot{L}}(\dot{v}) \Delta_{\mathbb{Q}}(R_2(\mu[\instantiation{u}{2}{\pi_r}]), R_2(\mu[\instantiation{v}{2}{\pi_r}])) =\\
									& = 2 \frac{8}{7} \frac{1}{8} \frac{1}{8} \{\Delta_{\mathbb{Q}}(R_2(\mu[\pi_s,\pi_r,\pi_s,\pi_s]), R_2(\mu[\pi_s,\pi_r,\pi_s,\pi_r])) +\\
									& + \Delta_{\mathbb{Q}}(R_2(\mu[\pi_s,\pi_r,\pi_s,\pi_s]), R_2(\mu[\pi_s,\pi_r,\pi_r,\pi_s])) +\\
									& \ \ \ \ \ \ \ \ \ \ \ \ \ \ \ \ \ \ \ \ \ \ \ \ \ \ \ \ \ \ \ \ \ \ \ \ \vdots\\
									& + \Delta_{\mathbb{Q}}(R_2(\mu[\pi_r,\pi_r,\pi_r,\pi_s]), R_2(\mu[\pi_r,\pi_r,\pi_r,\pi_r]))\} =\\
									& = 2 \frac{8}{7} \frac{1}{8} \frac{1}{8} \left\{28 \times 1\right\} = 1
\end{aligned}
\end{equation*}

For slot 3, also the result of these line-ups highly depends on the agents' positions. So:

\begin{equation*}
\begin{aligned}
RD_3(\pi_r,\Pi_o,w_{\dot{L}},\mu)	& = \eta_{\dot{L}^2} \sum_{\dot{u},\dot{v} \in \dot{L}^{N(\mu)}_{-3}(\Pi_o) | \dot{u} \neq \dot{v}} w_{\dot{L}}(\dot{u}) w_{\dot{L}}(\dot{v}) \Delta_{\mathbb{Q}}(R_3(\mu[\instantiation{u}{3}{\pi_r}]), R_3(\mu[\instantiation{v}{3}{\pi_r}])) =\\
									& = 2 \frac{8}{7} \frac{1}{8} \frac{1}{8} \{\Delta_{\mathbb{Q}}(R_3(\mu[\pi_s,\pi_s,\pi_r,\pi_s]), R_3(\mu[\pi_s,\pi_s,\pi_r,\pi_r])) +\\
									& + \Delta_{\mathbb{Q}}(R_3(\mu[\pi_s,\pi_s,\pi_r,\pi_s]), R_3(\mu[\pi_s,\pi_r,\pi_r,\pi_s])) +\\
									& \ \ \ \ \ \ \ \ \ \ \ \ \ \ \ \ \ \ \ \ \ \ \ \ \ \ \ \ \ \ \ \ \ \ \ \ \vdots\\
									& + \Delta_{\mathbb{Q}}(R_3(\mu[\pi_r,\pi_r,\pi_r,\pi_s]), R_3(\mu[\pi_r,\pi_r,\pi_r,\pi_r]))\} =\\
									& = 2 \frac{8}{7} \frac{1}{8} \frac{1}{8} \left\{28 \times 1\right\} = 1
\end{aligned}
\end{equation*}

And for slot 4, also the result of these line-ups highly depends on the agents' positions. So:

\begin{equation*}
\begin{aligned}
RD_4(\pi_r,\Pi_o,w_{\dot{L}},\mu)	& = \eta_{\dot{L}^2} \sum_{\dot{u},\dot{v} \in \dot{L}^{N(\mu)}_{-4}(\Pi_o) | \dot{u} \neq \dot{v}} w_{\dot{L}}(\dot{u}) w_{\dot{L}}(\dot{v}) \Delta_{\mathbb{Q}}(R_4(\mu[\instantiation{u}{4}{\pi_r}]), R_4(\mu[\instantiation{v}{4}{\pi_r}])) =\\
									& = 2 \frac{8}{7} \frac{1}{8} \frac{1}{8} \{\Delta_{\mathbb{Q}}(R_4(\mu[\pi_s,\pi_s,\pi_s,\pi_r]), R_4(\mu[\pi_s,\pi_s,\pi_r,\pi_r])) +\\
									& + \Delta_{\mathbb{Q}}(R_4(\mu[\pi_s,\pi_s,\pi_s,\pi_r]), R_4(\mu[\pi_s,\pi_r,\pi_s,\pi_r])) +\\
									& \ \ \ \ \ \ \ \ \ \ \ \ \ \ \ \ \ \ \ \ \ \ \ \ \ \ \ \ \ \ \ \ \ \ \ \ \vdots\\
									& + \Delta_{\mathbb{Q}}(R_4(\mu[\pi_r,\pi_r,\pi_s,\pi_r]), R_4(\mu[\pi_r,\pi_r,\pi_r,\pi_r]))\} =\\
									& = 2 \frac{8}{7} \frac{1}{8} \frac{1}{8} \left\{28 \times 1\right\} = 1
\end{aligned}
\end{equation*}

And finally, we weight over the slots:

\begin{equation*}
\begin{aligned}
RD(\Pi_e,w_{\Pi_e},\Pi_o,w_{\dot{L}},\mu,w_S)	& =	\sum_{i = 1}^{N(\mu)} w_S(i,\mu) RD_i(\Pi_e,w_{\Pi_e},\Pi_o,w_{\dot{L}},\mu) =\\
												& =	\sum_{i = 1}^{N(\mu)} w_S(i,\mu) RD_i(\pi_r,\Pi_o,w_{\dot{L}},\mu) =\\
												& = \frac{1}{4} \{RD_1(\pi_r,\Pi_o,w_{\dot{L}},\mu) + RD_2(\pi_r,\Pi_o,w_{\dot{L}},\mu) +\\
												& + RD_3(\pi_r,\Pi_o,w_{\dot{L}},\mu) + RD_4(\pi_r,\Pi_o,w_{\dot{L}},\mu)\} =\\
												& = \frac{1}{4} \left\{1 + 1 + 1 + 1\right\} = 1
\end{aligned}
\end{equation*}

Since $1$ is the highest possible value for the reward dependency property, therefore predator-prey has $General_{max} = 1$ for this property.
\end{conjecture}

\begin{proposition}
\label{prop:predator-prey_RD_left_max}
$Left_{max}$ for the reward dependency (RD) property is equal to $0$ for the predator-prey environment.

\begin{proof}
To find $Left_{max}$ (equation \ref{eq:left_max}), we need to find a pair $\left\langle\Pi_e,w_{\Pi_e}\right\rangle$ which maximises the property as much as possible while $\Pi_o$ minimises it. Using $\Pi_o = \{{\pi_d}_1,{\pi_d}_2\}$ (a $\pi_d$ agent always performs Down) we find this situation no matter which pair $\left\langle\Pi_e,w_{\Pi_e}\right\rangle$ we use.

Following definition \ref{def:RD}, we obtain the RD value for this $\left\langle\Pi_e,w_{\Pi_e},\Pi_o\right\rangle$ (where $\Pi_e$ and $w_{\Pi_e}$ are instantiated with any permitted values). Since the environment is not symmetric, we need to calculate this property for every slot. Following definition \ref{def:RD_set}, we can calculate its RD value for slot 1:

\begin{equation*}
RD_1(\Pi_e,w_{\Pi_e},\Pi_o,w_{\dot{L}},\mu) = \sum_{\pi \in \Pi_e} w_{\Pi_e}(\pi) RD_1(\pi,\Pi_o,w_{\dot{L}},\mu)
\end{equation*}

We do not know which $\Pi_e$ we have, but we know that we will need to evaluate $RD_1(\pi,\Pi_o,w_{\dot{L}},\mu)$ for all evaluated agent $\pi \in \Pi_e$. We follow definition \ref{def:RD_agent} to calculate this value for a figurative evaluated agent $\pi$ from $\Pi_e$:

\begin{equation*}
\begin{aligned}
RD_1(\pi,\Pi_o,w_{\dot{L}},\mu)	& = \eta_{\dot{L}^2} \sum_{\dot{u},\dot{v} \in \dot{L}^{N(\mu)}_{-1}(\Pi_o) | \dot{u} \neq \dot{v}} w_{\dot{L}}(\dot{u}) w_{\dot{L}}(\dot{v}) \Delta_{\mathbb{Q}}(R_1(\mu[\instantiation{u}{1}{\pi}]), R_1(\mu[\instantiation{v}{1}{\pi}])) =\\
								& = 2 \frac{8}{7} \frac{1}{8} \frac{1}{8} \{\Delta_{\mathbb{Q}}(R_1(\mu[\pi,{\pi_d}_1,{\pi_d}_1,{\pi_d}_1]), R_1(\mu[\pi,{\pi_d}_1,{\pi_d}_1,{\pi_d}_2])) +\\
								& + \Delta_{\mathbb{Q}}(R_1(\mu[\pi,{\pi_d}_1,{\pi_d}_1,{\pi_d}_1]), R_1(\mu[\pi,{\pi_d}_1,{\pi_d}_2,{\pi_d}_1])) +\\
								& \ \ \ \ \ \ \ \ \ \ \ \ \ \ \ \ \ \ \ \ \ \ \ \ \ \ \ \ \ \ \ \ \ \ \ \ \ \ \ \vdots\\
								& + \Delta_{\mathbb{Q}}(R_1(\mu[\pi,{\pi_d}_2,{\pi_d}_2,{\pi_d}_1]), R_1(\mu[\pi,{\pi_d}_2,{\pi_d}_2,{\pi_d}_2]))\}
\end{aligned}
\end{equation*}

\noindent Note that we avoided to calculate both $\Delta_{\mathbb{Q}}(a,b)$ and $\Delta_{\mathbb{Q}}(b,a)$ since they provide the same result, by calculating only $\Delta_{\mathbb{Q}}(a,b)$ and multiplying the result by $2$.

A $\pi_d$ agent will always perform Down, so for the 28 possible pairs of line-ups we obtain a situation where the agent in both slots 1 (any $\pi$) will be able to differentiate with which agents is interacting, so it will not be able to change its distribution of action sequences depending on the line-up, obtaining agent in both slots 1 (any $\pi$) the same expected average reward. So:

\begin{equation*}
RD_1(\pi,\Pi_o,w_{\dot{L}},\mu) = 2 \frac{8}{7} \frac{1}{8} \frac{1}{8} \left\{28 \times 0\right\} = 0
\end{equation*}

Therefore, no matter which agents are in $\Pi_e$ and their weights $w_{\Pi_e}$ we obtain:

\begin{equation*}
RD_1(\Pi_e,w_{\Pi_e},\Pi_o,w_{\dot{L}},\mu) = 0
\end{equation*}

For slot 2:

\begin{equation*}
RD_2(\Pi_e,w_{\Pi_e},\Pi_o,w_{\dot{L}},\mu) = \sum_{\pi \in \Pi_e} w_{\Pi_e}(\pi) RD_2(\pi,\Pi_o,w_{\dot{L}},\mu)
\end{equation*}

Again, we do not know which $\Pi_e$ we have, but we know that we will need to evaluate $RD_2(\pi,\Pi_o,w_{\dot{L}},\mu)$ for all evaluated agent $\pi \in \Pi_e$. We follow definition \ref{def:RD_agent} to calculate this value for a figurative evaluated agent $\pi$ from $\Pi_e$:

\begin{equation*}
\begin{aligned}
RD_2(\pi,\Pi_o,w_{\dot{L}},\mu)	& = \eta_{\dot{L}^2} \sum_{\dot{u},\dot{v} \in \dot{L}^{N(\mu)}_{-2}(\Pi_o) | \dot{u} \neq \dot{v}} w_{\dot{L}}(\dot{u}) w_{\dot{L}}(\dot{v}) \Delta_{\mathbb{Q}}(R_2(\mu[\instantiation{u}{2}{\pi}]), R_2(\mu[\instantiation{v}{2}{\pi}])) =\\
								& = 2 \frac{8}{7} \frac{1}{8} \frac{1}{8} \{\Delta_{\mathbb{Q}}(R_2(\mu[{\pi_d}_1,\pi,{\pi_d}_1,{\pi_d}_1]), R_2(\mu[{\pi_d}_1,\pi,{\pi_d}_1,{\pi_d}_2])) +\\
								& + \Delta_{\mathbb{Q}}(R_2(\mu[{\pi_d}_1,\pi,{\pi_d}_1,{\pi_d}_1]), R_2(\mu[{\pi_d}_1,\pi,{\pi_d}_2,{\pi_d}_1])) +\\
								& \ \ \ \ \ \ \ \ \ \ \ \ \ \ \ \ \ \ \ \ \ \ \ \ \ \ \ \ \ \ \ \ \ \ \ \ \ \ \ \vdots\\
								& + \Delta_{\mathbb{Q}}(R_2(\mu[{\pi_d}_2,\pi,{\pi_d}_2,{\pi_d}_1]), R_2(\mu[{\pi_d}_2,\pi,{\pi_d}_2,{\pi_d}_2]))\}
\end{aligned}
\end{equation*}

Again, a $\pi_d$ agent will always perform Down, so for the 28 possible pairs of line-ups we obtain a situation where the agent in both slots 2 (any $\pi$) will be able to differentiate with which agents is interacting, so it will not be able to change its distribution of action sequences depending on the line-up, obtaining agent in both slots 2 (any $\pi$) the same expected average reward. So:

\begin{equation*}
RD_2(\pi,\Pi_o,w_{\dot{L}},\mu) = 2 \frac{8}{7} \frac{1}{8} \frac{1}{8} \left\{28 \times 0\right\} = 0
\end{equation*}

Therefore, no matter which agents are in $\Pi_e$ and their weights $w_{\Pi_e}$ we obtain:

\begin{equation*}
RD_2(\Pi_e,w_{\Pi_e},\Pi_o,w_{\dot{L}},\mu) = 0
\end{equation*}

For slot 3:

\begin{equation*}
RD_3(\Pi_e,w_{\Pi_e},\Pi_o,w_{\dot{L}},\mu) = \sum_{\pi \in \Pi_e} w_{\Pi_e}(\pi) RD_3(\pi,\Pi_o,w_{\dot{L}},\mu)
\end{equation*}

Again, we do not know which $\Pi_e$ we have, but we know that we will need to evaluate $RD_3(\pi,\Pi_o,w_{\dot{L}},\mu)$ for all evaluated agent $\pi \in \Pi_e$. We follow definition \ref{def:RD_agent} to calculate this value for a figurative evaluated agent $\pi$ from $\Pi_e$:

\begin{equation*}
\begin{aligned}
RD_3(\pi,\Pi_o,w_{\dot{L}},\mu)	& = \eta_{\dot{L}^2} \sum_{\dot{u},\dot{v} \in \dot{L}^{N(\mu)}_{-3}(\Pi_o) | \dot{u} \neq \dot{v}} w_{\dot{L}}(\dot{u}) w_{\dot{L}}(\dot{v}) \Delta_{\mathbb{Q}}(R_3(\mu[\instantiation{u}{3}{\pi}]), R_3(\mu[\instantiation{v}{3}{\pi}])) =\\
								& = 2 \frac{8}{7} \frac{1}{8} \frac{1}{8} \{\Delta_{\mathbb{Q}}(R_3(\mu[{\pi_d}_1,{\pi_d}_1,\pi,{\pi_d}_1]), R_3(\mu[{\pi_d}_1,{\pi_d}_1,\pi,{\pi_d}_2])) +\\
								& + \Delta_{\mathbb{Q}}(R_3(\mu[{\pi_d}_1,{\pi_d}_1,\pi,{\pi_d}_1]), R_3(\mu[{\pi_d}_1,{\pi_d}_2,\pi,{\pi_d}_1])) +\\
								& \ \ \ \ \ \ \ \ \ \ \ \ \ \ \ \ \ \ \ \ \ \ \ \ \ \ \ \ \ \ \ \ \ \ \ \ \ \ \ \vdots\\
								& + \Delta_{\mathbb{Q}}(R_3(\mu[{\pi_d}_2,{\pi_d}_2,\pi,{\pi_d}_1]), R_3(\mu[{\pi_d}_2,{\pi_d}_2,\pi,{\pi_d}_2]))\}
\end{aligned}
\end{equation*}

Again, a $\pi_d$ agent will always perform Down, so for the 28 possible pairs of line-ups we obtain a situation where the agent in both slots 3 (any $\pi$) will be able to differentiate with which agents is interacting, so it will not be able to change its distribution of action sequences depending on the line-up, obtaining agent in both slots 3 (any $\pi$) the same expected average reward. So:

\begin{equation*}
RD_3(\pi,\Pi_o,w_{\dot{L}},\mu) = 2 \frac{8}{7} \frac{1}{8} \frac{1}{8} \left\{28 \times 0\right\} = 0
\end{equation*}

Therefore, no matter which agents are in $\Pi_e$ and their weights $w_{\Pi_e}$ we obtain:

\begin{equation*}
RD_3(\Pi_e,w_{\Pi_e},\Pi_o,w_{\dot{L}},\mu) = 0
\end{equation*}

And for slot 4:

\begin{equation*}
RD_4(\Pi_e,w_{\Pi_e},\Pi_o,w_{\dot{L}},\mu) = \sum_{\pi \in \Pi_e} w_{\Pi_e}(\pi) RD_4(\pi,\Pi_o,w_{\dot{L}},\mu)
\end{equation*}

Again, we do not know which $\Pi_e$ we have, but we know that we will need to evaluate $RD_4(\pi,\Pi_o,w_{\dot{L}},\mu)$ for all evaluated agent $\pi \in \Pi_e$. We follow definition \ref{def:RD_agent} to calculate this value for a figurative evaluated agent $\pi$ from $\Pi_e$:

\begin{equation*}
\begin{aligned}
RD_4(\pi,\Pi_o,w_{\dot{L}},\mu)	& = \eta_{\dot{L}^2} \sum_{\dot{u},\dot{v} \in \dot{L}^{N(\mu)}_{-4}(\Pi_o) | \dot{u} \neq \dot{v}} w_{\dot{L}}(\dot{u}) w_{\dot{L}}(\dot{v}) \Delta_{\mathbb{Q}}(R_4(\mu[\instantiation{u}{4}{\pi}]), R_4(\mu[\instantiation{v}{4}{\pi}])) =\\
								& = 2 \frac{8}{7} \frac{1}{8} \frac{1}{8} \{\Delta_{\mathbb{Q}}(R_4(\mu[{\pi_d}_1,{\pi_d}_1,{\pi_d}_1,\pi]), R_4(\mu[{\pi_d}_1,{\pi_d}_1,{\pi_d}_2,\pi])) +\\
								& + \Delta_{\mathbb{Q}}(R_4(\mu[{\pi_d}_1,{\pi_d}_1,{\pi_d}_1,\pi]), R_4(\mu[{\pi_d}_1,{\pi_d}_2,{\pi_d}_1,\pi])) +\\
								& \ \ \ \ \ \ \ \ \ \ \ \ \ \ \ \ \ \ \ \ \ \ \ \ \ \ \ \ \ \ \ \ \ \ \ \ \ \ \ \vdots\\
								& + \Delta_{\mathbb{Q}}(R_4(\mu[{\pi_d}_2,{\pi_d}_2,{\pi_d}_1,\pi]), R_4(\mu[{\pi_d}_2,{\pi_d}_2,{\pi_d}_2,\pi]))\}
\end{aligned}
\end{equation*}

Again, a $\pi_d$ agent will always perform Down, so for the 28 possible pairs of line-ups we obtain a situation where the agent in both slots 4 (any $\pi$) will be able to differentiate with which agents is interacting, so it will not be able to change its distribution of action sequences depending on the line-up, obtaining agent in both slots 4 (any $\pi$) the same expected average reward. So:

\begin{equation*}
RD_4(\pi,\Pi_o,w_{\dot{L}},\mu) = 2 \frac{8}{7} \frac{1}{8} \frac{1}{8} \left\{28 \times 0\right\} = 0
\end{equation*}

Therefore, no matter which agents are in $\Pi_e$ and their weights $w_{\Pi_e}$ we obtain:

\begin{equation*}
RD_4(\Pi_e,w_{\Pi_e},\Pi_o,w_{\dot{L}},\mu) = 0
\end{equation*}

And finally, we weight over the slots:

\begin{equation*}
\begin{aligned}
RD(\Pi_e,w_{\Pi_e},\Pi_o,w_{\dot{L}},\mu,w_S)	& =	\sum_{i = 1}^{N(\mu)} w_S(i,\mu) RD_i(\Pi_e,w_{\Pi_e},\Pi_o,w_{\dot{L}},\mu) =\\
												& = \frac{1}{4} \{RD_1(\Pi_e,w_{\Pi_e},\Pi_o,w_{\dot{L}},\mu) + RD_2(\Pi_e,w_{\Pi_e},\Pi_o,w_{\dot{L}},\mu) +\\
												& + RD_3(\Pi_e,w_{\Pi_e},\Pi_o,w_{\dot{L}},\mu) + RD_4(\Pi_e,w_{\Pi_e},\Pi_o,w_{\dot{L}},\mu)\} =\\
												& = \frac{1}{4} \left\{0 + 0 + 0 + 0\right\} = 0
\end{aligned}
\end{equation*}

So, for every pair $\left\langle\Pi_e,w_{\Pi_e}\right\rangle$ we obtain the same result:

\begin{equation*}
\forall \Pi_e,w_{\Pi_e} : RD(\Pi_e,w_{\Pi_e},\Pi_o,w_{\dot{L}},\mu,w_S) = 0
\end{equation*}

Therefore, predator-prey has $Left_{max} = 0$ for this property.
\end{proof}
\end{proposition}

\begin{approximation}
\label{approx:predator-prey_RD_right_min}
$Right_{min}$ for the reward dependency (RD) property is equal to $\frac{13}{28}$ (as a {\em lower} approximation) for the predator-prey environment.

\begin{proof}
To find $Right_{min}$ (equation \ref{eq:right_min}), we need to find a pair $\left\langle\Pi_e,w_{\Pi_e}\right\rangle$ which minimises the property as much as possible while $\Pi_o$ maximises it. Using $\Pi_e = \{\pi_{chase}\}$ with $w_{\Pi_e}(\pi_{chase}) = 1$ and $\Pi_o = \{\pi_{lose},\pi_{win}\}$ (a $\pi_{chase}$ agent always tries to be chased when playing as the prey and tries to chase when playing as a predator, a $\pi_{lose}$ agent always tries to be chased when playing as the prey and tries to do not chase when playing as the predator, and a $\pi_{win}$ agent always tries to not be chased when playing as the prey and tries to chase when playing as the predator) we find a {\em lower} approximation with this situation.

Following definition \ref{def:RD}, we obtain the RD value for this $\left\langle\Pi_e,w_{\Pi_e},\Pi_o\right\rangle$. Since the environment is not symmetric, we need to calculate this property for every slot. Following definition \ref{def:RD_set}, we could calculate its RD value for each slot but, since $\Pi_e$ has only one agent and its weight is equal to $1$, it is equivalent to use directly definition \ref{def:RD_agent}. We start with slot 1:

\begin{equation*}
\begin{aligned}
RD_1(\pi_{chase},\Pi_o,w_{\dot{L}},\mu)	& = \eta_{\dot{L}^2} \sum_{\dot{u},\dot{v} \in \dot{L}^{N(\mu)}_{-1}(\Pi_o) | \dot{u} \neq \dot{v}} w_{\dot{L}}(\dot{u}) w_{\dot{L}}(\dot{v}) \Delta_{\mathbb{Q}}(R_1(\mu[\instantiation{u}{1}{\pi_{chase}}]), R_1(\mu[\instantiation{v}{1}{\pi_{chase}}])) =\\
								& = 2 \frac{8}{7} \frac{1}{8} \frac{1}{8} \{\Delta_{\mathbb{Q}}(R_1(\mu[\pi_{chase},\pi_{lose},\pi_{lose},\pi_{lose}]), R_1(\mu[\pi_{chase},\pi_{lose},\pi_{lose},\pi_{win}])) +\\
								& + \Delta_{\mathbb{Q}}(R_1(\mu[\pi_{chase},\pi_{lose},\pi_{lose},\pi_{lose}]), R_1(\mu[\pi_{chase},\pi_{lose},\pi_{win},\pi_{lose}])) +\\
								& \ \ \ \ \ \ \ \ \ \ \ \ \ \ \ \ \ \ \ \ \ \ \ \ \ \ \ \ \ \ \ \ \ \ \ \ \ \ \ \ \ \ \ \ \ \ \ \ \ \ \vdots\\
								& + \Delta_{\mathbb{Q}}(R_1(\mu[\pi_{chase},\pi_{win},\pi_{win},\pi_{lose}]), R_1(\mu[\pi_{chase},\pi_{win},\pi_{win},\pi_{win}]))\}
\end{aligned}
\end{equation*}

\noindent Note that we avoided to calculate both $\Delta_{\mathbb{Q}}(a,b)$ and $\Delta_{\mathbb{Q}}(b,a)$ since they provide the same result, by calculating only $\Delta_{\mathbb{Q}}(a,b)$ and multiplying the result by $2$.

From the 28 possible pairs of line-ups that we obtained, $\pi_{chase}$ tries to make equal the maximum number of pairs, while $\pi_{win}$ and $\pi_{lose}$ try to diverge the maximum number of pairs. In this case, the agents from $\Pi_o$ can only assure that two line-up patterns obtain different results ($(*,\pi_{lose},\pi_{lose},\pi_{lose})$ and $(*,\pi_{win},\pi_{win},\pi_{win})$), therefore the agent from $\Pi_e$ ($\pi_{chase}$) can make equal seven of the eight line-ups\footnote{Note that only one predator trying to win is enough to chase a prey who wants to be chased.}. So:

\begin{equation*}
RD_1(\pi_{chase},\Pi_o,w_{\dot{L}},\mu) = 2 \frac{8}{7} \frac{1}{8} \frac{1}{8} \left\{7 \times 1 + 21 \times 0\right\} = \frac{1}{4}
\end{equation*}

For slot 2:

\begin{equation*}
\begin{aligned}
RD_2(\pi_{chase},\Pi_o,w_{\dot{L}},\mu)	& = \eta_{\dot{L}^2} \sum_{\dot{u},\dot{v} \in \dot{L}^{N(\mu)}_{-2}(\Pi_o) | \dot{u} \neq \dot{v}} w_{\dot{L}}(\dot{u}) w_{\dot{L}}(\dot{v}) \Delta_{\mathbb{Q}}(R_2(\mu[\instantiation{u}{2}{\pi_{chase}}]), R_2(\mu[\instantiation{v}{2}{\pi_{chase}}])) =\\
								& = 2 \frac{8}{7} \frac{1}{8} \frac{1}{8} \{\Delta_{\mathbb{Q}}(R_2(\mu[\pi_{lose},\pi_{chase},\pi_{lose},\pi_{lose}]), R_2(\mu[\pi_{lose},\pi_{chase},\pi_{lose},\pi_{win}])) +\\
								& + \Delta_{\mathbb{Q}}(R_2(\mu[\pi_{lose},\pi_{chase},\pi_{lose},\pi_{lose}]), R_2(\mu[\pi_{lose},\pi_{chase},\pi_{win},\pi_{lose}])) +\\
								& \ \ \ \ \ \ \ \ \ \ \ \ \ \ \ \ \ \ \ \ \ \ \ \ \ \ \ \ \ \ \ \ \ \ \ \ \ \ \ \ \ \ \ \ \ \ \ \ \ \ \vdots\\
								& + \Delta_{\mathbb{Q}}(R_2(\mu[\pi_{win},\pi_{chase},\pi_{win},\pi_{lose}]), R_2(\mu[\pi_{win},\pi_{chase},\pi_{win},\pi_{win}]))\}
\end{aligned}
\end{equation*}

From the 28 possible pairs of line-ups that we obtained, $\pi_{chase}$ tries to make equal the maximum number of pairs, while $\pi_{win}$ and $\pi_{lose}$ try to diverge the maximum number of pairs. In this case, the agents from $\Pi_o$ can assure that three line-up patterns obtain the same result ($(\pi_{lose},*,\pi_{lose},\pi_{win})$, $(\pi_{lose},*,\pi_{win},\pi_{lose})$ and $(\pi_{lose},*,\pi_{win},\pi_{win})$) and other three line-up patterns obtain a different result ($(\pi_{win},*,\pi_{lose},\pi_{lose})$, $(\pi_{win},*,\pi_{lose},\pi_{win})$ and $(\pi_{win},*,\pi_{win},\pi_{lose})$), therefore the agent from $\Pi_e$ ($\pi_{chase}$) can only make equal five of the eight line-ups. So:

\begin{equation*}
RD_2(\pi_{chase},\Pi_o,w_{\dot{L}},\mu) = 2 \frac{8}{7} \frac{1}{8} \frac{1}{8} \left\{15 \times 1 + 13 \times 0\right\} = \frac{15}{28}
\end{equation*}

For slot 3:

\begin{equation*}
\begin{aligned}
RD_3(\pi_{chase},\Pi_o,w_{\dot{L}},\mu)	& = \eta_{\dot{L}^2} \sum_{\dot{u},\dot{v} \in \dot{L}^{N(\mu)}_{-3}(\Pi_o) | \dot{u} \neq \dot{v}} w_{\dot{L}}(\dot{u}) w_{\dot{L}}(\dot{v}) \Delta_{\mathbb{Q}}(R_3(\mu[\instantiation{u}{3}{\pi_{chase}}]), R_3(\mu[\instantiation{v}{3}{\pi_{chase}}])) =\\
								& = 2 \frac{8}{7} \frac{1}{8} \frac{1}{8} \{\Delta_{\mathbb{Q}}(R_3(\mu[\pi_{lose},\pi_{lose},\pi_{chase},\pi_{lose}]), R_3(\mu[\pi_{lose},\pi_{lose},\pi_{chase},\pi_{win}])) +\\
								& + \Delta_{\mathbb{Q}}(R_3(\mu[\pi_{lose},\pi_{lose},\pi_{chase},\pi_{lose}]), R_3(\mu[\pi_{lose},\pi_{win},\pi_{chase},\pi_{lose}])) +\\
								& \ \ \ \ \ \ \ \ \ \ \ \ \ \ \ \ \ \ \ \ \ \ \ \ \ \ \ \ \ \ \ \ \ \ \ \ \ \ \ \ \ \ \ \ \ \ \ \ \ \ \vdots\\
								& + \Delta_{\mathbb{Q}}(R_3(\mu[\pi_{win},\pi_{win},\pi_{chase},\pi_{lose}]), R_3(\mu[\pi_{win},\pi_{win},\pi_{chase},\pi_{win}]))\}
\end{aligned}
\end{equation*}

From the 28 possible pairs of line-ups that we obtained, $\pi_{chase}$ tries to make equal the maximum number of pairs, while $\pi_{win}$ and $\pi_{lose}$ try to diverge the maximum number of pairs. In this case, the agents from $\Pi_o$ can assure that three line-up patterns obtain the same result ($(\pi_{lose},\pi_{lose},*,\pi_{win})$, $(\pi_{lose},\pi_{win},*,\pi_{lose})$ and $(\pi_{lose},\pi_{win},*,\pi_{win})$) and other three line-up patterns obtain a different result ($(\pi_{win},\pi_{lose},*,\pi_{lose})$, $(\pi_{win},\pi_{lose},*,\pi_{win})$ and $(\pi_{win},\pi_{win},*,\pi_{lose})$), therefore the agent from $\Pi_e$ ($\pi_{chase}$) can only make equal five of the eight line-ups. So:

\begin{equation*}
RD_3(\pi_{chase},\Pi_o,w_{\dot{L}},\mu) = 2 \frac{8}{7} \frac{1}{8} \frac{1}{8} \left\{15 \times 1 + 13 \times 0\right\} = \frac{15}{28}
\end{equation*}

And for slot 4:

\begin{equation*}
\begin{aligned}
RD_4(\pi_{chase},\Pi_o,w_{\dot{L}},\mu)	& = \eta_{\dot{L}^2} \sum_{\dot{u},\dot{v} \in \dot{L}^{N(\mu)}_{-4}(\Pi_o) | \dot{u} \neq \dot{v}} w_{\dot{L}}(\dot{u}) w_{\dot{L}}(\dot{v}) \Delta_{\mathbb{Q}}(R_4(\mu[\instantiation{u}{4}{\pi_{chase}}]), R_4(\mu[\instantiation{v}{4}{\pi_{chase}}])) =\\
								& = 2 \frac{8}{7} \frac{1}{8} \frac{1}{8} \{\Delta_{\mathbb{Q}}(R_4(\mu[\pi_{lose},\pi_{lose},\pi_{lose},\pi_{chase}]), R_4(\mu[\pi_{lose},\pi_{lose},\pi_{win},\pi_{chase}])) +\\
								& + \Delta_{\mathbb{Q}}(R_4(\mu[\pi_{lose},\pi_{lose},\pi_{lose},\pi_{chase}]), R_4(\mu[\pi_{lose},\pi_{win},\pi_{lose},\pi_{chase}])) +\\
								& \ \ \ \ \ \ \ \ \ \ \ \ \ \ \ \ \ \ \ \ \ \ \ \ \ \ \ \ \ \ \ \ \ \ \ \ \ \ \ \ \ \ \ \ \ \ \ \ \ \ \vdots\\
								& + \Delta_{\mathbb{Q}}(R_4(\mu[\pi_{win},\pi_{win},\pi_{lose},\pi_{chase}]), R_4(\mu[\pi_{win},\pi_{win},\pi_{win},\pi_{chase}]))\}
\end{aligned}
\end{equation*}

From the 28 possible pairs of line-ups that we obtained, $\pi_{chase}$ tries to make equal the maximum number of pairs, while $\pi_{win}$ and $\pi_{lose}$ try to diverge the maximum number of pairs. In this case, the agents from $\Pi_o$ can assure that three line-up patterns obtain the same result ($(\pi_{lose},\pi_{lose},\pi_{win},*)$, $(\pi_{lose},\pi_{win},\pi_{lose},*)$ and $(\pi_{lose},\pi_{win},\pi_{win},*)$) and other three line-up patterns obtain a different result ($(\pi_{win},\pi_{lose},\pi_{lose},*)$, $(\pi_{win},\pi_{lose},\pi_{win},*)$ and $(\pi_{win},\pi_{win},\pi_{lose},*)$), therefore the agent from $\Pi_e$ ($\pi_{chase}$) can only make equal five of the eight line-ups. So:

\begin{equation*}
RD_4(\pi_{chase},\Pi_o,w_{\dot{L}},\mu) = 2 \frac{8}{7} \frac{1}{8} \frac{1}{8} \left\{15 \times 1 + 13 \times 0\right\} = \frac{15}{28}
\end{equation*}

And finally, we weight over the slots:

\begin{equation*}
\begin{aligned}
RD(\Pi_e,w_{\Pi_e},\Pi_o,w_{\dot{L}},\mu,w_S)	& =	\sum_{i = 1}^{N(\mu)} w_S(i,\mu) RD_i(\Pi_e,w_{\Pi_e},\Pi_o,w_{\dot{L}},\mu) =\\
												& =	\sum_{i = 1}^{N(\mu)} w_S(i,\mu) RD_i(\pi_{chase},\Pi_o,w_{\dot{L}},\mu) =\\
												& = \frac{1}{4} \{RD_1(\pi_{chase},\Pi_o,w_{\dot{L}},\mu) + RD_2(\pi_{chase},\Pi_o,w_{\dot{L}},\mu) +\\
												& + RD_3(\pi_{chase},\Pi_o,w_{\dot{L}},\mu) + RD_4(\pi_{chase},\Pi_o,w_{\dot{L}},\mu)\} =\\
												& = \frac{1}{4} \left\{\frac{1}{4} + \frac{15}{28} + \frac{15}{28} + \frac{15}{28}\right\} = \frac{13}{28}
\end{aligned}
\end{equation*}

Therefore, predator-prey has $Right_{min} = \frac{13}{28}$ (as a {\em lower} approximation) for this property.
\end{proof}
\end{approximation}

\subsection{Fine Discrimination}
Now we move to the fine discrimination (FD) property. As given in section \ref{sec:FD}, we want to know if different evaluated agents obtain different expected average rewards when interacting in the environment. We use $\Delta_{\mathbb{Q}}(a,b) = 1$ if numbers $a$ and $b$ are equal and $0$ otherwise.

\begin{proposition}
\label{prop:predator-prey_FD_general_min}
$General_{min}$ for the fine discrimination (FD) property is equal to $0$ for the predator-prey environment.

\begin{proof}
To find $General_{min}$ (equation \ref{eq:general_min}), we need to find a trio $\left\langle\Pi_e,w_{\Pi_e},\Pi_o\right\rangle$ which minimises the property as much as possible. We can have this situation by selecting $\Pi_e = \{{\pi_u}_1,{\pi_u}_2\}$ with uniform weight for $w_{\Pi_e}$ and $\Pi_o = \{\pi_d\}$ (a $\pi_u$ agent always performs Up and a $\pi_d$ agent always performs Down).

Following definition \ref{def:FD}, we obtain the FD value for this $\left\langle\Pi_e,w_{\Pi_e},\Pi_o\right\rangle$. Since the environment is not symmetric, we need to calculate this property for every slot. Following definition \ref{def:FD_set}, we can calculate its FD value for each slot. We start with slot 1:

\begin{equation*}
\begin{aligned}
FD_1(\Pi_e,w_{\Pi_e},\Pi_o,w_{\dot{L}},\mu)	& = \eta_{\Pi^2} \sum_{\pi_1,\pi_2 \in \Pi_e | \pi_1 \neq \pi_2} w_{\Pi_e}(\pi_1) w_{\Pi_e}(\pi_2) FD_1(\pi_1,\pi_2,\Pi_o,w_{\dot{L}},\mu) =\\
											& = 2 \frac{2}{1} \frac{1}{2} \frac{1}{2} FD_1({\pi_u}_1,{\pi_u}_2,\Pi_o,w_{\dot{L}},\mu)
\end{aligned}
\end{equation*}

\noindent Note that we avoided to calculate both $FD_i(\pi_1,\pi_2,\Pi_o,w_{\dot{L}},\mu)\}$ and $FD_i(\pi_2,\pi_1,\Pi_o,w_{\dot{L}},\mu)\}$ since they provide the same result, by calculating only $FD_i(\pi_1,\pi_2,\Pi_o,w_{\dot{L}},\mu)\}$ and multiplying the result by $2$.

In this case, we only need to calculate $FD_1({\pi_u}_1,{\pi_u}_2,\Pi_o,w_{\dot{L}},\mu)$. We follow definition \ref{def:FD_agents} to calculate this value:

\begin{equation*}
\begin{aligned}
FD_1({\pi_u}_1,{\pi_u}_2,\Pi_o,w_{\dot{L}},\mu)	& = \sum_{\dot{l} \in \dot{L}^{N(\mu)}_{-1}(\Pi_o)} w_{\dot{L}}(\dot{l}) \Delta_{\mathbb{Q}}(R_1(\mu[\instantiation{l}{1}{{\pi_u}_1}]), R_1(\mu[\instantiation{l}{1}{{\pi_u}_2}])) =\\
												& = \Delta_{\mathbb{Q}}(R_1(\mu[{\pi_u}_1,\pi_d,\pi_d,\pi_d]), R_1(\mu[{\pi_u}_2,\pi_d,\pi_d,\pi_d]))
\end{aligned}
\end{equation*}

Here, a $\pi_u$ agent will always perform Up and $\pi_d$ will always perform Down, so both agents in slot 1 (${\pi_u}_1$ and ${\pi_u}_2$) will obtain the same expected average reward ($1$). So:

\begin{equation*}
FD_1({\pi_u}_1,{\pi_u}_2,\Pi_o,w_{\dot{L}},\mu) = 0
\end{equation*}

Therefore:

\begin{equation*}
FD_1(\Pi_e,w_{\Pi_e},\Pi_o,w_{\dot{L}},\mu) = 2 \frac{2}{1} \frac{1}{2} \frac{1}{2} 0 = 0
\end{equation*}

For slot 2:

\begin{equation*}
\begin{aligned}
FD_2(\Pi_e,w_{\Pi_e},\Pi_o,w_{\dot{L}},\mu)	& = \eta_{\Pi^2} \sum_{\pi_1,\pi_2 \in \Pi_e | \pi_1 \neq \pi_2} w_{\Pi_e}(\pi_1) w_{\Pi_e}(\pi_2) FD_2(\pi_1,\pi_2,\Pi_o,w_{\dot{L}},\mu) =\\
											& = 2 \frac{2}{1} \frac{1}{2} \frac{1}{2} FD_2({\pi_u}_1,{\pi_u}_2,\Pi_o,w_{\dot{L}},\mu)
\end{aligned}
\end{equation*}

Again, we only need to calculate $FD_2({\pi_u}_1,{\pi_u}_2,\Pi_o,w_{\dot{L}},\mu)$. We follow definition \ref{def:FD_agents} to calculate this value:

\begin{equation*}
\begin{aligned}
FD_2({\pi_u}_1,{\pi_u}_2,\Pi_o,w_{\dot{L}},\mu)	& = \sum_{\dot{l} \in \dot{L}^{N(\mu)}_{-2}(\Pi_o)} w_{\dot{L}}(\dot{l}) \Delta_{\mathbb{Q}}(R_2(\mu[\instantiation{l}{2}{{\pi_u}_1}]), R_2(\mu[\instantiation{l}{2}{{\pi_u}_2}])) =\\
												& = \Delta_{\mathbb{Q}}(R_2(\mu[\pi_d,{\pi_u}_1,\pi_d,\pi_d]), R_2(\mu[\pi_d,{\pi_u}_2,\pi_d,\pi_d]))
\end{aligned}
\end{equation*}

Again, a $\pi_u$ agent will always perform Up and $\pi_d$ will always perform Down, so both agents in slot 2 (${\pi_u}_1$ and ${\pi_u}_2$) will obtain the same expected average reward ($1$). So:

\begin{equation*}
FD_2({\pi_u}_1,{\pi_u}_2,\Pi_o,w_{\dot{L}},\mu) = 0
\end{equation*}

Therefore:

\begin{equation*}
FD_2(\Pi_e,w_{\Pi_e},\Pi_o,w_{\dot{L}},\mu) = 2 \frac{2}{1} \frac{1}{2} \frac{1}{2} 0 = 0
\end{equation*}

For slot 3:

\begin{equation*}
\begin{aligned}
FD_3(\Pi_e,w_{\Pi_e},\Pi_o,w_{\dot{L}},\mu)	& = \eta_{\Pi^2} \sum_{\pi_1,\pi_2 \in \Pi_e | \pi_1 \neq \pi_2} w_{\Pi_e}(\pi_1) w_{\Pi_e}(\pi_2) FD_3(\pi_1,\pi_2,\Pi_o,w_{\dot{L}},\mu) =\\
											& = 2 \frac{2}{1} \frac{1}{2} \frac{1}{2} FD_3({\pi_u}_1,{\pi_u}_2,\Pi_o,w_{\dot{L}},\mu)
\end{aligned}
\end{equation*}

Again, we only need to calculate $FD_3({\pi_u}_1,{\pi_u}_2,\Pi_o,w_{\dot{L}},\mu)$. We follow definition \ref{def:FD_agents} to calculate this value:

\begin{equation*}
\begin{aligned}
FD_3({\pi_u}_1,{\pi_u}_2,\Pi_o,w_{\dot{L}},\mu)	& = \sum_{\dot{l} \in \dot{L}^{N(\mu)}_{-3}(\Pi_o)} w_{\dot{L}}(\dot{l}) \Delta_{\mathbb{Q}}(R_3(\mu[\instantiation{l}{3}{{\pi_u}_1}]), R_3(\mu[\instantiation{l}{3}{{\pi_u}_2}])) =\\
												& = \Delta_{\mathbb{Q}}(R_3(\mu[\pi_d,\pi_d,{\pi_u}_1,\pi_d]), R_3(\mu[\pi_d,\pi_d,{\pi_u}_2,\pi_d]))
\end{aligned}
\end{equation*}

Again, a $\pi_u$ agent will always perform Up and $\pi_d$ will always perform Down, so both agents in slot 3 (${\pi_u}_1$ and ${\pi_u}_2$) will obtain the same expected average reward ($-1$). So:

\begin{equation*}
FD_3({\pi_u}_1,{\pi_u}_2,\Pi_o,w_{\dot{L}},\mu) = 0
\end{equation*}

Therefore:

\begin{equation*}
FD_3(\Pi_e,w_{\Pi_e},\Pi_o,w_{\dot{L}},\mu) = 2 \frac{2}{1} \frac{1}{2} \frac{1}{2} 0 = 0
\end{equation*}

And for slot 4:

\begin{equation*}
\begin{aligned}
FD_4(\Pi_e,w_{\Pi_e},\Pi_o,w_{\dot{L}},\mu)	& = \eta_{\Pi^2} \sum_{\pi_1,\pi_2 \in \Pi_e | \pi_1 \neq \pi_2} w_{\Pi_e}(\pi_1) w_{\Pi_e}(\pi_2) FD_4(\pi_1,\pi_2,\Pi_o,w_{\dot{L}},\mu) =\\
											& = 2 \frac{2}{1} \frac{1}{2} \frac{1}{2} FD_4({\pi_u}_1,{\pi_u}_2,\Pi_o,w_{\dot{L}},\mu)
\end{aligned}
\end{equation*}

Again, we only need to calculate $FD_4({\pi_u}_1,{\pi_u}_2,\Pi_o,w_{\dot{L}},\mu)$. We follow definition \ref{def:FD_agents} to calculate this value:

\begin{equation*}
\begin{aligned}
FD_4({\pi_u}_1,{\pi_u}_2,\Pi_o,w_{\dot{L}},\mu)	& = \sum_{\dot{l} \in \dot{L}^{N(\mu)}_{-4}(\Pi_o)} w_{\dot{L}}(\dot{l}) \Delta_{\mathbb{Q}}(R_4(\mu[\instantiation{l}{4}{{\pi_u}_1}]), R_4(\mu[\instantiation{l}{4}{{\pi_u}_2}])) =\\
												& = \Delta_{\mathbb{Q}}(R_4(\mu[\pi_d,\pi_d,\pi_d,{\pi_u}_1]), R_4(\mu[\pi_d,\pi_d,\pi_d,{\pi_u}_2]))
\end{aligned}
\end{equation*}

Again, a $\pi_u$ agent will always perform Up and $\pi_d$ will always perform Down, so both agents in slot 4 (${\pi_u}_1$ and ${\pi_u}_2$) will obtain the same expected average reward ($1$). So:

\begin{equation*}
FD_4({\pi_u}_1,{\pi_u}_2,\Pi_o,w_{\dot{L}},\mu) = 0
\end{equation*}

Therefore:

\begin{equation*}
FD_4(\Pi_e,w_{\Pi_e},\Pi_o,w_{\dot{L}},\mu) = 2 \frac{2}{1} \frac{1}{2} \frac{1}{2} 0 = 0
\end{equation*}

And finally, we weight over the slots:

\begin{equation*}
\begin{aligned}
FD(\Pi_e,w_{\Pi_e},\Pi_o,w_{\dot{L}},\mu,w_S)	& =	\sum_{i = 1}^{N(\mu)} w_S(i,\mu) FD_i(\Pi_e,w_{\Pi_e},\Pi_o,w_{\dot{L}},\mu) =\\
												& = \frac{1}{4} \{ FD_1(\Pi_e,w_{\Pi_e},\Pi_o,w_{\dot{L}},\mu) + FD_2(\Pi_e,w_{\Pi_e},\Pi_o,w_{\dot{L}},\mu) +\\
												& + FD_3(\Pi_e,w_{\Pi_e},\Pi_o,w_{\dot{L}},\mu) + FD_4(\Pi_e,w_{\Pi_e},\Pi_o,w_{\dot{L}},\mu)\} =\\
												& = \frac{1}{4} \left\{0 + 0 + 0 + 0\right\} = 0
\end{aligned}
\end{equation*}

Since $0$ is the lowest possible value for the fine discriminative property, therefore predator-prey has $General_{min} = 0$ for this property.
\end{proof}
\end{proposition}

\begin{proposition}
\label{prop:predator-prey_FD_general_max}
$General_{max}$ for the fine discrimination (FD) property is equal to $1$ for the predator-prey environment.

\begin{proof}
To find $General_{max}$ (equation \ref{eq:general_max}), we need to find a trio $\left\langle\Pi_e,w_{\Pi_e},\Pi_o\right\rangle$ which maximises the property as much as possible. We can have this situation by selecting $\Pi_e = \{\pi_s,\pi_r\}$ with uniform weight for $w_{\Pi_e}$ and $\Pi_o = \{\pi_s\}$ (a $\pi_r$ agent always acts randomly and a $\pi_s$ agent always stays in the same cell\footnote{Note that every cell has an action which is blocked by a block or a boundary, therefore an agent performing this action will stay at its current cell.}).

Following definition \ref{def:FD}, we obtain the FD value for this $\left\langle\Pi_e,w_{\Pi_e},\Pi_o\right\rangle$. Since the environment is not symmetric, we need to calculate this property for every slot. Following definition \ref{def:FD_set}, we can calculate its FD value for each slot. We start with slot 1:

\begin{equation*}
\begin{aligned}
FD_1(\Pi_e,w_{\Pi_e},\Pi_o,w_{\dot{L}},\mu)	& = \eta_{\Pi^2} \sum_{\pi_1,\pi_2 \in \Pi_e | \pi_1 \neq \pi_2} w_{\Pi_e}(\pi_1) w_{\Pi_e}(\pi_2) FD_1(\pi_1,\pi_2,\Pi_o,w_{\dot{L}},\mu) =\\
											& = 2 \frac{2}{1} \frac{1}{2} \frac{1}{2} FD_1(\pi_s,\pi_r,\Pi_o,w_{\dot{L}},\mu)
\end{aligned}
\end{equation*}

\noindent Note that we avoided to calculate both $FD_i(\pi_1,\pi_2,\Pi_o,w_{\dot{L}},\mu)\}$ and $FD_i(\pi_2,\pi_1,\Pi_o,w_{\dot{L}},\mu)\}$ since they provide the same result, by calculating only $FD_i(\pi_1,\pi_2,\Pi_o,w_{\dot{L}},\mu)\}$ and multiplying the result by $2$.

In this case, we only need to calculate $FD_1(\pi_s,\pi_r,\Pi_o,w_{\dot{L}},\mu)$. We follow definition \ref{def:FD_agents} to calculate this value:

\begin{equation*}
\begin{aligned}
FD_1(\pi_s,\pi_r,\Pi_o,w_{\dot{L}},\mu)	& = \sum_{\dot{l} \in \dot{L}^{N(\mu)}_{-1}(\Pi_o)} w_{\dot{L}}(\dot{l}) \Delta_{\mathbb{Q}}(R_1(\mu[\instantiation{l}{1}{\pi_s}]), R_1(\mu[\instantiation{l}{1}{\pi_r}])) =\\
												& = \Delta_{\mathbb{Q}}(R_1(\mu[\pi_s,\pi_s,\pi_s,\pi_s]), R_1(\mu[\pi_r,\pi_s,\pi_s,\pi_s]))
\end{aligned}
\end{equation*}

Here, a $\pi_s$ agent will always stay in the same cell and $\pi_r$ will always act randomly. In this case, $\pi_s$ will never been chased but $\pi_r$ will have a possibility to be chased, so they will obtain different expected average rewards. So:

\begin{equation*}
FD_1(\pi_s,\pi_r,\Pi_o,w_{\dot{L}},\mu) = 1
\end{equation*}

Therefore:

\begin{equation*}
FD_1(\Pi_e,w_{\Pi_e},\Pi_o,w_{\dot{L}},\mu) = 2 \frac{2}{1} \frac{1}{2} \frac{1}{2} 1 = 1
\end{equation*}

For slot 2:

\begin{equation*}
\begin{aligned}
FD_2(\Pi_e,w_{\Pi_e},\Pi_o,w_{\dot{L}},\mu)	& = \eta_{\Pi^2} \sum_{\pi_1,\pi_2 \in \Pi_e | \pi_1 \neq \pi_2} w_{\Pi_e}(\pi_1) w_{\Pi_e}(\pi_2) FD_2(\pi_1,\pi_2,\Pi_o,w_{\dot{L}},\mu) =\\
											& = 2 \frac{2}{1} \frac{1}{2} \frac{1}{2} FD_2(\pi_s,\pi_r,\Pi_o,w_{\dot{L}},\mu)
\end{aligned}
\end{equation*}

Again, we only need to calculate $FD_2(\pi_s,\pi_r,\Pi_o,w_{\dot{L}},\mu)$. We follow definition \ref{def:FD_agents} to calculate this value:

\begin{equation*}
\begin{aligned}
FD_2(\pi_s,\pi_r,\Pi_o,w_{\dot{L}},\mu)	& = \sum_{\dot{l} \in \dot{L}^{N(\mu)}_{-2}(\Pi_o)} w_{\dot{L}}(\dot{l}) \Delta_{\mathbb{Q}}(R_2(\mu[\instantiation{l}{2}{\pi_s}]), R_2(\mu[\instantiation{l}{2}{\pi_r}])) =\\
												& = \Delta_{\mathbb{Q}}(R_2(\mu[\pi_s,\pi_s,\pi_s,\pi_s]), R_2(\mu[\pi_s,\pi_r,\pi_s,\pi_s]))
\end{aligned}
\end{equation*}

Again, a $\pi_s$ agent will always stay in the same cell and $\pi_r$ will always act randomly. In this case, $\pi_s$ will never chase the prey but $\pi_r$ will have a possibility to chase the prey, so they will obtain different expected average rewards. So:

\begin{equation*}
FD_2(\pi_s,\pi_r,\Pi_o,w_{\dot{L}},\mu) = 1
\end{equation*}

Therefore:

\begin{equation*}
FD_2(\Pi_e,w_{\Pi_e},\Pi_o,w_{\dot{L}},\mu) = 2 \frac{2}{1} \frac{1}{2} \frac{1}{2} 1 = 1
\end{equation*}

For slot 3:

\begin{equation*}
\begin{aligned}
FD_3(\Pi_e,w_{\Pi_e},\Pi_o,w_{\dot{L}},\mu)	& = \eta_{\Pi^2} \sum_{\pi_1,\pi_2 \in \Pi_e | \pi_1 \neq \pi_2} w_{\Pi_e}(\pi_1) w_{\Pi_e}(\pi_2) FD_3(\pi_1,\pi_2,\Pi_o,w_{\dot{L}},\mu) =\\
											& = 2 \frac{2}{1} \frac{1}{2} \frac{1}{2} FD_3(\pi_s,\pi_r,\Pi_o,w_{\dot{L}},\mu)
\end{aligned}
\end{equation*}

Again, we only need to calculate $FD_3(\pi_s,\pi_r,\Pi_o,w_{\dot{L}},\mu)$. We follow definition \ref{def:FD_agents} to calculate this value:

\begin{equation*}
\begin{aligned}
FD_3(\pi_s,\pi_r,\Pi_o,w_{\dot{L}},\mu)	& = \sum_{\dot{l} \in \dot{L}^{N(\mu)}_{-3}(\Pi_o)} w_{\dot{L}}(\dot{l}) \Delta_{\mathbb{Q}}(R_3(\mu[\instantiation{l}{3}{\pi_s}]), R_3(\mu[\instantiation{l}{3}{\pi_r}])) =\\
												& = \Delta_{\mathbb{Q}}(R_3(\mu[\pi_s,\pi_s,\pi_s,\pi_s]), R_3(\mu[\pi_s,\pi_s,\pi_r,\pi_s]))
\end{aligned}
\end{equation*}

Again, a $\pi_s$ agent will always stay in the same cell and $\pi_r$ will always act randomly. In this case, $\pi_s$ will never chase the prey but $\pi_r$ will have a possibility to chase the prey, so they will obtain different expected average rewards. So:

\begin{equation*}
FD_3(\pi_s,\pi_r,\Pi_o,w_{\dot{L}},\mu) = 1
\end{equation*}

Therefore:

\begin{equation*}
FD_3(\Pi_e,w_{\Pi_e},\Pi_o,w_{\dot{L}},\mu) = 2 \frac{2}{1} \frac{1}{2} \frac{1}{2} 1 = 1
\end{equation*}

And for slot 4:

\begin{equation*}
\begin{aligned}
FD_4(\Pi_e,w_{\Pi_e},\Pi_o,w_{\dot{L}},\mu)	& = \eta_{\Pi^2} \sum_{\pi_1,\pi_2 \in \Pi_e | \pi_1 \neq \pi_2} w_{\Pi_e}(\pi_1) w_{\Pi_e}(\pi_2) FD_4(\pi_1,\pi_2,\Pi_o,w_{\dot{L}},\mu) =\\
											& = 2 \frac{2}{1} \frac{1}{2} \frac{1}{2} FD_4(\pi_s,\pi_r,\Pi_o,w_{\dot{L}},\mu)
\end{aligned}
\end{equation*}

Again, we only need to calculate $FD_4(\pi_s,\pi_r,\Pi_o,w_{\dot{L}},\mu)$. We follow definition \ref{def:FD_agents} to calculate this value:

\begin{equation*}
\begin{aligned}
FD_4(\pi_s,\pi_r,\Pi_o,w_{\dot{L}},\mu)	& = \sum_{\dot{l} \in \dot{L}^{N(\mu)}_{-4}(\Pi_o)} w_{\dot{L}}(\dot{l}) \Delta_{\mathbb{Q}}(R_4(\mu[\instantiation{l}{4}{\pi_s}]), R_4(\mu[\instantiation{l}{4}{\pi_r}])) =\\
												& = \Delta_{\mathbb{Q}}(R_4(\mu[\pi_s,\pi_s,\pi_s,\pi_s]), R_4(\mu[\pi_s,\pi_s,\pi_s,\pi_r]))
\end{aligned}
\end{equation*}

Again, a $\pi_s$ agent will always stay in the same cell and $\pi_r$ will always act randomly. In this case, $\pi_s$ will never chase the prey but $\pi_r$ will have a possibility to chase the prey, so they will obtain different expected average rewards. So:

\begin{equation*}
FD_4(\pi_s,\pi_r,\Pi_o,w_{\dot{L}},\mu) = 1
\end{equation*}

Therefore:

\begin{equation*}
FD_4(\Pi_e,w_{\Pi_e},\Pi_o,w_{\dot{L}},\mu) = 2 \frac{2}{1} \frac{1}{2} \frac{1}{2} 1 = 1
\end{equation*}

And finally, we weight over the slots:

\begin{equation*}
\begin{aligned}
FD(\Pi_e,w_{\Pi_e},\Pi_o,w_{\dot{L}},\mu,w_S)	& =	\sum_{i = 1}^{N(\mu)} w_S(i,\mu) FD_i(\Pi_e,w_{\Pi_e},\Pi_o,w_{\dot{L}},\mu) =\\
												& = \frac{1}{4} \{FD_1(\Pi_e,w_{\Pi_e},\Pi_o,w_{\dot{L}},\mu) + FD_2(\Pi_e,w_{\Pi_e},\Pi_o,w_{\dot{L}},\mu) +\\
												& + FD_3(\Pi_e,w_{\Pi_e},\Pi_o,w_{\dot{L}},\mu) + FD_4(\Pi_e,w_{\Pi_e},\Pi_o,w_{\dot{L}},\mu)\} =\\
												& = \frac{1}{4} \left\{1 + 1 + 1 + 1\right\} = 1
\end{aligned}
\end{equation*}

Since $1$ is the highest possible value for the fine discriminative property, therefore predator-prey has $General_{max} = 1$ for this property.
\end{proof}
\end{proposition}

\begin{proposition}
\label{prop:predator-prey_FD_left_max}
$Left_{max}$ for the fine discrimination (FD) property is equal to $0$ for the predator-prey environment.

\begin{proof}
To find $Left_{max}$ (equation \ref{eq:left_max}), we need to find a pair $\left\langle\Pi_e,w_{\Pi_e}\right\rangle$ which maximises the property as much as possible while $\Pi_o$ minimises it. Using $\Pi_o = \{\pi_{chase}\}$ (a $\pi_{chase}$ agent always tries to be chased when playing as the prey and tries to chase when playing as a predator) we find this situation no matter which pair $\left\langle\Pi_e,w_{\Pi_e}\right\rangle$ we use.

Following definition \ref{def:FD}, we obtain the FD value for this $\left\langle\Pi_e,w_{\Pi_e},\Pi_o\right\rangle$ (where $\Pi_e$ and $w_{\Pi_e}$ are instantiated with any permitted values). Since the environment is not symmetric, we need to calculate this property for every slot. Following definition \ref{def:FD_set}, we can calculate its FD value for each slot. We start with slot 1:

\begin{equation*}
FD_1(\Pi_e,w_{\Pi_e},\Pi_o,w_{\dot{L}},\mu) = \eta_{\Pi^2} \sum_{\pi_1,\pi_2 \in \Pi_e | \pi_1 \neq \pi_2} w_{\Pi_e}(\pi_1) w_{\Pi_e}(\pi_2) FD_1(\pi_1,\pi_2,\Pi_o,w_{\dot{L}},\mu)
\end{equation*}

We do not know which $\Pi_e$ we have, but we know that we will need to evaluate $FD_1(\pi_1,\pi_2,\Pi_o,w_{\dot{L}},\mu)$ for all pair of evaluated agents $\pi_1,\pi_2 \in \Pi_e | \pi_1 \neq \pi_2$. We follow definition \ref{def:FD_agents} to calculate this value for two figurative evaluated agents $\pi_1$ and $\pi_2$ from $\Pi_e$ such that $\pi_1 \neq \pi_2$:

\begin{equation*}
\begin{aligned}
FD_1(\pi_1,\pi_2,\Pi_o,w_{\dot{L}},\mu)	& = \sum_{\dot{l} \in \dot{L}^{N(\mu)}_{-1}(\Pi_o)} w_{\dot{L}}(\dot{l}) \Delta_{\mathbb{Q}}(R_1(\mu[\instantiation{l}{1}{\pi_1}]), R_1(\mu[\instantiation{l}{1}{\pi_2}])) =\\
										& = \Delta_{\mathbb{Q}}(R_1(\mu[\pi_1,\pi_{chase},\pi_{chase},\pi_{chase}]), R_1(\mu[\pi_2,\pi_{chase},\pi_{chase},\pi_{chase}]))
\end{aligned}
\end{equation*}

Here, the predators will coordinate to always chase the prey as seen in lemma \ref{lemma:predator-prey_always_chase}. So no matter which agents $\pi_1$ and $\pi_2$ are we obtain:

\begin{equation*}
FD_1(\pi_1,\pi_2,\Pi_o,w_{\dot{L}},\mu) = 0
\end{equation*}

Therefore:

\begin{equation*}
FD_1(\Pi_e,w_{\Pi_e},\Pi_o,w_{\dot{L}},\mu) = 0
\end{equation*}

For slot 2:

\begin{equation*}
FD_2(\Pi_e,w_{\Pi_e},\Pi_o,w_{\dot{L}},\mu) = \eta_{\Pi^2} \sum_{\pi_1,\pi_2 \in \Pi_e | \pi_1 \neq \pi_2} w_{\Pi_e}(\pi_1) w_{\Pi_e}(\pi_2) FD_2(\pi_1,\pi_2,\Pi_o,w_{\dot{L}},\mu)
\end{equation*}

We do not know which $\Pi_e$ we have, but we know that we will need to evaluate $FD_2(\pi_1,\pi_2,\Pi_o,w_{\dot{L}},\mu)$ for all pair of evaluated agents $\pi_1,\pi_2 \in \Pi_e | \pi_1 \neq \pi_2$. We follow definition \ref{def:FD_agents} to calculate this value for two figurative evaluated agents $\pi_1$ and $\pi_2$ from $\Pi_e$ such that $\pi_1 \neq \pi_2$:

\begin{equation*}
\begin{aligned}
FD_2(\pi_1,\pi_2,\Pi_o,w_{\dot{L}},\mu)	& = \sum_{\dot{l} \in \dot{L}^{N(\mu)}_{-2}(\Pi_o)} w_{\dot{L}}(\dot{l}) \Delta_{\mathbb{Q}}(R_2(\mu[\instantiation{l}{2}{\pi_1}]), R_2(\mu[\instantiation{l}{2}{\pi_2}])) =\\
										& = \Delta_{\mathbb{Q}}(R_2(\mu[\pi_{chase},\pi_1,\pi_{chase},\pi_{chase}]), R_2(\mu[\pi_{chase},\pi_2,\pi_{chase},\pi_{chase}]))
\end{aligned}
\end{equation*}

Here, the prey will always try to be chased, and also at least one predator is trying to chase they prey, therefore the prey will always been chased. So no matter which agents $\pi_1$ and $\pi_2$ are we obtain:

\begin{equation*}
FD_2(\pi_1,\pi_2,\Pi_o,w_{\dot{L}},\mu) = 0
\end{equation*}

Therefore:

\begin{equation*}
FD_2(\Pi_e,w_{\Pi_e},\Pi_o,w_{\dot{L}},\mu) = 0
\end{equation*}

For slot 3:

\begin{equation*}
FD_3(\Pi_e,w_{\Pi_e},\Pi_o,w_{\dot{L}},\mu) = \eta_{\Pi^2} \sum_{\pi_1,\pi_2 \in \Pi_e | \pi_1 \neq \pi_2} w_{\Pi_e}(\pi_1) w_{\Pi_e}(\pi_2) FD_3(\pi_1,\pi_2,\Pi_o,w_{\dot{L}},\mu)
\end{equation*}

We do not know which $\Pi_e$ we have, but we know that we will need to evaluate $FD_3(\pi_1,\pi_2,\Pi_o,w_{\dot{L}},\mu)$ for all pair of evaluated agents $\pi_1,\pi_2 \in \Pi_e | \pi_1 \neq \pi_2$. We follow definition \ref{def:FD_agents} to calculate this value for two figurative evaluated agents $\pi_1$ and $\pi_2$ from $\Pi_e$ such that $\pi_1 \neq \pi_2$:

\begin{equation*}
\begin{aligned}
FD_3(\pi_1,\pi_2,\Pi_o,w_{\dot{L}},\mu)	& = \sum_{\dot{l} \in \dot{L}^{N(\mu)}_{-3}(\Pi_o)} w_{\dot{L}}(\dot{l}) \Delta_{\mathbb{Q}}(R_3(\mu[\instantiation{l}{3}{\pi_1}]), R_3(\mu[\instantiation{l}{3}{\pi_2}])) =\\
										& = \Delta_{\mathbb{Q}}(R_3(\mu[\pi_{chase},\pi_{chase},\pi_1,\pi_{chase}]), R_3(\mu[\pi_{chase},\pi_{chase},\pi_2,\pi_{chase}]))
\end{aligned}
\end{equation*}

Here, the prey will always try to be chased, and also at least one predator is trying to chase they prey, therefore the prey will always been chased. So no matter which agents $\pi_1$ and $\pi_2$ are we obtain:

\begin{equation*}
FD_3(\pi_1,\pi_2,\Pi_o,w_{\dot{L}},\mu) = 0
\end{equation*}

Therefore:

\begin{equation*}
FD_3(\Pi_e,w_{\Pi_e},\Pi_o,w_{\dot{L}},\mu) = 0
\end{equation*}

And for slot 4:

\begin{equation*}
FD_4(\Pi_e,w_{\Pi_e},\Pi_o,w_{\dot{L}},\mu) = \eta_{\Pi^2} \sum_{\pi_1,\pi_2 \in \Pi_e | \pi_1 \neq \pi_2} w_{\Pi_e}(\pi_1) w_{\Pi_e}(\pi_2) FD_4(\pi_1,\pi_2,\Pi_o,w_{\dot{L}},\mu)
\end{equation*}

We do not know which $\Pi_e$ we have, but we know that we will need to evaluate $FD_4(\pi_1,\pi_2,\Pi_o,w_{\dot{L}},\mu)$ for all pair of evaluated agents $\pi_1,\pi_2 \in \Pi_e | \pi_1 \neq \pi_2$. We follow definition \ref{def:FD_agents} to calculate this value for two figurative evaluated agents $\pi_1$ and $\pi_2$ from $\Pi_e$ such that $\pi_1 \neq \pi_2$:

\begin{equation*}
\begin{aligned}
FD_4(\pi_1,\pi_2,\Pi_o,w_{\dot{L}},\mu)	& = \sum_{\dot{l} \in \dot{L}^{N(\mu)}_{-4}(\Pi_o)} w_{\dot{L}}(\dot{l}) \Delta_{\mathbb{Q}}(R_4(\mu[\instantiation{l}{4}{\pi_1}]), R_4(\mu[\instantiation{l}{4}{\pi_2}])) =\\
										& = \Delta_{\mathbb{Q}}(R_4(\mu[\pi_{chase},\pi_{chase},\pi_{chase},\pi_1]), R_4(\mu[\pi_{chase},\pi_{chase},\pi_{chase},\pi_2]))
\end{aligned}
\end{equation*}

Here, the prey will always try to be chased, and also at least one predator is trying to chase they prey, therefore the prey will always been chased. So no matter which agents $\pi_1$ and $\pi_2$ are we obtain:

\begin{equation*}
FD_4(\pi_1,\pi_2,\Pi_o,w_{\dot{L}},\mu) = 0
\end{equation*}

Therefore:

\begin{equation*}
FD_4(\Pi_e,w_{\Pi_e},\Pi_o,w_{\dot{L}},\mu) = 0
\end{equation*}

And finally, we weight over the slots:

\begin{equation*}
\begin{aligned}
FD(\Pi_e,w_{\Pi_e},\Pi_o,w_{\dot{L}},\mu,w_S)	& = \sum_{i=1}^{N(\mu)} w_S(i,\mu) FD_i(\Pi_e,w_{\Pi_e},\Pi_o,w_{\dot{L}},\mu)\\
												& = \frac{1}{4} \{FD_1(\Pi_e,w_{\Pi_e},\Pi_o,w_{\dot{L}},\mu) + FD_2(\Pi_e,w_{\Pi_e},\Pi_o,w_{\dot{L}},\mu) +\\
												& + FD_3(\Pi_e,w_{\Pi_e},\Pi_o,w_{\dot{L}},\mu) + FD_4(\Pi_e,w_{\Pi_e},\Pi_o,w_{\dot{L}},\mu)\} =\\
												& = \frac{1}{4} \left\{0 + 0 + 0 + 0\right\} = 0
\end{aligned}
\end{equation*}

So, for every pair $\left\langle\Pi_e,w_{\Pi_e}\right\rangle$ we obtain the same result:

\begin{equation*}
\forall \Pi_e,w_{\Pi_e} : FD(\Pi_e,w_{\Pi_e},\Pi_o,w_{\dot{L}},\mu,w_S) = 0
\end{equation*}

Therefore, predator-prey has $Left_{max} = 0$ for this property.
\end{proof}
\end{proposition}

\begin{proposition}
\label{prop:predator-prey_FD_right_min}
$Right_{min}$ for the fine discrimination (FD) property is equal to $0$ for the predator-prey environment.

\begin{proof}
To find $Right_{min}$ (equation \ref{eq:right_min}), we need to find a pair $\left\langle\Pi_e,w_{\Pi_e}\right\rangle$ which minimises the property as much as possible while $\Pi_o$ maximises it. Using $\Pi_e = \{{\pi_u}_1,{\pi_u}_2\}$ with uniform weight for $w_{\Pi_e}$ (a $\pi_u$ agent always performs Up) we find this situation no matter which $\Pi_o$ we use.

Following definition \ref{def:FD}, we obtain the FD value for this $\left\langle\Pi_e,w_{\Pi_e},\Pi_o\right\rangle$ (where $\Pi_o$ is instantiated with any permitted values). Since the environment is not symmetric, we need to calculate this property for every slot. Following definition \ref{def:FD_set}, we can calculate its FD value for each slot. We start with slot 1:

\begin{equation*}
\begin{aligned}
FD_1(\Pi_e,w_{\Pi_e},\Pi_o,w_{\dot{L}},\mu)	& = \eta_{\Pi^2} \sum_{\pi_1,\pi_2 \in \Pi_e | \pi_1 \neq \pi_2} w_{\Pi_e}(\pi_1) w_{\Pi_e}(\pi_2) FD_1(\pi_1,\pi_2,\Pi_o,w_{\dot{L}},\mu) =\\
											& = 2 \frac{2}{1} \frac{1}{2} \frac{1}{2} FD_1({\pi_u}_1,{\pi_u}_2,\Pi_o,w_{\dot{L}},\mu)
\end{aligned}
\end{equation*}

\noindent Note that we avoided to calculate both $FD_i(\pi_1,\pi_2,\Pi_o,w_{\dot{L}},\mu)\}$ and $FD_i(\pi_2,\pi_1,\Pi_o,w_{\dot{L}},\mu)\}$ since they provide the same result, by calculating only $FD_i(\pi_1,\pi_2,\Pi_o,w_{\dot{L}},\mu)\}$ and multiplying the result by $2$.

In this case, we only need to calculate $FD_1({\pi_u}_1,{\pi_u}_2,\Pi_o,w_{\dot{L}},\mu)$. We follow definition \ref{def:FD_agents} to calculate this value:

\begin{equation*}
FD_1({\pi_u}_1,{\pi_u}_2,\Pi_o,w_{\dot{L}},\mu) = \sum_{\dot{l} \in \dot{L}^{N(\mu)}_{-1}(\Pi_o)} w_{\dot{L}}(\dot{l}) \Delta_{\mathbb{Q}}(R_1(\mu[\instantiation{l}{1}{{\pi_u}_1}]), R_1(\mu[\instantiation{l}{1}{{\pi_u}_2}]))
\end{equation*}

We do not know which $\Pi_o$ we have, but we know that we will need to obtain a line-up pattern $\dot{l}$ from $\dot{L}^{N(\mu)}_{-1}(\Pi_o)$ to calculate $\Delta_{\mathbb{Q}}(R_1(\mu[\instantiation{l}{1}{{\pi_u}_1}]), R_1(\mu[\instantiation{l}{1}{{\pi_u}_2}]))$. We calculate this value for a figurative line-up pattern $\dot{l} = (*,\pi_1,\pi_2,\pi_3)$ from $\dot{L}^{N(\mu)}_{-1}(\Pi_o)$:

\begin{equation*}
\Delta_{\mathbb{Q}}(R_1(\mu[\instantiation{l}{1}{{\pi_u}_1}]), R_1(\mu[\instantiation{l}{1}{{\pi_u}_2}])) = \Delta_{\mathbb{Q}}(R_1(\mu[{\pi_u}_1,\pi_1,\pi_2,\pi_3]), R_1(\mu[{\pi_u}_2,\pi_1,\pi_2,\pi_3]))
\end{equation*}

A $\pi_u$ agent will always perform Up, so we obtain a situation where the other agents (any $\pi_1,\pi_2,\pi_3$) will be able to differentiate with which agent they are interacting, so they will not be able to change their distribution of action sequences depending on the behaviour of the agent in slot 1, obtaining both agents in slot 1 (${\pi_u}_1$ and ${\pi_u}_2$) the same expected average reward. So:

\begin{equation*}
FD_1({\pi_u}_1,{\pi_u}_2,\Pi_o,w_{\dot{L}},\mu) = 0
\end{equation*}

Therefore:

\begin{equation*}
FD_1(\Pi_e,w_{\Pi_e},\Pi_o,w_{\dot{L}},\mu) = 2 \frac{2}{1} \frac{1}{2} \frac{1}{2} 0 = 0
\end{equation*}

For slot 2:

\begin{equation*}
\begin{aligned}
FD_2(\Pi_e,w_{\Pi_e},\Pi_o,w_{\dot{L}},\mu)	& = \eta_{\Pi^2} \sum_{\pi_1,\pi_2 \in \Pi_e | \pi_1 \neq \pi_2} w_{\Pi_e}(\pi_1) w_{\Pi_e}(\pi_2) FD_2(\pi_1,\pi_2,\Pi_o,w_{\dot{L}},\mu) =\\
											& = 2 \frac{2}{1} \frac{1}{2} \frac{1}{2} FD_2({\pi_u}_1,{\pi_u}_2,\Pi_o,w_{\dot{L}},\mu)
\end{aligned}
\end{equation*}

In this case, we only need to calculate $FD_2({\pi_u}_1,{\pi_u}_2,\Pi_o,w_{\dot{L}},\mu)$. We follow definition \ref{def:FD_agents} to calculate this value:

\begin{equation*}
FD_2({\pi_u}_1,{\pi_u}_2,\Pi_o,w_{\dot{L}},\mu) = \sum_{\dot{l} \in \dot{L}^{N(\mu)}_{-2}(\Pi_o)} w_{\dot{L}}(\dot{l}) \Delta_{\mathbb{Q}}(R_2(\mu[\instantiation{l}{2}{{\pi_u}_1}]), R_2(\mu[\instantiation{l}{2}{{\pi_u}_2}]))
\end{equation*}

We do not know which $\Pi_o$ we have, but we know that we will need to obtain a line-up pattern $\dot{l}$ from $\dot{L}^{N(\mu)}_{-2}(\Pi_o)$ to calculate $\Delta_{\mathbb{Q}}(R_2(\mu[\instantiation{l}{2}{{\pi_u}_1}]), R_2(\mu[\instantiation{l}{2}{{\pi_u}_2}]))$. We calculate this value for a figurative line-up pattern $\dot{l} = (\pi_1,*,\pi_2,\pi_3)$ from $\dot{L}^{N(\mu)}_{-2}(\Pi_o)$:

\begin{equation*}
\Delta_{\mathbb{Q}}(R_2(\mu[\instantiation{l}{2}{{\pi_u}_1}]), R_2(\mu[\instantiation{l}{2}{{\pi_u}_2}])) = \Delta_{\mathbb{Q}}(R_2(\mu[\pi_1,{\pi_u}_1,\pi_2,\pi_3]), R_2(\mu[\pi_1,{\pi_u}_2,\pi_2,\pi_3]))
\end{equation*}

A $\pi_u$ agent will always perform Up, so we obtain a situation where the other agents (any $\pi_1,\pi_2,\pi_3$) will be able to differentiate with which agent they are interacting, so they will not be able to change their distribution of action sequences depending on the behaviour of the agent in slot 2, obtaining both agents in slot 2 (${\pi_u}_1$ and ${\pi_u}_2$) the same expected average reward. So:

\begin{equation*}
FD_2({\pi_u}_1,{\pi_u}_2,\Pi_o,w_{\dot{L}},\mu) = 0
\end{equation*}

Therefore:

\begin{equation*}
FD_2(\Pi_e,w_{\Pi_e},\Pi_o,w_{\dot{L}},\mu) = 2 \frac{2}{1} \frac{1}{2} \frac{1}{2} 0 = 0
\end{equation*}

For slot 3:

\begin{equation*}
\begin{aligned}
FD_3(\Pi_e,w_{\Pi_e},\Pi_o,w_{\dot{L}},\mu)	& = \eta_{\Pi^2} \sum_{\pi_1,\pi_2 \in \Pi_e | \pi_1 \neq \pi_2} w_{\Pi_e}(\pi_1) w_{\Pi_e}(\pi_2) FD_3(\pi_1,\pi_2,\Pi_o,w_{\dot{L}},\mu) =\\
											& = 2 \frac{2}{1} \frac{1}{2} \frac{1}{2} FD_3({\pi_u}_1,{\pi_u}_2,\Pi_o,w_{\dot{L}},\mu)
\end{aligned}
\end{equation*}

In this case, we only need to calculate $FD_3({\pi_u}_1,{\pi_u}_2,\Pi_o,w_{\dot{L}},\mu)$. We follow definition \ref{def:FD_agents} to calculate this value:

\begin{equation*}
FD_3({\pi_u}_1,{\pi_u}_2,\Pi_o,w_{\dot{L}},\mu) = \sum_{\dot{l} \in \dot{L}^{N(\mu)}_{-3}(\Pi_o)} w_{\dot{L}}(\dot{l}) \Delta_{\mathbb{Q}}(R_3(\mu[\instantiation{l}{3}{{\pi_u}_1}]), R_3(\mu[\instantiation{l}{3}{{\pi_u}_2}]))
\end{equation*}

We do not know which $\Pi_o$ we have, but we know that we will need to obtain a line-up pattern $\dot{l}$ from $\dot{L}^{N(\mu)}_{-3}(\Pi_o)$ to calculate $\Delta_{\mathbb{Q}}(R_3(\mu[\instantiation{l}{3}{{\pi_u}_1}]), R_3(\mu[\instantiation{l}{3}{{\pi_u}_2}]))$. We calculate this value for a figurative line-up pattern $\dot{l} = (\pi_1,\pi_2,*,\pi_3)$ from $\dot{L}^{N(\mu)}_{-3}(\Pi_o)$:

\begin{equation*}
\Delta_{\mathbb{Q}}(R_3(\mu[\instantiation{l}{3}{{\pi_u}_1}]), R_3(\mu[\instantiation{l}{3}{{\pi_u}_2}])) = \Delta_{\mathbb{Q}}(R_3(\mu[\pi_1,\pi_2,{\pi_u}_1,\pi_3]), R_3(\mu[\pi_1,\pi_2,{\pi_u}_2,\pi_3]))
\end{equation*}

A $\pi_u$ agent will always perform Up, so we obtain a situation where the other agents (any $\pi_1,\pi_2,\pi_3$) will be able to differentiate with which agent they are interacting, so they will not be able to change their distribution of action sequences depending on the behaviour of the agent in slot 3, obtaining both agents in slot 3 (${\pi_u}_1$ and ${\pi_u}_2$) the same expected average reward. So:

\begin{equation*}
FD_3({\pi_u}_1,{\pi_u}_2,\Pi_o,w_{\dot{L}},\mu) = 0
\end{equation*}

Therefore:

\begin{equation*}
FD_3(\Pi_e,w_{\Pi_e},\Pi_o,w_{\dot{L}},\mu) = 2 \frac{2}{1} \frac{1}{2} \frac{1}{2} 0 = 0
\end{equation*}

And for slot 4:

\begin{equation*}
\begin{aligned}
FD_4(\Pi_e,w_{\Pi_e},\Pi_o,w_{\dot{L}},\mu)	& = \eta_{\Pi^2} \sum_{\pi_1,\pi_2 \in \Pi_e | \pi_1 \neq \pi_2} w_{\Pi_e}(\pi_1) w_{\Pi_e}(\pi_2) FD_4(\pi_1,\pi_2,\Pi_o,w_{\dot{L}},\mu) =\\
											& = 2 \frac{2}{1} \frac{1}{2} \frac{1}{2} FD_4({\pi_u}_1,{\pi_u}_2,\Pi_o,w_{\dot{L}},\mu)
\end{aligned}
\end{equation*}

In this case, we only need to calculate $FD_4({\pi_u}_1,{\pi_u}_2,\Pi_o,w_{\dot{L}},\mu)$. We follow definition \ref{def:FD_agents} to calculate this value:

\begin{equation*}
FD_4({\pi_u}_1,{\pi_u}_2,\Pi_o,w_{\dot{L}},\mu) = \sum_{\dot{l} \in \dot{L}^{N(\mu)}_{-4}(\Pi_o)} w_{\dot{L}}(\dot{l}) \Delta_{\mathbb{Q}}(R_4(\mu[\instantiation{l}{4}{{\pi_u}_1}]), R_4(\mu[\instantiation{l}{4}{{\pi_u}_2}]))
\end{equation*}

We do not know which $\Pi_o$ we have, but we know that we will need to obtain a line-up pattern $\dot{l}$ from $\dot{L}^{N(\mu)}_{-4}(\Pi_o)$ to calculate $\Delta_{\mathbb{Q}}(R_4(\mu[\instantiation{l}{4}{{\pi_u}_1}]), R_4(\mu[\instantiation{l}{4}{{\pi_u}_2}]))$. We calculate this value for a figurative line-up pattern $\dot{l} = (\pi_1,\pi_2,\pi_3,*)$ from $\dot{L}^{N(\mu)}_{-4}(\Pi_o)$:

\begin{equation*}
\Delta_{\mathbb{Q}}(R_4(\mu[\instantiation{l}{4}{{\pi_u}_1}]), R_4(\mu[\instantiation{l}{4}{{\pi_u}_2}])) = \Delta_{\mathbb{Q}}(R_4(\mu[\pi_1,\pi_2,\pi_3,{\pi_u}_1]), R_4(\mu[\pi_1,\pi_2,\pi_3,{\pi_u}_2]))
\end{equation*}

A $\pi_u$ agent will always perform Up, so we obtain a situation where the other agents (any $\pi_1,\pi_2,\pi_3$) will be able to differentiate with which agent they are interacting, so they will not be able to change their distribution of action sequences depending on the behaviour of the agent in slot 4, obtaining both agents in slot 4 (${\pi_u}_1$ and ${\pi_u}_2$) the same expected average reward. So:

\begin{equation*}
FD_4({\pi_u}_1,{\pi_u}_2,\Pi_o,w_{\dot{L}},\mu) = 0
\end{equation*}

Therefore:

\begin{equation*}
FD_4(\Pi_e,w_{\Pi_e},\Pi_o,w_{\dot{L}},\mu) = 2 \frac{2}{1} \frac{1}{2} \frac{1}{2} 0 = 0
\end{equation*}

And finally, we weight over the slots:

\begin{equation*}
\begin{aligned}
FD(\Pi_e,w_{\Pi_e},\Pi_o,w_{\dot{L}},\mu,w_S)	& = \sum_{i=1}^{N(\mu)} w_S(i,\mu) FD_i(\Pi_e,w_{\Pi_e},\Pi_o,w_{\dot{L}},\mu)\\
												& = \frac{1}{4} \{FD_1(\Pi_e,w_{\Pi_e},\Pi_o,w_{\dot{L}},\mu) + FD_2(\Pi_e,w_{\Pi_e},\Pi_o,w_{\dot{L}},\mu) +\\
												& + FD_3(\Pi_e,w_{\Pi_e},\Pi_o,w_{\dot{L}},\mu) + FD_4(\Pi_e,w_{\Pi_e},\Pi_o,w_{\dot{L}},\mu)\} =\\
												& = \frac{1}{4} \left\{0 + 0 + 0 + 0\right\} = 0
\end{aligned}
\end{equation*}

So, for every $\Pi_o$ we obtain the same result:

\begin{equation*}
\forall \Pi_o : FD(\Pi_e,w_{\Pi_e},\Pi_o,w_{\dot{L}},\mu,w_S) = 0
\end{equation*}

Therefore, predator-prey has $Right_{min} = 0$ for this property.
\end{proof}
\end{proposition}

\subsection{Strict Total Grading}
We arrive to the strict total grading (STG) property. As given in section \ref{sec:STG}, we want to know if there exists a strict ordering between the evaluated agents when interacting in the environment.

To simplify the notation, we use the next table to represent the STO:
$R_i(\mu[\instantiation{l}{i,j}{\pi_1,\pi_2}]) < R_j(\mu[\instantiation{l}{i,j}{\pi_1,\pi_2}])$,
$R_i(\mu[\instantiation{l}{i,j}{\pi_2,\pi_3}]) < R_j(\mu[\instantiation{l}{i,j}{\pi_2,\pi_3}])$ and
$R_i(\mu[\instantiation{l}{i,j}{\pi_1,\pi_3}]) < R_j(\mu[\instantiation{l}{i,j}{\pi_1,\pi_3}])$.

\begin{center}
\begin{tabular}{c c c}
Slot i & & Slot j\\
\hline
$\pi_1$ & $<$ & $\pi_2$\\
$\pi_2$ & $<$ & $\pi_3$\\
$\pi_1$ & $<$ & $\pi_3$
\end{tabular}
\end{center}

\begin{proposition}
\label{prop:predator-prey_STG_general_min}
$General_{min}$ for the strict total grading (STG) property is equal to $0$ for the predator-prey environment.

\begin{proof}
To find $General_{min}$ (equation \ref{eq:general_min}), we need to find a trio $\left\langle\Pi_e,w_{\Pi_e},\Pi_o\right\rangle$ which minimises the property as much as possible. We can have this situation by selecting $\Pi_e = \{{\pi_s}_1,{\pi_s}_2,{\pi_s}_3\}$ with uniform weight for $w_{\Pi_e}$ and $\Pi_o = \{\pi_x\}$ (a $\pi_s$ agent always stays in the same cell\footnote{Note that every cell has an action which is blocked by a block or a boundary, therefore an agent performing this action will stay at its current cell.}, and a $\pi_x$ agent acts stochastically with a probability of $1/\sqrt{2}$ to do not reach the upper left corner and a probability of $1 - 1/\sqrt{2}$ to reach this corner).

Following definition \ref{def:STG}, we obtain the STG value for this $\left\langle\Pi_e,w_{\Pi_e},\Pi_o\right\rangle$. Since the environment is not symmetric, we need to calculate this property for every pair of slots. Following definition \ref{def:STG_set}, we can calculate its STG value for each pair of slots. We start with slots 1 and 2:

\begin{equation*}
\begin{aligned}
STG_{1,2}(\Pi_e,w_{\Pi_e},\Pi_o,w_{\dot{L}},\mu)	& = \eta_{\Pi^3} \sum_{\pi_1,\pi_2,\pi_3 \in \Pi_e | \pi_1 \neq \pi_2 \neq \pi_3} w_{\Pi_e}(\pi_1) w_{\Pi_e}(\pi_2) w_{\Pi_e}(\pi_3) STG_{1,2}(\pi_1,\pi_2,\pi_3,\Pi_o,w_{\dot{L}},\mu) =\\
													& = 6 \frac{9}{2} \frac{1}{3} \frac{1}{3} \frac{1}{3} STG_{1,2}({\pi_s}_1,{\pi_s}_2,{\pi_s}_3,\Pi_o,w_{\dot{L}},\mu)
\end{aligned}
\end{equation*}

\noindent Note that we avoided to calculate all the permutations of $\pi_1,\pi_2,\pi_3$ for $STG_{i,j}(\pi_1,\pi_2,\pi_3,\Pi_o,w_{\dot{L}},\mu)$ since they provide the same result, by calculating only one permutation and multiplying the result by the number of permutations $6$.

In this case, we only need to calculate $STG_{1,2}({\pi_s}_1,{\pi_s}_2,{\pi_s}_3,\Pi_o,w_{\dot{L}},\mu)$. We follow definition \ref{def:STG_agents} to calculate this value:

\begin{equation*}
\begin{aligned}
STG_{1,2}({\pi_s}_1,{\pi_s}_2,{\pi_s}_3,\Pi_o,w_{\dot{L}},\mu)	& = \sum_{\dot{l} \in \dot{L}^{N(\mu)}_{-1,2}(\Pi_o)} w_{\dot{L}}(\dot{l}) STO_{1,2}({\pi_s}_1,{\pi_s}_2,{\pi_s}_3,\dot{l},\mu) =\\
																& = STO_{1,2}({\pi_s}_1,{\pi_s}_2,{\pi_s}_3,(*,*,\pi_x,\pi_x),\mu)
\end{aligned}
\end{equation*}

The following table shows us $STO_{1,2}$ for all the permutations of ${\pi_s}_1,{\pi_s}_2,{\pi_s}_3$.

\begin{center}
\begin{tabular}{c c c | c c c | c c c}
Slot 1 & & Slot 2				& Slot 1 & & Slot 2					& Slot 1 & & Slot 2\\
\hline
${\pi_s}_1$ & $<$ & ${\pi_s}_2$	& ${\pi_s}_1$ & $<$ & ${\pi_s}_3$	& ${\pi_s}_2$ & $<$ & ${\pi_s}_1$\\
${\pi_s}_2$ & $<$ & ${\pi_s}_3$	& ${\pi_s}_3$ & $<$ & ${\pi_s}_2$	& ${\pi_s}_1$ & $<$ & ${\pi_s}_3$\\
${\pi_s}_1$ & $<$ & ${\pi_s}_3$	& ${\pi_s}_1$ & $<$ & ${\pi_s}_2$	& ${\pi_s}_2$ & $<$ & ${\pi_s}_3$
\end{tabular}
\begin{tabular}{c c c | c c c | c c c}
Slot 1 & & Slot 2				& Slot 1 & & Slot 2					& Slot 1 & & Slot 2\\
\hline
${\pi_s}_2$ & $<$ & ${\pi_s}_3$	& ${\pi_s}_3$ & $<$ & ${\pi_s}_1$	& ${\pi_s}_3$ & $<$ & ${\pi_s}_2$\\
${\pi_s}_3$ & $<$ & ${\pi_s}_1$	& ${\pi_s}_1$ & $<$ & ${\pi_s}_2$	& ${\pi_s}_2$ & $<$ & ${\pi_s}_1$\\
${\pi_s}_2$ & $<$ & ${\pi_s}_1$	& ${\pi_s}_3$ & $<$ & ${\pi_s}_2$	& ${\pi_s}_3$ & $<$ & ${\pi_s}_1$
\end{tabular}
\end{center}

But, it is not possible to find a STO, since for every permutation the agents in slots 3 ($\pi_x$) and 4 ($\pi_x$) have a probability of $(1/\sqrt{2}) \times (1/\sqrt{2}) = 1/2$ to do not chase the prey (any $\pi_s$) and the same probability $1 - 1/2 = 1/2$ to chase the prey, making for both agents in slots 1 (any $\pi_s$) and 2 (any $\pi_s$) to obtain the same expected average reward ($0$). So:

\begin{equation*}
STG_{1,2}({\pi_s}_1,{\pi_s}_2,{\pi_s}_3,\Pi_o,w_{\dot{L}},\mu) = 0
\end{equation*}

Therefore:

\begin{equation*}
STG_{1,2}(\Pi_e,w_{\Pi_e},\Pi_o,w_{\dot{L}},\mu) = 6 \frac{9}{2} \frac{1}{3} \frac{1}{3} \frac{1}{3} 0 = 0
\end{equation*}

For slots 1 and 3:

\begin{equation*}
\begin{aligned}
STG_{1,3}(\Pi_e,w_{\Pi_e},\Pi_o,w_{\dot{L}},\mu)	& = \eta_{\Pi^3} \sum_{\pi_1,\pi_2,\pi_3 \in \Pi_e | \pi_1 \neq \pi_2 \neq \pi_3} w_{\Pi_e}(\pi_1) w_{\Pi_e}(\pi_2) w_{\Pi_e}(\pi_3) STG_{1,3}(\pi_1,\pi_2,\pi_3,\Pi_o,w_{\dot{L}},\mu) =\\
													& = 6 \frac{9}{2} \frac{1}{3} \frac{1}{3} \frac{1}{3} STG_{1,3}({\pi_s}_1,{\pi_s}_2,{\pi_s}_3,\Pi_o,w_{\dot{L}},\mu)
\end{aligned}
\end{equation*}

In this case, we only need to calculate $STG_{1,3}({\pi_s}_1,{\pi_s}_2,{\pi_s}_3,\Pi_o,w_{\dot{L}},\mu)$. We follow definition \ref{def:STG_agents} to calculate this value:

\begin{equation*}
\begin{aligned}
STG_{1,3}({\pi_s}_1,{\pi_s}_2,{\pi_s}_3,\Pi_o,w_{\dot{L}},\mu)	& = \sum_{\dot{l} \in \dot{L}^{N(\mu)}_{-1,3}(\Pi_o)} w_{\dot{L}}(\dot{l}) STO_{1,3}({\pi_s}_1,{\pi_s}_2,{\pi_s}_3,\dot{l},\mu) =\\
																& = STO_{1,3}({\pi_s}_1,{\pi_s}_2,{\pi_s}_3,(*,\pi_x,*,\pi_x),\mu)
\end{aligned}
\end{equation*}

The following table shows us $STO_{1,3}$ for all the permutations of ${\pi_s}_1,{\pi_s}_2,{\pi_s}_3$.

\begin{center}
\begin{tabular}{c c c | c c c | c c c}
Slot 1 & & Slot 3				& Slot 1 & & Slot 3					& Slot 1 & & Slot 3\\
\hline
${\pi_s}_1$ & $<$ & ${\pi_s}_2$	& ${\pi_s}_1$ & $<$ & ${\pi_s}_3$	& ${\pi_s}_2$ & $<$ & ${\pi_s}_1$\\
${\pi_s}_2$ & $<$ & ${\pi_s}_3$	& ${\pi_s}_3$ & $<$ & ${\pi_s}_2$	& ${\pi_s}_1$ & $<$ & ${\pi_s}_3$\\
${\pi_s}_1$ & $<$ & ${\pi_s}_3$	& ${\pi_s}_1$ & $<$ & ${\pi_s}_2$	& ${\pi_s}_2$ & $<$ & ${\pi_s}_3$
\end{tabular}
\begin{tabular}{c c c | c c c | c c c}
Slot 1 & & Slot 3				& Slot 1 & & Slot 3					& Slot 1 & & Slot 3\\
\hline
${\pi_s}_2$ & $<$ & ${\pi_s}_3$	& ${\pi_s}_3$ & $<$ & ${\pi_s}_1$	& ${\pi_s}_3$ & $<$ & ${\pi_s}_2$\\
${\pi_s}_3$ & $<$ & ${\pi_s}_1$	& ${\pi_s}_1$ & $<$ & ${\pi_s}_2$	& ${\pi_s}_2$ & $<$ & ${\pi_s}_1$\\
${\pi_s}_2$ & $<$ & ${\pi_s}_1$	& ${\pi_s}_3$ & $<$ & ${\pi_s}_2$	& ${\pi_s}_3$ & $<$ & ${\pi_s}_1$
\end{tabular}
\end{center}

But, it is not possible to find a STO, since for every permutation the agents in slots 2 ($\pi_x$) and 4 ($\pi_x$) have a probability of $(1/\sqrt{2}) \times (1/\sqrt{2}) = 1/2$ to do not chase the prey (any $\pi_s$) and the same probability $1 - 1/2 = 1/2$ to chase the prey, making for both agents in slots 1 (any $\pi_s$) and 3 (any $\pi_s$) to obtain the same expected average reward ($0$). So:

\begin{equation*}
STG_{1,3}({\pi_s}_1,{\pi_s}_2,{\pi_s}_3,\Pi_o,w_{\dot{L}},\mu) = 0
\end{equation*}

Therefore:

\begin{equation*}
STG_{1,3}(\Pi_e,w_{\Pi_e},\Pi_o,w_{\dot{L}},\mu) = 6 \frac{9}{2} \frac{1}{3} \frac{1}{3} \frac{1}{3} 0 = 0
\end{equation*}

For slots 1 and 4:

\begin{equation*}
\begin{aligned}
STG_{1,4}(\Pi_e,w_{\Pi_e},\Pi_o,w_{\dot{L}},\mu)	& = \eta_{\Pi^3} \sum_{\pi_1,\pi_2,\pi_3 \in \Pi_e | \pi_1 \neq \pi_2 \neq \pi_3} w_{\Pi_e}(\pi_1) w_{\Pi_e}(\pi_2) w_{\Pi_e}(\pi_3) STG_{1,4}(\pi_1,\pi_2,\pi_3,\Pi_o,w_{\dot{L}},\mu) =\\
													& = 6 \frac{9}{2} \frac{1}{3} \frac{1}{3} \frac{1}{3} STG_{1,4}({\pi_s}_1,{\pi_s}_2,{\pi_s}_3,\Pi_o,w_{\dot{L}},\mu)
\end{aligned}
\end{equation*}

In this case, we only need to calculate $STG_{1,4}({\pi_s}_1,{\pi_s}_2,{\pi_s}_3,\Pi_o,w_{\dot{L}},\mu)$. We follow definition \ref{def:STG_agents} to calculate this value:

\begin{equation*}
\begin{aligned}
STG_{1,4}({\pi_s}_1,{\pi_s}_2,{\pi_s}_3,\Pi_o,w_{\dot{L}},\mu)	& = \sum_{\dot{l} \in \dot{L}^{N(\mu)}_{-1,4}(\Pi_o)} w_{\dot{L}}(\dot{l}) STO_{1,4}({\pi_s}_1,{\pi_s}_2,{\pi_s}_3,\dot{l},\mu) =\\
																& = STO_{1,4}({\pi_s}_1,{\pi_s}_2,{\pi_s}_3,(*,\pi_x,\pi_x,*),\mu)
\end{aligned}
\end{equation*}

The following table shows us $STO_{1,4}$ for all the permutations of ${\pi_s}_1,{\pi_s}_2,{\pi_s}_3$.

\begin{center}
\begin{tabular}{c c c | c c c | c c c}
Slot 1 & & Slot 4				& Slot 1 & & Slot 4					& Slot 1 & & Slot 4\\
\hline
${\pi_s}_1$ & $<$ & ${\pi_s}_2$	& ${\pi_s}_1$ & $<$ & ${\pi_s}_3$	& ${\pi_s}_2$ & $<$ & ${\pi_s}_1$\\
${\pi_s}_2$ & $<$ & ${\pi_s}_3$	& ${\pi_s}_3$ & $<$ & ${\pi_s}_2$	& ${\pi_s}_1$ & $<$ & ${\pi_s}_3$\\
${\pi_s}_1$ & $<$ & ${\pi_s}_3$	& ${\pi_s}_1$ & $<$ & ${\pi_s}_2$	& ${\pi_s}_2$ & $<$ & ${\pi_s}_3$
\end{tabular}
\begin{tabular}{c c c | c c c | c c c}
Slot 1 & & Slot 4				& Slot 1 & & Slot 4					& Slot 1 & & Slot 4\\
\hline
${\pi_s}_2$ & $<$ & ${\pi_s}_3$	& ${\pi_s}_3$ & $<$ & ${\pi_s}_1$	& ${\pi_s}_3$ & $<$ & ${\pi_s}_2$\\
${\pi_s}_3$ & $<$ & ${\pi_s}_1$	& ${\pi_s}_1$ & $<$ & ${\pi_s}_2$	& ${\pi_s}_2$ & $<$ & ${\pi_s}_1$\\
${\pi_s}_2$ & $<$ & ${\pi_s}_1$	& ${\pi_s}_3$ & $<$ & ${\pi_s}_2$	& ${\pi_s}_3$ & $<$ & ${\pi_s}_1$
\end{tabular}
\end{center}

But, it is not possible to find a STO, since for every permutation the agents in slots 2 ($\pi_x$) and 3 ($\pi_x$) have a probability of $(1/\sqrt{2}) \times (1/\sqrt{2}) = 1/2$ to do not chase the prey (any $\pi_s$) and the same probability $1 - 1/2 = 1/2$ to chase the prey, making for both agents in slots 1 (any $\pi_s$) and 4 (any $\pi_s$) to obtain the same expected average reward ($0$). So:

\begin{equation*}
STG_{1,4}({\pi_s}_1,{\pi_s}_2,{\pi_s}_3,\Pi_o,w_{\dot{L}},\mu) = 0
\end{equation*}

Therefore:

\begin{equation*}
STG_{1,4}(\Pi_e,w_{\Pi_e},\Pi_o,w_{\dot{L}},\mu) = 6 \frac{9}{2} \frac{1}{3} \frac{1}{3} \frac{1}{3} 0 = 0
\end{equation*}

For slots 2 and 1:

\begin{equation*}
\begin{aligned}
STG_{2,1}(\Pi_e,w_{\Pi_e},\Pi_o,w_{\dot{L}},\mu)	& = \eta_{\Pi^3} \sum_{\pi_1,\pi_2,\pi_3 \in \Pi_e | \pi_1 \neq \pi_2 \neq \pi_3} w_{\Pi_e}(\pi_1) w_{\Pi_e}(\pi_2) w_{\Pi_e}(\pi_3) STG_{2,1}(\pi_1,\pi_2,\pi_3,\Pi_o,w_{\dot{L}},\mu) =\\
													& = 6 \frac{9}{2} \frac{1}{3} \frac{1}{3} \frac{1}{3} STG_{2,1}({\pi_s}_1,{\pi_s}_2,{\pi_s}_3,\Pi_o,w_{\dot{L}},\mu)
\end{aligned}
\end{equation*}

In this case, we only need to calculate $STG_{2,1}({\pi_s}_1,{\pi_s}_2,{\pi_s}_3,\Pi_o,w_{\dot{L}},\mu)$. We follow definition \ref{def:STG_agents} to calculate this value:

\begin{equation*}
\begin{aligned}
STG_{2,1}({\pi_s}_1,{\pi_s}_2,{\pi_s}_3,\Pi_o,w_{\dot{L}},\mu)	& = \sum_{\dot{l} \in \dot{L}^{N(\mu)}_{-2,1}(\Pi_o)} w_{\dot{L}}(\dot{l}) STO_{2,1}({\pi_s}_1,{\pi_s}_2,{\pi_s}_3,\dot{l},\mu) =\\
																& = STO_{2,1}({\pi_s}_1,{\pi_s}_2,{\pi_s}_3,(*,*,\pi_x,\pi_x),\mu)
\end{aligned}
\end{equation*}

The following table shows us $STO_{2,1}$ for all the permutations of ${\pi_s}_1,{\pi_s}_2,{\pi_s}_3$.

\begin{center}
\begin{tabular}{c c c | c c c | c c c}
Slot 2 & & Slot 1				& Slot 2 & & Slot 1					& Slot 2 & & Slot 1\\
\hline
${\pi_s}_1$ & $<$ & ${\pi_s}_2$	& ${\pi_s}_1$ & $<$ & ${\pi_s}_3$	& ${\pi_s}_2$ & $<$ & ${\pi_s}_1$\\
${\pi_s}_2$ & $<$ & ${\pi_s}_3$	& ${\pi_s}_3$ & $<$ & ${\pi_s}_2$	& ${\pi_s}_1$ & $<$ & ${\pi_s}_3$\\
${\pi_s}_1$ & $<$ & ${\pi_s}_3$	& ${\pi_s}_1$ & $<$ & ${\pi_s}_2$	& ${\pi_s}_2$ & $<$ & ${\pi_s}_3$
\end{tabular}
\begin{tabular}{c c c | c c c | c c c}
Slot 2 & & Slot 1				& Slot 2 & & Slot 1					& Slot 2 & & Slot 1\\
\hline
${\pi_s}_2$ & $<$ & ${\pi_s}_3$	& ${\pi_s}_3$ & $<$ & ${\pi_s}_1$	& ${\pi_s}_3$ & $<$ & ${\pi_s}_2$\\
${\pi_s}_3$ & $<$ & ${\pi_s}_1$	& ${\pi_s}_1$ & $<$ & ${\pi_s}_2$	& ${\pi_s}_2$ & $<$ & ${\pi_s}_1$\\
${\pi_s}_2$ & $<$ & ${\pi_s}_1$	& ${\pi_s}_3$ & $<$ & ${\pi_s}_2$	& ${\pi_s}_3$ & $<$ & ${\pi_s}_1$
\end{tabular}
\end{center}

But, it is not possible to find a STO, since for every permutation the agents in slots 3 ($\pi_x$) and 4 ($\pi_x$) have a probability of $(1/\sqrt{2}) \times (1/\sqrt{2}) = 1/2$ to do not chase the prey (any $\pi_s$) and the same probability $1 - 1/2 = 1/2$ to chase the prey, making for both agents in slots 2 (any $\pi_s$) and 1 (any $\pi_s$) to obtain the same expected average reward ($0$). So:

\begin{equation*}
STG_{2,1}({\pi_s}_1,{\pi_s}_2,{\pi_s}_3,\Pi_o,w_{\dot{L}},\mu) = 0
\end{equation*}

Therefore:

\begin{equation*}
STG_{2,1}(\Pi_e,w_{\Pi_e},\Pi_o,w_{\dot{L}},\mu) = 6 \frac{9}{2} \frac{1}{3} \frac{1}{3} \frac{1}{3} 0 = 0
\end{equation*}

For slots 2 and 3:

\begin{equation*}
\begin{aligned}
STG_{2,3}(\Pi_e,w_{\Pi_e},\Pi_o,w_{\dot{L}},\mu)	& = \eta_{\Pi^3} \sum_{\pi_1,\pi_2,\pi_3 \in \Pi_e | \pi_1 \neq \pi_2 \neq \pi_3} w_{\Pi_e}(\pi_1) w_{\Pi_e}(\pi_2) w_{\Pi_e}(\pi_3) STG_{2,3}(\pi_1,\pi_2,\pi_3,\Pi_o,w_{\dot{L}},\mu) =\\
													& = 6 \frac{9}{2} \frac{1}{3} \frac{1}{3} \frac{1}{3} STG_{2,3}({\pi_s}_1,{\pi_s}_2,{\pi_s}_3,\Pi_o,w_{\dot{L}},\mu)
\end{aligned}
\end{equation*}

In this case, we only need to calculate $STG_{2,3}({\pi_s}_1,{\pi_s}_2,{\pi_s}_3,\Pi_o,w_{\dot{L}},\mu)$. We follow definition \ref{def:STG_agents} to calculate this value:

\begin{equation*}
\begin{aligned}
STG_{2,3}({\pi_s}_1,{\pi_s}_2,{\pi_s}_3,\Pi_o,w_{\dot{L}},\mu)	& = \sum_{\dot{l} \in \dot{L}^{N(\mu)}_{-2,3}(\Pi_o)} w_{\dot{L}}(\dot{l}) STO_{2,3}({\pi_s}_1,{\pi_s}_2,{\pi_s}_3,\dot{l},\mu) =\\
																& = STO_{2,3}({\pi_s}_1,{\pi_s}_2,{\pi_s}_3,(\pi_x,*,*,\pi_x),\mu)
\end{aligned}
\end{equation*}

The following table shows us $STO_{2,3}$ for all the permutations of ${\pi_s}_1,{\pi_s}_2,{\pi_s}_3$.

\begin{center}
\begin{tabular}{c c c | c c c | c c c}
Slot 2 & & Slot 3				& Slot 2 & & Slot 3					& Slot 2 & & Slot 3\\
\hline
${\pi_s}_1$ & $<$ & ${\pi_s}_2$	& ${\pi_s}_1$ & $<$ & ${\pi_s}_3$	& ${\pi_s}_2$ & $<$ & ${\pi_s}_1$\\
${\pi_s}_2$ & $<$ & ${\pi_s}_3$	& ${\pi_s}_3$ & $<$ & ${\pi_s}_2$	& ${\pi_s}_1$ & $<$ & ${\pi_s}_3$\\
${\pi_s}_1$ & $<$ & ${\pi_s}_3$	& ${\pi_s}_1$ & $<$ & ${\pi_s}_2$	& ${\pi_s}_2$ & $<$ & ${\pi_s}_3$
\end{tabular}
\begin{tabular}{c c c | c c c | c c c}
Slot 2 & & Slot 3				& Slot 2 & & Slot 3					& Slot 2 & & Slot 3\\
\hline
${\pi_s}_2$ & $<$ & ${\pi_s}_3$	& ${\pi_s}_3$ & $<$ & ${\pi_s}_1$	& ${\pi_s}_3$ & $<$ & ${\pi_s}_2$\\
${\pi_s}_3$ & $<$ & ${\pi_s}_1$	& ${\pi_s}_1$ & $<$ & ${\pi_s}_2$	& ${\pi_s}_2$ & $<$ & ${\pi_s}_1$\\
${\pi_s}_2$ & $<$ & ${\pi_s}_1$	& ${\pi_s}_3$ & $<$ & ${\pi_s}_2$	& ${\pi_s}_3$ & $<$ & ${\pi_s}_1$
\end{tabular}
\end{center}

But, it is not possible to find a STO, since for every permutation the agents in slots 2 (any $\pi_s$) and 3 (any $\pi_s$) share rewards (and expected average rewards as well) due they are in the same team. So:

\begin{equation*}
STG_{2,3}({\pi_s}_1,{\pi_s}_2,{\pi_s}_3,\Pi_o,w_{\dot{L}},\mu) = 0
\end{equation*}

Therefore:

\begin{equation*}
STG_{2,3}(\Pi_e,w_{\Pi_e},\Pi_o,w_{\dot{L}},\mu) = 6 \frac{9}{2} \frac{1}{3} \frac{1}{3} \frac{1}{3} 0 = 0
\end{equation*}

For slots 2 and 4:

\begin{equation*}
\begin{aligned}
STG_{2,4}(\Pi_e,w_{\Pi_e},\Pi_o,w_{\dot{L}},\mu)	& = \eta_{\Pi^3} \sum_{\pi_1,\pi_2,\pi_3 \in \Pi_e | \pi_1 \neq \pi_2 \neq \pi_3} w_{\Pi_e}(\pi_1) w_{\Pi_e}(\pi_2) w_{\Pi_e}(\pi_3) STG_{2,4}(\pi_1,\pi_2,\pi_3,\Pi_o,w_{\dot{L}},\mu) =\\
													& = 6 \frac{9}{2} \frac{1}{3} \frac{1}{3} \frac{1}{3} STG_{2,4}({\pi_s}_1,{\pi_s}_2,{\pi_s}_3,\Pi_o,w_{\dot{L}},\mu)
\end{aligned}
\end{equation*}

In this case, we only need to calculate $STG_{2,4}({\pi_s}_1,{\pi_s}_2,{\pi_s}_3,\Pi_o,w_{\dot{L}},\mu)$. We follow definition \ref{def:STG_agents} to calculate this value:

\begin{equation*}
\begin{aligned}
STG_{2,4}({\pi_s}_1,{\pi_s}_2,{\pi_s}_3,\Pi_o,w_{\dot{L}},\mu)	& = \sum_{\dot{l} \in \dot{L}^{N(\mu)}_{-2,4}(\Pi_o)} w_{\dot{L}}(\dot{l}) STO_{2,4}({\pi_s}_1,{\pi_s}_2,{\pi_s}_3,\dot{l},\mu) =\\
																& = STO_{2,4}({\pi_s}_1,{\pi_s}_2,{\pi_s}_3,(\pi_x,*,\pi_x,*),\mu)
\end{aligned}
\end{equation*}

The following table shows us $STO_{2,4}$ for all the permutations of ${\pi_s}_1,{\pi_s}_2,{\pi_s}_3$.

\begin{center}
\begin{tabular}{c c c | c c c | c c c}
Slot 2 & & Slot 4				& Slot 2 & & Slot 4					& Slot 2 & & Slot 4\\
\hline
${\pi_s}_1$ & $<$ & ${\pi_s}_2$	& ${\pi_s}_1$ & $<$ & ${\pi_s}_3$	& ${\pi_s}_2$ & $<$ & ${\pi_s}_1$\\
${\pi_s}_2$ & $<$ & ${\pi_s}_3$	& ${\pi_s}_3$ & $<$ & ${\pi_s}_2$	& ${\pi_s}_1$ & $<$ & ${\pi_s}_3$\\
${\pi_s}_1$ & $<$ & ${\pi_s}_3$	& ${\pi_s}_1$ & $<$ & ${\pi_s}_2$	& ${\pi_s}_2$ & $<$ & ${\pi_s}_3$
\end{tabular}
\begin{tabular}{c c c | c c c | c c c}
Slot 2 & & Slot 4				& Slot 2 & & Slot 4					& Slot 2 & & Slot 4\\
\hline
${\pi_s}_2$ & $<$ & ${\pi_s}_3$	& ${\pi_s}_3$ & $<$ & ${\pi_s}_1$	& ${\pi_s}_3$ & $<$ & ${\pi_s}_2$\\
${\pi_s}_3$ & $<$ & ${\pi_s}_1$	& ${\pi_s}_1$ & $<$ & ${\pi_s}_2$	& ${\pi_s}_2$ & $<$ & ${\pi_s}_1$\\
${\pi_s}_2$ & $<$ & ${\pi_s}_1$	& ${\pi_s}_3$ & $<$ & ${\pi_s}_2$	& ${\pi_s}_3$ & $<$ & ${\pi_s}_1$
\end{tabular}
\end{center}

But, it is not possible to find a STO, since for every permutation the agents in slots 2 (any $\pi_s$) and 4 (any $\pi_s$) share rewards (and expected average rewards as well) due they are in the same team. So:

\begin{equation*}
STG_{2,4}({\pi_s}_1,{\pi_s}_2,{\pi_s}_3,\Pi_o,w_{\dot{L}},\mu) = 0
\end{equation*}

Therefore:

\begin{equation*}
STG_{2,4}(\Pi_e,w_{\Pi_e},\Pi_o,w_{\dot{L}},\mu) = 6 \frac{9}{2} \frac{1}{3} \frac{1}{3} \frac{1}{3} 0 = 0
\end{equation*}

For slots 3 and 1:

\begin{equation*}
\begin{aligned}
STG_{3,1}(\Pi_e,w_{\Pi_e},\Pi_o,w_{\dot{L}},\mu)	& = \eta_{\Pi^3} \sum_{\pi_1,\pi_2,\pi_3 \in \Pi_e | \pi_1 \neq \pi_2 \neq \pi_3} w_{\Pi_e}(\pi_1) w_{\Pi_e}(\pi_2) w_{\Pi_e}(\pi_3) STG_{3,1}(\pi_1,\pi_2,\pi_3,\Pi_o,w_{\dot{L}},\mu) =\\
													& = 6 \frac{9}{2} \frac{1}{3} \frac{1}{3} \frac{1}{3} STG_{3,1}({\pi_s}_1,{\pi_s}_2,{\pi_s}_3,\Pi_o,w_{\dot{L}},\mu)
\end{aligned}
\end{equation*}

In this case, we only need to calculate $STG_{3,1}({\pi_s}_1,{\pi_s}_2,{\pi_s}_3,\Pi_o,w_{\dot{L}},\mu)$. We follow definition \ref{def:STG_agents} to calculate this value:

\begin{equation*}
\begin{aligned}
STG_{3,1}({\pi_s}_1,{\pi_s}_2,{\pi_s}_3,\Pi_o,w_{\dot{L}},\mu)	& = \sum_{\dot{l} \in \dot{L}^{N(\mu)}_{-3,1}(\Pi_o)} w_{\dot{L}}(\dot{l}) STO_{3,1}({\pi_s}_1,{\pi_s}_2,{\pi_s}_3,\dot{l},\mu) =\\
																& = STO_{3,1}({\pi_s}_1,{\pi_s}_2,{\pi_s}_3,(*,\pi_x,*,\pi_x),\mu)
\end{aligned}
\end{equation*}

The following table shows us $STO_{3,1}$ for all the permutations of ${\pi_s}_1,{\pi_s}_2,{\pi_s}_3$.

\begin{center}
\begin{tabular}{c c c | c c c | c c c}
Slot 3 & & Slot 1				& Slot 3 & & Slot 1					& Slot 3 & & Slot 1\\
\hline
${\pi_s}_1$ & $<$ & ${\pi_s}_2$	& ${\pi_s}_1$ & $<$ & ${\pi_s}_3$	& ${\pi_s}_2$ & $<$ & ${\pi_s}_1$\\
${\pi_s}_2$ & $<$ & ${\pi_s}_3$	& ${\pi_s}_3$ & $<$ & ${\pi_s}_2$	& ${\pi_s}_1$ & $<$ & ${\pi_s}_3$\\
${\pi_s}_1$ & $<$ & ${\pi_s}_3$	& ${\pi_s}_1$ & $<$ & ${\pi_s}_2$	& ${\pi_s}_2$ & $<$ & ${\pi_s}_3$
\end{tabular}
\begin{tabular}{c c c | c c c | c c c}
Slot 3 & & Slot 1				& Slot 3 & & Slot 1					& Slot 3 & & Slot 1\\
\hline
${\pi_s}_2$ & $<$ & ${\pi_s}_3$	& ${\pi_s}_3$ & $<$ & ${\pi_s}_1$	& ${\pi_s}_3$ & $<$ & ${\pi_s}_2$\\
${\pi_s}_3$ & $<$ & ${\pi_s}_1$	& ${\pi_s}_1$ & $<$ & ${\pi_s}_2$	& ${\pi_s}_2$ & $<$ & ${\pi_s}_1$\\
${\pi_s}_2$ & $<$ & ${\pi_s}_1$	& ${\pi_s}_3$ & $<$ & ${\pi_s}_2$	& ${\pi_s}_3$ & $<$ & ${\pi_s}_1$
\end{tabular}
\end{center}

But, it is not possible to find a STO, since for every permutation the agents in slots 2 ($\pi_x$) and 4 ($\pi_x$) have a probability of $(1/\sqrt{2}) \times (1/\sqrt{2}) = 1/2$ to do not chase the prey (any $\pi_s$) and the same probability $1 - 1/2 = 1/2$ to chase the prey, making for both agents in slots 3 (any $\pi_s$) and 1 (any $\pi_s$) to obtain the same expected average reward ($0$). So:

\begin{equation*}
STG_{3,1}({\pi_s}_1,{\pi_s}_2,{\pi_s}_3,\Pi_o,w_{\dot{L}},\mu) = 0
\end{equation*}

Therefore:

\begin{equation*}
STG_{3,1}(\Pi_e,w_{\Pi_e},\Pi_o,w_{\dot{L}},\mu) = 6 \frac{9}{2} \frac{1}{3} \frac{1}{3} \frac{1}{3} 0 = 0
\end{equation*}

For slots 3 and 2:

\begin{equation*}
\begin{aligned}
STG_{3,2}(\Pi_e,w_{\Pi_e},\Pi_o,w_{\dot{L}},\mu)	& = \eta_{\Pi^3} \sum_{\pi_1,\pi_2,\pi_3 \in \Pi_e | \pi_1 \neq \pi_2 \neq \pi_3} w_{\Pi_e}(\pi_1) w_{\Pi_e}(\pi_2) w_{\Pi_e}(\pi_3) STG_{3,2}(\pi_1,\pi_2,\pi_3,\Pi_o,w_{\dot{L}},\mu) =\\
													& = 6 \frac{9}{2} \frac{1}{3} \frac{1}{3} \frac{1}{3} STG_{3,2}({\pi_s}_1,{\pi_s}_2,{\pi_s}_3,\Pi_o,w_{\dot{L}},\mu)
\end{aligned}
\end{equation*}

In this case, we only need to calculate $STG_{3,2}({\pi_s}_1,{\pi_s}_2,{\pi_s}_3,\Pi_o,w_{\dot{L}},\mu)$. We follow definition \ref{def:STG_agents} to calculate this value:

\begin{equation*}
\begin{aligned}
STG_{3,2}({\pi_s}_1,{\pi_s}_2,{\pi_s}_3,\Pi_o,w_{\dot{L}},\mu)	& = \sum_{\dot{l} \in \dot{L}^{N(\mu)}_{-3,2}(\Pi_o)} w_{\dot{L}}(\dot{l}) STO_{3,2}({\pi_s}_1,{\pi_s}_2,{\pi_s}_3,\dot{l},\mu) =\\
																& = STO_{3,2}({\pi_s}_1,{\pi_s}_2,{\pi_s}_3,(\pi_x,*,*,\pi_x),\mu)
\end{aligned}
\end{equation*}

The following table shows us $STO_{3,2}$ for all the permutations of ${\pi_s}_1,{\pi_s}_2,{\pi_s}_3$.

\begin{center}
\begin{tabular}{c c c | c c c | c c c}
Slot 3 & & Slot 2				& Slot 3 & & Slot 2					& Slot 3 & & Slot 2\\
\hline
${\pi_s}_1$ & $<$ & ${\pi_s}_2$	& ${\pi_s}_1$ & $<$ & ${\pi_s}_3$	& ${\pi_s}_2$ & $<$ & ${\pi_s}_1$\\
${\pi_s}_2$ & $<$ & ${\pi_s}_3$	& ${\pi_s}_3$ & $<$ & ${\pi_s}_2$	& ${\pi_s}_1$ & $<$ & ${\pi_s}_3$\\
${\pi_s}_1$ & $<$ & ${\pi_s}_3$	& ${\pi_s}_1$ & $<$ & ${\pi_s}_2$	& ${\pi_s}_2$ & $<$ & ${\pi_s}_3$
\end{tabular}
\begin{tabular}{c c c | c c c | c c c}
Slot 3 & & Slot 2				& Slot 3 & & Slot 2					& Slot 3 & & Slot 2\\
\hline
${\pi_s}_2$ & $<$ & ${\pi_s}_3$	& ${\pi_s}_3$ & $<$ & ${\pi_s}_1$	& ${\pi_s}_3$ & $<$ & ${\pi_s}_2$\\
${\pi_s}_3$ & $<$ & ${\pi_s}_1$	& ${\pi_s}_1$ & $<$ & ${\pi_s}_2$	& ${\pi_s}_2$ & $<$ & ${\pi_s}_1$\\
${\pi_s}_2$ & $<$ & ${\pi_s}_1$	& ${\pi_s}_3$ & $<$ & ${\pi_s}_2$	& ${\pi_s}_3$ & $<$ & ${\pi_s}_1$
\end{tabular}
\end{center}

But, it is not possible to find a STO, since for every permutation the agents in slots 3 (any $\pi_s$) and 2 (any $\pi_s$) share rewards (and expected average rewards as well) due they are in the same team. So:

\begin{equation*}
STG_{3,2}({\pi_s}_1,{\pi_s}_2,{\pi_s}_3,\Pi_o,w_{\dot{L}},\mu) = 0
\end{equation*}

Therefore:

\begin{equation*}
STG_{3,2}(\Pi_e,w_{\Pi_e},\Pi_o,w_{\dot{L}},\mu) = 6 \frac{9}{2} \frac{1}{3} \frac{1}{3} \frac{1}{3} 0 = 0
\end{equation*}

For slots 3 and 4:

\begin{equation*}
\begin{aligned}
STG_{3,4}(\Pi_e,w_{\Pi_e},\Pi_o,w_{\dot{L}},\mu)	& = \eta_{\Pi^3} \sum_{\pi_1,\pi_2,\pi_3 \in \Pi_e | \pi_1 \neq \pi_2 \neq \pi_3} w_{\Pi_e}(\pi_1) w_{\Pi_e}(\pi_2) w_{\Pi_e}(\pi_3) STG_{3,4}(\pi_1,\pi_2,\pi_3,\Pi_o,w_{\dot{L}},\mu) =\\
													& = 6 \frac{9}{2} \frac{1}{3} \frac{1}{3} \frac{1}{3} STG_{3,4}({\pi_s}_1,{\pi_s}_2,{\pi_s}_3,\Pi_o,w_{\dot{L}},\mu)
\end{aligned}
\end{equation*}

In this case, we only need to calculate $STG_{3,4}({\pi_s}_1,{\pi_s}_2,{\pi_s}_3,\Pi_o,w_{\dot{L}},\mu)$. We follow definition \ref{def:STG_agents} to calculate this value:

\begin{equation*}
\begin{aligned}
STG_{3,4}({\pi_s}_1,{\pi_s}_2,{\pi_s}_3,\Pi_o,w_{\dot{L}},\mu)	& = \sum_{\dot{l} \in \dot{L}^{N(\mu)}_{-3,4}(\Pi_o)} w_{\dot{L}}(\dot{l}) STO_{3,4}({\pi_s}_1,{\pi_s}_2,{\pi_s}_3,\dot{l},\mu) =\\
																& = STO_{3,4}({\pi_s}_1,{\pi_s}_2,{\pi_s}_3,(\pi_x,\pi_x,*,*),\mu)
\end{aligned}
\end{equation*}

The following table shows us $STO_{3,4}$ for all the permutations of ${\pi_s}_1,{\pi_s}_2,{\pi_s}_3$.

\begin{center}
\begin{tabular}{c c c | c c c | c c c}
Slot 3 & & Slot 4				& Slot 3 & & Slot 4					& Slot 3 & & Slot 4\\
\hline
${\pi_s}_1$ & $<$ & ${\pi_s}_2$	& ${\pi_s}_1$ & $<$ & ${\pi_s}_3$	& ${\pi_s}_2$ & $<$ & ${\pi_s}_1$\\
${\pi_s}_2$ & $<$ & ${\pi_s}_3$	& ${\pi_s}_3$ & $<$ & ${\pi_s}_2$	& ${\pi_s}_1$ & $<$ & ${\pi_s}_3$\\
${\pi_s}_1$ & $<$ & ${\pi_s}_3$	& ${\pi_s}_1$ & $<$ & ${\pi_s}_2$	& ${\pi_s}_2$ & $<$ & ${\pi_s}_3$
\end{tabular}
\begin{tabular}{c c c | c c c | c c c}
Slot 3 & & Slot 4				& Slot 3 & & Slot 4					& Slot 3 & & Slot 4\\
\hline
${\pi_s}_2$ & $<$ & ${\pi_s}_3$	& ${\pi_s}_3$ & $<$ & ${\pi_s}_1$	& ${\pi_s}_3$ & $<$ & ${\pi_s}_2$\\
${\pi_s}_3$ & $<$ & ${\pi_s}_1$	& ${\pi_s}_1$ & $<$ & ${\pi_s}_2$	& ${\pi_s}_2$ & $<$ & ${\pi_s}_1$\\
${\pi_s}_2$ & $<$ & ${\pi_s}_1$	& ${\pi_s}_3$ & $<$ & ${\pi_s}_2$	& ${\pi_s}_3$ & $<$ & ${\pi_s}_1$
\end{tabular}
\end{center}

But, it is not possible to find a STO, since for every permutation the agents in slots 3 (any $\pi_s$) and 4 (any $\pi_s$) share rewards (and expected average rewards as well) due they are in the same team. So:

\begin{equation*}
STG_{3,4}({\pi_s}_1,{\pi_s}_2,{\pi_s}_3,\Pi_o,w_{\dot{L}},\mu) = 0
\end{equation*}

Therefore:

\begin{equation*}
STG_{3,4}(\Pi_e,w_{\Pi_e},\Pi_o,w_{\dot{L}},\mu) = 6 \frac{9}{2} \frac{1}{3} \frac{1}{3} \frac{1}{3} 0 = 0
\end{equation*}

For slots 4 and 1:

\begin{equation*}
\begin{aligned}
STG_{4,1}(\Pi_e,w_{\Pi_e},\Pi_o,w_{\dot{L}},\mu)	& = \eta_{\Pi^3} \sum_{\pi_1,\pi_2,\pi_3 \in \Pi_e | \pi_1 \neq \pi_2 \neq \pi_3} w_{\Pi_e}(\pi_1) w_{\Pi_e}(\pi_2) w_{\Pi_e}(\pi_3) STG_{4,1}(\pi_1,\pi_2,\pi_3,\Pi_o,w_{\dot{L}},\mu) =\\
													& = 6 \frac{9}{2} \frac{1}{3} \frac{1}{3} \frac{1}{3} STG_{4,1}({\pi_s}_1,{\pi_s}_2,{\pi_s}_3,\Pi_o,w_{\dot{L}},\mu)
\end{aligned}
\end{equation*}

In this case, we only need to calculate $STG_{4,1}({\pi_s}_1,{\pi_s}_2,{\pi_s}_3,\Pi_o,w_{\dot{L}},\mu)$. We follow definition \ref{def:STG_agents} to calculate this value:

\begin{equation*}
\begin{aligned}
STG_{4,1}({\pi_s}_1,{\pi_s}_2,{\pi_s}_3,\Pi_o,w_{\dot{L}},\mu)	& = \sum_{\dot{l} \in \dot{L}^{N(\mu)}_{-4,1}(\Pi_o)} w_{\dot{L}}(\dot{l}) STO_{4,1}({\pi_s}_1,{\pi_s}_2,{\pi_s}_3,\dot{l},\mu) =\\
																& = STO_{4,1}({\pi_s}_1,{\pi_s}_2,{\pi_s}_3,(*,\pi_x,\pi_x,*),\mu)
\end{aligned}
\end{equation*}

The following table shows us $STO_{4,1}$ for all the permutations of ${\pi_s}_1,{\pi_s}_2,{\pi_s}_3$.

\begin{center}
\begin{tabular}{c c c | c c c | c c c}
Slot 4 & & Slot 1				& Slot 4 & & Slot 1					& Slot 4 & & Slot 1\\
\hline
${\pi_s}_1$ & $<$ & ${\pi_s}_2$	& ${\pi_s}_1$ & $<$ & ${\pi_s}_3$	& ${\pi_s}_2$ & $<$ & ${\pi_s}_1$\\
${\pi_s}_2$ & $<$ & ${\pi_s}_3$	& ${\pi_s}_3$ & $<$ & ${\pi_s}_2$	& ${\pi_s}_1$ & $<$ & ${\pi_s}_3$\\
${\pi_s}_1$ & $<$ & ${\pi_s}_3$	& ${\pi_s}_1$ & $<$ & ${\pi_s}_2$	& ${\pi_s}_2$ & $<$ & ${\pi_s}_3$
\end{tabular}
\begin{tabular}{c c c | c c c | c c c}
Slot 4 & & Slot 1				& Slot 4 & & Slot 1					& Slot 4 & & Slot 1\\
\hline
${\pi_s}_2$ & $<$ & ${\pi_s}_3$	& ${\pi_s}_3$ & $<$ & ${\pi_s}_1$	& ${\pi_s}_3$ & $<$ & ${\pi_s}_2$\\
${\pi_s}_3$ & $<$ & ${\pi_s}_1$	& ${\pi_s}_1$ & $<$ & ${\pi_s}_2$	& ${\pi_s}_2$ & $<$ & ${\pi_s}_1$\\
${\pi_s}_2$ & $<$ & ${\pi_s}_1$	& ${\pi_s}_3$ & $<$ & ${\pi_s}_2$	& ${\pi_s}_3$ & $<$ & ${\pi_s}_1$
\end{tabular}
\end{center}

But, it is not possible to find a STO, since for every permutation the agents in slots 2 ($\pi_x$) and 3 ($\pi_x$) have a probability of $(1/\sqrt{2}) \times (1/\sqrt{2}) = 1/2$ to do not chase the prey (any $\pi_s$) and the same probability $1 - 1/2 = 1/2$ to chase the prey, making for both agents in slots 4 (any $\pi_s$) and 1 (any $\pi_s$) to obtain the same expected average reward ($0$). So:

\begin{equation*}
STG_{4,1}({\pi_s}_1,{\pi_s}_2,{\pi_s}_3,\Pi_o,w_{\dot{L}},\mu) = 0
\end{equation*}

Therefore:

\begin{equation*}
STG_{4,1}(\Pi_e,w_{\Pi_e},\Pi_o,w_{\dot{L}},\mu) = 6 \frac{9}{2} \frac{1}{3} \frac{1}{3} \frac{1}{3} 0 = 0
\end{equation*}

For slots 4 and 2:

\begin{equation*}
\begin{aligned}
STG_{4,2}(\Pi_e,w_{\Pi_e},\Pi_o,w_{\dot{L}},\mu)	& = \eta_{\Pi^3} \sum_{\pi_1,\pi_2,\pi_3 \in \Pi_e | \pi_1 \neq \pi_2 \neq \pi_3} w_{\Pi_e}(\pi_1) w_{\Pi_e}(\pi_2) w_{\Pi_e}(\pi_3) STG_{4,2}(\pi_1,\pi_2,\pi_3,\Pi_o,w_{\dot{L}},\mu) =\\
													& = 6 \frac{9}{2} \frac{1}{3} \frac{1}{3} \frac{1}{3} STG_{4,2}({\pi_s}_1,{\pi_s}_2,{\pi_s}_3,\Pi_o,w_{\dot{L}},\mu)
\end{aligned}
\end{equation*}

In this case, we only need to calculate $STG_{4,2}({\pi_s}_1,{\pi_s}_2,{\pi_s}_3,\Pi_o,w_{\dot{L}},\mu)$. We follow definition \ref{def:STG_agents} to calculate this value:

\begin{equation*}
\begin{aligned}
STG_{4,2}({\pi_s}_1,{\pi_s}_2,{\pi_s}_3,\Pi_o,w_{\dot{L}},\mu)	& = \sum_{\dot{l} \in \dot{L}^{N(\mu)}_{-4,2}(\Pi_o)} w_{\dot{L}}(\dot{l}) STO_{4,2}({\pi_s}_1,{\pi_s}_2,{\pi_s}_3,\dot{l},\mu) =\\
																& = STO_{4,2}({\pi_s}_1,{\pi_s}_2,{\pi_s}_3,(\pi_x,*,\pi_x,*),\mu)
\end{aligned}
\end{equation*}

The following table shows us $STO_{4,2}$ for all the permutations of ${\pi_s}_1,{\pi_s}_2,{\pi_s}_3$.

\begin{center}
\begin{tabular}{c c c | c c c | c c c}
Slot 4 & & Slot 2				& Slot 4 & & Slot 2					& Slot 4 & & Slot 2\\
\hline
${\pi_s}_1$ & $<$ & ${\pi_s}_2$	& ${\pi_s}_1$ & $<$ & ${\pi_s}_3$	& ${\pi_s}_2$ & $<$ & ${\pi_s}_1$\\
${\pi_s}_2$ & $<$ & ${\pi_s}_3$	& ${\pi_s}_3$ & $<$ & ${\pi_s}_2$	& ${\pi_s}_1$ & $<$ & ${\pi_s}_3$\\
${\pi_s}_1$ & $<$ & ${\pi_s}_3$	& ${\pi_s}_1$ & $<$ & ${\pi_s}_2$	& ${\pi_s}_2$ & $<$ & ${\pi_s}_3$
\end{tabular}
\begin{tabular}{c c c | c c c | c c c}
Slot 4 & & Slot 2				& Slot 4 & & Slot 2					& Slot 4 & & Slot 2\\
\hline
${\pi_s}_2$ & $<$ & ${\pi_s}_3$	& ${\pi_s}_3$ & $<$ & ${\pi_s}_1$	& ${\pi_s}_3$ & $<$ & ${\pi_s}_2$\\
${\pi_s}_3$ & $<$ & ${\pi_s}_1$	& ${\pi_s}_1$ & $<$ & ${\pi_s}_2$	& ${\pi_s}_2$ & $<$ & ${\pi_s}_1$\\
${\pi_s}_2$ & $<$ & ${\pi_s}_1$	& ${\pi_s}_3$ & $<$ & ${\pi_s}_2$	& ${\pi_s}_3$ & $<$ & ${\pi_s}_1$
\end{tabular}
\end{center}

But, it is not possible to find a STO, since for every permutation the agents in slots 4 (any $\pi_s$) and 2 (any $\pi_s$) share rewards (and expected average rewards as well) due they are in the same team. So:

\begin{equation*}
STG_{4,2}({\pi_s}_1,{\pi_s}_2,{\pi_s}_3,\Pi_o,w_{\dot{L}},\mu) = 0
\end{equation*}

Therefore:

\begin{equation*}
STG_{4,2}(\Pi_e,w_{\Pi_e},\Pi_o,w_{\dot{L}},\mu) = 6 \frac{9}{2} \frac{1}{3} \frac{1}{3} \frac{1}{3} 0 = 0
\end{equation*}

And for slots 4 and 3:

\begin{equation*}
\begin{aligned}
STG_{4,3}(\Pi_e,w_{\Pi_e},\Pi_o,w_{\dot{L}},\mu)	& = \eta_{\Pi^3} \sum_{\pi_1,\pi_2,\pi_3 \in \Pi_e | \pi_1 \neq \pi_2 \neq \pi_3} w_{\Pi_e}(\pi_1) w_{\Pi_e}(\pi_2) w_{\Pi_e}(\pi_3) STG_{4,3}(\pi_1,\pi_2,\pi_3,\Pi_o,w_{\dot{L}},\mu) =\\
													& = 6 \frac{9}{2} \frac{1}{3} \frac{1}{3} \frac{1}{3} STG_{4,3}({\pi_s}_1,{\pi_s}_2,{\pi_s}_3,\Pi_o,w_{\dot{L}},\mu)
\end{aligned}
\end{equation*}

In this case, we only need to calculate $STG_{4,3}({\pi_s}_1,{\pi_s}_2,{\pi_s}_3,\Pi_o,w_{\dot{L}},\mu)$. We follow definition \ref{def:STG_agents} to calculate this value:

\begin{equation*}
\begin{aligned}
STG_{4,3}({\pi_s}_1,{\pi_s}_2,{\pi_s}_3,\Pi_o,w_{\dot{L}},\mu)	& = \sum_{\dot{l} \in \dot{L}^{N(\mu)}_{-4,3}(\Pi_o)} w_{\dot{L}}(\dot{l}) STO_{4,3}({\pi_s}_1,{\pi_s}_2,{\pi_s}_3,\dot{l},\mu) =\\
																& = STO_{4,3}({\pi_s}_1,{\pi_s}_2,{\pi_s}_3,(\pi_x,\pi_x,*,*),\mu)
\end{aligned}
\end{equation*}

The following table shows us $STO_{4,3}$ for all the permutations of ${\pi_s}_1,{\pi_s}_2,{\pi_s}_3$.

\begin{center}
\begin{tabular}{c c c | c c c | c c c}
Slot 4 & & Slot 3				& Slot 4 & & Slot 3					& Slot 4 & & Slot 3\\
\hline
${\pi_s}_1$ & $<$ & ${\pi_s}_2$	& ${\pi_s}_1$ & $<$ & ${\pi_s}_3$	& ${\pi_s}_2$ & $<$ & ${\pi_s}_1$\\
${\pi_s}_2$ & $<$ & ${\pi_s}_3$	& ${\pi_s}_3$ & $<$ & ${\pi_s}_2$	& ${\pi_s}_1$ & $<$ & ${\pi_s}_3$\\
${\pi_s}_1$ & $<$ & ${\pi_s}_3$	& ${\pi_s}_1$ & $<$ & ${\pi_s}_2$	& ${\pi_s}_2$ & $<$ & ${\pi_s}_3$
\end{tabular}
\begin{tabular}{c c c | c c c | c c c}
Slot 4 & & Slot 3				& Slot 4 & & Slot 3					& Slot 4 & & Slot 3\\
\hline
${\pi_s}_2$ & $<$ & ${\pi_s}_3$	& ${\pi_s}_3$ & $<$ & ${\pi_s}_1$	& ${\pi_s}_3$ & $<$ & ${\pi_s}_2$\\
${\pi_s}_3$ & $<$ & ${\pi_s}_1$	& ${\pi_s}_1$ & $<$ & ${\pi_s}_2$	& ${\pi_s}_2$ & $<$ & ${\pi_s}_1$\\
${\pi_s}_2$ & $<$ & ${\pi_s}_1$	& ${\pi_s}_3$ & $<$ & ${\pi_s}_2$	& ${\pi_s}_3$ & $<$ & ${\pi_s}_1$
\end{tabular}
\end{center}

But, it is not possible to find a STO, since for every permutation the agents in slots 4 (any $\pi_s$) and 3 (any $\pi_s$) share rewards (and expected average rewards as well) due they are in the same team. So:

\begin{equation*}
STG_{4,3}({\pi_s}_1,{\pi_s}_2,{\pi_s}_3,\Pi_o,w_{\dot{L}},\mu) = 0
\end{equation*}

Therefore:

\begin{equation*}
STG_{4,3}(\Pi_e,w_{\Pi_e},\Pi_o,w_{\dot{L}},\mu) = 6 \frac{9}{2} \frac{1}{3} \frac{1}{3} \frac{1}{3} 0 = 0
\end{equation*}

And finally, we weight over the slots:

\begin{equation*}
\begin{aligned}
& STG(\Pi_e,w_{\Pi_e},\Pi_o,w_{\dot{L}},\mu,w_S) = \eta_{S_1^2} \sum_{i=1}^{N(\mu)} w_S(i,\mu) \times\\
& \times \left(\sum_{j=1}^{i-1} w_S(j,\mu) STG_{i,j}(\Pi_e,w_{\Pi_e},\Pi_o,w_{\dot{L}},\mu) + \sum_{j=i+1}^{N(\mu)} w_S(j,\mu) STG_{i,j}(\Pi_e,w_{\Pi_e},\Pi_o,w_{\dot{L}},\mu)\right) =\\
& \ \ \ \ \ \ \ \ \ \ \ \ \ \ \ \ \ \ \ \ \ \ \ \ \ \ \ \ \ \ \ \ \ \ \ \ \ \ = \frac{4}{3} \frac{1}{4} \frac{1}{4} \{STG_{1,2}(\Pi_e,w_{\Pi_e},\Pi_o,w_{\dot{L}},\mu) + STG_{1,3}(\Pi_e,w_{\Pi_e},\Pi_o,w_{\dot{L}},\mu) +\\
& \ \ \ \ \ \ \ \ \ \ \ \ \ \ \ \ \ \ \ \ \ \ \ \ \ \ \ \ \ \ \ \ \ \ \ \ \ \ + STG_{1,4}(\Pi_e,w_{\Pi_e},\Pi_o,w_{\dot{L}},\mu) + STG_{2,1}(\Pi_e,w_{\Pi_e},\Pi_o,w_{\dot{L}},\mu) +\\
& \ \ \ \ \ \ \ \ \ \ \ \ \ \ \ \ \ \ \ \ \ \ \ \ \ \ \ \ \ \ \ \ \ \ \ \ \ \ + STG_{2,3}(\Pi_e,w_{\Pi_e},\Pi_o,w_{\dot{L}},\mu) + STG_{2,4}(\Pi_e,w_{\Pi_e},\Pi_o,w_{\dot{L}},\mu) +\\
& \ \ \ \ \ \ \ \ \ \ \ \ \ \ \ \ \ \ \ \ \ \ \ \ \ \ \ \ \ \ \ \ \ \ \ \ \ \ + STG_{3,1}(\Pi_e,w_{\Pi_e},\Pi_o,w_{\dot{L}},\mu) + STG_{3,2}(\Pi_e,w_{\Pi_e},\Pi_o,w_{\dot{L}},\mu) +\\
& \ \ \ \ \ \ \ \ \ \ \ \ \ \ \ \ \ \ \ \ \ \ \ \ \ \ \ \ \ \ \ \ \ \ \ \ \ \ + STG_{3,4}(\Pi_e,w_{\Pi_e},\Pi_o,w_{\dot{L}},\mu) + STG_{4,1}(\Pi_e,w_{\Pi_e},\Pi_o,w_{\dot{L}},\mu) +\\
& \ \ \ \ \ \ \ \ \ \ \ \ \ \ \ \ \ \ \ \ \ \ \ \ \ \ \ \ \ \ \ \ \ \ \ \ \ \ + STG_{4,2}(\Pi_e,w_{\Pi_e},\Pi_o,w_{\dot{L}},\mu) + STG_{4,3}(\Pi_e,w_{\Pi_e},\Pi_o,w_{\dot{L}},\mu)\} =\\
& \ \ \ \ \ \ \ \ \ \ \ \ \ \ \ \ \ \ \ \ \ \ \ \ \ \ \ \ \ \ \ \ \ \ \ \ \ \ = \frac{4}{3} \frac{1}{4} \frac{1}{4} \left\{12 \times 0\right\} = 0
\end{aligned}
\end{equation*}

Since $0$ is the lowest possible value for the strict total grading property, therefore predator-prey has $General_{min} = 0$ for this property.
\end{proof}
\end{proposition}

\begin{proposition}
\label{prop:predator-prey_STG_general_max}
$General_{max}$ for the strict total grading (STG) property is equal to $\frac{1}{2}$ for the predator-prey environment.

\begin{proof}
To find $General_{max}$ (equation \ref{eq:general_max}), we need to find a trio $\left\langle\Pi_e,w_{\Pi_e},\Pi_o\right\rangle$ which maximises the property as much as possible. We can have this situation by selecting $\Pi_e = \{\pi_x,\pi_y,\pi_z\}$ with uniform weight for $w_{\Pi_e}$ and $\Pi_o = \{\pi_s\}$ (a $\pi_s$ agent always stays in the same cell\footnote{Note that every cell has an action which is blocked by a block or a boundary, therefore an agent performing this action will stay at its current cell.}).

$\pi_x$ behaves as shown in figure \ref{fig:predator-prey_STG_general_max_agent_x} when playing on each of the 4 slots.

\begin{figure}[!ht]
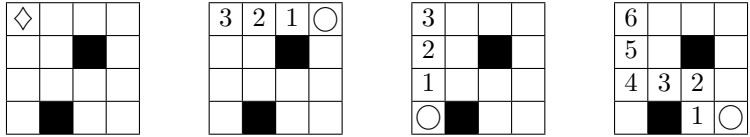

\newcolumntype{C}{>{\centering\arraybackslash}p{12px}}
\def \block {\cellcolor{black}}
\def \hunter {$\bigcirc$}
\def \prey {$\diamondsuit$}

\centering

\begin{minipage}{0.15\textwidth}
\centering
\setlength{\tabcolsep}{0px}

\begin{tabular}[c]{|C|C|C|C|}
\hline
\prey   &         &         &         \\
\hline
        &         & \block  &         \\
\hline
        &         &         &         \\
\hline
        & \block  &         &         \\
\hline
\end{tabular}
\end{minipage}
\begin{minipage}{0.15\textwidth}
\centering
\setlength{\tabcolsep}{0px}

\begin{tabular}[c]{|C|C|C|C|}
\hline
3       & 2       & 1       & \hunter \\
\hline
        &         & \block  &         \\
\hline
        &         &         &         \\
\hline
        & \block  &         &         \\
\hline
\end{tabular}
\end{minipage}
\begin{minipage}{0.15\textwidth}
\centering
\setlength{\tabcolsep}{0px}

\begin{tabular}[c]{|C|C|C|C|}
\hline
3       &         &         &         \\
\hline
2       &         & \block  &         \\
\hline
1       &         &         &         \\
\hline
\hunter & \block  &         &         \\
\hline
\end{tabular}
\end{minipage}
\begin{minipage}{0.15\textwidth}
\centering
\setlength{\tabcolsep}{0px}

\begin{tabular}[c]{|C|C|C|C|}
\hline
6       &         &         &         \\
\hline
5       &         & \block  &         \\
\hline
4       & 3       & 2       &         \\
\hline
        & \block  & 1       & \hunter \\
\hline
\end{tabular}
\end{minipage}

\caption{Behaviour of $\pi_x$ when playing on each of the slots. Numbers represent the position of $\pi_x$ during the iterations.}
\label{fig:predator-prey_STG_general_max_agent_x}
\end{figure}

$\pi_y$ behaves as shown in figure \ref{fig:predator-prey_STG_general_max_agent_y} when playing on each of the 4 slots.

\begin{figure}[!ht]
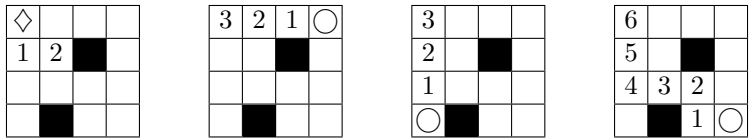

\newcolumntype{C}{>{\centering\arraybackslash}p{12px}}
\def \block {\cellcolor{black}}
\def \hunter {$\bigcirc$}
\def \prey {$\diamondsuit$}

\centering

\begin{minipage}{0.15\textwidth}
\centering
\setlength{\tabcolsep}{0px}

\begin{tabular}[c]{|C|C|C|C|}
\hline
\prey   &         &         &         \\
\hline
1       & 2       & \block  &         \\
\hline
        &         &         &         \\
\hline
        & \block  &         &         \\
\hline
\end{tabular}
\end{minipage}
\begin{minipage}{0.15\textwidth}
\centering
\setlength{\tabcolsep}{0px}

\begin{tabular}[c]{|C|C|C|C|}
\hline
3       & 2       & 1       & \hunter \\
\hline
        &         & \block  &         \\
\hline
        &         &         &         \\
\hline
        & \block  &         &         \\
\hline
\end{tabular}
\end{minipage}
\begin{minipage}{0.15\textwidth}
\centering
\setlength{\tabcolsep}{0px}

\begin{tabular}[c]{|C|C|C|C|}
\hline
3       &         &         &         \\
\hline
2       &         & \block  &         \\
\hline
1       &         &         &         \\
\hline
\hunter & \block  &         &         \\
\hline
\end{tabular}
\end{minipage}
\begin{minipage}{0.15\textwidth}
\centering
\setlength{\tabcolsep}{0px}

\begin{tabular}[c]{|C|C|C|C|}
\hline
6       &         &         &         \\
\hline
5       &         & \block  &         \\
\hline
4       & 3       & 2       &         \\
\hline
        & \block  & 1       & \hunter \\
\hline
\end{tabular}
\end{minipage}

\caption{Behaviour of $\pi_y$ when playing on each of the slots. Numbers represent the position of $\pi_y$ during the iterations.}
\label{fig:predator-prey_STG_general_max_agent_y}
\end{figure}

$\pi_z$ behaves as shown in figure \ref{fig:predator-prey_STG_general_max_agent_z} when playing on each of the 4 slots.

\begin{figure}[!ht]
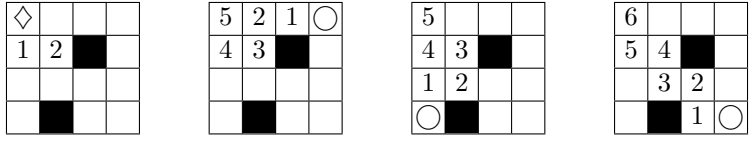

\newcolumntype{C}{>{\centering\arraybackslash}p{12px}}
\def \block {\cellcolor{black}}
\def \hunter {$\bigcirc$}
\def \prey {$\diamondsuit$}

\centering

\begin{minipage}{0.15\textwidth}
\centering
\setlength{\tabcolsep}{0px}

\begin{tabular}[c]{|C|C|C|C|}
\hline
\prey   &         &         &         \\
\hline
1       & 2       & \block  &         \\
\hline
        &         &         &         \\
\hline
        & \block  &         &         \\
\hline
\end{tabular}
\end{minipage}
\begin{minipage}{0.15\textwidth}
\centering
\setlength{\tabcolsep}{0px}

\begin{tabular}[c]{|C|C|C|C|}
\hline
5       & 2       & 1       & \hunter \\
\hline
4       & 3       & \block  &         \\
\hline
        &         &         &         \\
\hline
        & \block  &         &         \\
\hline
\end{tabular}
\end{minipage}
\begin{minipage}{0.15\textwidth}
\centering
\setlength{\tabcolsep}{0px}

\begin{tabular}[c]{|C|C|C|C|}
\hline
5       &         &         &         \\
\hline
4       & 3       & \block  &         \\
\hline
1       & 2       &         &         \\
\hline
\hunter & \block  &         &         \\
\hline
\end{tabular}
\end{minipage}
\begin{minipage}{0.15\textwidth}
\centering
\setlength{\tabcolsep}{0px}

\begin{tabular}[c]{|C|C|C|C|}
\hline
6       &         &         &         \\
\hline
5       & 4       & \block  &         \\
\hline
        & 3       & 2       &         \\
\hline
        & \block  & 1       & \hunter \\
\hline
\end{tabular}
\end{minipage}

\caption{Behaviour of $\pi_z$ when playing on each of the slots. Numbers represent the position of $\pi_z$ during the iterations.}
\label{fig:predator-prey_STG_general_max_agent_z}
\end{figure}

Following definition \ref{def:STG}, we obtain the STG value for this $\left\langle\Pi_e,w_{\Pi_e},\Pi_o\right\rangle$. Since the environment is not symmetric, we need to calculate this property for every pair of slots. Following definition \ref{def:STG_set}, we can calculate its STG value for each pair of slots. We start with slots 1 and 2:

\begin{equation*}
\begin{aligned}
STG_{1,2}(\Pi_e,w_{\Pi_e},\Pi_o,w_{\dot{L}},\mu)	& = \eta_{\Pi^3} \sum_{\pi_1,\pi_2,\pi_3 \in \Pi_e | \pi_1 \neq \pi_2 \neq \pi_3} w_{\Pi_e}(\pi_1) w_{\Pi_e}(\pi_2) w_{\Pi_e}(\pi_3) STG_{1,2}(\pi_1,\pi_2,\pi_3,\Pi_o,w_{\dot{L}},\mu) =\\
													& = 6 \frac{9}{2} \frac{1}{3} \frac{1}{3} \frac{1}{3} STG_{1,2}(\pi_x,\pi_y,\pi_z,\Pi_o,w_{\dot{L}},\mu)
\end{aligned}
\end{equation*}

\noindent Note that we avoided to calculate all the permutations of $\pi_1,\pi_2,\pi_3$ for $STG_{i,j}(\pi_1,\pi_2,\pi_3,\Pi_o,w_{\dot{L}},\mu)$ since they provide the same result, by calculating only one permutation and multiplying the result by the number of permutations $6$.

In this case, we only need to calculate $STG_{1,2}(\pi_x,\pi_y,\pi_z,\Pi_o,w_{\dot{L}},\mu)$. We follow definition \ref{def:STG_agents} to calculate this value:

\begin{equation*}
\begin{aligned}
STG_{1,2}(\pi_x,\pi_y,\pi_z,\Pi_o,w_{\dot{L}},\mu)	& = \sum_{\dot{l} \in \dot{L}^{N(\mu)}_{-1,2}(\Pi_o)} w_{\dot{L}}(\dot{l}) STO_{1,2}(\pi_x,\pi_y,\pi_z,\dot{l},\mu) =\\
													& = STO_{1,2}(\pi_x,\pi_y,\pi_z,(*,*,\pi_s,\pi_s),\mu)
\end{aligned}
\end{equation*}

The following table shows us $STO_{1,2}$ for all the permutations of $\pi_x,\pi_y,\pi_z$.

\begin{center}
\begin{tabular}{c c c | c c c | c c c}
Slot 1 & & Slot 2		& Slot 1 & & Slot 2			& Slot 1 & & Slot 2\\
\hline
$\pi_x$ & $<$ & $\pi_y$	& $\pi_x$ & $<$ & $\pi_z$	& $\pi_y$ & $<$ & $\pi_x$\\
$\pi_y$ & $<$ & $\pi_z$	& $\pi_z$ & $<$ & $\pi_y$	& $\pi_x$ & $<$ & $\pi_z$\\
$\pi_x$ & $<$ & $\pi_z$	& $\pi_x$ & $<$ & $\pi_y$	& $\pi_y$ & $<$ & $\pi_z$
\end{tabular}
\begin{tabular}{c c c | c c c | c c c}
Slot 1 & & Slot 2		& Slot 1 & & Slot 2			& Slot 1 & & Slot 2\\
\hline
$\pi_y$ & $<$ & $\pi_z$	& $\pi_z$ & $<$ & $\pi_x$	& $\pi_z$ & $<$ & $\pi_y$\\
$\pi_z$ & $<$ & $\pi_x$	& $\pi_x$ & $<$ & $\pi_y$	& $\pi_y$ & $<$ & $\pi_x$\\
$\pi_y$ & $<$ & $\pi_x$	& $\pi_z$ & $<$ & $\pi_y$	& $\pi_z$ & $<$ & $\pi_x$
\end{tabular}
\end{center}

It is possible to find a STO for the first permutation. In $\pi_x < \pi_y$, $\pi_x$ will stay at the upper left corner and $\pi_y$ will chase it in that cell in the 3rd iteration, so they will obtain an expected average reward of $-1$ and $1$ respectively. In $\pi_y < \pi_z$, $\pi_y$ will reach the 2nd row 2nd column cell and $\pi_z$ will chase it in that cell in the 3rd iteration, so they will obtain an expected average reward of $-1$ and $1$ respectively. In $\pi_x < \pi_z$, $\pi_x$ will stay at the upper left corner and $\pi_z$ will chase it in that cell in the 5th iteration, so they will obtain an expected average reward of $-1$ and $1$ respectively. So:

\begin{equation*}
STG_{1,2}(\pi_x,\pi_y,\pi_z,\Pi_o,w_{\dot{L}},\mu) = 1
\end{equation*}

Therefore:

\begin{equation*}
STG_{1,2}(\Pi_e,w_{\Pi_e},\Pi_o,w_{\dot{L}},\mu) = 6 \frac{9}{2} \frac{1}{3} \frac{1}{3} \frac{1}{3} 1 = 1
\end{equation*}

For slots 1 and 3:

\begin{equation*}
\begin{aligned}
STG_{1,3}(\Pi_e,w_{\Pi_e},\Pi_o,w_{\dot{L}},\mu)	& = \eta_{\Pi^3} \sum_{\pi_1,\pi_2,\pi_3 \in \Pi_e | \pi_1 \neq \pi_2 \neq \pi_3} w_{\Pi_e}(\pi_1) w_{\Pi_e}(\pi_2) w_{\Pi_e}(\pi_3) STG_{1,3}(\pi_1,\pi_2,\pi_3,\Pi_o,w_{\dot{L}},\mu) =\\
													& = 6 \frac{9}{2} \frac{1}{3} \frac{1}{3} \frac{1}{3} STG_{1,3}(\pi_x,\pi_y,\pi_z,\Pi_o,w_{\dot{L}},\mu)
\end{aligned}
\end{equation*}

In this case, we only need to calculate $STG_{1,3}(\pi_x,\pi_y,\pi_z,\Pi_o,w_{\dot{L}},\mu)$. We follow definition \ref{def:STG_agents} to calculate this value:

\begin{equation*}
\begin{aligned}
STG_{1,3}(\pi_x,\pi_y,\pi_z,\Pi_o,w_{\dot{L}},\mu)	& = \sum_{\dot{l} \in \dot{L}^{N(\mu)}_{-1,3}(\Pi_o)} w_{\dot{L}}(\dot{l}) STO_{1,3}(\pi_x,\pi_y,\pi_z,\dot{l},\mu) =\\
													& = STO_{1,3}(\pi_x,\pi_y,\pi_z,(*,\pi_s,*,\pi_s),\mu)
\end{aligned}
\end{equation*}

The following table shows us $STO_{1,3}$ for all the permutations of $\pi_x,\pi_y,\pi_z$.

\begin{center}
\begin{tabular}{c c c | c c c | c c c}
Slot 1 & & Slot 3		& Slot 1 & & Slot 3			& Slot 1 & & Slot 3\\
\hline
$\pi_x$ & $<$ & $\pi_y$	& $\pi_x$ & $<$ & $\pi_z$	& $\pi_y$ & $<$ & $\pi_x$\\
$\pi_y$ & $<$ & $\pi_z$	& $\pi_z$ & $<$ & $\pi_y$	& $\pi_x$ & $<$ & $\pi_z$\\
$\pi_x$ & $<$ & $\pi_z$	& $\pi_x$ & $<$ & $\pi_y$	& $\pi_y$ & $<$ & $\pi_z$
\end{tabular}
\begin{tabular}{c c c | c c c | c c c}
Slot 1 & & Slot 3		& Slot 1 & & Slot 3			& Slot 1 & & Slot 3\\
\hline
$\pi_y$ & $<$ & $\pi_z$	& $\pi_z$ & $<$ & $\pi_x$	& $\pi_z$ & $<$ & $\pi_y$\\
$\pi_z$ & $<$ & $\pi_x$	& $\pi_x$ & $<$ & $\pi_y$	& $\pi_y$ & $<$ & $\pi_x$\\
$\pi_y$ & $<$ & $\pi_x$	& $\pi_z$ & $<$ & $\pi_y$	& $\pi_z$ & $<$ & $\pi_x$
\end{tabular}
\end{center}

It is possible to find a STO for the first permutation. In $\pi_x < \pi_y$, $\pi_x$ will stay at the upper left corner and $\pi_y$ will chase it in that cell in the 3rd iteration, so they will obtain an expected average reward of $-1$ and $1$ respectively. In $\pi_y < \pi_z$, $\pi_y$ will reach the 2nd row 2nd column cell and $\pi_z$ will chase it in that cell in the 3rd iteration, so they will obtain an expected average reward of $-1$ and $1$ respectively. In $\pi_x < \pi_z$, $\pi_x$ will stay at the upper left corner and $\pi_z$ will chase it in that cell in the 5th iteration, so they will obtain an expected average reward of $-1$ and $1$ respectively. So:

\begin{equation*}
STG_{1,3}(\pi_x,\pi_y,\pi_z,\Pi_o,w_{\dot{L}},\mu) = 1
\end{equation*}

Therefore:

\begin{equation*}
STG_{1,3}(\Pi_e,w_{\Pi_e},\Pi_o,w_{\dot{L}},\mu) = 6 \frac{9}{2} \frac{1}{3} \frac{1}{3} \frac{1}{3} 1 = 1
\end{equation*}

For slots 1 and 4:

\begin{equation*}
\begin{aligned}
STG_{1,4}(\Pi_e,w_{\Pi_e},\Pi_o,w_{\dot{L}},\mu)	& = \eta_{\Pi^3} \sum_{\pi_1,\pi_2,\pi_3 \in \Pi_e | \pi_1 \neq \pi_2 \neq \pi_3} w_{\Pi_e}(\pi_1) w_{\Pi_e}(\pi_2) w_{\Pi_e}(\pi_3) STG_{1,4}(\pi_1,\pi_2,\pi_3,\Pi_o,w_{\dot{L}},\mu) =\\
													& = 6 \frac{9}{2} \frac{1}{3} \frac{1}{3} \frac{1}{3} STG_{1,4}(\pi_x,\pi_y,\pi_z,\Pi_o,w_{\dot{L}},\mu)
\end{aligned}
\end{equation*}

In this case, we only need to calculate $STG_{1,4}(\pi_x,\pi_y,\pi_z,\Pi_o,w_{\dot{L}},\mu)$. We follow definition \ref{def:STG_agents} to calculate this value:

\begin{equation*}
\begin{aligned}
STG_{1,4}(\pi_x,\pi_y,\pi_z,\Pi_o,w_{\dot{L}},\mu)	& = \sum_{\dot{l} \in \dot{L}^{N(\mu)}_{-1,4}(\Pi_o)} w_{\dot{L}}(\dot{l}) STO_{1,4}(\pi_x,\pi_y,\pi_z,\dot{l},\mu) =\\
													& = STO_{1,4}(\pi_x,\pi_y,\pi_z,(*,\pi_s,\pi_s,*),\mu)
\end{aligned}
\end{equation*}

The following table shows us $STO_{1,4}$ for all the permutations of $\pi_x,\pi_y,\pi_z$.

\begin{center}
\begin{tabular}{c c c | c c c | c c c}
Slot 1 & & Slot 4		& Slot 1 & & Slot 4			& Slot 1 & & Slot 4\\
\hline
$\pi_x$ & $<$ & $\pi_y$	& $\pi_x$ & $<$ & $\pi_z$	& $\pi_y$ & $<$ & $\pi_x$\\
$\pi_y$ & $<$ & $\pi_z$	& $\pi_z$ & $<$ & $\pi_y$	& $\pi_x$ & $<$ & $\pi_z$\\
$\pi_x$ & $<$ & $\pi_z$	& $\pi_x$ & $<$ & $\pi_y$	& $\pi_y$ & $<$ & $\pi_z$
\end{tabular}
\begin{tabular}{c c c | c c c | c c c}
Slot 1 & & Slot 4		& Slot 1 & & Slot 4			& Slot 1 & & Slot 4\\
\hline
$\pi_y$ & $<$ & $\pi_z$	& $\pi_z$ & $<$ & $\pi_x$	& $\pi_z$ & $<$ & $\pi_y$\\
$\pi_z$ & $<$ & $\pi_x$	& $\pi_x$ & $<$ & $\pi_y$	& $\pi_y$ & $<$ & $\pi_x$\\
$\pi_y$ & $<$ & $\pi_x$	& $\pi_z$ & $<$ & $\pi_y$	& $\pi_z$ & $<$ & $\pi_x$
\end{tabular}
\end{center}

It is possible to find a STO for the first permutation. In $\pi_x < \pi_y$, $\pi_x$ will stay at the upper left corner and $\pi_y$ will chase it in that cell in the 6th iteration, so they will obtain an expected average reward of $-1$ and $1$ respectively. In $\pi_y < \pi_z$, $\pi_y$ will reach the 2nd row 2nd column cell and $\pi_z$ will chase it in that cell in the 4th iteration, so they will obtain an expected average reward of $-1$ and $1$ respectively. In $\pi_x < \pi_z$, $\pi_x$ will stay at the upper left corner and $\pi_z$ will chase it in that cell in the 6th iteration, so they will obtain an expected average reward of $-1$ and $1$ respectively. So:

\begin{equation*}
STG_{1,4}(\pi_x,\pi_y,\pi_z,\Pi_o,w_{\dot{L}},\mu) = 1
\end{equation*}

Therefore:

\begin{equation*}
STG_{1,4}(\Pi_e,w_{\Pi_e},\Pi_o,w_{\dot{L}},\mu) = 6 \frac{9}{2} \frac{1}{3} \frac{1}{3} \frac{1}{3} 1 = 1
\end{equation*}

For slots 2 and 1:

\begin{equation*}
\begin{aligned}
STG_{2,1}(\Pi_e,w_{\Pi_e},\Pi_o,w_{\dot{L}},\mu)	& = \eta_{\Pi^3} \sum_{\pi_1,\pi_2,\pi_3 \in \Pi_e | \pi_1 \neq \pi_2 \neq \pi_3} w_{\Pi_e}(\pi_1) w_{\Pi_e}(\pi_2) w_{\Pi_e}(\pi_3) STG_{2,1}(\pi_1,\pi_2,\pi_3,\Pi_o,w_{\dot{L}},\mu) =\\
													& = 6 \frac{9}{2} \frac{1}{3} \frac{1}{3} \frac{1}{3} STG_{2,1}(\pi_x,\pi_y,\pi_z,\Pi_o,w_{\dot{L}},\mu)
\end{aligned}
\end{equation*}

In this case, we only need to calculate $STG_{2,1}(\pi_x,\pi_y,\pi_z,\Pi_o,w_{\dot{L}},\mu)$. We follow definition \ref{def:STG_agents} to calculate this value:

\begin{equation*}
\begin{aligned}
STG_{2,1}(\pi_x,\pi_y,\pi_z,\Pi_o,w_{\dot{L}},\mu)	& = \sum_{\dot{l} \in \dot{L}^{N(\mu)}_{-2,1}(\Pi_o)} w_{\dot{L}}(\dot{l}) STO_{2,1}(\pi_x,\pi_y,\pi_z,\dot{l},\mu) =\\
													& = STO_{2,1}(\pi_x,\pi_y,\pi_z,(*,*,\pi_s,\pi_s),\mu)
\end{aligned}
\end{equation*}

The following table shows us $STO_{2,1}$ for all the permutations of $\pi_x,\pi_y,\pi_z$.

\begin{center}
\begin{tabular}{c c c | c c c | c c c}
Slot 2 & & Slot 1		& Slot 2 & & Slot 1			& Slot 2 & & Slot 1\\
\hline
$\pi_x$ & $<$ & $\pi_y$	& $\pi_x$ & $<$ & $\pi_z$	& $\pi_y$ & $<$ & $\pi_x$\\
$\pi_y$ & $<$ & $\pi_z$	& $\pi_z$ & $<$ & $\pi_y$	& $\pi_x$ & $<$ & $\pi_z$\\
$\pi_x$ & $<$ & $\pi_z$	& $\pi_x$ & $<$ & $\pi_y$	& $\pi_y$ & $<$ & $\pi_z$
\end{tabular}
\begin{tabular}{c c c | c c c | c c c}
Slot 2 & & Slot 1		& Slot 2 & & Slot 1			& Slot 2 & & Slot 1\\
\hline
$\pi_y$ & $<$ & $\pi_z$	& $\pi_z$ & $<$ & $\pi_x$	& $\pi_z$ & $<$ & $\pi_y$\\
$\pi_z$ & $<$ & $\pi_x$	& $\pi_x$ & $<$ & $\pi_y$	& $\pi_y$ & $<$ & $\pi_x$\\
$\pi_y$ & $<$ & $\pi_x$	& $\pi_z$ & $<$ & $\pi_y$	& $\pi_z$ & $<$ & $\pi_x$
\end{tabular}
\end{center}

It is possible to find a STO for the first permutation. In $\pi_x < \pi_y$, $\pi_y$ will reach the 2nd row 2nd column cell and $\pi_x$ will never chase it in that cell, so they will obtain an expected average reward of $-1$ and $1$ respectively. In $\pi_y < \pi_z$, $\pi_z$ will reach the 2nd row 2nd column cell and $\pi_y$ will never chase it in that cell, so they will obtain an expected average reward of $-1$ and $1$ respectively. In $\pi_x < \pi_z$, $\pi_z$ will reach the 2nd row 2nd column cell and $\pi_x$ will never chase it in that cell, so they will obtain an expected average reward of $-1$ and $1$ respectively. So:

\begin{equation*}
STG_{2,1}(\pi_x,\pi_y,\pi_z,\Pi_o,w_{\dot{L}},\mu) = 1
\end{equation*}

Therefore:

\begin{equation*}
STG_{2,1}(\Pi_e,w_{\Pi_e},\Pi_o,w_{\dot{L}},\mu) = 6 \frac{9}{2} \frac{1}{3} \frac{1}{3} \frac{1}{3} 1 = 1
\end{equation*}

For slots 2 and 3:

\begin{equation*}
\begin{aligned}
STG_{2,3}(\Pi_e,w_{\Pi_e},\Pi_o,w_{\dot{L}},\mu)	& = \eta_{\Pi^3} \sum_{\pi_1,\pi_2,\pi_3 \in \Pi_e | \pi_1 \neq \pi_2 \neq \pi_3} w_{\Pi_e}(\pi_1) w_{\Pi_e}(\pi_2) w_{\Pi_e}(\pi_3) STG_{2,3}(\pi_1,\pi_2,\pi_3,\Pi_o,w_{\dot{L}},\mu) =\\
													& = 6 \frac{9}{2} \frac{1}{3} \frac{1}{3} \frac{1}{3} STG_{2,3}(\pi_x,\pi_y,\pi_z,\Pi_o,w_{\dot{L}},\mu)
\end{aligned}
\end{equation*}

In this case, we only need to calculate $STG_{2,3}(\pi_x,\pi_y,\pi_z,\Pi_o,w_{\dot{L}},\mu)$. We follow definition \ref{def:STG_agents} to calculate this value:

\begin{equation*}
\begin{aligned}
STG_{2,3}(\pi_x,\pi_y,\pi_z,\Pi_o,w_{\dot{L}},\mu)	& = \sum_{\dot{l} \in \dot{L}^{N(\mu)}_{-2,3}(\Pi_o)} w_{\dot{L}}(\dot{l}) STO_{2,3}(\pi_x,\pi_y,\pi_z,\dot{l},\mu) =\\
													& = STO_{2,3}(\pi_x,\pi_y,\pi_z,(\pi_s,*,*,\pi_s),\mu)
\end{aligned}
\end{equation*}

The following table shows us $STO_{2,3}$ for all the permutations of $\pi_x,\pi_y,\pi_z$.

\begin{center}
\begin{tabular}{c c c | c c c | c c c}
Slot 2 & & Slot 3		& Slot 2 & & Slot 3			& Slot 2 & & Slot 3\\
\hline
$\pi_x$ & $<$ & $\pi_y$	& $\pi_x$ & $<$ & $\pi_z$	& $\pi_y$ & $<$ & $\pi_x$\\
$\pi_y$ & $<$ & $\pi_z$	& $\pi_z$ & $<$ & $\pi_y$	& $\pi_x$ & $<$ & $\pi_z$\\
$\pi_x$ & $<$ & $\pi_z$	& $\pi_x$ & $<$ & $\pi_y$	& $\pi_y$ & $<$ & $\pi_z$
\end{tabular}
\begin{tabular}{c c c | c c c | c c c}
Slot 2 & & Slot 3		& Slot 2 & & Slot 3			& Slot 2 & & Slot 3\\
\hline
$\pi_y$ & $<$ & $\pi_z$	& $\pi_z$ & $<$ & $\pi_x$	& $\pi_z$ & $<$ & $\pi_y$\\
$\pi_z$ & $<$ & $\pi_x$	& $\pi_x$ & $<$ & $\pi_y$	& $\pi_y$ & $<$ & $\pi_x$\\
$\pi_y$ & $<$ & $\pi_x$	& $\pi_z$ & $<$ & $\pi_y$	& $\pi_z$ & $<$ & $\pi_x$
\end{tabular}
\end{center}

But, it is not possible to find a STO, since for every permutation the agents in slots 2 and 3 share rewards (and expected average rewards as well) due they are in the same team. So:

\begin{equation*}
STG_{2,3}(\pi_x,\pi_y,\pi_z,\Pi_o,w_{\dot{L}},\mu) = 0
\end{equation*}

Therefore:

\begin{equation*}
STG_{2,3}(\Pi_e,w_{\Pi_e},\Pi_o,w_{\dot{L}},\mu) = 6 \frac{9}{2} \frac{1}{3} \frac{1}{3} \frac{1}{3} 0 = 0
\end{equation*}

For slots 2 and 4:

\begin{equation*}
\begin{aligned}
STG_{2,4}(\Pi_e,w_{\Pi_e},\Pi_o,w_{\dot{L}},\mu)	& = \eta_{\Pi^3} \sum_{\pi_1,\pi_2,\pi_3 \in \Pi_e | \pi_1 \neq \pi_2 \neq \pi_3} w_{\Pi_e}(\pi_1) w_{\Pi_e}(\pi_2) w_{\Pi_e}(\pi_3) STG_{2,4}(\pi_1,\pi_2,\pi_3,\Pi_o,w_{\dot{L}},\mu) =\\
													& = 6 \frac{9}{2} \frac{1}{3} \frac{1}{3} \frac{1}{3} STG_{2,4}(\pi_x,\pi_y,\pi_z,\Pi_o,w_{\dot{L}},\mu)
\end{aligned}
\end{equation*}

In this case, we only need to calculate $STG_{2,4}(\pi_x,\pi_y,\pi_z,\Pi_o,w_{\dot{L}},\mu)$. We follow definition \ref{def:STG_agents} to calculate this value:

\begin{equation*}
\begin{aligned}
STG_{2,4}(\pi_x,\pi_y,\pi_z,\Pi_o,w_{\dot{L}},\mu)	& = \sum_{\dot{l} \in \dot{L}^{N(\mu)}_{-2,4}(\Pi_o)} w_{\dot{L}}(\dot{l}) STO_{2,4}(\pi_x,\pi_y,\pi_z,\dot{l},\mu) =\\
													& = STO_{2,4}(\pi_x,\pi_y,\pi_z,(\pi_s,*,\pi_s,*),\mu)
\end{aligned}
\end{equation*}

The following table shows us $STO_{2,4}$ for all the permutations of $\pi_x,\pi_y,\pi_z$.

\begin{center}
\begin{tabular}{c c c | c c c | c c c}
Slot 2 & & Slot 4		& Slot 2 & & Slot 4			& Slot 2 & & Slot 4\\
\hline
$\pi_x$ & $<$ & $\pi_y$	& $\pi_x$ & $<$ & $\pi_z$	& $\pi_y$ & $<$ & $\pi_x$\\
$\pi_y$ & $<$ & $\pi_z$	& $\pi_z$ & $<$ & $\pi_y$	& $\pi_x$ & $<$ & $\pi_z$\\
$\pi_x$ & $<$ & $\pi_z$	& $\pi_x$ & $<$ & $\pi_y$	& $\pi_y$ & $<$ & $\pi_z$
\end{tabular}
\begin{tabular}{c c c | c c c | c c c}
Slot 2 & & Slot 4		& Slot 2 & & Slot 4			& Slot 2 & & Slot 4\\
\hline
$\pi_y$ & $<$ & $\pi_z$	& $\pi_z$ & $<$ & $\pi_x$	& $\pi_z$ & $<$ & $\pi_y$\\
$\pi_z$ & $<$ & $\pi_x$	& $\pi_x$ & $<$ & $\pi_y$	& $\pi_y$ & $<$ & $\pi_x$\\
$\pi_y$ & $<$ & $\pi_x$	& $\pi_z$ & $<$ & $\pi_y$	& $\pi_z$ & $<$ & $\pi_x$
\end{tabular}
\end{center}

But, it is not possible to find a STO, since for every permutation the agents in slots 2 and 4 share rewards (and expected average rewards as well) due they are in the same team. So:

\begin{equation*}
STG_{2,4}(\pi_x,\pi_y,\pi_z,\Pi_o,w_{\dot{L}},\mu) = 0
\end{equation*}

Therefore:

\begin{equation*}
STG_{2,4}(\Pi_e,w_{\Pi_e},\Pi_o,w_{\dot{L}},\mu) = 6 \frac{9}{2} \frac{1}{3} \frac{1}{3} \frac{1}{3} 0 = 0
\end{equation*}

For slots 3 and 1:

\begin{equation*}
\begin{aligned}
STG_{3,1}(\Pi_e,w_{\Pi_e},\Pi_o,w_{\dot{L}},\mu)	& = \eta_{\Pi^3} \sum_{\pi_1,\pi_2,\pi_3 \in \Pi_e | \pi_1 \neq \pi_2 \neq \pi_3} w_{\Pi_e}(\pi_1) w_{\Pi_e}(\pi_2) w_{\Pi_e}(\pi_3) STG_{3,1}(\pi_1,\pi_2,\pi_3,\Pi_o,w_{\dot{L}},\mu) =\\
													& = 6 \frac{9}{2} \frac{1}{3} \frac{1}{3} \frac{1}{3} STG_{3,1}(\pi_x,\pi_y,\pi_z,\Pi_o,w_{\dot{L}},\mu)
\end{aligned}
\end{equation*}

In this case, we only need to calculate $STG_{3,1}(\pi_x,\pi_y,\pi_z,\Pi_o,w_{\dot{L}},\mu)$. We follow definition \ref{def:STG_agents} to calculate this value:

\begin{equation*}
\begin{aligned}
STG_{3,1}(\pi_x,\pi_y,\pi_z,\Pi_o,w_{\dot{L}},\mu)	& = \sum_{\dot{l} \in \dot{L}^{N(\mu)}_{-3,1}(\Pi_o)} w_{\dot{L}}(\dot{l}) STO_{3,1}(\pi_x,\pi_y,\pi_z,\dot{l},\mu) =\\
													& = STO_{3,1}(\pi_x,\pi_y,\pi_z,(*,\pi_s,*,\pi_s),\mu)
\end{aligned}
\end{equation*}

The following table shows us $STO_{3,1}$ for all the permutations of $\pi_x,\pi_y,\pi_z$.

\begin{center}
\begin{tabular}{c c c | c c c | c c c}
Slot 3 & & Slot 1		& Slot 3 & & Slot 1			& Slot 3 & & Slot 1\\
\hline
$\pi_x$ & $<$ & $\pi_y$	& $\pi_x$ & $<$ & $\pi_z$	& $\pi_y$ & $<$ & $\pi_x$\\
$\pi_y$ & $<$ & $\pi_z$	& $\pi_z$ & $<$ & $\pi_y$	& $\pi_x$ & $<$ & $\pi_z$\\
$\pi_x$ & $<$ & $\pi_z$	& $\pi_x$ & $<$ & $\pi_y$	& $\pi_y$ & $<$ & $\pi_z$
\end{tabular}
\begin{tabular}{c c c | c c c | c c c}
Slot 3 & & Slot 1		& Slot 3 & & Slot 1			& Slot 3 & & Slot 1\\
\hline
$\pi_y$ & $<$ & $\pi_z$	& $\pi_z$ & $<$ & $\pi_x$	& $\pi_z$ & $<$ & $\pi_y$\\
$\pi_z$ & $<$ & $\pi_x$	& $\pi_x$ & $<$ & $\pi_y$	& $\pi_y$ & $<$ & $\pi_x$\\
$\pi_y$ & $<$ & $\pi_x$	& $\pi_z$ & $<$ & $\pi_y$	& $\pi_z$ & $<$ & $\pi_x$
\end{tabular}
\end{center}

It is possible to find a STO for the first permutation. In $\pi_x < \pi_y$, $\pi_y$ will reach the 2nd row 2nd column cell and $\pi_x$ will never chase it in that cell, so they will obtain an expected average reward of $-1$ and $1$ respectively. In $\pi_y < \pi_z$, $\pi_z$ will reach the 2nd row 2nd column cell and $\pi_y$ will never chase it in that cell, so they will obtain an expected average reward of $-1$ and $1$ respectively. In $\pi_x < \pi_z$, $\pi_z$ will reach the 2nd row 2nd column cell and $\pi_x$ will never chase it in that cell, so they will obtain an expected average reward of $-1$ and $1$ respectively. So:

\begin{equation*}
STG_{3,1}(\pi_x,\pi_y,\pi_z,\Pi_o,w_{\dot{L}},\mu) = 1
\end{equation*}

Therefore:

\begin{equation*}
STG_{3,1}(\Pi_e,w_{\Pi_e},\Pi_o,w_{\dot{L}},\mu) = 6 \frac{9}{2} \frac{1}{3} \frac{1}{3} \frac{1}{3} 1 = 1
\end{equation*}

For slots 3 and 2:

\begin{equation*}
\begin{aligned}
STG_{3,2}(\Pi_e,w_{\Pi_e},\Pi_o,w_{\dot{L}},\mu)	& = \eta_{\Pi^3} \sum_{\pi_1,\pi_2,\pi_3 \in \Pi_e | \pi_1 \neq \pi_2 \neq \pi_3} w_{\Pi_e}(\pi_1) w_{\Pi_e}(\pi_2) w_{\Pi_e}(\pi_3) STG_{3,2}(\pi_1,\pi_2,\pi_3,\Pi_o,w_{\dot{L}},\mu) =\\
													& = 6 \frac{9}{2} \frac{1}{3} \frac{1}{3} \frac{1}{3} STG_{3,2}(\pi_x,\pi_y,\pi_z,\Pi_o,w_{\dot{L}},\mu)
\end{aligned}
\end{equation*}

In this case, we only need to calculate $STG_{3,2}(\pi_x,\pi_y,\pi_z,\Pi_o,w_{\dot{L}},\mu)$. We follow definition \ref{def:STG_agents} to calculate this value:

\begin{equation*}
\begin{aligned}
STG_{3,2}(\pi_x,\pi_y,\pi_z,\Pi_o,w_{\dot{L}},\mu)	& = \sum_{\dot{l} \in \dot{L}^{N(\mu)}_{-3,2}(\Pi_o)} w_{\dot{L}}(\dot{l}) STO_{3,2}(\pi_x,\pi_y,\pi_z,\dot{l},\mu) =\\
													& = STO_{3,2}(\pi_x,\pi_y,\pi_z,(\pi_s,*,*,\pi_s),\mu)
\end{aligned}
\end{equation*}

The following table shows us $STO_{3,2}$ for all the permutations of $\pi_x,\pi_y,\pi_z$.

\begin{center}
\begin{tabular}{c c c | c c c | c c c}
Slot 3 & & Slot 2		& Slot 3 & & Slot 2			& Slot 3 & & Slot 2\\
\hline
$\pi_x$ & $<$ & $\pi_y$	& $\pi_x$ & $<$ & $\pi_z$	& $\pi_y$ & $<$ & $\pi_x$\\
$\pi_y$ & $<$ & $\pi_z$	& $\pi_z$ & $<$ & $\pi_y$	& $\pi_x$ & $<$ & $\pi_z$\\
$\pi_x$ & $<$ & $\pi_z$	& $\pi_x$ & $<$ & $\pi_y$	& $\pi_y$ & $<$ & $\pi_z$
\end{tabular}
\begin{tabular}{c c c | c c c | c c c}
Slot 3 & & Slot 2		& Slot 3 & & Slot 2			& Slot 3 & & Slot 2\\
\hline
$\pi_y$ & $<$ & $\pi_z$	& $\pi_z$ & $<$ & $\pi_x$	& $\pi_z$ & $<$ & $\pi_y$\\
$\pi_z$ & $<$ & $\pi_x$	& $\pi_x$ & $<$ & $\pi_y$	& $\pi_y$ & $<$ & $\pi_x$\\
$\pi_y$ & $<$ & $\pi_x$	& $\pi_z$ & $<$ & $\pi_y$	& $\pi_z$ & $<$ & $\pi_x$
\end{tabular}
\end{center}

But, it is not possible to find a STO, since for every permutation the agents in slots 3 and 2 share rewards (and expected average rewards as well) due they are in the same team. So:

\begin{equation*}
STG_{3,2}(\pi_x,\pi_y,\pi_z,\Pi_o,w_{\dot{L}},\mu) = 0
\end{equation*}

Therefore:

\begin{equation*}
STG_{3,2}(\Pi_e,w_{\Pi_e},\Pi_o,w_{\dot{L}},\mu) = 6 \frac{9}{2} \frac{1}{3} \frac{1}{3} \frac{1}{3} 0 = 0
\end{equation*}

For slots 3 and 4:

\begin{equation*}
\begin{aligned}
STG_{3,4}(\Pi_e,w_{\Pi_e},\Pi_o,w_{\dot{L}},\mu)	& = \eta_{\Pi^3} \sum_{\pi_1,\pi_2,\pi_3 \in \Pi_e | \pi_1 \neq \pi_2 \neq \pi_3} w_{\Pi_e}(\pi_1) w_{\Pi_e}(\pi_2) w_{\Pi_e}(\pi_3) STG_{3,4}(\pi_1,\pi_2,\pi_3,\Pi_o,w_{\dot{L}},\mu) =\\
													& = 6 \frac{9}{2} \frac{1}{3} \frac{1}{3} \frac{1}{3} STG_{3,4}(\pi_x,\pi_y,\pi_z,\Pi_o,w_{\dot{L}},\mu)
\end{aligned}
\end{equation*}

In this case, we only need to calculate $STG_{3,4}(\pi_x,\pi_y,\pi_z,\Pi_o,w_{\dot{L}},\mu)$. We follow definition \ref{def:STG_agents} to calculate this value:

\begin{equation*}
\begin{aligned}
STG_{3,4}(\pi_x,\pi_y,\pi_z,\Pi_o,w_{\dot{L}},\mu)	& = \sum_{\dot{l} \in \dot{L}^{N(\mu)}_{-3,4}(\Pi_o)} w_{\dot{L}}(\dot{l}) STO_{3,4}(\pi_x,\pi_y,\pi_z,\dot{l},\mu) =\\
													& = STO_{3,4}(\pi_x,\pi_y,\pi_z,(\pi_s,\pi_s,*,*),\mu)
\end{aligned}
\end{equation*}

The following table shows us $STO_{3,4}$ for all the permutations of $\pi_x,\pi_y,\pi_z$.

\begin{center}
\begin{tabular}{c c c | c c c | c c c}
Slot 3 & & Slot 4		& Slot 3 & & Slot 4			& Slot 3 & & Slot 4\\
\hline
$\pi_x$ & $<$ & $\pi_y$	& $\pi_x$ & $<$ & $\pi_z$	& $\pi_y$ & $<$ & $\pi_x$\\
$\pi_y$ & $<$ & $\pi_z$	& $\pi_z$ & $<$ & $\pi_y$	& $\pi_x$ & $<$ & $\pi_z$\\
$\pi_x$ & $<$ & $\pi_z$	& $\pi_x$ & $<$ & $\pi_y$	& $\pi_y$ & $<$ & $\pi_z$
\end{tabular}
\begin{tabular}{c c c | c c c | c c c}
Slot 3 & & Slot 4		& Slot 3 & & Slot 4			& Slot 3 & & Slot 4\\
\hline
$\pi_y$ & $<$ & $\pi_z$	& $\pi_z$ & $<$ & $\pi_x$	& $\pi_z$ & $<$ & $\pi_y$\\
$\pi_z$ & $<$ & $\pi_x$	& $\pi_x$ & $<$ & $\pi_y$	& $\pi_y$ & $<$ & $\pi_x$\\
$\pi_y$ & $<$ & $\pi_x$	& $\pi_z$ & $<$ & $\pi_y$	& $\pi_z$ & $<$ & $\pi_x$
\end{tabular}
\end{center}

But, it is not possible to find a STO, since for every permutation the agents in slots 3 and 4 share rewards (and expected average rewards as well) due they are in the same team. So:

\begin{equation*}
STG_{3,4}(\pi_x,\pi_y,\pi_z,\Pi_o,w_{\dot{L}},\mu) = 0
\end{equation*}

Therefore:

\begin{equation*}
STG_{3,4}(\Pi_e,w_{\Pi_e},\Pi_o,w_{\dot{L}},\mu) = 6 \frac{9}{2} \frac{1}{3} \frac{1}{3} \frac{1}{3} 0 = 0
\end{equation*}

For slots 4 and 1:

\begin{equation*}
\begin{aligned}
STG_{4,1}(\Pi_e,w_{\Pi_e},\Pi_o,w_{\dot{L}},\mu)	& = \eta_{\Pi^3} \sum_{\pi_1,\pi_2,\pi_3 \in \Pi_e | \pi_1 \neq \pi_2 \neq \pi_3} w_{\Pi_e}(\pi_1) w_{\Pi_e}(\pi_2) w_{\Pi_e}(\pi_3) STG_{4,1}(\pi_1,\pi_2,\pi_3,\Pi_o,w_{\dot{L}},\mu) =\\
													& = 6 \frac{9}{2} \frac{1}{3} \frac{1}{3} \frac{1}{3} STG_{4,1}(\pi_x,\pi_y,\pi_z,\Pi_o,w_{\dot{L}},\mu)
\end{aligned}
\end{equation*}

In this case, we only need to calculate $STG_{4,1}(\pi_x,\pi_y,\pi_z,\Pi_o,w_{\dot{L}},\mu)$. We follow definition \ref{def:STG_agents} to calculate this value:

\begin{equation*}
\begin{aligned}
STG_{4,1}(\pi_x,\pi_y,\pi_z,\Pi_o,w_{\dot{L}},\mu)	& = \sum_{\dot{l} \in \dot{L}^{N(\mu)}_{-4,1}(\Pi_o)} w_{\dot{L}}(\dot{l}) STO_{4,1}(\pi_x,\pi_y,\pi_z,\dot{l},\mu) =\\
													& = STO_{4,1}(\pi_x,\pi_y,\pi_z,(*,\pi_s,\pi_s,*),\mu)
\end{aligned}
\end{equation*}

The following table shows us $STO_{4,1}$ for all the permutations of $\pi_x,\pi_y,\pi_z$.

\begin{center}
\begin{tabular}{c c c | c c c | c c c}
Slot 4 & & Slot 1		& Slot 4 & & Slot 1			& Slot 4 & & Slot 1\\
\hline
$\pi_x$ & $<$ & $\pi_y$	& $\pi_x$ & $<$ & $\pi_z$	& $\pi_y$ & $<$ & $\pi_x$\\
$\pi_y$ & $<$ & $\pi_z$	& $\pi_z$ & $<$ & $\pi_y$	& $\pi_x$ & $<$ & $\pi_z$\\
$\pi_x$ & $<$ & $\pi_z$	& $\pi_x$ & $<$ & $\pi_y$	& $\pi_y$ & $<$ & $\pi_z$
\end{tabular}
\begin{tabular}{c c c | c c c | c c c}
Slot 4 & & Slot 1		& Slot 4 & & Slot 1			& Slot 4 & & Slot 1\\
\hline
$\pi_y$ & $<$ & $\pi_z$	& $\pi_z$ & $<$ & $\pi_x$	& $\pi_z$ & $<$ & $\pi_y$\\
$\pi_z$ & $<$ & $\pi_x$	& $\pi_x$ & $<$ & $\pi_y$	& $\pi_y$ & $<$ & $\pi_x$\\
$\pi_y$ & $<$ & $\pi_x$	& $\pi_z$ & $<$ & $\pi_y$	& $\pi_z$ & $<$ & $\pi_x$
\end{tabular}
\end{center}

It is possible to find a STO for the first permutation. In $\pi_x < \pi_y$, $\pi_y$ will reach the 2nd row 2nd column cell and $\pi_x$ will never chase it in that cell, so they will obtain an expected average reward of $-1$ and $1$ respectively. In $\pi_y < \pi_z$, $\pi_z$ will reach the 2nd row 2nd column cell and $\pi_y$ will never chase it in that cell, so they will obtain an expected average reward of $-1$ and $1$ respectively. In $\pi_x < \pi_z$, $\pi_z$ will reach the 2nd row 2nd column cell and $\pi_x$ will never chase it in that cell, so they will obtain an expected average reward of $-1$ and $1$ respectively. So:

\begin{equation*}
STG_{4,1}(\pi_x,\pi_y,\pi_z,\Pi_o,w_{\dot{L}},\mu) = 1
\end{equation*}

Therefore:

\begin{equation*}
STG_{4,1}(\Pi_e,w_{\Pi_e},\Pi_o,w_{\dot{L}},\mu) = 6 \frac{9}{2} \frac{1}{3} \frac{1}{3} \frac{1}{3} 1 = 1
\end{equation*}

For slots 4 and 2:

\begin{equation*}
\begin{aligned}
STG_{4,2}(\Pi_e,w_{\Pi_e},\Pi_o,w_{\dot{L}},\mu)	& = \eta_{\Pi^3} \sum_{\pi_1,\pi_2,\pi_3 \in \Pi_e | \pi_1 \neq \pi_2 \neq \pi_3} w_{\Pi_e}(\pi_1) w_{\Pi_e}(\pi_2) w_{\Pi_e}(\pi_3) STG_{4,2}(\pi_1,\pi_2,\pi_3,\Pi_o,w_{\dot{L}},\mu) =\\
													& = 6 \frac{9}{2} \frac{1}{3} \frac{1}{3} \frac{1}{3} STG_{4,2}(\pi_x,\pi_y,\pi_z,\Pi_o,w_{\dot{L}},\mu)
\end{aligned}
\end{equation*}

In this case, we only need to calculate $STG_{4,2}(\pi_x,\pi_y,\pi_z,\Pi_o,w_{\dot{L}},\mu)$. We follow definition \ref{def:STG_agents} to calculate this value:

\begin{equation*}
\begin{aligned}
STG_{4,2}(\pi_x,\pi_y,\pi_z,\Pi_o,w_{\dot{L}},\mu)	& = \sum_{\dot{l} \in \dot{L}^{N(\mu)}_{-4,2}(\Pi_o)} w_{\dot{L}}(\dot{l}) STO_{4,2}(\pi_x,\pi_y,\pi_z,\dot{l},\mu) =\\
													& = STO_{4,2}(\pi_x,\pi_y,\pi_z,(\pi_s,*,\pi_s,*),\mu)
\end{aligned}
\end{equation*}

The following table shows us $STO_{4,2}$ for all the permutations of $\pi_x,\pi_y,\pi_z$.

\begin{center}
\begin{tabular}{c c c | c c c | c c c}
Slot 4 & & Slot 2		& Slot 4 & & Slot 2			& Slot 4 & & Slot 2\\
\hline
$\pi_x$ & $<$ & $\pi_y$	& $\pi_x$ & $<$ & $\pi_z$	& $\pi_y$ & $<$ & $\pi_x$\\
$\pi_y$ & $<$ & $\pi_z$	& $\pi_z$ & $<$ & $\pi_y$	& $\pi_x$ & $<$ & $\pi_z$\\
$\pi_x$ & $<$ & $\pi_z$	& $\pi_x$ & $<$ & $\pi_y$	& $\pi_y$ & $<$ & $\pi_z$
\end{tabular}
\begin{tabular}{c c c | c c c | c c c}
Slot 4 & & Slot 2		& Slot 4 & & Slot 2			& Slot 4 & & Slot 2\\
\hline
$\pi_y$ & $<$ & $\pi_z$	& $\pi_z$ & $<$ & $\pi_x$	& $\pi_z$ & $<$ & $\pi_y$\\
$\pi_z$ & $<$ & $\pi_x$	& $\pi_x$ & $<$ & $\pi_y$	& $\pi_y$ & $<$ & $\pi_x$\\
$\pi_y$ & $<$ & $\pi_x$	& $\pi_z$ & $<$ & $\pi_y$	& $\pi_z$ & $<$ & $\pi_x$
\end{tabular}
\end{center}

But, it is not possible to find a STO, since for every permutation the agents in slots 4 and 2 share rewards (and expected average rewards as well) due they are in the same team. So:

\begin{equation*}
STG_{4,2}(\pi_x,\pi_y,\pi_z,\Pi_o,w_{\dot{L}},\mu) = 0
\end{equation*}

Therefore:

\begin{equation*}
STG_{4,2}(\Pi_e,w_{\Pi_e},\Pi_o,w_{\dot{L}},\mu) = 6 \frac{9}{2} \frac{1}{3} \frac{1}{3} \frac{1}{3} 0 = 0
\end{equation*}

And for slots 4 and 3:

\begin{equation*}
\begin{aligned}
STG_{4,3}(\Pi_e,w_{\Pi_e},\Pi_o,w_{\dot{L}},\mu)	& = \eta_{\Pi^3} \sum_{\pi_1,\pi_2,\pi_3 \in \Pi_e | \pi_1 \neq \pi_2 \neq \pi_3} w_{\Pi_e}(\pi_1) w_{\Pi_e}(\pi_2) w_{\Pi_e}(\pi_3) STG_{4,3}(\pi_1,\pi_2,\pi_3,\Pi_o,w_{\dot{L}},\mu) =\\
													& = 6 \frac{9}{2} \frac{1}{3} \frac{1}{3} \frac{1}{3} STG_{4,3}(\pi_x,\pi_y,\pi_z,\Pi_o,w_{\dot{L}},\mu)
\end{aligned}
\end{equation*}

In this case, we only need to calculate $STG_{4,3}(\pi_x,\pi_y,\pi_z,\Pi_o,w_{\dot{L}},\mu)$. We follow definition \ref{def:STG_agents} to calculate this value:

\begin{equation*}
\begin{aligned}
STG_{4,3}(\pi_x,\pi_y,\pi_z,\Pi_o,w_{\dot{L}},\mu)	& = \sum_{\dot{l} \in \dot{L}^{N(\mu)}_{-4,3}(\Pi_o)} w_{\dot{L}}(\dot{l}) STO_{4,3}(\pi_x,\pi_y,\pi_z,\dot{l},\mu) =\\
													& = STO_{4,3}(\pi_x,\pi_y,\pi_z,(\pi_s,\pi_s,*,*),\mu)
\end{aligned}
\end{equation*}

The following table shows us $STO_{4,3}$ for all the permutations of $\pi_x,\pi_y,\pi_z$.

\begin{center}
\begin{tabular}{c c c | c c c | c c c}
Slot 4 & & Slot 3		& Slot 4 & & Slot 3			& Slot 4 & & Slot 3\\
\hline
$\pi_x$ & $<$ & $\pi_y$	& $\pi_x$ & $<$ & $\pi_z$	& $\pi_y$ & $<$ & $\pi_x$\\
$\pi_y$ & $<$ & $\pi_z$	& $\pi_z$ & $<$ & $\pi_y$	& $\pi_x$ & $<$ & $\pi_z$\\
$\pi_x$ & $<$ & $\pi_z$	& $\pi_x$ & $<$ & $\pi_y$	& $\pi_y$ & $<$ & $\pi_z$
\end{tabular}
\begin{tabular}{c c c | c c c | c c c}
Slot 4 & & Slot 3		& Slot 4 & & Slot 3			& Slot 4 & & Slot 3\\
\hline
$\pi_y$ & $<$ & $\pi_z$	& $\pi_z$ & $<$ & $\pi_x$	& $\pi_z$ & $<$ & $\pi_y$\\
$\pi_z$ & $<$ & $\pi_x$	& $\pi_x$ & $<$ & $\pi_y$	& $\pi_y$ & $<$ & $\pi_x$\\
$\pi_y$ & $<$ & $\pi_x$	& $\pi_z$ & $<$ & $\pi_y$	& $\pi_z$ & $<$ & $\pi_x$
\end{tabular}
\end{center}

But, it is not possible to find a STO, since for every permutation the agents in slots 4 and 3 share rewards (and expected average rewards as well) due they are in the same team. So:

\begin{equation*}
STG_{4,3}(\pi_x,\pi_y,\pi_z,\Pi_o,w_{\dot{L}},\mu) = 0
\end{equation*}

Therefore:

\begin{equation*}
STG_{4,3}(\Pi_e,w_{\Pi_e},\Pi_o,w_{\dot{L}},\mu) = 6 \frac{9}{2} \frac{1}{3} \frac{1}{3} \frac{1}{3} 0 = 0
\end{equation*}

And finally, we weight over the slots:

\begin{equation*}
\begin{aligned}
& STG(\Pi_e,w_{\Pi_e},\Pi_o,w_{\dot{L}},\mu,w_S) = \eta_{S_1^2} \sum_{i=1}^{N(\mu)} w_S(i,\mu) \times\\
& \times \left(\sum_{j=1}^{i-1} w_S(j,\mu) STG_{i,j}(\Pi_e,w_{\Pi_e},\Pi_o,w_{\dot{L}},\mu) + \sum_{j=i+1}^{N(\mu)} w_S(j,\mu) STG_{i,j}(\Pi_e,w_{\Pi_e},\Pi_o,w_{\dot{L}},\mu)\right) =\\
& \ \ \ \ \ \ \ \ \ \ \ \ \ \ \ \ \ \ \ \ \ \ \ \ \ \ \ \ \ \ \ \ \ \ \ \ \ \ = \frac{4}{3} \frac{1}{4} \frac{1}{4} \{STG_{1,2}(\Pi_e,w_{\Pi_e},\Pi_o,w_{\dot{L}},\mu) + STG_{1,3}(\Pi_e,w_{\Pi_e},\Pi_o,w_{\dot{L}},\mu) +\\
& \ \ \ \ \ \ \ \ \ \ \ \ \ \ \ \ \ \ \ \ \ \ \ \ \ \ \ \ \ \ \ \ \ \ \ \ \ \ + STG_{1,4}(\Pi_e,w_{\Pi_e},\Pi_o,w_{\dot{L}},\mu) + STG_{2,1}(\Pi_e,w_{\Pi_e},\Pi_o,w_{\dot{L}},\mu) +\\
& \ \ \ \ \ \ \ \ \ \ \ \ \ \ \ \ \ \ \ \ \ \ \ \ \ \ \ \ \ \ \ \ \ \ \ \ \ \ + STG_{2,3}(\Pi_e,w_{\Pi_e},\Pi_o,w_{\dot{L}},\mu) + STG_{2,4}(\Pi_e,w_{\Pi_e},\Pi_o,w_{\dot{L}},\mu) +\\
& \ \ \ \ \ \ \ \ \ \ \ \ \ \ \ \ \ \ \ \ \ \ \ \ \ \ \ \ \ \ \ \ \ \ \ \ \ \ + STG_{3,1}(\Pi_e,w_{\Pi_e},\Pi_o,w_{\dot{L}},\mu) + STG_{3,2}(\Pi_e,w_{\Pi_e},\Pi_o,w_{\dot{L}},\mu) +\\
& \ \ \ \ \ \ \ \ \ \ \ \ \ \ \ \ \ \ \ \ \ \ \ \ \ \ \ \ \ \ \ \ \ \ \ \ \ \ + STG_{3,4}(\Pi_e,w_{\Pi_e},\Pi_o,w_{\dot{L}},\mu) + STG_{4,1}(\Pi_e,w_{\Pi_e},\Pi_o,w_{\dot{L}},\mu) +\\
& \ \ \ \ \ \ \ \ \ \ \ \ \ \ \ \ \ \ \ \ \ \ \ \ \ \ \ \ \ \ \ \ \ \ \ \ \ \ + STG_{4,2}(\Pi_e,w_{\Pi_e},\Pi_o,w_{\dot{L}},\mu) + STG_{4,3}(\Pi_e,w_{\Pi_e},\Pi_o,w_{\dot{L}},\mu)\} =\\
& \ \ \ \ \ \ \ \ \ \ \ \ \ \ \ \ \ \ \ \ \ \ \ \ \ \ \ \ \ \ \ \ \ \ \ \ \ \ = \frac{4}{3} \frac{1}{4} \frac{1}{4} \left\{6 \times 1 + 6 \times 0\right\} = \frac{1}{2}
\end{aligned}
\end{equation*}

Since $\frac{1}{2}$ is the highest possible value that we can obtain for the strict total grading property, therefore predator-prey has $General_{max} = \frac{1}{2}$ for this property.
\end{proof}
\end{proposition}

\begin{approximation}
\label{approx:predator-prey_STG_left_max}
$Left_{max}$ for the strict total grading (STG) property is equal to $\frac{1}{4}$ (as a {\em lower} approximation) for the predator-prey environment.

\begin{proof}
To find $Left_{max}$ (equation \ref{eq:left_max}), we need to find a pair $\left\langle\Pi_e,w_{\Pi_e}\right\rangle$ which maximises the property as much as possible while $\Pi_o$ minimises it. Using $\Pi_e = \{{\pi_{chase}}_1,{\pi_{chase}}_2,{\pi_{chase}}_3\}$ with uniform weight for $w_{\Pi_e}$ (a $\pi_{chase}$ agent always tries to be chased when playing as the prey and tries to chase when playing as a predator) we find a {\em lower} approximation of this situation no matter which $\Pi_o$ we use.

Following definition \ref{def:STG}, we obtain the STG value for this $\left\langle\Pi_e,w_{\Pi_e},\Pi_o\right\rangle$ (where $\Pi_o$ is instantiated with any permitted value). Since the environment is not symmetric, we need to calculate this property for every pair of slots. Following definition \ref{def:STG_set}, we can calculate its STG value for each pair of slots. We start with slots 1 and 2:

\begin{equation*}
\begin{aligned}
STG_{1,2}(\Pi_e,w_{\Pi_e},\Pi_o,w_{\dot{L}},\mu)	& = \eta_{\Pi^3} \sum_{\pi_1,\pi_2,\pi_3 \in \Pi_e | \pi_1 \neq \pi_2 \neq \pi_3} w_{\Pi_e}(\pi_1) w_{\Pi_e}(\pi_2) w_{\Pi_e}(\pi_3) STG_{1,2}(\pi_1,\pi_2,\pi_3,\Pi_o,w_{\dot{L}},\mu) =\\
													& = 6 \frac{9}{2} \frac{1}{3} \frac{1}{3} \frac{1}{3} STG_{1,2}({\pi_{chase}}_1,{\pi_{chase}}_2,{\pi_{chase}}_3,\Pi_o,w_{\dot{L}},\mu)
\end{aligned}
\end{equation*}

\noindent Note that we avoided to calculate all the permutations of $\pi_1,\pi_2,\pi_3$ for $STG_{i,j}(\pi_1,\pi_2,\pi_3,\Pi_o,w_{\dot{L}},\mu)$ since they provide the same result, by calculating only one permutation and multiplying the result by the number of permutations $6$.

In this case, we only need to calculate $STG_{1,2}({\pi_{chase}}_1,{\pi_{chase}}_2,{\pi_{chase}}_3,\Pi_o,w_{\dot{L}},\mu)$. We follow definition \ref{def:STG_agents} to calculate this value:

\begin{equation*}
STG_{1,2}({\pi_{chase}}_1,{\pi_{chase}}_2,{\pi_{chase}}_3,\Pi_o,w_{\dot{L}},\mu) = \sum_{\dot{l} \in \dot{L}^{N(\mu)}_{-1,2}(\Pi_o)} w_{\dot{L}}(\dot{l}) STO_{1,2}({\pi_{chase}}_1,{\pi_{chase}}_2,{\pi_{chase}}_3,\dot{l},\mu)
\end{equation*}

We do not know which $\Pi_o$ we have, but we know that we will need to obtain a line-up pattern $\dot{l}$ from $\dot{L}^{N(\mu)}_{-1,2}(\Pi_o)$ to calculate $STO_{1,2}({\pi_{chase}}_1,{\pi_{chase}}_2,{\pi_{chase}}_3,\dot{l},\mu)$. We calculate this value for a figurative line-up pattern $\dot{l} = (*,*,\pi_1,\pi_2)$ from $\dot{L}^{N(\mu)}_{-1,2}(\Pi_o)$:

\begin{equation*}
STO_{1,2}({\pi_{chase}}_1,{\pi_{chase}}_2,{\pi_{chase}}_3,\dot{l},\mu) = STO_{1,2}({\pi_{chase}}_1,{\pi_{chase}}_2,{\pi_{chase}}_3,(*,*,\pi_1,\pi_2),\mu)
\end{equation*}

The following table shows us $STO_{1,2}$ for all the permutations of ${\pi_{chase}}_1,{\pi_{chase}}_2,{\pi_{chase}}_3$.

\begin{center}
\begin{tabular}{c c c | c c c | c c c}
Slot 1 & & Slot 2							& Slot 1 & & Slot 2								& Slot 1 & & Slot 2\\
\hline
${\pi_{chase}}_1$ & $<$ & ${\pi_{chase}}_2$	& ${\pi_{chase}}_1$ & $<$ & ${\pi_{chase}}_3$	& ${\pi_{chase}}_2$ & $<$ & ${\pi_{chase}}_1$\\
${\pi_{chase}}_2$ & $<$ & ${\pi_{chase}}_3$	& ${\pi_{chase}}_3$ & $<$ & ${\pi_{chase}}_2$	& ${\pi_{chase}}_1$ & $<$ & ${\pi_{chase}}_3$\\
${\pi_{chase}}_1$ & $<$ & ${\pi_{chase}}_3$	& ${\pi_{chase}}_1$ & $<$ & ${\pi_{chase}}_2$	& ${\pi_{chase}}_2$ & $<$ & ${\pi_{chase}}_3$
\end{tabular}
\begin{tabular}{c c c | c c c | c c c}
Slot 1 & & Slot 2							& Slot 1 & & Slot 2								& Slot 1 & & Slot 2\\
\hline
${\pi_{chase}}_2$ & $<$ & ${\pi_{chase}}_3$	& ${\pi_{chase}}_3$ & $<$ & ${\pi_{chase}}_1$	& ${\pi_{chase}}_3$ & $<$ & ${\pi_{chase}}_2$\\
${\pi_{chase}}_3$ & $<$ & ${\pi_{chase}}_1$	& ${\pi_{chase}}_1$ & $<$ & ${\pi_{chase}}_2$	& ${\pi_{chase}}_2$ & $<$ & ${\pi_{chase}}_1$\\
${\pi_{chase}}_2$ & $<$ & ${\pi_{chase}}_1$	& ${\pi_{chase}}_3$ & $<$ & ${\pi_{chase}}_2$	& ${\pi_{chase}}_3$ & $<$ & ${\pi_{chase}}_1$
\end{tabular}
\end{center}

It is possible to find a STO for every permutation, since we always have ${\pi_{chase}}_i < {\pi_{chase}}_j$, where a $\pi_{chase}$ agent always tries to be chased when playing as the prey and tries to chase when playing as a predator, so the agents in slots 1 and 2 will obtain an expected average reward of $-1$ and $1$ respectively. Note that the choice of $\Pi_o$ does not affect the result of $STO_{1,2}$, so no matter which agents are in $\Pi_o$ we obtain:

\begin{equation*}
STG_{1,2}({\pi_{chase}}_1,{\pi_{chase}}_2,{\pi_{chase}}_3,\Pi_o,w_{\dot{L}},\mu) = 1
\end{equation*}

Therefore:

\begin{equation*}
STG_{1,2}(\Pi_e,w_{\Pi_e},\Pi_o,w_{\dot{L}},\mu) = 6 \frac{9}{2} \frac{1}{3} \frac{1}{3} \frac{1}{3} 1 = 1
\end{equation*}

For slots 1 and 3:

\begin{equation*}
\begin{aligned}
STG_{1,3}(\Pi_e,w_{\Pi_e},\Pi_o,w_{\dot{L}},\mu)	& = \eta_{\Pi^3} \sum_{\pi_1,\pi_2,\pi_3 \in \Pi_e | \pi_1 \neq \pi_2 \neq \pi_3} w_{\Pi_e}(\pi_1) w_{\Pi_e}(\pi_2) w_{\Pi_e}(\pi_3) STG_{1,3}(\pi_1,\pi_2,\pi_3,\Pi_o,w_{\dot{L}},\mu) =\\
													& = 6 \frac{9}{2} \frac{1}{3} \frac{1}{3} \frac{1}{3} STG_{1,3}({\pi_{chase}}_1,{\pi_{chase}}_2,{\pi_{chase}}_3,\Pi_o,w_{\dot{L}},\mu)
\end{aligned}
\end{equation*}

In this case, we only need to calculate $STG_{1,3}({\pi_{chase}}_1,{\pi_{chase}}_2,{\pi_{chase}}_3,\Pi_o,w_{\dot{L}},\mu)$. We follow definition \ref{def:STG_agents} to calculate this value:

\begin{equation*}
STG_{1,3}({\pi_{chase}}_1,{\pi_{chase}}_2,{\pi_{chase}}_3,\Pi_o,w_{\dot{L}},\mu) = \sum_{\dot{l} \in \dot{L}^{N(\mu)}_{-1,3}(\Pi_o)} w_{\dot{L}}(\dot{l}) STO_{1,3}({\pi_{chase}}_1,{\pi_{chase}}_2,{\pi_{chase}}_3,\dot{l},\mu)
\end{equation*}

Again, we do not know which $\Pi_o$ we have, but we know that we will need to obtain a line-up pattern $\dot{l}$ from $\dot{L}^{N(\mu)}_{-1,3}(\Pi_o)$ to calculate $STO_{1,3}({\pi_{chase}}_1,{\pi_{chase}}_2,{\pi_{chase}}_3,\dot{l},\mu)$. We calculate this value for a figurative line-up pattern $\dot{l} = (*,\pi_1,*,\pi_2)$ from $\dot{L}^{N(\mu)}_{-1,3}(\Pi_o)$:

\begin{equation*}
STO_{1,3}({\pi_{chase}}_1,{\pi_{chase}}_2,{\pi_{chase}}_3,\dot{l},\mu) = STO_{1,3}({\pi_{chase}}_1,{\pi_{chase}}_2,{\pi_{chase}}_3,(*,\pi_1,*,\pi_2),\mu)
\end{equation*}

The following table shows us $STO_{1,3}$ for all the permutations of ${\pi_{chase}}_1,{\pi_{chase}}_2,{\pi_{chase}}_3$.

\begin{center}
\begin{tabular}{c c c | c c c | c c c}
Slot 1 & & Slot 3							& Slot 1 & & Slot 3								& Slot 1 & & Slot 3\\
\hline
${\pi_{chase}}_1$ & $<$ & ${\pi_{chase}}_2$	& ${\pi_{chase}}_1$ & $<$ & ${\pi_{chase}}_3$	& ${\pi_{chase}}_2$ & $<$ & ${\pi_{chase}}_1$\\
${\pi_{chase}}_2$ & $<$ & ${\pi_{chase}}_3$	& ${\pi_{chase}}_3$ & $<$ & ${\pi_{chase}}_2$	& ${\pi_{chase}}_1$ & $<$ & ${\pi_{chase}}_3$\\
${\pi_{chase}}_1$ & $<$ & ${\pi_{chase}}_3$	& ${\pi_{chase}}_1$ & $<$ & ${\pi_{chase}}_2$	& ${\pi_{chase}}_2$ & $<$ & ${\pi_{chase}}_3$
\end{tabular}
\begin{tabular}{c c c | c c c | c c c}
Slot 1 & & Slot 3							& Slot 1 & & Slot 3								& Slot 1 & & Slot 3\\
\hline
${\pi_{chase}}_2$ & $<$ & ${\pi_{chase}}_3$	& ${\pi_{chase}}_3$ & $<$ & ${\pi_{chase}}_1$	& ${\pi_{chase}}_3$ & $<$ & ${\pi_{chase}}_2$\\
${\pi_{chase}}_3$ & $<$ & ${\pi_{chase}}_1$	& ${\pi_{chase}}_1$ & $<$ & ${\pi_{chase}}_2$	& ${\pi_{chase}}_2$ & $<$ & ${\pi_{chase}}_1$\\
${\pi_{chase}}_2$ & $<$ & ${\pi_{chase}}_1$	& ${\pi_{chase}}_3$ & $<$ & ${\pi_{chase}}_2$	& ${\pi_{chase}}_3$ & $<$ & ${\pi_{chase}}_1$
\end{tabular}
\end{center}

Again, it is possible to find a STO for every permutation, since we always have ${\pi_{chase}}_i < {\pi_{chase}}_j$, where a $\pi_{chase}$ agent always tries to be chased when playing as the prey and tries to chase when playing as a predator, so the agents in slots 1 and 3 will obtain an expected average reward of $-1$ and $1$ respectively. Note that the choice of $\Pi_o$ does not affect the result of $STO_{1,3}$, so no matter which agents are in $\Pi_o$ we obtain:

\begin{equation*}
STG_{1,3}({\pi_{chase}}_1,{\pi_{chase}}_2,{\pi_{chase}}_3,\Pi_o,w_{\dot{L}},\mu) = 1
\end{equation*}

Therefore:

\begin{equation*}
STG_{1,3}(\Pi_e,w_{\Pi_e},\Pi_o,w_{\dot{L}},\mu) = 6 \frac{9}{2} \frac{1}{3} \frac{1}{3} \frac{1}{3} 1 = 1
\end{equation*}

For slots 1 and 4:

\begin{equation*}
\begin{aligned}
STG_{1,4}(\Pi_e,w_{\Pi_e},\Pi_o,w_{\dot{L}},\mu)	& = \eta_{\Pi^3} \sum_{\pi_1,\pi_2,\pi_3 \in \Pi_e | \pi_1 \neq \pi_2 \neq \pi_3} w_{\Pi_e}(\pi_1) w_{\Pi_e}(\pi_2) w_{\Pi_e}(\pi_3) STG_{1,4}(\pi_1,\pi_2,\pi_3,\Pi_o,w_{\dot{L}},\mu) =\\
													& = 6 \frac{9}{2} \frac{1}{3} \frac{1}{3} \frac{1}{3} STG_{1,4}({\pi_{chase}}_1,{\pi_{chase}}_2,{\pi_{chase}}_3,\Pi_o,w_{\dot{L}},\mu)
\end{aligned}
\end{equation*}

In this case, we only need to calculate $STG_{1,4}({\pi_{chase}}_1,{\pi_{chase}}_2,{\pi_{chase}}_3,\Pi_o,w_{\dot{L}},\mu)$. We follow definition \ref{def:STG_agents} to calculate this value:

\begin{equation*}
STG_{1,4}({\pi_{chase}}_1,{\pi_{chase}}_2,{\pi_{chase}}_3,\Pi_o,w_{\dot{L}},\mu) = \sum_{\dot{l} \in \dot{L}^{N(\mu)}_{-1,4}(\Pi_o)} w_{\dot{L}}(\dot{l}) STO_{1,4}({\pi_{chase}}_1,{\pi_{chase}}_2,{\pi_{chase}}_3,\dot{l},\mu)
\end{equation*}

Again, we do not know which $\Pi_o$ we have, but we know that we will need to obtain a line-up pattern $\dot{l}$ from $\dot{L}^{N(\mu)}_{-1,4}(\Pi_o)$ to calculate $STO_{1,4}({\pi_{chase}}_1,{\pi_{chase}}_2,{\pi_{chase}}_3,\dot{l},\mu)$. We calculate this value for a figurative line-up pattern $\dot{l} = (*,\pi_1,\pi_2,*)$ from $\dot{L}^{N(\mu)}_{-1,4}(\Pi_o)$:

\begin{equation*}
STO_{1,4}({\pi_{chase}}_1,{\pi_{chase}}_2,{\pi_{chase}}_3,\dot{l},\mu) = STO_{1,4}({\pi_{chase}}_1,{\pi_{chase}}_2,{\pi_{chase}}_3,(*,\pi_1,\pi_2,*),\mu)
\end{equation*}

The following table shows us $STO_{1,4}$ for all the permutations of ${\pi_{chase}}_1,{\pi_{chase}}_2,{\pi_{chase}}_3$.

\begin{center}
\begin{tabular}{c c c | c c c | c c c}
Slot 1 & & Slot 4							& Slot 1 & & Slot 4								& Slot 1 & & Slot 4\\
\hline
${\pi_{chase}}_1$ & $<$ & ${\pi_{chase}}_2$	& ${\pi_{chase}}_1$ & $<$ & ${\pi_{chase}}_3$	& ${\pi_{chase}}_2$ & $<$ & ${\pi_{chase}}_1$\\
${\pi_{chase}}_2$ & $<$ & ${\pi_{chase}}_3$	& ${\pi_{chase}}_3$ & $<$ & ${\pi_{chase}}_2$	& ${\pi_{chase}}_1$ & $<$ & ${\pi_{chase}}_3$\\
${\pi_{chase}}_1$ & $<$ & ${\pi_{chase}}_3$	& ${\pi_{chase}}_1$ & $<$ & ${\pi_{chase}}_2$	& ${\pi_{chase}}_2$ & $<$ & ${\pi_{chase}}_3$
\end{tabular}
\begin{tabular}{c c c | c c c | c c c}
Slot 1 & & Slot 4							& Slot 1 & & Slot 4								& Slot 1 & & Slot 4\\
\hline
${\pi_{chase}}_2$ & $<$ & ${\pi_{chase}}_3$	& ${\pi_{chase}}_3$ & $<$ & ${\pi_{chase}}_1$	& ${\pi_{chase}}_3$ & $<$ & ${\pi_{chase}}_2$\\
${\pi_{chase}}_3$ & $<$ & ${\pi_{chase}}_1$	& ${\pi_{chase}}_1$ & $<$ & ${\pi_{chase}}_2$	& ${\pi_{chase}}_2$ & $<$ & ${\pi_{chase}}_1$\\
${\pi_{chase}}_2$ & $<$ & ${\pi_{chase}}_1$	& ${\pi_{chase}}_3$ & $<$ & ${\pi_{chase}}_2$	& ${\pi_{chase}}_3$ & $<$ & ${\pi_{chase}}_1$
\end{tabular}
\end{center}

Again, it is possible to find a STO for every permutation, since we always have ${\pi_{chase}}_i < {\pi_{chase}}_j$, where a $\pi_{chase}$ agent always tries to be chased when playing as the prey and tries to chase when playing as a predator, so the agents in slots 1 and 4 will obtain an expected average reward of $-1$ and $1$ respectively. Note that the choice of $\Pi_o$ does not affect the result of $STO_{1,4}$, so no matter which agents are in $\Pi_o$ we obtain:

\begin{equation*}
STG_{1,4}({\pi_{chase}}_1,{\pi_{chase}}_2,{\pi_{chase}}_3,\Pi_o,w_{\dot{L}},\mu) = 1
\end{equation*}

Therefore:

\begin{equation*}
STG_{1,4}(\Pi_e,w_{\Pi_e},\Pi_o,w_{\dot{L}},\mu) = 6 \frac{9}{2} \frac{1}{3} \frac{1}{3} \frac{1}{3} 1 = 1
\end{equation*}

For slots 2 and 1:

\begin{equation*}
\begin{aligned}
STG_{2,1}(\Pi_e,w_{\Pi_e},\Pi_o,w_{\dot{L}},\mu)	& = \eta_{\Pi^3} \sum_{\pi_1,\pi_2,\pi_3 \in \Pi_e | \pi_1 \neq \pi_2 \neq \pi_3} w_{\Pi_e}(\pi_1) w_{\Pi_e}(\pi_2) w_{\Pi_e}(\pi_3) STG_{2,1}(\pi_1,\pi_2,\pi_3,\Pi_o,w_{\dot{L}},\mu) =\\
													& = 6 \frac{9}{2} \frac{1}{3} \frac{1}{3} \frac{1}{3} STG_{2,1}({\pi_{chase}}_1,{\pi_{chase}}_2,{\pi_{chase}}_3,\Pi_o,w_{\dot{L}},\mu)
\end{aligned}
\end{equation*}

In this case, we only need to calculate $STG_{2,1}({\pi_{chase}}_1,{\pi_{chase}}_2,{\pi_{chase}}_3,\Pi_o,w_{\dot{L}},\mu)$. We follow definition \ref{def:STG_agents} to calculate this value:

\begin{equation*}
STG_{2,1}({\pi_{chase}}_1,{\pi_{chase}}_2,{\pi_{chase}}_3,\Pi_o,w_{\dot{L}},\mu) = \sum_{\dot{l} \in \dot{L}^{N(\mu)}_{-2,1}(\Pi_o)} w_{\dot{L}}(\dot{l}) STO_{2,1}({\pi_{chase}}_1,{\pi_{chase}}_2,{\pi_{chase}}_3,\dot{l},\mu)
\end{equation*}

Again, we do not know which $\Pi_o$ we have, but we know that we will need to obtain a line-up pattern $\dot{l}$ from $\dot{L}^{N(\mu)}_{-2,1}(\Pi_o)$ to calculate $STO_{2,1}({\pi_{chase}}_1,{\pi_{chase}}_2,{\pi_{chase}}_3,\dot{l},\mu)$. We calculate this value for a figurative line-up pattern $\dot{l} = (*,*,\pi_1,\pi_2)$ from $\dot{L}^{N(\mu)}_{-2,1}(\Pi_o)$:

\begin{equation*}
STO_{2,1}({\pi_{chase}}_1,{\pi_{chase}}_2,{\pi_{chase}}_3,\dot{l},\mu) = STO_{2,1}({\pi_{chase}}_1,{\pi_{chase}}_2,{\pi_{chase}}_3,(*,*,\pi_1,\pi_2),\mu)
\end{equation*}

The following table shows us $STO_{2,1}$ for all the permutations of ${\pi_{chase}}_1,{\pi_{chase}}_2,{\pi_{chase}}_3$.

\begin{center}
\begin{tabular}{c c c | c c c | c c c}
Slot 2 & & Slot 1							& Slot 2 & & Slot 1								& Slot 2 & & Slot 1\\
\hline
${\pi_{chase}}_1$ & $<$ & ${\pi_{chase}}_2$	& ${\pi_{chase}}_1$ & $<$ & ${\pi_{chase}}_3$	& ${\pi_{chase}}_2$ & $<$ & ${\pi_{chase}}_1$\\
${\pi_{chase}}_2$ & $<$ & ${\pi_{chase}}_3$	& ${\pi_{chase}}_3$ & $<$ & ${\pi_{chase}}_2$	& ${\pi_{chase}}_1$ & $<$ & ${\pi_{chase}}_3$\\
${\pi_{chase}}_1$ & $<$ & ${\pi_{chase}}_3$	& ${\pi_{chase}}_1$ & $<$ & ${\pi_{chase}}_2$	& ${\pi_{chase}}_2$ & $<$ & ${\pi_{chase}}_3$
\end{tabular}
\begin{tabular}{c c c | c c c | c c c}
Slot 2 & & Slot 1							& Slot 2 & & Slot 1								& Slot 2 & & Slot 1\\
\hline
${\pi_{chase}}_2$ & $<$ & ${\pi_{chase}}_3$	& ${\pi_{chase}}_3$ & $<$ & ${\pi_{chase}}_1$	& ${\pi_{chase}}_3$ & $<$ & ${\pi_{chase}}_2$\\
${\pi_{chase}}_3$ & $<$ & ${\pi_{chase}}_1$	& ${\pi_{chase}}_1$ & $<$ & ${\pi_{chase}}_2$	& ${\pi_{chase}}_2$ & $<$ & ${\pi_{chase}}_1$\\
${\pi_{chase}}_2$ & $<$ & ${\pi_{chase}}_1$	& ${\pi_{chase}}_3$ & $<$ & ${\pi_{chase}}_2$	& ${\pi_{chase}}_3$ & $<$ & ${\pi_{chase}}_1$
\end{tabular}
\end{center}

It is not possible to find a STO for any permutation, since we always have ${\pi_{chase}}_i < {\pi_{chase}}_j$, where a $\pi_{chase}$ agent always tries to be chased when playing as the prey and tries to chase when playing as a predator, so the agents in slots 2 and 1 will obtain an expected average reward of $1$ and $-1$ respectively. Note that the choice of $\Pi_o$ does not affect the result of $STO_{2,1}$, so no matter which agents are in $\Pi_o$ we obtain:

\begin{equation*}
STG_{2,1}({\pi_{chase}}_1,{\pi_{chase}}_2,{\pi_{chase}}_3,\Pi_o,w_{\dot{L}},\mu) = 0
\end{equation*}

Therefore:

\begin{equation*}
STG_{2,1}(\Pi_e,w_{\Pi_e},\Pi_o,w_{\dot{L}},\mu) = 6 \frac{9}{2} \frac{1}{3} \frac{1}{3} \frac{1}{3} 0 = 0
\end{equation*}

For slots 2 and 3:

\begin{equation*}
\begin{aligned}
STG_{2,3}(\Pi_e,w_{\Pi_e},\Pi_o,w_{\dot{L}},\mu)	& = \eta_{\Pi^3} \sum_{\pi_1,\pi_2,\pi_3 \in \Pi_e | \pi_1 \neq \pi_2 \neq \pi_3} w_{\Pi_e}(\pi_1) w_{\Pi_e}(\pi_2) w_{\Pi_e}(\pi_3) STG_{2,3}(\pi_1,\pi_2,\pi_3,\Pi_o,w_{\dot{L}},\mu) =\\
													& = 6 \frac{9}{2} \frac{1}{3} \frac{1}{3} \frac{1}{3} STG_{2,3}({\pi_{chase}}_1,{\pi_{chase}}_2,{\pi_{chase}}_3,\Pi_o,w_{\dot{L}},\mu)
\end{aligned}
\end{equation*}

In this case, we only need to calculate $STG_{2,3}({\pi_{chase}}_1,{\pi_{chase}}_2,{\pi_{chase}}_3,\Pi_o,w_{\dot{L}},\mu)$. We follow definition \ref{def:STG_agents} to calculate this value:

\begin{equation*}
STG_{2,3}({\pi_{chase}}_1,{\pi_{chase}}_2,{\pi_{chase}}_3,\Pi_o,w_{\dot{L}},\mu) = \sum_{\dot{l} \in \dot{L}^{N(\mu)}_{-2,3}(\Pi_o)} w_{\dot{L}}(\dot{l}) STO_{2,3}({\pi_{chase}}_1,{\pi_{chase}}_2,{\pi_{chase}}_3,\dot{l},\mu)
\end{equation*}

Again, we do not know which $\Pi_o$ we have, but we know that we will need to obtain a line-up pattern $\dot{l}$ from $\dot{L}^{N(\mu)}_{-2,3}(\Pi_o)$ to calculate $STO_{2,3}({\pi_{chase}}_1,{\pi_{chase}}_2,{\pi_{chase}}_3,\dot{l},\mu)$. We calculate this value for a figurative line-up pattern $\dot{l} = (\pi_1,*,*,\pi_2)$ from $\dot{L}^{N(\mu)}_{-2,3}(\Pi_o)$:

\begin{equation*}
STO_{2,3}({\pi_{chase}}_1,{\pi_{chase}}_2,{\pi_{chase}}_3,\dot{l},\mu) = STO_{2,3}({\pi_{chase}}_1,{\pi_{chase}}_2,{\pi_{chase}}_3,(\pi_1,*,*,\pi_2),\mu)
\end{equation*}

The following table shows us $STO_{2,3}$ for all the permutations of ${\pi_{chase}}_1,{\pi_{chase}}_2,{\pi_{chase}}_3$.

\begin{center}
\begin{tabular}{c c c | c c c | c c c}
Slot 2 & & Slot 3							& Slot 2 & & Slot 3								& Slot 2 & & Slot 3\\
\hline
${\pi_{chase}}_1$ & $<$ & ${\pi_{chase}}_2$	& ${\pi_{chase}}_1$ & $<$ & ${\pi_{chase}}_3$	& ${\pi_{chase}}_2$ & $<$ & ${\pi_{chase}}_1$\\
${\pi_{chase}}_2$ & $<$ & ${\pi_{chase}}_3$	& ${\pi_{chase}}_3$ & $<$ & ${\pi_{chase}}_2$	& ${\pi_{chase}}_1$ & $<$ & ${\pi_{chase}}_3$\\
${\pi_{chase}}_1$ & $<$ & ${\pi_{chase}}_3$	& ${\pi_{chase}}_1$ & $<$ & ${\pi_{chase}}_2$	& ${\pi_{chase}}_2$ & $<$ & ${\pi_{chase}}_3$
\end{tabular}
\begin{tabular}{c c c | c c c | c c c}
Slot 2 & & Slot 3							& Slot 2 & & Slot 3								& Slot 2 & & Slot 3\\
\hline
${\pi_{chase}}_2$ & $<$ & ${\pi_{chase}}_3$	& ${\pi_{chase}}_3$ & $<$ & ${\pi_{chase}}_1$	& ${\pi_{chase}}_3$ & $<$ & ${\pi_{chase}}_2$\\
${\pi_{chase}}_3$ & $<$ & ${\pi_{chase}}_1$	& ${\pi_{chase}}_1$ & $<$ & ${\pi_{chase}}_2$	& ${\pi_{chase}}_2$ & $<$ & ${\pi_{chase}}_1$\\
${\pi_{chase}}_2$ & $<$ & ${\pi_{chase}}_1$	& ${\pi_{chase}}_3$ & $<$ & ${\pi_{chase}}_2$	& ${\pi_{chase}}_3$ & $<$ & ${\pi_{chase}}_1$
\end{tabular}
\end{center}

But, it is not possible to find a STO, since for every permutation the agents in slots 2 and 3 share rewards (and expected average rewards as well) due they are in the same team. So no matter which agents are in $\Pi_o$ we obtain:

\begin{equation*}
STG_{2,3}({\pi_{chase}}_1,{\pi_{chase}}_2,{\pi_{chase}}_3,\Pi_o,w_{\dot{L}},\mu) = 0
\end{equation*}

Therefore:

\begin{equation*}
STG_{2,3}(\Pi_e,w_{\Pi_e},\Pi_o,w_{\dot{L}},\mu) = 6 \frac{9}{2} \frac{1}{3} \frac{1}{3} \frac{1}{3} 0 = 0
\end{equation*}

For slots 2 and 4:

\begin{equation*}
\begin{aligned}
STG_{2,4}(\Pi_e,w_{\Pi_e},\Pi_o,w_{\dot{L}},\mu)	& = \eta_{\Pi^3} \sum_{\pi_1,\pi_2,\pi_3 \in \Pi_e | \pi_1 \neq \pi_2 \neq \pi_3} w_{\Pi_e}(\pi_1) w_{\Pi_e}(\pi_2) w_{\Pi_e}(\pi_3) STG_{2,4}(\pi_1,\pi_2,\pi_3,\Pi_o,w_{\dot{L}},\mu) =\\
													& = 6 \frac{9}{2} \frac{1}{3} \frac{1}{3} \frac{1}{3} STG_{2,4}({\pi_{chase}}_1,{\pi_{chase}}_2,{\pi_{chase}}_3,\Pi_o,w_{\dot{L}},\mu)
\end{aligned}
\end{equation*}

In this case, we only need to calculate $STG_{2,4}({\pi_{chase}}_1,{\pi_{chase}}_2,{\pi_{chase}}_3,\Pi_o,w_{\dot{L}},\mu)$. We follow definition \ref{def:STG_agents} to calculate this value:

\begin{equation*}
STG_{2,4}({\pi_{chase}}_1,{\pi_{chase}}_2,{\pi_{chase}}_3,\Pi_o,w_{\dot{L}},\mu) = \sum_{\dot{l} \in \dot{L}^{N(\mu)}_{-2,4}(\Pi_o)} w_{\dot{L}}(\dot{l}) STO_{2,4}({\pi_{chase}}_1,{\pi_{chase}}_2,{\pi_{chase}}_3,\dot{l},\mu)
\end{equation*}

Again, we do not know which $\Pi_o$ we have, but we know that we will need to obtain a line-up pattern $\dot{l}$ from $\dot{L}^{N(\mu)}_{-2,4}(\Pi_o)$ to calculate $STO_{2,4}({\pi_{chase}}_1,{\pi_{chase}}_2,{\pi_{chase}}_3,\dot{l},\mu)$. We calculate this value for a figurative line-up pattern $\dot{l} = (\pi_1,*,\pi_2,*)$ from $\dot{L}^{N(\mu)}_{-2,4}(\Pi_o)$:

\begin{equation*}
STO_{2,4}({\pi_{chase}}_1,{\pi_{chase}}_2,{\pi_{chase}}_3,\dot{l},\mu) = STO_{2,4}({\pi_{chase}}_1,{\pi_{chase}}_2,{\pi_{chase}}_3,(\pi_1,*,\pi_2,*),\mu)
\end{equation*}

The following table shows us $STO_{2,4}$ for all the permutations of ${\pi_{chase}}_1,{\pi_{chase}}_2,{\pi_{chase}}_3$.

\begin{center}
\begin{tabular}{c c c | c c c | c c c}
Slot 2 & & Slot 4							& Slot 2 & & Slot 4								& Slot 2 & & Slot 4\\
\hline
${\pi_{chase}}_1$ & $<$ & ${\pi_{chase}}_2$	& ${\pi_{chase}}_1$ & $<$ & ${\pi_{chase}}_3$	& ${\pi_{chase}}_2$ & $<$ & ${\pi_{chase}}_1$\\
${\pi_{chase}}_2$ & $<$ & ${\pi_{chase}}_3$	& ${\pi_{chase}}_3$ & $<$ & ${\pi_{chase}}_2$	& ${\pi_{chase}}_1$ & $<$ & ${\pi_{chase}}_3$\\
${\pi_{chase}}_1$ & $<$ & ${\pi_{chase}}_3$	& ${\pi_{chase}}_1$ & $<$ & ${\pi_{chase}}_2$	& ${\pi_{chase}}_2$ & $<$ & ${\pi_{chase}}_3$
\end{tabular}
\begin{tabular}{c c c | c c c | c c c}
Slot 2 & & Slot 4							& Slot 2 & & Slot 4								& Slot 2 & & Slot 4\\
\hline
${\pi_{chase}}_2$ & $<$ & ${\pi_{chase}}_3$	& ${\pi_{chase}}_3$ & $<$ & ${\pi_{chase}}_1$	& ${\pi_{chase}}_3$ & $<$ & ${\pi_{chase}}_2$\\
${\pi_{chase}}_3$ & $<$ & ${\pi_{chase}}_1$	& ${\pi_{chase}}_1$ & $<$ & ${\pi_{chase}}_2$	& ${\pi_{chase}}_2$ & $<$ & ${\pi_{chase}}_1$\\
${\pi_{chase}}_2$ & $<$ & ${\pi_{chase}}_1$	& ${\pi_{chase}}_3$ & $<$ & ${\pi_{chase}}_2$	& ${\pi_{chase}}_3$ & $<$ & ${\pi_{chase}}_1$
\end{tabular}
\end{center}

But, it is not possible to find a STO, since for every permutation the agents in slots 2 and 4 share rewards (and expected average rewards as well) due they are in the same team. So no matter which agents are in $\Pi_o$ we obtain:

\begin{equation*}
STG_{2,4}({\pi_{chase}}_1,{\pi_{chase}}_2,{\pi_{chase}}_3,\Pi_o,w_{\dot{L}},\mu) = 0
\end{equation*}

Therefore:

\begin{equation*}
STG_{2,4}(\Pi_e,w_{\Pi_e},\Pi_o,w_{\dot{L}},\mu) = 6 \frac{9}{2} \frac{1}{3} \frac{1}{3} \frac{1}{3} 0 = 0
\end{equation*}

For slots 3 and 1:

\begin{equation*}
\begin{aligned}
STG_{3,1}(\Pi_e,w_{\Pi_e},\Pi_o,w_{\dot{L}},\mu)	& = \eta_{\Pi^3} \sum_{\pi_1,\pi_2,\pi_3 \in \Pi_e | \pi_1 \neq \pi_2 \neq \pi_3} w_{\Pi_e}(\pi_1) w_{\Pi_e}(\pi_2) w_{\Pi_e}(\pi_3) STG_{3,1}(\pi_1,\pi_2,\pi_3,\Pi_o,w_{\dot{L}},\mu) =\\
													& = 6 \frac{9}{2} \frac{1}{3} \frac{1}{3} \frac{1}{3} STG_{3,1}({\pi_{chase}}_1,{\pi_{chase}}_2,{\pi_{chase}}_3,\Pi_o,w_{\dot{L}},\mu)
\end{aligned}
\end{equation*}

In this case, we only need to calculate $STG_{3,1}({\pi_{chase}}_1,{\pi_{chase}}_2,{\pi_{chase}}_3,\Pi_o,w_{\dot{L}},\mu)$. We follow definition \ref{def:STG_agents} to calculate this value:

\begin{equation*}
STG_{3,1}({\pi_{chase}}_1,{\pi_{chase}}_2,{\pi_{chase}}_3,\Pi_o,w_{\dot{L}},\mu) = \sum_{\dot{l} \in \dot{L}^{N(\mu)}_{-3,1}(\Pi_o)} w_{\dot{L}}(\dot{l}) STO_{3,1}({\pi_{chase}}_1,{\pi_{chase}}_2,{\pi_{chase}}_3,\dot{l},\mu)
\end{equation*}

Again, we do not know which $\Pi_o$ we have, but we know that we will need to obtain a line-up pattern $\dot{l}$ from $\dot{L}^{N(\mu)}_{-3,1}(\Pi_o)$ to calculate $STO_{3,1}({\pi_{chase}}_1,{\pi_{chase}}_2,{\pi_{chase}}_3,\dot{l},\mu)$. We calculate this value for a figurative line-up pattern $\dot{l} = (*,\pi_1,*,\pi_2)$ from $\dot{L}^{N(\mu)}_{-3,1}(\Pi_o)$:

\begin{equation*}
STO_{3,1}({\pi_{chase}}_1,{\pi_{chase}}_2,{\pi_{chase}}_3,\dot{l},\mu) = STO_{3,1}({\pi_{chase}}_1,{\pi_{chase}}_2,{\pi_{chase}}_3,(*,\pi_1,*,\pi_2),\mu)
\end{equation*}

The following table shows us $STO_{3,1}$ for all the permutations of ${\pi_{chase}}_1,{\pi_{chase}}_2,{\pi_{chase}}_3$.

\begin{center}
\begin{tabular}{c c c | c c c | c c c}
Slot 3 & & Slot 1							& Slot 3 & & Slot 1								& Slot 3 & & Slot 1\\
\hline
${\pi_{chase}}_1$ & $<$ & ${\pi_{chase}}_2$	& ${\pi_{chase}}_1$ & $<$ & ${\pi_{chase}}_3$	& ${\pi_{chase}}_2$ & $<$ & ${\pi_{chase}}_1$\\
${\pi_{chase}}_2$ & $<$ & ${\pi_{chase}}_3$	& ${\pi_{chase}}_3$ & $<$ & ${\pi_{chase}}_2$	& ${\pi_{chase}}_1$ & $<$ & ${\pi_{chase}}_3$\\
${\pi_{chase}}_1$ & $<$ & ${\pi_{chase}}_3$	& ${\pi_{chase}}_1$ & $<$ & ${\pi_{chase}}_2$	& ${\pi_{chase}}_2$ & $<$ & ${\pi_{chase}}_3$
\end{tabular}
\begin{tabular}{c c c | c c c | c c c}
Slot 3 & & Slot 1							& Slot 3 & & Slot 1								& Slot 3 & & Slot 1\\
\hline
${\pi_{chase}}_2$ & $<$ & ${\pi_{chase}}_3$	& ${\pi_{chase}}_3$ & $<$ & ${\pi_{chase}}_1$	& ${\pi_{chase}}_3$ & $<$ & ${\pi_{chase}}_2$\\
${\pi_{chase}}_3$ & $<$ & ${\pi_{chase}}_1$	& ${\pi_{chase}}_1$ & $<$ & ${\pi_{chase}}_2$	& ${\pi_{chase}}_2$ & $<$ & ${\pi_{chase}}_1$\\
${\pi_{chase}}_2$ & $<$ & ${\pi_{chase}}_1$	& ${\pi_{chase}}_3$ & $<$ & ${\pi_{chase}}_2$	& ${\pi_{chase}}_3$ & $<$ & ${\pi_{chase}}_1$
\end{tabular}
\end{center}

Again, it is not possible to find a STO for any permutation, since we always have ${\pi_{chase}}_i < {\pi_{chase}}_j$, where a $\pi_{chase}$ agent always tries to be chased when playing as the prey and tries to chase when playing as a predator, so the agents in slots 3 and 1 will obtain an expected average reward of $1$ and $-1$ respectively. Note that the choice of $\Pi_o$ does not affect the result of $STO_{3,1}$, so no matter which agents are in $\Pi_o$ we obtain:

\begin{equation*}
STG_{3,1}({\pi_{chase}}_1,{\pi_{chase}}_2,{\pi_{chase}}_3,\Pi_o,w_{\dot{L}},\mu) = 0
\end{equation*}

Therefore:

\begin{equation*}
STG_{3,1}(\Pi_e,w_{\Pi_e},\Pi_o,w_{\dot{L}},\mu) = 6 \frac{9}{2} \frac{1}{3} \frac{1}{3} \frac{1}{3} 0 = 0
\end{equation*}

For slots 3 and 2:

\begin{equation*}
\begin{aligned}
STG_{3,2}(\Pi_e,w_{\Pi_e},\Pi_o,w_{\dot{L}},\mu)	& = \eta_{\Pi^3} \sum_{\pi_1,\pi_2,\pi_3 \in \Pi_e | \pi_1 \neq \pi_2 \neq \pi_3} w_{\Pi_e}(\pi_1) w_{\Pi_e}(\pi_2) w_{\Pi_e}(\pi_3) STG_{3,2}(\pi_1,\pi_2,\pi_3,\Pi_o,w_{\dot{L}},\mu) =\\
													& = 6 \frac{9}{2} \frac{1}{3} \frac{1}{3} \frac{1}{3} STG_{3,2}({\pi_{chase}}_1,{\pi_{chase}}_2,{\pi_{chase}}_3,\Pi_o,w_{\dot{L}},\mu)
\end{aligned}
\end{equation*}

In this case, we only need to calculate $STG_{3,2}({\pi_{chase}}_1,{\pi_{chase}}_2,{\pi_{chase}}_3,\Pi_o,w_{\dot{L}},\mu)$. We follow definition \ref{def:STG_agents} to calculate this value:

\begin{equation*}
STG_{3,2}({\pi_{chase}}_1,{\pi_{chase}}_2,{\pi_{chase}}_3,\Pi_o,w_{\dot{L}},\mu) = \sum_{\dot{l} \in \dot{L}^{N(\mu)}_{-3,2}(\Pi_o)} w_{\dot{L}}(\dot{l}) STO_{3,2}({\pi_{chase}}_1,{\pi_{chase}}_2,{\pi_{chase}}_3,\dot{l},\mu)
\end{equation*}

Again, we do not know which $\Pi_o$ we have, but we know that we will need to obtain a line-up pattern $\dot{l}$ from $\dot{L}^{N(\mu)}_{-3,2}(\Pi_o)$ to calculate $STO_{3,2}({\pi_{chase}}_1,{\pi_{chase}}_2,{\pi_{chase}}_3,\dot{l},\mu)$. We calculate this value for a figurative line-up pattern $\dot{l} = (\pi_1,*,*,\pi_2)$ from $\dot{L}^{N(\mu)}_{-3,2}(\Pi_o)$:

\begin{equation*}
STO_{3,2}({\pi_{chase}}_1,{\pi_{chase}}_2,{\pi_{chase}}_3,\dot{l},\mu) = STO_{3,2}({\pi_{chase}}_1,{\pi_{chase}}_2,{\pi_{chase}}_3,(\pi_1,*,*,\pi_2),\mu)
\end{equation*}

The following table shows us $STO_{3,2}$ for all the permutations of ${\pi_{chase}}_1,{\pi_{chase}}_2,{\pi_{chase}}_3$.

\begin{center}
\begin{tabular}{c c c | c c c | c c c}
Slot 3 & & Slot 2							& Slot 3 & & Slot 2								& Slot 3 & & Slot 2\\
\hline
${\pi_{chase}}_1$ & $<$ & ${\pi_{chase}}_2$	& ${\pi_{chase}}_1$ & $<$ & ${\pi_{chase}}_3$	& ${\pi_{chase}}_2$ & $<$ & ${\pi_{chase}}_1$\\
${\pi_{chase}}_2$ & $<$ & ${\pi_{chase}}_3$	& ${\pi_{chase}}_3$ & $<$ & ${\pi_{chase}}_2$	& ${\pi_{chase}}_1$ & $<$ & ${\pi_{chase}}_3$\\
${\pi_{chase}}_1$ & $<$ & ${\pi_{chase}}_3$	& ${\pi_{chase}}_1$ & $<$ & ${\pi_{chase}}_2$	& ${\pi_{chase}}_2$ & $<$ & ${\pi_{chase}}_3$
\end{tabular}
\begin{tabular}{c c c | c c c | c c c}
Slot 3 & & Slot 2							& Slot 3 & & Slot 2								& Slot 3 & & Slot 2\\
\hline
${\pi_{chase}}_2$ & $<$ & ${\pi_{chase}}_3$	& ${\pi_{chase}}_3$ & $<$ & ${\pi_{chase}}_1$	& ${\pi_{chase}}_3$ & $<$ & ${\pi_{chase}}_2$\\
${\pi_{chase}}_3$ & $<$ & ${\pi_{chase}}_1$	& ${\pi_{chase}}_1$ & $<$ & ${\pi_{chase}}_2$	& ${\pi_{chase}}_2$ & $<$ & ${\pi_{chase}}_1$\\
${\pi_{chase}}_2$ & $<$ & ${\pi_{chase}}_1$	& ${\pi_{chase}}_3$ & $<$ & ${\pi_{chase}}_2$	& ${\pi_{chase}}_3$ & $<$ & ${\pi_{chase}}_1$
\end{tabular}
\end{center}

But, it is not possible to find a STO, since for every permutation the agents in slots 3 and 2 share rewards (and expected average rewards as well) due they are in the same team. So no matter which agents are in $\Pi_o$ we obtain:

\begin{equation*}
STG_{3,2}({\pi_{chase}}_1,{\pi_{chase}}_2,{\pi_{chase}}_3,\Pi_o,w_{\dot{L}},\mu) = 0
\end{equation*}

Therefore:

\begin{equation*}
STG_{3,2}(\Pi_e,w_{\Pi_e},\Pi_o,w_{\dot{L}},\mu) = 6 \frac{9}{2} \frac{1}{3} \frac{1}{3} \frac{1}{3} 0 = 0
\end{equation*}

For slots 3 and 4:

\begin{equation*}
\begin{aligned}
STG_{3,4}(\Pi_e,w_{\Pi_e},\Pi_o,w_{\dot{L}},\mu)	& = \eta_{\Pi^3} \sum_{\pi_1,\pi_2,\pi_3 \in \Pi_e | \pi_1 \neq \pi_2 \neq \pi_3} w_{\Pi_e}(\pi_1) w_{\Pi_e}(\pi_2) w_{\Pi_e}(\pi_3) STG_{3,4}(\pi_1,\pi_2,\pi_3,\Pi_o,w_{\dot{L}},\mu) =\\
													& = 6 \frac{9}{2} \frac{1}{3} \frac{1}{3} \frac{1}{3} STG_{3,4}({\pi_{chase}}_1,{\pi_{chase}}_2,{\pi_{chase}}_3,\Pi_o,w_{\dot{L}},\mu)
\end{aligned}
\end{equation*}

In this case, we only need to calculate $STG_{3,4}({\pi_{chase}}_1,{\pi_{chase}}_2,{\pi_{chase}}_3,\Pi_o,w_{\dot{L}},\mu)$. We follow definition \ref{def:STG_agents} to calculate this value:

\begin{equation*}
STG_{3,4}({\pi_{chase}}_1,{\pi_{chase}}_2,{\pi_{chase}}_3,\Pi_o,w_{\dot{L}},\mu) = \sum_{\dot{l} \in \dot{L}^{N(\mu)}_{-3,4}(\Pi_o)} w_{\dot{L}}(\dot{l}) STO_{3,4}({\pi_{chase}}_1,{\pi_{chase}}_2,{\pi_{chase}}_3,\dot{l},\mu)
\end{equation*}

Again, we do not know which $\Pi_o$ we have, but we know that we will need to obtain a line-up pattern $\dot{l}$ from $\dot{L}^{N(\mu)}_{-3,4}(\Pi_o)$ to calculate $STO_{3,4}({\pi_{chase}}_1,{\pi_{chase}}_2,{\pi_{chase}}_3,\dot{l},\mu)$. We calculate this value for a figurative line-up pattern $\dot{l} = (\pi_1,\pi_2,*,*)$ from $\dot{L}^{N(\mu)}_{-3,4}(\Pi_o)$:

\begin{equation*}
STO_{3,4}({\pi_{chase}}_1,{\pi_{chase}}_2,{\pi_{chase}}_3,\dot{l},\mu) = STO_{3,4}({\pi_{chase}}_1,{\pi_{chase}}_2,{\pi_{chase}}_3,(\pi_1,\pi_2,*,*),\mu)
\end{equation*}

The following table shows us $STO_{3,4}$ for all the permutations of ${\pi_{chase}}_1,{\pi_{chase}}_2,{\pi_{chase}}_3$.

\begin{center}
\begin{tabular}{c c c | c c c | c c c}
Slot 3 & & Slot 4							& Slot 3 & & Slot 4								& Slot 3 & & Slot 4\\
\hline
${\pi_{chase}}_1$ & $<$ & ${\pi_{chase}}_2$	& ${\pi_{chase}}_1$ & $<$ & ${\pi_{chase}}_3$	& ${\pi_{chase}}_2$ & $<$ & ${\pi_{chase}}_1$\\
${\pi_{chase}}_2$ & $<$ & ${\pi_{chase}}_3$	& ${\pi_{chase}}_3$ & $<$ & ${\pi_{chase}}_2$	& ${\pi_{chase}}_1$ & $<$ & ${\pi_{chase}}_3$\\
${\pi_{chase}}_1$ & $<$ & ${\pi_{chase}}_3$	& ${\pi_{chase}}_1$ & $<$ & ${\pi_{chase}}_2$	& ${\pi_{chase}}_2$ & $<$ & ${\pi_{chase}}_3$
\end{tabular}
\begin{tabular}{c c c | c c c | c c c}
Slot 3 & & Slot 4							& Slot 3 & & Slot 4								& Slot 3 & & Slot 4\\
\hline
${\pi_{chase}}_2$ & $<$ & ${\pi_{chase}}_3$	& ${\pi_{chase}}_3$ & $<$ & ${\pi_{chase}}_1$	& ${\pi_{chase}}_3$ & $<$ & ${\pi_{chase}}_2$\\
${\pi_{chase}}_3$ & $<$ & ${\pi_{chase}}_1$	& ${\pi_{chase}}_1$ & $<$ & ${\pi_{chase}}_2$	& ${\pi_{chase}}_2$ & $<$ & ${\pi_{chase}}_1$\\
${\pi_{chase}}_2$ & $<$ & ${\pi_{chase}}_1$	& ${\pi_{chase}}_3$ & $<$ & ${\pi_{chase}}_2$	& ${\pi_{chase}}_3$ & $<$ & ${\pi_{chase}}_1$
\end{tabular}
\end{center}

But, it is not possible to find a STO, since for every permutation the agents in slots 3 and 4 share rewards (and expected average rewards as well) due they are in the same team. So no matter which agents are in $\Pi_o$ we obtain:

\begin{equation*}
STG_{3,4}({\pi_{chase}}_1,{\pi_{chase}}_2,{\pi_{chase}}_3,\Pi_o,w_{\dot{L}},\mu) = 0
\end{equation*}

Therefore:

\begin{equation*}
STG_{3,4}(\Pi_e,w_{\Pi_e},\Pi_o,w_{\dot{L}},\mu) = 6 \frac{9}{2} \frac{1}{3} \frac{1}{3} \frac{1}{3} 0 = 0
\end{equation*}

For slots 4 and 1:

\begin{equation*}
\begin{aligned}
STG_{4,1}(\Pi_e,w_{\Pi_e},\Pi_o,w_{\dot{L}},\mu)	& = \eta_{\Pi^3} \sum_{\pi_1,\pi_2,\pi_3 \in \Pi_e | \pi_1 \neq \pi_2 \neq \pi_3} w_{\Pi_e}(\pi_1) w_{\Pi_e}(\pi_2) w_{\Pi_e}(\pi_3) STG_{4,1}(\pi_1,\pi_2,\pi_3,\Pi_o,w_{\dot{L}},\mu) =\\
													& = 6 \frac{9}{2} \frac{1}{3} \frac{1}{3} \frac{1}{3} STG_{4,1}({\pi_{chase}}_1,{\pi_{chase}}_2,{\pi_{chase}}_3,\Pi_o,w_{\dot{L}},\mu)
\end{aligned}
\end{equation*}

In this case, we only need to calculate $STG_{4,1}({\pi_{chase}}_1,{\pi_{chase}}_2,{\pi_{chase}}_3,\Pi_o,w_{\dot{L}},\mu)$. We follow definition \ref{def:STG_agents} to calculate this value:

\begin{equation*}
STG_{4,1}({\pi_{chase}}_1,{\pi_{chase}}_2,{\pi_{chase}}_3,\Pi_o,w_{\dot{L}},\mu) = \sum_{\dot{l} \in \dot{L}^{N(\mu)}_{-4,1}(\Pi_o)} w_{\dot{L}}(\dot{l}) STO_{4,1}({\pi_{chase}}_1,{\pi_{chase}}_2,{\pi_{chase}}_3,\dot{l},\mu)
\end{equation*}

Again, we do not know which $\Pi_o$ we have, but we know that we will need to obtain a line-up pattern $\dot{l}$ from $\dot{L}^{N(\mu)}_{-4,1}(\Pi_o)$ to calculate $STO_{4,1}({\pi_{chase}}_1,{\pi_{chase}}_2,{\pi_{chase}}_3,\dot{l},\mu)$. We calculate this value for a figurative line-up pattern $\dot{l} = (*,\pi_1,\pi_2,*)$ from $\dot{L}^{N(\mu)}_{-4,1}(\Pi_o)$:

\begin{equation*}
STO_{4,1}({\pi_{chase}}_1,{\pi_{chase}}_2,{\pi_{chase}}_3,\dot{l},\mu) = STO_{4,1}({\pi_{chase}}_1,{\pi_{chase}}_2,{\pi_{chase}}_3,(*,\pi_1,\pi_2,*),\mu)
\end{equation*}

The following table shows us $STO_{4,1}$ for all the permutations of ${\pi_{chase}}_1,{\pi_{chase}}_2,{\pi_{chase}}_3$.

\begin{center}
\begin{tabular}{c c c | c c c | c c c}
Slot 4 & & Slot 1							& Slot 4 & & Slot 1								& Slot 4 & & Slot 1\\
\hline
${\pi_{chase}}_1$ & $<$ & ${\pi_{chase}}_2$	& ${\pi_{chase}}_1$ & $<$ & ${\pi_{chase}}_3$	& ${\pi_{chase}}_2$ & $<$ & ${\pi_{chase}}_1$\\
${\pi_{chase}}_2$ & $<$ & ${\pi_{chase}}_3$	& ${\pi_{chase}}_3$ & $<$ & ${\pi_{chase}}_2$	& ${\pi_{chase}}_1$ & $<$ & ${\pi_{chase}}_3$\\
${\pi_{chase}}_1$ & $<$ & ${\pi_{chase}}_3$	& ${\pi_{chase}}_1$ & $<$ & ${\pi_{chase}}_2$	& ${\pi_{chase}}_2$ & $<$ & ${\pi_{chase}}_3$
\end{tabular}
\begin{tabular}{c c c | c c c | c c c}
Slot 4 & & Slot 1							& Slot 4 & & Slot 1								& Slot 4 & & Slot 1\\
\hline
${\pi_{chase}}_2$ & $<$ & ${\pi_{chase}}_3$	& ${\pi_{chase}}_3$ & $<$ & ${\pi_{chase}}_1$	& ${\pi_{chase}}_3$ & $<$ & ${\pi_{chase}}_2$\\
${\pi_{chase}}_3$ & $<$ & ${\pi_{chase}}_1$	& ${\pi_{chase}}_1$ & $<$ & ${\pi_{chase}}_2$	& ${\pi_{chase}}_2$ & $<$ & ${\pi_{chase}}_1$\\
${\pi_{chase}}_2$ & $<$ & ${\pi_{chase}}_1$	& ${\pi_{chase}}_3$ & $<$ & ${\pi_{chase}}_2$	& ${\pi_{chase}}_3$ & $<$ & ${\pi_{chase}}_1$
\end{tabular}
\end{center}

Again, it is not possible to find a STO for any permutation, since we always have ${\pi_{chase}}_i < {\pi_{chase}}_j$, where a $\pi_{chase}$ agent always tries to be chased when playing as the prey and tries to chase when playing as a predator, so the agents in slots 4 and 1 will obtain an expected average reward of $1$ and $-1$ respectively. Note that the choice of $\Pi_o$ does not affect the result of $STO_{4,1}$, so no matter which agents are in $\Pi_o$ we obtain:

\begin{equation*}
STG_{4,1}({\pi_{chase}}_1,{\pi_{chase}}_2,{\pi_{chase}}_3,\Pi_o,w_{\dot{L}},\mu) = 0
\end{equation*}

Therefore:

\begin{equation*}
STG_{4,1}(\Pi_e,w_{\Pi_e},\Pi_o,w_{\dot{L}},\mu) = 6 \frac{9}{2} \frac{1}{3} \frac{1}{3} \frac{1}{3} 0 = 0
\end{equation*}

For slots 4 and 2:

\begin{equation*}
\begin{aligned}
STG_{4,2}(\Pi_e,w_{\Pi_e},\Pi_o,w_{\dot{L}},\mu)	& = \eta_{\Pi^3} \sum_{\pi_1,\pi_2,\pi_3 \in \Pi_e | \pi_1 \neq \pi_2 \neq \pi_3} w_{\Pi_e}(\pi_1) w_{\Pi_e}(\pi_2) w_{\Pi_e}(\pi_3) STG_{4,2}(\pi_1,\pi_2,\pi_3,\Pi_o,w_{\dot{L}},\mu) =\\
													& = 6 \frac{9}{2} \frac{1}{3} \frac{1}{3} \frac{1}{3} STG_{4,2}({\pi_{chase}}_1,{\pi_{chase}}_2,{\pi_{chase}}_3,\Pi_o,w_{\dot{L}},\mu)
\end{aligned}
\end{equation*}

In this case, we only need to calculate $STG_{4,2}({\pi_{chase}}_1,{\pi_{chase}}_2,{\pi_{chase}}_3,\Pi_o,w_{\dot{L}},\mu)$. We follow definition \ref{def:STG_agents} to calculate this value:

\begin{equation*}
STG_{4,2}({\pi_{chase}}_1,{\pi_{chase}}_2,{\pi_{chase}}_3,\Pi_o,w_{\dot{L}},\mu) = \sum_{\dot{l} \in \dot{L}^{N(\mu)}_{-4,2}(\Pi_o)} w_{\dot{L}}(\dot{l}) STO_{4,2}({\pi_{chase}}_1,{\pi_{chase}}_2,{\pi_{chase}}_3,\dot{l},\mu)
\end{equation*}

Again, we do not know which $\Pi_o$ we have, but we know that we will need to obtain a line-up pattern $\dot{l}$ from $\dot{L}^{N(\mu)}_{-4,2}(\Pi_o)$ to calculate $STO_{4,2}({\pi_{chase}}_1,{\pi_{chase}}_2,{\pi_{chase}}_3,\dot{l},\mu)$. We calculate this value for a figurative line-up pattern $\dot{l} = (\pi_1,*,\pi_2,*)$ from $\dot{L}^{N(\mu)}_{-4,2}(\Pi_o)$:

\begin{equation*}
STO_{4,2}({\pi_{chase}}_1,{\pi_{chase}}_2,{\pi_{chase}}_3,\dot{l},\mu) = STO_{4,2}({\pi_{chase}}_1,{\pi_{chase}}_2,{\pi_{chase}}_3,(\pi_1,*,\pi_2,*),\mu)
\end{equation*}

The following table shows us $STO_{4,2}$ for all the permutations of ${\pi_{chase}}_1,{\pi_{chase}}_2,{\pi_{chase}}_3$.

\begin{center}
\begin{tabular}{c c c | c c c | c c c}
Slot 4 & & Slot 2							& Slot 4 & & Slot 2								& Slot 4 & & Slot 2\\
\hline
${\pi_{chase}}_1$ & $<$ & ${\pi_{chase}}_2$	& ${\pi_{chase}}_1$ & $<$ & ${\pi_{chase}}_3$	& ${\pi_{chase}}_2$ & $<$ & ${\pi_{chase}}_1$\\
${\pi_{chase}}_2$ & $<$ & ${\pi_{chase}}_3$	& ${\pi_{chase}}_3$ & $<$ & ${\pi_{chase}}_2$	& ${\pi_{chase}}_1$ & $<$ & ${\pi_{chase}}_3$\\
${\pi_{chase}}_1$ & $<$ & ${\pi_{chase}}_3$	& ${\pi_{chase}}_1$ & $<$ & ${\pi_{chase}}_2$	& ${\pi_{chase}}_2$ & $<$ & ${\pi_{chase}}_3$
\end{tabular}
\begin{tabular}{c c c | c c c | c c c}
Slot 4 & & Slot 2							& Slot 4 & & Slot 2								& Slot 4 & & Slot 2\\
\hline
${\pi_{chase}}_2$ & $<$ & ${\pi_{chase}}_3$	& ${\pi_{chase}}_3$ & $<$ & ${\pi_{chase}}_1$	& ${\pi_{chase}}_3$ & $<$ & ${\pi_{chase}}_2$\\
${\pi_{chase}}_3$ & $<$ & ${\pi_{chase}}_1$	& ${\pi_{chase}}_1$ & $<$ & ${\pi_{chase}}_2$	& ${\pi_{chase}}_2$ & $<$ & ${\pi_{chase}}_1$\\
${\pi_{chase}}_2$ & $<$ & ${\pi_{chase}}_1$	& ${\pi_{chase}}_3$ & $<$ & ${\pi_{chase}}_2$	& ${\pi_{chase}}_3$ & $<$ & ${\pi_{chase}}_1$
\end{tabular}
\end{center}

But, it is not possible to find a STO, since for every permutation the agents in slots 4 and 2 share rewards (and expected average rewards as well) due they are in the same team. So no matter which agents are in $\Pi_o$ we obtain:

\begin{equation*}
STG_{4,2}({\pi_{chase}}_1,{\pi_{chase}}_2,{\pi_{chase}}_3,\Pi_o,w_{\dot{L}},\mu) = 0
\end{equation*}

Therefore:

\begin{equation*}
STG_{4,2}(\Pi_e,w_{\Pi_e},\Pi_o,w_{\dot{L}},\mu) = 6 \frac{9}{2} \frac{1}{3} \frac{1}{3} \frac{1}{3} 0 = 0
\end{equation*}

And for slots 4 and 3:

\begin{equation*}
\begin{aligned}
STG_{4,3}(\Pi_e,w_{\Pi_e},\Pi_o,w_{\dot{L}},\mu)	& = \eta_{\Pi^3} \sum_{\pi_1,\pi_2,\pi_3 \in \Pi_e | \pi_1 \neq \pi_2 \neq \pi_3} w_{\Pi_e}(\pi_1) w_{\Pi_e}(\pi_2) w_{\Pi_e}(\pi_3) STG_{4,3}(\pi_1,\pi_2,\pi_3,\Pi_o,w_{\dot{L}},\mu) =\\
													& = 6 \frac{9}{2} \frac{1}{3} \frac{1}{3} \frac{1}{3} STG_{4,3}({\pi_{chase}}_1,{\pi_{chase}}_2,{\pi_{chase}}_3,\Pi_o,w_{\dot{L}},\mu)
\end{aligned}
\end{equation*}

In this case, we only need to calculate $STG_{4,3}({\pi_{chase}}_1,{\pi_{chase}}_2,{\pi_{chase}}_3,\Pi_o,w_{\dot{L}},\mu)$. We follow definition \ref{def:STG_agents} to calculate this value:

\begin{equation*}
STG_{4,3}({\pi_{chase}}_1,{\pi_{chase}}_2,{\pi_{chase}}_3,\Pi_o,w_{\dot{L}},\mu) = \sum_{\dot{l} \in \dot{L}^{N(\mu)}_{-4,3}(\Pi_o)} w_{\dot{L}}(\dot{l}) STO_{4,3}({\pi_{chase}}_1,{\pi_{chase}}_2,{\pi_{chase}}_3,\dot{l},\mu)
\end{equation*}

Again, we do not know which $\Pi_o$ we have, but we know that we will need to obtain a line-up pattern $\dot{l}$ from $\dot{L}^{N(\mu)}_{-4,3}(\Pi_o)$ to calculate $STO_{4,3}({\pi_{chase}}_1,{\pi_{chase}}_2,{\pi_{chase}}_3,\dot{l},\mu)$. We calculate this value for a figurative line-up pattern $\dot{l} = (\pi_1,\pi_2,*,*)$ from $\dot{L}^{N(\mu)}_{-4,3}(\Pi_o)$:

\begin{equation*}
STO_{4,3}({\pi_{chase}}_1,{\pi_{chase}}_2,{\pi_{chase}}_3,\dot{l},\mu) = STO_{4,3}({\pi_{chase}}_1,{\pi_{chase}}_2,{\pi_{chase}}_3,(\pi_1,\pi_2,*,*),\mu)
\end{equation*}

The following table shows us $STO_{4,3}$ for all the permutations of ${\pi_{chase}}_1,{\pi_{chase}}_2,{\pi_{chase}}_3$.

\begin{center}
\begin{tabular}{c c c | c c c | c c c}
Slot 4 & & Slot 3							& Slot 4 & & Slot 3								& Slot 4 & & Slot 3\\
\hline
${\pi_{chase}}_1$ & $<$ & ${\pi_{chase}}_2$	& ${\pi_{chase}}_1$ & $<$ & ${\pi_{chase}}_3$	& ${\pi_{chase}}_2$ & $<$ & ${\pi_{chase}}_1$\\
${\pi_{chase}}_2$ & $<$ & ${\pi_{chase}}_3$	& ${\pi_{chase}}_3$ & $<$ & ${\pi_{chase}}_2$	& ${\pi_{chase}}_1$ & $<$ & ${\pi_{chase}}_3$\\
${\pi_{chase}}_1$ & $<$ & ${\pi_{chase}}_3$	& ${\pi_{chase}}_1$ & $<$ & ${\pi_{chase}}_2$	& ${\pi_{chase}}_2$ & $<$ & ${\pi_{chase}}_3$
\end{tabular}
\begin{tabular}{c c c | c c c | c c c}
Slot 4 & & Slot 3							& Slot 4 & & Slot 3								& Slot 4 & & Slot 3\\
\hline
${\pi_{chase}}_2$ & $<$ & ${\pi_{chase}}_3$	& ${\pi_{chase}}_3$ & $<$ & ${\pi_{chase}}_1$	& ${\pi_{chase}}_3$ & $<$ & ${\pi_{chase}}_2$\\
${\pi_{chase}}_3$ & $<$ & ${\pi_{chase}}_1$	& ${\pi_{chase}}_1$ & $<$ & ${\pi_{chase}}_2$	& ${\pi_{chase}}_2$ & $<$ & ${\pi_{chase}}_1$\\
${\pi_{chase}}_2$ & $<$ & ${\pi_{chase}}_1$	& ${\pi_{chase}}_3$ & $<$ & ${\pi_{chase}}_2$	& ${\pi_{chase}}_3$ & $<$ & ${\pi_{chase}}_1$
\end{tabular}
\end{center}

But, it is not possible to find a STO, since for every permutation the agents in slots 4 and 3 share rewards (and expected average rewards as well) due they are in the same team. So no matter which agents are in $\Pi_o$ we obtain:

\begin{equation*}
STG_{4,3}({\pi_{chase}}_1,{\pi_{chase}}_2,{\pi_{chase}}_3,\Pi_o,w_{\dot{L}},\mu) = 0
\end{equation*}

Therefore:

\begin{equation*}
STG_{4,3}(\Pi_e,w_{\Pi_e},\Pi_o,w_{\dot{L}},\mu) = 6 \frac{9}{2} \frac{1}{3} \frac{1}{3} \frac{1}{3} 0 = 0
\end{equation*}

And finally, we weight over the slots:

\begin{equation*}
\begin{aligned}
& STG(\Pi_e,w_{\Pi_e},\Pi_o,w_{\dot{L}},\mu,w_S) = \eta_{S_1^2} \sum_{i=1}^{N(\mu)} w_S(i,\mu) \times\\
& \times \left(\sum_{j=1}^{i-1} w_S(j,\mu) STG_{i,j}(\Pi_e,w_{\Pi_e},\Pi_o,w_{\dot{L}},\mu) + \sum_{j=i+1}^{N(\mu)} w_S(j,\mu) STG_{i,j}(\Pi_e,w_{\Pi_e},\Pi_o,w_{\dot{L}},\mu)\right) =\\
& \ \ \ \ \ \ \ \ \ \ \ \ \ \ \ \ \ \ \ \ \ \ \ \ \ \ \ \ \ \ \ \ \ \ \ \ \ \ = \frac{4}{3} \frac{1}{4} \frac{1}{4} \{STG_{1,2}(\Pi_e,w_{\Pi_e},\Pi_o,w_{\dot{L}},\mu) + STG_{1,3}(\Pi_e,w_{\Pi_e},\Pi_o,w_{\dot{L}},\mu) +\\
& \ \ \ \ \ \ \ \ \ \ \ \ \ \ \ \ \ \ \ \ \ \ \ \ \ \ \ \ \ \ \ \ \ \ \ \ \ \ + STG_{1,4}(\Pi_e,w_{\Pi_e},\Pi_o,w_{\dot{L}},\mu) + STG_{2,1}(\Pi_e,w_{\Pi_e},\Pi_o,w_{\dot{L}},\mu) +\\
& \ \ \ \ \ \ \ \ \ \ \ \ \ \ \ \ \ \ \ \ \ \ \ \ \ \ \ \ \ \ \ \ \ \ \ \ \ \ + STG_{2,3}(\Pi_e,w_{\Pi_e},\Pi_o,w_{\dot{L}},\mu) + STG_{2,4}(\Pi_e,w_{\Pi_e},\Pi_o,w_{\dot{L}},\mu) +\\
& \ \ \ \ \ \ \ \ \ \ \ \ \ \ \ \ \ \ \ \ \ \ \ \ \ \ \ \ \ \ \ \ \ \ \ \ \ \ + STG_{3,1}(\Pi_e,w_{\Pi_e},\Pi_o,w_{\dot{L}},\mu) + STG_{3,2}(\Pi_e,w_{\Pi_e},\Pi_o,w_{\dot{L}},\mu) +\\
& \ \ \ \ \ \ \ \ \ \ \ \ \ \ \ \ \ \ \ \ \ \ \ \ \ \ \ \ \ \ \ \ \ \ \ \ \ \ + STG_{3,4}(\Pi_e,w_{\Pi_e},\Pi_o,w_{\dot{L}},\mu) + STG_{4,1}(\Pi_e,w_{\Pi_e},\Pi_o,w_{\dot{L}},\mu) +\\
& \ \ \ \ \ \ \ \ \ \ \ \ \ \ \ \ \ \ \ \ \ \ \ \ \ \ \ \ \ \ \ \ \ \ \ \ \ \ + STG_{4,2}(\Pi_e,w_{\Pi_e},\Pi_o,w_{\dot{L}},\mu) + STG_{4,3}(\Pi_e,w_{\Pi_e},\Pi_o,w_{\dot{L}},\mu)\} =\\
& \ \ \ \ \ \ \ \ \ \ \ \ \ \ \ \ \ \ \ \ \ \ \ \ \ \ \ \ \ \ \ \ \ \ \ \ \ \ = \frac{4}{3} \frac{1}{4} \frac{1}{4} \left\{3 \times 1 + 9 \times 0\right\} = \frac{1}{4}
\end{aligned}
\end{equation*}

So, for every $\Pi_o$ we obtain the same result:

\begin{equation*}
\forall \Pi_o : STG(\Pi_e,w_{\Pi_e},\Pi_o,w_{\dot{L}},\mu,w_S) = \frac{1}{4}
\end{equation*}

Therefore, predator-prey has $Left_{max} = \frac{1}{4}$ (as a {\em lower} approximation) for this property.
\end{proof}
\end{approximation}

\begin{approximation}
\label{approx:predator-prey_STG_right_min}
$Right_{min}$ for the strict total grading (STG) property is equal to $\frac{1}{4}$ (as a {\em higher} approximation) for the predator-prey environment.

\begin{proof}
To find $Right_{min}$ (equation \ref{eq:right_min}), we need to find a pair $\left\langle\Pi_e,w_{\Pi_e}\right\rangle$ which minimises the property as much as possible while $\Pi_o$ maximises it. Using $\Pi_e = \{{\pi_{chase}}_1,{\pi_{chase}}_2,{\pi_{chase}}_3\}$ with uniform weight for $w_{\Pi_e}$ (a $\pi_{chase}$ agent always tries to be chased when playing as the prey and tries to chase when playing as a predator) we find a {\em higher} approximation of this situation no matter which $\Pi_o$ we use.

Following definition \ref{def:STG}, we obtain the STG value for this $\left\langle\Pi_e,w_{\Pi_e},\Pi_o\right\rangle$ (where $\Pi_o$ is instantiated with any permitted value). Since the environment is not symmetric, we need to calculate this property for every pair of slots. Following definition \ref{def:STG_set}, we can calculate its STG value for each pair of slots. We start with slots 1 and 2:

\begin{equation*}
\begin{aligned}
STG_{1,2}(\Pi_e,w_{\Pi_e},\Pi_o,w_{\dot{L}},\mu)	& = \eta_{\Pi^3} \sum_{\pi_1,\pi_2,\pi_3 \in \Pi_e | \pi_1 \neq \pi_2 \neq \pi_3} w_{\Pi_e}(\pi_1) w_{\Pi_e}(\pi_2) w_{\Pi_e}(\pi_3) STG_{1,2}(\pi_1,\pi_2,\pi_3,\Pi_o,w_{\dot{L}},\mu) =\\
													& = 6 \frac{9}{2} \frac{1}{3} \frac{1}{3} \frac{1}{3} STG_{1,2}({\pi_{chase}}_1,{\pi_{chase}}_2,{\pi_{chase}}_3,\Pi_o,w_{\dot{L}},\mu)
\end{aligned}
\end{equation*}

\noindent Note that we avoided to calculate all the permutations of $\pi_1,\pi_2,\pi_3$ for $STG_{i,j}(\pi_1,\pi_2,\pi_3,\Pi_o,w_{\dot{L}},\mu)$ since they provide the same result, by calculating only one permutation and multiplying the result by the number of permutations $6$.

In this case, we only need to calculate $STG_{1,2}({\pi_{chase}}_1,{\pi_{chase}}_2,{\pi_{chase}}_3,\Pi_o,w_{\dot{L}},\mu)$. We follow definition \ref{def:STG_agents} to calculate this value:

\begin{equation*}
STG_{1,2}({\pi_{chase}}_1,{\pi_{chase}}_2,{\pi_{chase}}_3,\Pi_o,w_{\dot{L}},\mu) = \sum_{\dot{l} \in \dot{L}^{N(\mu)}_{-1,2}(\Pi_o)} w_{\dot{L}}(\dot{l}) STO_{1,2}({\pi_{chase}}_1,{\pi_{chase}}_2,{\pi_{chase}}_3,\dot{l},\mu)
\end{equation*}

We do not know which $\Pi_o$ we have, but we know that we will need to obtain a line-up pattern $\dot{l}$ from $\dot{L}^{N(\mu)}_{-1,2}(\Pi_o)$ to calculate $STO_{1,2}({\pi_{chase}}_1,{\pi_{chase}}_2,{\pi_{chase}}_3,\dot{l},\mu)$. We calculate this value for a figurative line-up pattern $\dot{l} = (*,*,\pi_1,\pi_2)$ from $\dot{L}^{N(\mu)}_{-1,2}(\Pi_o)$:

\begin{equation*}
STO_{1,2}({\pi_{chase}}_1,{\pi_{chase}}_2,{\pi_{chase}}_3,\dot{l},\mu) = STO_{1,2}({\pi_{chase}}_1,{\pi_{chase}}_2,{\pi_{chase}}_3,(*,*,\pi_1,\pi_2),\mu)
\end{equation*}

The following table shows us $STO_{1,2}$ for all the permutations of ${\pi_{chase}}_1,{\pi_{chase}}_2,{\pi_{chase}}_3$.

\begin{center}
\begin{tabular}{c c c | c c c | c c c}
Slot 1 & & Slot 2							& Slot 1 & & Slot 2								& Slot 1 & & Slot 2\\
\hline
${\pi_{chase}}_1$ & $<$ & ${\pi_{chase}}_2$	& ${\pi_{chase}}_1$ & $<$ & ${\pi_{chase}}_3$	& ${\pi_{chase}}_2$ & $<$ & ${\pi_{chase}}_1$\\
${\pi_{chase}}_2$ & $<$ & ${\pi_{chase}}_3$	& ${\pi_{chase}}_3$ & $<$ & ${\pi_{chase}}_2$	& ${\pi_{chase}}_1$ & $<$ & ${\pi_{chase}}_3$\\
${\pi_{chase}}_1$ & $<$ & ${\pi_{chase}}_3$	& ${\pi_{chase}}_1$ & $<$ & ${\pi_{chase}}_2$	& ${\pi_{chase}}_2$ & $<$ & ${\pi_{chase}}_3$
\end{tabular}
\begin{tabular}{c c c | c c c | c c c}
Slot 1 & & Slot 2							& Slot 1 & & Slot 2								& Slot 1 & & Slot 2\\
\hline
${\pi_{chase}}_2$ & $<$ & ${\pi_{chase}}_3$	& ${\pi_{chase}}_3$ & $<$ & ${\pi_{chase}}_1$	& ${\pi_{chase}}_3$ & $<$ & ${\pi_{chase}}_2$\\
${\pi_{chase}}_3$ & $<$ & ${\pi_{chase}}_1$	& ${\pi_{chase}}_1$ & $<$ & ${\pi_{chase}}_2$	& ${\pi_{chase}}_2$ & $<$ & ${\pi_{chase}}_1$\\
${\pi_{chase}}_2$ & $<$ & ${\pi_{chase}}_1$	& ${\pi_{chase}}_3$ & $<$ & ${\pi_{chase}}_2$	& ${\pi_{chase}}_3$ & $<$ & ${\pi_{chase}}_1$
\end{tabular}
\end{center}

It is possible to find a STO for every permutation, since we always have ${\pi_{chase}}_i < {\pi_{chase}}_j$, where a $\pi_{chase}$ agent always tries to be chased when playing as the prey and tries to chase when playing as a predator, so the agents in slots 1 and 2 will obtain an expected average reward of $-1$ and $1$ respectively. Note that the choice of $\Pi_o$ does not affect the result of $STO_{1,2}$, so no matter which agents are in $\Pi_o$ we obtain:

\begin{equation*}
STG_{1,2}({\pi_{chase}}_1,{\pi_{chase}}_2,{\pi_{chase}}_3,\Pi_o,w_{\dot{L}},\mu) = 1
\end{equation*}

Therefore:

\begin{equation*}
STG_{1,2}(\Pi_e,w_{\Pi_e},\Pi_o,w_{\dot{L}},\mu) = 6 \frac{9}{2} \frac{1}{3} \frac{1}{3} \frac{1}{3} 1 = 1
\end{equation*}

For slots 1 and 3:

\begin{equation*}
\begin{aligned}
STG_{1,3}(\Pi_e,w_{\Pi_e},\Pi_o,w_{\dot{L}},\mu)	& = \eta_{\Pi^3} \sum_{\pi_1,\pi_2,\pi_3 \in \Pi_e | \pi_1 \neq \pi_2 \neq \pi_3} w_{\Pi_e}(\pi_1) w_{\Pi_e}(\pi_2) w_{\Pi_e}(\pi_3) STG_{1,3}(\pi_1,\pi_2,\pi_3,\Pi_o,w_{\dot{L}},\mu) =\\
													& = 6 \frac{9}{2} \frac{1}{3} \frac{1}{3} \frac{1}{3} STG_{1,3}({\pi_{chase}}_1,{\pi_{chase}}_2,{\pi_{chase}}_3,\Pi_o,w_{\dot{L}},\mu)
\end{aligned}
\end{equation*}

In this case, we only need to calculate $STG_{1,3}({\pi_{chase}}_1,{\pi_{chase}}_2,{\pi_{chase}}_3,\Pi_o,w_{\dot{L}},\mu)$. We follow definition \ref{def:STG_agents} to calculate this value:

\begin{equation*}
STG_{1,3}({\pi_{chase}}_1,{\pi_{chase}}_2,{\pi_{chase}}_3,\Pi_o,w_{\dot{L}},\mu) = \sum_{\dot{l} \in \dot{L}^{N(\mu)}_{-1,3}(\Pi_o)} w_{\dot{L}}(\dot{l}) STO_{1,3}({\pi_{chase}}_1,{\pi_{chase}}_2,{\pi_{chase}}_3,\dot{l},\mu)
\end{equation*}

Again, we do not know which $\Pi_o$ we have, but we know that we will need to obtain a line-up pattern $\dot{l}$ from $\dot{L}^{N(\mu)}_{-1,3}(\Pi_o)$ to calculate $STO_{1,3}({\pi_{chase}}_1,{\pi_{chase}}_2,{\pi_{chase}}_3,\dot{l},\mu)$. We calculate this value for a figurative line-up pattern $\dot{l} = (*,\pi_1,*,\pi_2)$ from $\dot{L}^{N(\mu)}_{-1,3}(\Pi_o)$:

\begin{equation*}
STO_{1,3}({\pi_{chase}}_1,{\pi_{chase}}_2,{\pi_{chase}}_3,\dot{l},\mu) = STO_{1,3}({\pi_{chase}}_1,{\pi_{chase}}_2,{\pi_{chase}}_3,(*,\pi_1,*,\pi_2),\mu)
\end{equation*}

The following table shows us $STO_{1,3}$ for all the permutations of ${\pi_{chase}}_1,{\pi_{chase}}_2,{\pi_{chase}}_3$.

\begin{center}
\begin{tabular}{c c c | c c c | c c c}
Slot 1 & & Slot 3							& Slot 1 & & Slot 3								& Slot 1 & & Slot 3\\
\hline
${\pi_{chase}}_1$ & $<$ & ${\pi_{chase}}_2$	& ${\pi_{chase}}_1$ & $<$ & ${\pi_{chase}}_3$	& ${\pi_{chase}}_2$ & $<$ & ${\pi_{chase}}_1$\\
${\pi_{chase}}_2$ & $<$ & ${\pi_{chase}}_3$	& ${\pi_{chase}}_3$ & $<$ & ${\pi_{chase}}_2$	& ${\pi_{chase}}_1$ & $<$ & ${\pi_{chase}}_3$\\
${\pi_{chase}}_1$ & $<$ & ${\pi_{chase}}_3$	& ${\pi_{chase}}_1$ & $<$ & ${\pi_{chase}}_2$	& ${\pi_{chase}}_2$ & $<$ & ${\pi_{chase}}_3$
\end{tabular}
\begin{tabular}{c c c | c c c | c c c}
Slot 1 & & Slot 3							& Slot 1 & & Slot 3								& Slot 1 & & Slot 3\\
\hline
${\pi_{chase}}_2$ & $<$ & ${\pi_{chase}}_3$	& ${\pi_{chase}}_3$ & $<$ & ${\pi_{chase}}_1$	& ${\pi_{chase}}_3$ & $<$ & ${\pi_{chase}}_2$\\
${\pi_{chase}}_3$ & $<$ & ${\pi_{chase}}_1$	& ${\pi_{chase}}_1$ & $<$ & ${\pi_{chase}}_2$	& ${\pi_{chase}}_2$ & $<$ & ${\pi_{chase}}_1$\\
${\pi_{chase}}_2$ & $<$ & ${\pi_{chase}}_1$	& ${\pi_{chase}}_3$ & $<$ & ${\pi_{chase}}_2$	& ${\pi_{chase}}_3$ & $<$ & ${\pi_{chase}}_1$
\end{tabular}
\end{center}

Again, it is possible to find a STO for every permutation, since we always have ${\pi_{chase}}_i < {\pi_{chase}}_j$, where a $\pi_{chase}$ agent always tries to be chased when playing as the prey and tries to chase when playing as a predator, so the agents in slots 1 and 3 will obtain an expected average reward of $-1$ and $1$ respectively. Note that the choice of $\Pi_o$ does not affect the result of $STO_{1,3}$, so no matter which agents are in $\Pi_o$ we obtain:

\begin{equation*}
STG_{1,3}({\pi_{chase}}_1,{\pi_{chase}}_2,{\pi_{chase}}_3,\Pi_o,w_{\dot{L}},\mu) = 1
\end{equation*}

Therefore:

\begin{equation*}
STG_{1,3}(\Pi_e,w_{\Pi_e},\Pi_o,w_{\dot{L}},\mu) = 6 \frac{9}{2} \frac{1}{3} \frac{1}{3} \frac{1}{3} 1 = 1
\end{equation*}

For slots 1 and 4:

\begin{equation*}
\begin{aligned}
STG_{1,4}(\Pi_e,w_{\Pi_e},\Pi_o,w_{\dot{L}},\mu)	& = \eta_{\Pi^3} \sum_{\pi_1,\pi_2,\pi_3 \in \Pi_e | \pi_1 \neq \pi_2 \neq \pi_3} w_{\Pi_e}(\pi_1) w_{\Pi_e}(\pi_2) w_{\Pi_e}(\pi_3) STG_{1,4}(\pi_1,\pi_2,\pi_3,\Pi_o,w_{\dot{L}},\mu) =\\
													& = 6 \frac{9}{2} \frac{1}{3} \frac{1}{3} \frac{1}{3} STG_{1,4}({\pi_{chase}}_1,{\pi_{chase}}_2,{\pi_{chase}}_3,\Pi_o,w_{\dot{L}},\mu)
\end{aligned}
\end{equation*}

In this case, we only need to calculate $STG_{1,4}({\pi_{chase}}_1,{\pi_{chase}}_2,{\pi_{chase}}_3,\Pi_o,w_{\dot{L}},\mu)$. We follow definition \ref{def:STG_agents} to calculate this value:

\begin{equation*}
STG_{1,4}({\pi_{chase}}_1,{\pi_{chase}}_2,{\pi_{chase}}_3,\Pi_o,w_{\dot{L}},\mu) = \sum_{\dot{l} \in \dot{L}^{N(\mu)}_{-1,4}(\Pi_o)} w_{\dot{L}}(\dot{l}) STO_{1,4}({\pi_{chase}}_1,{\pi_{chase}}_2,{\pi_{chase}}_3,\dot{l},\mu)
\end{equation*}

Again, we do not know which $\Pi_o$ we have, but we know that we will need to obtain a line-up pattern $\dot{l}$ from $\dot{L}^{N(\mu)}_{-1,4}(\Pi_o)$ to calculate $STO_{1,4}({\pi_{chase}}_1,{\pi_{chase}}_2,{\pi_{chase}}_3,\dot{l},\mu)$. We calculate this value for a figurative line-up pattern $\dot{l} = (*,\pi_1,\pi_2,*)$ from $\dot{L}^{N(\mu)}_{-1,4}(\Pi_o)$:

\begin{equation*}
STO_{1,4}({\pi_{chase}}_1,{\pi_{chase}}_2,{\pi_{chase}}_3,\dot{l},\mu) = STO_{1,4}({\pi_{chase}}_1,{\pi_{chase}}_2,{\pi_{chase}}_3,(*,\pi_1,\pi_2,*),\mu)
\end{equation*}

The following table shows us $STO_{1,4}$ for all the permutations of ${\pi_{chase}}_1,{\pi_{chase}}_2,{\pi_{chase}}_3$.

\begin{center}
\begin{tabular}{c c c | c c c | c c c}
Slot 1 & & Slot 4							& Slot 1 & & Slot 4								& Slot 1 & & Slot 4\\
\hline
${\pi_{chase}}_1$ & $<$ & ${\pi_{chase}}_2$	& ${\pi_{chase}}_1$ & $<$ & ${\pi_{chase}}_3$	& ${\pi_{chase}}_2$ & $<$ & ${\pi_{chase}}_1$\\
${\pi_{chase}}_2$ & $<$ & ${\pi_{chase}}_3$	& ${\pi_{chase}}_3$ & $<$ & ${\pi_{chase}}_2$	& ${\pi_{chase}}_1$ & $<$ & ${\pi_{chase}}_3$\\
${\pi_{chase}}_1$ & $<$ & ${\pi_{chase}}_3$	& ${\pi_{chase}}_1$ & $<$ & ${\pi_{chase}}_2$	& ${\pi_{chase}}_2$ & $<$ & ${\pi_{chase}}_3$
\end{tabular}
\begin{tabular}{c c c | c c c | c c c}
Slot 1 & & Slot 4							& Slot 1 & & Slot 4								& Slot 1 & & Slot 4\\
\hline
${\pi_{chase}}_2$ & $<$ & ${\pi_{chase}}_3$	& ${\pi_{chase}}_3$ & $<$ & ${\pi_{chase}}_1$	& ${\pi_{chase}}_3$ & $<$ & ${\pi_{chase}}_2$\\
${\pi_{chase}}_3$ & $<$ & ${\pi_{chase}}_1$	& ${\pi_{chase}}_1$ & $<$ & ${\pi_{chase}}_2$	& ${\pi_{chase}}_2$ & $<$ & ${\pi_{chase}}_1$\\
${\pi_{chase}}_2$ & $<$ & ${\pi_{chase}}_1$	& ${\pi_{chase}}_3$ & $<$ & ${\pi_{chase}}_2$	& ${\pi_{chase}}_3$ & $<$ & ${\pi_{chase}}_1$
\end{tabular}
\end{center}

Again, it is possible to find a STO for every permutation, since we always have ${\pi_{chase}}_i < {\pi_{chase}}_j$, where a $\pi_{chase}$ agent always tries to be chased when playing as the prey and tries to chase when playing as a predator, so the agents in slots 1 and 4 will obtain an expected average reward of $-1$ and $1$ respectively. Note that the choice of $\Pi_o$ does not affect the result of $STO_{1,4}$, so no matter which agents are in $\Pi_o$ we obtain:

\begin{equation*}
STG_{1,4}({\pi_{chase}}_1,{\pi_{chase}}_2,{\pi_{chase}}_3,\Pi_o,w_{\dot{L}},\mu) = 1
\end{equation*}

Therefore:

\begin{equation*}
STG_{1,4}(\Pi_e,w_{\Pi_e},\Pi_o,w_{\dot{L}},\mu) = 6 \frac{9}{2} \frac{1}{3} \frac{1}{3} \frac{1}{3} 1 = 1
\end{equation*}

For slots 2 and 1:

\begin{equation*}
\begin{aligned}
STG_{2,1}(\Pi_e,w_{\Pi_e},\Pi_o,w_{\dot{L}},\mu)	& = \eta_{\Pi^3} \sum_{\pi_1,\pi_2,\pi_3 \in \Pi_e | \pi_1 \neq \pi_2 \neq \pi_3} w_{\Pi_e}(\pi_1) w_{\Pi_e}(\pi_2) w_{\Pi_e}(\pi_3) STG_{2,1}(\pi_1,\pi_2,\pi_3,\Pi_o,w_{\dot{L}},\mu) =\\
													& = 6 \frac{9}{2} \frac{1}{3} \frac{1}{3} \frac{1}{3} STG_{2,1}({\pi_{chase}}_1,{\pi_{chase}}_2,{\pi_{chase}}_3,\Pi_o,w_{\dot{L}},\mu)
\end{aligned}
\end{equation*}

In this case, we only need to calculate $STG_{2,1}({\pi_{chase}}_1,{\pi_{chase}}_2,{\pi_{chase}}_3,\Pi_o,w_{\dot{L}},\mu)$. We follow definition \ref{def:STG_agents} to calculate this value:

\begin{equation*}
STG_{2,1}({\pi_{chase}}_1,{\pi_{chase}}_2,{\pi_{chase}}_3,\Pi_o,w_{\dot{L}},\mu) = \sum_{\dot{l} \in \dot{L}^{N(\mu)}_{-2,1}(\Pi_o)} w_{\dot{L}}(\dot{l}) STO_{2,1}({\pi_{chase}}_1,{\pi_{chase}}_2,{\pi_{chase}}_3,\dot{l},\mu)
\end{equation*}

Again, we do not know which $\Pi_o$ we have, but we know that we will need to obtain a line-up pattern $\dot{l}$ from $\dot{L}^{N(\mu)}_{-2,1}(\Pi_o)$ to calculate $STO_{2,1}({\pi_{chase}}_1,{\pi_{chase}}_2,{\pi_{chase}}_3,\dot{l},\mu)$. We calculate this value for a figurative line-up pattern $\dot{l} = (*,*,\pi_1,\pi_2)$ from $\dot{L}^{N(\mu)}_{-2,1}(\Pi_o)$:

\begin{equation*}
STO_{2,1}({\pi_{chase}}_1,{\pi_{chase}}_2,{\pi_{chase}}_3,\dot{l},\mu) = STO_{2,1}({\pi_{chase}}_1,{\pi_{chase}}_2,{\pi_{chase}}_3,(*,*,\pi_1,\pi_2),\mu)
\end{equation*}

The following table shows us $STO_{2,1}$ for all the permutations of ${\pi_{chase}}_1,{\pi_{chase}}_2,{\pi_{chase}}_3$.

\begin{center}
\begin{tabular}{c c c | c c c | c c c}
Slot 2 & & Slot 1							& Slot 2 & & Slot 1								& Slot 2 & & Slot 1\\
\hline
${\pi_{chase}}_1$ & $<$ & ${\pi_{chase}}_2$	& ${\pi_{chase}}_1$ & $<$ & ${\pi_{chase}}_3$	& ${\pi_{chase}}_2$ & $<$ & ${\pi_{chase}}_1$\\
${\pi_{chase}}_2$ & $<$ & ${\pi_{chase}}_3$	& ${\pi_{chase}}_3$ & $<$ & ${\pi_{chase}}_2$	& ${\pi_{chase}}_1$ & $<$ & ${\pi_{chase}}_3$\\
${\pi_{chase}}_1$ & $<$ & ${\pi_{chase}}_3$	& ${\pi_{chase}}_1$ & $<$ & ${\pi_{chase}}_2$	& ${\pi_{chase}}_2$ & $<$ & ${\pi_{chase}}_3$
\end{tabular}
\begin{tabular}{c c c | c c c | c c c}
Slot 2 & & Slot 1							& Slot 2 & & Slot 1								& Slot 2 & & Slot 1\\
\hline
${\pi_{chase}}_2$ & $<$ & ${\pi_{chase}}_3$	& ${\pi_{chase}}_3$ & $<$ & ${\pi_{chase}}_1$	& ${\pi_{chase}}_3$ & $<$ & ${\pi_{chase}}_2$\\
${\pi_{chase}}_3$ & $<$ & ${\pi_{chase}}_1$	& ${\pi_{chase}}_1$ & $<$ & ${\pi_{chase}}_2$	& ${\pi_{chase}}_2$ & $<$ & ${\pi_{chase}}_1$\\
${\pi_{chase}}_2$ & $<$ & ${\pi_{chase}}_1$	& ${\pi_{chase}}_3$ & $<$ & ${\pi_{chase}}_2$	& ${\pi_{chase}}_3$ & $<$ & ${\pi_{chase}}_1$
\end{tabular}
\end{center}

It is not possible to find a STO for any permutation, since we always have ${\pi_{chase}}_i < {\pi_{chase}}_j$, where a $\pi_{chase}$ agent always tries to be chased when playing as the prey and tries to chase when playing as a predator, so the agents in slots 2 and 1 will obtain an expected average reward of $1$ and $-1$ respectively. Note that the choice of $\Pi_o$ does not affect the result of $STO_{2,1}$, so no matter which agents are in $\Pi_o$ we obtain:

\begin{equation*}
STG_{2,1}({\pi_{chase}}_1,{\pi_{chase}}_2,{\pi_{chase}}_3,\Pi_o,w_{\dot{L}},\mu) = 0
\end{equation*}

Therefore:

\begin{equation*}
STG_{2,1}(\Pi_e,w_{\Pi_e},\Pi_o,w_{\dot{L}},\mu) = 6 \frac{9}{2} \frac{1}{3} \frac{1}{3} \frac{1}{3} 0 = 0
\end{equation*}

For slots 2 and 3:

\begin{equation*}
\begin{aligned}
STG_{2,3}(\Pi_e,w_{\Pi_e},\Pi_o,w_{\dot{L}},\mu)	& = \eta_{\Pi^3} \sum_{\pi_1,\pi_2,\pi_3 \in \Pi_e | \pi_1 \neq \pi_2 \neq \pi_3} w_{\Pi_e}(\pi_1) w_{\Pi_e}(\pi_2) w_{\Pi_e}(\pi_3) STG_{2,3}(\pi_1,\pi_2,\pi_3,\Pi_o,w_{\dot{L}},\mu) =\\
													& = 6 \frac{9}{2} \frac{1}{3} \frac{1}{3} \frac{1}{3} STG_{2,3}({\pi_{chase}}_1,{\pi_{chase}}_2,{\pi_{chase}}_3,\Pi_o,w_{\dot{L}},\mu)
\end{aligned}
\end{equation*}

In this case, we only need to calculate $STG_{2,3}({\pi_{chase}}_1,{\pi_{chase}}_2,{\pi_{chase}}_3,\Pi_o,w_{\dot{L}},\mu)$. We follow definition \ref{def:STG_agents} to calculate this value:

\begin{equation*}
STG_{2,3}({\pi_{chase}}_1,{\pi_{chase}}_2,{\pi_{chase}}_3,\Pi_o,w_{\dot{L}},\mu) = \sum_{\dot{l} \in \dot{L}^{N(\mu)}_{-2,3}(\Pi_o)} w_{\dot{L}}(\dot{l}) STO_{2,3}({\pi_{chase}}_1,{\pi_{chase}}_2,{\pi_{chase}}_3,\dot{l},\mu)
\end{equation*}

Again, we do not know which $\Pi_o$ we have, but we know that we will need to obtain a line-up pattern $\dot{l}$ from $\dot{L}^{N(\mu)}_{-2,3}(\Pi_o)$ to calculate $STO_{2,3}({\pi_{chase}}_1,{\pi_{chase}}_2,{\pi_{chase}}_3,\dot{l},\mu)$. We calculate this value for a figurative line-up pattern $\dot{l} = (\pi_1,*,*,\pi_2)$ from $\dot{L}^{N(\mu)}_{-2,3}(\Pi_o)$:

\begin{equation*}
STO_{2,3}({\pi_{chase}}_1,{\pi_{chase}}_2,{\pi_{chase}}_3,\dot{l},\mu) = STO_{2,3}({\pi_{chase}}_1,{\pi_{chase}}_2,{\pi_{chase}}_3,(\pi_1,*,*,\pi_2),\mu)
\end{equation*}

The following table shows us $STO_{2,3}$ for all the permutations of ${\pi_{chase}}_1,{\pi_{chase}}_2,{\pi_{chase}}_3$.

\begin{center}
\begin{tabular}{c c c | c c c | c c c}
Slot 2 & & Slot 3							& Slot 2 & & Slot 3								& Slot 2 & & Slot 3\\
\hline
${\pi_{chase}}_1$ & $<$ & ${\pi_{chase}}_2$	& ${\pi_{chase}}_1$ & $<$ & ${\pi_{chase}}_3$	& ${\pi_{chase}}_2$ & $<$ & ${\pi_{chase}}_1$\\
${\pi_{chase}}_2$ & $<$ & ${\pi_{chase}}_3$	& ${\pi_{chase}}_3$ & $<$ & ${\pi_{chase}}_2$	& ${\pi_{chase}}_1$ & $<$ & ${\pi_{chase}}_3$\\
${\pi_{chase}}_1$ & $<$ & ${\pi_{chase}}_3$	& ${\pi_{chase}}_1$ & $<$ & ${\pi_{chase}}_2$	& ${\pi_{chase}}_2$ & $<$ & ${\pi_{chase}}_3$
\end{tabular}
\begin{tabular}{c c c | c c c | c c c}
Slot 2 & & Slot 3							& Slot 2 & & Slot 3								& Slot 2 & & Slot 3\\
\hline
${\pi_{chase}}_2$ & $<$ & ${\pi_{chase}}_3$	& ${\pi_{chase}}_3$ & $<$ & ${\pi_{chase}}_1$	& ${\pi_{chase}}_3$ & $<$ & ${\pi_{chase}}_2$\\
${\pi_{chase}}_3$ & $<$ & ${\pi_{chase}}_1$	& ${\pi_{chase}}_1$ & $<$ & ${\pi_{chase}}_2$	& ${\pi_{chase}}_2$ & $<$ & ${\pi_{chase}}_1$\\
${\pi_{chase}}_2$ & $<$ & ${\pi_{chase}}_1$	& ${\pi_{chase}}_3$ & $<$ & ${\pi_{chase}}_2$	& ${\pi_{chase}}_3$ & $<$ & ${\pi_{chase}}_1$
\end{tabular}
\end{center}

But, it is not possible to find a STO, since for every permutation the agents in slots 2 and 3 share rewards (and expected average rewards as well) due they are in the same team. So no matter which agents are in $\Pi_o$ we obtain:

\begin{equation*}
STG_{2,3}({\pi_{chase}}_1,{\pi_{chase}}_2,{\pi_{chase}}_3,\Pi_o,w_{\dot{L}},\mu) = 0
\end{equation*}

Therefore:

\begin{equation*}
STG_{2,3}(\Pi_e,w_{\Pi_e},\Pi_o,w_{\dot{L}},\mu) = 6 \frac{9}{2} \frac{1}{3} \frac{1}{3} \frac{1}{3} 0 = 0
\end{equation*}

For slots 2 and 4:

\begin{equation*}
\begin{aligned}
STG_{2,4}(\Pi_e,w_{\Pi_e},\Pi_o,w_{\dot{L}},\mu)	& = \eta_{\Pi^3} \sum_{\pi_1,\pi_2,\pi_3 \in \Pi_e | \pi_1 \neq \pi_2 \neq \pi_3} w_{\Pi_e}(\pi_1) w_{\Pi_e}(\pi_2) w_{\Pi_e}(\pi_3) STG_{2,4}(\pi_1,\pi_2,\pi_3,\Pi_o,w_{\dot{L}},\mu) =\\
													& = 6 \frac{9}{2} \frac{1}{3} \frac{1}{3} \frac{1}{3} STG_{2,4}({\pi_{chase}}_1,{\pi_{chase}}_2,{\pi_{chase}}_3,\Pi_o,w_{\dot{L}},\mu)
\end{aligned}
\end{equation*}

In this case, we only need to calculate $STG_{2,4}({\pi_{chase}}_1,{\pi_{chase}}_2,{\pi_{chase}}_3,\Pi_o,w_{\dot{L}},\mu)$. We follow definition \ref{def:STG_agents} to calculate this value:

\begin{equation*}
STG_{2,4}({\pi_{chase}}_1,{\pi_{chase}}_2,{\pi_{chase}}_3,\Pi_o,w_{\dot{L}},\mu) = \sum_{\dot{l} \in \dot{L}^{N(\mu)}_{-2,4}(\Pi_o)} w_{\dot{L}}(\dot{l}) STO_{2,4}({\pi_{chase}}_1,{\pi_{chase}}_2,{\pi_{chase}}_3,\dot{l},\mu)
\end{equation*}

Again, we do not know which $\Pi_o$ we have, but we know that we will need to obtain a line-up pattern $\dot{l}$ from $\dot{L}^{N(\mu)}_{-2,4}(\Pi_o)$ to calculate $STO_{2,4}({\pi_{chase}}_1,{\pi_{chase}}_2,{\pi_{chase}}_3,\dot{l},\mu)$. We calculate this value for a figurative line-up pattern $\dot{l} = (\pi_1,*,\pi_2,*)$ from $\dot{L}^{N(\mu)}_{-2,4}(\Pi_o)$:

\begin{equation*}
STO_{2,4}({\pi_{chase}}_1,{\pi_{chase}}_2,{\pi_{chase}}_3,\dot{l},\mu) = STO_{2,4}({\pi_{chase}}_1,{\pi_{chase}}_2,{\pi_{chase}}_3,(\pi_1,*,\pi_2,*),\mu)
\end{equation*}

The following table shows us $STO_{2,4}$ for all the permutations of ${\pi_{chase}}_1,{\pi_{chase}}_2,{\pi_{chase}}_3$.

\begin{center}
\begin{tabular}{c c c | c c c | c c c}
Slot 2 & & Slot 4							& Slot 2 & & Slot 4								& Slot 2 & & Slot 4\\
\hline
${\pi_{chase}}_1$ & $<$ & ${\pi_{chase}}_2$	& ${\pi_{chase}}_1$ & $<$ & ${\pi_{chase}}_3$	& ${\pi_{chase}}_2$ & $<$ & ${\pi_{chase}}_1$\\
${\pi_{chase}}_2$ & $<$ & ${\pi_{chase}}_3$	& ${\pi_{chase}}_3$ & $<$ & ${\pi_{chase}}_2$	& ${\pi_{chase}}_1$ & $<$ & ${\pi_{chase}}_3$\\
${\pi_{chase}}_1$ & $<$ & ${\pi_{chase}}_3$	& ${\pi_{chase}}_1$ & $<$ & ${\pi_{chase}}_2$	& ${\pi_{chase}}_2$ & $<$ & ${\pi_{chase}}_3$
\end{tabular}
\begin{tabular}{c c c | c c c | c c c}
Slot 2 & & Slot 4							& Slot 2 & & Slot 4								& Slot 2 & & Slot 4\\
\hline
${\pi_{chase}}_2$ & $<$ & ${\pi_{chase}}_3$	& ${\pi_{chase}}_3$ & $<$ & ${\pi_{chase}}_1$	& ${\pi_{chase}}_3$ & $<$ & ${\pi_{chase}}_2$\\
${\pi_{chase}}_3$ & $<$ & ${\pi_{chase}}_1$	& ${\pi_{chase}}_1$ & $<$ & ${\pi_{chase}}_2$	& ${\pi_{chase}}_2$ & $<$ & ${\pi_{chase}}_1$\\
${\pi_{chase}}_2$ & $<$ & ${\pi_{chase}}_1$	& ${\pi_{chase}}_3$ & $<$ & ${\pi_{chase}}_2$	& ${\pi_{chase}}_3$ & $<$ & ${\pi_{chase}}_1$
\end{tabular}
\end{center}

But, it is not possible to find a STO, since for every permutation the agents in slots 2 and 4 share rewards (and expected average rewards as well) due they are in the same team. So no matter which agents are in $\Pi_o$ we obtain:

\begin{equation*}
STG_{2,4}({\pi_{chase}}_1,{\pi_{chase}}_2,{\pi_{chase}}_3,\Pi_o,w_{\dot{L}},\mu) = 0
\end{equation*}

Therefore:

\begin{equation*}
STG_{2,4}(\Pi_e,w_{\Pi_e},\Pi_o,w_{\dot{L}},\mu) = 6 \frac{9}{2} \frac{1}{3} \frac{1}{3} \frac{1}{3} 0 = 0
\end{equation*}

For slots 3 and 1:

\begin{equation*}
\begin{aligned}
STG_{3,1}(\Pi_e,w_{\Pi_e},\Pi_o,w_{\dot{L}},\mu)	& = \eta_{\Pi^3} \sum_{\pi_1,\pi_2,\pi_3 \in \Pi_e | \pi_1 \neq \pi_2 \neq \pi_3} w_{\Pi_e}(\pi_1) w_{\Pi_e}(\pi_2) w_{\Pi_e}(\pi_3) STG_{3,1}(\pi_1,\pi_2,\pi_3,\Pi_o,w_{\dot{L}},\mu) =\\
													& = 6 \frac{9}{2} \frac{1}{3} \frac{1}{3} \frac{1}{3} STG_{3,1}({\pi_{chase}}_1,{\pi_{chase}}_2,{\pi_{chase}}_3,\Pi_o,w_{\dot{L}},\mu)
\end{aligned}
\end{equation*}

In this case, we only need to calculate $STG_{3,1}({\pi_{chase}}_1,{\pi_{chase}}_2,{\pi_{chase}}_3,\Pi_o,w_{\dot{L}},\mu)$. We follow definition \ref{def:STG_agents} to calculate this value:

\begin{equation*}
STG_{3,1}({\pi_{chase}}_1,{\pi_{chase}}_2,{\pi_{chase}}_3,\Pi_o,w_{\dot{L}},\mu) = \sum_{\dot{l} \in \dot{L}^{N(\mu)}_{-3,1}(\Pi_o)} w_{\dot{L}}(\dot{l}) STO_{3,1}({\pi_{chase}}_1,{\pi_{chase}}_2,{\pi_{chase}}_3,\dot{l},\mu)
\end{equation*}

Again, we do not know which $\Pi_o$ we have, but we know that we will need to obtain a line-up pattern $\dot{l}$ from $\dot{L}^{N(\mu)}_{-3,1}(\Pi_o)$ to calculate $STO_{3,1}({\pi_{chase}}_1,{\pi_{chase}}_2,{\pi_{chase}}_3,\dot{l},\mu)$. We calculate this value for a figurative line-up pattern $\dot{l} = (*,\pi_1,*,\pi_2)$ from $\dot{L}^{N(\mu)}_{-3,1}(\Pi_o)$:

\begin{equation*}
STO_{3,1}({\pi_{chase}}_1,{\pi_{chase}}_2,{\pi_{chase}}_3,\dot{l},\mu) = STO_{3,1}({\pi_{chase}}_1,{\pi_{chase}}_2,{\pi_{chase}}_3,(*,\pi_1,*,\pi_2),\mu)
\end{equation*}

The following table shows us $STO_{3,1}$ for all the permutations of ${\pi_{chase}}_1,{\pi_{chase}}_2,{\pi_{chase}}_3$.

\begin{center}
\begin{tabular}{c c c | c c c | c c c}
Slot 3 & & Slot 1							& Slot 3 & & Slot 1								& Slot 3 & & Slot 1\\
\hline
${\pi_{chase}}_1$ & $<$ & ${\pi_{chase}}_2$	& ${\pi_{chase}}_1$ & $<$ & ${\pi_{chase}}_3$	& ${\pi_{chase}}_2$ & $<$ & ${\pi_{chase}}_1$\\
${\pi_{chase}}_2$ & $<$ & ${\pi_{chase}}_3$	& ${\pi_{chase}}_3$ & $<$ & ${\pi_{chase}}_2$	& ${\pi_{chase}}_1$ & $<$ & ${\pi_{chase}}_3$\\
${\pi_{chase}}_1$ & $<$ & ${\pi_{chase}}_3$	& ${\pi_{chase}}_1$ & $<$ & ${\pi_{chase}}_2$	& ${\pi_{chase}}_2$ & $<$ & ${\pi_{chase}}_3$
\end{tabular}
\begin{tabular}{c c c | c c c | c c c}
Slot 3 & & Slot 1							& Slot 3 & & Slot 1								& Slot 3 & & Slot 1\\
\hline
${\pi_{chase}}_2$ & $<$ & ${\pi_{chase}}_3$	& ${\pi_{chase}}_3$ & $<$ & ${\pi_{chase}}_1$	& ${\pi_{chase}}_3$ & $<$ & ${\pi_{chase}}_2$\\
${\pi_{chase}}_3$ & $<$ & ${\pi_{chase}}_1$	& ${\pi_{chase}}_1$ & $<$ & ${\pi_{chase}}_2$	& ${\pi_{chase}}_2$ & $<$ & ${\pi_{chase}}_1$\\
${\pi_{chase}}_2$ & $<$ & ${\pi_{chase}}_1$	& ${\pi_{chase}}_3$ & $<$ & ${\pi_{chase}}_2$	& ${\pi_{chase}}_3$ & $<$ & ${\pi_{chase}}_1$
\end{tabular}
\end{center}

Again, it is not possible to find a STO for any permutation, since we always have ${\pi_{chase}}_i < {\pi_{chase}}_j$, where a $\pi_{chase}$ agent always tries to be chased when playing as the prey and tries to chase when playing as a predator, so the agents in slots 3 and 1 will obtain an expected average reward of $1$ and $-1$ respectively. Note that the choice of $\Pi_o$ does not affect the result of $STO_{3,1}$, so no matter which agents are in $\Pi_o$ we obtain:

\begin{equation*}
STG_{3,1}({\pi_{chase}}_1,{\pi_{chase}}_2,{\pi_{chase}}_3,\Pi_o,w_{\dot{L}},\mu) = 0
\end{equation*}

Therefore:

\begin{equation*}
STG_{3,1}(\Pi_e,w_{\Pi_e},\Pi_o,w_{\dot{L}},\mu) = 6 \frac{9}{2} \frac{1}{3} \frac{1}{3} \frac{1}{3} 0 = 0
\end{equation*}

For slots 3 and 2:

\begin{equation*}
\begin{aligned}
STG_{3,2}(\Pi_e,w_{\Pi_e},\Pi_o,w_{\dot{L}},\mu)	& = \eta_{\Pi^3} \sum_{\pi_1,\pi_2,\pi_3 \in \Pi_e | \pi_1 \neq \pi_2 \neq \pi_3} w_{\Pi_e}(\pi_1) w_{\Pi_e}(\pi_2) w_{\Pi_e}(\pi_3) STG_{3,2}(\pi_1,\pi_2,\pi_3,\Pi_o,w_{\dot{L}},\mu) =\\
													& = 6 \frac{9}{2} \frac{1}{3} \frac{1}{3} \frac{1}{3} STG_{3,2}({\pi_{chase}}_1,{\pi_{chase}}_2,{\pi_{chase}}_3,\Pi_o,w_{\dot{L}},\mu)
\end{aligned}
\end{equation*}

In this case, we only need to calculate $STG_{3,2}({\pi_{chase}}_1,{\pi_{chase}}_2,{\pi_{chase}}_3,\Pi_o,w_{\dot{L}},\mu)$. We follow definition \ref{def:STG_agents} to calculate this value:

\begin{equation*}
STG_{3,2}({\pi_{chase}}_1,{\pi_{chase}}_2,{\pi_{chase}}_3,\Pi_o,w_{\dot{L}},\mu) = \sum_{\dot{l} \in \dot{L}^{N(\mu)}_{-3,2}(\Pi_o)} w_{\dot{L}}(\dot{l}) STO_{3,2}({\pi_{chase}}_1,{\pi_{chase}}_2,{\pi_{chase}}_3,\dot{l},\mu)
\end{equation*}

Again, we do not know which $\Pi_o$ we have, but we know that we will need to obtain a line-up pattern $\dot{l}$ from $\dot{L}^{N(\mu)}_{-3,2}(\Pi_o)$ to calculate $STO_{3,2}({\pi_{chase}}_1,{\pi_{chase}}_2,{\pi_{chase}}_3,\dot{l},\mu)$. We calculate this value for a figurative line-up pattern $\dot{l} = (\pi_1,*,*,\pi_2)$ from $\dot{L}^{N(\mu)}_{-3,2}(\Pi_o)$:

\begin{equation*}
STO_{3,2}({\pi_{chase}}_1,{\pi_{chase}}_2,{\pi_{chase}}_3,\dot{l},\mu) = STO_{3,2}({\pi_{chase}}_1,{\pi_{chase}}_2,{\pi_{chase}}_3,(\pi_1,*,*,\pi_2),\mu)
\end{equation*}

The following table shows us $STO_{3,2}$ for all the permutations of ${\pi_{chase}}_1,{\pi_{chase}}_2,{\pi_{chase}}_3$.

\begin{center}
\begin{tabular}{c c c | c c c | c c c}
Slot 3 & & Slot 2							& Slot 3 & & Slot 2								& Slot 3 & & Slot 2\\
\hline
${\pi_{chase}}_1$ & $<$ & ${\pi_{chase}}_2$	& ${\pi_{chase}}_1$ & $<$ & ${\pi_{chase}}_3$	& ${\pi_{chase}}_2$ & $<$ & ${\pi_{chase}}_1$\\
${\pi_{chase}}_2$ & $<$ & ${\pi_{chase}}_3$	& ${\pi_{chase}}_3$ & $<$ & ${\pi_{chase}}_2$	& ${\pi_{chase}}_1$ & $<$ & ${\pi_{chase}}_3$\\
${\pi_{chase}}_1$ & $<$ & ${\pi_{chase}}_3$	& ${\pi_{chase}}_1$ & $<$ & ${\pi_{chase}}_2$	& ${\pi_{chase}}_2$ & $<$ & ${\pi_{chase}}_3$
\end{tabular}
\begin{tabular}{c c c | c c c | c c c}
Slot 3 & & Slot 2							& Slot 3 & & Slot 2								& Slot 3 & & Slot 2\\
\hline
${\pi_{chase}}_2$ & $<$ & ${\pi_{chase}}_3$	& ${\pi_{chase}}_3$ & $<$ & ${\pi_{chase}}_1$	& ${\pi_{chase}}_3$ & $<$ & ${\pi_{chase}}_2$\\
${\pi_{chase}}_3$ & $<$ & ${\pi_{chase}}_1$	& ${\pi_{chase}}_1$ & $<$ & ${\pi_{chase}}_2$	& ${\pi_{chase}}_2$ & $<$ & ${\pi_{chase}}_1$\\
${\pi_{chase}}_2$ & $<$ & ${\pi_{chase}}_1$	& ${\pi_{chase}}_3$ & $<$ & ${\pi_{chase}}_2$	& ${\pi_{chase}}_3$ & $<$ & ${\pi_{chase}}_1$
\end{tabular}
\end{center}

But, it is not possible to find a STO, since for every permutation the agents in slots 3 and 2 share rewards (and expected average rewards as well) due they are in the same team. So no matter which agents are in $\Pi_o$ we obtain:

\begin{equation*}
STG_{3,2}({\pi_{chase}}_1,{\pi_{chase}}_2,{\pi_{chase}}_3,\Pi_o,w_{\dot{L}},\mu) = 0
\end{equation*}

Therefore:

\begin{equation*}
STG_{3,2}(\Pi_e,w_{\Pi_e},\Pi_o,w_{\dot{L}},\mu) = 6 \frac{9}{2} \frac{1}{3} \frac{1}{3} \frac{1}{3} 0 = 0
\end{equation*}

For slots 3 and 4:

\begin{equation*}
\begin{aligned}
STG_{3,4}(\Pi_e,w_{\Pi_e},\Pi_o,w_{\dot{L}},\mu)	& = \eta_{\Pi^3} \sum_{\pi_1,\pi_2,\pi_3 \in \Pi_e | \pi_1 \neq \pi_2 \neq \pi_3} w_{\Pi_e}(\pi_1) w_{\Pi_e}(\pi_2) w_{\Pi_e}(\pi_3) STG_{3,4}(\pi_1,\pi_2,\pi_3,\Pi_o,w_{\dot{L}},\mu) =\\
													& = 6 \frac{9}{2} \frac{1}{3} \frac{1}{3} \frac{1}{3} STG_{3,4}({\pi_{chase}}_1,{\pi_{chase}}_2,{\pi_{chase}}_3,\Pi_o,w_{\dot{L}},\mu)
\end{aligned}
\end{equation*}

In this case, we only need to calculate $STG_{3,4}({\pi_{chase}}_1,{\pi_{chase}}_2,{\pi_{chase}}_3,\Pi_o,w_{\dot{L}},\mu)$. We follow definition \ref{def:STG_agents} to calculate this value:

\begin{equation*}
STG_{3,4}({\pi_{chase}}_1,{\pi_{chase}}_2,{\pi_{chase}}_3,\Pi_o,w_{\dot{L}},\mu) = \sum_{\dot{l} \in \dot{L}^{N(\mu)}_{-3,4}(\Pi_o)} w_{\dot{L}}(\dot{l}) STO_{3,4}({\pi_{chase}}_1,{\pi_{chase}}_2,{\pi_{chase}}_3,\dot{l},\mu)
\end{equation*}

Again, we do not know which $\Pi_o$ we have, but we know that we will need to obtain a line-up pattern $\dot{l}$ from $\dot{L}^{N(\mu)}_{-3,4}(\Pi_o)$ to calculate $STO_{3,4}({\pi_{chase}}_1,{\pi_{chase}}_2,{\pi_{chase}}_3,\dot{l},\mu)$. We calculate this value for a figurative line-up pattern $\dot{l} = (\pi_1,\pi_2,*,*)$ from $\dot{L}^{N(\mu)}_{-3,4}(\Pi_o)$:

\begin{equation*}
STO_{3,4}({\pi_{chase}}_1,{\pi_{chase}}_2,{\pi_{chase}}_3,\dot{l},\mu) = STO_{3,4}({\pi_{chase}}_1,{\pi_{chase}}_2,{\pi_{chase}}_3,(\pi_1,\pi_2,*,*),\mu)
\end{equation*}

The following table shows us $STO_{3,4}$ for all the permutations of ${\pi_{chase}}_1,{\pi_{chase}}_2,{\pi_{chase}}_3$.

\begin{center}
\begin{tabular}{c c c | c c c | c c c}
Slot 3 & & Slot 4							& Slot 3 & & Slot 4								& Slot 3 & & Slot 4\\
\hline
${\pi_{chase}}_1$ & $<$ & ${\pi_{chase}}_2$	& ${\pi_{chase}}_1$ & $<$ & ${\pi_{chase}}_3$	& ${\pi_{chase}}_2$ & $<$ & ${\pi_{chase}}_1$\\
${\pi_{chase}}_2$ & $<$ & ${\pi_{chase}}_3$	& ${\pi_{chase}}_3$ & $<$ & ${\pi_{chase}}_2$	& ${\pi_{chase}}_1$ & $<$ & ${\pi_{chase}}_3$\\
${\pi_{chase}}_1$ & $<$ & ${\pi_{chase}}_3$	& ${\pi_{chase}}_1$ & $<$ & ${\pi_{chase}}_2$	& ${\pi_{chase}}_2$ & $<$ & ${\pi_{chase}}_3$
\end{tabular}
\begin{tabular}{c c c | c c c | c c c}
Slot 3 & & Slot 4							& Slot 3 & & Slot 4								& Slot 3 & & Slot 4\\
\hline
${\pi_{chase}}_2$ & $<$ & ${\pi_{chase}}_3$	& ${\pi_{chase}}_3$ & $<$ & ${\pi_{chase}}_1$	& ${\pi_{chase}}_3$ & $<$ & ${\pi_{chase}}_2$\\
${\pi_{chase}}_3$ & $<$ & ${\pi_{chase}}_1$	& ${\pi_{chase}}_1$ & $<$ & ${\pi_{chase}}_2$	& ${\pi_{chase}}_2$ & $<$ & ${\pi_{chase}}_1$\\
${\pi_{chase}}_2$ & $<$ & ${\pi_{chase}}_1$	& ${\pi_{chase}}_3$ & $<$ & ${\pi_{chase}}_2$	& ${\pi_{chase}}_3$ & $<$ & ${\pi_{chase}}_1$
\end{tabular}
\end{center}

But, it is not possible to find a STO, since for every permutation the agents in slots 3 and 4 share rewards (and expected average rewards as well) due they are in the same team. So no matter which agents are in $\Pi_o$ we obtain:

\begin{equation*}
STG_{3,4}({\pi_{chase}}_1,{\pi_{chase}}_2,{\pi_{chase}}_3,\Pi_o,w_{\dot{L}},\mu) = 0
\end{equation*}

Therefore:

\begin{equation*}
STG_{3,4}(\Pi_e,w_{\Pi_e},\Pi_o,w_{\dot{L}},\mu) = 6 \frac{9}{2} \frac{1}{3} \frac{1}{3} \frac{1}{3} 0 = 0
\end{equation*}

For slots 4 and 1:

\begin{equation*}
\begin{aligned}
STG_{4,1}(\Pi_e,w_{\Pi_e},\Pi_o,w_{\dot{L}},\mu)	& = \eta_{\Pi^3} \sum_{\pi_1,\pi_2,\pi_3 \in \Pi_e | \pi_1 \neq \pi_2 \neq \pi_3} w_{\Pi_e}(\pi_1) w_{\Pi_e}(\pi_2) w_{\Pi_e}(\pi_3) STG_{4,1}(\pi_1,\pi_2,\pi_3,\Pi_o,w_{\dot{L}},\mu) =\\
													& = 6 \frac{9}{2} \frac{1}{3} \frac{1}{3} \frac{1}{3} STG_{4,1}({\pi_{chase}}_1,{\pi_{chase}}_2,{\pi_{chase}}_3,\Pi_o,w_{\dot{L}},\mu)
\end{aligned}
\end{equation*}

In this case, we only need to calculate $STG_{4,1}({\pi_{chase}}_1,{\pi_{chase}}_2,{\pi_{chase}}_3,\Pi_o,w_{\dot{L}},\mu)$. We follow definition \ref{def:STG_agents} to calculate this value:

\begin{equation*}
STG_{4,1}({\pi_{chase}}_1,{\pi_{chase}}_2,{\pi_{chase}}_3,\Pi_o,w_{\dot{L}},\mu) = \sum_{\dot{l} \in \dot{L}^{N(\mu)}_{-4,1}(\Pi_o)} w_{\dot{L}}(\dot{l}) STO_{4,1}({\pi_{chase}}_1,{\pi_{chase}}_2,{\pi_{chase}}_3,\dot{l},\mu)
\end{equation*}

Again, we do not know which $\Pi_o$ we have, but we know that we will need to obtain a line-up pattern $\dot{l}$ from $\dot{L}^{N(\mu)}_{-4,1}(\Pi_o)$ to calculate $STO_{4,1}({\pi_{chase}}_1,{\pi_{chase}}_2,{\pi_{chase}}_3,\dot{l},\mu)$. We calculate this value for a figurative line-up pattern $\dot{l} = (*,\pi_1,\pi_2,*)$ from $\dot{L}^{N(\mu)}_{-4,1}(\Pi_o)$:

\begin{equation*}
STO_{4,1}({\pi_{chase}}_1,{\pi_{chase}}_2,{\pi_{chase}}_3,\dot{l},\mu) = STO_{4,1}({\pi_{chase}}_1,{\pi_{chase}}_2,{\pi_{chase}}_3,(*,\pi_1,\pi_2,*),\mu)
\end{equation*}

The following table shows us $STO_{4,1}$ for all the permutations of ${\pi_{chase}}_1,{\pi_{chase}}_2,{\pi_{chase}}_3$.

\begin{center}
\begin{tabular}{c c c | c c c | c c c}
Slot 4 & & Slot 1							& Slot 4 & & Slot 1								& Slot 4 & & Slot 1\\
\hline
${\pi_{chase}}_1$ & $<$ & ${\pi_{chase}}_2$	& ${\pi_{chase}}_1$ & $<$ & ${\pi_{chase}}_3$	& ${\pi_{chase}}_2$ & $<$ & ${\pi_{chase}}_1$\\
${\pi_{chase}}_2$ & $<$ & ${\pi_{chase}}_3$	& ${\pi_{chase}}_3$ & $<$ & ${\pi_{chase}}_2$	& ${\pi_{chase}}_1$ & $<$ & ${\pi_{chase}}_3$\\
${\pi_{chase}}_1$ & $<$ & ${\pi_{chase}}_3$	& ${\pi_{chase}}_1$ & $<$ & ${\pi_{chase}}_2$	& ${\pi_{chase}}_2$ & $<$ & ${\pi_{chase}}_3$
\end{tabular}
\begin{tabular}{c c c | c c c | c c c}
Slot 4 & & Slot 1							& Slot 4 & & Slot 1								& Slot 4 & & Slot 1\\
\hline
${\pi_{chase}}_2$ & $<$ & ${\pi_{chase}}_3$	& ${\pi_{chase}}_3$ & $<$ & ${\pi_{chase}}_1$	& ${\pi_{chase}}_3$ & $<$ & ${\pi_{chase}}_2$\\
${\pi_{chase}}_3$ & $<$ & ${\pi_{chase}}_1$	& ${\pi_{chase}}_1$ & $<$ & ${\pi_{chase}}_2$	& ${\pi_{chase}}_2$ & $<$ & ${\pi_{chase}}_1$\\
${\pi_{chase}}_2$ & $<$ & ${\pi_{chase}}_1$	& ${\pi_{chase}}_3$ & $<$ & ${\pi_{chase}}_2$	& ${\pi_{chase}}_3$ & $<$ & ${\pi_{chase}}_1$
\end{tabular}
\end{center}

Again, it is not possible to find a STO for any permutation, since we always have ${\pi_{chase}}_i < {\pi_{chase}}_j$, where a $\pi_{chase}$ agent always tries to be chased when playing as the prey and tries to chase when playing as a predator, so the agents in slots 4 and 1 will obtain an expected average reward of $1$ and $-1$ respectively. Note that the choice of $\Pi_o$ does not affect the result of $STO_{4,1}$, so no matter which agents are in $\Pi_o$ we obtain:

\begin{equation*}
STG_{4,1}({\pi_{chase}}_1,{\pi_{chase}}_2,{\pi_{chase}}_3,\Pi_o,w_{\dot{L}},\mu) = 0
\end{equation*}

Therefore:

\begin{equation*}
STG_{4,1}(\Pi_e,w_{\Pi_e},\Pi_o,w_{\dot{L}},\mu) = 6 \frac{9}{2} \frac{1}{3} \frac{1}{3} \frac{1}{3} 0 = 0
\end{equation*}

For slots 4 and 2:

\begin{equation*}
\begin{aligned}
STG_{4,2}(\Pi_e,w_{\Pi_e},\Pi_o,w_{\dot{L}},\mu)	& = \eta_{\Pi^3} \sum_{\pi_1,\pi_2,\pi_3 \in \Pi_e | \pi_1 \neq \pi_2 \neq \pi_3} w_{\Pi_e}(\pi_1) w_{\Pi_e}(\pi_2) w_{\Pi_e}(\pi_3) STG_{4,2}(\pi_1,\pi_2,\pi_3,\Pi_o,w_{\dot{L}},\mu) =\\
													& = 6 \frac{9}{2} \frac{1}{3} \frac{1}{3} \frac{1}{3} STG_{4,2}({\pi_{chase}}_1,{\pi_{chase}}_2,{\pi_{chase}}_3,\Pi_o,w_{\dot{L}},\mu)
\end{aligned}
\end{equation*}

In this case, we only need to calculate $STG_{4,2}({\pi_{chase}}_1,{\pi_{chase}}_2,{\pi_{chase}}_3,\Pi_o,w_{\dot{L}},\mu)$. We follow definition \ref{def:STG_agents} to calculate this value:

\begin{equation*}
STG_{4,2}({\pi_{chase}}_1,{\pi_{chase}}_2,{\pi_{chase}}_3,\Pi_o,w_{\dot{L}},\mu) = \sum_{\dot{l} \in \dot{L}^{N(\mu)}_{-4,2}(\Pi_o)} w_{\dot{L}}(\dot{l}) STO_{4,2}({\pi_{chase}}_1,{\pi_{chase}}_2,{\pi_{chase}}_3,\dot{l},\mu)
\end{equation*}

Again, we do not know which $\Pi_o$ we have, but we know that we will need to obtain a line-up pattern $\dot{l}$ from $\dot{L}^{N(\mu)}_{-4,2}(\Pi_o)$ to calculate $STO_{4,2}({\pi_{chase}}_1,{\pi_{chase}}_2,{\pi_{chase}}_3,\dot{l},\mu)$. We calculate this value for a figurative line-up pattern $\dot{l} = (\pi_1,*,\pi_2,*)$ from $\dot{L}^{N(\mu)}_{-4,2}(\Pi_o)$:

\begin{equation*}
STO_{4,2}({\pi_{chase}}_1,{\pi_{chase}}_2,{\pi_{chase}}_3,\dot{l},\mu) = STO_{4,2}({\pi_{chase}}_1,{\pi_{chase}}_2,{\pi_{chase}}_3,(\pi_1,*,\pi_2,*),\mu)
\end{equation*}

The following table shows us $STO_{4,2}$ for all the permutations of ${\pi_{chase}}_1,{\pi_{chase}}_2,{\pi_{chase}}_3$.

\begin{center}
\begin{tabular}{c c c | c c c | c c c}
Slot 4 & & Slot 2							& Slot 4 & & Slot 2								& Slot 4 & & Slot 2\\
\hline
${\pi_{chase}}_1$ & $<$ & ${\pi_{chase}}_2$	& ${\pi_{chase}}_1$ & $<$ & ${\pi_{chase}}_3$	& ${\pi_{chase}}_2$ & $<$ & ${\pi_{chase}}_1$\\
${\pi_{chase}}_2$ & $<$ & ${\pi_{chase}}_3$	& ${\pi_{chase}}_3$ & $<$ & ${\pi_{chase}}_2$	& ${\pi_{chase}}_1$ & $<$ & ${\pi_{chase}}_3$\\
${\pi_{chase}}_1$ & $<$ & ${\pi_{chase}}_3$	& ${\pi_{chase}}_1$ & $<$ & ${\pi_{chase}}_2$	& ${\pi_{chase}}_2$ & $<$ & ${\pi_{chase}}_3$
\end{tabular}
\begin{tabular}{c c c | c c c | c c c}
Slot 4 & & Slot 2							& Slot 4 & & Slot 2								& Slot 4 & & Slot 2\\
\hline
${\pi_{chase}}_2$ & $<$ & ${\pi_{chase}}_3$	& ${\pi_{chase}}_3$ & $<$ & ${\pi_{chase}}_1$	& ${\pi_{chase}}_3$ & $<$ & ${\pi_{chase}}_2$\\
${\pi_{chase}}_3$ & $<$ & ${\pi_{chase}}_1$	& ${\pi_{chase}}_1$ & $<$ & ${\pi_{chase}}_2$	& ${\pi_{chase}}_2$ & $<$ & ${\pi_{chase}}_1$\\
${\pi_{chase}}_2$ & $<$ & ${\pi_{chase}}_1$	& ${\pi_{chase}}_3$ & $<$ & ${\pi_{chase}}_2$	& ${\pi_{chase}}_3$ & $<$ & ${\pi_{chase}}_1$
\end{tabular}
\end{center}

But, it is not possible to find a STO, since for every permutation the agents in slots 4 and 2 share rewards (and expected average rewards as well) due they are in the same team. So no matter which agents are in $\Pi_o$ we obtain:

\begin{equation*}
STG_{4,2}({\pi_{chase}}_1,{\pi_{chase}}_2,{\pi_{chase}}_3,\Pi_o,w_{\dot{L}},\mu) = 0
\end{equation*}

Therefore:

\begin{equation*}
STG_{4,2}(\Pi_e,w_{\Pi_e},\Pi_o,w_{\dot{L}},\mu) = 6 \frac{9}{2} \frac{1}{3} \frac{1}{3} \frac{1}{3} 0 = 0
\end{equation*}

And for slots 4 and 3:

\begin{equation*}
\begin{aligned}
STG_{4,3}(\Pi_e,w_{\Pi_e},\Pi_o,w_{\dot{L}},\mu)	& = \eta_{\Pi^3} \sum_{\pi_1,\pi_2,\pi_3 \in \Pi_e | \pi_1 \neq \pi_2 \neq \pi_3} w_{\Pi_e}(\pi_1) w_{\Pi_e}(\pi_2) w_{\Pi_e}(\pi_3) STG_{4,3}(\pi_1,\pi_2,\pi_3,\Pi_o,w_{\dot{L}},\mu) =\\
													& = 6 \frac{9}{2} \frac{1}{3} \frac{1}{3} \frac{1}{3} STG_{4,3}({\pi_{chase}}_1,{\pi_{chase}}_2,{\pi_{chase}}_3,\Pi_o,w_{\dot{L}},\mu)
\end{aligned}
\end{equation*}

In this case, we only need to calculate $STG_{4,3}({\pi_{chase}}_1,{\pi_{chase}}_2,{\pi_{chase}}_3,\Pi_o,w_{\dot{L}},\mu)$. We follow definition \ref{def:STG_agents} to calculate this value:

\begin{equation*}
STG_{4,3}({\pi_{chase}}_1,{\pi_{chase}}_2,{\pi_{chase}}_3,\Pi_o,w_{\dot{L}},\mu) = \sum_{\dot{l} \in \dot{L}^{N(\mu)}_{-4,3}(\Pi_o)} w_{\dot{L}}(\dot{l}) STO_{4,3}({\pi_{chase}}_1,{\pi_{chase}}_2,{\pi_{chase}}_3,\dot{l},\mu)
\end{equation*}

Again, we do not know which $\Pi_o$ we have, but we know that we will need to obtain a line-up pattern $\dot{l}$ from $\dot{L}^{N(\mu)}_{-4,3}(\Pi_o)$ to calculate $STO_{4,3}({\pi_{chase}}_1,{\pi_{chase}}_2,{\pi_{chase}}_3,\dot{l},\mu)$. We calculate this value for a figurative line-up pattern $\dot{l} = (\pi_1,\pi_2,*,*)$ from $\dot{L}^{N(\mu)}_{-4,3}(\Pi_o)$:

\begin{equation*}
STO_{4,3}({\pi_{chase}}_1,{\pi_{chase}}_2,{\pi_{chase}}_3,\dot{l},\mu) = STO_{4,3}({\pi_{chase}}_1,{\pi_{chase}}_2,{\pi_{chase}}_3,(\pi_1,\pi_2,*,*),\mu)
\end{equation*}

The following table shows us $STO_{4,3}$ for all the permutations of ${\pi_{chase}}_1,{\pi_{chase}}_2,{\pi_{chase}}_3$.

\begin{center}
\begin{tabular}{c c c | c c c | c c c}
Slot 4 & & Slot 3							& Slot 4 & & Slot 3								& Slot 4 & & Slot 3\\
\hline
${\pi_{chase}}_1$ & $<$ & ${\pi_{chase}}_2$	& ${\pi_{chase}}_1$ & $<$ & ${\pi_{chase}}_3$	& ${\pi_{chase}}_2$ & $<$ & ${\pi_{chase}}_1$\\
${\pi_{chase}}_2$ & $<$ & ${\pi_{chase}}_3$	& ${\pi_{chase}}_3$ & $<$ & ${\pi_{chase}}_2$	& ${\pi_{chase}}_1$ & $<$ & ${\pi_{chase}}_3$\\
${\pi_{chase}}_1$ & $<$ & ${\pi_{chase}}_3$	& ${\pi_{chase}}_1$ & $<$ & ${\pi_{chase}}_2$	& ${\pi_{chase}}_2$ & $<$ & ${\pi_{chase}}_3$
\end{tabular}
\begin{tabular}{c c c | c c c | c c c}
Slot 4 & & Slot 3							& Slot 4 & & Slot 3								& Slot 4 & & Slot 3\\
\hline
${\pi_{chase}}_2$ & $<$ & ${\pi_{chase}}_3$	& ${\pi_{chase}}_3$ & $<$ & ${\pi_{chase}}_1$	& ${\pi_{chase}}_3$ & $<$ & ${\pi_{chase}}_2$\\
${\pi_{chase}}_3$ & $<$ & ${\pi_{chase}}_1$	& ${\pi_{chase}}_1$ & $<$ & ${\pi_{chase}}_2$	& ${\pi_{chase}}_2$ & $<$ & ${\pi_{chase}}_1$\\
${\pi_{chase}}_2$ & $<$ & ${\pi_{chase}}_1$	& ${\pi_{chase}}_3$ & $<$ & ${\pi_{chase}}_2$	& ${\pi_{chase}}_3$ & $<$ & ${\pi_{chase}}_1$
\end{tabular}
\end{center}

But, it is not possible to find a STO, since for every permutation the agents in slots 4 and 3 share rewards (and expected average rewards as well) due they are in the same team. So no matter which agents are in $\Pi_o$ we obtain:

\begin{equation*}
STG_{4,3}({\pi_{chase}}_1,{\pi_{chase}}_2,{\pi_{chase}}_3,\Pi_o,w_{\dot{L}},\mu) = 0
\end{equation*}

Therefore:

\begin{equation*}
STG_{4,3}(\Pi_e,w_{\Pi_e},\Pi_o,w_{\dot{L}},\mu) = 6 \frac{9}{2} \frac{1}{3} \frac{1}{3} \frac{1}{3} 0 = 0
\end{equation*}

And finally, we weight over the slots:

\begin{equation*}
\begin{aligned}
& STG(\Pi_e,w_{\Pi_e},\Pi_o,w_{\dot{L}},\mu,w_S) = \eta_{S_1^2} \sum_{i=1}^{N(\mu)} w_S(i,\mu) \times\\
& \times \left(\sum_{j=1}^{i-1} w_S(j,\mu) STG_{i,j}(\Pi_e,w_{\Pi_e},\Pi_o,w_{\dot{L}},\mu) + \sum_{j=i+1}^{N(\mu)} w_S(j,\mu) STG_{i,j}(\Pi_e,w_{\Pi_e},\Pi_o,w_{\dot{L}},\mu)\right) =\\
& \ \ \ \ \ \ \ \ \ \ \ \ \ \ \ \ \ \ \ \ \ \ \ \ \ \ \ \ \ \ \ \ \ \ \ \ \ \ = \frac{4}{3} \frac{1}{4} \frac{1}{4} \{STG_{1,2}(\Pi_e,w_{\Pi_e},\Pi_o,w_{\dot{L}},\mu) + STG_{1,3}(\Pi_e,w_{\Pi_e},\Pi_o,w_{\dot{L}},\mu) +\\
& \ \ \ \ \ \ \ \ \ \ \ \ \ \ \ \ \ \ \ \ \ \ \ \ \ \ \ \ \ \ \ \ \ \ \ \ \ \ + STG_{1,4}(\Pi_e,w_{\Pi_e},\Pi_o,w_{\dot{L}},\mu) + STG_{2,1}(\Pi_e,w_{\Pi_e},\Pi_o,w_{\dot{L}},\mu) +\\
& \ \ \ \ \ \ \ \ \ \ \ \ \ \ \ \ \ \ \ \ \ \ \ \ \ \ \ \ \ \ \ \ \ \ \ \ \ \ + STG_{2,3}(\Pi_e,w_{\Pi_e},\Pi_o,w_{\dot{L}},\mu) + STG_{2,4}(\Pi_e,w_{\Pi_e},\Pi_o,w_{\dot{L}},\mu) +\\
& \ \ \ \ \ \ \ \ \ \ \ \ \ \ \ \ \ \ \ \ \ \ \ \ \ \ \ \ \ \ \ \ \ \ \ \ \ \ + STG_{3,1}(\Pi_e,w_{\Pi_e},\Pi_o,w_{\dot{L}},\mu) + STG_{3,2}(\Pi_e,w_{\Pi_e},\Pi_o,w_{\dot{L}},\mu) +\\
& \ \ \ \ \ \ \ \ \ \ \ \ \ \ \ \ \ \ \ \ \ \ \ \ \ \ \ \ \ \ \ \ \ \ \ \ \ \ + STG_{3,4}(\Pi_e,w_{\Pi_e},\Pi_o,w_{\dot{L}},\mu) + STG_{4,1}(\Pi_e,w_{\Pi_e},\Pi_o,w_{\dot{L}},\mu) +\\
& \ \ \ \ \ \ \ \ \ \ \ \ \ \ \ \ \ \ \ \ \ \ \ \ \ \ \ \ \ \ \ \ \ \ \ \ \ \ + STG_{4,2}(\Pi_e,w_{\Pi_e},\Pi_o,w_{\dot{L}},\mu) + STG_{4,3}(\Pi_e,w_{\Pi_e},\Pi_o,w_{\dot{L}},\mu)\} =\\
& \ \ \ \ \ \ \ \ \ \ \ \ \ \ \ \ \ \ \ \ \ \ \ \ \ \ \ \ \ \ \ \ \ \ \ \ \ \ = \frac{4}{3} \frac{1}{4} \frac{1}{4} \left\{3 \times 1 + 9 \times 0\right\} = \frac{1}{4}
\end{aligned}
\end{equation*}

So, for every $\Pi_o$ we obtain the same result:

\begin{equation*}
\forall \Pi_o : STG(\Pi_e,w_{\Pi_e},\Pi_o,w_{\dot{L}},\mu,w_S) = \frac{1}{4}
\end{equation*}

Therefore, predator-prey has $Right_{min} = \frac{1}{4}$ (as a {\em higher} approximation) for this property.
\end{proof}
\end{approximation}

\subsection{Partial Grading}
Now we arrive to the partial grading (PG) property. As given in section \ref{sec:PG}, we want to know if there exists a partial ordering between the evaluated agents when interacting in the environment.

To simplify the notation, we use the next table to represent the PO:
$R_i(\mu[\instantiation{l}{i,j}{\pi_1,\pi_2}]) \leq R_j(\mu[\instantiation{l}{i,j}{\pi_1,\pi_2}])$,
$R_i(\mu[\instantiation{l}{i,j}{\pi_2,\pi_3}]) \leq R_j(\mu[\instantiation{l}{i,j}{\pi_2,\pi_3}])$ and
$R_i(\mu[\instantiation{l}{i,j}{\pi_1,\pi_3}]) \leq R_j(\mu[\instantiation{l}{i,j}{\pi_1,\pi_3}])$.

\begin{center}
\begin{tabular}{c c c}
Slot i & & Slot j\\
\hline
$\pi_1$ & $\leq$ & $\pi_2$\\
$\pi_2$ & $\leq$ & $\pi_3$\\
$\pi_1$ & $\leq$ & $\pi_3$
\end{tabular}
\end{center}

\begin{proposition}
\label{prop:predator-prey_PG_general_min}
$General_{min}$ for the partial grading (PG) property is equal to $\frac{1}{2}$ for the predator-prey environment.

\begin{proof}
To find $General_{min}$ (equation \ref{eq:general_min}), we need to find a trio $\left\langle\Pi_e,w_{\Pi_e},\Pi_o\right\rangle$ which minimises the property as much as possible. We can have this situation by selecting $\Pi_e = \{\pi_x,\pi_y,\pi_z\}$ with uniform weight for $w_{\Pi_e}$ and $\Pi_o = \{\pi_s\}$ (a $\pi_s$ agent always stays in the same cell\footnote{Note that every cell has an action which is blocked by a block or a boundary, therefore an agent performing this action will stay at its current cell.}).

$\pi_x$ behaves as shown in figure \ref{fig:predator-prey_PG_general_min_agent_x} when playing on each of the 4 slots.

\begin{figure}[!ht]
\newcolumntype{C}{>{\centering\arraybackslash}p{12px}}
\def \block {\cellcolor{black}}
\def \hunter {$\bigcirc$}
\def \prey {$\diamondsuit$}

\centering

\begin{minipage}{0.15\textwidth}
\centering
\setlength{\tabcolsep}{0px}

\begin{tabular}[c]{|C|C|C|C|}
\hline
\prey   &         &         &         \\
\hline
1       & 2       & \block  &         \\
\hline
        &         &         &         \\
\hline
        & \block  &         &         \\
\hline
\end{tabular}
\end{minipage}
\begin{minipage}{0.15\textwidth}
\centering
\setlength{\tabcolsep}{0px}

\begin{tabular}[c]{|C|C|C|C|}
\hline
        & 2       & 1       & \hunter \\
\hline
4       & 3       & \block  &         \\
\hline
5       &         &         &         \\
\hline
        & \block  &         &         \\
\hline
\end{tabular}
\end{minipage}
\begin{minipage}{0.15\textwidth}
\centering
\setlength{\tabcolsep}{0px}

\begin{tabular}[c]{|C|C|C|C|}
\hline
        &         &         &         \\
\hline
        &         & \block  &         \\
\hline
1       &         &         &         \\
\hline
\hunter & \block  &         &         \\
\hline
\end{tabular}
\end{minipage}
\begin{minipage}{0.15\textwidth}
\centering
\setlength{\tabcolsep}{0px}

\begin{tabular}[c]{|C|C|C|C|}
\hline
        &         &         &         \\
\hline
        &         & \block  &         \\
\hline
4       & 3       & 2       &         \\
\hline
        & \block  & 1       & \hunter \\
\hline
\end{tabular}
\end{minipage}

\caption{Behaviour of $\pi_x$ when playing on each of the slots. Numbers represent the position of $\pi_x$ during the iterations.}
\label{fig:predator-prey_PG_general_min_agent_x}
\end{figure}

$\pi_y$ behaves as shown in figure \ref{fig:predator-prey_PG_general_min_agent_y} when playing on each of the 4 slots.

\begin{figure}[!ht]
\newcolumntype{C}{>{\centering\arraybackslash}p{12px}}
\def \block {\cellcolor{black}}
\def \hunter {$\bigcirc$}
\def \prey {$\diamondsuit$}

\centering

\begin{minipage}{0.15\textwidth}
\centering
\setlength{\tabcolsep}{0px}

\begin{tabular}[c]{|C|C|C|C|}
\hline
\prey   &         &         &         \\
\hline
        &         & \block  &         \\
\hline
        &         &         &         \\
\hline
        & \block  &         &         \\
\hline
\end{tabular}
\end{minipage}
\begin{minipage}{0.15\textwidth}
\centering
\setlength{\tabcolsep}{0px}

\begin{tabular}[c]{|C|C|C|C|}
\hline
5       & 2       & 1       & \hunter \\
\hline
4       & 3       & \block  &         \\
\hline
        &         &         &         \\
\hline
        & \block  &         &         \\
\hline
\end{tabular}
\end{minipage}
\begin{minipage}{0.15\textwidth}
\centering
\setlength{\tabcolsep}{0px}

\begin{tabular}[c]{|C|C|C|C|}
\hline
5       & 4       &         &         \\
\hline
        & 3       & \block  &         \\
\hline
1       & 2       &         &         \\
\hline
\hunter & \block  &         &         \\
\hline
\end{tabular}
\end{minipage}
\begin{minipage}{0.15\textwidth}
\centering
\setlength{\tabcolsep}{0px}

\begin{tabular}[c]{|C|C|C|C|}
\hline
6       &         &         &         \\
\hline
5       & 4       & \block  &         \\
\hline
        & 3       & 2       &         \\
\hline
        & \block  & 1       & \hunter \\
\hline
\end{tabular}
\end{minipage}

\caption{Behaviour of $\pi_y$ when playing on each of the slots. Numbers represent the position of $\pi_y$ during the iterations.}
\label{fig:predator-prey_PG_general_min_agent_y}
\end{figure}

$\pi_z$ behaves as shown in figure \ref{fig:predator-prey_PG_general_min_agent_z} when playing on each of the 4 slots.

\begin{figure}[!ht]
\newcolumntype{C}{>{\centering\arraybackslash}p{12px}}
\def \block {\cellcolor{black}}
\def \hunter {$\bigcirc$}
\def \prey {$\diamondsuit$}

\centering

\begin{minipage}{0.15\textwidth}
\centering
\setlength{\tabcolsep}{0px}

\begin{tabular}[c]{|C|C|C|C|}
\hline
\prey   & 1       &         &         \\
\hline
3       & 2       & \block  &         \\
\hline
4       &         &         &         \\
\hline
        & \block  &         &         \\
\hline
\end{tabular}
\end{minipage}
\begin{minipage}{0.15\textwidth}
\centering
\setlength{\tabcolsep}{0px}

\begin{tabular}[c]{|C|C|C|C|}
\hline
3       & 2       & 1       & \hunter \\
\hline
        &         & \block  &         \\
\hline
        &         &         &         \\
\hline
        & \block  &         &         \\
\hline
\end{tabular}
\end{minipage}
\begin{minipage}{0.15\textwidth}
\centering
\setlength{\tabcolsep}{0px}

\begin{tabular}[c]{|C|C|C|C|}
\hline
3       &         &         &         \\
\hline
2       &         & \block  &         \\
\hline
1       &         &         &         \\
\hline
\hunter & \block  &         &         \\
\hline
\end{tabular}
\end{minipage}
\begin{minipage}{0.15\textwidth}
\centering
\setlength{\tabcolsep}{0px}

\begin{tabular}[c]{|C|C|C|C|}
\hline
6       &         &         &         \\
\hline
5       &         & \block  &         \\
\hline
4       & 3       & 2       &         \\
\hline
        & \block  & 1       & \hunter \\
\hline
\end{tabular}
\end{minipage}

\caption{Behaviour of $\pi_z$ when playing on each of the slots. Numbers represent the position of $\pi_z$ during the iterations.}
\label{fig:predator-prey_PG_general_min_agent_z}
\end{figure}

Following definition \ref{def:PG}, we obtain the PG value for this $\left\langle\Pi_e,w_{\Pi_e},\Pi_o\right\rangle$. Since the environment is not symmetric, we need to calculate this property for every pair of slots. Following definition \ref{def:STG_set} (for PG), we can calculate its PG value for each pair of slots. We start with slots 1 and 2:

\begin{equation*}
\begin{aligned}
PG_{1,2}(\Pi_e,w_{\Pi_e},\Pi_o,w_{\dot{L}},\mu)	& = \eta_{\Pi^3} \sum_{\pi_1,\pi_2,\pi_3 \in \Pi_e | \pi_1 \neq \pi_2 \neq \pi_3} w_{\Pi_e}(\pi_1) w_{\Pi_e}(\pi_2) w_{\Pi_e}(\pi_3) PG_{1,2}(\pi_1,\pi_2,\pi_3,\Pi_o,w_{\dot{L}},\mu) =\\
												& = 6 \frac{9}{2} \frac{1}{3} \frac{1}{3} \frac{1}{3} PG_{1,2}(\pi_x,\pi_y,\pi_z,\Pi_o,w_{\dot{L}},\mu)
\end{aligned}
\end{equation*}

\noindent Note that we avoided to calculate all the permutations of $\pi_1,\pi_2,\pi_3$ for $PG_{i,j}(\pi_1,\pi_2,\pi_3,\Pi_o,w_{\dot{L}},\mu)$ since they provide the same result, by calculating only one permutation and multiplying the result by the number of permutations $6$.

In this case, we only need to calculate $PG_{1,2}(\pi_x,\pi_y,\pi_z,\Pi_o,w_{\dot{L}},\mu)$. We follow definition \ref{def:STG_agents} (for PG) to calculate this value:

\begin{equation*}
\begin{aligned}
PG_{1,2}(\pi_x,\pi_y,\pi_z,\Pi_o,w_{\dot{L}},\mu)	& = \sum_{\dot{l} \in \dot{L}^{N(\mu)}_{-1,2}(\Pi_o)} w_{\dot{L}}(\dot{l}) PO_{1,2}(\pi_x,\pi_y,\pi_z,\dot{l},\mu) =\\
													& = PO_{1,2}(\pi_x,\pi_y,\pi_z,(*,*,\pi_s,\pi_s),\mu)
\end{aligned}
\end{equation*}

The following table shows us $PO_{1,2}$ for all the permutations of $\pi_x,\pi_y,\pi_z$.

\begin{center}
\begin{tabular}{c c c | c c c | c c c}
Slot 1 & & Slot 2			& Slot 1 & & Slot 2				& Slot 1 & & Slot 2\\
\hline
$\pi_x$ & $\leq$ & $\pi_y$	& $\pi_x$ & $\leq$ & $\pi_z$	& $\pi_y$ & $\leq$ & $\pi_x$\\
$\pi_y$ & $\leq$ & $\pi_z$	& $\pi_z$ & $\leq$ & $\pi_y$	& $\pi_x$ & $\leq$ & $\pi_z$\\
$\pi_x$ & $\leq$ & $\pi_z$	& $\pi_x$ & $\leq$ & $\pi_y$	& $\pi_y$ & $\leq$ & $\pi_z$
\end{tabular}
\begin{tabular}{c c c | c c c | c c c}
Slot 1 & & Slot 2			& Slot 1 & & Slot 2				& Slot 1 & & Slot 2\\
\hline
$\pi_y$ & $\leq$ & $\pi_z$	& $\pi_z$ & $\leq$ & $\pi_x$	& $\pi_z$ & $\leq$ & $\pi_y$\\
$\pi_z$ & $\leq$ & $\pi_x$	& $\pi_x$ & $\leq$ & $\pi_y$	& $\pi_y$ & $\leq$ & $\pi_x$\\
$\pi_y$ & $\leq$ & $\pi_x$	& $\pi_z$ & $\leq$ & $\pi_y$	& $\pi_z$ & $\leq$ & $\pi_x$
\end{tabular}
\end{center}

But, it is not possible to find a PO for any permutation, since $\pi_x > \pi_z$, $\pi_y > \pi_x$ and $\pi_z > \pi_y$. So:

\begin{equation*}
PG_{1,2}(\pi_x,\pi_y,\pi_z,\Pi_o,w_{\dot{L}},\mu) = 0
\end{equation*}

Therefore:

\begin{equation*}
PG_{1,2}(\Pi_e,w_{\Pi_e},\Pi_o,w_{\dot{L}},\mu) = 6 \frac{9}{2} \frac{1}{3} \frac{1}{3} \frac{1}{3} 0 = 0
\end{equation*}

For slots 1 and 3:

\begin{equation*}
\begin{aligned}
PG_{1,3}(\Pi_e,w_{\Pi_e},\Pi_o,w_{\dot{L}},\mu)	& = \eta_{\Pi^3} \sum_{\pi_1,\pi_2,\pi_3 \in \Pi_e | \pi_1 \neq \pi_2 \neq \pi_3} w_{\Pi_e}(\pi_1) w_{\Pi_e}(\pi_2) w_{\Pi_e}(\pi_3) PG_{1,3}(\pi_1,\pi_2,\pi_3,\Pi_o,w_{\dot{L}},\mu) =\\
												& = 6 \frac{9}{2} \frac{1}{3} \frac{1}{3} \frac{1}{3} PG_{1,3}(\pi_x,\pi_y,\pi_z,\Pi_o,w_{\dot{L}},\mu)
\end{aligned}
\end{equation*}

In this case, we only need to calculate $PG_{1,3}(\pi_x,\pi_y,\pi_z,\Pi_o,w_{\dot{L}},\mu)$. We follow definition \ref{def:STG_agents} (for PG) to calculate this value:

\begin{equation*}
\begin{aligned}
PG_{1,3}(\pi_x,\pi_y,\pi_z,\Pi_o,w_{\dot{L}},\mu)	& = \sum_{\dot{l} \in \dot{L}^{N(\mu)}_{-1,3}(\Pi_o)} w_{\dot{L}}(\dot{l}) PO_{1,3}(\pi_x,\pi_y,\pi_z,\dot{l},\mu) =\\
													& = PO_{1,3}(\pi_x,\pi_y,\pi_z,(*,\pi_s,*,\pi_s),\mu)
\end{aligned}
\end{equation*}

The following table shows us $PO_{1,3}$ for all the permutations of $\pi_x,\pi_y,\pi_z$.

\begin{center}
\begin{tabular}{c c c | c c c | c c c}
Slot 1 & & Slot 3			& Slot 1 & & Slot 3				& Slot 1 & & Slot 3\\
\hline
$\pi_x$ & $\leq$ & $\pi_y$	& $\pi_x$ & $\leq$ & $\pi_z$	& $\pi_y$ & $\leq$ & $\pi_x$\\
$\pi_y$ & $\leq$ & $\pi_z$	& $\pi_z$ & $\leq$ & $\pi_y$	& $\pi_x$ & $\leq$ & $\pi_z$\\
$\pi_x$ & $\leq$ & $\pi_z$	& $\pi_x$ & $\leq$ & $\pi_y$	& $\pi_y$ & $\leq$ & $\pi_z$
\end{tabular}
\begin{tabular}{c c c | c c c | c c c}
Slot 1 & & Slot 3			& Slot 1 & & Slot 3				& Slot 1 & & Slot 3\\
\hline
$\pi_y$ & $\leq$ & $\pi_z$	& $\pi_z$ & $\leq$ & $\pi_x$	& $\pi_z$ & $\leq$ & $\pi_y$\\
$\pi_z$ & $\leq$ & $\pi_x$	& $\pi_x$ & $\leq$ & $\pi_y$	& $\pi_y$ & $\leq$ & $\pi_x$\\
$\pi_y$ & $\leq$ & $\pi_x$	& $\pi_z$ & $\leq$ & $\pi_y$	& $\pi_z$ & $\leq$ & $\pi_x$
\end{tabular}
\end{center}

But, it is not possible to find a PO for any permutation, since $\pi_x > \pi_z$, $\pi_y > \pi_x$ and $\pi_z > \pi_y$. So:

\begin{equation*}
PG_{1,3}(\pi_x,\pi_y,\pi_z,\Pi_o,w_{\dot{L}},\mu) = 0
\end{equation*}

Therefore:

\begin{equation*}
PG_{1,3}(\Pi_e,w_{\Pi_e},\Pi_o,w_{\dot{L}},\mu) = 6 \frac{9}{2} \frac{1}{3} \frac{1}{3} \frac{1}{3} 0 = 0
\end{equation*}

For slots 1 and 4:

\begin{equation*}
\begin{aligned}
PG_{1,4}(\Pi_e,w_{\Pi_e},\Pi_o,w_{\dot{L}},\mu)	& = \eta_{\Pi^3} \sum_{\pi_1,\pi_2,\pi_3 \in \Pi_e | \pi_1 \neq \pi_2 \neq \pi_3} w_{\Pi_e}(\pi_1) w_{\Pi_e}(\pi_2) w_{\Pi_e}(\pi_3) PG_{1,4}(\pi_1,\pi_2,\pi_3,\Pi_o,w_{\dot{L}},\mu) =\\
												& = 6 \frac{9}{2} \frac{1}{3} \frac{1}{3} \frac{1}{3} PG_{1,4}(\pi_x,\pi_y,\pi_z,\Pi_o,w_{\dot{L}},\mu)
\end{aligned}
\end{equation*}

In this case, we only need to calculate $PG_{1,4}(\pi_x,\pi_y,\pi_z,\Pi_o,w_{\dot{L}},\mu)$. We follow definition \ref{def:STG_agents} (for PG) to calculate this value:

\begin{equation*}
\begin{aligned}
PG_{1,4}(\pi_x,\pi_y,\pi_z,\Pi_o,w_{\dot{L}},\mu)	& = \sum_{\dot{l} \in \dot{L}^{N(\mu)}_{-1,4}(\Pi_o)} w_{\dot{L}}(\dot{l}) PO_{1,4}(\pi_x,\pi_y,\pi_z,\dot{l},\mu) =\\
													& = PO_{1,4}(\pi_x,\pi_y,\pi_z,(*,\pi_s,\pi_s,*),\mu)
\end{aligned}
\end{equation*}

The following table shows us $PO_{1,4}$ for all the permutations of $\pi_x,\pi_y,\pi_z$.

\begin{center}
\begin{tabular}{c c c | c c c | c c c}
Slot 1 & & Slot 4			& Slot 1 & & Slot 4				& Slot 1 & & Slot 4\\
\hline
$\pi_x$ & $\leq$ & $\pi_y$	& $\pi_x$ & $\leq$ & $\pi_z$	& $\pi_y$ & $\leq$ & $\pi_x$\\
$\pi_y$ & $\leq$ & $\pi_z$	& $\pi_z$ & $\leq$ & $\pi_y$	& $\pi_x$ & $\leq$ & $\pi_z$\\
$\pi_x$ & $\leq$ & $\pi_z$	& $\pi_x$ & $\leq$ & $\pi_y$	& $\pi_y$ & $\leq$ & $\pi_z$
\end{tabular}
\begin{tabular}{c c c | c c c | c c c}
Slot 1 & & Slot 4			& Slot 1 & & Slot 4				& Slot 1 & & Slot 4\\
\hline
$\pi_y$ & $\leq$ & $\pi_z$	& $\pi_z$ & $\leq$ & $\pi_x$	& $\pi_z$ & $\leq$ & $\pi_y$\\
$\pi_z$ & $\leq$ & $\pi_x$	& $\pi_x$ & $\leq$ & $\pi_y$	& $\pi_y$ & $\leq$ & $\pi_x$\\
$\pi_y$ & $\leq$ & $\pi_x$	& $\pi_z$ & $\leq$ & $\pi_y$	& $\pi_z$ & $\leq$ & $\pi_x$
\end{tabular}
\end{center}

But, it is not possible to find a PO for any permutation, since $\pi_x > \pi_z$, $\pi_y > \pi_x$ and $\pi_z > \pi_y$. So:

\begin{equation*}
PG_{1,4}(\pi_x,\pi_y,\pi_z,\Pi_o,w_{\dot{L}},\mu) = 0
\end{equation*}

Therefore:

\begin{equation*}
PG_{1,4}(\Pi_e,w_{\Pi_e},\Pi_o,w_{\dot{L}},\mu) = 6 \frac{9}{2} \frac{1}{3} \frac{1}{3} \frac{1}{3} 0 = 0
\end{equation*}

For slots 2 and 1:

\begin{equation*}
\begin{aligned}
PG_{2,1}(\Pi_e,w_{\Pi_e},\Pi_o,w_{\dot{L}},\mu)	& = \eta_{\Pi^3} \sum_{\pi_1,\pi_2,\pi_3 \in \Pi_e | \pi_1 \neq \pi_2 \neq \pi_3} w_{\Pi_e}(\pi_1) w_{\Pi_e}(\pi_2) w_{\Pi_e}(\pi_3) PG_{2,1}(\pi_1,\pi_2,\pi_3,\Pi_o,w_{\dot{L}},\mu) =\\
												& = 6 \frac{9}{2} \frac{1}{3} \frac{1}{3} \frac{1}{3} PG_{2,1}(\pi_x,\pi_y,\pi_z,\Pi_o,w_{\dot{L}},\mu)
\end{aligned}
\end{equation*}

In this case, we only need to calculate $PG_{2,1}(\pi_x,\pi_y,\pi_z,\Pi_o,w_{\dot{L}},\mu)$. We follow definition \ref{def:STG_agents} (for PG) to calculate this value:

\begin{equation*}
\begin{aligned}
PG_{2,1}(\pi_x,\pi_y,\pi_z,\Pi_o,w_{\dot{L}},\mu)	& = \sum_{\dot{l} \in \dot{L}^{N(\mu)}_{-2,1}(\Pi_o)} w_{\dot{L}}(\dot{l}) PO_{2,1}(\pi_x,\pi_y,\pi_z,\dot{l},\mu) =\\
													& = PO_{2,1}(\pi_x,\pi_y,\pi_z,(*,*,\pi_s,\pi_s),\mu)
\end{aligned}
\end{equation*}

The following table shows us $PO_{2,1}$ for all the permutations of $\pi_x,\pi_y,\pi_z$.

\begin{center}
\begin{tabular}{c c c | c c c | c c c}
Slot 2 & & Slot 1			& Slot 2 & & Slot 1				& Slot 2 & & Slot 1\\
\hline
$\pi_x$ & $\leq$ & $\pi_y$	& $\pi_x$ & $\leq$ & $\pi_z$	& $\pi_y$ & $\leq$ & $\pi_x$\\
$\pi_y$ & $\leq$ & $\pi_z$	& $\pi_z$ & $\leq$ & $\pi_y$	& $\pi_x$ & $\leq$ & $\pi_z$\\
$\pi_x$ & $\leq$ & $\pi_z$	& $\pi_x$ & $\leq$ & $\pi_y$	& $\pi_y$ & $\leq$ & $\pi_z$
\end{tabular}
\begin{tabular}{c c c | c c c | c c c}
Slot 2 & & Slot 1			& Slot 2 & & Slot 1				& Slot 2 & & Slot 1\\
\hline
$\pi_y$ & $\leq$ & $\pi_z$	& $\pi_z$ & $\leq$ & $\pi_x$	& $\pi_z$ & $\leq$ & $\pi_y$\\
$\pi_z$ & $\leq$ & $\pi_x$	& $\pi_x$ & $\leq$ & $\pi_y$	& $\pi_y$ & $\leq$ & $\pi_x$\\
$\pi_y$ & $\leq$ & $\pi_x$	& $\pi_z$ & $\leq$ & $\pi_y$	& $\pi_z$ & $\leq$ & $\pi_x$
\end{tabular}
\end{center}

But, it is not possible to find a PO for any permutation, since $\pi_x > \pi_z$, $\pi_y > \pi_x$ and $\pi_z > \pi_y$. So:

\begin{equation*}
PG_{2,1}(\pi_x,\pi_y,\pi_z,\Pi_o,w_{\dot{L}},\mu) = 0
\end{equation*}

Therefore:

\begin{equation*}
PG_{2,1}(\Pi_e,w_{\Pi_e},\Pi_o,w_{\dot{L}},\mu) = 6 \frac{9}{2} \frac{1}{3} \frac{1}{3} \frac{1}{3} 0 = 0
\end{equation*}

For slots 2 and 3:

\begin{equation*}
\begin{aligned}
PG_{2,3}(\Pi_e,w_{\Pi_e},\Pi_o,w_{\dot{L}},\mu)	& = \eta_{\Pi^3} \sum_{\pi_1,\pi_2,\pi_3 \in \Pi_e | \pi_1 \neq \pi_2 \neq \pi_3} w_{\Pi_e}(\pi_1) w_{\Pi_e}(\pi_2) w_{\Pi_e}(\pi_3) PG_{2,3}(\pi_1,\pi_2,\pi_3,\Pi_o,w_{\dot{L}},\mu) =\\
												& = 6 \frac{9}{2} \frac{1}{3} \frac{1}{3} \frac{1}{3} PG_{2,3}(\pi_x,\pi_y,\pi_z,\Pi_o,w_{\dot{L}},\mu)
\end{aligned}
\end{equation*}

In this case, we only need to calculate $PG_{2,3}(\pi_x,\pi_y,\pi_z,\Pi_o,w_{\dot{L}},\mu)$. We follow definition \ref{def:STG_agents} (for PG) to calculate this value:

\begin{equation*}
\begin{aligned}
PG_{2,3}(\pi_x,\pi_y,\pi_z,\Pi_o,w_{\dot{L}},\mu)	& = \sum_{\dot{l} \in \dot{L}^{N(\mu)}_{-2,3}(\Pi_o)} w_{\dot{L}}(\dot{l}) PO_{2,3}(\pi_x,\pi_y,\pi_z,\dot{l},\mu) =\\
													& = PO_{2,3}(\pi_x,\pi_y,\pi_z,(\pi_s,*,*,\pi_s),\mu)
\end{aligned}
\end{equation*}

The following table shows us $PO_{2,3}$ for all the permutations of $\pi_x,\pi_y,\pi_z$.

\begin{center}
\begin{tabular}{c c c | c c c | c c c}
Slot 2 & & Slot 3			& Slot 2 & & Slot 3				& Slot 2 & & Slot 3\\
\hline
$\pi_x$ & $\leq$ & $\pi_y$	& $\pi_x$ & $\leq$ & $\pi_z$	& $\pi_y$ & $\leq$ & $\pi_x$\\
$\pi_y$ & $\leq$ & $\pi_z$	& $\pi_z$ & $\leq$ & $\pi_y$	& $\pi_x$ & $\leq$ & $\pi_z$\\
$\pi_x$ & $\leq$ & $\pi_z$	& $\pi_x$ & $\leq$ & $\pi_y$	& $\pi_y$ & $\leq$ & $\pi_z$
\end{tabular}
\begin{tabular}{c c c | c c c | c c c}
Slot 2 & & Slot 3			& Slot 2 & & Slot 3				& Slot 2 & & Slot 3\\
\hline
$\pi_y$ & $\leq$ & $\pi_z$	& $\pi_z$ & $\leq$ & $\pi_x$	& $\pi_z$ & $\leq$ & $\pi_y$\\
$\pi_z$ & $\leq$ & $\pi_x$	& $\pi_x$ & $\leq$ & $\pi_y$	& $\pi_y$ & $\leq$ & $\pi_x$\\
$\pi_y$ & $\leq$ & $\pi_x$	& $\pi_z$ & $\leq$ & $\pi_y$	& $\pi_z$ & $\leq$ & $\pi_x$
\end{tabular}
\end{center}

It is possible to find a PO for every permutation, since the agents in slots 2 and 3 share rewards (and expected average rewards as well) due they are in the same team. So:

\begin{equation*}
PG_{2,3}(\pi_x,\pi_y,\pi_z,\Pi_o,w_{\dot{L}},\mu) = 1
\end{equation*}

Therefore:

\begin{equation*}
PG_{2,3}(\Pi_e,w_{\Pi_e},\Pi_o,w_{\dot{L}},\mu) = 6 \frac{9}{2} \frac{1}{3} \frac{1}{3} \frac{1}{3} 1 = 1
\end{equation*}

For slots 2 and 4:

\begin{equation*}
\begin{aligned}
PG_{2,4}(\Pi_e,w_{\Pi_e},\Pi_o,w_{\dot{L}},\mu)	& = \eta_{\Pi^3} \sum_{\pi_1,\pi_2,\pi_3 \in \Pi_e | \pi_1 \neq \pi_2 \neq \pi_3} w_{\Pi_e}(\pi_1) w_{\Pi_e}(\pi_2) w_{\Pi_e}(\pi_3) PG_{2,4}(\pi_1,\pi_2,\pi_3,\Pi_o,w_{\dot{L}},\mu) =\\
												& = 6 \frac{9}{2} \frac{1}{3} \frac{1}{3} \frac{1}{3} PG_{2,4}(\pi_x,\pi_y,\pi_z,\Pi_o,w_{\dot{L}},\mu)
\end{aligned}
\end{equation*}

In this case, we only need to calculate $PG_{2,4}(\pi_x,\pi_y,\pi_z,\Pi_o,w_{\dot{L}},\mu)$. We follow definition \ref{def:STG_agents} (for PG) to calculate this value:

\begin{equation*}
\begin{aligned}
PG_{2,4}(\pi_x,\pi_y,\pi_z,\Pi_o,w_{\dot{L}},\mu)	& = \sum_{\dot{l} \in \dot{L}^{N(\mu)}_{-2,4}(\Pi_o)} w_{\dot{L}}(\dot{l}) PO_{2,4}(\pi_x,\pi_y,\pi_z,\dot{l},\mu) =\\
													& = PO_{2,4}(\pi_x,\pi_y,\pi_z,(\pi_s,*,\pi_s,*),\mu)
\end{aligned}
\end{equation*}

The following table shows us $PO_{2,4}$ for all the permutations of $\pi_x,\pi_y,\pi_z$.

\begin{center}
\begin{tabular}{c c c | c c c | c c c}
Slot 2 & & Slot 4			& Slot 2 & & Slot 4				& Slot 2 & & Slot 4\\
\hline
$\pi_x$ & $\leq$ & $\pi_y$	& $\pi_x$ & $\leq$ & $\pi_z$	& $\pi_y$ & $\leq$ & $\pi_x$\\
$\pi_y$ & $\leq$ & $\pi_z$	& $\pi_z$ & $\leq$ & $\pi_y$	& $\pi_x$ & $\leq$ & $\pi_z$\\
$\pi_x$ & $\leq$ & $\pi_z$	& $\pi_x$ & $\leq$ & $\pi_y$	& $\pi_y$ & $\leq$ & $\pi_z$
\end{tabular}
\begin{tabular}{c c c | c c c | c c c}
Slot 2 & & Slot 4			& Slot 2 & & Slot 4				& Slot 2 & & Slot 4\\
\hline
$\pi_y$ & $\leq$ & $\pi_z$	& $\pi_z$ & $\leq$ & $\pi_x$	& $\pi_z$ & $\leq$ & $\pi_y$\\
$\pi_z$ & $\leq$ & $\pi_x$	& $\pi_x$ & $\leq$ & $\pi_y$	& $\pi_y$ & $\leq$ & $\pi_x$\\
$\pi_y$ & $\leq$ & $\pi_x$	& $\pi_z$ & $\leq$ & $\pi_y$	& $\pi_z$ & $\leq$ & $\pi_x$
\end{tabular}
\end{center}

It is possible to find a PO for every permutation, since the agents in slots 2 and 4 share rewards (and expected average rewards as well) due they are in the same team. So:

\begin{equation*}
PG_{2,4}(\pi_x,\pi_y,\pi_z,\Pi_o,w_{\dot{L}},\mu) = 1
\end{equation*}

Therefore:

\begin{equation*}
PG_{2,4}(\Pi_e,w_{\Pi_e},\Pi_o,w_{\dot{L}},\mu) = 6 \frac{9}{2} \frac{1}{3} \frac{1}{3} \frac{1}{3} 1 = 1
\end{equation*}

For slots 3 and 1:

\begin{equation*}
\begin{aligned}
PG_{3,1}(\Pi_e,w_{\Pi_e},\Pi_o,w_{\dot{L}},\mu)	& = \eta_{\Pi^3} \sum_{\pi_1,\pi_2,\pi_3 \in \Pi_e | \pi_1 \neq \pi_2 \neq \pi_3} w_{\Pi_e}(\pi_1) w_{\Pi_e}(\pi_2) w_{\Pi_e}(\pi_3) PG_{3,1}(\pi_1,\pi_2,\pi_3,\Pi_o,w_{\dot{L}},\mu) =\\
												& = 6 \frac{9}{2} \frac{1}{3} \frac{1}{3} \frac{1}{3} PG_{3,1}(\pi_x,\pi_y,\pi_z,\Pi_o,w_{\dot{L}},\mu)
\end{aligned}
\end{equation*}

In this case, we only need to calculate $PG_{3,1}(\pi_x,\pi_y,\pi_z,\Pi_o,w_{\dot{L}},\mu)$. We follow definition \ref{def:STG_agents} (for PG) to calculate this value:

\begin{equation*}
\begin{aligned}
PG_{3,1}(\pi_x,\pi_y,\pi_z,\Pi_o,w_{\dot{L}},\mu)	& = \sum_{\dot{l} \in \dot{L}^{N(\mu)}_{-3,1}(\Pi_o)} w_{\dot{L}}(\dot{l}) PO_{3,1}(\pi_x,\pi_y,\pi_z,\dot{l},\mu) =\\
													& = PO_{3,1}(\pi_x,\pi_y,\pi_z,(*,\pi_s,*,\pi_s),\mu)
\end{aligned}
\end{equation*}

The following table shows us $PO_{3,1}$ for all the permutations of $\pi_x,\pi_y,\pi_z$.

\begin{center}
\begin{tabular}{c c c | c c c | c c c}
Slot 3 & & Slot 1			& Slot 3 & & Slot 1				& Slot 3 & & Slot 1\\
\hline
$\pi_x$ & $\leq$ & $\pi_y$	& $\pi_x$ & $\leq$ & $\pi_z$	& $\pi_y$ & $\leq$ & $\pi_x$\\
$\pi_y$ & $\leq$ & $\pi_z$	& $\pi_z$ & $\leq$ & $\pi_y$	& $\pi_x$ & $\leq$ & $\pi_z$\\
$\pi_x$ & $\leq$ & $\pi_z$	& $\pi_x$ & $\leq$ & $\pi_y$	& $\pi_y$ & $\leq$ & $\pi_z$
\end{tabular}
\begin{tabular}{c c c | c c c | c c c}
Slot 3 & & Slot 1			& Slot 3 & & Slot 1				& Slot 3 & & Slot 1\\
\hline
$\pi_y$ & $\leq$ & $\pi_z$	& $\pi_z$ & $\leq$ & $\pi_x$	& $\pi_z$ & $\leq$ & $\pi_y$\\
$\pi_z$ & $\leq$ & $\pi_x$	& $\pi_x$ & $\leq$ & $\pi_y$	& $\pi_y$ & $\leq$ & $\pi_x$\\
$\pi_y$ & $\leq$ & $\pi_x$	& $\pi_z$ & $\leq$ & $\pi_y$	& $\pi_z$ & $\leq$ & $\pi_x$
\end{tabular}
\end{center}

But, it is not possible to find a PO for any permutation, since $\pi_x > \pi_z$, $\pi_y > \pi_x$ and $\pi_z > \pi_y$. So:

\begin{equation*}
PG_{3,1}(\pi_x,\pi_y,\pi_z,\Pi_o,w_{\dot{L}},\mu) = 0
\end{equation*}

Therefore:

\begin{equation*}
PG_{3,1}(\Pi_e,w_{\Pi_e},\Pi_o,w_{\dot{L}},\mu) = 6 \frac{9}{2} \frac{1}{3} \frac{1}{3} \frac{1}{3} 0 = 0
\end{equation*}

For slots 3 and 2:

\begin{equation*}
\begin{aligned}
PG_{3,2}(\Pi_e,w_{\Pi_e},\Pi_o,w_{\dot{L}},\mu)	& = \eta_{\Pi^3} \sum_{\pi_1,\pi_2,\pi_3 \in \Pi_e | \pi_1 \neq \pi_2 \neq \pi_3} w_{\Pi_e}(\pi_1) w_{\Pi_e}(\pi_2) w_{\Pi_e}(\pi_3) PG_{3,2}(\pi_1,\pi_2,\pi_3,\Pi_o,w_{\dot{L}},\mu) =\\
												& = 6 \frac{9}{2} \frac{1}{3} \frac{1}{3} \frac{1}{3} PG_{3,2}(\pi_x,\pi_y,\pi_z,\Pi_o,w_{\dot{L}},\mu)
\end{aligned}
\end{equation*}

In this case, we only need to calculate $PG_{3,2}(\pi_x,\pi_y,\pi_z,\Pi_o,w_{\dot{L}},\mu)$. We follow definition \ref{def:STG_agents} (for PG) to calculate this value:

\begin{equation*}
\begin{aligned}
PG_{3,2}(\pi_x,\pi_y,\pi_z,\Pi_o,w_{\dot{L}},\mu)	& = \sum_{\dot{l} \in \dot{L}^{N(\mu)}_{-3,2}(\Pi_o)} w_{\dot{L}}(\dot{l}) PO_{3,2}(\pi_x,\pi_y,\pi_z,\dot{l},\mu) =\\
													& = PO_{3,2}(\pi_x,\pi_y,\pi_z,(\pi_s,*,*,\pi_s),\mu)
\end{aligned}
\end{equation*}

The following table shows us $PO_{3,2}$ for all the permutations of $\pi_x,\pi_y,\pi_z$.

\begin{center}
\begin{tabular}{c c c | c c c | c c c}
Slot 3 & & Slot 2			& Slot 3 & & Slot 2				& Slot 3 & & Slot 2\\
\hline
$\pi_x$ & $\leq$ & $\pi_y$	& $\pi_x$ & $\leq$ & $\pi_z$	& $\pi_y$ & $\leq$ & $\pi_x$\\
$\pi_y$ & $\leq$ & $\pi_z$	& $\pi_z$ & $\leq$ & $\pi_y$	& $\pi_x$ & $\leq$ & $\pi_z$\\
$\pi_x$ & $\leq$ & $\pi_z$	& $\pi_x$ & $\leq$ & $\pi_y$	& $\pi_y$ & $\leq$ & $\pi_z$
\end{tabular}
\begin{tabular}{c c c | c c c | c c c}
Slot 3 & & Slot 2			& Slot 3 & & Slot 2				& Slot 3 & & Slot 2\\
\hline
$\pi_y$ & $\leq$ & $\pi_z$	& $\pi_z$ & $\leq$ & $\pi_x$	& $\pi_z$ & $\leq$ & $\pi_y$\\
$\pi_z$ & $\leq$ & $\pi_x$	& $\pi_x$ & $\leq$ & $\pi_y$	& $\pi_y$ & $\leq$ & $\pi_x$\\
$\pi_y$ & $\leq$ & $\pi_x$	& $\pi_z$ & $\leq$ & $\pi_y$	& $\pi_z$ & $\leq$ & $\pi_x$
\end{tabular}
\end{center}

It is possible to find a PO for every permutation, since the agents in slots 3 and 2 share rewards (and expected average rewards as well) due they are in the same team. So:

\begin{equation*}
PG_{3,2}(\pi_x,\pi_y,\pi_z,\Pi_o,w_{\dot{L}},\mu) = 1
\end{equation*}

Therefore:

\begin{equation*}
PG_{3,2}(\Pi_e,w_{\Pi_e},\Pi_o,w_{\dot{L}},\mu) = 6 \frac{9}{2} \frac{1}{3} \frac{1}{3} \frac{1}{3} 1 = 1
\end{equation*}

For slots 3 and 4:

\begin{equation*}
\begin{aligned}
PG_{3,4}(\Pi_e,w_{\Pi_e},\Pi_o,w_{\dot{L}},\mu)	& = \eta_{\Pi^3} \sum_{\pi_1,\pi_2,\pi_3 \in \Pi_e | \pi_1 \neq \pi_2 \neq \pi_3} w_{\Pi_e}(\pi_1) w_{\Pi_e}(\pi_2) w_{\Pi_e}(\pi_3) PG_{3,4}(\pi_1,\pi_2,\pi_3,\Pi_o,w_{\dot{L}},\mu) =\\
												& = 6 \frac{9}{2} \frac{1}{3} \frac{1}{3} \frac{1}{3} PG_{3,4}(\pi_x,\pi_y,\pi_z,\Pi_o,w_{\dot{L}},\mu)
\end{aligned}
\end{equation*}

In this case, we only need to calculate $PG_{3,4}(\pi_x,\pi_y,\pi_z,\Pi_o,w_{\dot{L}},\mu)$. We follow definition \ref{def:STG_agents} (for PG) to calculate this value:

\begin{equation*}
\begin{aligned}
PG_{3,4}(\pi_x,\pi_y,\pi_z,\Pi_o,w_{\dot{L}},\mu)	& = \sum_{\dot{l} \in \dot{L}^{N(\mu)}_{-3,4}(\Pi_o)} w_{\dot{L}}(\dot{l}) PO_{3,4}(\pi_x,\pi_y,\pi_z,\dot{l},\mu) =\\
													& = PO_{3,4}(\pi_x,\pi_y,\pi_z,(\pi_s,\pi_s,*,*),\mu)
\end{aligned}
\end{equation*}

The following table shows us $PO_{3,4}$ for all the permutations of $\pi_x,\pi_y,\pi_z$.

\begin{center}
\begin{tabular}{c c c | c c c | c c c}
Slot 3 & & Slot 4			& Slot 3 & & Slot 4				& Slot 3 & & Slot 4\\
\hline
$\pi_x$ & $\leq$ & $\pi_y$	& $\pi_x$ & $\leq$ & $\pi_z$	& $\pi_y$ & $\leq$ & $\pi_x$\\
$\pi_y$ & $\leq$ & $\pi_z$	& $\pi_z$ & $\leq$ & $\pi_y$	& $\pi_x$ & $\leq$ & $\pi_z$\\
$\pi_x$ & $\leq$ & $\pi_z$	& $\pi_x$ & $\leq$ & $\pi_y$	& $\pi_y$ & $\leq$ & $\pi_z$
\end{tabular}
\begin{tabular}{c c c | c c c | c c c}
Slot 3 & & Slot 4			& Slot 3 & & Slot 4				& Slot 3 & & Slot 4\\
\hline
$\pi_y$ & $\leq$ & $\pi_z$	& $\pi_z$ & $\leq$ & $\pi_x$	& $\pi_z$ & $\leq$ & $\pi_y$\\
$\pi_z$ & $\leq$ & $\pi_x$	& $\pi_x$ & $\leq$ & $\pi_y$	& $\pi_y$ & $\leq$ & $\pi_x$\\
$\pi_y$ & $\leq$ & $\pi_x$	& $\pi_z$ & $\leq$ & $\pi_y$	& $\pi_z$ & $\leq$ & $\pi_x$
\end{tabular}
\end{center}

It is possible to find a PO for every permutation, since the agents in slots 3 and 4 share rewards (and expected average rewards as well) due they are in the same team. So:

\begin{equation*}
PG_{3,4}(\pi_x,\pi_y,\pi_z,\Pi_o,w_{\dot{L}},\mu) = 1
\end{equation*}

Therefore:

\begin{equation*}
PG_{3,4}(\Pi_e,w_{\Pi_e},\Pi_o,w_{\dot{L}},\mu) = 6 \frac{9}{2} \frac{1}{3} \frac{1}{3} \frac{1}{3} 1 = 1
\end{equation*}

For slots 4 and 1:

\begin{equation*}
\begin{aligned}
PG_{4,1}(\Pi_e,w_{\Pi_e},\Pi_o,w_{\dot{L}},\mu)	& = \eta_{\Pi^3} \sum_{\pi_1,\pi_2,\pi_3 \in \Pi_e | \pi_1 \neq \pi_2 \neq \pi_3} w_{\Pi_e}(\pi_1) w_{\Pi_e}(\pi_2) w_{\Pi_e}(\pi_3) PG_{4,1}(\pi_1,\pi_2,\pi_3,\Pi_o,w_{\dot{L}},\mu) =\\
												& = 6 \frac{9}{2} \frac{1}{3} \frac{1}{3} \frac{1}{3} PG_{4,1}(\pi_x,\pi_y,\pi_z,\Pi_o,w_{\dot{L}},\mu)
\end{aligned}
\end{equation*}

In this case, we only need to calculate $PG_{4,1}(\pi_x,\pi_y,\pi_z,\Pi_o,w_{\dot{L}},\mu)$. We follow definition \ref{def:STG_agents} (for PG) to calculate this value:

\begin{equation*}
\begin{aligned}
PG_{4,1}(\pi_x,\pi_y,\pi_z,\Pi_o,w_{\dot{L}},\mu)	& = \sum_{\dot{l} \in \dot{L}^{N(\mu)}_{-4,1}(\Pi_o)} w_{\dot{L}}(\dot{l}) PO_{4,1}(\pi_x,\pi_y,\pi_z,\dot{l},\mu) =\\
													& = PO_{4,1}(\pi_x,\pi_y,\pi_z,(*,\pi_s,\pi_s,*),\mu)
\end{aligned}
\end{equation*}

The following table shows us $PO_{4,1}$ for all the permutations of $\pi_x,\pi_y,\pi_z$.

\begin{center}
\begin{tabular}{c c c | c c c | c c c}
Slot 4 & & Slot 1			& Slot 4 & & Slot 1				& Slot 4 & & Slot 1\\
\hline
$\pi_x$ & $\leq$ & $\pi_y$	& $\pi_x$ & $\leq$ & $\pi_z$	& $\pi_y$ & $\leq$ & $\pi_x$\\
$\pi_y$ & $\leq$ & $\pi_z$	& $\pi_z$ & $\leq$ & $\pi_y$	& $\pi_x$ & $\leq$ & $\pi_z$\\
$\pi_x$ & $\leq$ & $\pi_z$	& $\pi_x$ & $\leq$ & $\pi_y$	& $\pi_y$ & $\leq$ & $\pi_z$
\end{tabular}
\begin{tabular}{c c c | c c c | c c c}
Slot 4 & & Slot 1			& Slot 4 & & Slot 1				& Slot 4 & & Slot 1\\
\hline
$\pi_y$ & $\leq$ & $\pi_z$	& $\pi_z$ & $\leq$ & $\pi_x$	& $\pi_z$ & $\leq$ & $\pi_y$\\
$\pi_z$ & $\leq$ & $\pi_x$	& $\pi_x$ & $\leq$ & $\pi_y$	& $\pi_y$ & $\leq$ & $\pi_x$\\
$\pi_y$ & $\leq$ & $\pi_x$	& $\pi_z$ & $\leq$ & $\pi_y$	& $\pi_z$ & $\leq$ & $\pi_x$
\end{tabular}
\end{center}

But, it is not possible to find a PO for any permutation, since $\pi_x > \pi_z$, $\pi_y > \pi_x$ and $\pi_z > \pi_y$. So:

\begin{equation*}
PG_{4,1}(\pi_x,\pi_y,\pi_z,\Pi_o,w_{\dot{L}},\mu) = 0
\end{equation*}

Therefore:

\begin{equation*}
PG_{4,1}(\Pi_e,w_{\Pi_e},\Pi_o,w_{\dot{L}},\mu) = 6 \frac{9}{2} \frac{1}{3} \frac{1}{3} \frac{1}{3} 0 = 0
\end{equation*}

For slots 4 and 2:

\begin{equation*}
\begin{aligned}
PG_{4,2}(\Pi_e,w_{\Pi_e},\Pi_o,w_{\dot{L}},\mu)	& = \eta_{\Pi^3} \sum_{\pi_1,\pi_2,\pi_3 \in \Pi_e | \pi_1 \neq \pi_2 \neq \pi_3} w_{\Pi_e}(\pi_1) w_{\Pi_e}(\pi_2) w_{\Pi_e}(\pi_3) PG_{4,2}(\pi_1,\pi_2,\pi_3,\Pi_o,w_{\dot{L}},\mu) =\\
												& = 6 \frac{9}{2} \frac{1}{3} \frac{1}{3} \frac{1}{3} PG_{4,2}(\pi_x,\pi_y,\pi_z,\Pi_o,w_{\dot{L}},\mu)
\end{aligned}
\end{equation*}

In this case, we only need to calculate $PG_{4,2}(\pi_x,\pi_y,\pi_z,\Pi_o,w_{\dot{L}},\mu)$. We follow definition \ref{def:STG_agents} (for PG) to calculate this value:

\begin{equation*}
\begin{aligned}
PG_{4,2}(\pi_x,\pi_y,\pi_z,\Pi_o,w_{\dot{L}},\mu)	& = \sum_{\dot{l} \in \dot{L}^{N(\mu)}_{-4,2}(\Pi_o)} w_{\dot{L}}(\dot{l}) PO_{4,2}(\pi_x,\pi_y,\pi_z,\dot{l},\mu) =\\
													& = PO_{4,2}(\pi_x,\pi_y,\pi_z,(\pi_s,*,\pi_s,*),\mu)
\end{aligned}
\end{equation*}

The following table shows us $PO_{4,2}$ for all the permutations of $\pi_x,\pi_y,\pi_z$.

\begin{center}
\begin{tabular}{c c c | c c c | c c c}
Slot 4 & & Slot 2			& Slot 4 & & Slot 2				& Slot 4 & & Slot 2\\
\hline
$\pi_x$ & $\leq$ & $\pi_y$	& $\pi_x$ & $\leq$ & $\pi_z$	& $\pi_y$ & $\leq$ & $\pi_x$\\
$\pi_y$ & $\leq$ & $\pi_z$	& $\pi_z$ & $\leq$ & $\pi_y$	& $\pi_x$ & $\leq$ & $\pi_z$\\
$\pi_x$ & $\leq$ & $\pi_z$	& $\pi_x$ & $\leq$ & $\pi_y$	& $\pi_y$ & $\leq$ & $\pi_z$
\end{tabular}
\begin{tabular}{c c c | c c c | c c c}
Slot 4 & & Slot 2			& Slot 4 & & Slot 2				& Slot 4 & & Slot 2\\
\hline
$\pi_y$ & $\leq$ & $\pi_z$	& $\pi_z$ & $\leq$ & $\pi_x$	& $\pi_z$ & $\leq$ & $\pi_y$\\
$\pi_z$ & $\leq$ & $\pi_x$	& $\pi_x$ & $\leq$ & $\pi_y$	& $\pi_y$ & $\leq$ & $\pi_x$\\
$\pi_y$ & $\leq$ & $\pi_x$	& $\pi_z$ & $\leq$ & $\pi_y$	& $\pi_z$ & $\leq$ & $\pi_x$
\end{tabular}
\end{center}

It is possible to find a PO for every permutation, since the agents in slots 4 and 2 share rewards (and expected average rewards as well) due they are in the same team. So:

\begin{equation*}
PG_{4,2}(\pi_x,\pi_y,\pi_z,\Pi_o,w_{\dot{L}},\mu) = 1
\end{equation*}

Therefore:

\begin{equation*}
PG_{4,2}(\Pi_e,w_{\Pi_e},\Pi_o,w_{\dot{L}},\mu) = 6 \frac{9}{2} \frac{1}{3} \frac{1}{3} \frac{1}{3} 1 = 1
\end{equation*}

And for slots 4 and 3:

\begin{equation*}
\begin{aligned}
PG_{4,3}(\Pi_e,w_{\Pi_e},\Pi_o,w_{\dot{L}},\mu)	& = \eta_{\Pi^3} \sum_{\pi_1,\pi_2,\pi_3 \in \Pi_e | \pi_1 \neq \pi_2 \neq \pi_3} w_{\Pi_e}(\pi_1) w_{\Pi_e}(\pi_2) w_{\Pi_e}(\pi_3) PG_{4,3}(\pi_1,\pi_2,\pi_3,\Pi_o,w_{\dot{L}},\mu) =\\
												& = 6 \frac{9}{2} \frac{1}{3} \frac{1}{3} \frac{1}{3} PG_{4,3}(\pi_x,\pi_y,\pi_z,\Pi_o,w_{\dot{L}},\mu)
\end{aligned}
\end{equation*}

In this case, we only need to calculate $PG_{4,3}(\pi_x,\pi_y,\pi_z,\Pi_o,w_{\dot{L}},\mu)$. We follow definition \ref{def:STG_agents} (for PG) to calculate this value:

\begin{equation*}
\begin{aligned}
PG_{4,3}(\pi_x,\pi_y,\pi_z,\Pi_o,w_{\dot{L}},\mu)	& = \sum_{\dot{l} \in \dot{L}^{N(\mu)}_{-4,3}(\Pi_o)} w_{\dot{L}}(\dot{l}) PO_{4,3}(\pi_x,\pi_y,\pi_z,\dot{l},\mu) =\\
													& = PO_{4,3}(\pi_x,\pi_y,\pi_z,(\pi_s,\pi_s,*,*),\mu)
\end{aligned}
\end{equation*}

The following table shows us $PO_{4,3}$ for all the permutations of $\pi_x,\pi_y,\pi_z$.

\begin{center}
\begin{tabular}{c c c | c c c | c c c}
Slot 4 & & Slot 3			& Slot 4 & & Slot 3				& Slot 4 & & Slot 3\\
\hline
$\pi_x$ & $\leq$ & $\pi_y$	& $\pi_x$ & $\leq$ & $\pi_z$	& $\pi_y$ & $\leq$ & $\pi_x$\\
$\pi_y$ & $\leq$ & $\pi_z$	& $\pi_z$ & $\leq$ & $\pi_y$	& $\pi_x$ & $\leq$ & $\pi_z$\\
$\pi_x$ & $\leq$ & $\pi_z$	& $\pi_x$ & $\leq$ & $\pi_y$	& $\pi_y$ & $\leq$ & $\pi_z$
\end{tabular}
\begin{tabular}{c c c | c c c | c c c}
Slot 4 & & Slot 3			& Slot 4 & & Slot 3				& Slot 4 & & Slot 3\\
\hline
$\pi_y$ & $\leq$ & $\pi_z$	& $\pi_z$ & $\leq$ & $\pi_x$	& $\pi_z$ & $\leq$ & $\pi_y$\\
$\pi_z$ & $\leq$ & $\pi_x$	& $\pi_x$ & $\leq$ & $\pi_y$	& $\pi_y$ & $\leq$ & $\pi_x$\\
$\pi_y$ & $\leq$ & $\pi_x$	& $\pi_z$ & $\leq$ & $\pi_y$	& $\pi_z$ & $\leq$ & $\pi_x$
\end{tabular}
\end{center}

It is possible to find a PO for every permutation, since the agents in slots 4 and 3 share rewards (and expected average rewards as well) due they are in the same team. So:

\begin{equation*}
PG_{4,3}(\pi_x,\pi_y,\pi_z,\Pi_o,w_{\dot{L}},\mu) = 1
\end{equation*}

Therefore:

\begin{equation*}
PG_{4,3}(\Pi_e,w_{\Pi_e},\Pi_o,w_{\dot{L}},\mu) = 6 \frac{9}{2} \frac{1}{3} \frac{1}{3} \frac{1}{3} 1 = 1
\end{equation*}

And finally, we weight over the slots:

\begin{equation*}
\begin{aligned}
& PG(\Pi_e,w_{\Pi_e},\Pi_o,w_{\dot{L}},\mu,w_S) = \eta_{S_1^2} \sum_{i=1}^{N(\mu)} w_S(i,\mu) \times\\
& \times \left(\sum_{j=1}^{i-1} w_S(j,\mu) PG_{i,j}(\Pi_e,w_{\Pi_e},\Pi_o,w_{\dot{L}},\mu) + \sum_{j=i+1}^{N(\mu)} w_S(j,\mu) PG_{i,j}(\Pi_e,w_{\Pi_e},\Pi_o,w_{\dot{L}},\mu)\right) =\\
& \ \ \ \ \ \ \ \ \ \ \ \ \ \ \ \ \ \ \ \ \ \ \ \ \ \ \ \ \ \ \ \ \ \ \ \ = \frac{4}{3} \frac{1}{4} \frac{1}{4} \{PG_{1,2}(\Pi_e,w_{\Pi_e},\Pi_o,w_{\dot{L}},\mu) + PG_{1,3}(\Pi_e,w_{\Pi_e},\Pi_o,w_{\dot{L}},\mu) +\\
& \ \ \ \ \ \ \ \ \ \ \ \ \ \ \ \ \ \ \ \ \ \ \ \ \ \ \ \ \ \ \ \ \ \ \ \ + PG_{1,4}(\Pi_e,w_{\Pi_e},\Pi_o,w_{\dot{L}},\mu) + PG_{2,1}(\Pi_e,w_{\Pi_e},\Pi_o,w_{\dot{L}},\mu) +\\
& \ \ \ \ \ \ \ \ \ \ \ \ \ \ \ \ \ \ \ \ \ \ \ \ \ \ \ \ \ \ \ \ \ \ \ \ + PG_{2,3}(\Pi_e,w_{\Pi_e},\Pi_o,w_{\dot{L}},\mu) + PG_{2,4}(\Pi_e,w_{\Pi_e},\Pi_o,w_{\dot{L}},\mu) +\\
& \ \ \ \ \ \ \ \ \ \ \ \ \ \ \ \ \ \ \ \ \ \ \ \ \ \ \ \ \ \ \ \ \ \ \ \ + PG_{3,1}(\Pi_e,w_{\Pi_e},\Pi_o,w_{\dot{L}},\mu) + PG_{3,2}(\Pi_e,w_{\Pi_e},\Pi_o,w_{\dot{L}},\mu) +\\
& \ \ \ \ \ \ \ \ \ \ \ \ \ \ \ \ \ \ \ \ \ \ \ \ \ \ \ \ \ \ \ \ \ \ \ \ + PG_{3,4}(\Pi_e,w_{\Pi_e},\Pi_o,w_{\dot{L}},\mu) + PG_{4,1}(\Pi_e,w_{\Pi_e},\Pi_o,w_{\dot{L}},\mu) +\\
& \ \ \ \ \ \ \ \ \ \ \ \ \ \ \ \ \ \ \ \ \ \ \ \ \ \ \ \ \ \ \ \ \ \ \ \ + PG_{4,2}(\Pi_e,w_{\Pi_e},\Pi_o,w_{\dot{L}},\mu) + PG_{4,3}(\Pi_e,w_{\Pi_e},\Pi_o,w_{\dot{L}},\mu)\} =\\
& \ \ \ \ \ \ \ \ \ \ \ \ \ \ \ \ \ \ \ \ \ \ \ \ \ \ \ \ \ \ \ \ \ \ \ \ = \frac{4}{3} \frac{1}{4} \frac{1}{4} \left\{6 \times 1 + 6 \times 0\right\} = \frac{1}{2}
\end{aligned}
\end{equation*}

Since $\frac{1}{2}$ is the lowest possible value that we can obtain for the partial grading property, therefore predator-prey has $General_{min} = \frac{1}{2}$ for this property.
\end{proof}
\end{proposition}

\begin{proposition}
\label{prop:predator-prey_PG_general_max}
$General_{max}$ for the partial grading (PG) property is equal to $1$ for the predator-prey environment.

\begin{proof}
To find $General_{max}$ (equation \ref{eq:general_max}), we need to find a trio $\left\langle\Pi_e,w_{\Pi_e},\Pi_o\right\rangle$ which maximises the property as much as possible. We can have this situation by selecting $\Pi_e = \{{\pi_s}_1,{\pi_s}_2,{\pi_s}_3\}$ with uniform weight for $w_{\Pi_e}$ and $\Pi_o = \{\pi_x\}$ (a $\pi_s$ agent always stays in the same cell\footnote{Note that every cell has an action which is blocked by a block or a boundary, therefore an agent performing this action will stay at its current cell.}, and a $\pi_x$ agent acts stochastically with a probability of $1/\sqrt{2}$ to do not reach the upper left corner and a probability of $1 - 1/\sqrt{2}$ to reach this corner).

Following definition \ref{def:PG}, we obtain the PG value for this $\left\langle\Pi_e,w_{\Pi_e},\Pi_o\right\rangle$. Since the environment is not symmetric, we need to calculate this property for every pair of slots. Following definition \ref{def:STG_set} (for PG), we can calculate its PG value for each pair of slots. We start with slots 1 and 2:

\begin{equation*}
\begin{aligned}
PG_{1,2}(\Pi_e,w_{\Pi_e},\Pi_o,w_{\dot{L}},\mu)	& = \eta_{\Pi^3} \sum_{\pi_1,\pi_2,\pi_3 \in \Pi_e | \pi_1 \neq \pi_2 \neq \pi_3} w_{\Pi_e}(\pi_1) w_{\Pi_e}(\pi_2) w_{\Pi_e}(\pi_3) PG_{1,2}(\pi_1,\pi_2,\pi_3,\Pi_o,w_{\dot{L}},\mu) =\\
												& = 6 \frac{9}{2} \frac{1}{3} \frac{1}{3} \frac{1}{3} PG_{1,2}({\pi_s}_1,{\pi_s}_2,{\pi_s}_3,\Pi_o,w_{\dot{L}},\mu)
\end{aligned}
\end{equation*}

\noindent Note that we avoided to calculate all the permutations of $\pi_1,\pi_2,\pi_3$ for $PG_{i,j}(\pi_1,\pi_2,\pi_3,\Pi_o,w_{\dot{L}},\mu)$ since they provide the same result, by calculating only one permutation and multiplying the result by the number of permutations $6$.

In this case, we only need to calculate $PG_{1,2}({\pi_s}_1,{\pi_s}_2,{\pi_s}_3,\Pi_o,w_{\dot{L}},\mu)$. We follow definition \ref{def:STG_agents} (for PG) to calculate this value:

\begin{equation*}
\begin{aligned}
PG_{1,2}({\pi_s}_1,{\pi_s}_2,{\pi_s}_3,\Pi_o,w_{\dot{L}},\mu)	& = \sum_{\dot{l} \in \dot{L}^{N(\mu)}_{-1,2}(\Pi_o)} w_{\dot{L}}(\dot{l}) PO_{1,2}({\pi_s}_1,{\pi_s}_2,{\pi_s}_3,\dot{l},\mu) =\\
																& = PO_{1,2}({\pi_s}_1,{\pi_s}_2,{\pi_s}_3,(*,*,\pi_x,\pi_x),\mu)
\end{aligned}
\end{equation*}

The following table shows us $PO_{1,2}$ for all the permutations of ${\pi_s}_1,{\pi_s}_2,{\pi_s}_3$.

\begin{center}
\begin{tabular}{c c c | c c c | c c c}
Slot 1 & & Slot 2					& Slot 1 & & Slot 2						& Slot 1 & & Slot 2\\
\hline
${\pi_s}_1$ & $\leq$ & ${\pi_s}_2$	& ${\pi_s}_1$ & $\leq$ & ${\pi_s}_3$	& ${\pi_s}_2$ & $\leq$ & ${\pi_s}_1$\\
${\pi_s}_2$ & $\leq$ & ${\pi_s}_3$	& ${\pi_s}_3$ & $\leq$ & ${\pi_s}_2$	& ${\pi_s}_1$ & $\leq$ & ${\pi_s}_3$\\
${\pi_s}_1$ & $\leq$ & ${\pi_s}_3$	& ${\pi_s}_1$ & $\leq$ & ${\pi_s}_2$	& ${\pi_s}_2$ & $\leq$ & ${\pi_s}_3$
\end{tabular}
\begin{tabular}{c c c | c c c | c c c}
Slot 1 & & Slot 2					& Slot 1 & & Slot 2						& Slot 1 & & Slot 2\\
\hline
${\pi_s}_2$ & $\leq$ & ${\pi_s}_3$	& ${\pi_s}_3$ & $\leq$ & ${\pi_s}_1$	& ${\pi_s}_3$ & $\leq$ & ${\pi_s}_2$\\
${\pi_s}_3$ & $\leq$ & ${\pi_s}_1$	& ${\pi_s}_1$ & $\leq$ & ${\pi_s}_2$	& ${\pi_s}_2$ & $\leq$ & ${\pi_s}_1$\\
${\pi_s}_2$ & $\leq$ & ${\pi_s}_1$	& ${\pi_s}_3$ & $\leq$ & ${\pi_s}_2$	& ${\pi_s}_3$ & $\leq$ & ${\pi_s}_1$
\end{tabular}
\end{center}

It is possible to find a PO for every permutation, since the agents in slots 3 ($\pi_x$) and 4 ($\pi_x$) have a probability of $(1/\sqrt{2}) \times (1/\sqrt{2}) = 1/2$ to do not chase the prey (any $\pi_s$) and the same probability $1 - 1/2 = 1/2$ to chase the prey, making for both agents in slots 1 (any $\pi_s$) and 2 (any $\pi_s$) to obtain the same expected average reward ($0$). So:

\begin{equation*}
PG_{1,2}({\pi_s}_1,{\pi_s}_2,{\pi_s}_3,\Pi_o,w_{\dot{L}},\mu) = 1
\end{equation*}

Therefore:

\begin{equation*}
PG_{1,2}(\Pi_e,w_{\Pi_e},\Pi_o,w_{\dot{L}},\mu) = 6 \frac{9}{2} \frac{1}{3} \frac{1}{3} \frac{1}{3} 1 = 1
\end{equation*}

For slots 1 and 3:

\begin{equation*}
\begin{aligned}
PG_{1,3}(\Pi_e,w_{\Pi_e},\Pi_o,w_{\dot{L}},\mu)	& = \eta_{\Pi^3} \sum_{\pi_1,\pi_2,\pi_3 \in \Pi_e | \pi_1 \neq \pi_2 \neq \pi_3} w_{\Pi_e}(\pi_1) w_{\Pi_e}(\pi_2) w_{\Pi_e}(\pi_3) PG_{1,3}(\pi_1,\pi_2,\pi_3,\Pi_o,w_{\dot{L}},\mu) =\\
												& = 6 \frac{9}{2} \frac{1}{3} \frac{1}{3} \frac{1}{3} PG_{1,3}({\pi_s}_1,{\pi_s}_2,{\pi_s}_3,\Pi_o,w_{\dot{L}},\mu)
\end{aligned}
\end{equation*}

In this case, we only need to calculate $PG_{1,3}({\pi_s}_1,{\pi_s}_2,{\pi_s}_3,\Pi_o,w_{\dot{L}},\mu)$. We follow definition \ref{def:STG_agents} (for PG) to calculate this value:

\begin{equation*}
\begin{aligned}
PG_{1,3}({\pi_s}_1,{\pi_s}_2,{\pi_s}_3,\Pi_o,w_{\dot{L}},\mu)	& = \sum_{\dot{l} \in \dot{L}^{N(\mu)}_{-1,3}(\Pi_o)} w_{\dot{L}}(\dot{l}) PO_{1,3}({\pi_s}_1,{\pi_s}_2,{\pi_s}_3,\dot{l},\mu) =\\
																& = PO_{1,3}({\pi_s}_1,{\pi_s}_2,{\pi_s}_3,(*,\pi_x,*,\pi_x),\mu)
\end{aligned}
\end{equation*}

The following table shows us $PO_{1,3}$ for all the permutations of ${\pi_s}_1,{\pi_s}_2,{\pi_s}_3$.

\begin{center}
\begin{tabular}{c c c | c c c | c c c}
Slot 1 & & Slot 3					& Slot 1 & & Slot 3						& Slot 1 & & Slot 3\\
\hline
${\pi_s}_1$ & $\leq$ & ${\pi_s}_2$	& ${\pi_s}_1$ & $\leq$ & ${\pi_s}_3$	& ${\pi_s}_2$ & $\leq$ & ${\pi_s}_1$\\
${\pi_s}_2$ & $\leq$ & ${\pi_s}_3$	& ${\pi_s}_3$ & $\leq$ & ${\pi_s}_2$	& ${\pi_s}_1$ & $\leq$ & ${\pi_s}_3$\\
${\pi_s}_1$ & $\leq$ & ${\pi_s}_3$	& ${\pi_s}_1$ & $\leq$ & ${\pi_s}_2$	& ${\pi_s}_2$ & $\leq$ & ${\pi_s}_3$
\end{tabular}
\begin{tabular}{c c c | c c c | c c c}
Slot 1 & & Slot 3					& Slot 1 & & Slot 3						& Slot 1 & & Slot 3\\
\hline
${\pi_s}_2$ & $\leq$ & ${\pi_s}_3$	& ${\pi_s}_3$ & $\leq$ & ${\pi_s}_1$	& ${\pi_s}_3$ & $\leq$ & ${\pi_s}_2$\\
${\pi_s}_3$ & $\leq$ & ${\pi_s}_1$	& ${\pi_s}_1$ & $\leq$ & ${\pi_s}_2$	& ${\pi_s}_2$ & $\leq$ & ${\pi_s}_1$\\
${\pi_s}_2$ & $\leq$ & ${\pi_s}_1$	& ${\pi_s}_3$ & $\leq$ & ${\pi_s}_2$	& ${\pi_s}_3$ & $\leq$ & ${\pi_s}_1$
\end{tabular}
\end{center}

It is possible to find a PO for every permutation, since the agents in slots 2 ($\pi_x$) and 4 ($\pi_x$) have a probability of $(1/\sqrt{2}) \times (1/\sqrt{2}) = 1/2$ to do not chase the prey (any $\pi_s$) and the same probability $1 - 1/2 = 1/2$ to chase the prey, making for both agents in slots 1 (any $\pi_s$) and 3 (any $\pi_s$) to obtain the same expected average reward ($0$). So:

\begin{equation*}
PG_{1,3}({\pi_s}_1,{\pi_s}_2,{\pi_s}_3,\Pi_o,w_{\dot{L}},\mu) = 1
\end{equation*}

Therefore:

\begin{equation*}
PG_{1,3}(\Pi_e,w_{\Pi_e},\Pi_o,w_{\dot{L}},\mu) = 6 \frac{9}{2} \frac{1}{3} \frac{1}{3} \frac{1}{3} 1 = 1
\end{equation*}

For slots 1 and 4:

\begin{equation*}
\begin{aligned}
PG_{1,4}(\Pi_e,w_{\Pi_e},\Pi_o,w_{\dot{L}},\mu)	& = \eta_{\Pi^3} \sum_{\pi_1,\pi_2,\pi_3 \in \Pi_e | \pi_1 \neq \pi_2 \neq \pi_3} w_{\Pi_e}(\pi_1) w_{\Pi_e}(\pi_2) w_{\Pi_e}(\pi_3) PG_{1,4}(\pi_1,\pi_2,\pi_3,\Pi_o,w_{\dot{L}},\mu) =\\
												& = 6 \frac{9}{2} \frac{1}{3} \frac{1}{3} \frac{1}{3} PG_{1,4}({\pi_s}_1,{\pi_s}_2,{\pi_s}_3,\Pi_o,w_{\dot{L}},\mu)
\end{aligned}
\end{equation*}

In this case, we only need to calculate $PG_{1,4}({\pi_s}_1,{\pi_s}_2,{\pi_s}_3,\Pi_o,w_{\dot{L}},\mu)$. We follow definition \ref{def:STG_agents} (for PG) to calculate this value:

\begin{equation*}
\begin{aligned}
PG_{1,4}({\pi_s}_1,{\pi_s}_2,{\pi_s}_3,\Pi_o,w_{\dot{L}},\mu)	& = \sum_{\dot{l} \in \dot{L}^{N(\mu)}_{-1,4}(\Pi_o)} w_{\dot{L}}(\dot{l}) PO_{1,4}({\pi_s}_1,{\pi_s}_2,{\pi_s}_3,\dot{l},\mu) =\\
																& = PO_{1,4}({\pi_s}_1,{\pi_s}_2,{\pi_s}_3,(*,\pi_x,\pi_x,*),\mu)
\end{aligned}
\end{equation*}

The following table shows us $PO_{1,4}$ for all the permutations of ${\pi_s}_1,{\pi_s}_2,{\pi_s}_3$.

\begin{center}
\begin{tabular}{c c c | c c c | c c c}
Slot 1 & & Slot 4					& Slot 1 & & Slot 4						& Slot 1 & & Slot 4\\
\hline
${\pi_s}_1$ & $\leq$ & ${\pi_s}_2$	& ${\pi_s}_1$ & $\leq$ & ${\pi_s}_3$	& ${\pi_s}_2$ & $\leq$ & ${\pi_s}_1$\\
${\pi_s}_2$ & $\leq$ & ${\pi_s}_3$	& ${\pi_s}_3$ & $\leq$ & ${\pi_s}_2$	& ${\pi_s}_1$ & $\leq$ & ${\pi_s}_3$\\
${\pi_s}_1$ & $\leq$ & ${\pi_s}_3$	& ${\pi_s}_1$ & $\leq$ & ${\pi_s}_2$	& ${\pi_s}_2$ & $\leq$ & ${\pi_s}_3$
\end{tabular}
\begin{tabular}{c c c | c c c | c c c}
Slot 1 & & Slot 4					& Slot 1 & & Slot 4						& Slot 1 & & Slot 4\\
\hline
${\pi_s}_2$ & $\leq$ & ${\pi_s}_3$	& ${\pi_s}_3$ & $\leq$ & ${\pi_s}_1$	& ${\pi_s}_3$ & $\leq$ & ${\pi_s}_2$\\
${\pi_s}_3$ & $\leq$ & ${\pi_s}_1$	& ${\pi_s}_1$ & $\leq$ & ${\pi_s}_2$	& ${\pi_s}_2$ & $\leq$ & ${\pi_s}_1$\\
${\pi_s}_2$ & $\leq$ & ${\pi_s}_1$	& ${\pi_s}_3$ & $\leq$ & ${\pi_s}_2$	& ${\pi_s}_3$ & $\leq$ & ${\pi_s}_1$
\end{tabular}
\end{center}

It is possible to find a PO for every permutation, since the agents in slots 2 ($\pi_x$) and 3 ($\pi_x$) have a probability of $(1/\sqrt{2}) \times (1/\sqrt{2}) = 1/2$ to do not chase the prey (any $\pi_s$) and the same probability $1 - 1/2 = 1/2$ to chase the prey, making for both agents in slots 1 (any $\pi_s$) and 4 (any $\pi_s$) to obtain the same expected average reward ($0$). So:

\begin{equation*}
PG_{1,4}({\pi_s}_1,{\pi_s}_2,{\pi_s}_3,\Pi_o,w_{\dot{L}},\mu) = 1
\end{equation*}

Therefore:

\begin{equation*}
PG_{1,4}(\Pi_e,w_{\Pi_e},\Pi_o,w_{\dot{L}},\mu) = 6 \frac{9}{2} \frac{1}{3} \frac{1}{3} \frac{1}{3} 1 = 1
\end{equation*}

For slots 2 and 1:

\begin{equation*}
\begin{aligned}
PG_{2,1}(\Pi_e,w_{\Pi_e},\Pi_o,w_{\dot{L}},\mu)	& = \eta_{\Pi^3} \sum_{\pi_1,\pi_2,\pi_3 \in \Pi_e | \pi_1 \neq \pi_2 \neq \pi_3} w_{\Pi_e}(\pi_1) w_{\Pi_e}(\pi_2) w_{\Pi_e}(\pi_3) PG_{2,1}(\pi_1,\pi_2,\pi_3,\Pi_o,w_{\dot{L}},\mu) =\\
												& = 6 \frac{9}{2} \frac{1}{3} \frac{1}{3} \frac{1}{3} PG_{2,1}({\pi_s}_1,{\pi_s}_2,{\pi_s}_3,\Pi_o,w_{\dot{L}},\mu)
\end{aligned}
\end{equation*}

In this case, we only need to calculate $PG_{2,1}({\pi_s}_1,{\pi_s}_2,{\pi_s}_3,\Pi_o,w_{\dot{L}},\mu)$. We follow definition \ref{def:STG_agents} (for PG) to calculate this value:

\begin{equation*}
\begin{aligned}
PG_{2,1}({\pi_s}_1,{\pi_s}_2,{\pi_s}_3,\Pi_o,w_{\dot{L}},\mu)	& = \sum_{\dot{l} \in \dot{L}^{N(\mu)}_{-2,1}(\Pi_o)} w_{\dot{L}}(\dot{l}) PO_{2,1}({\pi_s}_1,{\pi_s}_2,{\pi_s}_3,\dot{l},\mu) =\\
																& = PO_{2,1}({\pi_s}_1,{\pi_s}_2,{\pi_s}_3,(*,*,\pi_x,\pi_x),\mu)
\end{aligned}
\end{equation*}

The following table shows us $PO_{2,1}$ for all the permutations of ${\pi_s}_1,{\pi_s}_2,{\pi_s}_3$.

\begin{center}
\begin{tabular}{c c c | c c c | c c c}
Slot 2 & & Slot 1					& Slot 2 & & Slot 1						& Slot 2 & & Slot 1\\
\hline
${\pi_s}_1$ & $\leq$ & ${\pi_s}_2$	& ${\pi_s}_1$ & $\leq$ & ${\pi_s}_3$	& ${\pi_s}_2$ & $\leq$ & ${\pi_s}_1$\\
${\pi_s}_2$ & $\leq$ & ${\pi_s}_3$	& ${\pi_s}_3$ & $\leq$ & ${\pi_s}_2$	& ${\pi_s}_1$ & $\leq$ & ${\pi_s}_3$\\
${\pi_s}_1$ & $\leq$ & ${\pi_s}_3$	& ${\pi_s}_1$ & $\leq$ & ${\pi_s}_2$	& ${\pi_s}_2$ & $\leq$ & ${\pi_s}_3$
\end{tabular}
\begin{tabular}{c c c | c c c | c c c}
Slot 2 & & Slot 1					& Slot 2 & & Slot 1						& Slot 2 & & Slot 1\\
\hline
${\pi_s}_2$ & $\leq$ & ${\pi_s}_3$	& ${\pi_s}_3$ & $\leq$ & ${\pi_s}_1$	& ${\pi_s}_3$ & $\leq$ & ${\pi_s}_2$\\
${\pi_s}_3$ & $\leq$ & ${\pi_s}_1$	& ${\pi_s}_1$ & $\leq$ & ${\pi_s}_2$	& ${\pi_s}_2$ & $\leq$ & ${\pi_s}_1$\\
${\pi_s}_2$ & $\leq$ & ${\pi_s}_1$	& ${\pi_s}_3$ & $\leq$ & ${\pi_s}_2$	& ${\pi_s}_3$ & $\leq$ & ${\pi_s}_1$
\end{tabular}
\end{center}

It is possible to find a PO for every permutation, since the agents in slots 3 ($\pi_x$) and 4 ($\pi_x$) have a probability of $(1/\sqrt{2}) \times (1/\sqrt{2}) = 1/2$ to do not chase the prey (any $\pi_s$) and the same probability $1 - 1/2 = 1/2$ to chase the prey, making for both agents in slots 2 (any $\pi_s$) and 1 (any $\pi_s$) to obtain the same expected average reward ($0$). So:

\begin{equation*}
PG_{2,1}({\pi_s}_1,{\pi_s}_2,{\pi_s}_3,\Pi_o,w_{\dot{L}},\mu) = 1
\end{equation*}

Therefore:

\begin{equation*}
PG_{2,1}(\Pi_e,w_{\Pi_e},\Pi_o,w_{\dot{L}},\mu) = 6 \frac{9}{2} \frac{1}{3} \frac{1}{3} \frac{1}{3} 1 = 1
\end{equation*}

For slots 2 and 3:

\begin{equation*}
\begin{aligned}
PG_{2,3}(\Pi_e,w_{\Pi_e},\Pi_o,w_{\dot{L}},\mu)	& = \eta_{\Pi^3} \sum_{\pi_1,\pi_2,\pi_3 \in \Pi_e | \pi_1 \neq \pi_2 \neq \pi_3} w_{\Pi_e}(\pi_1) w_{\Pi_e}(\pi_2) w_{\Pi_e}(\pi_3) PG_{2,3}(\pi_1,\pi_2,\pi_3,\Pi_o,w_{\dot{L}},\mu) =\\
												& = 6 \frac{9}{2} \frac{1}{3} \frac{1}{3} \frac{1}{3} PG_{2,3}({\pi_s}_1,{\pi_s}_2,{\pi_s}_3,\Pi_o,w_{\dot{L}},\mu)
\end{aligned}
\end{equation*}

In this case, we only need to calculate $PG_{2,3}({\pi_s}_1,{\pi_s}_2,{\pi_s}_3,\Pi_o,w_{\dot{L}},\mu)$. We follow definition \ref{def:STG_agents} (for PG) to calculate this value:

\begin{equation*}
\begin{aligned}
PG_{2,3}({\pi_s}_1,{\pi_s}_2,{\pi_s}_3,\Pi_o,w_{\dot{L}},\mu)	& = \sum_{\dot{l} \in \dot{L}^{N(\mu)}_{-2,3}(\Pi_o)} w_{\dot{L}}(\dot{l}) PO_{2,3}({\pi_s}_1,{\pi_s}_2,{\pi_s}_3,\dot{l},\mu) =\\
																& = PO_{2,3}({\pi_s}_1,{\pi_s}_2,{\pi_s}_3,(\pi_x,*,*,\pi_x),\mu)
\end{aligned}
\end{equation*}

The following table shows us $PO_{2,3}$ for all the permutations of ${\pi_s}_1,{\pi_s}_2,{\pi_s}_3$.

\begin{center}
\begin{tabular}{c c c | c c c | c c c}
Slot 2 & & Slot 3					& Slot 2 & & Slot 3						& Slot 2 & & Slot 3\\
\hline
${\pi_s}_1$ & $\leq$ & ${\pi_s}_2$	& ${\pi_s}_1$ & $\leq$ & ${\pi_s}_3$	& ${\pi_s}_2$ & $\leq$ & ${\pi_s}_1$\\
${\pi_s}_2$ & $\leq$ & ${\pi_s}_3$	& ${\pi_s}_3$ & $\leq$ & ${\pi_s}_2$	& ${\pi_s}_1$ & $\leq$ & ${\pi_s}_3$\\
${\pi_s}_1$ & $\leq$ & ${\pi_s}_3$	& ${\pi_s}_1$ & $\leq$ & ${\pi_s}_2$	& ${\pi_s}_2$ & $\leq$ & ${\pi_s}_3$
\end{tabular}
\begin{tabular}{c c c | c c c | c c c}
Slot 2 & & Slot 3					& Slot 2 & & Slot 3						& Slot 2 & & Slot 3\\
\hline
${\pi_s}_2$ & $\leq$ & ${\pi_s}_3$	& ${\pi_s}_3$ & $\leq$ & ${\pi_s}_1$	& ${\pi_s}_3$ & $\leq$ & ${\pi_s}_2$\\
${\pi_s}_3$ & $\leq$ & ${\pi_s}_1$	& ${\pi_s}_1$ & $\leq$ & ${\pi_s}_2$	& ${\pi_s}_2$ & $\leq$ & ${\pi_s}_1$\\
${\pi_s}_2$ & $\leq$ & ${\pi_s}_1$	& ${\pi_s}_3$ & $\leq$ & ${\pi_s}_2$	& ${\pi_s}_3$ & $\leq$ & ${\pi_s}_1$
\end{tabular}
\end{center}

It is possible to find a PO for every permutation, since the agents in slots 2 and 3 share rewards (and expected average rewards as well) due they are in the same team. So:

\begin{equation*}
PG_{2,3}({\pi_s}_1,{\pi_s}_2,{\pi_s}_3,\Pi_o,w_{\dot{L}},\mu) = 1
\end{equation*}

Therefore:

\begin{equation*}
PG_{2,3}(\Pi_e,w_{\Pi_e},\Pi_o,w_{\dot{L}},\mu) = 6 \frac{9}{2} \frac{1}{3} \frac{1}{3} \frac{1}{3} 1 = 1
\end{equation*}

For slots 2 and 4:

\begin{equation*}
\begin{aligned}
PG_{2,4}(\Pi_e,w_{\Pi_e},\Pi_o,w_{\dot{L}},\mu)	& = \eta_{\Pi^3} \sum_{\pi_1,\pi_2,\pi_3 \in \Pi_e | \pi_1 \neq \pi_2 \neq \pi_3} w_{\Pi_e}(\pi_1) w_{\Pi_e}(\pi_2) w_{\Pi_e}(\pi_3) PG_{2,4}(\pi_1,\pi_2,\pi_3,\Pi_o,w_{\dot{L}},\mu) =\\
												& = 6 \frac{9}{2} \frac{1}{3} \frac{1}{3} \frac{1}{3} PG_{2,4}({\pi_s}_1,{\pi_s}_2,{\pi_s}_3,\Pi_o,w_{\dot{L}},\mu)
\end{aligned}
\end{equation*}

In this case, we only need to calculate $PG_{2,4}({\pi_s}_1,{\pi_s}_2,{\pi_s}_3,\Pi_o,w_{\dot{L}},\mu)$. We follow definition \ref{def:STG_agents} (for PG) to calculate this value:

\begin{equation*}
\begin{aligned}
PG_{2,4}({\pi_s}_1,{\pi_s}_2,{\pi_s}_3,\Pi_o,w_{\dot{L}},\mu)	& = \sum_{\dot{l} \in \dot{L}^{N(\mu)}_{-2,4}(\Pi_o)} w_{\dot{L}}(\dot{l}) PO_{2,4}({\pi_s}_1,{\pi_s}_2,{\pi_s}_3,\dot{l},\mu) =\\
																& = PO_{2,4}({\pi_s}_1,{\pi_s}_2,{\pi_s}_3,(\pi_x,*,\pi_x,*),\mu)
\end{aligned}
\end{equation*}

The following table shows us $PO_{2,4}$ for all the permutations of ${\pi_s}_1,{\pi_s}_2,{\pi_s}_3$.

\begin{center}
\begin{tabular}{c c c | c c c | c c c}
Slot 2 & & Slot 4					& Slot 2 & & Slot 4						& Slot 2 & & Slot 4\\
\hline
${\pi_s}_1$ & $\leq$ & ${\pi_s}_2$	& ${\pi_s}_1$ & $\leq$ & ${\pi_s}_3$	& ${\pi_s}_2$ & $\leq$ & ${\pi_s}_1$\\
${\pi_s}_2$ & $\leq$ & ${\pi_s}_3$	& ${\pi_s}_3$ & $\leq$ & ${\pi_s}_2$	& ${\pi_s}_1$ & $\leq$ & ${\pi_s}_3$\\
${\pi_s}_1$ & $\leq$ & ${\pi_s}_3$	& ${\pi_s}_1$ & $\leq$ & ${\pi_s}_2$	& ${\pi_s}_2$ & $\leq$ & ${\pi_s}_3$
\end{tabular}
\begin{tabular}{c c c | c c c | c c c}
Slot 2 & & Slot 4					& Slot 2 & & Slot 4						& Slot 2 & & Slot 4\\
\hline
${\pi_s}_2$ & $\leq$ & ${\pi_s}_3$	& ${\pi_s}_3$ & $\leq$ & ${\pi_s}_1$	& ${\pi_s}_3$ & $\leq$ & ${\pi_s}_2$\\
${\pi_s}_3$ & $\leq$ & ${\pi_s}_1$	& ${\pi_s}_1$ & $\leq$ & ${\pi_s}_2$	& ${\pi_s}_2$ & $\leq$ & ${\pi_s}_1$\\
${\pi_s}_2$ & $\leq$ & ${\pi_s}_1$	& ${\pi_s}_3$ & $\leq$ & ${\pi_s}_2$	& ${\pi_s}_3$ & $\leq$ & ${\pi_s}_1$
\end{tabular}
\end{center}

It is possible to find a PO for every permutation, since the agents in slots 2 and 4 share rewards (and expected average rewards as well) due they are in the same team. So:

\begin{equation*}
PG_{2,4}({\pi_s}_1,{\pi_s}_2,{\pi_s}_3,\Pi_o,w_{\dot{L}},\mu) = 1
\end{equation*}

Therefore:

\begin{equation*}
PG_{2,4}(\Pi_e,w_{\Pi_e},\Pi_o,w_{\dot{L}},\mu) = 6 \frac{9}{2} \frac{1}{3} \frac{1}{3} \frac{1}{3} 1 = 1
\end{equation*}

For slots 3 and 1:

\begin{equation*}
\begin{aligned}
PG_{3,1}(\Pi_e,w_{\Pi_e},\Pi_o,w_{\dot{L}},\mu)	& = \eta_{\Pi^3} \sum_{\pi_1,\pi_2,\pi_3 \in \Pi_e | \pi_1 \neq \pi_2 \neq \pi_3} w_{\Pi_e}(\pi_1) w_{\Pi_e}(\pi_2) w_{\Pi_e}(\pi_3) PG_{3,1}(\pi_1,\pi_2,\pi_3,\Pi_o,w_{\dot{L}},\mu) =\\
												& = 6 \frac{9}{2} \frac{1}{3} \frac{1}{3} \frac{1}{3} PG_{3,1}({\pi_s}_1,{\pi_s}_2,{\pi_s}_3,\Pi_o,w_{\dot{L}},\mu)
\end{aligned}
\end{equation*}

In this case, we only need to calculate $PG_{3,1}({\pi_s}_1,{\pi_s}_2,{\pi_s}_3,\Pi_o,w_{\dot{L}},\mu)$. We follow definition \ref{def:STG_agents} (for PG) to calculate this value:

\begin{equation*}
\begin{aligned}
PG_{3,1}({\pi_s}_1,{\pi_s}_2,{\pi_s}_3,\Pi_o,w_{\dot{L}},\mu)	& = \sum_{\dot{l} \in \dot{L}^{N(\mu)}_{-3,1}(\Pi_o)} w_{\dot{L}}(\dot{l}) PO_{3,1}({\pi_s}_1,{\pi_s}_2,{\pi_s}_3,\dot{l},\mu) =\\
																& = PO_{3,1}({\pi_s}_1,{\pi_s}_2,{\pi_s}_3,(*,\pi_x,*,\pi_x),\mu)
\end{aligned}
\end{equation*}

The following table shows us $PO_{3,1}$ for all the permutations of ${\pi_s}_1,{\pi_s}_2,{\pi_s}_3$.

\begin{center}
\begin{tabular}{c c c | c c c | c c c}
Slot 3 & & Slot 1					& Slot 3 & & Slot 1						& Slot 3 & & Slot 1\\
\hline
${\pi_s}_1$ & $\leq$ & ${\pi_s}_2$	& ${\pi_s}_1$ & $\leq$ & ${\pi_s}_3$	& ${\pi_s}_2$ & $\leq$ & ${\pi_s}_1$\\
${\pi_s}_2$ & $\leq$ & ${\pi_s}_3$	& ${\pi_s}_3$ & $\leq$ & ${\pi_s}_2$	& ${\pi_s}_1$ & $\leq$ & ${\pi_s}_3$\\
${\pi_s}_1$ & $\leq$ & ${\pi_s}_3$	& ${\pi_s}_1$ & $\leq$ & ${\pi_s}_2$	& ${\pi_s}_2$ & $\leq$ & ${\pi_s}_3$
\end{tabular}
\begin{tabular}{c c c | c c c | c c c}
Slot 3 & & Slot 1					& Slot 3 & & Slot 1						& Slot 3 & & Slot 1\\
\hline
${\pi_s}_2$ & $\leq$ & ${\pi_s}_3$	& ${\pi_s}_3$ & $\leq$ & ${\pi_s}_1$	& ${\pi_s}_3$ & $\leq$ & ${\pi_s}_2$\\
${\pi_s}_3$ & $\leq$ & ${\pi_s}_1$	& ${\pi_s}_1$ & $\leq$ & ${\pi_s}_2$	& ${\pi_s}_2$ & $\leq$ & ${\pi_s}_1$\\
${\pi_s}_2$ & $\leq$ & ${\pi_s}_1$	& ${\pi_s}_3$ & $\leq$ & ${\pi_s}_2$	& ${\pi_s}_3$ & $\leq$ & ${\pi_s}_1$
\end{tabular}
\end{center}

It is possible to find a PO for every permutation, since the agents in slots 2 ($\pi_x$) and 4 ($\pi_x$) have a probability of $(1/\sqrt{2}) \times (1/\sqrt{2}) = 1/2$ to do not chase the prey (any $\pi_s$) and the same probability $1 - 1/2 = 1/2$ to chase the prey, making for both agents in slots 3 (any $\pi_s$) and 1 (any $\pi_s$) to obtain the same expected average reward ($0$). So:

\begin{equation*}
PG_{3,1}({\pi_s}_1,{\pi_s}_2,{\pi_s}_3,\Pi_o,w_{\dot{L}},\mu) = 1
\end{equation*}

Therefore:

\begin{equation*}
PG_{3,1}(\Pi_e,w_{\Pi_e},\Pi_o,w_{\dot{L}},\mu) = 6 \frac{9}{2} \frac{1}{3} \frac{1}{3} \frac{1}{3} 1 = 1
\end{equation*}

For slots 3 and 2:

\begin{equation*}
\begin{aligned}
PG_{3,2}(\Pi_e,w_{\Pi_e},\Pi_o,w_{\dot{L}},\mu)	& = \eta_{\Pi^3} \sum_{\pi_1,\pi_2,\pi_3 \in \Pi_e | \pi_1 \neq \pi_2 \neq \pi_3} w_{\Pi_e}(\pi_1) w_{\Pi_e}(\pi_2) w_{\Pi_e}(\pi_3) PG_{3,2}(\pi_1,\pi_2,\pi_3,\Pi_o,w_{\dot{L}},\mu) =\\
												& = 6 \frac{9}{2} \frac{1}{3} \frac{1}{3} \frac{1}{3} PG_{3,2}({\pi_s}_1,{\pi_s}_2,{\pi_s}_3,\Pi_o,w_{\dot{L}},\mu)
\end{aligned}
\end{equation*}

In this case, we only need to calculate $PG_{3,2}({\pi_s}_1,{\pi_s}_2,{\pi_s}_3,\Pi_o,w_{\dot{L}},\mu)$. We follow definition \ref{def:STG_agents} (for PG) to calculate this value:

\begin{equation*}
\begin{aligned}
PG_{3,2}({\pi_s}_1,{\pi_s}_2,{\pi_s}_3,\Pi_o,w_{\dot{L}},\mu)	& = \sum_{\dot{l} \in \dot{L}^{N(\mu)}_{-3,2}(\Pi_o)} w_{\dot{L}}(\dot{l}) PO_{3,2}({\pi_s}_1,{\pi_s}_2,{\pi_s}_3,\dot{l},\mu) =\\
																& = PO_{3,2}({\pi_s}_1,{\pi_s}_2,{\pi_s}_3,(\pi_x,*,*,\pi_x),\mu)
\end{aligned}
\end{equation*}

The following table shows us $PO_{3,2}$ for all the permutations of ${\pi_s}_1,{\pi_s}_2,{\pi_s}_3$.

\begin{center}
\begin{tabular}{c c c | c c c | c c c}
Slot 3 & & Slot 2					& Slot 3 & & Slot 2						& Slot 3 & & Slot 2\\
\hline
${\pi_s}_1$ & $\leq$ & ${\pi_s}_2$	& ${\pi_s}_1$ & $\leq$ & ${\pi_s}_3$	& ${\pi_s}_2$ & $\leq$ & ${\pi_s}_1$\\
${\pi_s}_2$ & $\leq$ & ${\pi_s}_3$	& ${\pi_s}_3$ & $\leq$ & ${\pi_s}_2$	& ${\pi_s}_1$ & $\leq$ & ${\pi_s}_3$\\
${\pi_s}_1$ & $\leq$ & ${\pi_s}_3$	& ${\pi_s}_1$ & $\leq$ & ${\pi_s}_2$	& ${\pi_s}_2$ & $\leq$ & ${\pi_s}_3$
\end{tabular}
\begin{tabular}{c c c | c c c | c c c}
Slot 3 & & Slot 2					& Slot 3 & & Slot 2						& Slot 3 & & Slot 2\\
\hline
${\pi_s}_2$ & $\leq$ & ${\pi_s}_3$	& ${\pi_s}_3$ & $\leq$ & ${\pi_s}_1$	& ${\pi_s}_3$ & $\leq$ & ${\pi_s}_2$\\
${\pi_s}_3$ & $\leq$ & ${\pi_s}_1$	& ${\pi_s}_1$ & $\leq$ & ${\pi_s}_2$	& ${\pi_s}_2$ & $\leq$ & ${\pi_s}_1$\\
${\pi_s}_2$ & $\leq$ & ${\pi_s}_1$	& ${\pi_s}_3$ & $\leq$ & ${\pi_s}_2$	& ${\pi_s}_3$ & $\leq$ & ${\pi_s}_1$
\end{tabular}
\end{center}

It is possible to find a PO for every permutation, since the agents in slots 3 and 2 share rewards (and expected average rewards as well) due they are in the same team. So:

\begin{equation*}
PG_{3,2}({\pi_s}_1,{\pi_s}_2,{\pi_s}_3,\Pi_o,w_{\dot{L}},\mu) = 1
\end{equation*}

Therefore:

\begin{equation*}
PG_{3,2}(\Pi_e,w_{\Pi_e},\Pi_o,w_{\dot{L}},\mu) = 6 \frac{9}{2} \frac{1}{3} \frac{1}{3} \frac{1}{3} 1 = 1
\end{equation*}

For slots 3 and 4:

\begin{equation*}
\begin{aligned}
PG_{3,4}(\Pi_e,w_{\Pi_e},\Pi_o,w_{\dot{L}},\mu)	& = \eta_{\Pi^3} \sum_{\pi_1,\pi_2,\pi_3 \in \Pi_e | \pi_1 \neq \pi_2 \neq \pi_3} w_{\Pi_e}(\pi_1) w_{\Pi_e}(\pi_2) w_{\Pi_e}(\pi_3) PG_{3,4}(\pi_1,\pi_2,\pi_3,\Pi_o,w_{\dot{L}},\mu) =\\
												& = 6 \frac{9}{2} \frac{1}{3} \frac{1}{3} \frac{1}{3} PG_{3,4}({\pi_s}_1,{\pi_s}_2,{\pi_s}_3,\Pi_o,w_{\dot{L}},\mu)
\end{aligned}
\end{equation*}

In this case, we only need to calculate $PG_{3,4}({\pi_s}_1,{\pi_s}_2,{\pi_s}_3,\Pi_o,w_{\dot{L}},\mu)$. We follow definition \ref{def:STG_agents} (for PG) to calculate this value:

\begin{equation*}
\begin{aligned}
PG_{3,4}({\pi_s}_1,{\pi_s}_2,{\pi_s}_3,\Pi_o,w_{\dot{L}},\mu)	& = \sum_{\dot{l} \in \dot{L}^{N(\mu)}_{-3,4}(\Pi_o)} w_{\dot{L}}(\dot{l}) PO_{3,4}({\pi_s}_1,{\pi_s}_2,{\pi_s}_3,\dot{l},\mu) =\\
																& = PO_{3,4}({\pi_s}_1,{\pi_s}_2,{\pi_s}_3,(\pi_x,\pi_x,*,*),\mu)
\end{aligned}
\end{equation*}

The following table shows us $PO_{3,4}$ for all the permutations of ${\pi_s}_1,{\pi_s}_2,{\pi_s}_3$.

\begin{center}
\begin{tabular}{c c c | c c c | c c c}
Slot 3 & & Slot 4					& Slot 3 & & Slot 4						& Slot 3 & & Slot 4\\
\hline
${\pi_s}_1$ & $\leq$ & ${\pi_s}_2$	& ${\pi_s}_1$ & $\leq$ & ${\pi_s}_3$	& ${\pi_s}_2$ & $\leq$ & ${\pi_s}_1$\\
${\pi_s}_2$ & $\leq$ & ${\pi_s}_3$	& ${\pi_s}_3$ & $\leq$ & ${\pi_s}_2$	& ${\pi_s}_1$ & $\leq$ & ${\pi_s}_3$\\
${\pi_s}_1$ & $\leq$ & ${\pi_s}_3$	& ${\pi_s}_1$ & $\leq$ & ${\pi_s}_2$	& ${\pi_s}_2$ & $\leq$ & ${\pi_s}_3$
\end{tabular}
\begin{tabular}{c c c | c c c | c c c}
Slot 3 & & Slot 4					& Slot 3 & & Slot 4						& Slot 3 & & Slot 4\\
\hline
${\pi_s}_2$ & $\leq$ & ${\pi_s}_3$	& ${\pi_s}_3$ & $\leq$ & ${\pi_s}_1$	& ${\pi_s}_3$ & $\leq$ & ${\pi_s}_2$\\
${\pi_s}_3$ & $\leq$ & ${\pi_s}_1$	& ${\pi_s}_1$ & $\leq$ & ${\pi_s}_2$	& ${\pi_s}_2$ & $\leq$ & ${\pi_s}_1$\\
${\pi_s}_2$ & $\leq$ & ${\pi_s}_1$	& ${\pi_s}_3$ & $\leq$ & ${\pi_s}_2$	& ${\pi_s}_3$ & $\leq$ & ${\pi_s}_1$
\end{tabular}
\end{center}

It is possible to find a PO for every permutation, since the agents in slots 3 and 4 share rewards (and expected average rewards as well) due they are in the same team. So:

\begin{equation*}
PG_{3,4}({\pi_s}_1,{\pi_s}_2,{\pi_s}_3,\Pi_o,w_{\dot{L}},\mu) = 1
\end{equation*}

Therefore:

\begin{equation*}
PG_{3,4}(\Pi_e,w_{\Pi_e},\Pi_o,w_{\dot{L}},\mu) = 6 \frac{9}{2} \frac{1}{3} \frac{1}{3} \frac{1}{3} 1 = 1
\end{equation*}

For slots 4 and 1:

\begin{equation*}
\begin{aligned}
PG_{4,1}(\Pi_e,w_{\Pi_e},\Pi_o,w_{\dot{L}},\mu)	& = \eta_{\Pi^3} \sum_{\pi_1,\pi_2,\pi_3 \in \Pi_e | \pi_1 \neq \pi_2 \neq \pi_3} w_{\Pi_e}(\pi_1) w_{\Pi_e}(\pi_2) w_{\Pi_e}(\pi_3) PG_{4,1}(\pi_1,\pi_2,\pi_3,\Pi_o,w_{\dot{L}},\mu) =\\
												& = 6 \frac{9}{2} \frac{1}{3} \frac{1}{3} \frac{1}{3} PG_{4,1}({\pi_s}_1,{\pi_s}_2,{\pi_s}_3,\Pi_o,w_{\dot{L}},\mu)
\end{aligned}
\end{equation*}

In this case, we only need to calculate $PG_{4,1}({\pi_s}_1,{\pi_s}_2,{\pi_s}_3,\Pi_o,w_{\dot{L}},\mu)$. We follow definition \ref{def:STG_agents} (for PG) to calculate this value:

\begin{equation*}
\begin{aligned}
PG_{4,1}({\pi_s}_1,{\pi_s}_2,{\pi_s}_3,\Pi_o,w_{\dot{L}},\mu)	& = \sum_{\dot{l} \in \dot{L}^{N(\mu)}_{-4,1}(\Pi_o)} w_{\dot{L}}(\dot{l}) PO_{4,1}({\pi_s}_1,{\pi_s}_2,{\pi_s}_3,\dot{l},\mu) =\\
																& = PO_{4,1}({\pi_s}_1,{\pi_s}_2,{\pi_s}_3,(*,\pi_x,\pi_x,*),\mu)
\end{aligned}
\end{equation*}

The following table shows us $PO_{4,1}$ for all the permutations of ${\pi_s}_1,{\pi_s}_2,{\pi_s}_3$.

\begin{center}
\begin{tabular}{c c c | c c c | c c c}
Slot 4 & & Slot 1					& Slot 4 & & Slot 1						& Slot 4 & & Slot 1\\
\hline
${\pi_s}_1$ & $\leq$ & ${\pi_s}_2$	& ${\pi_s}_1$ & $\leq$ & ${\pi_s}_3$	& ${\pi_s}_2$ & $\leq$ & ${\pi_s}_1$\\
${\pi_s}_2$ & $\leq$ & ${\pi_s}_3$	& ${\pi_s}_3$ & $\leq$ & ${\pi_s}_2$	& ${\pi_s}_1$ & $\leq$ & ${\pi_s}_3$\\
${\pi_s}_1$ & $\leq$ & ${\pi_s}_3$	& ${\pi_s}_1$ & $\leq$ & ${\pi_s}_2$	& ${\pi_s}_2$ & $\leq$ & ${\pi_s}_3$
\end{tabular}
\begin{tabular}{c c c | c c c | c c c}
Slot 4 & & Slot 1					& Slot 4 & & Slot 1						& Slot 4 & & Slot 1\\
\hline
${\pi_s}_2$ & $\leq$ & ${\pi_s}_3$	& ${\pi_s}_3$ & $\leq$ & ${\pi_s}_1$	& ${\pi_s}_3$ & $\leq$ & ${\pi_s}_2$\\
${\pi_s}_3$ & $\leq$ & ${\pi_s}_1$	& ${\pi_s}_1$ & $\leq$ & ${\pi_s}_2$	& ${\pi_s}_2$ & $\leq$ & ${\pi_s}_1$\\
${\pi_s}_2$ & $\leq$ & ${\pi_s}_1$	& ${\pi_s}_3$ & $\leq$ & ${\pi_s}_2$	& ${\pi_s}_3$ & $\leq$ & ${\pi_s}_1$
\end{tabular}
\end{center}

It is possible to find a PO for every permutation, since the agents in slots 2 ($\pi_x$) and 3 ($\pi_x$) have a probability of $(1/\sqrt{2}) \times (1/\sqrt{2}) = 1/2$ to do not chase the prey (any $\pi_s$) and the same probability $1 - 1/2 = 1/2$ to chase the prey, making for both agents in slots 4 (any $\pi_s$) and 1 (any $\pi_s$) to obtain the same expected average reward ($0$). So:

\begin{equation*}
PG_{4,1}({\pi_s}_1,{\pi_s}_2,{\pi_s}_3,\Pi_o,w_{\dot{L}},\mu) = 1
\end{equation*}

Therefore:

\begin{equation*}
PG_{4,1}(\Pi_e,w_{\Pi_e},\Pi_o,w_{\dot{L}},\mu) = 6 \frac{9}{2} \frac{1}{3} \frac{1}{3} \frac{1}{3} 1 = 1
\end{equation*}

For slots 4 and 2:

\begin{equation*}
\begin{aligned}
PG_{4,2}(\Pi_e,w_{\Pi_e},\Pi_o,w_{\dot{L}},\mu)	& = \eta_{\Pi^3} \sum_{\pi_1,\pi_2,\pi_3 \in \Pi_e | \pi_1 \neq \pi_2 \neq \pi_3} w_{\Pi_e}(\pi_1) w_{\Pi_e}(\pi_2) w_{\Pi_e}(\pi_3) PG_{4,2}(\pi_1,\pi_2,\pi_3,\Pi_o,w_{\dot{L}},\mu) =\\
												& = 6 \frac{9}{2} \frac{1}{3} \frac{1}{3} \frac{1}{3} PG_{4,2}({\pi_s}_1,{\pi_s}_2,{\pi_s}_3,\Pi_o,w_{\dot{L}},\mu)
\end{aligned}
\end{equation*}

In this case, we only need to calculate $PG_{4,2}({\pi_s}_1,{\pi_s}_2,{\pi_s}_3,\Pi_o,w_{\dot{L}},\mu)$. We follow definition \ref{def:STG_agents} (for PG) to calculate this value:

\begin{equation*}
\begin{aligned}
PG_{4,2}({\pi_s}_1,{\pi_s}_2,{\pi_s}_3,\Pi_o,w_{\dot{L}},\mu)	& = \sum_{\dot{l} \in \dot{L}^{N(\mu)}_{-4,2}(\Pi_o)} w_{\dot{L}}(\dot{l}) PO_{4,2}({\pi_s}_1,{\pi_s}_2,{\pi_s}_3,\dot{l},\mu) =\\
																& = PO_{4,2}({\pi_s}_1,{\pi_s}_2,{\pi_s}_3,(\pi_x,*,\pi_x,*),\mu)
\end{aligned}
\end{equation*}

The following table shows us $PO_{4,2}$ for all the permutations of ${\pi_s}_1,{\pi_s}_2,{\pi_s}_3$.

\begin{center}
\begin{tabular}{c c c | c c c | c c c}
Slot 4 & & Slot 2					& Slot 4 & & Slot 2						& Slot 4 & & Slot 2\\
\hline
${\pi_s}_1$ & $\leq$ & ${\pi_s}_2$	& ${\pi_s}_1$ & $\leq$ & ${\pi_s}_3$	& ${\pi_s}_2$ & $\leq$ & ${\pi_s}_1$\\
${\pi_s}_2$ & $\leq$ & ${\pi_s}_3$	& ${\pi_s}_3$ & $\leq$ & ${\pi_s}_2$	& ${\pi_s}_1$ & $\leq$ & ${\pi_s}_3$\\
${\pi_s}_1$ & $\leq$ & ${\pi_s}_3$	& ${\pi_s}_1$ & $\leq$ & ${\pi_s}_2$	& ${\pi_s}_2$ & $\leq$ & ${\pi_s}_3$
\end{tabular}
\begin{tabular}{c c c | c c c | c c c}
Slot 4 & & Slot 2					& Slot 4 & & Slot 2						& Slot 4 & & Slot 2\\
\hline
${\pi_s}_2$ & $\leq$ & ${\pi_s}_3$	& ${\pi_s}_3$ & $\leq$ & ${\pi_s}_1$	& ${\pi_s}_3$ & $\leq$ & ${\pi_s}_2$\\
${\pi_s}_3$ & $\leq$ & ${\pi_s}_1$	& ${\pi_s}_1$ & $\leq$ & ${\pi_s}_2$	& ${\pi_s}_2$ & $\leq$ & ${\pi_s}_1$\\
${\pi_s}_2$ & $\leq$ & ${\pi_s}_1$	& ${\pi_s}_3$ & $\leq$ & ${\pi_s}_2$	& ${\pi_s}_3$ & $\leq$ & ${\pi_s}_1$
\end{tabular}
\end{center}

It is possible to find a PO for every permutation, since the agents in slots 4 and 2 share rewards (and expected average rewards as well) due they are in the same team. So:

\begin{equation*}
PG_{4,2}({\pi_s}_1,{\pi_s}_2,{\pi_s}_3,\Pi_o,w_{\dot{L}},\mu) = 1
\end{equation*}

Therefore:

\begin{equation*}
PG_{4,2}(\Pi_e,w_{\Pi_e},\Pi_o,w_{\dot{L}},\mu) = 6 \frac{9}{2} \frac{1}{3} \frac{1}{3} \frac{1}{3} 1 = 1
\end{equation*}

And for slots 4 and 3:

\begin{equation*}
\begin{aligned}
PG_{4,3}(\Pi_e,w_{\Pi_e},\Pi_o,w_{\dot{L}},\mu)	& = \eta_{\Pi^3} \sum_{\pi_1,\pi_2,\pi_3 \in \Pi_e | \pi_1 \neq \pi_2 \neq \pi_3} w_{\Pi_e}(\pi_1) w_{\Pi_e}(\pi_2) w_{\Pi_e}(\pi_3) PG_{4,3}(\pi_1,\pi_2,\pi_3,\Pi_o,w_{\dot{L}},\mu) =\\
												& = 6 \frac{9}{2} \frac{1}{3} \frac{1}{3} \frac{1}{3} PG_{4,3}({\pi_s}_1,{\pi_s}_2,{\pi_s}_3,\Pi_o,w_{\dot{L}},\mu)
\end{aligned}
\end{equation*}

In this case, we only need to calculate $PG_{4,3}({\pi_s}_1,{\pi_s}_2,{\pi_s}_3,\Pi_o,w_{\dot{L}},\mu)$. We follow definition \ref{def:STG_agents} (for PG) to calculate this value:

\begin{equation*}
\begin{aligned}
PG_{4,3}({\pi_s}_1,{\pi_s}_2,{\pi_s}_3,\Pi_o,w_{\dot{L}},\mu)	& = \sum_{\dot{l} \in \dot{L}^{N(\mu)}_{-4,3}(\Pi_o)} w_{\dot{L}}(\dot{l}) PO_{4,3}({\pi_s}_1,{\pi_s}_2,{\pi_s}_3,\dot{l},\mu) =\\
																& = PO_{4,3}({\pi_s}_1,{\pi_s}_2,{\pi_s}_3,(\pi_x,\pi_x,*,*),\mu)
\end{aligned}
\end{equation*}

The following table shows us $PO_{4,3}$ for all the permutations of ${\pi_s}_1,{\pi_s}_2,{\pi_s}_3$.

\begin{center}
\begin{tabular}{c c c | c c c | c c c}
Slot 4 & & Slot 3					& Slot 4 & & Slot 3						& Slot 4 & & Slot 3\\
\hline
${\pi_s}_1$ & $\leq$ & ${\pi_s}_2$	& ${\pi_s}_1$ & $\leq$ & ${\pi_s}_3$	& ${\pi_s}_2$ & $\leq$ & ${\pi_s}_1$\\
${\pi_s}_2$ & $\leq$ & ${\pi_s}_3$	& ${\pi_s}_3$ & $\leq$ & ${\pi_s}_2$	& ${\pi_s}_1$ & $\leq$ & ${\pi_s}_3$\\
${\pi_s}_1$ & $\leq$ & ${\pi_s}_3$	& ${\pi_s}_1$ & $\leq$ & ${\pi_s}_2$	& ${\pi_s}_2$ & $\leq$ & ${\pi_s}_3$
\end{tabular}
\begin{tabular}{c c c | c c c | c c c}
Slot 4 & & Slot 3					& Slot 4 & & Slot 3						& Slot 4 & & Slot 3\\
\hline
${\pi_s}_2$ & $\leq$ & ${\pi_s}_3$	& ${\pi_s}_3$ & $\leq$ & ${\pi_s}_1$	& ${\pi_s}_3$ & $\leq$ & ${\pi_s}_2$\\
${\pi_s}_3$ & $\leq$ & ${\pi_s}_1$	& ${\pi_s}_1$ & $\leq$ & ${\pi_s}_2$	& ${\pi_s}_2$ & $\leq$ & ${\pi_s}_1$\\
${\pi_s}_2$ & $\leq$ & ${\pi_s}_1$	& ${\pi_s}_3$ & $\leq$ & ${\pi_s}_2$	& ${\pi_s}_3$ & $\leq$ & ${\pi_s}_1$
\end{tabular}
\end{center}

It is possible to find a PO for every permutation, since the agents in slots 4 and 3 share rewards (and expected average rewards as well) due they are in the same team. So:

\begin{equation*}
PG_{4,3}({\pi_s}_1,{\pi_s}_2,{\pi_s}_3,\Pi_o,w_{\dot{L}},\mu) = 1
\end{equation*}

Therefore:

\begin{equation*}
PG_{4,3}(\Pi_e,w_{\Pi_e},\Pi_o,w_{\dot{L}},\mu) = 6 \frac{9}{2} \frac{1}{3} \frac{1}{3} \frac{1}{3} 1 = 1
\end{equation*}

And finally, we weight over the slots:

\begin{equation*}
\begin{aligned}
& PG(\Pi_e,w_{\Pi_e},\Pi_o,w_{\dot{L}},\mu,w_S) = \eta_{S_1^2} \sum_{i=1}^{N(\mu)} w_S(i,\mu) \times\\
& \times \left(\sum_{j=1}^{i-1} w_S(j,\mu) PG_{i,j}(\Pi_e,w_{\Pi_e},\Pi_o,w_{\dot{L}},\mu) + \sum_{j=i+1}^{N(\mu)} w_S(j,\mu) PG_{i,j}(\Pi_e,w_{\Pi_e},\Pi_o,w_{\dot{L}},\mu)\right) =\\
& \ \ \ \ \ \ \ \ \ \ \ \ \ \ \ \ \ \ \ \ \ \ \ \ \ \ \ \ \ \ \ \ \ \ \ \ = \frac{4}{3} \frac{1}{4} \frac{1}{4} \{PG_{1,2}(\Pi_e,w_{\Pi_e},\Pi_o,w_{\dot{L}},\mu) + PG_{1,3}(\Pi_e,w_{\Pi_e},\Pi_o,w_{\dot{L}},\mu) +\\
& \ \ \ \ \ \ \ \ \ \ \ \ \ \ \ \ \ \ \ \ \ \ \ \ \ \ \ \ \ \ \ \ \ \ \ \ + PG_{1,4}(\Pi_e,w_{\Pi_e},\Pi_o,w_{\dot{L}},\mu) + PG_{2,1}(\Pi_e,w_{\Pi_e},\Pi_o,w_{\dot{L}},\mu) +\\
& \ \ \ \ \ \ \ \ \ \ \ \ \ \ \ \ \ \ \ \ \ \ \ \ \ \ \ \ \ \ \ \ \ \ \ \ + PG_{2,3}(\Pi_e,w_{\Pi_e},\Pi_o,w_{\dot{L}},\mu) + PG_{2,4}(\Pi_e,w_{\Pi_e},\Pi_o,w_{\dot{L}},\mu) +\\
& \ \ \ \ \ \ \ \ \ \ \ \ \ \ \ \ \ \ \ \ \ \ \ \ \ \ \ \ \ \ \ \ \ \ \ \ + PG_{3,1}(\Pi_e,w_{\Pi_e},\Pi_o,w_{\dot{L}},\mu) + PG_{3,2}(\Pi_e,w_{\Pi_e},\Pi_o,w_{\dot{L}},\mu) +\\
& \ \ \ \ \ \ \ \ \ \ \ \ \ \ \ \ \ \ \ \ \ \ \ \ \ \ \ \ \ \ \ \ \ \ \ \ + PG_{3,4}(\Pi_e,w_{\Pi_e},\Pi_o,w_{\dot{L}},\mu) + PG_{4,1}(\Pi_e,w_{\Pi_e},\Pi_o,w_{\dot{L}},\mu) +\\
& \ \ \ \ \ \ \ \ \ \ \ \ \ \ \ \ \ \ \ \ \ \ \ \ \ \ \ \ \ \ \ \ \ \ \ \ + PG_{4,2}(\Pi_e,w_{\Pi_e},\Pi_o,w_{\dot{L}},\mu) + PG_{4,3}(\Pi_e,w_{\Pi_e},\Pi_o,w_{\dot{L}},\mu)\} =\\
& \ \ \ \ \ \ \ \ \ \ \ \ \ \ \ \ \ \ \ \ \ \ \ \ \ \ \ \ \ \ \ \ \ \ \ \ = \frac{4}{3} \frac{1}{4} \frac{1}{4} \left\{12 \times 1\right\} = 1
\end{aligned}
\end{equation*}

Since $1$ is the highest possible value for the partial grading property, therefore predator-prey has $General_{max} = 1$ for this property.
\end{proof}
\end{proposition}

\begin{approximation}
\label{approx:predator-prey_PG_left_max}
$Left_{max}$ for the partial grading (PG) property is equal to $\frac{3}{4}$ (as a {\em lower} approximation) for the predator-prey environment.

\begin{proof}
To find $Left_{max}$ (equation \ref{eq:left_max}), we need to find a pair $\left\langle\Pi_e,w_{\Pi_e}\right\rangle$ which maximises the property as much as possible while $\Pi_o$ minimises it. Using $\Pi_e = \{{\pi_{chase}}_1,{\pi_{chase}}_2,{\pi_{chase}}_3\}$ with uniform weight for $w_{\Pi_e}$ (a $\pi_{chase}$ agent always tries to be chased when playing as the prey and tries to chase when playing as a predator) we find a {\em lower} approximation of this situation no matter which $\Pi_o$ we use.

Following definition \ref{def:PG}, we obtain the PG value for this $\left\langle\Pi_e,w_{\Pi_e},\Pi_o\right\rangle$ (where $\Pi_o$ is instantiated with any permitted value). Since the environment is not symmetric, we need to calculate this property for every pair of slots. Following definition \ref{def:STG_set} (for PG), we can calculate its PG value for each pair of slots. We start with slots 1 and 2:

\begin{equation*}
\begin{aligned}
PG_{1,2}(\Pi_e,w_{\Pi_e},\Pi_o,w_{\dot{L}},\mu)	& = \eta_{\Pi^3} \sum_{\pi_1,\pi_2,\pi_3 \in \Pi_e | \pi_1 \neq \pi_2 \neq \pi_3} w_{\Pi_e}(\pi_1) w_{\Pi_e}(\pi_2) w_{\Pi_e}(\pi_3) PG_{1,2}(\pi_1,\pi_2,\pi_3,\Pi_o,w_{\dot{L}},\mu) =\\
												& = 6 \frac{9}{2} \frac{1}{3} \frac{1}{3} \frac{1}{3} PG_{1,2}({\pi_{chase}}_1,{\pi_{chase}}_2,{\pi_{chase}}_3,\Pi_o,w_{\dot{L}},\mu)
\end{aligned}
\end{equation*}

\noindent Note that we avoided to calculate all the permutations of $\pi_1,\pi_2,\pi_3$ for $PG_{i,j}(\pi_1,\pi_2,\pi_3,\Pi_o,w_{\dot{L}},\mu)$ since they provide the same result, by calculating only one permutation and multiplying the result by the number of permutations $6$.

In this case, we only need to calculate $PG_{1,2}({\pi_{chase}}_1,{\pi_{chase}}_2,{\pi_{chase}}_3,\Pi_o,w_{\dot{L}},\mu)$. We follow definition \ref{def:STG_agents} (for PG) to calculate this value:

\begin{equation*}
PG_{1,2}({\pi_{chase}}_1,{\pi_{chase}}_2,{\pi_{chase}}_3,\Pi_o,w_{\dot{L}},\mu) = \sum_{\dot{l} \in \dot{L}^{N(\mu)}_{-1,2}(\Pi_o)} w_{\dot{L}}(\dot{l}) PO_{1,2}({\pi_{chase}}_1,{\pi_{chase}}_2,{\pi_{chase}}_3,\dot{l},\mu)
\end{equation*}

We do not know which $\Pi_o$ we have, but we know that we will need to obtain a line-up pattern $\dot{l}$ from $\dot{L}^{N(\mu)}_{-1,2}(\Pi_o)$ to calculate $PO_{1,2}({\pi_{chase}}_1,{\pi_{chase}}_2,{\pi_{chase}}_3,\dot{l},\mu)$. We calculate this value for a figurative line-up pattern $\dot{l} = (*,*,\pi_1,\pi_2)$ from $\dot{L}^{N(\mu)}_{-1,2}(\Pi_o)$:

\begin{equation*}
PO_{1,2}({\pi_{chase}}_1,{\pi_{chase}}_2,{\pi_{chase}}_3,\dot{l},\mu) = PO_{1,2}({\pi_{chase}}_1,{\pi_{chase}}_2,{\pi_{chase}}_3,(*,*,\pi_1,\pi_2),\mu)
\end{equation*}

The following table shows us $PO_{1,2}$ for all the permutations of ${\pi_{chase}}_1,{\pi_{chase}}_2,{\pi_{chase}}_3$.

\begin{center}
\begin{tabular}{c c c | c c c | c c c}
Slot 1 & & Slot 2								& Slot 1 & & Slot 2									& Slot 1 & & Slot 2\\
\hline
${\pi_{chase}}_1$ & $\leq$ & ${\pi_{chase}}_2$	& ${\pi_{chase}}_1$ & $\leq$ & ${\pi_{chase}}_3$	& ${\pi_{chase}}_2$ & $\leq$ & ${\pi_{chase}}_1$\\
${\pi_{chase}}_2$ & $\leq$ & ${\pi_{chase}}_3$	& ${\pi_{chase}}_3$ & $\leq$ & ${\pi_{chase}}_2$	& ${\pi_{chase}}_1$ & $\leq$ & ${\pi_{chase}}_3$\\
${\pi_{chase}}_1$ & $\leq$ & ${\pi_{chase}}_3$	& ${\pi_{chase}}_1$ & $\leq$ & ${\pi_{chase}}_2$	& ${\pi_{chase}}_2$ & $\leq$ & ${\pi_{chase}}_3$
\end{tabular}
\begin{tabular}{c c c | c c c | c c c}
Slot 1 & & Slot 2								& Slot 1 & & Slot 2									& Slot 1 & & Slot 2\\
\hline
${\pi_{chase}}_2$ & $\leq$ & ${\pi_{chase}}_3$	& ${\pi_{chase}}_3$ & $\leq$ & ${\pi_{chase}}_1$	& ${\pi_{chase}}_3$ & $\leq$ & ${\pi_{chase}}_2$\\
${\pi_{chase}}_3$ & $\leq$ & ${\pi_{chase}}_1$	& ${\pi_{chase}}_1$ & $\leq$ & ${\pi_{chase}}_2$	& ${\pi_{chase}}_2$ & $\leq$ & ${\pi_{chase}}_1$\\
${\pi_{chase}}_2$ & $\leq$ & ${\pi_{chase}}_1$	& ${\pi_{chase}}_3$ & $\leq$ & ${\pi_{chase}}_2$	& ${\pi_{chase}}_3$ & $\leq$ & ${\pi_{chase}}_1$
\end{tabular}
\end{center}

It is possible to find a PO for every permutation, since we always have ${\pi_{chase}}_i \leq {\pi_{chase}}_j$, where a $\pi_{chase}$ agent always tries to be chased when playing as the prey and tries to chase when playing as a predator, so the agents in slots 1 and 2 will obtain an expected average reward of $-1$ and $1$ respectively. Note that the choice of $\Pi_o$ does not affect the result of $PO_{1,2}$, so no matter which agents are in $\Pi_o$ we obtain:

\begin{equation*}
PG_{1,2}({\pi_{chase}}_1,{\pi_{chase}}_2,{\pi_{chase}}_3,\Pi_o,w_{\dot{L}},\mu) = 1
\end{equation*}

Therefore:

\begin{equation*}
PG_{1,2}(\Pi_e,w_{\Pi_e},\Pi_o,w_{\dot{L}},\mu) = 6 \frac{9}{2} \frac{1}{3} \frac{1}{3} \frac{1}{3} 1 = 1
\end{equation*}

For slots 1 and 3:

\begin{equation*}
\begin{aligned}
PG_{1,3}(\Pi_e,w_{\Pi_e},\Pi_o,w_{\dot{L}},\mu)	& = \eta_{\Pi^3} \sum_{\pi_1,\pi_2,\pi_3 \in \Pi_e | \pi_1 \neq \pi_2 \neq \pi_3} w_{\Pi_e}(\pi_1) w_{\Pi_e}(\pi_2) w_{\Pi_e}(\pi_3) PG_{1,3}(\pi_1,\pi_2,\pi_3,\Pi_o,w_{\dot{L}},\mu) =\\
												& = 6 \frac{9}{2} \frac{1}{3} \frac{1}{3} \frac{1}{3} PG_{1,3}({\pi_{chase}}_1,{\pi_{chase}}_2,{\pi_{chase}}_3,\Pi_o,w_{\dot{L}},\mu)
\end{aligned}
\end{equation*}

In this case, we only need to calculate $PG_{1,3}({\pi_{chase}}_1,{\pi_{chase}}_2,{\pi_{chase}}_3,\Pi_o,w_{\dot{L}},\mu)$. We follow definition \ref{def:STG_agents} (for PG) to calculate this value:

\begin{equation*}
PG_{1,3}({\pi_{chase}}_1,{\pi_{chase}}_2,{\pi_{chase}}_3,\Pi_o,w_{\dot{L}},\mu) = \sum_{\dot{l} \in \dot{L}^{N(\mu)}_{-1,3}(\Pi_o)} w_{\dot{L}}(\dot{l}) PO_{1,3}({\pi_{chase}}_1,{\pi_{chase}}_2,{\pi_{chase}}_3,\dot{l},\mu)
\end{equation*}

Again, we do not know which $\Pi_o$ we have, but we know that we will need to obtain a line-up pattern $\dot{l}$ from $\dot{L}^{N(\mu)}_{-1,3}(\Pi_o)$ to calculate $PO_{1,3}({\pi_{chase}}_1,{\pi_{chase}}_2,{\pi_{chase}}_3,\dot{l},\mu)$. We calculate this value for a figurative line-up pattern $\dot{l} = (*,\pi_1,*,\pi_2)$ from $\dot{L}^{N(\mu)}_{-1,3}(\Pi_o)$:

\begin{equation*}
PO_{1,3}({\pi_{chase}}_1,{\pi_{chase}}_2,{\pi_{chase}}_3,\dot{l},\mu) = PO_{1,3}({\pi_{chase}}_1,{\pi_{chase}}_2,{\pi_{chase}}_3,(*,\pi_1,*,\pi_2),\mu)
\end{equation*}

The following table shows us $PO_{1,3}$ for all the permutations of ${\pi_{chase}}_1,{\pi_{chase}}_2,{\pi_{chase}}_3$.

\begin{center}
\begin{tabular}{c c c | c c c | c c c}
Slot 1 & & Slot 3								& Slot 1 & & Slot 3									& Slot 1 & & Slot 3\\
\hline
${\pi_{chase}}_1$ & $\leq$ & ${\pi_{chase}}_2$	& ${\pi_{chase}}_1$ & $\leq$ & ${\pi_{chase}}_3$	& ${\pi_{chase}}_2$ & $\leq$ & ${\pi_{chase}}_1$\\
${\pi_{chase}}_2$ & $\leq$ & ${\pi_{chase}}_3$	& ${\pi_{chase}}_3$ & $\leq$ & ${\pi_{chase}}_2$	& ${\pi_{chase}}_1$ & $\leq$ & ${\pi_{chase}}_3$\\
${\pi_{chase}}_1$ & $\leq$ & ${\pi_{chase}}_3$	& ${\pi_{chase}}_1$ & $\leq$ & ${\pi_{chase}}_2$	& ${\pi_{chase}}_2$ & $\leq$ & ${\pi_{chase}}_3$
\end{tabular}
\begin{tabular}{c c c | c c c | c c c}
Slot 1 & & Slot 3								& Slot 1 & & Slot 3									& Slot 1 & & Slot 3\\
\hline
${\pi_{chase}}_2$ & $\leq$ & ${\pi_{chase}}_3$	& ${\pi_{chase}}_3$ & $\leq$ & ${\pi_{chase}}_1$	& ${\pi_{chase}}_3$ & $\leq$ & ${\pi_{chase}}_2$\\
${\pi_{chase}}_3$ & $\leq$ & ${\pi_{chase}}_1$	& ${\pi_{chase}}_1$ & $\leq$ & ${\pi_{chase}}_2$	& ${\pi_{chase}}_2$ & $\leq$ & ${\pi_{chase}}_1$\\
${\pi_{chase}}_2$ & $\leq$ & ${\pi_{chase}}_1$	& ${\pi_{chase}}_3$ & $\leq$ & ${\pi_{chase}}_2$	& ${\pi_{chase}}_3$ & $\leq$ & ${\pi_{chase}}_1$
\end{tabular}
\end{center}

Again, it is possible to find a PO for every permutation, since we always have ${\pi_{chase}}_i \leq {\pi_{chase}}_j$, where a $\pi_{chase}$ agent always tries to be chased when playing as the prey and tries to chase when playing as a predator, so the agents in slots 1 and 3 will obtain an expected average reward of $-1$ and $1$ respectively. Note that the choice of $\Pi_o$ does not affect the result of $PO_{1,3}$, so no matter which agents are in $\Pi_o$ we obtain:

\begin{equation*}
PG_{1,3}({\pi_{chase}}_1,{\pi_{chase}}_2,{\pi_{chase}}_3,\Pi_o,w_{\dot{L}},\mu) = 1
\end{equation*}

Therefore:

\begin{equation*}
PG_{1,3}(\Pi_e,w_{\Pi_e},\Pi_o,w_{\dot{L}},\mu) = 6 \frac{9}{2} \frac{1}{3} \frac{1}{3} \frac{1}{3} 1 = 1
\end{equation*}

For slots 1 and 4:

\begin{equation*}
\begin{aligned}
PG_{1,4}(\Pi_e,w_{\Pi_e},\Pi_o,w_{\dot{L}},\mu)	& = \eta_{\Pi^3} \sum_{\pi_1,\pi_2,\pi_3 \in \Pi_e | \pi_1 \neq \pi_2 \neq \pi_3} w_{\Pi_e}(\pi_1) w_{\Pi_e}(\pi_2) w_{\Pi_e}(\pi_3) PG_{1,4}(\pi_1,\pi_2,\pi_3,\Pi_o,w_{\dot{L}},\mu) =\\
												& = 6 \frac{9}{2} \frac{1}{3} \frac{1}{3} \frac{1}{3} PG_{1,4}({\pi_{chase}}_1,{\pi_{chase}}_2,{\pi_{chase}}_3,\Pi_o,w_{\dot{L}},\mu)
\end{aligned}
\end{equation*}

In this case, we only need to calculate $PG_{1,4}({\pi_{chase}}_1,{\pi_{chase}}_2,{\pi_{chase}}_3,\Pi_o,w_{\dot{L}},\mu)$. We follow definition \ref{def:STG_agents} (for PG) to calculate this value:

\begin{equation*}
PG_{1,4}({\pi_{chase}}_1,{\pi_{chase}}_2,{\pi_{chase}}_3,\Pi_o,w_{\dot{L}},\mu) = \sum_{\dot{l} \in \dot{L}^{N(\mu)}_{-1,4}(\Pi_o)} w_{\dot{L}}(\dot{l}) PO_{1,4}({\pi_{chase}}_1,{\pi_{chase}}_2,{\pi_{chase}}_3,\dot{l},\mu)
\end{equation*}

Again, we do not know which $\Pi_o$ we have, but we know that we will need to obtain a line-up pattern $\dot{l}$ from $\dot{L}^{N(\mu)}_{-1,4}(\Pi_o)$ to calculate $PO_{1,4}({\pi_{chase}}_1,{\pi_{chase}}_2,{\pi_{chase}}_3,\dot{l},\mu)$. We calculate this value for a figurative line-up pattern $\dot{l} = (*,\pi_1,\pi_2,*)$ from $\dot{L}^{N(\mu)}_{-1,4}(\Pi_o)$:

\begin{equation*}
PO_{1,4}({\pi_{chase}}_1,{\pi_{chase}}_2,{\pi_{chase}}_3,\dot{l},\mu) = PO_{1,4}({\pi_{chase}}_1,{\pi_{chase}}_2,{\pi_{chase}}_3,(*,\pi_1,\pi_2,*),\mu)
\end{equation*}

The following table shows us $PO_{1,4}$ for all the permutations of ${\pi_{chase}}_1,{\pi_{chase}}_2,{\pi_{chase}}_3$.

\begin{center}
\begin{tabular}{c c c | c c c | c c c}
Slot 1 & & Slot 4								& Slot 1 & & Slot 4									& Slot 1 & & Slot 4\\
\hline
${\pi_{chase}}_1$ & $\leq$ & ${\pi_{chase}}_2$	& ${\pi_{chase}}_1$ & $\leq$ & ${\pi_{chase}}_3$	& ${\pi_{chase}}_2$ & $\leq$ & ${\pi_{chase}}_1$\\
${\pi_{chase}}_2$ & $\leq$ & ${\pi_{chase}}_3$	& ${\pi_{chase}}_3$ & $\leq$ & ${\pi_{chase}}_2$	& ${\pi_{chase}}_1$ & $\leq$ & ${\pi_{chase}}_3$\\
${\pi_{chase}}_1$ & $\leq$ & ${\pi_{chase}}_3$	& ${\pi_{chase}}_1$ & $\leq$ & ${\pi_{chase}}_2$	& ${\pi_{chase}}_2$ & $\leq$ & ${\pi_{chase}}_3$
\end{tabular}
\begin{tabular}{c c c | c c c | c c c}
Slot 1 & & Slot 4								& Slot 1 & & Slot 4									& Slot 1 & & Slot 4\\
\hline
${\pi_{chase}}_2$ & $\leq$ & ${\pi_{chase}}_3$	& ${\pi_{chase}}_3$ & $\leq$ & ${\pi_{chase}}_1$	& ${\pi_{chase}}_3$ & $\leq$ & ${\pi_{chase}}_2$\\
${\pi_{chase}}_3$ & $\leq$ & ${\pi_{chase}}_1$	& ${\pi_{chase}}_1$ & $\leq$ & ${\pi_{chase}}_2$	& ${\pi_{chase}}_2$ & $\leq$ & ${\pi_{chase}}_1$\\
${\pi_{chase}}_2$ & $\leq$ & ${\pi_{chase}}_1$	& ${\pi_{chase}}_3$ & $\leq$ & ${\pi_{chase}}_2$	& ${\pi_{chase}}_3$ & $\leq$ & ${\pi_{chase}}_1$
\end{tabular}
\end{center}

Again, it is possible to find a PO for every permutation, since we always have ${\pi_{chase}}_i \leq {\pi_{chase}}_j$, where a $\pi_{chase}$ agent always tries to be chased when playing as the prey and tries to chase when playing as a predator, so the agents in slots 1 and 4 will obtain an expected average reward of $-1$ and $1$ respectively. Note that the choice of $\Pi_o$ does not affect the result of $PO_{1,4}$, so no matter which agents are in $\Pi_o$ we obtain:

\begin{equation*}
PG_{1,4}({\pi_{chase}}_1,{\pi_{chase}}_2,{\pi_{chase}}_3,\Pi_o,w_{\dot{L}},\mu) = 1
\end{equation*}

Therefore:

\begin{equation*}
PG_{1,4}(\Pi_e,w_{\Pi_e},\Pi_o,w_{\dot{L}},\mu) = 6 \frac{9}{2} \frac{1}{3} \frac{1}{3} \frac{1}{3} 1 = 1
\end{equation*}

For slots 2 and 1:

\begin{equation*}
\begin{aligned}
PG_{2,1}(\Pi_e,w_{\Pi_e},\Pi_o,w_{\dot{L}},\mu)	& = \eta_{\Pi^3} \sum_{\pi_1,\pi_2,\pi_3 \in \Pi_e | \pi_1 \neq \pi_2 \neq \pi_3} w_{\Pi_e}(\pi_1) w_{\Pi_e}(\pi_2) w_{\Pi_e}(\pi_3) PG_{2,1}(\pi_1,\pi_2,\pi_3,\Pi_o,w_{\dot{L}},\mu) =\\
												& = 6 \frac{9}{2} \frac{1}{3} \frac{1}{3} \frac{1}{3} PG_{2,1}({\pi_{chase}}_1,{\pi_{chase}}_2,{\pi_{chase}}_3,\Pi_o,w_{\dot{L}},\mu)
\end{aligned}
\end{equation*}

In this case, we only need to calculate $PG_{2,1}({\pi_{chase}}_1,{\pi_{chase}}_2,{\pi_{chase}}_3,\Pi_o,w_{\dot{L}},\mu)$. We follow definition \ref{def:STG_agents} (for PG) to calculate this value:

\begin{equation*}
PG_{2,1}({\pi_{chase}}_1,{\pi_{chase}}_2,{\pi_{chase}}_3,\Pi_o,w_{\dot{L}},\mu) = \sum_{\dot{l} \in \dot{L}^{N(\mu)}_{-2,1}(\Pi_o)} w_{\dot{L}}(\dot{l}) PO_{2,1}({\pi_{chase}}_1,{\pi_{chase}}_2,{\pi_{chase}}_3,\dot{l},\mu)
\end{equation*}

Again, we do not know which $\Pi_o$ we have, but we know that we will need to obtain a line-up pattern $\dot{l}$ from $\dot{L}^{N(\mu)}_{-2,1}(\Pi_o)$ to calculate $PO_{2,1}({\pi_{chase}}_1,{\pi_{chase}}_2,{\pi_{chase}}_3,\dot{l},\mu)$. We calculate this value for a figurative line-up pattern $\dot{l} = (*,*,\pi_1,\pi_2)$ from $\dot{L}^{N(\mu)}_{-2,1}(\Pi_o)$:

\begin{equation*}
PO_{2,1}({\pi_{chase}}_1,{\pi_{chase}}_2,{\pi_{chase}}_3,\dot{l},\mu) = PO_{2,1}({\pi_{chase}}_1,{\pi_{chase}}_2,{\pi_{chase}}_3,(*,*,\pi_1,\pi_2),\mu)
\end{equation*}

The following table shows us $PO_{2,1}$ for all the permutations of ${\pi_{chase}}_1,{\pi_{chase}}_2,{\pi_{chase}}_3$.

\begin{center}
\begin{tabular}{c c c | c c c | c c c}
Slot 2 & & Slot 1								& Slot 2 & & Slot 1									& Slot 2 & & Slot 1\\
\hline
${\pi_{chase}}_1$ & $\leq$ & ${\pi_{chase}}_2$	& ${\pi_{chase}}_1$ & $\leq$ & ${\pi_{chase}}_3$	& ${\pi_{chase}}_2$ & $\leq$ & ${\pi_{chase}}_1$\\
${\pi_{chase}}_2$ & $\leq$ & ${\pi_{chase}}_3$	& ${\pi_{chase}}_3$ & $\leq$ & ${\pi_{chase}}_2$	& ${\pi_{chase}}_1$ & $\leq$ & ${\pi_{chase}}_3$\\
${\pi_{chase}}_1$ & $\leq$ & ${\pi_{chase}}_3$	& ${\pi_{chase}}_1$ & $\leq$ & ${\pi_{chase}}_2$	& ${\pi_{chase}}_2$ & $\leq$ & ${\pi_{chase}}_3$
\end{tabular}
\begin{tabular}{c c c | c c c | c c c}
Slot 2 & & Slot 1								& Slot 2 & & Slot 1									& Slot 2 & & Slot 1\\
\hline
${\pi_{chase}}_2$ & $\leq$ & ${\pi_{chase}}_3$	& ${\pi_{chase}}_3$ & $\leq$ & ${\pi_{chase}}_1$	& ${\pi_{chase}}_3$ & $\leq$ & ${\pi_{chase}}_2$\\
${\pi_{chase}}_3$ & $\leq$ & ${\pi_{chase}}_1$	& ${\pi_{chase}}_1$ & $\leq$ & ${\pi_{chase}}_2$	& ${\pi_{chase}}_2$ & $\leq$ & ${\pi_{chase}}_1$\\
${\pi_{chase}}_2$ & $\leq$ & ${\pi_{chase}}_1$	& ${\pi_{chase}}_3$ & $\leq$ & ${\pi_{chase}}_2$	& ${\pi_{chase}}_3$ & $\leq$ & ${\pi_{chase}}_1$
\end{tabular}
\end{center}

It is not possible to find a PO for any permutation, since we always have ${\pi_{chase}}_i \leq {\pi_{chase}}_j$, where a $\pi_{chase}$ agent always tries to be chased when playing as the prey and tries to chase when playing as a predator, so the agents in slots 2 and 1 will obtain an expected average reward of $1$ and $-1$ respectively. Note that the choice of $\Pi_o$ does not affect the result of $PO_{2,1}$, so no matter which agents are in $\Pi_o$ we obtain:

\begin{equation*}
PG_{2,1}({\pi_{chase}}_1,{\pi_{chase}}_2,{\pi_{chase}}_3,\Pi_o,w_{\dot{L}},\mu) = 0
\end{equation*}

Therefore:

\begin{equation*}
PG_{2,1}(\Pi_e,w_{\Pi_e},\Pi_o,w_{\dot{L}},\mu) = 6 \frac{9}{2} \frac{1}{3} \frac{1}{3} \frac{1}{3} 0 = 0
\end{equation*}

For slots 2 and 3:

\begin{equation*}
\begin{aligned}
PG_{2,3}(\Pi_e,w_{\Pi_e},\Pi_o,w_{\dot{L}},\mu)	& = \eta_{\Pi^3} \sum_{\pi_1,\pi_2,\pi_3 \in \Pi_e | \pi_1 \neq \pi_2 \neq \pi_3} w_{\Pi_e}(\pi_1) w_{\Pi_e}(\pi_2) w_{\Pi_e}(\pi_3) PG_{2,3}(\pi_1,\pi_2,\pi_3,\Pi_o,w_{\dot{L}},\mu) =\\
												& = 6 \frac{9}{2} \frac{1}{3} \frac{1}{3} \frac{1}{3} PG_{2,3}({\pi_{chase}}_1,{\pi_{chase}}_2,{\pi_{chase}}_3,\Pi_o,w_{\dot{L}},\mu)
\end{aligned}
\end{equation*}

In this case, we only need to calculate $PG_{2,3}({\pi_{chase}}_1,{\pi_{chase}}_2,{\pi_{chase}}_3,\Pi_o,w_{\dot{L}},\mu)$. We follow definition \ref{def:STG_agents} (for PG) to calculate this value:

\begin{equation*}
PG_{2,3}({\pi_{chase}}_1,{\pi_{chase}}_2,{\pi_{chase}}_3,\Pi_o,w_{\dot{L}},\mu) = \sum_{\dot{l} \in \dot{L}^{N(\mu)}_{-2,3}(\Pi_o)} w_{\dot{L}}(\dot{l}) PO_{2,3}({\pi_{chase}}_1,{\pi_{chase}}_2,{\pi_{chase}}_3,\dot{l},\mu)
\end{equation*}

Again, we do not know which $\Pi_o$ we have, but we know that we will need to obtain a line-up pattern $\dot{l}$ from $\dot{L}^{N(\mu)}_{-2,3}(\Pi_o)$ to calculate $PO_{2,3}({\pi_{chase}}_1,{\pi_{chase}}_2,{\pi_{chase}}_3,\dot{l},\mu)$. We calculate this value for a figurative line-up pattern $\dot{l} = (\pi_1,*,*,\pi_2)$ from $\dot{L}^{N(\mu)}_{-2,3}(\Pi_o)$:

\begin{equation*}
PO_{2,3}({\pi_{chase}}_1,{\pi_{chase}}_2,{\pi_{chase}}_3,\dot{l},\mu) = PO_{2,3}({\pi_{chase}}_1,{\pi_{chase}}_2,{\pi_{chase}}_3,(\pi_1,*,*,\pi_2),\mu)
\end{equation*}

The following table shows us $PO_{2,3}$ for all the permutations of ${\pi_{chase}}_1,{\pi_{chase}}_2,{\pi_{chase}}_3$.

\begin{center}
\begin{tabular}{c c c | c c c | c c c}
Slot 2 & & Slot 3								& Slot 2 & & Slot 3									& Slot 2 & & Slot 3\\
\hline
${\pi_{chase}}_1$ & $\leq$ & ${\pi_{chase}}_2$	& ${\pi_{chase}}_1$ & $\leq$ & ${\pi_{chase}}_3$	& ${\pi_{chase}}_2$ & $\leq$ & ${\pi_{chase}}_1$\\
${\pi_{chase}}_2$ & $\leq$ & ${\pi_{chase}}_3$	& ${\pi_{chase}}_3$ & $\leq$ & ${\pi_{chase}}_2$	& ${\pi_{chase}}_1$ & $\leq$ & ${\pi_{chase}}_3$\\
${\pi_{chase}}_1$ & $\leq$ & ${\pi_{chase}}_3$	& ${\pi_{chase}}_1$ & $\leq$ & ${\pi_{chase}}_2$	& ${\pi_{chase}}_2$ & $\leq$ & ${\pi_{chase}}_3$
\end{tabular}
\begin{tabular}{c c c | c c c | c c c}
Slot 2 & & Slot 3								& Slot 2 & & Slot 3									& Slot 2 & & Slot 3\\
\hline
${\pi_{chase}}_2$ & $\leq$ & ${\pi_{chase}}_3$	& ${\pi_{chase}}_3$ & $\leq$ & ${\pi_{chase}}_1$	& ${\pi_{chase}}_3$ & $\leq$ & ${\pi_{chase}}_2$\\
${\pi_{chase}}_3$ & $\leq$ & ${\pi_{chase}}_1$	& ${\pi_{chase}}_1$ & $\leq$ & ${\pi_{chase}}_2$	& ${\pi_{chase}}_2$ & $\leq$ & ${\pi_{chase}}_1$\\
${\pi_{chase}}_2$ & $\leq$ & ${\pi_{chase}}_1$	& ${\pi_{chase}}_3$ & $\leq$ & ${\pi_{chase}}_2$	& ${\pi_{chase}}_3$ & $\leq$ & ${\pi_{chase}}_1$
\end{tabular}
\end{center}

It is possible to find a PO for every permutation, since the agents in slots 2 and 3 share rewards (and expected average rewards as well) due they are in the same team. So no matter which agents are in $\Pi_o$ we obtain:

\begin{equation*}
PG_{2,3}({\pi_{chase}}_1,{\pi_{chase}}_2,{\pi_{chase}}_3,\Pi_o,w_{\dot{L}},\mu) = 1
\end{equation*}

Therefore:

\begin{equation*}
PG_{2,3}(\Pi_e,w_{\Pi_e},\Pi_o,w_{\dot{L}},\mu) = 6 \frac{9}{2} \frac{1}{3} \frac{1}{3} \frac{1}{3} 1 = 1
\end{equation*}

For slots 2 and 4:

\begin{equation*}
\begin{aligned}
PG_{2,4}(\Pi_e,w_{\Pi_e},\Pi_o,w_{\dot{L}},\mu)	& = \eta_{\Pi^3} \sum_{\pi_1,\pi_2,\pi_3 \in \Pi_e | \pi_1 \neq \pi_2 \neq \pi_3} w_{\Pi_e}(\pi_1) w_{\Pi_e}(\pi_2) w_{\Pi_e}(\pi_3) PG_{2,4}(\pi_1,\pi_2,\pi_3,\Pi_o,w_{\dot{L}},\mu) =\\
												& = 6 \frac{9}{2} \frac{1}{3} \frac{1}{3} \frac{1}{3} PG_{2,4}({\pi_{chase}}_1,{\pi_{chase}}_2,{\pi_{chase}}_3,\Pi_o,w_{\dot{L}},\mu)
\end{aligned}
\end{equation*}

In this case, we only need to calculate $PG_{2,4}({\pi_{chase}}_1,{\pi_{chase}}_2,{\pi_{chase}}_3,\Pi_o,w_{\dot{L}},\mu)$. We follow definition \ref{def:STG_agents} (for PG) to calculate this value:

\begin{equation*}
PG_{2,4}({\pi_{chase}}_1,{\pi_{chase}}_2,{\pi_{chase}}_3,\Pi_o,w_{\dot{L}},\mu) = \sum_{\dot{l} \in \dot{L}^{N(\mu)}_{-2,4}(\Pi_o)} w_{\dot{L}}(\dot{l}) PO_{2,4}({\pi_{chase}}_1,{\pi_{chase}}_2,{\pi_{chase}}_3,\dot{l},\mu)
\end{equation*}

Again, we do not know which $\Pi_o$ we have, but we know that we will need to obtain a line-up pattern $\dot{l}$ from $\dot{L}^{N(\mu)}_{-2,4}(\Pi_o)$ to calculate $PO_{2,4}({\pi_{chase}}_1,{\pi_{chase}}_2,{\pi_{chase}}_3,\dot{l},\mu)$. We calculate this value for a figurative line-up pattern $\dot{l} = (\pi_1,*,\pi_2,*)$ from $\dot{L}^{N(\mu)}_{-2,4}(\Pi_o)$:

\begin{equation*}
PO_{2,4}({\pi_{chase}}_1,{\pi_{chase}}_2,{\pi_{chase}}_3,\dot{l},\mu) = PO_{2,4}({\pi_{chase}}_1,{\pi_{chase}}_2,{\pi_{chase}}_3,(\pi_1,*,\pi_2,*),\mu)
\end{equation*}

The following table shows us $PO_{2,4}$ for all the permutations of ${\pi_{chase}}_1,{\pi_{chase}}_2,{\pi_{chase}}_3$.

\begin{center}
\begin{tabular}{c c c | c c c | c c c}
Slot 2 & & Slot 4								& Slot 2 & & Slot 4									& Slot 2 & & Slot 4\\
\hline
${\pi_{chase}}_1$ & $\leq$ & ${\pi_{chase}}_2$	& ${\pi_{chase}}_1$ & $\leq$ & ${\pi_{chase}}_3$	& ${\pi_{chase}}_2$ & $\leq$ & ${\pi_{chase}}_1$\\
${\pi_{chase}}_2$ & $\leq$ & ${\pi_{chase}}_3$	& ${\pi_{chase}}_3$ & $\leq$ & ${\pi_{chase}}_2$	& ${\pi_{chase}}_1$ & $\leq$ & ${\pi_{chase}}_3$\\
${\pi_{chase}}_1$ & $\leq$ & ${\pi_{chase}}_3$	& ${\pi_{chase}}_1$ & $\leq$ & ${\pi_{chase}}_2$	& ${\pi_{chase}}_2$ & $\leq$ & ${\pi_{chase}}_3$
\end{tabular}
\begin{tabular}{c c c | c c c | c c c}
Slot 2 & & Slot 4								& Slot 2 & & Slot 4									& Slot 2 & & Slot 4\\
\hline
${\pi_{chase}}_2$ & $\leq$ & ${\pi_{chase}}_3$	& ${\pi_{chase}}_3$ & $\leq$ & ${\pi_{chase}}_1$	& ${\pi_{chase}}_3$ & $\leq$ & ${\pi_{chase}}_2$\\
${\pi_{chase}}_3$ & $\leq$ & ${\pi_{chase}}_1$	& ${\pi_{chase}}_1$ & $\leq$ & ${\pi_{chase}}_2$	& ${\pi_{chase}}_2$ & $\leq$ & ${\pi_{chase}}_1$\\
${\pi_{chase}}_2$ & $\leq$ & ${\pi_{chase}}_1$	& ${\pi_{chase}}_3$ & $\leq$ & ${\pi_{chase}}_2$	& ${\pi_{chase}}_3$ & $\leq$ & ${\pi_{chase}}_1$
\end{tabular}
\end{center}

It is possible to find a PO for every permutation, since the agents in slots 2 and 4 share rewards (and expected average rewards as well) due they are in the same team. So no matter which agents are in $\Pi_o$ we obtain:

\begin{equation*}
PG_{2,4}({\pi_{chase}}_1,{\pi_{chase}}_2,{\pi_{chase}}_3,\Pi_o,w_{\dot{L}},\mu) = 1
\end{equation*}

Therefore:

\begin{equation*}
PG_{2,4}(\Pi_e,w_{\Pi_e},\Pi_o,w_{\dot{L}},\mu) = 6 \frac{9}{2} \frac{1}{3} \frac{1}{3} \frac{1}{3} 1 = 1
\end{equation*}

For slots 3 and 1:

\begin{equation*}
\begin{aligned}
PG_{3,1}(\Pi_e,w_{\Pi_e},\Pi_o,w_{\dot{L}},\mu)	& = \eta_{\Pi^3} \sum_{\pi_1,\pi_2,\pi_3 \in \Pi_e | \pi_1 \neq \pi_2 \neq \pi_3} w_{\Pi_e}(\pi_1) w_{\Pi_e}(\pi_2) w_{\Pi_e}(\pi_3) PG_{3,1}(\pi_1,\pi_2,\pi_3,\Pi_o,w_{\dot{L}},\mu) =\\
												& = 6 \frac{9}{2} \frac{1}{3} \frac{1}{3} \frac{1}{3} PG_{3,1}({\pi_{chase}}_1,{\pi_{chase}}_2,{\pi_{chase}}_3,\Pi_o,w_{\dot{L}},\mu)
\end{aligned}
\end{equation*}

In this case, we only need to calculate $PG_{3,1}({\pi_{chase}}_1,{\pi_{chase}}_2,{\pi_{chase}}_3,\Pi_o,w_{\dot{L}},\mu)$. We follow definition \ref{def:STG_agents} (for PG) to calculate this value:

\begin{equation*}
PG_{3,1}({\pi_{chase}}_1,{\pi_{chase}}_2,{\pi_{chase}}_3,\Pi_o,w_{\dot{L}},\mu) = \sum_{\dot{l} \in \dot{L}^{N(\mu)}_{-3,1}(\Pi_o)} w_{\dot{L}}(\dot{l}) PO_{3,1}({\pi_{chase}}_1,{\pi_{chase}}_2,{\pi_{chase}}_3,\dot{l},\mu)
\end{equation*}

Again, we do not know which $\Pi_o$ we have, but we know that we will need to obtain a line-up pattern $\dot{l}$ from $\dot{L}^{N(\mu)}_{-3,1}(\Pi_o)$ to calculate $PO_{3,1}({\pi_{chase}}_1,{\pi_{chase}}_2,{\pi_{chase}}_3,\dot{l},\mu)$. We calculate this value for a figurative line-up pattern $\dot{l} = (*,\pi_1,*,\pi_2)$ from $\dot{L}^{N(\mu)}_{-3,1}(\Pi_o)$:

\begin{equation*}
PO_{3,1}({\pi_{chase}}_1,{\pi_{chase}}_2,{\pi_{chase}}_3,\dot{l},\mu) = PO_{3,1}({\pi_{chase}}_1,{\pi_{chase}}_2,{\pi_{chase}}_3,(*,\pi_1,*,\pi_2),\mu)
\end{equation*}

The following table shows us $PO_{3,1}$ for all the permutations of ${\pi_{chase}}_1,{\pi_{chase}}_2,{\pi_{chase}}_3$.

\begin{center}
\begin{tabular}{c c c | c c c | c c c}
Slot 3 & & Slot 1								& Slot 3 & & Slot 1									& Slot 3 & & Slot 1\\
\hline
${\pi_{chase}}_1$ & $\leq$ & ${\pi_{chase}}_2$	& ${\pi_{chase}}_1$ & $\leq$ & ${\pi_{chase}}_3$	& ${\pi_{chase}}_2$ & $\leq$ & ${\pi_{chase}}_1$\\
${\pi_{chase}}_2$ & $\leq$ & ${\pi_{chase}}_3$	& ${\pi_{chase}}_3$ & $\leq$ & ${\pi_{chase}}_2$	& ${\pi_{chase}}_1$ & $\leq$ & ${\pi_{chase}}_3$\\
${\pi_{chase}}_1$ & $\leq$ & ${\pi_{chase}}_3$	& ${\pi_{chase}}_1$ & $\leq$ & ${\pi_{chase}}_2$	& ${\pi_{chase}}_2$ & $\leq$ & ${\pi_{chase}}_3$
\end{tabular}
\begin{tabular}{c c c | c c c | c c c}
Slot 3 & & Slot 1								& Slot 3 & & Slot 1									& Slot 3 & & Slot 1\\
\hline
${\pi_{chase}}_2$ & $\leq$ & ${\pi_{chase}}_3$	& ${\pi_{chase}}_3$ & $\leq$ & ${\pi_{chase}}_1$	& ${\pi_{chase}}_3$ & $\leq$ & ${\pi_{chase}}_2$\\
${\pi_{chase}}_3$ & $\leq$ & ${\pi_{chase}}_1$	& ${\pi_{chase}}_1$ & $\leq$ & ${\pi_{chase}}_2$	& ${\pi_{chase}}_2$ & $\leq$ & ${\pi_{chase}}_1$\\
${\pi_{chase}}_2$ & $\leq$ & ${\pi_{chase}}_1$	& ${\pi_{chase}}_3$ & $\leq$ & ${\pi_{chase}}_2$	& ${\pi_{chase}}_3$ & $\leq$ & ${\pi_{chase}}_1$
\end{tabular}
\end{center}

Again, it is not possible to find a PO for any permutation, since we always have ${\pi_{chase}}_i \leq {\pi_{chase}}_j$, where a $\pi_{chase}$ agent always tries to be chased when playing as the prey and tries to chase when playing as a predator, so the agents in slots 3 and 1 will obtain an expected average reward of $1$ and $-1$ respectively. Note that the choice of $\Pi_o$ does not affect the result of $PO_{3,1}$, so no matter which agents are in $\Pi_o$ we obtain:

\begin{equation*}
PG_{3,1}({\pi_{chase}}_1,{\pi_{chase}}_2,{\pi_{chase}}_3,\Pi_o,w_{\dot{L}},\mu) = 0
\end{equation*}

Therefore:

\begin{equation*}
PG_{3,1}(\Pi_e,w_{\Pi_e},\Pi_o,w_{\dot{L}},\mu) = 6 \frac{9}{2} \frac{1}{3} \frac{1}{3} \frac{1}{3} 1 = 0
\end{equation*}

For slots 3 and 2:

\begin{equation*}
\begin{aligned}
PG_{3,2}(\Pi_e,w_{\Pi_e},\Pi_o,w_{\dot{L}},\mu)	& = \eta_{\Pi^3} \sum_{\pi_1,\pi_2,\pi_3 \in \Pi_e | \pi_1 \neq \pi_2 \neq \pi_3} w_{\Pi_e}(\pi_1) w_{\Pi_e}(\pi_2) w_{\Pi_e}(\pi_3) PG_{3,2}(\pi_1,\pi_2,\pi_3,\Pi_o,w_{\dot{L}},\mu) =\\
												& = 6 \frac{9}{2} \frac{1}{3} \frac{1}{3} \frac{1}{3} PG_{3,2}({\pi_{chase}}_1,{\pi_{chase}}_2,{\pi_{chase}}_3,\Pi_o,w_{\dot{L}},\mu)
\end{aligned}
\end{equation*}

In this case, we only need to calculate $PG_{3,2}({\pi_{chase}}_1,{\pi_{chase}}_2,{\pi_{chase}}_3,\Pi_o,w_{\dot{L}},\mu)$. We follow definition \ref{def:STG_agents} (for PG) to calculate this value:

\begin{equation*}
PG_{3,2}({\pi_{chase}}_1,{\pi_{chase}}_2,{\pi_{chase}}_3,\Pi_o,w_{\dot{L}},\mu) = \sum_{\dot{l} \in \dot{L}^{N(\mu)}_{-3,2}(\Pi_o)} w_{\dot{L}}(\dot{l}) PO_{3,2}({\pi_{chase}}_1,{\pi_{chase}}_2,{\pi_{chase}}_3,\dot{l},\mu)
\end{equation*}

Again, we do not know which $\Pi_o$ we have, but we know that we will need to obtain a line-up pattern $\dot{l}$ from $\dot{L}^{N(\mu)}_{-3,2}(\Pi_o)$ to calculate $PO_{3,2}({\pi_{chase}}_1,{\pi_{chase}}_2,{\pi_{chase}}_3,\dot{l},\mu)$. We calculate this value for a figurative line-up pattern $\dot{l} = (\pi_1,*,*,\pi_2)$ from $\dot{L}^{N(\mu)}_{-3,2}(\Pi_o)$:

\begin{equation*}
PO_{3,2}({\pi_{chase}}_1,{\pi_{chase}}_2,{\pi_{chase}}_3,\dot{l},\mu) = PO_{3,2}({\pi_{chase}}_1,{\pi_{chase}}_2,{\pi_{chase}}_3,(\pi_1,*,*,\pi_2),\mu)
\end{equation*}

The following table shows us $PO_{3,2}$ for all the permutations of ${\pi_{chase}}_1,{\pi_{chase}}_2,{\pi_{chase}}_3$.

\begin{center}
\begin{tabular}{c c c | c c c | c c c}
Slot 3 & & Slot 2								& Slot 3 & & Slot 2									& Slot 3 & & Slot 2\\
\hline
${\pi_{chase}}_1$ & $\leq$ & ${\pi_{chase}}_2$	& ${\pi_{chase}}_1$ & $\leq$ & ${\pi_{chase}}_3$	& ${\pi_{chase}}_2$ & $\leq$ & ${\pi_{chase}}_1$\\
${\pi_{chase}}_2$ & $\leq$ & ${\pi_{chase}}_3$	& ${\pi_{chase}}_3$ & $\leq$ & ${\pi_{chase}}_2$	& ${\pi_{chase}}_1$ & $\leq$ & ${\pi_{chase}}_3$\\
${\pi_{chase}}_1$ & $\leq$ & ${\pi_{chase}}_3$	& ${\pi_{chase}}_1$ & $\leq$ & ${\pi_{chase}}_2$	& ${\pi_{chase}}_2$ & $\leq$ & ${\pi_{chase}}_3$
\end{tabular}
\begin{tabular}{c c c | c c c | c c c}
Slot 3 & & Slot 2								& Slot 3 & & Slot 2									& Slot 3 & & Slot 2\\
\hline
${\pi_{chase}}_2$ & $\leq$ & ${\pi_{chase}}_3$	& ${\pi_{chase}}_3$ & $\leq$ & ${\pi_{chase}}_1$	& ${\pi_{chase}}_3$ & $\leq$ & ${\pi_{chase}}_2$\\
${\pi_{chase}}_3$ & $\leq$ & ${\pi_{chase}}_1$	& ${\pi_{chase}}_1$ & $\leq$ & ${\pi_{chase}}_2$	& ${\pi_{chase}}_2$ & $\leq$ & ${\pi_{chase}}_1$\\
${\pi_{chase}}_2$ & $\leq$ & ${\pi_{chase}}_1$	& ${\pi_{chase}}_3$ & $\leq$ & ${\pi_{chase}}_2$	& ${\pi_{chase}}_3$ & $\leq$ & ${\pi_{chase}}_1$
\end{tabular}
\end{center}

It is possible to find a PO for every permutation, since the agents in slots 3 and 2 share rewards (and expected average rewards as well) due they are in the same team. So no matter which agents are in $\Pi_o$ we obtain:

\begin{equation*}
PG_{3,2}({\pi_{chase}}_1,{\pi_{chase}}_2,{\pi_{chase}}_3,\Pi_o,w_{\dot{L}},\mu) = 1
\end{equation*}

Therefore:

\begin{equation*}
PG_{3,2}(\Pi_e,w_{\Pi_e},\Pi_o,w_{\dot{L}},\mu) = 6 \frac{9}{2} \frac{1}{3} \frac{1}{3} \frac{1}{3} 1 = 1
\end{equation*}

For slots 3 and 4:

\begin{equation*}
\begin{aligned}
PG_{3,4}(\Pi_e,w_{\Pi_e},\Pi_o,w_{\dot{L}},\mu)	& = \eta_{\Pi^3} \sum_{\pi_1,\pi_2,\pi_3 \in \Pi_e | \pi_1 \neq \pi_2 \neq \pi_3} w_{\Pi_e}(\pi_1) w_{\Pi_e}(\pi_2) w_{\Pi_e}(\pi_3) PG_{3,4}(\pi_1,\pi_2,\pi_3,\Pi_o,w_{\dot{L}},\mu) =\\
												& = 6 \frac{9}{2} \frac{1}{3} \frac{1}{3} \frac{1}{3} PG_{3,4}({\pi_{chase}}_1,{\pi_{chase}}_2,{\pi_{chase}}_3,\Pi_o,w_{\dot{L}},\mu)
\end{aligned}
\end{equation*}

In this case, we only need to calculate $PG_{3,4}({\pi_{chase}}_1,{\pi_{chase}}_2,{\pi_{chase}}_3,\Pi_o,w_{\dot{L}},\mu)$. We follow definition \ref{def:STG_agents} (for PG) to calculate this value:

\begin{equation*}
PG_{3,4}({\pi_{chase}}_1,{\pi_{chase}}_2,{\pi_{chase}}_3,\Pi_o,w_{\dot{L}},\mu) = \sum_{\dot{l} \in \dot{L}^{N(\mu)}_{-3,4}(\Pi_o)} w_{\dot{L}}(\dot{l}) PO_{3,4}({\pi_{chase}}_1,{\pi_{chase}}_2,{\pi_{chase}}_3,\dot{l},\mu)
\end{equation*}

Again, we do not know which $\Pi_o$ we have, but we know that we will need to obtain a line-up pattern $\dot{l}$ from $\dot{L}^{N(\mu)}_{-3,4}(\Pi_o)$ to calculate $PO_{3,4}({\pi_{chase}}_1,{\pi_{chase}}_2,{\pi_{chase}}_3,\dot{l},\mu)$. We calculate this value for a figurative line-up pattern $\dot{l} = (\pi_1,\pi_2,*,*)$ from $\dot{L}^{N(\mu)}_{-3,4}(\Pi_o)$:

\begin{equation*}
PO_{3,4}({\pi_{chase}}_1,{\pi_{chase}}_2,{\pi_{chase}}_3,\dot{l},\mu) = PO_{3,4}({\pi_{chase}}_1,{\pi_{chase}}_2,{\pi_{chase}}_3,(\pi_1,\pi_2,*,*),\mu)
\end{equation*}

The following table shows us $PO_{3,4}$ for all the permutations of ${\pi_{chase}}_1,{\pi_{chase}}_2,{\pi_{chase}}_3$.

\begin{center}
\begin{tabular}{c c c | c c c | c c c}
Slot 3 & & Slot 4								& Slot 3 & & Slot 4									& Slot 3 & & Slot 4\\
\hline
${\pi_{chase}}_1$ & $\leq$ & ${\pi_{chase}}_2$	& ${\pi_{chase}}_1$ & $\leq$ & ${\pi_{chase}}_3$	& ${\pi_{chase}}_2$ & $\leq$ & ${\pi_{chase}}_1$\\
${\pi_{chase}}_2$ & $\leq$ & ${\pi_{chase}}_3$	& ${\pi_{chase}}_3$ & $\leq$ & ${\pi_{chase}}_2$	& ${\pi_{chase}}_1$ & $\leq$ & ${\pi_{chase}}_3$\\
${\pi_{chase}}_1$ & $\leq$ & ${\pi_{chase}}_3$	& ${\pi_{chase}}_1$ & $\leq$ & ${\pi_{chase}}_2$	& ${\pi_{chase}}_2$ & $\leq$ & ${\pi_{chase}}_3$
\end{tabular}
\begin{tabular}{c c c | c c c | c c c}
Slot 3 & & Slot 4								& Slot 3 & & Slot 4									& Slot 3 & & Slot 4\\
\hline
${\pi_{chase}}_2$ & $\leq$ & ${\pi_{chase}}_3$	& ${\pi_{chase}}_3$ & $\leq$ & ${\pi_{chase}}_1$	& ${\pi_{chase}}_3$ & $\leq$ & ${\pi_{chase}}_2$\\
${\pi_{chase}}_3$ & $\leq$ & ${\pi_{chase}}_1$	& ${\pi_{chase}}_1$ & $\leq$ & ${\pi_{chase}}_2$	& ${\pi_{chase}}_2$ & $\leq$ & ${\pi_{chase}}_1$\\
${\pi_{chase}}_2$ & $\leq$ & ${\pi_{chase}}_1$	& ${\pi_{chase}}_3$ & $\leq$ & ${\pi_{chase}}_2$	& ${\pi_{chase}}_3$ & $\leq$ & ${\pi_{chase}}_1$
\end{tabular}
\end{center}

It is possible to find a PO for every permutation, since the agents in slots 3 and 4 share rewards (and expected average rewards as well) due they are in the same team. So no matter which agents are in $\Pi_o$ we obtain:

\begin{equation*}
PG_{3,4}({\pi_{chase}}_1,{\pi_{chase}}_2,{\pi_{chase}}_3,\Pi_o,w_{\dot{L}},\mu) = 1
\end{equation*}

Therefore:

\begin{equation*}
PG_{3,4}(\Pi_e,w_{\Pi_e},\Pi_o,w_{\dot{L}},\mu) = 6 \frac{9}{2} \frac{1}{3} \frac{1}{3} \frac{1}{3} 1 = 1
\end{equation*}

For slots 4 and 1:

\begin{equation*}
\begin{aligned}
PG_{4,1}(\Pi_e,w_{\Pi_e},\Pi_o,w_{\dot{L}},\mu)	& = \eta_{\Pi^3} \sum_{\pi_1,\pi_2,\pi_3 \in \Pi_e | \pi_1 \neq \pi_2 \neq \pi_3} w_{\Pi_e}(\pi_1) w_{\Pi_e}(\pi_2) w_{\Pi_e}(\pi_3) PG_{4,1}(\pi_1,\pi_2,\pi_3,\Pi_o,w_{\dot{L}},\mu) =\\
												& = 6 \frac{9}{2} \frac{1}{3} \frac{1}{3} \frac{1}{3} PG_{4,1}({\pi_{chase}}_1,{\pi_{chase}}_2,{\pi_{chase}}_3,\Pi_o,w_{\dot{L}},\mu)
\end{aligned}
\end{equation*}

In this case, we only need to calculate $PG_{4,1}({\pi_{chase}}_1,{\pi_{chase}}_2,{\pi_{chase}}_3,\Pi_o,w_{\dot{L}},\mu)$. We follow definition \ref{def:STG_agents} (for PG) to calculate this value:

\begin{equation*}
PG_{4,1}({\pi_{chase}}_1,{\pi_{chase}}_2,{\pi_{chase}}_3,\Pi_o,w_{\dot{L}},\mu) = \sum_{\dot{l} \in \dot{L}^{N(\mu)}_{-4,1}(\Pi_o)} w_{\dot{L}}(\dot{l}) PO_{4,1}({\pi_{chase}}_1,{\pi_{chase}}_2,{\pi_{chase}}_3,\dot{l},\mu)
\end{equation*}

Again, we do not know which $\Pi_o$ we have, but we know that we will need to obtain a line-up pattern $\dot{l}$ from $\dot{L}^{N(\mu)}_{-4,1}(\Pi_o)$ to calculate $PO_{4,1}({\pi_{chase}}_1,{\pi_{chase}}_2,{\pi_{chase}}_3,\dot{l},\mu)$. We calculate this value for a figurative line-up pattern $\dot{l} = (*,\pi_1,\pi_2,*)$ from $\dot{L}^{N(\mu)}_{-4,1}(\Pi_o)$:

\begin{equation*}
PO_{4,1}({\pi_{chase}}_1,{\pi_{chase}}_2,{\pi_{chase}}_3,\dot{l},\mu) = PO_{4,1}({\pi_{chase}}_1,{\pi_{chase}}_2,{\pi_{chase}}_3,(*,\pi_1,\pi_2,*),\mu)
\end{equation*}

The following table shows us $PO_{4,1}$ for all the permutations of ${\pi_{chase}}_1,{\pi_{chase}}_2,{\pi_{chase}}_3$.

\begin{center}
\begin{tabular}{c c c | c c c | c c c}
Slot 4 & & Slot 1								& Slot 4 & & Slot 1									& Slot 4 & & Slot 1\\
\hline
${\pi_{chase}}_1$ & $\leq$ & ${\pi_{chase}}_2$	& ${\pi_{chase}}_1$ & $\leq$ & ${\pi_{chase}}_3$	& ${\pi_{chase}}_2$ & $\leq$ & ${\pi_{chase}}_1$\\
${\pi_{chase}}_2$ & $\leq$ & ${\pi_{chase}}_3$	& ${\pi_{chase}}_3$ & $\leq$ & ${\pi_{chase}}_2$	& ${\pi_{chase}}_1$ & $\leq$ & ${\pi_{chase}}_3$\\
${\pi_{chase}}_1$ & $\leq$ & ${\pi_{chase}}_3$	& ${\pi_{chase}}_1$ & $\leq$ & ${\pi_{chase}}_2$	& ${\pi_{chase}}_2$ & $\leq$ & ${\pi_{chase}}_3$
\end{tabular}
\begin{tabular}{c c c | c c c | c c c}
Slot 4 & & Slot 1								& Slot 4 & & Slot 1									& Slot 4 & & Slot 1\\
\hline
${\pi_{chase}}_2$ & $\leq$ & ${\pi_{chase}}_3$	& ${\pi_{chase}}_3$ & $\leq$ & ${\pi_{chase}}_1$	& ${\pi_{chase}}_3$ & $\leq$ & ${\pi_{chase}}_2$\\
${\pi_{chase}}_3$ & $\leq$ & ${\pi_{chase}}_1$	& ${\pi_{chase}}_1$ & $\leq$ & ${\pi_{chase}}_2$	& ${\pi_{chase}}_2$ & $\leq$ & ${\pi_{chase}}_1$\\
${\pi_{chase}}_2$ & $\leq$ & ${\pi_{chase}}_1$	& ${\pi_{chase}}_3$ & $\leq$ & ${\pi_{chase}}_2$	& ${\pi_{chase}}_3$ & $\leq$ & ${\pi_{chase}}_1$
\end{tabular}
\end{center}

Again, it is not possible to find a PO for any permutation, since we always have ${\pi_{chase}}_i \leq {\pi_{chase}}_j$, where a $\pi_{chase}$ agent always tries to be chased when playing as the prey and tries to chase when playing as a predator, so the agents in slots 4 and 1 will obtain an expected average reward of $1$ and $-1$ respectively. Note that the choice of $\Pi_o$ does not affect the result of $PO_{4,1}$, so no matter which agents are in $\Pi_o$ we obtain:

\begin{equation*}
PG_{4,1}({\pi_{chase}}_1,{\pi_{chase}}_2,{\pi_{chase}}_3,\Pi_o,w_{\dot{L}},\mu) = 0
\end{equation*}

Therefore:

\begin{equation*}
PG_{4,1}(\Pi_e,w_{\Pi_e},\Pi_o,w_{\dot{L}},\mu) = 6 \frac{9}{2} \frac{1}{3} \frac{1}{3} \frac{1}{3} 1 = 0
\end{equation*}

For slots 4 and 2:

\begin{equation*}
\begin{aligned}
PG_{4,2}(\Pi_e,w_{\Pi_e},\Pi_o,w_{\dot{L}},\mu)	& = \eta_{\Pi^3} \sum_{\pi_1,\pi_2,\pi_3 \in \Pi_e | \pi_1 \neq \pi_2 \neq \pi_3} w_{\Pi_e}(\pi_1) w_{\Pi_e}(\pi_2) w_{\Pi_e}(\pi_3) PG_{4,2}(\pi_1,\pi_2,\pi_3,\Pi_o,w_{\dot{L}},\mu) =\\
												& = 6 \frac{9}{2} \frac{1}{3} \frac{1}{3} \frac{1}{3} PG_{4,2}({\pi_{chase}}_1,{\pi_{chase}}_2,{\pi_{chase}}_3,\Pi_o,w_{\dot{L}},\mu)
\end{aligned}
\end{equation*}

In this case, we only need to calculate $PG_{4,2}({\pi_{chase}}_1,{\pi_{chase}}_2,{\pi_{chase}}_3,\Pi_o,w_{\dot{L}},\mu)$. We follow definition \ref{def:STG_agents} (for PG) to calculate this value:

\begin{equation*}
PG_{4,2}({\pi_{chase}}_1,{\pi_{chase}}_2,{\pi_{chase}}_3,\Pi_o,w_{\dot{L}},\mu) = \sum_{\dot{l} \in \dot{L}^{N(\mu)}_{-4,2}(\Pi_o)} w_{\dot{L}}(\dot{l}) PO_{4,2}({\pi_{chase}}_1,{\pi_{chase}}_2,{\pi_{chase}}_3,\dot{l},\mu)
\end{equation*}

Again, we do not know which $\Pi_o$ we have, but we know that we will need to obtain a line-up pattern $\dot{l}$ from $\dot{L}^{N(\mu)}_{-4,2}(\Pi_o)$ to calculate $PO_{4,2}({\pi_{chase}}_1,{\pi_{chase}}_2,{\pi_{chase}}_3,\dot{l},\mu)$. We calculate this value for a figurative line-up pattern $\dot{l} = (\pi_1,*,\pi_2,*)$ from $\dot{L}^{N(\mu)}_{-4,2}(\Pi_o)$:

\begin{equation*}
PO_{4,2}({\pi_{chase}}_1,{\pi_{chase}}_2,{\pi_{chase}}_3,\dot{l},\mu) = PO_{4,2}({\pi_{chase}}_1,{\pi_{chase}}_2,{\pi_{chase}}_3,(\pi_1,*,\pi_2,*),\mu)
\end{equation*}

The following table shows us $PO_{4,2}$ for all the permutations of ${\pi_{chase}}_1,{\pi_{chase}}_2,{\pi_{chase}}_3$.

\begin{center}
\begin{tabular}{c c c | c c c | c c c}
Slot 4 & & Slot 2								& Slot 4 & & Slot 2									& Slot 4 & & Slot 2\\
\hline
${\pi_{chase}}_1$ & $\leq$ & ${\pi_{chase}}_2$	& ${\pi_{chase}}_1$ & $\leq$ & ${\pi_{chase}}_3$	& ${\pi_{chase}}_2$ & $\leq$ & ${\pi_{chase}}_1$\\
${\pi_{chase}}_2$ & $\leq$ & ${\pi_{chase}}_3$	& ${\pi_{chase}}_3$ & $\leq$ & ${\pi_{chase}}_2$	& ${\pi_{chase}}_1$ & $\leq$ & ${\pi_{chase}}_3$\\
${\pi_{chase}}_1$ & $\leq$ & ${\pi_{chase}}_3$	& ${\pi_{chase}}_1$ & $\leq$ & ${\pi_{chase}}_2$	& ${\pi_{chase}}_2$ & $\leq$ & ${\pi_{chase}}_3$
\end{tabular}
\begin{tabular}{c c c | c c c | c c c}
Slot 4 & & Slot 2								& Slot 4 & & Slot 2									& Slot 4 & & Slot 2\\
\hline
${\pi_{chase}}_2$ & $\leq$ & ${\pi_{chase}}_3$	& ${\pi_{chase}}_3$ & $\leq$ & ${\pi_{chase}}_1$	& ${\pi_{chase}}_3$ & $\leq$ & ${\pi_{chase}}_2$\\
${\pi_{chase}}_3$ & $\leq$ & ${\pi_{chase}}_1$	& ${\pi_{chase}}_1$ & $\leq$ & ${\pi_{chase}}_2$	& ${\pi_{chase}}_2$ & $\leq$ & ${\pi_{chase}}_1$\\
${\pi_{chase}}_2$ & $\leq$ & ${\pi_{chase}}_1$	& ${\pi_{chase}}_3$ & $\leq$ & ${\pi_{chase}}_2$	& ${\pi_{chase}}_3$ & $\leq$ & ${\pi_{chase}}_1$
\end{tabular}
\end{center}

It is possible to find a PO for every permutation, since the agents in slots 4 and 2 share rewards (and expected average rewards as well) due they are in the same team. So no matter which agents are in $\Pi_o$ we obtain:

\begin{equation*}
PG_{4,2}({\pi_{chase}}_1,{\pi_{chase}}_2,{\pi_{chase}}_3,\Pi_o,w_{\dot{L}},\mu) = 1
\end{equation*}

Therefore:

\begin{equation*}
PG_{4,2}(\Pi_e,w_{\Pi_e},\Pi_o,w_{\dot{L}},\mu) = 6 \frac{9}{2} \frac{1}{3} \frac{1}{3} \frac{1}{3} 1 = 1
\end{equation*}

And for slots 4 and 3:

\begin{equation*}
\begin{aligned}
PG_{4,3}(\Pi_e,w_{\Pi_e},\Pi_o,w_{\dot{L}},\mu)	& = \eta_{\Pi^3} \sum_{\pi_1,\pi_2,\pi_3 \in \Pi_e | \pi_1 \neq \pi_2 \neq \pi_3} w_{\Pi_e}(\pi_1) w_{\Pi_e}(\pi_2) w_{\Pi_e}(\pi_3) PG_{4,3}(\pi_1,\pi_2,\pi_3,\Pi_o,w_{\dot{L}},\mu) =\\
												& = 6 \frac{9}{2} \frac{1}{3} \frac{1}{3} \frac{1}{3} PG_{4,3}({\pi_{chase}}_1,{\pi_{chase}}_2,{\pi_{chase}}_3,\Pi_o,w_{\dot{L}},\mu)
\end{aligned}
\end{equation*}

In this case, we only need to calculate $PG_{4,3}({\pi_{chase}}_1,{\pi_{chase}}_2,{\pi_{chase}}_3,\Pi_o,w_{\dot{L}},\mu)$. We follow definition \ref{def:STG_agents} (for PG) to calculate this value:

\begin{equation*}
PG_{4,3}({\pi_{chase}}_1,{\pi_{chase}}_2,{\pi_{chase}}_3,\Pi_o,w_{\dot{L}},\mu) = \sum_{\dot{l} \in \dot{L}^{N(\mu)}_{-4,3}(\Pi_o)} w_{\dot{L}}(\dot{l}) PO_{4,3}({\pi_{chase}}_1,{\pi_{chase}}_2,{\pi_{chase}}_3,\dot{l},\mu)
\end{equation*}

Again, we do not know which $\Pi_o$ we have, but we know that we will need to obtain a line-up pattern $\dot{l}$ from $\dot{L}^{N(\mu)}_{-4,3}(\Pi_o)$ to calculate $PO_{4,3}({\pi_{chase}}_1,{\pi_{chase}}_2,{\pi_{chase}}_3,\dot{l},\mu)$. We calculate this value for a figurative line-up pattern $\dot{l} = (\pi_1,\pi_2,*,*)$ from $\dot{L}^{N(\mu)}_{-4,3}(\Pi_o)$:

\begin{equation*}
PO_{4,3}({\pi_{chase}}_1,{\pi_{chase}}_2,{\pi_{chase}}_3,\dot{l},\mu) = PO_{4,3}({\pi_{chase}}_1,{\pi_{chase}}_2,{\pi_{chase}}_3,(\pi_1,\pi_2,*,*),\mu)
\end{equation*}

The following table shows us $PO_{4,3}$ for all the permutations of ${\pi_{chase}}_1,{\pi_{chase}}_2,{\pi_{chase}}_3$.

\begin{center}
\begin{tabular}{c c c | c c c | c c c}
Slot 4 & & Slot 3								& Slot 4 & & Slot 3									& Slot 4 & & Slot 3\\
\hline
${\pi_{chase}}_1$ & $\leq$ & ${\pi_{chase}}_2$	& ${\pi_{chase}}_1$ & $\leq$ & ${\pi_{chase}}_3$	& ${\pi_{chase}}_2$ & $\leq$ & ${\pi_{chase}}_1$\\
${\pi_{chase}}_2$ & $\leq$ & ${\pi_{chase}}_3$	& ${\pi_{chase}}_3$ & $\leq$ & ${\pi_{chase}}_2$	& ${\pi_{chase}}_1$ & $\leq$ & ${\pi_{chase}}_3$\\
${\pi_{chase}}_1$ & $\leq$ & ${\pi_{chase}}_3$	& ${\pi_{chase}}_1$ & $\leq$ & ${\pi_{chase}}_2$	& ${\pi_{chase}}_2$ & $\leq$ & ${\pi_{chase}}_3$
\end{tabular}
\begin{tabular}{c c c | c c c | c c c}
Slot 4 & & Slot 3								& Slot 4 & & Slot 3									& Slot 4 & & Slot 3\\
\hline
${\pi_{chase}}_2$ & $\leq$ & ${\pi_{chase}}_3$	& ${\pi_{chase}}_3$ & $\leq$ & ${\pi_{chase}}_1$	& ${\pi_{chase}}_3$ & $\leq$ & ${\pi_{chase}}_2$\\
${\pi_{chase}}_3$ & $\leq$ & ${\pi_{chase}}_1$	& ${\pi_{chase}}_1$ & $\leq$ & ${\pi_{chase}}_2$	& ${\pi_{chase}}_2$ & $\leq$ & ${\pi_{chase}}_1$\\
${\pi_{chase}}_2$ & $\leq$ & ${\pi_{chase}}_1$	& ${\pi_{chase}}_3$ & $\leq$ & ${\pi_{chase}}_2$	& ${\pi_{chase}}_3$ & $\leq$ & ${\pi_{chase}}_1$
\end{tabular}
\end{center}

It is possible to find a PO for every permutation, since the agents in slots 4 and 3 share rewards (and expected average rewards as well) due they are in the same team. So no matter which agents are in $\Pi_o$ we obtain:

\begin{equation*}
PG_{4,3}({\pi_{chase}}_1,{\pi_{chase}}_2,{\pi_{chase}}_3,\Pi_o,w_{\dot{L}},\mu) = 1
\end{equation*}

Therefore:

\begin{equation*}
PG_{4,3}(\Pi_e,w_{\Pi_e},\Pi_o,w_{\dot{L}},\mu) = 6 \frac{9}{2} \frac{1}{3} \frac{1}{3} \frac{1}{3} 1 = 1
\end{equation*}

And finally, we weight over the slots:

\begin{equation*}
\begin{aligned}
& PG(\Pi_e,w_{\Pi_e},\Pi_o,w_{\dot{L}},\mu,w_S) = \eta_{S_1^2} \sum_{i=1}^{N(\mu)} w_S(i,\mu) \times\\
& \times \left(\sum_{j=1}^{i-1} w_S(j,\mu) PG_{i,j}(\Pi_e,w_{\Pi_e},\Pi_o,w_{\dot{L}},\mu) + \sum_{j=i+1}^{N(\mu)} w_S(j,\mu) PG_{i,j}(\Pi_e,w_{\Pi_e},\Pi_o,w_{\dot{L}},\mu)\right) =\\
& \ \ \ \ \ \ \ \ \ \ \ \ \ \ \ \ \ \ \ \ \ \ \ \ \ \ \ \ \ \ \ \ \ \ \ \ = \frac{4}{3} \frac{1}{4} \frac{1}{4} \{PG_{1,2}(\Pi_e,w_{\Pi_e},\Pi_o,w_{\dot{L}},\mu) + PG_{1,3}(\Pi_e,w_{\Pi_e},\Pi_o,w_{\dot{L}},\mu) +\\
& \ \ \ \ \ \ \ \ \ \ \ \ \ \ \ \ \ \ \ \ \ \ \ \ \ \ \ \ \ \ \ \ \ \ \ \ + PG_{1,4}(\Pi_e,w_{\Pi_e},\Pi_o,w_{\dot{L}},\mu) + PG_{2,1}(\Pi_e,w_{\Pi_e},\Pi_o,w_{\dot{L}},\mu) +\\
& \ \ \ \ \ \ \ \ \ \ \ \ \ \ \ \ \ \ \ \ \ \ \ \ \ \ \ \ \ \ \ \ \ \ \ \ + PG_{2,3}(\Pi_e,w_{\Pi_e},\Pi_o,w_{\dot{L}},\mu) + PG_{2,4}(\Pi_e,w_{\Pi_e},\Pi_o,w_{\dot{L}},\mu) +\\
& \ \ \ \ \ \ \ \ \ \ \ \ \ \ \ \ \ \ \ \ \ \ \ \ \ \ \ \ \ \ \ \ \ \ \ \ + PG_{3,1}(\Pi_e,w_{\Pi_e},\Pi_o,w_{\dot{L}},\mu) + PG_{3,2}(\Pi_e,w_{\Pi_e},\Pi_o,w_{\dot{L}},\mu) +\\
& \ \ \ \ \ \ \ \ \ \ \ \ \ \ \ \ \ \ \ \ \ \ \ \ \ \ \ \ \ \ \ \ \ \ \ \ + PG_{3,4}(\Pi_e,w_{\Pi_e},\Pi_o,w_{\dot{L}},\mu) + PG_{4,1}(\Pi_e,w_{\Pi_e},\Pi_o,w_{\dot{L}},\mu) +\\
& \ \ \ \ \ \ \ \ \ \ \ \ \ \ \ \ \ \ \ \ \ \ \ \ \ \ \ \ \ \ \ \ \ \ \ \ + PG_{4,2}(\Pi_e,w_{\Pi_e},\Pi_o,w_{\dot{L}},\mu) + PG_{4,3}(\Pi_e,w_{\Pi_e},\Pi_o,w_{\dot{L}},\mu)\} =\\
& \ \ \ \ \ \ \ \ \ \ \ \ \ \ \ \ \ \ \ \ \ \ \ \ \ \ \ \ \ \ \ \ \ \ \ \ = \frac{4}{3} \frac{1}{4} \frac{1}{4} \left\{9 \times 1 + 3 \times 0\right\} = \frac{3}{4}
\end{aligned}
\end{equation*}

So, for every $\Pi_o$ we obtain the same result:

\begin{equation*}
\forall \Pi_o : PG(\Pi_e,w_{\Pi_e},\Pi_o,w_{\dot{L}},\mu,w_S) = \frac{3}{4}
\end{equation*}

Therefore, predator-prey has $Left_{max} = \frac{3}{4}$ (as a {\em lower} approximation) for this property.
\end{proof}
\end{approximation}

\begin{approximation}
\label{approx:predator-prey_PG_right_min}
$Right_{min}$ for the partial grading (PG) property is equal to $\frac{3}{4}$ (as a {\em higher} approximation) for the predator-prey environment.

\begin{proof}
To find $Right_{min}$ (equation \ref{eq:right_min}), we need to find a pair $\left\langle\Pi_e,w_{\Pi_e}\right\rangle$ which minimises the property as much as possible while $\Pi_o$ maximises it. Using $\Pi_e = \{{\pi_{chase}}_1,{\pi_{chase}}_2,{\pi_{chase}}_3\}$ with uniform weight for $w_{\Pi_e}$ (a $\pi_{chase}$ agent always tries to be chased when playing as the prey and tries to chase when playing as a predator) we find a {\em higher} approximation of this situation no matter which $\Pi_o$ we use.

Following definition \ref{def:PG}, we obtain the PG value for this $\left\langle\Pi_e,w_{\Pi_e},\Pi_o\right\rangle$ (where $\Pi_o$ is instantiated with any permitted value). Since the environment is not symmetric, we need to calculate this property for every pair of slots. Following definition \ref{def:STG_set} (for PG), we can calculate its PG value for each pair of slots. We start with slots 1 and 2:

\begin{equation*}
\begin{aligned}
PG_{1,2}(\Pi_e,w_{\Pi_e},\Pi_o,w_{\dot{L}},\mu)	& = \eta_{\Pi^3} \sum_{\pi_1,\pi_2,\pi_3 \in \Pi_e | \pi_1 \neq \pi_2 \neq \pi_3} w_{\Pi_e}(\pi_1) w_{\Pi_e}(\pi_2) w_{\Pi_e}(\pi_3) PG_{1,2}(\pi_1,\pi_2,\pi_3,\Pi_o,w_{\dot{L}},\mu) =\\
												& = 6 \frac{9}{2} \frac{1}{3} \frac{1}{3} \frac{1}{3} PG_{1,2}({\pi_{chase}}_1,{\pi_{chase}}_2,{\pi_{chase}}_3,\Pi_o,w_{\dot{L}},\mu)
\end{aligned}
\end{equation*}

\noindent Note that we avoided to calculate all the permutations of $\pi_1,\pi_2,\pi_3$ for $PG_{i,j}(\pi_1,\pi_2,\pi_3,\Pi_o,w_{\dot{L}},\mu)$ since they provide the same result, by calculating only one permutation and multiplying the result by the number of permutations $6$.

In this case, we only need to calculate $PG_{1,2}({\pi_{chase}}_1,{\pi_{chase}}_2,{\pi_{chase}}_3,\Pi_o,w_{\dot{L}},\mu)$. We follow definition \ref{def:STG_agents} (for PG) to calculate this value:

\begin{equation*}
PG_{1,2}({\pi_{chase}}_1,{\pi_{chase}}_2,{\pi_{chase}}_3,\Pi_o,w_{\dot{L}},\mu) = \sum_{\dot{l} \in \dot{L}^{N(\mu)}_{-1,2}(\Pi_o)} w_{\dot{L}}(\dot{l}) PO_{1,2}({\pi_{chase}}_1,{\pi_{chase}}_2,{\pi_{chase}}_3,\dot{l},\mu)
\end{equation*}

We do not know which $\Pi_o$ we have, but we know that we will need to obtain a line-up pattern $\dot{l}$ from $\dot{L}^{N(\mu)}_{-1,2}(\Pi_o)$ to calculate $PO_{1,2}({\pi_{chase}}_1,{\pi_{chase}}_2,{\pi_{chase}}_3,\dot{l},\mu)$. We calculate this value for a figurative line-up pattern $\dot{l} = (*,*,\pi_1,\pi_2)$ from $\dot{L}^{N(\mu)}_{-1,2}(\Pi_o)$:

\begin{equation*}
PO_{1,2}({\pi_{chase}}_1,{\pi_{chase}}_2,{\pi_{chase}}_3,\dot{l},\mu) = PO_{1,2}({\pi_{chase}}_1,{\pi_{chase}}_2,{\pi_{chase}}_3,(*,*,\pi_1,\pi_2),\mu)
\end{equation*}

The following table shows us $PO_{1,2}$ for all the permutations of ${\pi_{chase}}_1,{\pi_{chase}}_2,{\pi_{chase}}_3$.

\begin{center}
\begin{tabular}{c c c | c c c | c c c}
Slot 1 & & Slot 2								& Slot 1 & & Slot 2									& Slot 1 & & Slot 2\\
\hline
${\pi_{chase}}_1$ & $\leq$ & ${\pi_{chase}}_2$	& ${\pi_{chase}}_1$ & $\leq$ & ${\pi_{chase}}_3$	& ${\pi_{chase}}_2$ & $\leq$ & ${\pi_{chase}}_1$\\
${\pi_{chase}}_2$ & $\leq$ & ${\pi_{chase}}_3$	& ${\pi_{chase}}_3$ & $\leq$ & ${\pi_{chase}}_2$	& ${\pi_{chase}}_1$ & $\leq$ & ${\pi_{chase}}_3$\\
${\pi_{chase}}_1$ & $\leq$ & ${\pi_{chase}}_3$	& ${\pi_{chase}}_1$ & $\leq$ & ${\pi_{chase}}_2$	& ${\pi_{chase}}_2$ & $\leq$ & ${\pi_{chase}}_3$
\end{tabular}
\begin{tabular}{c c c | c c c | c c c}
Slot 1 & & Slot 2								& Slot 1 & & Slot 2									& Slot 1 & & Slot 2\\
\hline
${\pi_{chase}}_2$ & $\leq$ & ${\pi_{chase}}_3$	& ${\pi_{chase}}_3$ & $\leq$ & ${\pi_{chase}}_1$	& ${\pi_{chase}}_3$ & $\leq$ & ${\pi_{chase}}_2$\\
${\pi_{chase}}_3$ & $\leq$ & ${\pi_{chase}}_1$	& ${\pi_{chase}}_1$ & $\leq$ & ${\pi_{chase}}_2$	& ${\pi_{chase}}_2$ & $\leq$ & ${\pi_{chase}}_1$\\
${\pi_{chase}}_2$ & $\leq$ & ${\pi_{chase}}_1$	& ${\pi_{chase}}_3$ & $\leq$ & ${\pi_{chase}}_2$	& ${\pi_{chase}}_3$ & $\leq$ & ${\pi_{chase}}_1$
\end{tabular}
\end{center}

It is possible to find a PO for every permutation, since we always have ${\pi_{chase}}_i \leq {\pi_{chase}}_j$, where a $\pi_{chase}$ agent always tries to be chased when playing as the prey and tries to chase when playing as a predator, so the agents in slots 1 and 2 will obtain an expected average reward of $-1$ and $1$ respectively. Note that the choice of $\Pi_o$ does not affect the result of $PO_{1,2}$, so no matter which agents are in $\Pi_o$ we obtain:

\begin{equation*}
PG_{1,2}({\pi_{chase}}_1,{\pi_{chase}}_2,{\pi_{chase}}_3,\Pi_o,w_{\dot{L}},\mu) = 1
\end{equation*}

Therefore:

\begin{equation*}
PG_{1,2}(\Pi_e,w_{\Pi_e},\Pi_o,w_{\dot{L}},\mu) = 6 \frac{9}{2} \frac{1}{3} \frac{1}{3} \frac{1}{3} 1 = 1
\end{equation*}

For slots 1 and 3:

\begin{equation*}
\begin{aligned}
PG_{1,3}(\Pi_e,w_{\Pi_e},\Pi_o,w_{\dot{L}},\mu)	& = \eta_{\Pi^3} \sum_{\pi_1,\pi_2,\pi_3 \in \Pi_e | \pi_1 \neq \pi_2 \neq \pi_3} w_{\Pi_e}(\pi_1) w_{\Pi_e}(\pi_2) w_{\Pi_e}(\pi_3) PG_{1,3}(\pi_1,\pi_2,\pi_3,\Pi_o,w_{\dot{L}},\mu) =\\
												& = 6 \frac{9}{2} \frac{1}{3} \frac{1}{3} \frac{1}{3} PG_{1,3}({\pi_{chase}}_1,{\pi_{chase}}_2,{\pi_{chase}}_3,\Pi_o,w_{\dot{L}},\mu)
\end{aligned}
\end{equation*}

In this case, we only need to calculate $PG_{1,3}({\pi_{chase}}_1,{\pi_{chase}}_2,{\pi_{chase}}_3,\Pi_o,w_{\dot{L}},\mu)$. We follow definition \ref{def:STG_agents} (for PG) to calculate this value:

\begin{equation*}
PG_{1,3}({\pi_{chase}}_1,{\pi_{chase}}_2,{\pi_{chase}}_3,\Pi_o,w_{\dot{L}},\mu) = \sum_{\dot{l} \in \dot{L}^{N(\mu)}_{-1,3}(\Pi_o)} w_{\dot{L}}(\dot{l}) PO_{1,3}({\pi_{chase}}_1,{\pi_{chase}}_2,{\pi_{chase}}_3,\dot{l},\mu)
\end{equation*}

Again, we do not know which $\Pi_o$ we have, but we know that we will need to obtain a line-up pattern $\dot{l}$ from $\dot{L}^{N(\mu)}_{-1,3}(\Pi_o)$ to calculate $PO_{1,3}({\pi_{chase}}_1,{\pi_{chase}}_2,{\pi_{chase}}_3,\dot{l},\mu)$. We calculate this value for a figurative line-up pattern $\dot{l} = (*,\pi_1,*,\pi_2)$ from $\dot{L}^{N(\mu)}_{-1,3}(\Pi_o)$:

\begin{equation*}
PO_{1,3}({\pi_{chase}}_1,{\pi_{chase}}_2,{\pi_{chase}}_3,\dot{l},\mu) = PO_{1,3}({\pi_{chase}}_1,{\pi_{chase}}_2,{\pi_{chase}}_3,(*,\pi_1,*,\pi_2),\mu)
\end{equation*}

The following table shows us $PO_{1,3}$ for all the permutations of ${\pi_{chase}}_1,{\pi_{chase}}_2,{\pi_{chase}}_3$.

\begin{center}
\begin{tabular}{c c c | c c c | c c c}
Slot 1 & & Slot 3								& Slot 1 & & Slot 3									& Slot 1 & & Slot 3\\
\hline
${\pi_{chase}}_1$ & $\leq$ & ${\pi_{chase}}_2$	& ${\pi_{chase}}_1$ & $\leq$ & ${\pi_{chase}}_3$	& ${\pi_{chase}}_2$ & $\leq$ & ${\pi_{chase}}_1$\\
${\pi_{chase}}_2$ & $\leq$ & ${\pi_{chase}}_3$	& ${\pi_{chase}}_3$ & $\leq$ & ${\pi_{chase}}_2$	& ${\pi_{chase}}_1$ & $\leq$ & ${\pi_{chase}}_3$\\
${\pi_{chase}}_1$ & $\leq$ & ${\pi_{chase}}_3$	& ${\pi_{chase}}_1$ & $\leq$ & ${\pi_{chase}}_2$	& ${\pi_{chase}}_2$ & $\leq$ & ${\pi_{chase}}_3$
\end{tabular}
\begin{tabular}{c c c | c c c | c c c}
Slot 1 & & Slot 3								& Slot 1 & & Slot 3									& Slot 1 & & Slot 3\\
\hline
${\pi_{chase}}_2$ & $\leq$ & ${\pi_{chase}}_3$	& ${\pi_{chase}}_3$ & $\leq$ & ${\pi_{chase}}_1$	& ${\pi_{chase}}_3$ & $\leq$ & ${\pi_{chase}}_2$\\
${\pi_{chase}}_3$ & $\leq$ & ${\pi_{chase}}_1$	& ${\pi_{chase}}_1$ & $\leq$ & ${\pi_{chase}}_2$	& ${\pi_{chase}}_2$ & $\leq$ & ${\pi_{chase}}_1$\\
${\pi_{chase}}_2$ & $\leq$ & ${\pi_{chase}}_1$	& ${\pi_{chase}}_3$ & $\leq$ & ${\pi_{chase}}_2$	& ${\pi_{chase}}_3$ & $\leq$ & ${\pi_{chase}}_1$
\end{tabular}
\end{center}

Again, it is possible to find a PO for every permutation, since we always have ${\pi_{chase}}_i \leq {\pi_{chase}}_j$, where a $\pi_{chase}$ agent always tries to be chased when playing as the prey and tries to chase when playing as a predator, so the agents in slots 1 and 3 will obtain an expected average reward of $-1$ and $1$ respectively. Note that the choice of $\Pi_o$ does not affect the result of $PO_{1,3}$, so no matter which agents are in $\Pi_o$ we obtain:

\begin{equation*}
PG_{1,3}({\pi_{chase}}_1,{\pi_{chase}}_2,{\pi_{chase}}_3,\Pi_o,w_{\dot{L}},\mu) = 1
\end{equation*}

Therefore:

\begin{equation*}
PG_{1,3}(\Pi_e,w_{\Pi_e},\Pi_o,w_{\dot{L}},\mu) = 6 \frac{9}{2} \frac{1}{3} \frac{1}{3} \frac{1}{3} 1 = 1
\end{equation*}

For slots 1 and 4:

\begin{equation*}
\begin{aligned}
PG_{1,4}(\Pi_e,w_{\Pi_e},\Pi_o,w_{\dot{L}},\mu)	& = \eta_{\Pi^3} \sum_{\pi_1,\pi_2,\pi_3 \in \Pi_e | \pi_1 \neq \pi_2 \neq \pi_3} w_{\Pi_e}(\pi_1) w_{\Pi_e}(\pi_2) w_{\Pi_e}(\pi_3) PG_{1,4}(\pi_1,\pi_2,\pi_3,\Pi_o,w_{\dot{L}},\mu) =\\
												& = 6 \frac{9}{2} \frac{1}{3} \frac{1}{3} \frac{1}{3} PG_{1,4}({\pi_{chase}}_1,{\pi_{chase}}_2,{\pi_{chase}}_3,\Pi_o,w_{\dot{L}},\mu)
\end{aligned}
\end{equation*}

In this case, we only need to calculate $PG_{1,4}({\pi_{chase}}_1,{\pi_{chase}}_2,{\pi_{chase}}_3,\Pi_o,w_{\dot{L}},\mu)$. We follow definition \ref{def:STG_agents} (for PG) to calculate this value:

\begin{equation*}
PG_{1,4}({\pi_{chase}}_1,{\pi_{chase}}_2,{\pi_{chase}}_3,\Pi_o,w_{\dot{L}},\mu) = \sum_{\dot{l} \in \dot{L}^{N(\mu)}_{-1,4}(\Pi_o)} w_{\dot{L}}(\dot{l}) PO_{1,4}({\pi_{chase}}_1,{\pi_{chase}}_2,{\pi_{chase}}_3,\dot{l},\mu)
\end{equation*}

Again, we do not know which $\Pi_o$ we have, but we know that we will need to obtain a line-up pattern $\dot{l}$ from $\dot{L}^{N(\mu)}_{-1,4}(\Pi_o)$ to calculate $PO_{1,4}({\pi_{chase}}_1,{\pi_{chase}}_2,{\pi_{chase}}_3,\dot{l},\mu)$. We calculate this value for a figurative line-up pattern $\dot{l} = (*,\pi_1,\pi_2,*)$ from $\dot{L}^{N(\mu)}_{-1,4}(\Pi_o)$:

\begin{equation*}
PO_{1,4}({\pi_{chase}}_1,{\pi_{chase}}_2,{\pi_{chase}}_3,\dot{l},\mu) = PO_{1,4}({\pi_{chase}}_1,{\pi_{chase}}_2,{\pi_{chase}}_3,(*,\pi_1,\pi_2,*),\mu)
\end{equation*}

The following table shows us $PO_{1,4}$ for all the permutations of ${\pi_{chase}}_1,{\pi_{chase}}_2,{\pi_{chase}}_3$.

\begin{center}
\begin{tabular}{c c c | c c c | c c c}
Slot 1 & & Slot 4								& Slot 1 & & Slot 4									& Slot 1 & & Slot 4\\
\hline
${\pi_{chase}}_1$ & $\leq$ & ${\pi_{chase}}_2$	& ${\pi_{chase}}_1$ & $\leq$ & ${\pi_{chase}}_3$	& ${\pi_{chase}}_2$ & $\leq$ & ${\pi_{chase}}_1$\\
${\pi_{chase}}_2$ & $\leq$ & ${\pi_{chase}}_3$	& ${\pi_{chase}}_3$ & $\leq$ & ${\pi_{chase}}_2$	& ${\pi_{chase}}_1$ & $\leq$ & ${\pi_{chase}}_3$\\
${\pi_{chase}}_1$ & $\leq$ & ${\pi_{chase}}_3$	& ${\pi_{chase}}_1$ & $\leq$ & ${\pi_{chase}}_2$	& ${\pi_{chase}}_2$ & $\leq$ & ${\pi_{chase}}_3$
\end{tabular}
\begin{tabular}{c c c | c c c | c c c}
Slot 1 & & Slot 4								& Slot 1 & & Slot 4									& Slot 1 & & Slot 4\\
\hline
${\pi_{chase}}_2$ & $\leq$ & ${\pi_{chase}}_3$	& ${\pi_{chase}}_3$ & $\leq$ & ${\pi_{chase}}_1$	& ${\pi_{chase}}_3$ & $\leq$ & ${\pi_{chase}}_2$\\
${\pi_{chase}}_3$ & $\leq$ & ${\pi_{chase}}_1$	& ${\pi_{chase}}_1$ & $\leq$ & ${\pi_{chase}}_2$	& ${\pi_{chase}}_2$ & $\leq$ & ${\pi_{chase}}_1$\\
${\pi_{chase}}_2$ & $\leq$ & ${\pi_{chase}}_1$	& ${\pi_{chase}}_3$ & $\leq$ & ${\pi_{chase}}_2$	& ${\pi_{chase}}_3$ & $\leq$ & ${\pi_{chase}}_1$
\end{tabular}
\end{center}

Again, it is possible to find a PO for every permutation, since we always have ${\pi_{chase}}_i \leq {\pi_{chase}}_j$, where a $\pi_{chase}$ agent always tries to be chased when playing as the prey and tries to chase when playing as a predator, so the agents in slots 1 and 4 will obtain an expected average reward of $-1$ and $1$ respectively. Note that the choice of $\Pi_o$ does not affect the result of $PO_{1,4}$, so no matter which agents are in $\Pi_o$ we obtain:

\begin{equation*}
PG_{1,4}({\pi_{chase}}_1,{\pi_{chase}}_2,{\pi_{chase}}_3,\Pi_o,w_{\dot{L}},\mu) = 1
\end{equation*}

Therefore:

\begin{equation*}
PG_{1,4}(\Pi_e,w_{\Pi_e},\Pi_o,w_{\dot{L}},\mu) = 6 \frac{9}{2} \frac{1}{3} \frac{1}{3} \frac{1}{3} 1 = 1
\end{equation*}

For slots 2 and 1:

\begin{equation*}
\begin{aligned}
PG_{2,1}(\Pi_e,w_{\Pi_e},\Pi_o,w_{\dot{L}},\mu)	& = \eta_{\Pi^3} \sum_{\pi_1,\pi_2,\pi_3 \in \Pi_e | \pi_1 \neq \pi_2 \neq \pi_3} w_{\Pi_e}(\pi_1) w_{\Pi_e}(\pi_2) w_{\Pi_e}(\pi_3) PG_{2,1}(\pi_1,\pi_2,\pi_3,\Pi_o,w_{\dot{L}},\mu) =\\
												& = 6 \frac{9}{2} \frac{1}{3} \frac{1}{3} \frac{1}{3} PG_{2,1}({\pi_{chase}}_1,{\pi_{chase}}_2,{\pi_{chase}}_3,\Pi_o,w_{\dot{L}},\mu)
\end{aligned}
\end{equation*}

In this case, we only need to calculate $PG_{2,1}({\pi_{chase}}_1,{\pi_{chase}}_2,{\pi_{chase}}_3,\Pi_o,w_{\dot{L}},\mu)$. We follow definition \ref{def:STG_agents} (for PG) to calculate this value:

\begin{equation*}
PG_{2,1}({\pi_{chase}}_1,{\pi_{chase}}_2,{\pi_{chase}}_3,\Pi_o,w_{\dot{L}},\mu) = \sum_{\dot{l} \in \dot{L}^{N(\mu)}_{-2,1}(\Pi_o)} w_{\dot{L}}(\dot{l}) PO_{2,1}({\pi_{chase}}_1,{\pi_{chase}}_2,{\pi_{chase}}_3,\dot{l},\mu)
\end{equation*}

Again, we do not know which $\Pi_o$ we have, but we know that we will need to obtain a line-up pattern $\dot{l}$ from $\dot{L}^{N(\mu)}_{-2,1}(\Pi_o)$ to calculate $PO_{2,1}({\pi_{chase}}_1,{\pi_{chase}}_2,{\pi_{chase}}_3,\dot{l},\mu)$. We calculate this value for a figurative line-up pattern $\dot{l} = (*,*,\pi_1,\pi_2)$ from $\dot{L}^{N(\mu)}_{-2,1}(\Pi_o)$:

\begin{equation*}
PO_{2,1}({\pi_{chase}}_1,{\pi_{chase}}_2,{\pi_{chase}}_3,\dot{l},\mu) = PO_{2,1}({\pi_{chase}}_1,{\pi_{chase}}_2,{\pi_{chase}}_3,(*,*,\pi_1,\pi_2),\mu)
\end{equation*}

The following table shows us $PO_{2,1}$ for all the permutations of ${\pi_{chase}}_1,{\pi_{chase}}_2,{\pi_{chase}}_3$.

\begin{center}
\begin{tabular}{c c c | c c c | c c c}
Slot 2 & & Slot 1								& Slot 2 & & Slot 1									& Slot 2 & & Slot 1\\
\hline
${\pi_{chase}}_1$ & $\leq$ & ${\pi_{chase}}_2$	& ${\pi_{chase}}_1$ & $\leq$ & ${\pi_{chase}}_3$	& ${\pi_{chase}}_2$ & $\leq$ & ${\pi_{chase}}_1$\\
${\pi_{chase}}_2$ & $\leq$ & ${\pi_{chase}}_3$	& ${\pi_{chase}}_3$ & $\leq$ & ${\pi_{chase}}_2$	& ${\pi_{chase}}_1$ & $\leq$ & ${\pi_{chase}}_3$\\
${\pi_{chase}}_1$ & $\leq$ & ${\pi_{chase}}_3$	& ${\pi_{chase}}_1$ & $\leq$ & ${\pi_{chase}}_2$	& ${\pi_{chase}}_2$ & $\leq$ & ${\pi_{chase}}_3$
\end{tabular}
\begin{tabular}{c c c | c c c | c c c}
Slot 2 & & Slot 1								& Slot 2 & & Slot 1									& Slot 2 & & Slot 1\\
\hline
${\pi_{chase}}_2$ & $\leq$ & ${\pi_{chase}}_3$	& ${\pi_{chase}}_3$ & $\leq$ & ${\pi_{chase}}_1$	& ${\pi_{chase}}_3$ & $\leq$ & ${\pi_{chase}}_2$\\
${\pi_{chase}}_3$ & $\leq$ & ${\pi_{chase}}_1$	& ${\pi_{chase}}_1$ & $\leq$ & ${\pi_{chase}}_2$	& ${\pi_{chase}}_2$ & $\leq$ & ${\pi_{chase}}_1$\\
${\pi_{chase}}_2$ & $\leq$ & ${\pi_{chase}}_1$	& ${\pi_{chase}}_3$ & $\leq$ & ${\pi_{chase}}_2$	& ${\pi_{chase}}_3$ & $\leq$ & ${\pi_{chase}}_1$
\end{tabular}
\end{center}

It is not possible to find a PO for any permutation, since we always have ${\pi_{chase}}_i \leq {\pi_{chase}}_j$, where a $\pi_{chase}$ agent always tries to be chased when playing as the prey and tries to chase when playing as a predator, so the agents in slots 2 and 1 will obtain an expected average reward of $1$ and $-1$ respectively. Note that the choice of $\Pi_o$ does not affect the result of $PO_{2,1}$, so no matter which agents are in $\Pi_o$ we obtain:

\begin{equation*}
PG_{2,1}({\pi_{chase}}_1,{\pi_{chase}}_2,{\pi_{chase}}_3,\Pi_o,w_{\dot{L}},\mu) = 0
\end{equation*}

Therefore:

\begin{equation*}
PG_{2,1}(\Pi_e,w_{\Pi_e},\Pi_o,w_{\dot{L}},\mu) = 6 \frac{9}{2} \frac{1}{3} \frac{1}{3} \frac{1}{3} 0 = 0
\end{equation*}

For slots 2 and 3:

\begin{equation*}
\begin{aligned}
PG_{2,3}(\Pi_e,w_{\Pi_e},\Pi_o,w_{\dot{L}},\mu)	& = \eta_{\Pi^3} \sum_{\pi_1,\pi_2,\pi_3 \in \Pi_e | \pi_1 \neq \pi_2 \neq \pi_3} w_{\Pi_e}(\pi_1) w_{\Pi_e}(\pi_2) w_{\Pi_e}(\pi_3) PG_{2,3}(\pi_1,\pi_2,\pi_3,\Pi_o,w_{\dot{L}},\mu) =\\
												& = 6 \frac{9}{2} \frac{1}{3} \frac{1}{3} \frac{1}{3} PG_{2,3}({\pi_{chase}}_1,{\pi_{chase}}_2,{\pi_{chase}}_3,\Pi_o,w_{\dot{L}},\mu)
\end{aligned}
\end{equation*}

In this case, we only need to calculate $PG_{2,3}({\pi_{chase}}_1,{\pi_{chase}}_2,{\pi_{chase}}_3,\Pi_o,w_{\dot{L}},\mu)$. We follow definition \ref{def:STG_agents} (for PG) to calculate this value:

\begin{equation*}
PG_{2,3}({\pi_{chase}}_1,{\pi_{chase}}_2,{\pi_{chase}}_3,\Pi_o,w_{\dot{L}},\mu) = \sum_{\dot{l} \in \dot{L}^{N(\mu)}_{-2,3}(\Pi_o)} w_{\dot{L}}(\dot{l}) PO_{2,3}({\pi_{chase}}_1,{\pi_{chase}}_2,{\pi_{chase}}_3,\dot{l},\mu)
\end{equation*}

Again, we do not know which $\Pi_o$ we have, but we know that we will need to obtain a line-up pattern $\dot{l}$ from $\dot{L}^{N(\mu)}_{-2,3}(\Pi_o)$ to calculate $PO_{2,3}({\pi_{chase}}_1,{\pi_{chase}}_2,{\pi_{chase}}_3,\dot{l},\mu)$. We calculate this value for a figurative line-up pattern $\dot{l} = (\pi_1,*,*,\pi_2)$ from $\dot{L}^{N(\mu)}_{-2,3}(\Pi_o)$:

\begin{equation*}
PO_{2,3}({\pi_{chase}}_1,{\pi_{chase}}_2,{\pi_{chase}}_3,\dot{l},\mu) = PO_{2,3}({\pi_{chase}}_1,{\pi_{chase}}_2,{\pi_{chase}}_3,(\pi_1,*,*,\pi_2),\mu)
\end{equation*}

The following table shows us $PO_{2,3}$ for all the permutations of ${\pi_{chase}}_1,{\pi_{chase}}_2,{\pi_{chase}}_3$.

\begin{center}
\begin{tabular}{c c c | c c c | c c c}
Slot 2 & & Slot 3								& Slot 2 & & Slot 3									& Slot 2 & & Slot 3\\
\hline
${\pi_{chase}}_1$ & $\leq$ & ${\pi_{chase}}_2$	& ${\pi_{chase}}_1$ & $\leq$ & ${\pi_{chase}}_3$	& ${\pi_{chase}}_2$ & $\leq$ & ${\pi_{chase}}_1$\\
${\pi_{chase}}_2$ & $\leq$ & ${\pi_{chase}}_3$	& ${\pi_{chase}}_3$ & $\leq$ & ${\pi_{chase}}_2$	& ${\pi_{chase}}_1$ & $\leq$ & ${\pi_{chase}}_3$\\
${\pi_{chase}}_1$ & $\leq$ & ${\pi_{chase}}_3$	& ${\pi_{chase}}_1$ & $\leq$ & ${\pi_{chase}}_2$	& ${\pi_{chase}}_2$ & $\leq$ & ${\pi_{chase}}_3$
\end{tabular}
\begin{tabular}{c c c | c c c | c c c}
Slot 2 & & Slot 3								& Slot 2 & & Slot 3									& Slot 2 & & Slot 3\\
\hline
${\pi_{chase}}_2$ & $\leq$ & ${\pi_{chase}}_3$	& ${\pi_{chase}}_3$ & $\leq$ & ${\pi_{chase}}_1$	& ${\pi_{chase}}_3$ & $\leq$ & ${\pi_{chase}}_2$\\
${\pi_{chase}}_3$ & $\leq$ & ${\pi_{chase}}_1$	& ${\pi_{chase}}_1$ & $\leq$ & ${\pi_{chase}}_2$	& ${\pi_{chase}}_2$ & $\leq$ & ${\pi_{chase}}_1$\\
${\pi_{chase}}_2$ & $\leq$ & ${\pi_{chase}}_1$	& ${\pi_{chase}}_3$ & $\leq$ & ${\pi_{chase}}_2$	& ${\pi_{chase}}_3$ & $\leq$ & ${\pi_{chase}}_1$
\end{tabular}
\end{center}

It is possible to find a PO for every permutation, since the agents in slots 2 and 3 share rewards (and expected average rewards as well) due they are in the same team. So no matter which agents are in $\Pi_o$ we obtain:

\begin{equation*}
PG_{2,3}({\pi_{chase}}_1,{\pi_{chase}}_2,{\pi_{chase}}_3,\Pi_o,w_{\dot{L}},\mu) = 1
\end{equation*}

Therefore:

\begin{equation*}
PG_{2,3}(\Pi_e,w_{\Pi_e},\Pi_o,w_{\dot{L}},\mu) = 6 \frac{9}{2} \frac{1}{3} \frac{1}{3} \frac{1}{3} 1 = 1
\end{equation*}

For slots 2 and 4:

\begin{equation*}
\begin{aligned}
PG_{2,4}(\Pi_e,w_{\Pi_e},\Pi_o,w_{\dot{L}},\mu)	& = \eta_{\Pi^3} \sum_{\pi_1,\pi_2,\pi_3 \in \Pi_e | \pi_1 \neq \pi_2 \neq \pi_3} w_{\Pi_e}(\pi_1) w_{\Pi_e}(\pi_2) w_{\Pi_e}(\pi_3) PG_{2,4}(\pi_1,\pi_2,\pi_3,\Pi_o,w_{\dot{L}},\mu) =\\
												& = 6 \frac{9}{2} \frac{1}{3} \frac{1}{3} \frac{1}{3} PG_{2,4}({\pi_{chase}}_1,{\pi_{chase}}_2,{\pi_{chase}}_3,\Pi_o,w_{\dot{L}},\mu)
\end{aligned}
\end{equation*}

In this case, we only need to calculate $PG_{2,4}({\pi_{chase}}_1,{\pi_{chase}}_2,{\pi_{chase}}_3,\Pi_o,w_{\dot{L}},\mu)$. We follow definition \ref{def:STG_agents} (for PG) to calculate this value:

\begin{equation*}
PG_{2,4}({\pi_{chase}}_1,{\pi_{chase}}_2,{\pi_{chase}}_3,\Pi_o,w_{\dot{L}},\mu) = \sum_{\dot{l} \in \dot{L}^{N(\mu)}_{-2,4}(\Pi_o)} w_{\dot{L}}(\dot{l}) PO_{2,4}({\pi_{chase}}_1,{\pi_{chase}}_2,{\pi_{chase}}_3,\dot{l},\mu)
\end{equation*}

Again, we do not know which $\Pi_o$ we have, but we know that we will need to obtain a line-up pattern $\dot{l}$ from $\dot{L}^{N(\mu)}_{-2,4}(\Pi_o)$ to calculate $PO_{2,4}({\pi_{chase}}_1,{\pi_{chase}}_2,{\pi_{chase}}_3,\dot{l},\mu)$. We calculate this value for a figurative line-up pattern $\dot{l} = (\pi_1,*,\pi_2,*)$ from $\dot{L}^{N(\mu)}_{-2,4}(\Pi_o)$:

\begin{equation*}
PO_{2,4}({\pi_{chase}}_1,{\pi_{chase}}_2,{\pi_{chase}}_3,\dot{l},\mu) = PO_{2,4}({\pi_{chase}}_1,{\pi_{chase}}_2,{\pi_{chase}}_3,(\pi_1,*,\pi_2,*),\mu)
\end{equation*}

The following table shows us $PO_{2,4}$ for all the permutations of ${\pi_{chase}}_1,{\pi_{chase}}_2,{\pi_{chase}}_3$.

\begin{center}
\begin{tabular}{c c c | c c c | c c c}
Slot 2 & & Slot 4								& Slot 2 & & Slot 4									& Slot 2 & & Slot 4\\
\hline
${\pi_{chase}}_1$ & $\leq$ & ${\pi_{chase}}_2$	& ${\pi_{chase}}_1$ & $\leq$ & ${\pi_{chase}}_3$	& ${\pi_{chase}}_2$ & $\leq$ & ${\pi_{chase}}_1$\\
${\pi_{chase}}_2$ & $\leq$ & ${\pi_{chase}}_3$	& ${\pi_{chase}}_3$ & $\leq$ & ${\pi_{chase}}_2$	& ${\pi_{chase}}_1$ & $\leq$ & ${\pi_{chase}}_3$\\
${\pi_{chase}}_1$ & $\leq$ & ${\pi_{chase}}_3$	& ${\pi_{chase}}_1$ & $\leq$ & ${\pi_{chase}}_2$	& ${\pi_{chase}}_2$ & $\leq$ & ${\pi_{chase}}_3$
\end{tabular}
\begin{tabular}{c c c | c c c | c c c}
Slot 2 & & Slot 4								& Slot 2 & & Slot 4									& Slot 2 & & Slot 4\\
\hline
${\pi_{chase}}_2$ & $\leq$ & ${\pi_{chase}}_3$	& ${\pi_{chase}}_3$ & $\leq$ & ${\pi_{chase}}_1$	& ${\pi_{chase}}_3$ & $\leq$ & ${\pi_{chase}}_2$\\
${\pi_{chase}}_3$ & $\leq$ & ${\pi_{chase}}_1$	& ${\pi_{chase}}_1$ & $\leq$ & ${\pi_{chase}}_2$	& ${\pi_{chase}}_2$ & $\leq$ & ${\pi_{chase}}_1$\\
${\pi_{chase}}_2$ & $\leq$ & ${\pi_{chase}}_1$	& ${\pi_{chase}}_3$ & $\leq$ & ${\pi_{chase}}_2$	& ${\pi_{chase}}_3$ & $\leq$ & ${\pi_{chase}}_1$
\end{tabular}
\end{center}

It is possible to find a PO for every permutation, since the agents in slots 2 and 4 share rewards (and expected average rewards as well) due they are in the same team. So no matter which agents are in $\Pi_o$ we obtain:

\begin{equation*}
PG_{2,4}({\pi_{chase}}_1,{\pi_{chase}}_2,{\pi_{chase}}_3,\Pi_o,w_{\dot{L}},\mu) = 1
\end{equation*}

Therefore:

\begin{equation*}
PG_{2,4}(\Pi_e,w_{\Pi_e},\Pi_o,w_{\dot{L}},\mu) = 6 \frac{9}{2} \frac{1}{3} \frac{1}{3} \frac{1}{3} 1 = 1
\end{equation*}

For slots 3 and 1:

\begin{equation*}
\begin{aligned}
PG_{3,1}(\Pi_e,w_{\Pi_e},\Pi_o,w_{\dot{L}},\mu)	& = \eta_{\Pi^3} \sum_{\pi_1,\pi_2,\pi_3 \in \Pi_e | \pi_1 \neq \pi_2 \neq \pi_3} w_{\Pi_e}(\pi_1) w_{\Pi_e}(\pi_2) w_{\Pi_e}(\pi_3) PG_{3,1}(\pi_1,\pi_2,\pi_3,\Pi_o,w_{\dot{L}},\mu) =\\
												& = 6 \frac{9}{2} \frac{1}{3} \frac{1}{3} \frac{1}{3} PG_{3,1}({\pi_{chase}}_1,{\pi_{chase}}_2,{\pi_{chase}}_3,\Pi_o,w_{\dot{L}},\mu)
\end{aligned}
\end{equation*}

In this case, we only need to calculate $PG_{3,1}({\pi_{chase}}_1,{\pi_{chase}}_2,{\pi_{chase}}_3,\Pi_o,w_{\dot{L}},\mu)$. We follow definition \ref{def:STG_agents} (for PG) to calculate this value:

\begin{equation*}
PG_{3,1}({\pi_{chase}}_1,{\pi_{chase}}_2,{\pi_{chase}}_3,\Pi_o,w_{\dot{L}},\mu) = \sum_{\dot{l} \in \dot{L}^{N(\mu)}_{-3,1}(\Pi_o)} w_{\dot{L}}(\dot{l}) PO_{3,1}({\pi_{chase}}_1,{\pi_{chase}}_2,{\pi_{chase}}_3,\dot{l},\mu)
\end{equation*}

Again, we do not know which $\Pi_o$ we have, but we know that we will need to obtain a line-up pattern $\dot{l}$ from $\dot{L}^{N(\mu)}_{-3,1}(\Pi_o)$ to calculate $PO_{3,1}({\pi_{chase}}_1,{\pi_{chase}}_2,{\pi_{chase}}_3,\dot{l},\mu)$. We calculate this value for a figurative line-up pattern $\dot{l} = (*,\pi_1,*,\pi_2)$ from $\dot{L}^{N(\mu)}_{-3,1}(\Pi_o)$:

\begin{equation*}
PO_{3,1}({\pi_{chase}}_1,{\pi_{chase}}_2,{\pi_{chase}}_3,\dot{l},\mu) = PO_{3,1}({\pi_{chase}}_1,{\pi_{chase}}_2,{\pi_{chase}}_3,(*,\pi_1,*,\pi_2),\mu)
\end{equation*}

The following table shows us $PO_{3,1}$ for all the permutations of ${\pi_{chase}}_1,{\pi_{chase}}_2,{\pi_{chase}}_3$.

\begin{center}
\begin{tabular}{c c c | c c c | c c c}
Slot 3 & & Slot 1								& Slot 3 & & Slot 1									& Slot 3 & & Slot 1\\
\hline
${\pi_{chase}}_1$ & $\leq$ & ${\pi_{chase}}_2$	& ${\pi_{chase}}_1$ & $\leq$ & ${\pi_{chase}}_3$	& ${\pi_{chase}}_2$ & $\leq$ & ${\pi_{chase}}_1$\\
${\pi_{chase}}_2$ & $\leq$ & ${\pi_{chase}}_3$	& ${\pi_{chase}}_3$ & $\leq$ & ${\pi_{chase}}_2$	& ${\pi_{chase}}_1$ & $\leq$ & ${\pi_{chase}}_3$\\
${\pi_{chase}}_1$ & $\leq$ & ${\pi_{chase}}_3$	& ${\pi_{chase}}_1$ & $\leq$ & ${\pi_{chase}}_2$	& ${\pi_{chase}}_2$ & $\leq$ & ${\pi_{chase}}_3$
\end{tabular}
\begin{tabular}{c c c | c c c | c c c}
Slot 3 & & Slot 1								& Slot 3 & & Slot 1									& Slot 3 & & Slot 1\\
\hline
${\pi_{chase}}_2$ & $\leq$ & ${\pi_{chase}}_3$	& ${\pi_{chase}}_3$ & $\leq$ & ${\pi_{chase}}_1$	& ${\pi_{chase}}_3$ & $\leq$ & ${\pi_{chase}}_2$\\
${\pi_{chase}}_3$ & $\leq$ & ${\pi_{chase}}_1$	& ${\pi_{chase}}_1$ & $\leq$ & ${\pi_{chase}}_2$	& ${\pi_{chase}}_2$ & $\leq$ & ${\pi_{chase}}_1$\\
${\pi_{chase}}_2$ & $\leq$ & ${\pi_{chase}}_1$	& ${\pi_{chase}}_3$ & $\leq$ & ${\pi_{chase}}_2$	& ${\pi_{chase}}_3$ & $\leq$ & ${\pi_{chase}}_1$
\end{tabular}
\end{center}

Again, it is not possible to find a PO for any permutation, since we always have ${\pi_{chase}}_i \leq {\pi_{chase}}_j$, where a $\pi_{chase}$ agent always tries to be chased when playing as the prey and tries to chase when playing as a predator, so the agents in slots 3 and 1 will obtain an expected average reward of $1$ and $-1$ respectively. Note that the choice of $\Pi_o$ does not affect the result of $PO_{3,1}$, so no matter which agents are in $\Pi_o$ we obtain:

\begin{equation*}
PG_{3,1}({\pi_{chase}}_1,{\pi_{chase}}_2,{\pi_{chase}}_3,\Pi_o,w_{\dot{L}},\mu) = 0
\end{equation*}

Therefore:

\begin{equation*}
PG_{3,1}(\Pi_e,w_{\Pi_e},\Pi_o,w_{\dot{L}},\mu) = 6 \frac{9}{2} \frac{1}{3} \frac{1}{3} \frac{1}{3} 0 = 0
\end{equation*}

For slots 3 and 2:

\begin{equation*}
\begin{aligned}
PG_{3,2}(\Pi_e,w_{\Pi_e},\Pi_o,w_{\dot{L}},\mu)	& = \eta_{\Pi^3} \sum_{\pi_1,\pi_2,\pi_3 \in \Pi_e | \pi_1 \neq \pi_2 \neq \pi_3} w_{\Pi_e}(\pi_1) w_{\Pi_e}(\pi_2) w_{\Pi_e}(\pi_3) PG_{3,2}(\pi_1,\pi_2,\pi_3,\Pi_o,w_{\dot{L}},\mu) =\\
												& = 6 \frac{9}{2} \frac{1}{3} \frac{1}{3} \frac{1}{3} PG_{3,2}({\pi_{chase}}_1,{\pi_{chase}}_2,{\pi_{chase}}_3,\Pi_o,w_{\dot{L}},\mu)
\end{aligned}
\end{equation*}

In this case, we only need to calculate $PG_{3,2}({\pi_{chase}}_1,{\pi_{chase}}_2,{\pi_{chase}}_3,\Pi_o,w_{\dot{L}},\mu)$. We follow definition \ref{def:STG_agents} (for PG) to calculate this value:

\begin{equation*}
PG_{3,2}({\pi_{chase}}_1,{\pi_{chase}}_2,{\pi_{chase}}_3,\Pi_o,w_{\dot{L}},\mu) = \sum_{\dot{l} \in \dot{L}^{N(\mu)}_{-3,2}(\Pi_o)} w_{\dot{L}}(\dot{l}) PO_{3,2}({\pi_{chase}}_1,{\pi_{chase}}_2,{\pi_{chase}}_3,\dot{l},\mu)
\end{equation*}

Again, we do not know which $\Pi_o$ we have, but we know that we will need to obtain a line-up pattern $\dot{l}$ from $\dot{L}^{N(\mu)}_{-3,2}(\Pi_o)$ to calculate $PO_{3,2}({\pi_{chase}}_1,{\pi_{chase}}_2,{\pi_{chase}}_3,\dot{l},\mu)$. We calculate this value for a figurative line-up pattern $\dot{l} = (\pi_1,*,*,\pi_2)$ from $\dot{L}^{N(\mu)}_{-3,2}(\Pi_o)$:

\begin{equation*}
PO_{3,2}({\pi_{chase}}_1,{\pi_{chase}}_2,{\pi_{chase}}_3,\dot{l},\mu) = PO_{3,2}({\pi_{chase}}_1,{\pi_{chase}}_2,{\pi_{chase}}_3,(\pi_1,*,*,\pi_2),\mu)
\end{equation*}

The following table shows us $PO_{3,2}$ for all the permutations of ${\pi_{chase}}_1,{\pi_{chase}}_2,{\pi_{chase}}_3$.

\begin{center}
\begin{tabular}{c c c | c c c | c c c}
Slot 3 & & Slot 2								& Slot 3 & & Slot 2									& Slot 3 & & Slot 2\\
\hline
${\pi_{chase}}_1$ & $\leq$ & ${\pi_{chase}}_2$	& ${\pi_{chase}}_1$ & $\leq$ & ${\pi_{chase}}_3$	& ${\pi_{chase}}_2$ & $\leq$ & ${\pi_{chase}}_1$\\
${\pi_{chase}}_2$ & $\leq$ & ${\pi_{chase}}_3$	& ${\pi_{chase}}_3$ & $\leq$ & ${\pi_{chase}}_2$	& ${\pi_{chase}}_1$ & $\leq$ & ${\pi_{chase}}_3$\\
${\pi_{chase}}_1$ & $\leq$ & ${\pi_{chase}}_3$	& ${\pi_{chase}}_1$ & $\leq$ & ${\pi_{chase}}_2$	& ${\pi_{chase}}_2$ & $\leq$ & ${\pi_{chase}}_3$
\end{tabular}
\begin{tabular}{c c c | c c c | c c c}
Slot 3 & & Slot 2								& Slot 3 & & Slot 2									& Slot 3 & & Slot 2\\
\hline
${\pi_{chase}}_2$ & $\leq$ & ${\pi_{chase}}_3$	& ${\pi_{chase}}_3$ & $\leq$ & ${\pi_{chase}}_1$	& ${\pi_{chase}}_3$ & $\leq$ & ${\pi_{chase}}_2$\\
${\pi_{chase}}_3$ & $\leq$ & ${\pi_{chase}}_1$	& ${\pi_{chase}}_1$ & $\leq$ & ${\pi_{chase}}_2$	& ${\pi_{chase}}_2$ & $\leq$ & ${\pi_{chase}}_1$\\
${\pi_{chase}}_2$ & $\leq$ & ${\pi_{chase}}_1$	& ${\pi_{chase}}_3$ & $\leq$ & ${\pi_{chase}}_2$	& ${\pi_{chase}}_3$ & $\leq$ & ${\pi_{chase}}_1$
\end{tabular}
\end{center}

It is possible to find a PO for every permutation, since the agents in slots 3 and 2 share rewards (and expected average rewards as well) due they are in the same team. So no matter which agents are in $\Pi_o$ we obtain:

\begin{equation*}
PG_{3,2}({\pi_{chase}}_1,{\pi_{chase}}_2,{\pi_{chase}}_3,\Pi_o,w_{\dot{L}},\mu) = 1
\end{equation*}

Therefore:

\begin{equation*}
PG_{3,2}(\Pi_e,w_{\Pi_e},\Pi_o,w_{\dot{L}},\mu) = 6 \frac{9}{2} \frac{1}{3} \frac{1}{3} \frac{1}{3} 1 = 1
\end{equation*}

For slots 3 and 4:

\begin{equation*}
\begin{aligned}
PG_{3,4}(\Pi_e,w_{\Pi_e},\Pi_o,w_{\dot{L}},\mu)	& = \eta_{\Pi^3} \sum_{\pi_1,\pi_2,\pi_3 \in \Pi_e | \pi_1 \neq \pi_2 \neq \pi_3} w_{\Pi_e}(\pi_1) w_{\Pi_e}(\pi_2) w_{\Pi_e}(\pi_3) PG_{3,4}(\pi_1,\pi_2,\pi_3,\Pi_o,w_{\dot{L}},\mu) =\\
												& = 6 \frac{9}{2} \frac{1}{3} \frac{1}{3} \frac{1}{3} PG_{3,4}({\pi_{chase}}_1,{\pi_{chase}}_2,{\pi_{chase}}_3,\Pi_o,w_{\dot{L}},\mu)
\end{aligned}
\end{equation*}

In this case, we only need to calculate $PG_{3,4}({\pi_{chase}}_1,{\pi_{chase}}_2,{\pi_{chase}}_3,\Pi_o,w_{\dot{L}},\mu)$. We follow definition \ref{def:STG_agents} (for PG) to calculate this value:

\begin{equation*}
PG_{3,4}({\pi_{chase}}_1,{\pi_{chase}}_2,{\pi_{chase}}_3,\Pi_o,w_{\dot{L}},\mu) = \sum_{\dot{l} \in \dot{L}^{N(\mu)}_{-3,4}(\Pi_o)} w_{\dot{L}}(\dot{l}) PO_{3,4}({\pi_{chase}}_1,{\pi_{chase}}_2,{\pi_{chase}}_3,\dot{l},\mu)
\end{equation*}

Again, we do not know which $\Pi_o$ we have, but we know that we will need to obtain a line-up pattern $\dot{l}$ from $\dot{L}^{N(\mu)}_{-3,4}(\Pi_o)$ to calculate $PO_{3,4}({\pi_{chase}}_1,{\pi_{chase}}_2,{\pi_{chase}}_3,\dot{l},\mu)$. We calculate this value for a figurative line-up pattern $\dot{l} = (\pi_1,\pi_2,*,*)$ from $\dot{L}^{N(\mu)}_{-3,4}(\Pi_o)$:

\begin{equation*}
PO_{3,4}({\pi_{chase}}_1,{\pi_{chase}}_2,{\pi_{chase}}_3,\dot{l},\mu) = PO_{3,4}({\pi_{chase}}_1,{\pi_{chase}}_2,{\pi_{chase}}_3,(\pi_1,\pi_2,*,*),\mu)
\end{equation*}

The following table shows us $PO_{3,4}$ for all the permutations of ${\pi_{chase}}_1,{\pi_{chase}}_2,{\pi_{chase}}_3$.

\begin{center}
\begin{tabular}{c c c | c c c | c c c}
Slot 3 & & Slot 4								& Slot 3 & & Slot 4									& Slot 3 & & Slot 4\\
\hline
${\pi_{chase}}_1$ & $\leq$ & ${\pi_{chase}}_2$	& ${\pi_{chase}}_1$ & $\leq$ & ${\pi_{chase}}_3$	& ${\pi_{chase}}_2$ & $\leq$ & ${\pi_{chase}}_1$\\
${\pi_{chase}}_2$ & $\leq$ & ${\pi_{chase}}_3$	& ${\pi_{chase}}_3$ & $\leq$ & ${\pi_{chase}}_2$	& ${\pi_{chase}}_1$ & $\leq$ & ${\pi_{chase}}_3$\\
${\pi_{chase}}_1$ & $\leq$ & ${\pi_{chase}}_3$	& ${\pi_{chase}}_1$ & $\leq$ & ${\pi_{chase}}_2$	& ${\pi_{chase}}_2$ & $\leq$ & ${\pi_{chase}}_3$
\end{tabular}
\begin{tabular}{c c c | c c c | c c c}
Slot 3 & & Slot 4								& Slot 3 & & Slot 4									& Slot 3 & & Slot 4\\
\hline
${\pi_{chase}}_2$ & $\leq$ & ${\pi_{chase}}_3$	& ${\pi_{chase}}_3$ & $\leq$ & ${\pi_{chase}}_1$	& ${\pi_{chase}}_3$ & $\leq$ & ${\pi_{chase}}_2$\\
${\pi_{chase}}_3$ & $\leq$ & ${\pi_{chase}}_1$	& ${\pi_{chase}}_1$ & $\leq$ & ${\pi_{chase}}_2$	& ${\pi_{chase}}_2$ & $\leq$ & ${\pi_{chase}}_1$\\
${\pi_{chase}}_2$ & $\leq$ & ${\pi_{chase}}_1$	& ${\pi_{chase}}_3$ & $\leq$ & ${\pi_{chase}}_2$	& ${\pi_{chase}}_3$ & $\leq$ & ${\pi_{chase}}_1$
\end{tabular}
\end{center}

It is possible to find a PO for every permutation, since the agents in slots 3 and 4 share rewards (and expected average rewards as well) due they are in the same team. So no matter which agents are in $\Pi_o$ we obtain:

\begin{equation*}
PG_{3,4}({\pi_{chase}}_1,{\pi_{chase}}_2,{\pi_{chase}}_3,\Pi_o,w_{\dot{L}},\mu) = 1
\end{equation*}

Therefore:

\begin{equation*}
PG_{3,4}(\Pi_e,w_{\Pi_e},\Pi_o,w_{\dot{L}},\mu) = 6 \frac{9}{2} \frac{1}{3} \frac{1}{3} \frac{1}{3} 1 = 1
\end{equation*}

For slots 4 and 1:

\begin{equation*}
\begin{aligned}
PG_{4,1}(\Pi_e,w_{\Pi_e},\Pi_o,w_{\dot{L}},\mu)	& = \eta_{\Pi^3} \sum_{\pi_1,\pi_2,\pi_3 \in \Pi_e | \pi_1 \neq \pi_2 \neq \pi_3} w_{\Pi_e}(\pi_1) w_{\Pi_e}(\pi_2) w_{\Pi_e}(\pi_3) PG_{4,1}(\pi_1,\pi_2,\pi_3,\Pi_o,w_{\dot{L}},\mu) =\\
												& = 6 \frac{9}{2} \frac{1}{3} \frac{1}{3} \frac{1}{3} PG_{4,1}({\pi_{chase}}_1,{\pi_{chase}}_2,{\pi_{chase}}_3,\Pi_o,w_{\dot{L}},\mu)
\end{aligned}
\end{equation*}

In this case, we only need to calculate $PG_{4,1}({\pi_{chase}}_1,{\pi_{chase}}_2,{\pi_{chase}}_3,\Pi_o,w_{\dot{L}},\mu)$. We follow definition \ref{def:STG_agents} (for PG) to calculate this value:

\begin{equation*}
PG_{4,1}({\pi_{chase}}_1,{\pi_{chase}}_2,{\pi_{chase}}_3,\Pi_o,w_{\dot{L}},\mu) = \sum_{\dot{l} \in \dot{L}^{N(\mu)}_{-4,1}(\Pi_o)} w_{\dot{L}}(\dot{l}) PO_{4,1}({\pi_{chase}}_1,{\pi_{chase}}_2,{\pi_{chase}}_3,\dot{l},\mu)
\end{equation*}

Again, we do not know which $\Pi_o$ we have, but we know that we will need to obtain a line-up pattern $\dot{l}$ from $\dot{L}^{N(\mu)}_{-4,1}(\Pi_o)$ to calculate $PO_{4,1}({\pi_{chase}}_1,{\pi_{chase}}_2,{\pi_{chase}}_3,\dot{l},\mu)$. We calculate this value for a figurative line-up pattern $\dot{l} = (*,\pi_1,\pi_2,*)$ from $\dot{L}^{N(\mu)}_{-4,1}(\Pi_o)$:

\begin{equation*}
PO_{4,1}({\pi_{chase}}_1,{\pi_{chase}}_2,{\pi_{chase}}_3,\dot{l},\mu) = PO_{4,1}({\pi_{chase}}_1,{\pi_{chase}}_2,{\pi_{chase}}_3,(*,\pi_1,\pi_2,*),\mu)
\end{equation*}

The following table shows us $PO_{4,1}$ for all the permutations of ${\pi_{chase}}_1,{\pi_{chase}}_2,{\pi_{chase}}_3$.

\begin{center}
\begin{tabular}{c c c | c c c | c c c}
Slot 4 & & Slot 1								& Slot 4 & & Slot 1									& Slot 4 & & Slot 1\\
\hline
${\pi_{chase}}_1$ & $\leq$ & ${\pi_{chase}}_2$	& ${\pi_{chase}}_1$ & $\leq$ & ${\pi_{chase}}_3$	& ${\pi_{chase}}_2$ & $\leq$ & ${\pi_{chase}}_1$\\
${\pi_{chase}}_2$ & $\leq$ & ${\pi_{chase}}_3$	& ${\pi_{chase}}_3$ & $\leq$ & ${\pi_{chase}}_2$	& ${\pi_{chase}}_1$ & $\leq$ & ${\pi_{chase}}_3$\\
${\pi_{chase}}_1$ & $\leq$ & ${\pi_{chase}}_3$	& ${\pi_{chase}}_1$ & $\leq$ & ${\pi_{chase}}_2$	& ${\pi_{chase}}_2$ & $\leq$ & ${\pi_{chase}}_3$
\end{tabular}
\begin{tabular}{c c c | c c c | c c c}
Slot 4 & & Slot 1								& Slot 4 & & Slot 1									& Slot 4 & & Slot 1\\
\hline
${\pi_{chase}}_2$ & $\leq$ & ${\pi_{chase}}_3$	& ${\pi_{chase}}_3$ & $\leq$ & ${\pi_{chase}}_1$	& ${\pi_{chase}}_3$ & $\leq$ & ${\pi_{chase}}_2$\\
${\pi_{chase}}_3$ & $\leq$ & ${\pi_{chase}}_1$	& ${\pi_{chase}}_1$ & $\leq$ & ${\pi_{chase}}_2$	& ${\pi_{chase}}_2$ & $\leq$ & ${\pi_{chase}}_1$\\
${\pi_{chase}}_2$ & $\leq$ & ${\pi_{chase}}_1$	& ${\pi_{chase}}_3$ & $\leq$ & ${\pi_{chase}}_2$	& ${\pi_{chase}}_3$ & $\leq$ & ${\pi_{chase}}_1$
\end{tabular}
\end{center}

Again, it is not possible to find a PO for any permutation, since we always have ${\pi_{chase}}_i \leq {\pi_{chase}}_j$, where a $\pi_{chase}$ agent always tries to be chased when playing as the prey and tries to chase when playing as a predator, so the agents in slots 4 and 1 will obtain an expected average reward of $1$ and $-1$ respectively. Note that the choice of $\Pi_o$ does not affect the result of $PO_{4,1}$, so no matter which agents are in $\Pi_o$ we obtain:

\begin{equation*}
PG_{4,1}({\pi_{chase}}_1,{\pi_{chase}}_2,{\pi_{chase}}_3,\Pi_o,w_{\dot{L}},\mu) = 0
\end{equation*}

Therefore:

\begin{equation*}
PG_{4,1}(\Pi_e,w_{\Pi_e},\Pi_o,w_{\dot{L}},\mu) = 6 \frac{9}{2} \frac{1}{3} \frac{1}{3} \frac{1}{3} 0 = 0
\end{equation*}

For slots 4 and 2:

\begin{equation*}
\begin{aligned}
PG_{4,2}(\Pi_e,w_{\Pi_e},\Pi_o,w_{\dot{L}},\mu)	& = \eta_{\Pi^3} \sum_{\pi_1,\pi_2,\pi_3 \in \Pi_e | \pi_1 \neq \pi_2 \neq \pi_3} w_{\Pi_e}(\pi_1) w_{\Pi_e}(\pi_2) w_{\Pi_e}(\pi_3) PG_{4,2}(\pi_1,\pi_2,\pi_3,\Pi_o,w_{\dot{L}},\mu) =\\
												& = 6 \frac{9}{2} \frac{1}{3} \frac{1}{3} \frac{1}{3} PG_{4,2}({\pi_{chase}}_1,{\pi_{chase}}_2,{\pi_{chase}}_3,\Pi_o,w_{\dot{L}},\mu)
\end{aligned}
\end{equation*}

In this case, we only need to calculate $PG_{4,2}({\pi_{chase}}_1,{\pi_{chase}}_2,{\pi_{chase}}_3,\Pi_o,w_{\dot{L}},\mu)$. We follow definition \ref{def:STG_agents} (for PG) to calculate this value:

\begin{equation*}
PG_{4,2}({\pi_{chase}}_1,{\pi_{chase}}_2,{\pi_{chase}}_3,\Pi_o,w_{\dot{L}},\mu) = \sum_{\dot{l} \in \dot{L}^{N(\mu)}_{-4,2}(\Pi_o)} w_{\dot{L}}(\dot{l}) PO_{4,2}({\pi_{chase}}_1,{\pi_{chase}}_2,{\pi_{chase}}_3,\dot{l},\mu)
\end{equation*}

Again, we do not know which $\Pi_o$ we have, but we know that we will need to obtain a line-up pattern $\dot{l}$ from $\dot{L}^{N(\mu)}_{-4,2}(\Pi_o)$ to calculate $PO_{4,2}({\pi_{chase}}_1,{\pi_{chase}}_2,{\pi_{chase}}_3,\dot{l},\mu)$. We calculate this value for a figurative line-up pattern $\dot{l} = (\pi_1,*,\pi_2,*)$ from $\dot{L}^{N(\mu)}_{-4,2}(\Pi_o)$:

\begin{equation*}
PO_{4,2}({\pi_{chase}}_1,{\pi_{chase}}_2,{\pi_{chase}}_3,\dot{l},\mu) = PO_{4,2}({\pi_{chase}}_1,{\pi_{chase}}_2,{\pi_{chase}}_3,(\pi_1,*,\pi_2,*),\mu)
\end{equation*}

The following table shows us $PO_{4,2}$ for all the permutations of ${\pi_{chase}}_1,{\pi_{chase}}_2,{\pi_{chase}}_3$.

\begin{center}
\begin{tabular}{c c c | c c c | c c c}
Slot 4 & & Slot 2								& Slot 4 & & Slot 2									& Slot 4 & & Slot 2\\
\hline
${\pi_{chase}}_1$ & $\leq$ & ${\pi_{chase}}_2$	& ${\pi_{chase}}_1$ & $\leq$ & ${\pi_{chase}}_3$	& ${\pi_{chase}}_2$ & $\leq$ & ${\pi_{chase}}_1$\\
${\pi_{chase}}_2$ & $\leq$ & ${\pi_{chase}}_3$	& ${\pi_{chase}}_3$ & $\leq$ & ${\pi_{chase}}_2$	& ${\pi_{chase}}_1$ & $\leq$ & ${\pi_{chase}}_3$\\
${\pi_{chase}}_1$ & $\leq$ & ${\pi_{chase}}_3$	& ${\pi_{chase}}_1$ & $\leq$ & ${\pi_{chase}}_2$	& ${\pi_{chase}}_2$ & $\leq$ & ${\pi_{chase}}_3$
\end{tabular}
\begin{tabular}{c c c | c c c | c c c}
Slot 4 & & Slot 2								& Slot 4 & & Slot 2									& Slot 4 & & Slot 2\\
\hline
${\pi_{chase}}_2$ & $\leq$ & ${\pi_{chase}}_3$	& ${\pi_{chase}}_3$ & $\leq$ & ${\pi_{chase}}_1$	& ${\pi_{chase}}_3$ & $\leq$ & ${\pi_{chase}}_2$\\
${\pi_{chase}}_3$ & $\leq$ & ${\pi_{chase}}_1$	& ${\pi_{chase}}_1$ & $\leq$ & ${\pi_{chase}}_2$	& ${\pi_{chase}}_2$ & $\leq$ & ${\pi_{chase}}_1$\\
${\pi_{chase}}_2$ & $\leq$ & ${\pi_{chase}}_1$	& ${\pi_{chase}}_3$ & $\leq$ & ${\pi_{chase}}_2$	& ${\pi_{chase}}_3$ & $\leq$ & ${\pi_{chase}}_1$
\end{tabular}
\end{center}

It is possible to find a PO for every permutation, since the agents in slots 4 and 2 share rewards (and expected average rewards as well) due they are in the same team. So no matter which agents are in $\Pi_o$ we obtain:

\begin{equation*}
PG_{4,2}({\pi_{chase}}_1,{\pi_{chase}}_2,{\pi_{chase}}_3,\Pi_o,w_{\dot{L}},\mu) = 1
\end{equation*}

Therefore:

\begin{equation*}
PG_{4,2}(\Pi_e,w_{\Pi_e},\Pi_o,w_{\dot{L}},\mu) = 6 \frac{9}{2} \frac{1}{3} \frac{1}{3} \frac{1}{3} 1 = 1
\end{equation*}

And for slots 4 and 3:

\begin{equation*}
\begin{aligned}
PG_{4,3}(\Pi_e,w_{\Pi_e},\Pi_o,w_{\dot{L}},\mu)	& = \eta_{\Pi^3} \sum_{\pi_1,\pi_2,\pi_3 \in \Pi_e | \pi_1 \neq \pi_2 \neq \pi_3} w_{\Pi_e}(\pi_1) w_{\Pi_e}(\pi_2) w_{\Pi_e}(\pi_3) PG_{4,3}(\pi_1,\pi_2,\pi_3,\Pi_o,w_{\dot{L}},\mu) =\\
												& = 6 \frac{9}{2} \frac{1}{3} \frac{1}{3} \frac{1}{3} PG_{4,3}({\pi_{chase}}_1,{\pi_{chase}}_2,{\pi_{chase}}_3,\Pi_o,w_{\dot{L}},\mu)
\end{aligned}
\end{equation*}

In this case, we only need to calculate $PG_{4,3}({\pi_{chase}}_1,{\pi_{chase}}_2,{\pi_{chase}}_3,\Pi_o,w_{\dot{L}},\mu)$. We follow definition \ref{def:STG_agents} (for PG) to calculate this value:

\begin{equation*}
PG_{4,3}({\pi_{chase}}_1,{\pi_{chase}}_2,{\pi_{chase}}_3,\Pi_o,w_{\dot{L}},\mu) = \sum_{\dot{l} \in \dot{L}^{N(\mu)}_{-4,3}(\Pi_o)} w_{\dot{L}}(\dot{l}) PO_{4,3}({\pi_{chase}}_1,{\pi_{chase}}_2,{\pi_{chase}}_3,\dot{l},\mu)
\end{equation*}

Again, we do not know which $\Pi_o$ we have, but we know that we will need to obtain a line-up pattern $\dot{l}$ from $\dot{L}^{N(\mu)}_{-4,3}(\Pi_o)$ to calculate $PO_{4,3}({\pi_{chase}}_1,{\pi_{chase}}_2,{\pi_{chase}}_3,\dot{l},\mu)$. We calculate this value for a figurative line-up pattern $\dot{l} = (\pi_1,\pi_2,*,*)$ from $\dot{L}^{N(\mu)}_{-4,3}(\Pi_o)$:

\begin{equation*}
PO_{4,3}({\pi_{chase}}_1,{\pi_{chase}}_2,{\pi_{chase}}_3,\dot{l},\mu) = PO_{4,3}({\pi_{chase}}_1,{\pi_{chase}}_2,{\pi_{chase}}_3,(\pi_1,\pi_2,*,*),\mu)
\end{equation*}

The following table shows us $PO_{4,3}$ for all the permutations of ${\pi_{chase}}_1,{\pi_{chase}}_2,{\pi_{chase}}_3$.

\begin{center}
\begin{tabular}{c c c | c c c | c c c}
Slot 4 & & Slot 3								& Slot 4 & & Slot 3									& Slot 4 & & Slot 3\\
\hline
${\pi_{chase}}_1$ & $\leq$ & ${\pi_{chase}}_2$	& ${\pi_{chase}}_1$ & $\leq$ & ${\pi_{chase}}_3$	& ${\pi_{chase}}_2$ & $\leq$ & ${\pi_{chase}}_1$\\
${\pi_{chase}}_2$ & $\leq$ & ${\pi_{chase}}_3$	& ${\pi_{chase}}_3$ & $\leq$ & ${\pi_{chase}}_2$	& ${\pi_{chase}}_1$ & $\leq$ & ${\pi_{chase}}_3$\\
${\pi_{chase}}_1$ & $\leq$ & ${\pi_{chase}}_3$	& ${\pi_{chase}}_1$ & $\leq$ & ${\pi_{chase}}_2$	& ${\pi_{chase}}_2$ & $\leq$ & ${\pi_{chase}}_3$
\end{tabular}
\begin{tabular}{c c c | c c c | c c c}
Slot 4 & & Slot 3								& Slot 4 & & Slot 3									& Slot 4 & & Slot 3\\
\hline
${\pi_{chase}}_2$ & $\leq$ & ${\pi_{chase}}_3$	& ${\pi_{chase}}_3$ & $\leq$ & ${\pi_{chase}}_1$	& ${\pi_{chase}}_3$ & $\leq$ & ${\pi_{chase}}_2$\\
${\pi_{chase}}_3$ & $\leq$ & ${\pi_{chase}}_1$	& ${\pi_{chase}}_1$ & $\leq$ & ${\pi_{chase}}_2$	& ${\pi_{chase}}_2$ & $\leq$ & ${\pi_{chase}}_1$\\
${\pi_{chase}}_2$ & $\leq$ & ${\pi_{chase}}_1$	& ${\pi_{chase}}_3$ & $\leq$ & ${\pi_{chase}}_2$	& ${\pi_{chase}}_3$ & $\leq$ & ${\pi_{chase}}_1$
\end{tabular}
\end{center}

It is possible to find a PO for every permutation, since the agents in slots 4 and 3 share rewards (and expected average rewards as well) due they are in the same team. So no matter which agents are in $\Pi_o$ we obtain:

\begin{equation*}
PG_{4,3}({\pi_{chase}}_1,{\pi_{chase}}_2,{\pi_{chase}}_3,\Pi_o,w_{\dot{L}},\mu) = 1
\end{equation*}

Therefore:

\begin{equation*}
PG_{4,3}(\Pi_e,w_{\Pi_e},\Pi_o,w_{\dot{L}},\mu) = 6 \frac{9}{2} \frac{1}{3} \frac{1}{3} \frac{1}{3} 1 = 1
\end{equation*}

And finally, we weight over the slots:

\begin{equation*}
\begin{aligned}
& PG(\Pi_e,w_{\Pi_e},\Pi_o,w_{\dot{L}},\mu,w_S) = \eta_{S_1^2} \sum_{i=1}^{N(\mu)} w_S(i,\mu) \times\\
& \times \left(\sum_{j=1}^{i-1} w_S(j,\mu) PG_{i,j}(\Pi_e,w_{\Pi_e},\Pi_o,w_{\dot{L}},\mu) + \sum_{j=i+1}^{N(\mu)} w_S(j,\mu) PG_{i,j}(\Pi_e,w_{\Pi_e},\Pi_o,w_{\dot{L}},\mu)\right) =\\
& \ \ \ \ \ \ \ \ \ \ \ \ \ \ \ \ \ \ \ \ \ \ \ \ \ \ \ \ \ \ \ \ \ \ \ \ = \frac{4}{3} \frac{1}{4} \frac{1}{4} \{PG_{1,2}(\Pi_e,w_{\Pi_e},\Pi_o,w_{\dot{L}},\mu) + PG_{1,3}(\Pi_e,w_{\Pi_e},\Pi_o,w_{\dot{L}},\mu) +\\
& \ \ \ \ \ \ \ \ \ \ \ \ \ \ \ \ \ \ \ \ \ \ \ \ \ \ \ \ \ \ \ \ \ \ \ \ + PG_{1,4}(\Pi_e,w_{\Pi_e},\Pi_o,w_{\dot{L}},\mu) + PG_{2,1}(\Pi_e,w_{\Pi_e},\Pi_o,w_{\dot{L}},\mu) +\\
& \ \ \ \ \ \ \ \ \ \ \ \ \ \ \ \ \ \ \ \ \ \ \ \ \ \ \ \ \ \ \ \ \ \ \ \ + PG_{2,3}(\Pi_e,w_{\Pi_e},\Pi_o,w_{\dot{L}},\mu) + PG_{2,4}(\Pi_e,w_{\Pi_e},\Pi_o,w_{\dot{L}},\mu) +\\
& \ \ \ \ \ \ \ \ \ \ \ \ \ \ \ \ \ \ \ \ \ \ \ \ \ \ \ \ \ \ \ \ \ \ \ \ + PG_{3,1}(\Pi_e,w_{\Pi_e},\Pi_o,w_{\dot{L}},\mu) + PG_{3,2}(\Pi_e,w_{\Pi_e},\Pi_o,w_{\dot{L}},\mu) +\\
& \ \ \ \ \ \ \ \ \ \ \ \ \ \ \ \ \ \ \ \ \ \ \ \ \ \ \ \ \ \ \ \ \ \ \ \ + PG_{3,4}(\Pi_e,w_{\Pi_e},\Pi_o,w_{\dot{L}},\mu) + PG_{4,1}(\Pi_e,w_{\Pi_e},\Pi_o,w_{\dot{L}},\mu) +\\
& \ \ \ \ \ \ \ \ \ \ \ \ \ \ \ \ \ \ \ \ \ \ \ \ \ \ \ \ \ \ \ \ \ \ \ \ + PG_{4,2}(\Pi_e,w_{\Pi_e},\Pi_o,w_{\dot{L}},\mu) + PG_{4,3}(\Pi_e,w_{\Pi_e},\Pi_o,w_{\dot{L}},\mu)\} =\\
& \ \ \ \ \ \ \ \ \ \ \ \ \ \ \ \ \ \ \ \ \ \ \ \ \ \ \ \ \ \ \ \ \ \ \ \ = \frac{4}{3} \frac{1}{4} \frac{1}{4} \left\{9 \times 1 + 3 \times 0\right\} = \frac{3}{4}
\end{aligned}
\end{equation*}

So, for every $\Pi_o$ we obtain the same result:

\begin{equation*}
\forall \Pi_o : PG(\Pi_e,w_{\Pi_e},\Pi_o,w_{\dot{L}},\mu,w_S) = \frac{3}{4}
\end{equation*}

Therefore, predator-prey has $Right_{min} = \frac{3}{4}$ (as a {\em higher} approximation) for this property.
\end{proof}
\end{approximation}

\subsection{Slot Reward Dependency}
Next we see the slot reward dependency (SRD) property. As given in section \ref{sec:SRD}, we want to know how much competitive or cooperative the environment is.

\begin{proposition}
\label{prop:predator-prey_SRD_general_range}
$General$ range for the slot reward dependency (SRD) property is equal to $[0,0]$ for the predator-prey environment.

\begin{proof}
Following definition \ref{def:SRD}, we obtain the SRD value for any $\left\langle\Pi_e,w_{\Pi_e},\Pi_o\right\rangle$ (where $\Pi_e, w_{\Pi_e}$ and $\Pi_o$ are instantiated with any permitted values). Since the environment is not symmetric, we need to calculate this property for every pair of slots. Following definition \ref{def:SRD_set}, we can calculate its SRD value for each pair of slots. We start with slots 1 and 2:

\begin{equation*}
SRD_{1,2}(\Pi_e,w_{\Pi_e},\Pi_o,w_{\dot{L}},\mu) = \sum_{\pi \in \Pi_e} w_{\Pi_e}(\pi) SRD_{1,2}(\pi,\Pi_o,w_{\dot{L}},\mu)
\end{equation*}

We do not know which $\Pi_e$ we have, but we know that we will need to evaluate $SRD_{1,2}(\pi,\Pi_o,w_{\dot{L}},\mu)$ for all evaluated agent $\pi \in \Pi_e$. We follow definition \ref{def:SRD_agent} to calculate this value for a figurative evaluated agent $\pi_1$ from $\Pi_e$:

\begin{equation*}
SRD_{1,2}(\pi_1,\Pi_o,w_{\dot{L}},\mu) = corr_{\dot{l} \in \dot{L}^{N(\mu)}_{-1}(\Pi_o)}[w_{\dot{L}}(\dot{l})](R_1(\mu[\instantiation{l}{1}{\pi_1}]), R_2(\mu[\instantiation{l}{1}{\pi_1}]))
\end{equation*}

We do not know which $\Pi_o$ we have, but we know that we will need to obtain a line-up pattern $\dot{l}$ from $\dot{L}^{N(\mu)}_{-1}(\Pi_o)$ to calculate $corr(R_1(\mu[\instantiation{l}{1}{\pi_1}]), R_2(\mu[\instantiation{l}{1}{\pi_1}]))$. We calculate this value for a figurative line-up pattern $\dot{l} = (*,\pi_2,\pi_3,\pi_4)$ from $\dot{L}^{N(\mu)}_{-1}(\Pi_o)$:

\begin{equation*}
corr(R_1(\mu[\instantiation{l}{1}{\pi_1}]), R_2(\mu[\instantiation{l}{1}{\pi_1}])) = corr(R_1(\mu[\pi_1,\pi_2,\pi_3,\pi_4]), R_2(\mu[\pi_1,\pi_2,\pi_3,\pi_4]))
\end{equation*}

From the predator-prey's payoff matrix (figure \ref{fig:predator-prey_payoff}), we can see that when the agent in slot 1 (any $\pi_1$) obtains a reward $r$ the agent in slot 2 (any $\pi_2$) obtains $-r$ as reward, and this relation is propagated to expected average rewards as well. Since we use a correlation function between the expected average rewards, and the agents in slots 1 and 2 always obtain opposite expected average reward, then the correlation function will always obtain the same value\footnote{Provided there is at least one game which is not a tie.} of $-1$. So:

\begin{equation*}
SRD_{1,2}(\pi_1,\Pi_o,w_{\dot{L}},\mu) = -1
\end{equation*}

Therefore:

\begin{equation*}
SRD_{1,2}(\Pi_e,w_{\Pi_e},\Pi_o,w_{\dot{L}},\mu) = -1
\end{equation*}

For slots 1 and 3:

\begin{equation*}
SRD_{1,3}(\Pi_e,w_{\Pi_e},\Pi_o,w_{\dot{L}},\mu) = \sum_{\pi \in \Pi_e} w_{\Pi_e}(\pi) SRD_{1,3}(\pi,\Pi_o,w_{\dot{L}},\mu)
\end{equation*}

We do not know which $\Pi_e$ we have, but we know that we will need to evaluate $SRD_{1,3}(\pi,\Pi_o,w_{\dot{L}},\mu)$ for all evaluated agent $\pi \in \Pi_e$. We follow definition \ref{def:SRD_agent} to calculate this value for a figurative evaluated agent $\pi_1$ from $\Pi_e$:

\begin{equation*}
SRD_{1,3}(\pi_1,\Pi_o,w_{\dot{L}},\mu) = corr_{\dot{l} \in \dot{L}^{N(\mu)}_{-1}(\Pi_o)}[w_{\dot{L}}(\dot{l})](R_1(\mu[\instantiation{l}{1}{\pi_1}]), R_3(\mu[\instantiation{l}{1}{\pi_1}]))
\end{equation*}

We do not know which $\Pi_o$ we have, but we know that we will need to obtain a line-up pattern $\dot{l}$ from $\dot{L}^{N(\mu)}_{-1}(\Pi_o)$ to calculate $corr(R_1(\mu[\instantiation{l}{1}{\pi_1}]), R_3(\mu[\instantiation{l}{1}{\pi_1}]))$. We calculate this value for a figurative line-up pattern $\dot{l} = (*,\pi_2,\pi_3,\pi_4)$ from $\dot{L}^{N(\mu)}_{-1}(\Pi_o)$:

\begin{equation*}
corr(R_1(\mu[\instantiation{l}{1}{\pi_1}]), R_3(\mu[\instantiation{l}{1}{\pi_1}])) = corr(R_1(\mu[\pi_1,\pi_2,\pi_3,\pi_4]), R_3(\mu[\pi_1,\pi_2,\pi_3,\pi_4]))
\end{equation*}

From the predator-prey's payoff matrix (figure \ref{fig:predator-prey_payoff}), we can see that when the agent in slot 1 (any $\pi_1$) obtains a reward $r$ the agent in slot 3 (any $\pi_3$) obtains $-r$ as reward, and this relation is propagated to expected average rewards as well. Since we use a correlation function between the expected average rewards, and the agents in slots 1 and 3 always obtain opposite expected average reward, then the correlation function will always obtain the same value of $-1$. So:

\begin{equation*}
SRD_{1,3}(\pi_1,\Pi_o,w_{\dot{L}},\mu) = -1
\end{equation*}

Therefore:

\begin{equation*}
SRD_{1,3}(\Pi_e,w_{\Pi_e},\Pi_o,w_{\dot{L}},\mu) = -1
\end{equation*}

For slots 1 and 4:

\begin{equation*}
SRD_{1,4}(\Pi_e,w_{\Pi_e},\Pi_o,w_{\dot{L}},\mu) = \sum_{\pi \in \Pi_e} w_{\Pi_e}(\pi) SRD_{1,4}(\pi,\Pi_o,w_{\dot{L}},\mu)
\end{equation*}

We do not know which $\Pi_e$ we have, but we know that we will need to evaluate $SRD_{1,4}(\pi,\Pi_o,w_{\dot{L}},\mu)$ for all evaluated agent $\pi \in \Pi_e$. We follow definition \ref{def:SRD_agent} to calculate this value for a figurative evaluated agent $\pi_1$ from $\Pi_e$:

\begin{equation*}
SRD_{1,4}(\pi_1,\Pi_o,w_{\dot{L}},\mu) = corr_{\dot{l} \in \dot{L}^{N(\mu)}_{-1}(\Pi_o)}[w_{\dot{L}}(\dot{l})](R_1(\mu[\instantiation{l}{1}{\pi_1}]), R_4(\mu[\instantiation{l}{1}{\pi_1}]))
\end{equation*}

We do not know which $\Pi_o$ we have, but we know that we will need to obtain a line-up pattern $\dot{l}$ from $\dot{L}^{N(\mu)}_{-1}(\Pi_o)$ to calculate $corr(R_1(\mu[\instantiation{l}{1}{\pi_1}]), R_4(\mu[\instantiation{l}{1}{\pi_1}]))$. We calculate this value for a figurative line-up pattern $\dot{l} = (*,\pi_2,\pi_3,\pi_4)$ from $\dot{L}^{N(\mu)}_{-1}(\Pi_o)$:

\begin{equation*}
corr(R_1(\mu[\instantiation{l}{1}{\pi_1}]), R_4(\mu[\instantiation{l}{1}{\pi_1}])) = corr(R_1(\mu[\pi_1,\pi_2,\pi_3,\pi_4]), R_4(\mu[\pi_1,\pi_2,\pi_3,\pi_4]))
\end{equation*}

From the predator-prey's payoff matrix (figure \ref{fig:predator-prey_payoff}), we can see that when the agent in slot 1 (any $\pi_1$) obtains a reward $r$ the agent in slot 4 (any $\pi_4$) obtains $-r$ as reward, and this relation is propagated to expected average rewards as well. Since we use a correlation function between the expected average rewards, and the agents in slots 1 and 4 always obtain opposite expected average reward, then the correlation function will always obtain the same value of $-1$. So:

\begin{equation*}
SRD_{1,4}(\pi_1,\Pi_o,w_{\dot{L}},\mu) = -1
\end{equation*}

Therefore:

\begin{equation*}
SRD_{1,4}(\Pi_e,w_{\Pi_e},\Pi_o,w_{\dot{L}},\mu) = -1
\end{equation*}

For slots 2 and 1:

\begin{equation*}
SRD_{2,1}(\Pi_e,w_{\Pi_e},\Pi_o,w_{\dot{L}},\mu) = \sum_{\pi \in \Pi_e} w_{\Pi_e}(\pi) SRD_{2,1}(\pi,\Pi_o,w_{\dot{L}},\mu)
\end{equation*}

We do not know which $\Pi_e$ we have, but we know that we will need to evaluate $SRD_{2,1}(\pi,\Pi_o,w_{\dot{L}},\mu)$ for all evaluated agent $\pi \in \Pi_e$. We follow definition \ref{def:SRD_agent} to calculate this value for a figurative evaluated agent $\pi_1$ from $\Pi_e$:

\begin{equation*}
SRD_{2,1}(\pi_1,\Pi_o,w_{\dot{L}},\mu) = corr_{\dot{l} \in \dot{L}^{N(\mu)}_{-2}(\Pi_o)}[w_{\dot{L}}(\dot{l})](R_2(\mu[\instantiation{l}{2}{\pi_1}]), R_1(\mu[\instantiation{l}{2}{\pi_1}]))
\end{equation*}

We do not know which $\Pi_o$ we have, but we know that we will need to obtain a line-up pattern $\dot{l}$ from $\dot{L}^{N(\mu)}_{-2}(\Pi_o)$ to calculate $corr(R_2(\mu[\instantiation{l}{2}{\pi_1}]), R_1(\mu[\instantiation{l}{2}{\pi_1}]))$. We calculate this value for a figurative line-up pattern $\dot{l} = (\pi_2,*,\pi_3,\pi_4)$ from $\dot{L}^{N(\mu)}_{-2}(\Pi_o)$:

\begin{equation*}
corr(R_2(\mu[\instantiation{l}{2}{\pi_1}]), R_1(\mu[\instantiation{l}{2}{\pi_1}])) = corr(R_2(\mu[\pi_2,\pi_1,\pi_3,\pi_4]), R_1(\mu[\pi_2,\pi_1,\pi_3,\pi_4]))
\end{equation*}

From the predator-prey's payoff matrix (figure \ref{fig:predator-prey_payoff}), we can see that when the agent in slot 2 (any $\pi_1$) obtains a reward $r$ the agent in slot 1 (any $\pi_2$) obtains $-r$ as reward, and this relation is propagated to expected average rewards as well. Since we use a correlation function between the expected average rewards, and the agents in slots 2 and 1 always obtain opposite expected average reward, then the correlation function will always obtain the same value of $-1$. So:

\begin{equation*}
SRD_{2,1}(\pi_1,\Pi_o,w_{\dot{L}},\mu) = -1
\end{equation*}

Therefore:

\begin{equation*}
SRD_{2,1}(\Pi_e,w_{\Pi_e},\Pi_o,w_{\dot{L}},\mu) = -1
\end{equation*}

For slots 2 and 3:

\begin{equation*}
SRD_{2,3}(\Pi_e,w_{\Pi_e},\Pi_o,w_{\dot{L}},\mu) = \sum_{\pi \in \Pi_e} w_{\Pi_e}(\pi) SRD_{2,3}(\pi,\Pi_o,w_{\dot{L}},\mu)
\end{equation*}

We do not know which $\Pi_e$ we have, but we know that we will need to evaluate $SRD_{2,3}(\pi,\Pi_o,w_{\dot{L}},\mu)$ for all evaluated agent $\pi \in \Pi_e$. We follow definition \ref{def:SRD_agent} to calculate this value for a figurative evaluated agent $\pi_1$ from $\Pi_e$:

\begin{equation*}
SRD_{2,3}(\pi_1,\Pi_o,w_{\dot{L}},\mu) = corr_{\dot{l} \in \dot{L}^{N(\mu)}_{-2}(\Pi_o)}[w_{\dot{L}}(\dot{l})](R_2(\mu[\instantiation{l}{2}{\pi_1}]), R_3(\mu[\instantiation{l}{2}{\pi_1}]))
\end{equation*}

We do not know which $\Pi_o$ we have, but we know that we will need to obtain a line-up pattern $\dot{l}$ from $\dot{L}^{N(\mu)}_{-2}(\Pi_o)$ to calculate $corr(R_2(\mu[\instantiation{l}{2}{\pi_1}]), R_3(\mu[\instantiation{l}{2}{\pi_1}]))$. We calculate this value for a figurative line-up pattern $\dot{l} = (\pi_2,*,\pi_3,\pi_4)$ from $\dot{L}^{N(\mu)}_{-2}(\Pi_o)$:

\begin{equation*}
corr(R_2(\mu[\instantiation{l}{2}{\pi_1}]), R_3(\mu[\instantiation{l}{2}{\pi_1}])) = corr(R_2(\mu[\pi_2,\pi_1,\pi_3,\pi_4]), R_3(\mu[\pi_2,\pi_1,\pi_3,\pi_4]))
\end{equation*}

From the predator-prey's payoff matrix (figure \ref{fig:predator-prey_payoff}), we can see that when the agent in slot 2 (any $\pi_1$) obtains a reward $r$ the agent in slot 3 (any $\pi_3$) obtains the same value $r$ as reward (since they are in the same team), and this relation is propagated to expected average rewards as well. Since we use a correlation function between the expected average rewards, and the agents in slots 2 and 3 always obtain the same expected average reward, then the correlation function will always obtain the same value of $1$. So:

\begin{equation*}
SRD_{2,3}(\pi_1,\Pi_o,w_{\dot{L}},\mu) = 1
\end{equation*}

Therefore:

\begin{equation*}
SRD_{2,3}(\Pi_e,w_{\Pi_e},\Pi_o,w_{\dot{L}},\mu) = 1
\end{equation*}

For slots 2 and 4:

\begin{equation*}
SRD_{2,4}(\Pi_e,w_{\Pi_e},\Pi_o,w_{\dot{L}},\mu) = \sum_{\pi \in \Pi_e} w_{\Pi_e}(\pi) SRD_{2,4}(\pi,\Pi_o,w_{\dot{L}},\mu)
\end{equation*}

We do not know which $\Pi_e$ we have, but we know that we will need to evaluate $SRD_{2,4}(\pi,\Pi_o,w_{\dot{L}},\mu)$ for all evaluated agent $\pi \in \Pi_e$. We follow definition \ref{def:SRD_agent} to calculate this value for a figurative evaluated agent $\pi_1$ from $\Pi_e$:

\begin{equation*}
SRD_{2,4}(\pi_1,\Pi_o,w_{\dot{L}},\mu) = corr_{\dot{l} \in \dot{L}^{N(\mu)}_{-2}(\Pi_o)}[w_{\dot{L}}(\dot{l})](R_2(\mu[\instantiation{l}{2}{\pi_1}]), R_4(\mu[\instantiation{l}{2}{\pi_1}]))
\end{equation*}

We do not know which $\Pi_o$ we have, but we know that we will need to obtain a line-up pattern $\dot{l}$ from $\dot{L}^{N(\mu)}_{-2}(\Pi_o)$ to calculate $corr(R_2(\mu[\instantiation{l}{2}{\pi_1}]), R_4(\mu[\instantiation{l}{2}{\pi_1}]))$. We calculate this value for a figurative line-up pattern $\dot{l} = (\pi_2,*,\pi_3,\pi_4)$ from $\dot{L}^{N(\mu)}_{-2}(\Pi_o)$:

\begin{equation*}
corr(R_2(\mu[\instantiation{l}{2}{\pi_1}]), R_4(\mu[\instantiation{l}{2}{\pi_1}])) = corr(R_2(\mu[\pi_2,\pi_1,\pi_3,\pi_4]), R_4(\mu[\pi_2,\pi_1,\pi_3,\pi_4]))
\end{equation*}

From the predator-prey's payoff matrix (figure \ref{fig:predator-prey_payoff}), we can see that when the agent in slot 2 (any $\pi_1$) obtains a reward $r$ the agent in slot 4 (any $\pi_4$) obtains the same value $r$ as reward (since they are in the same team), and this relation is propagated to expected average rewards as well. Since we use a correlation function between the expected average rewards, and the agents in slots 2 and 4 always obtain the same expected average reward, then the correlation function will always obtain the same value of $1$. So:

\begin{equation*}
SRD_{2,4}(\pi_1,\Pi_o,w_{\dot{L}},\mu) = 1
\end{equation*}

Therefore:

\begin{equation*}
SRD_{2,4}(\Pi_e,w_{\Pi_e},\Pi_o,w_{\dot{L}},\mu) = 1
\end{equation*}

For slots 3 and 1:

\begin{equation*}
SRD_{3,1}(\Pi_e,w_{\Pi_e},\Pi_o,w_{\dot{L}},\mu) = \sum_{\pi \in \Pi_e} w_{\Pi_e}(\pi) SRD_{3,1}(\pi,\Pi_o,w_{\dot{L}},\mu)
\end{equation*}

We do not know which $\Pi_e$ we have, but we know that we will need to evaluate $SRD_{3,1}(\pi,\Pi_o,w_{\dot{L}},\mu)$ for all evaluated agent $\pi \in \Pi_e$. We follow definition \ref{def:SRD_agent} to calculate this value for a figurative evaluated agent $\pi_1$ from $\Pi_e$:

\begin{equation*}
SRD_{3,1}(\pi_1,\Pi_o,w_{\dot{L}},\mu) = corr_{\dot{l} \in \dot{L}^{N(\mu)}_{-3}(\Pi_o)}[w_{\dot{L}}(\dot{l})](R_3(\mu[\instantiation{l}{3}{\pi_1}]), R_1(\mu[\instantiation{l}{3}{\pi_1}]))
\end{equation*}

We do not know which $\Pi_o$ we have, but we know that we will need to obtain a line-up pattern $\dot{l}$ from $\dot{L}^{N(\mu)}_{-3}(\Pi_o)$ to calculate $corr(R_3(\mu[\instantiation{l}{3}{\pi_1}]), R_1(\mu[\instantiation{l}{3}{\pi_1}]))$. We calculate this value for a figurative line-up pattern $\dot{l} = (\pi_2,\pi_3,*,\pi_4)$ from $\dot{L}^{N(\mu)}_{-3}(\Pi_o)$:

\begin{equation*}
corr(R_3(\mu[\instantiation{l}{3}{\pi_1}]), R_1(\mu[\instantiation{l}{3}{\pi_1}])) = corr(R_3(\mu[\pi_2,\pi_3,\pi_1,\pi_4]), R_1(\mu[\pi_2,\pi_3,\pi_1,\pi_4]))
\end{equation*}

From the predator-prey's payoff matrix (figure \ref{fig:predator-prey_payoff}), we can see that when the agent in slot 3 (any $\pi_1$) obtains a reward $r$ the agent in slot 1 (any $\pi_2$) obtains $-r$ as reward, and this relation is propagated to expected average rewards as well. Since we use a correlation function between the expected average rewards, and the agents in slots 3 and 1 always obtain opposite expected average reward, then the correlation function will always obtain the same value of $-1$. So:

\begin{equation*}
SRD_{3,1}(\pi_1,\Pi_o,w_{\dot{L}},\mu) = -1
\end{equation*}

Therefore:

\begin{equation*}
SRD_{3,1}(\Pi_e,w_{\Pi_e},\Pi_o,w_{\dot{L}},\mu) = -1
\end{equation*}

For slots 3 and 2:

\begin{equation*}
SRD_{3,2}(\Pi_e,w_{\Pi_e},\Pi_o,w_{\dot{L}},\mu) = \sum_{\pi \in \Pi_e} w_{\Pi_e}(\pi) SRD_{3,2}(\pi,\Pi_o,w_{\dot{L}},\mu)
\end{equation*}

We do not know which $\Pi_e$ we have, but we know that we will need to evaluate $SRD_{3,2}(\pi,\Pi_o,w_{\dot{L}},\mu)$ for all evaluated agent $\pi \in \Pi_e$. We follow definition \ref{def:SRD_agent} to calculate this value for a figurative evaluated agent $\pi_1$ from $\Pi_e$:

\begin{equation*}
SRD_{3,2}(\pi_1,\Pi_o,w_{\dot{L}},\mu) = corr_{\dot{l} \in \dot{L}^{N(\mu)}_{-3}(\Pi_o)}[w_{\dot{L}}(\dot{l})](R_3(\mu[\instantiation{l}{3}{\pi_1}]), R_2(\mu[\instantiation{l}{3}{\pi_1}]))
\end{equation*}

We do not know which $\Pi_o$ we have, but we know that we will need to obtain a line-up pattern $\dot{l}$ from $\dot{L}^{N(\mu)}_{-3}(\Pi_o)$ to calculate $corr(R_3(\mu[\instantiation{l}{3}{\pi_1}]), R_2(\mu[\instantiation{l}{3}{\pi_1}]))$. We calculate this value for a figurative line-up pattern $\dot{l} = (\pi_2,\pi_3,*,\pi_4)$ from $\dot{L}^{N(\mu)}_{-3}(\Pi_o)$:

\begin{equation*}
corr(R_3(\mu[\instantiation{l}{3}{\pi_1}]), R_2(\mu[\instantiation{l}{3}{\pi_1}])) = corr(R_3(\mu[\pi_2,\pi_3,\pi_1,\pi_4]), R_2(\mu[\pi_2,\pi_3,\pi_1,\pi_4]))
\end{equation*}

From the predator-prey's payoff matrix (figure \ref{fig:predator-prey_payoff}), we can see that when the agent in slot 3 (any $\pi_1$) obtains a reward $r$ the agent in slot 2 (any $\pi_3$) obtains the same value $r$ as reward (since they are in the same team), and this relation is propagated to expected average rewards as well. Since we use a correlation function between the expected average rewards, and the agents in slots 3 and 2 always obtain the same expected average reward, then the correlation function will always obtain the same value of $1$. So:

\begin{equation*}
SRD_{3,2}(\pi_1,\Pi_o,w_{\dot{L}},\mu) = 1
\end{equation*}

Therefore:

\begin{equation*}
SRD_{3,2}(\Pi_e,w_{\Pi_e},\Pi_o,w_{\dot{L}},\mu) = 1
\end{equation*}

For slots 3 and 4:

\begin{equation*}
SRD_{3,4}(\Pi_e,w_{\Pi_e},\Pi_o,w_{\dot{L}},\mu) = \sum_{\pi \in \Pi_e} w_{\Pi_e}(\pi) SRD_{3,4}(\pi,\Pi_o,w_{\dot{L}},\mu)
\end{equation*}

We do not know which $\Pi_e$ we have, but we know that we will need to evaluate $SRD_{3,4}(\pi,\Pi_o,w_{\dot{L}},\mu)$ for all evaluated agent $\pi \in \Pi_e$. We follow definition \ref{def:SRD_agent} to calculate this value for a figurative evaluated agent $\pi_1$ from $\Pi_e$:

\begin{equation*}
SRD_{3,4}(\pi_1,\Pi_o,w_{\dot{L}},\mu) = corr_{\dot{l} \in \dot{L}^{N(\mu)}_{-3}(\Pi_o)}[w_{\dot{L}}(\dot{l})](R_3(\mu[\instantiation{l}{3}{\pi_1}]), R_4(\mu[\instantiation{l}{3}{\pi_1}]))
\end{equation*}

We do not know which $\Pi_o$ we have, but we know that we will need to obtain a line-up pattern $\dot{l}$ from $\dot{L}^{N(\mu)}_{-3}(\Pi_o)$ to calculate $corr(R_3(\mu[\instantiation{l}{3}{\pi_1}]), R_4(\mu[\instantiation{l}{3}{\pi_1}]))$. We calculate this value for a figurative line-up pattern $\dot{l} = (\pi_2,\pi_3,*,\pi_4)$ from $\dot{L}^{N(\mu)}_{-3}(\Pi_o)$:

\begin{equation*}
corr(R_3(\mu[\instantiation{l}{3}{\pi_1}]), R_4(\mu[\instantiation{l}{3}{\pi_1}])) = corr(R_3(\mu[\pi_2,\pi_3,\pi_1,\pi_4]), R_4(\mu[\pi_2,\pi_3,\pi_1,\pi_4]))
\end{equation*}

From the predator-prey's payoff matrix (figure \ref{fig:predator-prey_payoff}), we can see that when the agent in slot 3 (any $\pi_1$) obtains a reward $r$ the agent in slot 4 (any $\pi_4$) obtains the same value $r$ as reward (since they are in the same team), and this relation is propagated to expected average rewards as well. Since we use a correlation function between the expected average rewards, and the agents in slots 3 and 4 always obtain the same expected average reward, then the correlation function will always obtain the same value of $1$. So:

\begin{equation*}
SRD_{3,4}(\pi_1,\Pi_o,w_{\dot{L}},\mu) = 1
\end{equation*}

Therefore:

\begin{equation*}
SRD_{3,4}(\Pi_e,w_{\Pi_e},\Pi_o,w_{\dot{L}},\mu) = 1
\end{equation*}

For slots 4 and 1:

\begin{equation*}
SRD_{4,1}(\Pi_e,w_{\Pi_e},\Pi_o,w_{\dot{L}},\mu) = \sum_{\pi \in \Pi_e} w_{\Pi_e}(\pi) SRD_{4,1}(\pi,\Pi_o,w_{\dot{L}},\mu)
\end{equation*}

We do not know which $\Pi_e$ we have, but we know that we will need to evaluate $SRD_{4,1}(\pi,\Pi_o,w_{\dot{L}},\mu)$ for all evaluated agent $\pi \in \Pi_e$. We follow definition \ref{def:SRD_agent} to calculate this value for a figurative evaluated agent $\pi_1$ from $\Pi_e$:

\begin{equation*}
SRD_{4,1}(\pi_1,\Pi_o,w_{\dot{L}},\mu) = corr_{\dot{l} \in \dot{L}^{N(\mu)}_{-4}(\Pi_o)}[w_{\dot{L}}(\dot{l})](R_4(\mu[\instantiation{l}{4}{\pi_1}]), R_1(\mu[\instantiation{l}{4}{\pi_1}]))
\end{equation*}

We do not know which $\Pi_o$ we have, but we know that we will need to obtain a line-up pattern $\dot{l}$ from $\dot{L}^{N(\mu)}_{-4}(\Pi_o)$ to calculate $corr(R_4(\mu[\instantiation{l}{4}{\pi_1}]), R_1(\mu[\instantiation{l}{4}{\pi_1}]))$. We calculate this value for a figurative line-up pattern $\dot{l} = (\pi_2,\pi_3,\pi_4,*)$ from $\dot{L}^{N(\mu)}_{-4}(\Pi_o)$:

\begin{equation*}
corr(R_4(\mu[\instantiation{l}{4}{\pi_1}]), R_1(\mu[\instantiation{l}{4}{\pi_1}])) = corr(R_4(\mu[\pi_2,\pi_3,\pi_4,\pi_1]), R_1(\mu[\pi_2,\pi_3,\pi_4,\pi_1]))
\end{equation*}

From the predator-prey's payoff matrix (figure \ref{fig:predator-prey_payoff}), we can see that when the agent in slot 4 (any $\pi_1$) obtains a reward $r$ the agent in slot 1 (any $\pi_2$) obtains $-r$ as reward, and this relation is propagated to expected average rewards as well. Since we use a correlation function between the expected average rewards, and the agents in slots 4 and 1 always obtain opposite expected average reward, then the correlation function will always obtain the same value of $-1$. So:

\begin{equation*}
SRD_{4,1}(\pi_1,\Pi_o,w_{\dot{L}},\mu) = -1
\end{equation*}

Therefore:

\begin{equation*}
SRD_{4,1}(\Pi_e,w_{\Pi_e},\Pi_o,w_{\dot{L}},\mu) = -1
\end{equation*}

For slots 4 and 2:

\begin{equation*}
SRD_{4,2}(\Pi_e,w_{\Pi_e},\Pi_o,w_{\dot{L}},\mu) = \sum_{\pi \in \Pi_e} w_{\Pi_e}(\pi) SRD_{4,2}(\pi,\Pi_o,w_{\dot{L}},\mu)
\end{equation*}

We do not know which $\Pi_e$ we have, but we know that we will need to evaluate $SRD_{4,2}(\pi,\Pi_o,w_{\dot{L}},\mu)$ for all evaluated agent $\pi \in \Pi_e$. We follow definition \ref{def:SRD_agent} to calculate this value for a figurative evaluated agent $\pi_1$ from $\Pi_e$:

\begin{equation*}
SRD_{4,2}(\pi_1,\Pi_o,w_{\dot{L}},\mu) = corr_{\dot{l} \in \dot{L}^{N(\mu)}_{-4}(\Pi_o)}[w_{\dot{L}}(\dot{l})](R_4(\mu[\instantiation{l}{4}{\pi_1}]), R_2(\mu[\instantiation{l}{4}{\pi_1}]))
\end{equation*}

We do not know which $\Pi_o$ we have, but we know that we will need to obtain a line-up pattern $\dot{l}$ from $\dot{L}^{N(\mu)}_{-4}(\Pi_o)$ to calculate $corr(R_4(\mu[\instantiation{l}{4}{\pi_1}]), R_2(\mu[\instantiation{l}{4}{\pi_1}]))$. We calculate this value for a figurative line-up pattern $\dot{l} = (\pi_2,\pi_3,\pi_4,*)$ from $\dot{L}^{N(\mu)}_{-4}(\Pi_o)$:

\begin{equation*}
corr(R_4(\mu[\instantiation{l}{4}{\pi_1}]), R_2(\mu[\instantiation{l}{4}{\pi_1}])) = corr(R_4(\mu[\pi_2,\pi_3,\pi_4,\pi_1]), R_2(\mu[\pi_2,\pi_3,\pi_4,\pi_1]))
\end{equation*}

From the predator-prey's payoff matrix (figure \ref{fig:predator-prey_payoff}), we can see that when the agent in slot 4 (any $\pi_1$) obtains a reward $r$ the agent in slot 2 (any $\pi_3$) obtains the same value $r$ as reward (since they are in the same team), and this relation is propagated to expected average rewards as well. Since we use a correlation function between the expected average rewards, and the agents in slots 4 and 2 always obtain the same expected average reward, then the correlation function will always obtain the same value of $1$. So:

\begin{equation*}
SRD_{4,2}(\pi_1,\Pi_o,w_{\dot{L}},\mu) = 1
\end{equation*}

Therefore:

\begin{equation*}
SRD_{4,2}(\Pi_e,w_{\Pi_e},\Pi_o,w_{\dot{L}},\mu) = 1
\end{equation*}

And for slots 4 and 3:

\begin{equation*}
SRD_{4,3}(\Pi_e,w_{\Pi_e},\Pi_o,w_{\dot{L}},\mu) = \sum_{\pi \in \Pi_e} w_{\Pi_e}(\pi) SRD_{4,3}(\pi,\Pi_o,w_{\dot{L}},\mu)
\end{equation*}

We do not know which $\Pi_e$ we have, but we know that we will need to evaluate $SRD_{4,3}(\pi,\Pi_o,w_{\dot{L}},\mu)$ for all evaluated agent $\pi \in \Pi_e$. We follow definition \ref{def:SRD_agent} to calculate this value for a figurative evaluated agent $\pi_1$ from $\Pi_e$:

\begin{equation*}
SRD_{4,3}(\pi_1,\Pi_o,w_{\dot{L}},\mu) = corr_{\dot{l} \in \dot{L}^{N(\mu)}_{-4}(\Pi_o)}[w_{\dot{L}}(\dot{l})](R_4(\mu[\instantiation{l}{4}{\pi_1}]), R_3(\mu[\instantiation{l}{4}{\pi_1}]))
\end{equation*}

We do not know which $\Pi_o$ we have, but we know that we will need to obtain a line-up pattern $\dot{l}$ from $\dot{L}^{N(\mu)}_{-4}(\Pi_o)$ to calculate $corr(R_4(\mu[\instantiation{l}{4}{\pi_1}]), R_3(\mu[\instantiation{l}{4}{\pi_1}]))$. We calculate this value for a figurative line-up pattern $\dot{l} = (\pi_2,\pi_3,\pi_4,*)$ from $\dot{L}^{N(\mu)}_{-4}(\Pi_o)$:

\begin{equation*}
corr(R_4(\mu[\instantiation{l}{4}{\pi_1}]), R_3(\mu[\instantiation{l}{4}{\pi_1}])) = corr(R_4(\mu[\pi_2,\pi_3,\pi_4,\pi_1]), R_3(\mu[\pi_2,\pi_3,\pi_4,\pi_1]))
\end{equation*}

From the predator-prey's payoff matrix (figure \ref{fig:predator-prey_payoff}), we can see that when the agent in slot 4 (any $\pi_1$) obtains a reward $r$ the agent in slot 3 (any $\pi_4$) obtains the same value $r$ as reward (since they are in the same team), and this relation is propagated to expected average rewards as well. Since we use a correlation function between the expected average rewards, and the agents in slots 4 and 3 always obtain the same expected average reward, then the correlation function will always obtain the same value of $1$. So:

\begin{equation*}
SRD_{4,3}(\pi_1,\Pi_o,w_{\dot{L}},\mu) = 1
\end{equation*}

Therefore:

\begin{equation*}
SRD_{4,3}(\Pi_e,w_{\Pi_e},\Pi_o,w_{\dot{L}},\mu) = 1
\end{equation*}

And finally, we weight over the slots:

\begin{equation*}
\begin{aligned}
& SRD(\Pi_e,w_{\Pi_e},\Pi_o,w_{\dot{L}},\mu,w_S) = \eta_{S_1^2} \sum_{i=1}^{N(\mu)} w_S(i,\mu) \times\\
& \times \left(\sum_{j=1}^{i-1} w_S(j,\mu) SRD_{i,j}(\Pi_e,w_{\Pi_e},\Pi_o,w_{\dot{L}},\mu) + \sum_{j=i+1}^{N(\mu)} w_S(j,\mu) SRD_{i,j}(\Pi_e,w_{\Pi_e},\Pi_o,w_{\dot{L}},\mu)\right) =\\
& \ \ \ \ \ \ \ \ \ \ \ \ \ \ \ \ \ \ \ \ \ \ \ \ \ \ \ \ \ \ \ \ \ \ \ \ \ \ = \frac{8}{3} \frac{1}{4} \frac{1}{4} \{SRD_{1,2}(\Pi_e,w_{\Pi_e},\Pi_o,w_{\dot{L}},\mu) + SRD_{1,3}(\Pi_e,w_{\Pi_e},\Pi_o,w_{\dot{L}},\mu) +\\
& \ \ \ \ \ \ \ \ \ \ \ \ \ \ \ \ \ \ \ \ \ \ \ \ \ \ \ \ \ \ \ \ \ \ \ \ \ \ + SRD_{1,4}(\Pi_e,w_{\Pi_e},\Pi_o,w_{\dot{L}},\mu) + SRD_{2,1}(\Pi_e,w_{\Pi_e},\Pi_o,w_{\dot{L}},\mu) +\\
& \ \ \ \ \ \ \ \ \ \ \ \ \ \ \ \ \ \ \ \ \ \ \ \ \ \ \ \ \ \ \ \ \ \ \ \ \ \ + SRD_{2,3}(\Pi_e,w_{\Pi_e},\Pi_o,w_{\dot{L}},\mu) + SRD_{2,4}(\Pi_e,w_{\Pi_e},\Pi_o,w_{\dot{L}},\mu) +\\
& \ \ \ \ \ \ \ \ \ \ \ \ \ \ \ \ \ \ \ \ \ \ \ \ \ \ \ \ \ \ \ \ \ \ \ \ \ \ + SRD_{3,1}(\Pi_e,w_{\Pi_e},\Pi_o,w_{\dot{L}},\mu) + SRD_{3,2}(\Pi_e,w_{\Pi_e},\Pi_o,w_{\dot{L}},\mu) +\\
& \ \ \ \ \ \ \ \ \ \ \ \ \ \ \ \ \ \ \ \ \ \ \ \ \ \ \ \ \ \ \ \ \ \ \ \ \ \ + SRD_{3,4}(\Pi_e,w_{\Pi_e},\Pi_o,w_{\dot{L}},\mu) + SRD_{4,1}(\Pi_e,w_{\Pi_e},\Pi_o,w_{\dot{L}},\mu) +\\
& \ \ \ \ \ \ \ \ \ \ \ \ \ \ \ \ \ \ \ \ \ \ \ \ \ \ \ \ \ \ \ \ \ \ \ \ \ \ + SRD_{4,2}(\Pi_e,w_{\Pi_e},\Pi_o,w_{\dot{L}},\mu) + SRD_{4,3}(\Pi_e,w_{\Pi_e},\Pi_o,w_{\dot{L}},\mu)\} =\\
& \ \ \ \ \ \ \ \ \ \ \ \ \ \ \ \ \ \ \ \ \ \ \ \ \ \ \ \ \ \ \ \ \ \ \ \ \ \ = \frac{4}{3} \frac{1}{4} \frac{1}{4} \left\{6 \times (-1) + 6 \times 1\right\} = 0
\end{aligned}
\end{equation*}

So, for every trio $\left\langle\Pi_e,w_{\Pi_e},\Pi_o\right\rangle$ we obtain the same result:

\begin{equation*}
\forall \Pi_e,w_{\Pi_e},\Pi_o : SRD(\Pi_e,w_{\Pi_e},\Pi_o,w_{\dot{L}},\mu,w_S) = 0
\end{equation*}

Therefore, predator-prey has $General = [0,0]$ for this property.
\end{proof}
\end{proposition}

\subsection{Competitive Anticipation}
Then, we follow with the competitive anticipation (AComp) property. As given in section \ref{sec:AComp}, we want to know how much benefit the evaluated agents obtain when they anticipate competing agents.

\begin{proposition}
\label{prop:predator-prey_AComp_general_min}
$General_{min}$ for the competitive anticipation (AComp) property is equal to $-1$ for the predator-prey environment.

\begin{proof}
To find $General_{min}$ (equation \ref{eq:general_min}), we need to find a trio $\left\langle\Pi_e,w_{\Pi_e},\Pi_o\right\rangle$ which minimises the property as much as possible. We can have this situation by selecting $\Pi_e = \{\pi_{win/win'}\}$ with $w_{\Pi_e}(\pi_{win/win'}) = 1$ and $\Pi_o = \{\pi_{win}\}$ (a $\pi_{win}$ agent always tries to not be chased when playing as the prey and tries to chase when playing as the predator, and a $\pi_{win/win'}$ agent always tries to not be chased when playing as the prey and tries to chase when playing as the predator but from the fifth iteration stops chasing the prey).

Following definition \ref{def:AComp}, we obtain the AComp value for this $\left\langle\Pi_e,w_{\Pi_e},\Pi_o\right\rangle$. Since the environment is not symmetric, we need to calculate this property for every pair of slots in different teams. Following definition \ref{def:AComp_set}, we could calculate its AComp value for each pair of slots but, since $\Pi_e$ has only one agent, its weight is equal to $1$ and $\Pi_o$ also has only one agent, it is equivalent to use directly definition \ref{def:AComp_agents}. We start with slots 1 and 2:

\begin{equation*}
\begin{aligned}
AComp_{1,2}(\pi_{win/win'},\pi_{win},\Pi_o,w_{\dot{L}},\mu)	& = \sum_{\dot{l} \in \dot{L}^{N(\mu)}_{-1,2}(\Pi_o)} w_{\dot{L}}(\dot{l}) \frac{1}{2} \left(R_1(\mu[\instantiation{l}{1,2}{\pi_{win/win'},\pi_{win}}]) - R_1(\mu[\instantiation{l}{1,2}{\pi_{win/win'},\pi_r}])\right) =\\
															& = \frac{1}{2} \left(R_1(\mu[\pi_{win/win'},\pi_{win},\pi_{win},\pi_{win}]) - R_1(\mu[\pi_{win/win'},\pi_r,\pi_{win},\pi_{win}])\right)
\end{aligned}
\end{equation*}

We know from lemma \ref{lemma:predator-prey_always_chase} that three predators correctly coordinating will always chase the prey. In line-up $(\pi_{win/win'},\pi_{win},\pi_{win},\pi_{win})$ the agent in slot 1 ($\pi_{win/win'}$) will obtain an expected average reward of $-1$, while in line-up $(\pi_{win/win'},\pi_r,\pi_{win},\pi_{win})$ the agent in slot 1 ($\pi_{win/win'}$), where the prey will almost always be able to avoid the predators, will almost obtain an expected average reward of $1$. So:

\begin{equation*}
AComp_{1,2}(\pi_{win/win'},\pi_{win},\Pi_o,w_{\dot{L}},\mu) = \frac{1}{2} \left((-1) - 1\right) = -1
\end{equation*}

For slots 1 and 3:

\begin{equation*}
\begin{aligned}
AComp_{1,3}(\pi_{win/win'},\pi_{win},\Pi_o,w_{\dot{L}},\mu)	& = \sum_{\dot{l} \in \dot{L}^{N(\mu)}_{-1,3}(\Pi_o)} w_{\dot{L}}(\dot{l}) \frac{1}{2} \left(R_1(\mu[\instantiation{l}{1,3}{\pi_{win/win'},\pi_{win}}]) - R_1(\mu[\instantiation{l}{1,3}{\pi_{win/win'},\pi_r}])\right) =\\
															& = \frac{1}{2} \left(R_1(\mu[\pi_{win/win'},\pi_{win},\pi_{win},\pi_{win}]) - R_1(\mu[\pi_{win/win'},\pi_{win},\pi_r,\pi_{win}])\right)
\end{aligned}
\end{equation*}

We know from lemma \ref{lemma:predator-prey_always_chase} that three predators correctly coordinating will always chase the prey. In line-up $(\pi_{win/win'},\pi_{win},\pi_{win},\pi_{win})$ the agent in slot 1 ($\pi_{win/win'}$) will obtain an expected average reward of $-1$, while in line-up $(\pi_{win/win'},\pi_{win},\pi_r,\pi_{win})$ the agent in slot 1 ($\pi_{win/win'}$), where the prey will almost always be able to avoid the predators, will almost obtain an expected average reward of $1$. So:

\begin{equation*}
AComp_{1,3}(\pi_{win/win'},\pi_{win},\Pi_o,w_{\dot{L}},\mu) = \frac{1}{2} \left((-1) - 1\right) = -1
\end{equation*}

For slots 1 and 4:

\begin{equation*}
\begin{aligned}
AComp_{1,4}(\pi_{win/win'},\pi_{win},\Pi_o,w_{\dot{L}},\mu)	& = \sum_{\dot{l} \in \dot{L}^{N(\mu)}_{-1,4}(\Pi_o)} w_{\dot{L}}(\dot{l}) \frac{1}{2} \left(R_1(\mu[\instantiation{l}{1,4}{\pi_{win/win'},\pi_{win}}]) - R_1(\mu[\instantiation{l}{1,4}{\pi_{win/win'},\pi_r}])\right) =\\
															& = \frac{1}{2} \left(R_1(\mu[\pi_{win/win'},\pi_{win},\pi_{win},\pi_{win}]) - R_1(\mu[\pi_{win/win'},\pi_{win},\pi_{win},\pi_r])\right)
\end{aligned}
\end{equation*}

We know from lemma \ref{lemma:predator-prey_always_chase} that three predators correctly coordinating will always chase the prey. In line-up $(\pi_{win/win'},\pi_{win},\pi_{win},\pi_{win})$ the agent in slot 1 ($\pi_{win/win'}$) will obtain an expected average reward of $-1$, while in line-up $(\pi_{win/win'},\pi_{win},\pi_{win},\pi_r)$ the agent in slot 1 ($\pi_{win/win'}$), where the prey will almost always be able to avoid the predators, will almost obtain an expected average reward of $1$. So:

\begin{equation*}
AComp_{1,4}(\pi_{win/win'},\pi_{win},\Pi_o,w_{\dot{L}},\mu) = \frac{1}{2} \left((-1) - 1\right) = -1
\end{equation*}

For slots 2 and 1:

\begin{equation*}
\begin{aligned}
AComp_{2,1}(\pi_{win/win'},\pi_{win},\Pi_o,w_{\dot{L}},\mu)	& = \sum_{\dot{l} \in \dot{L}^{N(\mu)}_{-2,1}(\Pi_o)} w_{\dot{L}}(\dot{l}) \frac{1}{2} \left(R_2(\mu[\instantiation{l}{2,1}{\pi_{win/win'},\pi_{win}}]) - R_2(\mu[\instantiation{l}{2,1}{\pi_{win/win'},\pi_r}])\right) =\\
															& = \frac{1}{2} \left(R_2(\mu[\pi_{win},\pi_{win/win'},\pi_{win},\pi_{win}]) - R_2(\mu[\pi_r,\pi_{win/win'},\pi_{win},\pi_{win}])\right)
\end{aligned}
\end{equation*}

We know from lemma \ref{lemma:predator-prey_always_chase} that three predators correctly coordinating will always chase the prey. In line-up $(\pi_{win},\pi_{win/win'},\pi_{win},\pi_{win})$ the agent in slot 2 ($\pi_{win/win'}$), where they prey will not be chased due to the miss-coordination of $\pi_{win/win'}$ in the last iterations, will obtain an expected average reward of $-1$, while in line-up $(\pi_r,\pi_{win/win'},\pi_{win},\pi_{win})$ the agent in slot 2 ($\pi_{win/win'}$), where the prey will almost always be chased by the predators, will almost obtain an expected average reward of $1$. So:

\begin{equation*}
AComp_{2,1}(\pi_{win/win'},\pi_{win},\Pi_o,w_{\dot{L}},\mu) = \frac{1}{2} \left((-1) - 1\right) = -1
\end{equation*}

For slots 3 and 1:

\begin{equation*}
\begin{aligned}
AComp_{3,1}(\pi_{win/win'},\pi_{win},\Pi_o,w_{\dot{L}},\mu)	& = \sum_{\dot{l} \in \dot{L}^{N(\mu)}_{-3,1}(\Pi_o)} w_{\dot{L}}(\dot{l}) \frac{1}{2} \left(R_3(\mu[\instantiation{l}{3,1}{\pi_{win/win'},\pi_{win}}]) - R_3(\mu[\instantiation{l}{3,1}{\pi_{win/win'},\pi_r}])\right) =\\
															& = \frac{1}{2} \left(R_3(\mu[\pi_{win},\pi_{win},\pi_{win/win'},\pi_{win}]) - R_3(\mu[\pi_r,\pi_{win},\pi_{win/win'},\pi_{win}])\right)
\end{aligned}
\end{equation*}

We know from lemma \ref{lemma:predator-prey_always_chase} that three predators correctly coordinating will always chase the prey. In line-up $(\pi_{win},\pi_{win},\pi_{win/win'},\pi_{win})$ the agent in slot 3 ($\pi_{win/win'}$), where they prey will not be chased due to the miss-coordination of $\pi_{win/win'}$ in the last iterations, will obtain an expected average reward of $-1$, while in line-up $(\pi_r,\pi_{win},\pi_{win/win'},\pi_{win})$ the agent in slot 3 ($\pi_{win/win'}$), where the prey will almost always be chased by the predators, will almost obtain an expected average reward of $1$. So:

\begin{equation*}
AComp_{3,1}(\pi_{win/win'},\pi_{win},\Pi_o,w_{\dot{L}},\mu) = \frac{1}{2} \left((-1) - 1\right) = -1
\end{equation*}

And for slots 4 and 1:

\begin{equation*}
\begin{aligned}
AComp_{4,1}(\pi_{win/win'},\pi_{win},\Pi_o,w_{\dot{L}},\mu)	& = \sum_{\dot{l} \in \dot{L}^{N(\mu)}_{-4,1}(\Pi_o)} w_{\dot{L}}(\dot{l}) \frac{1}{2} \left(R_4(\mu[\instantiation{l}{4,1}{\pi_{win/win'},\pi_{win}}]) - R_4(\mu[\instantiation{l}{4,1}{\pi_{win/win'},\pi_r}])\right) =\\
															& = \frac{1}{2} \left(R_4(\mu[\pi_{win},\pi_{win},\pi_{win},\pi_{win/win'}]) - R_4(\mu[\pi_r,\pi_{win},\pi_{win},\pi_{win/win'}])\right)
\end{aligned}
\end{equation*}

We know from lemma \ref{lemma:predator-prey_always_chase} that three predators correctly coordinating will always chase the prey. In line-up $(\pi_{win},\pi_{win},\pi_{win},\pi_{win/win'})$ the agent in slot 4 ($\pi_{win/win'}$), where they prey will not be chased due to the miss-coordination of $\pi_{win/win'}$ in the last iterations, will obtain an expected average reward of $-1$, while in line-up $(\pi_r,\pi_{win},\pi_{win},\pi_{win/win'})$ the agent in slot 4 ($\pi_{win/win'}$), where the prey will almost always be chased by the predators, will almost obtain an expected average reward of $1$. So:

\begin{equation*}
AComp_{4,1}(\pi_{win/win'},\pi_{win},\Pi_o,w_{\dot{L}},\mu) = \frac{1}{2} \left((-1) - 1\right) = -1
\end{equation*}

And finally, we weight over the slots:

\begin{equation*}
\begin{aligned}
AComp(\Pi_e,w_{\Pi_e},\Pi_o,w_{\dot{L}},\mu,w_S)	& = \eta_{S_2^2} \sum_{t_1,t_2 \in \tau | t_1 \neq t_2} \sum_{i \in t_1} w_S(i,\mu) \sum_{j \in t_2} w_S(j,\mu) AComp_{i,j}(\Pi_e,w_{\Pi_e},\Pi_o,w_{\dot{L}},\mu) =\\
													& = \frac{8}{3} \frac{1}{4} \frac{1}{4} \{AComp_{1,2}(\Pi_e,w_{\Pi_e},\Pi_o,w_{\dot{L}},\mu) + AComp_{1,3}(\Pi_e,w_{\Pi_e},\Pi_o,w_{\dot{L}},\mu) +\\
													& + AComp_{1,4}(\Pi_e,w_{\Pi_e},\Pi_o,w_{\dot{L}},\mu) + AComp_{2,1}(\Pi_e,w_{\Pi_e},\Pi_o,w_{\dot{L}},\mu) +\\
													& + AComp_{3,1}(\Pi_e,w_{\Pi_e},\Pi_o,w_{\dot{L}},\mu) + AComp_{4,1}(\Pi_e,w_{\Pi_e},\Pi_o,w_{\dot{L}},\mu)\} =\\
													& = \frac{8}{3} \frac{1}{4} \frac{1}{4} \{AComp_{1,2}(\pi_{win/win'},\pi_{win},\Pi_o,w_{\dot{L}},\mu) + AComp_{1,3}(\pi_{win/win'},\pi_{win},\Pi_o,w_{\dot{L}},\mu) +\\
													& + AComp_{1,4}(\pi_{win/win'},\pi_{win},\Pi_o,w_{\dot{L}},\mu) + AComp_{2,1}(\pi_{win/win'},\pi_{win},\Pi_o,w_{\dot{L}},\mu) +\\
													& + AComp_{3,1}(\pi_{win/win'},\pi_{win},\Pi_o,w_{\dot{L}},\mu) + AComp_{4,1}(\pi_{win/win'},\pi_{win},\Pi_o,w_{\dot{L}},\mu)\} =\\
													& = \frac{8}{3} \frac{1}{4} \frac{1}{4} \left\{6 \times (-1)\right\} = -1
\end{aligned}
\end{equation*}

Since $-1$ is the lowest possible value for the competitive anticipation property, therefore predator-prey has $General_{min} = -1$ for this property.
\end{proof}
\end{proposition}

\begin{proposition}
\label{prop:predator-prey_AComp_general_max}
$General_{max}$ for the competitive anticipation (AComp) property is equal to $\frac{1}{2}$ for the predator-prey environment.

\begin{proof}
To find $General_{max}$ (equation \ref{eq:general_max}), we need to find a trio $\left\langle\Pi_e,w_{\Pi_e},\Pi_o\right\rangle$ which maximises the property as much as possible. We can have this situation by selecting $\Pi_e = \{\pi_{tl/br}\}$ with $w_{\Pi_e}(\pi_{tl/br}) = 1$ and $\Pi_o = \{\pi_{br}\}$ (a $\pi_{tl/br}$ agent always stays in the top left corner when playing as the prey and always goes to the bottom right corner when playing as a predator, and a $\pi_{br}$ agent always goes to the bottom right corner).

Following definition \ref{def:AComp}, we obtain the AComp value for this $\left\langle\Pi_e,w_{\Pi_e},\Pi_o\right\rangle$. Since the environment is not symmetric, we need to calculate this property for every pair of slots in different teams. Following definition \ref{def:AComp_set}, we could calculate its AComp value for each pair of slots but, since $\Pi_e$ has only one agent, its weight is equal to $1$ and $\Pi_o$ also has only one agent, it is equivalent to use directly definition \ref{def:AComp_agents}. We start with slots 1 and 2:

\begin{equation*}
\begin{aligned}
AComp_{1,2}(\pi_{tl/br},\pi_{br},\Pi_o,w_{\dot{L}},\mu)	& = \sum_{\dot{l} \in \dot{L}^{N(\mu)}_{-1,2}(\Pi_o)} w_{\dot{L}}(\dot{l}) \frac{1}{2} \left(R_1(\mu[\instantiation{l}{1,2}{\pi_{tl/br},\pi_{br}}]) - R_1(\mu[\instantiation{l}{1,2}{\pi_{tl/br},\pi_r}])\right) =\\
														& = \frac{1}{2} \left(R_1(\mu[\pi_{tl/br},\pi_{br},\pi_{br},\pi_{br}]) - R_1(\mu[\pi_{tl/br},\pi_r,\pi_{br},\pi_{br}])\right)
\end{aligned}
\end{equation*}

In line-up $(\pi_{tl/br},\pi_{br},\pi_{br},\pi_{br})$ the agent in slot 1 ($\pi_{tl/br}$) will obtain an expected average reward of $1$, while in line-up $(\pi_{tl/br},\pi_r,\pi_{br},\pi_{br})$ the agent in slot 1 ($\pi_{tl/br}$), where the prey will almost always be able to avoid the predators, will almost obtain an expected average reward of $1$\footnote{It is arguably that some behaviours for the prey could try to be chased by the random agent, providing a greater value for this property, but this probability of been chased will still remain too low, which would not increase too much the value.}. So:

\begin{equation*}
AComp_{1,2}(\pi_{tl/br},\pi_{br},\Pi_o,w_{\dot{L}},\mu) = \frac{1}{2} \left(1 - 1\right) = 0
\end{equation*}

For slots 1 and 3:

\begin{equation*}
\begin{aligned}
AComp_{1,3}(\pi_{tl/br},\pi_{br},\Pi_o,w_{\dot{L}},\mu)	& = \sum_{\dot{l} \in \dot{L}^{N(\mu)}_{-1,3}(\Pi_o)} w_{\dot{L}}(\dot{l}) \frac{1}{2} \left(R_1(\mu[\instantiation{l}{1,3}{\pi_{tl/br},\pi_{br}}]) - R_1(\mu[\instantiation{l}{1,3}{\pi_{tl/br},\pi_r}])\right) =\\
														& = \frac{1}{2} \left(R_1(\mu[\pi_{tl/br},\pi_{br},\pi_{br},\pi_{br}]) - R_1(\mu[\pi_{tl/br},\pi_{br},\pi_r,\pi_{br}])\right)
\end{aligned}
\end{equation*}

In line-up $(\pi_{tl/br},\pi_{br},\pi_{br},\pi_{br})$ the agent in slot 1 ($\pi_{tl/br}$) will obtain an expected average reward of $1$, while in line-up $(\pi_{tl/br},\pi_{br},\pi_r,\pi_{br})$ the agent in slot 1 ($\pi_{tl/br}$), where the prey will almost always be able to avoid the predators, will almost obtain an expected average reward of $1$. So:

\begin{equation*}
AComp_{1,3}(\pi_{tl/br},\pi_{br},\Pi_o,w_{\dot{L}},\mu) = \frac{1}{2} (1 - 1) = 0
\end{equation*}

For slots 1 and 4:

\begin{equation*}
\begin{aligned}
AComp_{1,4}(\pi_{tl/br},\pi_{br},\Pi_o,w_{\dot{L}},\mu)	& = \sum_{\dot{l} \in \dot{L}^{N(\mu)}_{-1,4}(\Pi_o)} w_{\dot{L}}(\dot{l}) \frac{1}{2} \left(R_1(\mu[\instantiation{l}{1,4}{\pi_{tl/br},\pi_{br}}]) - R_1(\mu[\instantiation{l}{1,4}{\pi_{tl/br},\pi_r}])\right) =\\
														& = \frac{1}{2} \left(R_1(\mu[\pi_{tl/br},\pi_{br},\pi_{br},\pi_{br}]) - R_1(\mu[\pi_{tl/br},\pi_{br},\pi_{br},\pi_r])\right)
\end{aligned}
\end{equation*}

In line-up $(\pi_{tl/br},\pi_{br},\pi_{br},\pi_{br})$ the agent in slot 1 ($\pi_{tl/br}$) will obtain an expected average reward of $1$, while in line-up $(\pi_{tl/br},\pi_{br},\pi_{br},\pi_r)$ the agent in slot 1 ($\pi_{tl/br}$), where the prey will almost always be able to avoid the predators, will almost obtain an expected average reward of $1$. So:

\begin{equation*}
AComp_{1,4}(\pi_{tl/br},\pi_{br},\Pi_o,w_{\dot{L}},\mu) = \frac{1}{2} \left(1 - 1\right) = 0
\end{equation*}

For slots 2 and 1:

\begin{equation*}
\begin{aligned}
AComp_{2,1}(\pi_{tl/br},\pi_{br},\Pi_o,w_{\dot{L}},\mu)	& = \sum_{\dot{l} \in \dot{L}^{N(\mu)}_{-2,1}(\Pi_o)} w_{\dot{L}}(\dot{l}) \frac{1}{2} \left(R_2(\mu[\instantiation{l}{2,1}{\pi_{tl/br},\pi_{br}}]) - R_2(\mu[\instantiation{l}{2,1}{\pi_{tl/br},\pi_r}])\right) =\\
														& = \frac{1}{2} \left(R_2(\mu[\pi_{br},\pi_{tl/br},\pi_{br},\pi_{br}]) - R_2(\mu[\pi_r,\pi_{tl/br},\pi_{br},\pi_{br}])\right)
\end{aligned}
\end{equation*}

In line-up $(\pi_{br},\pi_{tl/br},\pi_{br},\pi_{br})$, where $\pi_{tl/br}$ and $\pi_{br}$ will always go to the bottom right corner, the agent in slot 2 ($\pi_{tl/br}$) will obtain an expected average reward of $1$, since the prey will always be chased. In line-up $(\pi_r,\pi_{tl/br},\pi_{br},\pi_{br})$, where $\pi_{tl/br}$ and $\pi_{br}$ will always go to the bottom right corner and $\pi_r$ will act randomly, the agent in slot 2 ($\pi_{tl/br}$) will obtain an expected average reward of $-1$, since the prey will rarely be chased. So:

\begin{equation*}
AComp_{2,1}(\pi_{tl/br},\pi_{br},\Pi_o,w_{\dot{L}},\mu) = \frac{1}{2} \left(1 - (-1)\right) = 1
\end{equation*}

For slots 3 and 1:

\begin{equation*}
\begin{aligned}
AComp_{3,1}(\pi_{tl/br},\pi_{br},\Pi_o,w_{\dot{L}},\mu)	& = \sum_{\dot{l} \in \dot{L}^{N(\mu)}_{-3,1}(\Pi_o)} w_{\dot{L}}(\dot{l}) \frac{1}{2} \left(R_3(\mu[\instantiation{l}{3,1}{\pi_{tl/br},\pi_{br}}]) - R_3(\mu[\instantiation{l}{3,1}{\pi_{tl/br},\pi_r}])\right) =\\
														& = \frac{1}{2} \left(R_3(\mu[\pi_{br},\pi_{br},\pi_{tl/br},\pi_{br}]) - R_3(\mu[\pi_r,\pi_{br},\pi_{tl/br},\pi_{br}])\right)
\end{aligned}
\end{equation*}

In line-up $(\pi_{br},\pi_{br},\pi_{tl/br},\pi_{br})$, where $\pi_{tl/br}$ and $\pi_{br}$ will always go to the bottom right corner, the agent in slot 3 ($\pi_{tl/br}$) will obtain an expected average reward of $1$, since the prey will always be chased. In line-up $(\pi_r,\pi_{br},\pi_{tl/br},\pi_{br})$, where $\pi_{tl/br}$ and $\pi_{br}$ will always go to the bottom right corner and $\pi_r$ will act randomly, the agent in slot 3 ($\pi_{tl/br}$) will obtain an expected average reward of $-1$, since the prey will rarely be chased. So:

\begin{equation*}
AComp_{3,1}(\pi_{tl/br},\pi_{br},\Pi_o,w_{\dot{L}},\mu) = \frac{1}{2} \left(1 - (-1)\right) = 1
\end{equation*}

And for slots 4 and 1:

\begin{equation*}
\begin{aligned}
AComp_{4,1}(\pi_{tl/br},\pi_{br},\Pi_o,w_{\dot{L}},\mu)	& = \sum_{\dot{l} \in \dot{L}^{N(\mu)}_{-4,1}(\Pi_o)} w_{\dot{L}}(\dot{l}) \frac{1}{2} \left(R_4(\mu[\instantiation{l}{4,1}{\pi_{tl/br},\pi_{br}}]) - R_4(\mu[\instantiation{l}{4,1}{\pi_{tl/br},\pi_r}])\right) =\\
												& = \frac{1}{2} \left(R_4(\mu[\pi_{br},\pi_{br},\pi_{br},\pi_{tl/br}]) - R_4(\mu[\pi_r,\pi_{br},\pi_{br},\pi_{tl/br}])\right)
\end{aligned}
\end{equation*}

In line-up $(\pi_{br},\pi_{br},\pi_{br},\pi_{tl/br})$, where $\pi_{tl/br}$ and $\pi_{br}$ will always go to the bottom right corner, the agent in slot 4 ($\pi_{tl/br}$) will obtain an expected average reward of $1$, since the prey will always be chased. In line-up $(\pi_r,\pi_{br},\pi_{br},\pi_{tl/br})$, where $\pi_{tl/br}$ and $\pi_{br}$ will always go to the bottom right corner and $\pi_r$ will act randomly, the agent in slot 4 ($\pi_{tl/br}$) will obtain an expected average reward of $-1$, since the prey will rarely be chased. So:

\begin{equation*}
AComp_{4,1}(\pi_{tl/br},\pi_{br},\Pi_o,w_{\dot{L}},\mu) = \frac{1}{2} \left(1 - (-1)\right) = 1
\end{equation*}

And finally, we weight over the slots:

\begin{equation*}
\begin{aligned}
AComp(\Pi_e,w_{\Pi_e},\Pi_o,w_{\dot{L}},\mu,w_S)	& = \eta_{S_2^2} \sum_{t_1,t_2 \in \tau | t_1 \neq t_2} \sum_{i \in t_1} w_S(i,\mu) \sum_{j \in t_2} w_S(j,\mu) AComp_{i,j}(\Pi_e,w_{\Pi_e},\Pi_o,w_{\dot{L}},\mu) =\\
													& = \frac{8}{3} \frac{1}{4} \frac{1}{4} \{AComp_{1,2}(\Pi_e,w_{\Pi_e},\Pi_o,w_{\dot{L}},\mu) + AComp_{1,3}(\Pi_e,w_{\Pi_e},\Pi_o,w_{\dot{L}},\mu) +\\
													& + AComp_{1,4}(\Pi_e,w_{\Pi_e},\Pi_o,w_{\dot{L}},\mu) + AComp_{2,1}(\Pi_e,w_{\Pi_e},\Pi_o,w_{\dot{L}},\mu) +\\
													& + AComp_{3,1}(\Pi_e,w_{\Pi_e},\Pi_o,w_{\dot{L}},\mu) + AComp_{4,1}(\Pi_e,w_{\Pi_e},\Pi_o,w_{\dot{L}},\mu)\} =\\
													& = \frac{8}{3} \frac{1}{4} \frac{1}{4} \{AComp_{1,2}(\pi_{tl/br},\pi_{br},\Pi_o,w_{\dot{L}},\mu) + AComp_{1,3}(\pi_{tl/br},\pi_{br},\Pi_o,w_{\dot{L}},\mu) +\\
													& + AComp_{1,4}(\pi_{tl/br},\pi_{br},\Pi_o,w_{\dot{L}},\mu) + AComp_{2,1}(\pi_{tl/br},\pi_{br},\Pi_o,w_{\dot{L}},\mu) +\\
													& + AComp_{3,1}(\pi_{tl/br},\pi_{br},\Pi_o,w_{\dot{L}},\mu) + AComp_{4,1}(\pi_{tl/br},\pi_{br},\Pi_o,w_{\dot{L}},\mu)\} =\\
													& = \frac{8}{3} \frac{1}{4} \frac{1}{4} \left\{3 \times 0 + 3 \times 1\right\} = \frac{1}{2}
\end{aligned}
\end{equation*}

Since $\frac{1}{2}$ is the highest possible value that we can obtain for the competitive anticipation property, therefore predator-prey has $General_{max} = \frac{1}{2}$ for this property.
\end{proof}
\end{proposition}

\subsection{Cooperative Anticipation}
Finally, we follow with the cooperative anticipation (ACoop) property. As given in section \ref{sec:ACoop}, we want to know how much benefit the evaluated agents obtain when they anticipate cooperating agents.

\begin{proposition}
\label{prop:predator-prey_ACoop_general_min}
$General_{min}$ for the cooperative anticipation (ACoop) property is equal to $-1$ for the predator-prey environment.

\begin{proof}
To find $General_{min}$ (equation \ref{eq:general_min}), we need to find a trio $\left\langle\Pi_e,w_{\Pi_e},\Pi_o\right\rangle$ which minimises the property as much as possible. We can have this situation by selecting $\Pi_e = \{\pi_{br'}\}$ with $w_{\Pi_e}(\pi_{br'}) = 1$ and $\Pi_o = \{\pi_{33/br'}\}$ (a $\pi_{br'}$ agent always goes to the bottom right corner, but if it notices that not all the predators are going directly to this corner, then it will go to the cell in the 3rd row and 3rd column, and a $\pi_{33/br'}$ agent always goes to the cell in the 3rd row and 3rd column when playing as the prey and will go directly to the bottom right corner when playing as a predator, but if it notices that not all the predators are going directly to this corner, then it will go to the cell in the 3rd row and 3rd column).

Following definition \ref{def:ACoop}, we obtain the ACoop value for this $\left\langle\Pi_e,w_{\Pi_e},\Pi_o\right\rangle$. Since the environment is not symmetric, we need to calculate this property for every pair of different slots in the same team. Following definition \ref{def:ACoop_set}, we could calculate its ACoop value for each pair of slots but, since $\Pi_e$ has only one agent, its weight is equal to $1$ and $\Pi_o$ also has only one agent, it is equivalent to use directly definition \ref{def:ACoop_agents}. We start with slots 2 and 3:

\begin{equation*}
\begin{aligned}
& ACoop_{2,3}(\pi_{br'},\pi_{33/br'},\Pi_o,w_{\dot{L}},\mu) = \sum_{\dot{l} \in \dot{L}^{N(\mu)}_{-2,3}(\Pi_o)} w_{\dot{L}}(\dot{l}) \frac{1}{4} \times\\
& \times \left(R_2(\mu[\instantiation{l}{2,3}{\pi_{br'},\pi_{33/br'}}]) + R_3(\mu[\instantiation{l}{2,3}{\pi_{br'},\pi_{33/br'}}]) - R_2(\mu[\instantiation{l}{2,3}{\pi_{br'},\pi_r}]) - R_3(\mu[\instantiation{l}{2,3}{\pi_r,\pi_{33/br'}}])\right) =\\
& \ \ \ \ \ \ \ \ \ \ \ \ \ \ \ \ \ \ \ \ \ \ \ \ \ \ \ \ \ \ \ \ \ \ \ \ \ \ \ \ \ \ \ = \frac{1}{4} \Big(R_2(\mu[\pi_{33/br'},\pi_{br'},\pi_{33/br'},\pi_{33/br'}]) + R_3(\mu[\pi_{33/br'},\pi_{br'},\pi_{33/br'},\pi_{33/br'}]) -\\
& \ \ \ \ \ \ \ \ \ \ \ \ \ \ \ \ \ \ \ \ \ \ \ \ \ \ \ \ \ \ \ \ \ \ \ \ \ \ \ \ \ \ \ - R_2(\mu[\pi_{33/br'},\pi_{br'},\pi_r,\pi_{33/br'}]) - R_3(\mu[\pi_{33/br'},\pi_r,\pi_{33/br'},\pi_{33/br'}])\Big)
\end{aligned}
\end{equation*}

In both line-ups $(\pi_{33/br'},\pi_{br'},\pi_{33/br'},\pi_{33/br'})$, where the predators will go directly to the bottom right cell and the prey will go to the cell in the 3rd row and 3rd column, the agents in slot 2 ($\pi_{br'}$) and slot 3 ($\pi_{33/br'}$) will both obtain an expected average reward of $-1$. In line-up $(\pi_{33/br'},\pi_{br'},\pi_r,\pi_{33/br'})$, where $\pi_r$ will act randomly so the predators will almost always notice this random movement and then they will go to the cell in the 3rd row and 3rd column, the agent in slot 2 ($\pi_{br'}$) will almost obtain an expected average reward of $1$. In line-up $(\pi_{33/br'},\pi_r,\pi_{33/br'},\pi_{33/br'})$, where $\pi_r$ will act randomly so the predators will almost always notice this random movement and then they will go to the cell in the 3rd row and 3rd column, the agent in slot 3 ($\pi_{33/br'}$) will almost obtain an expected average reward of $1$. So:

\begin{equation*}
ACoop_{2,3}(\pi_{br'},\pi_{33/br'},\Pi_o,w_{\dot{L}},\mu) = \frac{1}{4} \left((-1) + (-1) - 1 - 1\right) = -1
\end{equation*}

For slots 2 and 4:

\begin{equation*}
\begin{aligned}
& ACoop_{2,4}(\pi_{br'},\pi_{33/br'},\Pi_o,w_{\dot{L}},\mu) = \sum_{\dot{l} \in \dot{L}^{N(\mu)}_{-2,4}(\Pi_o)} w_{\dot{L}}(\dot{l}) \frac{1}{4} \times\\
& \times \left(R_2(\mu[\instantiation{l}{2,4}{\pi_{br'},\pi_{33/br'}}]) + R_4(\mu[\instantiation{l}{2,4}{\pi_{br'},\pi_{33/br'}}]) - R_2(\mu[\instantiation{l}{2,4}{\pi_{br'},\pi_r}]) - R_4(\mu[\instantiation{l}{2,4}{\pi_r,\pi_{33/br'}}])\right) =\\
& \ \ \ \ \ \ \ \ \ \ \ \ \ \ \ \ \ \ \ \ \ \ \ \ \ \ \ \ \ \ \ \ \ \ \ \ \ \ \ \ \ \ \ = \frac{1}{4} \Big(R_2(\mu[\pi_{33/br'},\pi_{br'},\pi_{33/br'},\pi_{33/br'}]) + R_4(\mu[\pi_{33/br'},\pi_{br'},\pi_{33/br'},\pi_{33/br'}]) -\\
& \ \ \ \ \ \ \ \ \ \ \ \ \ \ \ \ \ \ \ \ \ \ \ \ \ \ \ \ \ \ \ \ \ \ \ \ \ \ \ \ \ \ \ - R_2(\mu[\pi_{33/br'},\pi_{br'},\pi_{33/br'},\pi_r]) - R_4(\mu[\pi_{33/br'},\pi_r,\pi_{33/br'},\pi_{33/br'}])\Big)
\end{aligned}
\end{equation*}

In both line-ups $(\pi_{33/br'},\pi_{br'},\pi_{33/br'},\pi_{33/br'})$, where the predators will go directly to the bottom right cell and the prey will go to the cell in the 3rd row and 3rd column, the agents in slot 2 ($\pi_{br'}$) and slot 4 ($\pi_{33/br'}$) will both obtain an expected average reward of $-1$. In line-up $(\pi_{33/br'},\pi_{br'},\pi_{33/br'},\pi_r)$, where $\pi_r$ will act randomly so the predators will almost always notice this random movement and then they will go to the cell in the 3rd row and 3rd column, the agent in slot 2 ($\pi_{br'}$) will almost obtain an expected average reward of $1$. In line-up $(\pi_{33/br'},\pi_r,\pi_{33/br'},\pi_{33/br'})$, where $\pi_r$ will act randomly so the predators will almost always notice this random movement and then they will go to the cell in the 3rd row and 3rd column, the agent in slot 4 ($\pi_{33/br'}$) will almost obtain an expected average reward of $1$. So:

\begin{equation*}
ACoop_{2,4}(\pi_{br'},\pi_{33/br'},\Pi_o,w_{\dot{L}},\mu) = \frac{1}{4} \left((-1) + (-1) - 1 - 1\right) = -1
\end{equation*}

For slots 3 and 2:

\begin{equation*}
\begin{aligned}
& ACoop_{3,2}(\pi_{br'},\pi_{33/br'},\Pi_o,w_{\dot{L}},\mu) = \sum_{\dot{l} \in \dot{L}^{N(\mu)}_{-3,2}(\Pi_o)} w_{\dot{L}}(\dot{l}) \frac{1}{4} \times\\
& \times \left(R_3(\mu[\instantiation{l}{3,2}{\pi_{br'},\pi_{33/br'}}]) + R_2(\mu[\instantiation{l}{3,2}{\pi_{br'},\pi_{33/br'}}]) - R_3(\mu[\instantiation{l}{3,2}{\pi_{br'},\pi_r}]) - R_2(\mu[\instantiation{l}{3,2}{\pi_r,\pi_{33/br'}}])\right) =\\
& \ \ \ \ \ \ \ \ \ \ \ \ \ \ \ \ \ \ \ \ \ \ \ \ \ \ \ \ \ \ \ \ \ \ \ \ \ \ \ \ \ \ \ = \frac{1}{4} \Big(R_3(\mu[\pi_{33/br'},\pi_{33/br'},\pi_{br'},\pi_{33/br'}]) + R_2(\mu[\pi_{33/br'},\pi_{33/br'},\pi_{br'},\pi_{33/br'}]) -\\
& \ \ \ \ \ \ \ \ \ \ \ \ \ \ \ \ \ \ \ \ \ \ \ \ \ \ \ \ \ \ \ \ \ \ \ \ \ \ \ \ \ \ \ - R_3(\mu[\pi_{33/br'},\pi_r,\pi_{br'},\pi_{33/br'}]) - R_2(\mu[\pi_{33/br'},\pi_{33/br'},\pi_r,\pi_{33/br'}])\Big)
\end{aligned}
\end{equation*}

In both line-ups $(\pi_{33/br'},\pi_{33/br'},\pi_{br'},\pi_{33/br'})$, where the predators will go directly to the bottom right cell and the prey will go to the cell in the 3rd row and 3rd column, the agents in slot 3 ($\pi_{br'}$) and slot 2 ($\pi_{33/br'}$) will both obtain an expected average reward of $-1$. In line-up $(\pi_{33/br'},\pi_r,\pi_{br'},\pi_{33/br'})$, where $\pi_r$ will act randomly so the predators will almost always notice this random movement and then they will go to the cell in the 3rd row and 3rd column, the agent in slot 3 ($\pi_{br'}$) will almost obtain an expected average reward of $1$. In line-up $(\pi_{33/br'},\pi_{33/br'},\pi_r,\pi_{33/br'})$, where $\pi_r$ will act randomly so the predators will almost always notice this random movement and then they will go to the cell in the 3rd row and 3rd column, the agent in slot 2 ($\pi_{33/br'}$) will almost obtain an expected average reward of $1$. So:

\begin{equation*}
ACoop_{3,2}(\pi_{br'},\pi_{33/br'},\Pi_o,w_{\dot{L}},\mu) = \frac{1}{4} \left((-1) + (-1) - 1 - 1\right) = -1
\end{equation*}

For slots 3 and 4:

\begin{equation*}
\begin{aligned}
& ACoop_{3,4}(\pi_{br'},\pi_{33/br'},\Pi_o,w_{\dot{L}},\mu) = \sum_{\dot{l} \in \dot{L}^{N(\mu)}_{-3,4}(\Pi_o)} w_{\dot{L}}(\dot{l}) \frac{1}{4} \times\\
& \times \left(R_3(\mu[\instantiation{l}{3,4}{\pi_{br'},\pi_{33/br'}}]) + R_4(\mu[\instantiation{l}{3,4}{\pi_{br'},\pi_{33/br'}}]) - R_3(\mu[\instantiation{l}{3,4}{\pi_{br'},\pi_r}]) - R_4(\mu[\instantiation{l}{3,4}{\pi_r,\pi_{33/br'}}])\right) =\\
& \ \ \ \ \ \ \ \ \ \ \ \ \ \ \ \ \ \ \ \ \ \ \ \ \ \ \ \ \ \ \ \ \ \ \ \ \ \ \ \ \ \ \ = \frac{1}{4} \Big(R_3(\mu[\pi_{33/br'},\pi_{33/br'},\pi_{br'},\pi_{33/br'}]) + R_4(\mu[\pi_{33/br'},\pi_{33/br'},\pi_{br'},\pi_{33/br'}]) -\\
& \ \ \ \ \ \ \ \ \ \ \ \ \ \ \ \ \ \ \ \ \ \ \ \ \ \ \ \ \ \ \ \ \ \ \ \ \ \ \ \ \ \ \ - R_3(\mu[\pi_{33/br'},\pi_{33/br'},\pi_{br'},\pi_r]) - R_4(\mu[\pi_{33/br'},\pi_{33/br'},\pi_r,\pi_{33/br'}])\Big)
\end{aligned}
\end{equation*}

In both line-ups $(\pi_{33/br'},\pi_{33/br'},\pi_{br'},\pi_{33/br'})$, where the predators will go directly to the bottom right cell and the prey will go to the cell in the 3rd row and 3rd column, the agents in slot 3 ($\pi_{br'}$) and slot 4 ($\pi_{33/br'}$) will both obtain an expected average reward of $-1$. In line-up $(\pi_{33/br'},\pi_{33/br'},\pi_{br'},\pi_r)$, where $\pi_r$ will act randomly so the predators will almost always notice this random movement and then they will go to the cell in the 3rd row and 3rd column, the agent in slot 3 ($\pi_{br'}$) will almost obtain an expected average reward of $1$. In line-up $(\pi_{33/br'},\pi_{33/br'},\pi_r,\pi_{33/br'})$, where $\pi_r$ will act randomly so the predators will almost always notice this random movement and then they will go to the cell in the 3rd row and 3rd column, the agent in slot 4 ($\pi_{33/br'}$) will almost obtain an expected average reward of $1$. So:

\begin{equation*}
ACoop_{3,4}(\pi_{br'},\pi_{33/br'},\Pi_o,w_{\dot{L}},\mu) = \frac{1}{4} \left((-1) + (-1) - 1 - 1\right) = -1
\end{equation*}

For slots 4 and 2:

\begin{equation*}
\begin{aligned}
& ACoop_{4,2}(\pi_{br'},\pi_{33/br'},\Pi_o,w_{\dot{L}},\mu) = \sum_{\dot{l} \in \dot{L}^{N(\mu)}_{-4,2}(\Pi_o)} w_{\dot{L}}(\dot{l}) \frac{1}{4} \times\\
& \times \left(R_4(\mu[\instantiation{l}{4,2}{\pi_{br'},\pi_{33/br'}}]) + R_2(\mu[\instantiation{l}{4,2}{\pi_{br'},\pi_{33/br'}}]) - R_4(\mu[\instantiation{l}{4,2}{\pi_{br'},\pi_r}]) - R_2(\mu[\instantiation{l}{4,2}{\pi_r,\pi_{33/br'}}])\right) =\\
& \ \ \ \ \ \ \ \ \ \ \ \ \ \ \ \ \ \ \ \ \ \ \ \ \ \ \ \ \ \ \ \ \ \ \ \ \ \ \ \ \ \ \ = \frac{1}{4} \Big(R_4(\mu[\pi_{33/br'},\pi_{33/br'},\pi_{33/br'},\pi_{br'}]) + R_2(\mu[\pi_{33/br'},\pi_{33/br'},\pi_{33/br'},\pi_{br'}]) -\\
& \ \ \ \ \ \ \ \ \ \ \ \ \ \ \ \ \ \ \ \ \ \ \ \ \ \ \ \ \ \ \ \ \ \ \ \ \ \ \ \ \ \ \ - R_4(\mu[\pi_{33/br'},\pi_r,\pi_{33/br'},\pi_{br'}]) - R_2(\mu[\pi_{33/br'},\pi_{33/br'},\pi_{33/br'},\pi_r])\Big)
\end{aligned}
\end{equation*}

In both line-ups $(\pi_{33/br'},\pi_{33/br'},\pi_{33/br'},\pi_{br'})$, where the predators will go directly to the bottom right cell and the prey will go to the cell in the 3rd row and 3rd column, the agents in slot 4 ($\pi_{br'}$) and slot 2 ($\pi_{33/br'}$) will both obtain an expected average reward of $-1$. In line-up $(\pi_{33/br'},\pi_r,\pi_{33/br'},\pi_{br'})$, where $\pi_r$ will act randomly so the predators will almost always notice this random movement and then they will go to the cell in the 3rd row and 3rd column, the agent in slot 4 ($\pi_{br'}$) will almost obtain an expected average reward of $1$. In line-up $(\pi_{33/br'},\pi_{33/br'},\pi_{33/br'},\pi_r)$, where $\pi_r$ will act randomly so the predators will almost always notice this random movement and then they will go to the cell in the 3rd row and 3rd column, the agent in slot 2 ($\pi_{33/br'}$) will almost obtain an expected average reward of $1$. So:

\begin{equation*}
ACoop_{4,2}(\pi_{br'},\pi_{33/br'},\Pi_o,w_{\dot{L}},\mu) = \frac{1}{4} \left((-1) + (-1) - 1 - 1\right) = -1
\end{equation*}

And for slots 4 and 3:

\begin{equation*}
\begin{aligned}
& ACoop_{4,3}(\pi_{br'},\pi_{33/br'},\Pi_o,w_{\dot{L}},\mu) = \sum_{\dot{l} \in \dot{L}^{N(\mu)}_{-4,3}(\Pi_o)} w_{\dot{L}}(\dot{l}) \frac{1}{4} \times\\
& \times \left(R_4(\mu[\instantiation{l}{4,3}{\pi_{br'},\pi_{33/br'}}]) + R_3(\mu[\instantiation{l}{4,3}{\pi_{br'},\pi_{33/br'}}]) - R_4(\mu[\instantiation{l}{4,3}{\pi_{br'},\pi_r}]) - R_3(\mu[\instantiation{l}{4,3}{\pi_r,\pi_{33/br'}}])\right) =\\
& \ \ \ \ \ \ \ \ \ \ \ \ \ \ \ \ \ \ \ \ \ \ \ \ \ \ \ \ \ \ \ \ \ \ \ \ \ \ \ \ \ \ \ = \frac{1}{4} \Big(R_4(\mu[\pi_{33/br'},\pi_{33/br'},\pi_{33/br'},\pi_{br'}]) + R_3(\mu[\pi_{33/br'},\pi_{33/br'},\pi_{33/br'},\pi_{br'}]) -\\
& \ \ \ \ \ \ \ \ \ \ \ \ \ \ \ \ \ \ \ \ \ \ \ \ \ \ \ \ \ \ \ \ \ \ \ \ \ \ \ \ \ \ \ - R_4(\mu[\pi_{33/br'},\pi_{33/br'},\pi_r,\pi_{br'}]) - R_3(\mu[\pi_{33/br'},\pi_{33/br'},\pi_{33/br'},\pi_r])\Big)
\end{aligned}
\end{equation*}

In both line-ups $(\pi_{33/br'},\pi_{33/br'},\pi_{33/br'},\pi_{br'})$, where the predators will go directly to the bottom right cell and the prey will go to the cell in the 3rd row and 3rd column, the agents in slot 4 ($\pi_{br'}$) and slot 3 ($\pi_{33/br'}$) will both obtain an expected average reward of $-1$. In line-up $(\pi_{33/br'},\pi_{33/br'},\pi_r,\pi_{br'})$, where $\pi_r$ will act randomly so the predators will almost always notice this random movement and then they will go to the cell in the 3rd row and 3rd column, the agent in slot 4 ($\pi_{br'}$) will almost obtain an expected average reward of $1$. In line-up $(\pi_{33/br'},\pi_{33/br'},\pi_{33/br'},\pi_r)$, where $\pi_r$ will act randomly so the predators will almost always notice this random movement and then they will go to the cell in the 3rd row and 3rd column, the agent in slot 3 ($\pi_{33/br'}$) will almost obtain an expected average reward of $1$. So:

\begin{equation*}
ACoop_{4,3}(\pi_{br'},\pi_{33/br'},\Pi_o,w_{\dot{L}},\mu) = \frac{1}{4} \left((-1) + (-1) - 1 - 1\right) = -1
\end{equation*}

And finally, we weight over the slots:

\begin{equation*}
\begin{aligned}
ACoop(\Pi_e,w_{\Pi_e},\Pi_o,w_{\dot{L}},\mu,w_S)	& = \eta_{S_3^2} \sum_{t \in \tau} \sum_{i,j \in t | i \neq j} w_S(i,\mu) w_S(j,\mu) ACoop_{i,j}(\Pi_e,w_{\Pi_e},\Pi_o,w_{\dot{L}},\mu) +\\
													& + \sum_{t_1,t_2,t_3 \in \tau | t_1 \neq t_2 \neq t_3} \sum_{i \in t_1} w_S(i,\mu) \sum_{j \in t_2} w_S(j,\mu) ACoop_{i,j}(\Pi_e,w_{\Pi_e},\Pi_o,w_{\dot{L}},\mu) =\\
													& = \frac{8}{3} \frac{1}{4} \frac{1}{4} \{ACoop_{2,3}(\Pi_e,w_{\Pi_e},\Pi_o,w_{\dot{L}},\mu) + ACoop_{2,4}(\Pi_e,w_{\Pi_e},\Pi_o,w_{\dot{L}},\mu) +\\
													& + ACoop_{3,2}(\Pi_e,w_{\Pi_e},\Pi_o,w_{\dot{L}},\mu) + ACoop_{3,4}(\Pi_e,w_{\Pi_e},\Pi_o,w_{\dot{L}},\mu) +\\
													& + ACoop_{4,2}(\Pi_e,w_{\Pi_e},\Pi_o,w_{\dot{L}},\mu) + ACoop_{4,3}(\Pi_e,w_{\Pi_e},\Pi_o,w_{\dot{L}},\mu)\} =\\
													& = \frac{8}{3} \frac{1}{4} \frac{1}{4} \{ACoop_{2,3}(\pi_{br'},\pi_{33/br'},\Pi_o,w_{\dot{L}},\mu) + ACoop_{2,4}(\pi_{br'},\pi_{33/br'},\Pi_o,w_{\dot{L}},\mu) +\\
													& + ACoop_{3,2}(\pi_{br'},\pi_{33/br'},\Pi_o,w_{\dot{L}},\mu) + ACoop_{3,4}(\pi_{br'},\pi_{33/br'},\Pi_o,w_{\dot{L}},\mu) +\\
													& + ACoop_{4,2}(\pi_{br'},\pi_{33/br'},\Pi_o,w_{\dot{L}},\mu) + ACoop_{4,3}(\pi_{br'},\pi_{33/br'},\Pi_o,w_{\dot{L}},\mu)\} =\\
													& = \frac{8}{3} \frac{1}{4} \frac{1}{4} \left\{6 \times (-1)\right\} = -1
\end{aligned}
\end{equation*}

Since $-1$ is the lowest possible value for the cooperative anticipation property, therefore predator-prey has $General_{min} = -1$ for this property.
\end{proof}
\end{proposition}

\begin{proposition}
\label{prop:predator-prey_ACoop_general_max}
$General_{max}$ for the cooperative anticipation (ACoop) property is equal to $1$ for the predator-prey environment.

\begin{proof}
To find $General_{max}$ (equation \ref{eq:general_max}), we need to find a trio $\left\langle\Pi_e,w_{\Pi_e},\Pi_o\right\rangle$ which maximises the property as much as possible. We can have this situation by selecting $\Pi_e = \{\pi_{win}\}$ with $w_{\Pi_e}(\pi_{win}) = 1$ and $\Pi_o = \{\pi_{win}\}$ (a $\pi_{win}$ agent always tries to not be chased when playing as the prey and tries to chase when playing as the predator).

Following definition \ref{def:ACoop}, we obtain the ACoop value for this $\left\langle\Pi_e,w_{\Pi_e},\Pi_o\right\rangle$. Since the environment is not symmetric, we need to calculate this property for every pair of different slots in the same team. Following definition \ref{def:ACoop_set}, we could calculate its ACoop value for each pair of slots but, since $\Pi_e$ has only one agent, its weight is equal to $1$ and $\Pi_o$ also has only one agent, it is equivalent to use directly definition \ref{def:ACoop_agents}. We start with slots 2 and 3:

\begin{equation*}
\begin{aligned}
& ACoop_{2,3}(\pi_{win},\pi_{win},\Pi_o,w_{\dot{L}},\mu) = \sum_{\dot{l} \in \dot{L}^{N(\mu)}_{-2,3}(\Pi_o)} w_{\dot{L}}(\dot{l}) \frac{1}{4} \times\\
& \times \left(R_2(\mu[\instantiation{l}{2,3}{\pi_{win},\pi_{win}}]) + R_3(\mu[\instantiation{l}{2,3}{\pi_{win},\pi_{win}}]) - R_2(\mu[\instantiation{l}{2,3}{\pi_{win},\pi_r}]) - R_3(\mu[\instantiation{l}{2,3}{\pi_r,\pi_{win}}])\right) =\\
& \ \ \ \ \ \ \ \ \ \ \ \ \ \ \ \ \ \ \ \ \ \ \ \ \ \ \ \ \ \ \ \ \ \ \ \ \ \ \ \ \ = \frac{1}{4} \Big(R_2(\mu[\pi_{win},\pi_{win},\pi_{win},\pi_{win}]) + R_3(\mu[\pi_{win},\pi_{win},\pi_{win},\pi_{win}]) -\\
& \ \ \ \ \ \ \ \ \ \ \ \ \ \ \ \ \ \ \ \ \ \ \ \ \ \ \ \ \ \ \ \ \ \ \ \ \ \ \ \ \ - R_2(\mu[\pi_{win},\pi_{win},\pi_r,\pi_{win}]) - R_3(\mu[\pi_{win},\pi_r,\pi_{win},\pi_{win}])\Big)
\end{aligned}
\end{equation*}

In both line-ups $(\pi_{win},\pi_{win},\pi_{win},\pi_{win})$, where the predators will coordinate to always chase the prey as seen in lemma \ref{lemma:predator-prey_always_chase}, the agents in slot 2 ($\pi_{win}$) and slot 3 ($\pi_{win}$) will both obtain an expected average reward of $1$. In line-up $(\pi_{win},\pi_{win},\pi_r,\pi_{win})$, where the random agent will almost never coordinate with the other predators so the prey will almost always survive, the agent in slot 2 ($\pi_{win}$) will almost obtain an expected average reward of $-1$. In line-up $(\pi_{win},\pi_r,\pi_{win},\pi_{win})$, where the random agent will almost never coordinate with the other predators so the prey will almost always survive, the agent in slot 3 ($\pi_{win}$) will almost obtain an expected average reward of $-1$. So:

\begin{equation*}
ACoop_{2,3}(\pi_{win},\pi_{win},\Pi_o,w_{\dot{L}},\mu) = \frac{1}{4} \left(1 + 1 - (-1) - (-1)\right) = 1
\end{equation*}

For slots 2 and 4:

\begin{equation*}
\begin{aligned}
& ACoop_{2,4}(\pi_{win},\pi_{win},\Pi_o,w_{\dot{L}},\mu) = \sum_{\dot{l} \in \dot{L}^{N(\mu)}_{-2,4}(\Pi_o)} w_{\dot{L}}(\dot{l}) \frac{1}{4} \times\\
& \times \left(R_2(\mu[\instantiation{l}{2,4}{\pi_{win},\pi_{win}}]) + R_4(\mu[\instantiation{l}{2,4}{\pi_{win},\pi_{win}}]) - R_2(\mu[\instantiation{l}{2,4}{\pi_{win},\pi_r}]) - R_4(\mu[\instantiation{l}{2,4}{\pi_r,\pi_{win}}])\right) =\\
& \ \ \ \ \ \ \ \ \ \ \ \ \ \ \ \ \ \ \ \ \ \ \ \ \ \ \ \ \ \ \ \ \ \ \ \ \ \ \ \ \ = \frac{1}{4} \Big(R_2(\mu[\pi_{win},\pi_{win},\pi_{win},\pi_{win}]) + R_4(\mu[\pi_{win},\pi_{win},\pi_{win},\pi_{win}]) -\\
& \ \ \ \ \ \ \ \ \ \ \ \ \ \ \ \ \ \ \ \ \ \ \ \ \ \ \ \ \ \ \ \ \ \ \ \ \ \ \ \ \ - R_2(\mu[\pi_{win},\pi_{win},\pi_{win},\pi_r]) - R_4(\mu[\pi_{win},\pi_r,\pi_{win},\pi_{win}])\Big)
\end{aligned}
\end{equation*}

In both line-ups $(\pi_{win},\pi_{win},\pi_{win},\pi_{win})$, where the predators will coordinate to always chase the prey as seen in lemma \ref{lemma:predator-prey_always_chase}, the agents in slot 2 ($\pi_{win}$) and slot 4 ($\pi_{win}$) will both obtain an expected average reward of $1$. In line-up $(\pi_{win},\pi_{win},\pi_{win},\pi_r)$, where the random agent will almost never coordinate with the other predators so the prey will almost always survive, the agent in slot 2 ($\pi_{win}$) will almost obtain an expected average reward of $-1$. In line-up $(\pi_{win},\pi_r,\pi_{win},\pi_{win})$, where the random agent will almost never coordinate with the other predators so the prey will almost always survive, the agent in slot 4 ($\pi_{win}$) will almost obtain an expected average reward of $-1$. So:

\begin{equation*}
ACoop_{2,4}(\pi_{win},\pi_{win},\Pi_o,w_{\dot{L}},\mu) = \frac{1}{4} \left(1 + 1 - (-1) - (-1)\right) = 1
\end{equation*}

For slots 3 and 2:

\begin{equation*}
\begin{aligned}
& ACoop_{3,2}(\pi_{win},\pi_{win},\Pi_o,w_{\dot{L}},\mu) = \sum_{\dot{l} \in \dot{L}^{N(\mu)}_{-3,2}(\Pi_o)} w_{\dot{L}}(\dot{l}) \frac{1}{4} \times\\
& \times \left(R_3(\mu[\instantiation{l}{3,2}{\pi_{win},\pi_{win}}]) + R_2(\mu[\instantiation{l}{3,2}{\pi_{win},\pi_{win}}]) - R_3(\mu[\instantiation{l}{3,2}{\pi_{win},\pi_r}]) - R_2(\mu[\instantiation{l}{3,2}{\pi_r,\pi_{win}}])\right) =\\
& \ \ \ \ \ \ \ \ \ \ \ \ \ \ \ \ \ \ \ \ \ \ \ \ \ \ \ \ \ \ \ \ \ \ \ \ \ \ \ \ \ = \frac{1}{4} \Big(R_3(\mu[\pi_{win},\pi_{win},\pi_{win},\pi_{win}]) + R_2(\mu[\pi_{win},\pi_{win},\pi_{win},\pi_{win}]) -\\
& \ \ \ \ \ \ \ \ \ \ \ \ \ \ \ \ \ \ \ \ \ \ \ \ \ \ \ \ \ \ \ \ \ \ \ \ \ \ \ \ \ - R_3(\mu[\pi_{win},\pi_r,\pi_{win},\pi_{win}]) - R_2(\mu[\pi_{win},\pi_{win},\pi_r,\pi_{win}])\Big)
\end{aligned}
\end{equation*}

In both line-ups $(\pi_{win},\pi_{win},\pi_{win},\pi_{win})$, where the predators will coordinate to always chase the prey as seen in lemma \ref{lemma:predator-prey_always_chase}, the agents in slot 3 ($\pi_{win}$) and slot 2 ($\pi_{win}$) will both obtain an expected average reward of $1$. In line-up $(\pi_{win},\pi_r,\pi_{win},\pi_{win})$, where the random agent will almost never coordinate with the other predators so the prey will almost always survive, the agent in slot 3 ($\pi_{win}$) will almost obtain an expected average reward of $-1$. In line-up $(\pi_{win},\pi_{win},\pi_r,\pi_{win})$, where the random agent will almost never coordinate with the other predators so the prey will almost always survive, the agent in slot 2 ($\pi_{win}$) will almost obtain an expected average reward of $-1$. So:

\begin{equation*}
ACoop_{3,2}(\pi_{win},\pi_{win},\Pi_o,w_{\dot{L}},\mu) = \frac{1}{4} \left(1 + 1 - (-1) - (-1)\right) = 1
\end{equation*}

For slots 3 and 4:

\begin{equation*}
\begin{aligned}
& ACoop_{3,4}(\pi_{win},\pi_{win},\Pi_o,w_{\dot{L}},\mu) = \sum_{\dot{l} \in \dot{L}^{N(\mu)}_{-3,4}(\Pi_o)} w_{\dot{L}}(\dot{l}) \frac{1}{4} \times\\
& \times \left(R_3(\mu[\instantiation{l}{3,4}{\pi_{win},\pi_{win}}]) + R_4(\mu[\instantiation{l}{3,4}{\pi_{win},\pi_{win}}]) - R_3(\mu[\instantiation{l}{3,4}{\pi_{win},\pi_r}]) - R_4(\mu[\instantiation{l}{3,4}{\pi_r,\pi_{win}}])\right) =\\
& \ \ \ \ \ \ \ \ \ \ \ \ \ \ \ \ \ \ \ \ \ \ \ \ \ \ \ \ \ \ \ \ \ \ \ \ \ \ \ \ \ = \frac{1}{4} \Big(R_3(\mu[\pi_{win},\pi_{win},\pi_{win},\pi_{win}]) + R_4(\mu[\pi_{win},\pi_{win},\pi_{win},\pi_{win}]) -\\
& \ \ \ \ \ \ \ \ \ \ \ \ \ \ \ \ \ \ \ \ \ \ \ \ \ \ \ \ \ \ \ \ \ \ \ \ \ \ \ \ \ - R_3(\mu[\pi_{win},\pi_{win},\pi_{win},\pi_r]) - R_4(\mu[\pi_{win},\pi_{win},\pi_r,\pi_{win}])\Big)
\end{aligned}
\end{equation*}

In both line-ups $(\pi_{win},\pi_{win},\pi_{win},\pi_{win})$, where the predators will coordinate to always chase the prey as seen in lemma \ref{lemma:predator-prey_always_chase}, the agents in slot 3 ($\pi_{win}$) and slot 4 ($\pi_{win}$) will both obtain an expected average reward of $1$. In line-up $(\pi_{win},\pi_{win},\pi_{win},\pi_r)$, where the random agent will almost never coordinate with the other predators so the prey will almost always survive, the agent in slot 3 ($\pi_{win}$) will almost obtain an expected average reward of $-1$. In line-up $(\pi_{win},\pi_{win},\pi_r,\pi_{win})$, where the random agent will almost never coordinate with the other predators so the prey will almost always survive, the agent in slot 4 ($\pi_{win}$) will almost obtain an expected average reward of $-1$. So:

\begin{equation*}
ACoop_{3,4}(\pi_{win},\pi_{win},\Pi_o,w_{\dot{L}},\mu) = \frac{1}{4} \left(1 + 1 - (-1) - (-1)\right) = 1
\end{equation*}

For slots 4 and 2:

\begin{equation*}
\begin{aligned}
& ACoop_{4,2}(\pi_{win},\pi_{win},\Pi_o,w_{\dot{L}},\mu) = \sum_{\dot{l} \in \dot{L}^{N(\mu)}_{-4,2}(\Pi_o)} w_{\dot{L}}(\dot{l}) \frac{1}{4} \times\\
& \times \left(R_4(\mu[\instantiation{l}{4,2}{\pi_{win},\pi_{win}}]) + R_2(\mu[\instantiation{l}{4,2}{\pi_{win},\pi_{win}}]) - R_4(\mu[\instantiation{l}{4,2}{\pi_{win},\pi_r}]) - R_2(\mu[\instantiation{l}{4,2}{\pi_r,\pi_{win}}])\right) =\\
& \ \ \ \ \ \ \ \ \ \ \ \ \ \ \ \ \ \ \ \ \ \ \ \ \ \ \ \ \ \ \ \ \  \ \ \ \ \ \ \ \ = \frac{1}{4} \Big(R_4(\mu[\pi_{win},\pi_{win},\pi_{win},\pi_{win}]) + R_2(\mu[\pi_{win},\pi_{win},\pi_{win},\pi_{win}]) -\\
& \ \ \ \ \ \ \ \ \ \ \ \ \ \ \ \ \ \ \ \ \ \ \ \ \ \ \ \ \ \ \ \ \  \ \ \ \ \ \ \ \ - R_4(\mu[\pi_{win},\pi_r,\pi_{win},\pi_{win}]) - R_2(\mu[\pi_{win},\pi_{win},\pi_{win},\pi_r])\Big)
\end{aligned}
\end{equation*}

In both line-ups $(\pi_{win},\pi_{win},\pi_{win},\pi_{win})$, where the predators will coordinate to always chase the prey as seen in lemma \ref{lemma:predator-prey_always_chase}, the agents in slot 4 ($\pi_{win}$) and slot 2 ($\pi_{win}$) will both obtain an expected average reward of $1$. In line-up $(\pi_{win},\pi_r,\pi_{win},\pi_{win})$, where the random agent will almost never coordinate with the other predators so the prey will almost always survive, the agent in slot 4 ($\pi_{win}$) will almost obtain an expected average reward of $-1$. In line-up $(\pi_{win},\pi_{win},\pi_{win},\pi_r)$, where the random agent will almost never coordinate with the other predators so the prey will almost always survive, the agent in slot 2 ($\pi_{win}$) will almost obtain an expected average reward of $-1$. So:

\begin{equation*}
ACoop_{4,2}(\pi_{win},\pi_{win},\Pi_o,w_{\dot{L}},\mu) = \frac{1}{4} \left(1 + 1 - (-1) - (-1)\right) = 1
\end{equation*}

And for slots 4 and 3:

\begin{equation*}
\begin{aligned}
& ACoop_{4,3}(\pi_{win},\pi_{win},\Pi_o,w_{\dot{L}},\mu) = \sum_{\dot{l} \in \dot{L}^{N(\mu)}_{-4,3}(\Pi_o)} w_{\dot{L}}(\dot{l}) \frac{1}{4} \times\\
& \times \left(R_4(\mu[\instantiation{l}{4,3}{\pi_{win},\pi_{win}}]) + R_3(\mu[\instantiation{l}{4,3}{\pi_{win},\pi_{win}}]) - R_4(\mu[\instantiation{l}{4,3}{\pi_{win},\pi_r}]) - R_3(\mu[\instantiation{l}{4,3}{\pi_r,\pi_{win}}])\right) =\\
& \ \ \ \ \ \ \ \ \ \ \ \ \ \ \ \ \ \ \ \ \ \ \ \ \ \ \ \ \ \ \ \ \ \ \ \ \ \ \ \ \ = \frac{1}{4} \Big(R_4(\mu[\pi_{win},\pi_{win},\pi_{win},\pi_{win}]) + R_3(\mu[\pi_{win},\pi_{win},\pi_{win},\pi_{win}]) -\\
& \ \ \ \ \ \ \ \ \ \ \ \ \ \ \ \ \ \ \ \ \ \ \ \ \ \ \ \ \ \ \ \ \ \ \ \ \ \ \ \ \ - R_4(\mu[\pi_{win},\pi_{win},\pi_r,\pi_{win}]) - R_3(\mu[\pi_{win},\pi_{win},\pi_{win},\pi_r])\Big)
\end{aligned}
\end{equation*}

In both line-ups $(\pi_{win},\pi_{win},\pi_{win},\pi_{win})$, where the predators will coordinate to always chase the prey as seen in lemma \ref{lemma:predator-prey_always_chase}, the agents in slot 4 ($\pi_{win}$) and slot 3 ($\pi_{win}$) will both obtain an expected average reward of $1$. In line-up $(\pi_{win},\pi_{win},\pi_{win},\pi_r)$, where the random agent will almost never coordinate with the other predators so the prey will almost always survive, the agent in slot 4 ($\pi_{win}$) will almost obtain an expected average reward of $-1$. In line-up $(\pi_{win},\pi_{win},\pi_r,\pi_{win})$, where the random agent will almost never coordinate with the other predators so the prey will almost always survive, the agent in slot 3 ($\pi_{win}$) will almost obtain an expected average reward of $-1$. So:

\begin{equation*}
ACoop_{4,3}(\pi_{win},\pi_{win},\Pi_o,w_{\dot{L}},\mu) = \frac{1}{4} \left(1 + 1 - (-1) - (-1)\right) = 1
\end{equation*}

And finally, we weight over the slots:

\begin{equation*}
\begin{aligned}
ACoop(\Pi_e,w_{\Pi_e},\Pi_o,w_{\dot{L}},\mu,w_S)	& = \eta_{S_3^2} \sum_{t \in \tau} \sum_{i,j \in t | i \neq j} w_S(i,\mu) w_S(j,\mu) ACoop_{i,j}(\Pi_e,w_{\Pi_e},\Pi_o,w_{\dot{L}},\mu) +\\
													& + \sum_{t_1,t_2,t_3 \in \tau | t_1 \neq t_2 \neq t_3} \sum_{i \in t_1} w_S(i,\mu) \sum_{j \in t_2} w_S(j,\mu) ACoop_{i,j}(\Pi_e,w_{\Pi_e},\Pi_o,w_{\dot{L}},\mu) =\\
													& = \frac{8}{3} \frac{1}{4} \frac{1}{4} \{ACoop_{2,3}(\Pi_e,w_{\Pi_e},\Pi_o,w_{\dot{L}},\mu) + ACoop_{2,4}(\Pi_e,w_{\Pi_e},\Pi_o,w_{\dot{L}},\mu) +\\
													& + ACoop_{3,2}(\Pi_e,w_{\Pi_e},\Pi_o,w_{\dot{L}},\mu) + ACoop_{3,4}(\Pi_e,w_{\Pi_e},\Pi_o,w_{\dot{L}},\mu) +\\
													& + ACoop_{4,2}(\Pi_e,w_{\Pi_e},\Pi_o,w_{\dot{L}},\mu) + ACoop_{4,3}(\Pi_e,w_{\Pi_e},\Pi_o,w_{\dot{L}},\mu)\} =\\
													& = \frac{8}{3} \frac{1}{4} \frac{1}{4} \{ACoop_{2,3}(\pi_{win},\pi_{win},\Pi_o,w_{\dot{L}},\mu) + ACoop_{2,4}(\pi_{win},\pi_{win},\Pi_o,w_{\dot{L}},\mu) +\\
													& + ACoop_{3,2}(\pi_{win},\pi_{win},\Pi_o,w_{\dot{L}},\mu) + ACoop_{3,4}(\pi_{win},\pi_{win},\Pi_o,w_{\dot{L}},\mu) +\\
													& + ACoop_{4,2}(\pi_{win},\pi_{win},\Pi_o,w_{\dot{L}},\mu) + ACoop_{4,3}(\pi_{win},\pi_{win},\Pi_o,w_{\dot{L}},\mu)\} =\\
													& = \frac{8}{3} \frac{1}{4} \frac{1}{4} \left\{6 \times 1\right\} = 1
\end{aligned}
\end{equation*}

Since $1$ is the highest possible value for the cooperative anticipation property, therefore predator-prey has $General_{max} = 1$ for this property.
\end{proof}
\end{proposition}

\end{document}